\documentclass[12pt]{article}
\title{Analytical Methods in Physics}
\author{Yi-Zen Chu}
\date{}

\usepackage{graphicx}
\usepackage{multicol}
\usepackage{amsmath}
\usepackage{slashed}
\usepackage{young}
\usepackage{amsfonts}
\usepackage{amssymb}
\usepackage[usenames,dvipsnames]{color}
\usepackage{amsthm}
\usepackage{hyperref}

\theoremstyle{definition}

\numberwithin{equation}{subsection}
\newtheorem{myP}{Problem}[section]

\newcommand{\mpl}{M_{\rm pl}}
\newcommand{\GN}{G_{\rm N}}
\newcommand{\gb}{\bar{g}}
\newcommand{\Db}{\overline{\nabla}}

\newcommand{\Tr}[1]{\text{Tr}\left[ #1 \right]}

\newcommand{\dd}{\text{d}}

\newcommand{\epud}[2]{\varepsilon^{#1}_{\phantom{#1}#2}}
\newcommand{\epdu}[2]{\varepsilon_{#1}^{\phantom{#1}#2}}

\newcommand{\dbeinud}[2]{\varepsilon^{\widehat{#1}}_{\phantom{\widehat{#1}}#2}}
\newcommand{\dbeindu}[2]{\varepsilon_{\widehat{#1}}^{\phantom{\widehat{#1}}#2}}
\newcommand{\dbeinuu}[2]{\varepsilon^{\widehat{#1}#2}}
\newcommand{\dbeindd}[2]{\varepsilon_{\widehat{#1}#2}}

\newcommand{\ket}[1]{\left\vert #1 \right\rangle}
\newcommand{\bra}[1]{\left\langle #1 \right\vert}
\newcommand{\braket}[2]{\left.\left\langle #1 \right\vert #2 \right\rangle}
\newcommand{\ketbra}[2]{\left\vert #1 \right\rangle \left\langle #2 \right\vert}
\newcommand{\braOket}[3]{\left\langle #1 \left\vert #2 \right\vert #3 \right\rangle}
\newcommand{\ketk}{\vert \vec{k} \rangle}

\newcommand{\sbar}{\bar{\sigma}}
\newcommand{\LTud}[2]{\Lambda^{#1}_{\phantom{#1}#2}}

\newcommand{\Christ}[3]{\Gamma^{#1}_{\phantom{#1}#2#3}}

\begin{document}

\maketitle

\tableofcontents

\newpage

\section{Preface}

This work constitutes the free textbook project I initiated towards the end of Summer 2015, while preparing for the Fall 2015 {\it Analytical Methods in Physics} course I taught to upper level (mostly 2nd and 3rd year) undergraduates here at the University of Minnesota Duluth. During Fall 2017, I taught the graduate-level {\it Differential Geometry and Physics in Curved Spacetimes} here at National Central University, Taiwan; this has allowed me to further expand the text.

I assumed that the reader has taken the first three semesters of calculus, i.e., up to multi-variable calculus, as well as a first course in Linear Algebra and ordinary differential equations. (These are typical prerequisites for the Physics major within the US college curriculum.) My primary goal was to impart a good working knowledge of the mathematical tools that underlie fundamental physics -- quantum mechanics and electromagnetism, in particular. This meant that Linear Algebra in its abstract formulation had to take a central role in these notes.\footnote{That the textbook originally assigned for this course relegated the axioms of Linear Algebra towards the very end of the discussion was one major reason why I decided to write these notes. This same book also cost nearly two hundred (US) dollars -- a fine example of exorbitant textbook prices these days -- so I am glad I saved my students quite a bit of their educational expenses that semester.} To this end, I first reviewed complex numbers and matrix algebra. The middle chapters cover calculus beyond the first three semesters: complex analysis and special/approximation/asymptotic methods. The latter, I feel, is not taught widely enough in the undergraduate setting. The final chapter is meant to give a solid introduction to the topic of linear partial differential equations (PDEs), which is crucial to the study of electromagnetism, linearized gravitation and quantum mechanics/field theory. But before tackling PDEs, I feel that having a good grounding in the basic elements of differential geometry not only helps streamlines one's fluency in multi-variable calculus; it also provides a stepping stone to the discussion of curved spacetime wave equations. 

Some of the other distinctive features of this free textbook project are as follows.

Index notation and Einstein summation convention is widely used throughout the physics literature, so I have not shied away from introducing it early on, starting in \S \eqref{Chapter_MatrixAlgebra} on matrix algebra. In a similar spirit, I have phrased the abstract formulation of Linear Algebra in \S \eqref{Chapter_LinearAlgebra} entirely in terms of P.A.M. Dirac's bra-ket notation. When discussing inner products, I do make a brief comparison of Dirac's notation against the one commonly found in math textbooks.

I made no pretense at making the material mathematically rigorous, but I strived to make the flow coherent, so that the reader comes away with a firm conceptual grasp of the overall structure of each major topic. For instance, while the full fledged study of continuous (as opposed to discrete) vector spaces can take up a whole math class of its own, I feel the physicist should be exposed to it right after learning the discrete case. For, the basics are not only accessible, the Fourier transform is in fact a physically important application of the continuous space spanned by the position eigenkets $\{ \vert \vec{x} \rangle \}$. One key difference between Hermitian operators in discrete versus continuous vector spaces is the need to impose appropriate boundary conditions in the latter; this is highlighted in the Linear Algebra chapter as a prelude to the PDE chapter \S \eqref{Chapter_LinearPDE}, where the Laplacian and its spectrum plays a significant role. Additionally, while the Linear Algebra chapter was heavily inspired by the first chapter of Sakurai's {\it Modern Quantum Mechanics}, I have taken effort to emphasize that quantum mechanics is merely a very important application of the framework; for e.g., even the famous commutation relation $[X^i, P_j] = i \delta^i_j$ is not necessarily a quantum mechanical statement. This emphasis is based on the belief that the power of a given mathematical tool is very much tied to its versatility -- this issue arises again in the JWKB discussion within \S \eqref{Chapter_SpecialTechniquesAndAsymptotics}, where I highlight it is not merely some ``semi-classical" limit of quantum mechanical problems, but really a general technique for solving differential equations.

Much of \S \eqref{Chapter_ComplexCalculus} is a standard introduction to calculus on the complex plane and the theory of complex analytic functions. However, the Fourier transform application section gave me the chance to introduce the concept of the Green's function; specifically, that of the ordinary differential equation describing the damped harmonic oscillator. This (retarded) Green's function can be computed via the theory of residues -- and through its key role in the initial value formulation of the ODE solution, allows the two linearly independent solutions to the associated homogeneous equation to be obtained for any value of the damping parameter.

Differential geometry may appear to be an advanced topic to many, but it really is not. From a practical standpoint, it cannot be overemphasized that most vector calculus operations can be readily carried out and the curved space(time) Laplacian/wave operator computed once the relevant metric is specified explicitly. I wrote much of \S \eqref{Chapter_DifferentialGeometry_CurvedSpaces} in this ``practical physicist" spirit. Although it deals primarily with curved spaces, teaching {\it Physics in Curved Spacetimes} during Fall 2017 at National Central University, Taiwan, gave me the opportunity to add its curved spacetime sequel, \S \eqref{Chapter_DifferentialGeometry_CurvedSpacetimes}, where I elaborated upon geometric concepts -- the emergence of the Riemann tensor from parallel transporting a vector around an infinitesimal parallelogram, for instance -- deliberately glossed over in \S \eqref{Chapter_DifferentialGeometry_CurvedSpaces}. It is my hope that \S \eqref{Chapter_DifferentialGeometry_CurvedSpaces} and \S \eqref{Chapter_DifferentialGeometry_CurvedSpacetimes} can be used to build the differential geometric tools one could then employ to understand General Relativity, Einstein's field equations for gravitation.

In \S \eqref{Chapter_LinearPDE} on PDEs, I begin with the Poisson equation in curved space, followed by the enumeration of the eigensystem of the Laplacian in different flat spaces. By imposing Dirichlet or periodic boundary conditions for the most part, I view the development there as the culmination of the Linear Algebra of continuous spaces. The spectrum of the Laplacian also finds important applications in the solution of the heat and wave equations. I have deliberately discussed the heat instead of the Schr\"{o}dinger equation because the two are similar enough, I hope when the reader learns about the latter in her/his quantum mechanics course, it will only serve to enrich her/his understanding when she/he compares it with the discourse here. Finally, the wave equation in Minkowski spacetime -- {\it the} basis of electromagnetism and linearized gravitation -- is discussed from both the position/real and Fourier/reciprocal space perspectives. The retarded Green's function plays a central role here, and I spend significant effort exploring different means of computing it. The tail effect is also highlighted there: classical waves associated with massless particles transmit physical information {\it within} the null cone in $(1+1)$D and all odd dimensions. Wave solutions are examined from different perspectives: in real/position space; in frequency space; in the non-relativistic/static limits; and with the multipole-expansion employed to extract leading order features. The final section contains a brief introduction to the variational principle for the classical field theories of the Poisson and wave equations.

Finally, I have interspersed problems throughout each chapter because this is how I personally like to engage with new material -- read and ``doodle" along the way, to make sure I am properly following the details. My hope is that these notes are concise but accessible enough that anyone can work through both the main text as well as the problems along the way; and discover they have indeed acquired a new set of mathematical tools to tackle physical problems.

One glaring omission is the subject of Group Theory. It can easily take up a whole course of its own, but I have tried to sprinkle problems and a discussion or two throughout these notes that allude to it. By making this material available online, I view it as an ongoing project: I plan to update and add new material whenever time permits, so Group Theory, as well as illustrations/figures accompanying the main text, may show up at some point down the road. The most updated version can be found at the following URL:
\begin{quotation}
\url{http://www.stargazing.net/yizen/AnalyticalMethods_YZChu.pdf}
\end{quotation}
I would very much welcome suggestions, questions, comments, error reports, etc.; please feel free to contact me at yizen [dot] chu @ gmail [dot] com.

\begin{flushright}
	-- Yi-Zen Chu
\end{flushright}

\newpage

\section{Complex Numbers and Functions}

\footnote{Some of the material in this section is based on James Nearing's \href{http://www.physics.miami.edu/~nearing/mathmethods/}{{\it Mathematical Tools for Physics}}.}The motivational introduction to complex numbers, in particular the number $i$,\footnote{Engineers use $j$.} is the solution to the equation
\begin{align}
\label{Defi}
i^2 = -1.
\end{align}
That is, ``what's the square root of $-1$?" For us, we will simply take eq. \eqref{Defi} as the {\it defining} equation for the algebra obeyed by $i$. A general complex number $z$ can then be expressed as 
\begin{align}
z = x+iy
\end{align}
where $x$ and $y$ are real numbers. The $x$ is called the {\it real part} ($\equiv$ Re$(z)$) and $y$ the {\it imaginary part} of $z$ ($\equiv$ Im$(z)$). 

{\bf Geometrically speaking} $z$ is a vector $(x,y)$ on the 2-dimensional plane spanned by the real axis (the $x$ part of $z$) and the imaginary axis (the $iy$ part of $z$). Moreover, you may recall from (perhaps) multi-variable calculus, that if $r$ is the distance between the origin and the point $(x,y)$ and $\phi$ is the angle between the vector joining $(0,0)$ to $(x,y)$ and the positive horizontal axis -- then
\begin{align}
(x,y) = (r \cos\phi,r \sin\phi) .
\end{align}
Therefore a complex number must be expressible as
\begin{align}
z = x+iy = r(\cos\phi + i \sin\phi) .
\end{align}
This actually takes a compact form using the exponential:
\begin{align}
z = x+iy = r(\cos\phi + i \sin\phi) = r e^{i\phi}, \qquad r \geq 0, \ 0 \leq \phi < 2\pi.
\end{align}
Some words on notation. The distance $r$ between $(0,0)$ and $(x,y)$ in the complex number context is written as an absolute value, i.e.,
\begin{align}
|z| = |x+iy| = r = \sqrt{x^2+y^2} ,
\end{align}
where the final equality follows from Pythagoras' Theorem. The angle $\phi$ is denoted as
\begin{align}
\text{arg}(z) = \text{arg}(r e^{i\phi}) = \phi .
\end{align}
The symbol $\mathbb{C}$ is often used to represent the 2D space of complex numbers.
\begin{align}
z = |z| e^{i \text{arg}(z)} \in \mathbb{C} .
\end{align}
\begin{myP}
{\it Euler's formula.} \qquad Assuming $\exp z$ can be defined through its Taylor series for any complex $z$, prove by Taylor expansion and eq. \eqref{Defi} that
\begin{align}
\label{EulerFormula}
e^{i\phi} = \cos(\phi) + i \sin(\phi), \qquad \phi \in \mathbb{R} .
\end{align}
\end{myP}
{\bf Arithmetic} \qquad Addition and subtraction of complex numbers take place component-by-component, just like adding/subtracting 2D real vectors; for example, if
\begin{align}
z_1 = x_1 + i y_1 \qquad \text{ and } \qquad z_2 = x_2 + i y_2 ,
\end{align}
then
\begin{align}
\label{Complex_Addition}
z_1 \pm z_2 = (x_1 \pm x_2) + i (y_1 \pm y_2) .
\end{align}
Multiplication is more easily done in polar coordinates: if $z_1 = r_1 e^{i\phi_1}$ and $z_2 = r_2 e^{i\phi_2}$, their product amounts to adding their phases and multiplying their radii, namely
\begin{align}
\label{Complex_Multiplication}
z_1 z_2 = r_1 r_2 e^{i(\phi_1 + \phi_2)} .
\end{align}
To summarize: \qquad \begin{quotation}
	Complex numbers $\{ z = x+iy = r e^{i\phi} \vert x,y \in \mathbb{R}; r \geq 0, \phi \in \mathbb{R} \}$ are 2D real vectors as far as addition/subtraction goes -- Cartesian coordinates are useful here (cf. \eqref{Complex_Addition}). It is their multiplication that the additional ingredient/algebra $i^2 \equiv -1$ comes into play. In particular, using polar coordinates to multiply two complex numbers (cf. \eqref{Complex_Multiplication}) allows us to see the result is a combination of a re-scaling of their radii plus a rotation.
\end{quotation}
\begin{myP}
\qquad If $z=x+iy$ what is $z^2$ in terms of $x$ and $y$? \qed
\end{myP}
\begin{myP}
\qquad Explain why multiplying a complex number $z=x+iy$ by $i$ amounts to rotating the vector $(x,y)$ on the complex plane counter-clockwise by $\pi/2$. Hint: first write $i$ in polar coordinates. \qed
\end{myP}
\begin{myP}
\qquad Describe the points on the complex $z$-plane satisfying $|z - z_0| < R$, where $z_0$ is some fixed complex number and $R > 0$ is a real number.
\end{myP}
\begin{myP}
\qquad Use the polar form of the complex number to proof that multiplication of complex numbers is associative, i.e., $z_1 z_2 z_3 = z_1 (z_2 z_3) = (z_1 z_2) z_3$. \qed
\end{myP}
{\bf Complex conjugation} \qquad Taking the complex conjugate of $z=x+iy$ means we flip the sign of its imaginary part, i.e.,
\begin{align}
z^* = x-iy ;
\end{align}
it is also denoted as $\bar{z}$. In polar coordinates, if $z=re^{i\phi}=r(\cos\phi + i \sin\phi)$ then $z^*=re^{-i\phi}$ because
\begin{align}
e^{-i\phi} = \cos(-\phi) + i \sin(-\phi) = \cos\phi - i\sin\phi .
\end{align} 
The $\sin\phi \to -\sin\phi$ is what brings us from $x+iy$ to $x-iy$. Now
\begin{align}
z^* z = z z^* = (x+iy)(x-iy) = x^2 + y^2 = |z|^2 .
\end{align}
When we take the ratio of complex numbers, it is possible to ensure that the imaginary number $i$ appears only in the numerator, by multiplying the numerator and denominator by the complex conjugate of the denominator. For $x$, $y$, $a$ and $b$ all real,
\begin{align}
\frac{x + i y}{a + i b} = \frac{(a-ib)(x+iy)}{a^2+b^2} = \frac{(ax+by) + i (ay-bx)}{a^2+b^2} .
\end{align}
\begin{myP}
\qquad Is $(z_1 z_2)^* = z_1^* z_2^*$, i.e., is the complex conjugate of the product of 2 complex numbers equal to the product of their complex conjugates? What about $(z_1/z_2)^* = z_1^*/z_2^*$? Is $|z_1 z_2| = |z_1||z_2|$? What about $|z_1/z_2| = |z_1|/|z_2|$? Also show that arg$(z_1 \cdot z_2) = \text{arg}(z_1) + \text{arg}(z_2)$. Strictly speaking, $\arg(z)$ is well defined only up to an additive multiple of $2\pi$. Can you explain why? Hint: polar coordinates are very useful in this problem. \qed
\end{myP}
\begin{myP}
\qquad Show that $z$ is real if and only if $z=z^*$. Show that $z$ is purely imaginary if and only if $z=-z^*$. Show that $z + z^* = 2 \text{Re}(z)$ and $z - z^* = 2 i \text{Im}(z)$. Hint: use Cartesian coordinates. \qed
\end{myP}
\begin{myP}
\qquad Prove that the roots of a polynomial with real coefficients 
\begin{align}
P_N(z) \equiv c_0 + c_1 z + c_2 z^2 + \dots + c_N z^N, \qquad\qquad \{c_i \in \mathbb{R}\},
\end{align}
come in complex conjugate pairs; i.e., if $z$ is a root then so is $z^*$. \qed
\end{myP}
{\bf Trigonometric, hyperbolic and exponential functions} \qquad Complex numbers allow us to connect trigonometric, hyperbolic and exponential (exp) functions. Start from
\begin{align}
e^{\pm i\phi} = \cos\phi \pm i\sin\phi .
\end{align}
These two equations can be added and subtracted to yield
\begin{align}
\label{TrigFromExp}
\cos(z) = \frac{e^{iz} + e^{-iz}}{2}, \qquad
\sin(z) = \frac{e^{iz} - e^{-iz}}{2i}, \qquad \tan(z) = \frac{\sin(z)}{\cos(z)} .
\end{align}
We have made the replacement $\phi \to z$. This change is cosmetic if $0 \leq z < 2\pi$, but we can in fact now use eq. \eqref{TrigFromExp} to {\it define} the trigonometric functions in terms of the exp function for any complex $z$.

Trigonometric identities can be readily obtained from their exponential definitions. For example, the addition formulas would now begin from
\begin{align}
e^{i(\theta_1 + \theta_2)} = e^{i\theta_1} e^{i\theta_2} .
\end{align}
Applying Euler's formula (eq. \eqref{EulerFormula}) on both sides,
\begin{align}
\cos(\theta_1+\theta_2) + i \sin(\theta_1+\theta_2)
&= (\cos\theta_1 + i \sin\theta_1)(\cos\theta_2 + i \sin\theta_2) \\
&= (\cos\theta_1 \cos\theta_2 - \sin\theta_1 \sin\theta_2) + i (\sin\theta_1 \cos\theta_2 + \sin\theta_2 \cos\theta_1) . \nonumber
\end{align}
If we suppose $\theta_{1,2}$ are real angles, equating the real and imaginary parts of the left-hand-side and the last line tell us
\begin{align}
\cos(\theta_1+\theta_2) &= \cos\theta_1 \cos\theta_2 - \sin\theta_1 \sin\theta_2, \\
\sin(\theta_1+\theta_2) &= \sin\theta_1 \cos\theta_2 + \sin\theta_2 \cos\theta_1 .
\end{align}
\begin{myP}
	\qquad You are probably familiar with the hyperbolic functions, now defined as
\begin{align}
\label{Complex_HyperbolicFunctions}
\cosh(z) = \frac{e^z + e^{-z}}{2}, \qquad \sinh(z) = \frac{e^z - e^{-z}}{2}, \qquad \tanh(z) = \frac{\sinh(z)}{\cosh(z)} ,
\end{align}
for any complex $z$. Show that
\begin{align}
\cosh(iz) = \cos(z), \qquad \sinh(iz) = i \sin(z),
\cos(iz) = \cosh(z), \qquad \sin(iz) = i \sinh(z) .
\end{align}\qed
\end{myP}
\begin{myP}
	\qquad Calculate, for real $\theta$ and positive integer $N$:
\begin{align}
\cos(\theta) + \cos(2\theta) + \cos(3\theta) + \dots + \cos(N\theta) &= ? \\
\sin(\theta) + \sin(2\theta) + \sin(3\theta) + \dots + \sin(N\theta) &= ?
\end{align}
Hint: consider the geometric series $e^{i\theta} + e^{2i\theta} + \dots + e^{N i\theta}$. \qed
\end{myP}
\begin{myP}
\qquad Starting from $(e^{i\theta})^n$, for arbitrary integer $n$, re-write $\cos(n\theta)$ and $\sin(n\theta)$ as a sum involving products/powers of $\sin\theta$ and $\cos\theta$. Hint: if the arbitrary $n$ case is confusing at first, start with $n=1,2,3$ first. \qed
\end{myP}
{\bf Roots of unity} \qquad In polar coordinates, circling the origin $n$ times bring us back to the same point,
\begin{align}
\label{Polar2PiN}
z = r e^{i\theta + i 2\pi n},  \qquad n = 0,\pm 1,\pm 2,\pm 3,\dots.
\end{align}
This observation is useful for the following problem: what is $m$th root of 1, when $m$ is a positive integer? Of course, 1 is an answer, but so are
\begin{align}
1^{1/m} = e^{i 2\pi n/m}, \qquad n = 0, 1, \dots, m-1.
\end{align}
The terms repeat themselves for $n \geq m$; the negative integers $n$ do not give new solutions for $m$ integer. If we replace $1/m$ with $a/b$ where $a$ and $b$ are integers that do not share any common factors, then
\begin{align}
1^{a/b} = e^{i 2\pi n(a/b)} \qquad \text{ for } \qquad n = 0, 1, \dots, b-1 ,
\end{align}
since when $n=b$ we will get back $1$. If we replaced $(a/b)$ with say $1/\pi$, 
\begin{align}
1^{1/\pi} = e^{i2 \pi n/\pi} = e^{i 2 n},
\end{align}
then there will be infinite number of solutions, because $1/\pi$ cannot be expressed as a ratio of integers -- there is no way to get $2n = 2\pi n'$, for $n'$ integer.

In general, when you are finding the $m$th root of a complex number $z$, you are actually solving for $w$ in the polynomial equation $w^m = z$. The fundamental theorem of algebra tells us, if $m$ is a positive integer, you are guaranteed $m$ solutions -- although not all of them may be distinct.

{\it Square root of $-1$} \qquad What is $\sqrt{-1}$? Since $-1 = e^{i(\pi + 2\pi n)}$ for any integer $n$,
\begin{align}
(e^{i(\pi + 2\pi n)})^{1/2} = e^{i\pi/2 + i \pi n} = \pm i. \qquad n = 0, 1.
\end{align}
\begin{myP}
	\qquad Find all the solutions to $\sqrt{1-i}$. \qed
\end{myP}
{\bf Logarithm and powers} \qquad As we have just seen, whenever we take the root of some complex number $z$, we really have a multi-valued function. The inverse of the exponential is another such function. For $w = x+iy$, where $x$ and $y$ are real, we may consider
\begin{align}
e^w = e^x e^{i(y + 2\pi n)}, \qquad n = 0,\pm 1,\pm 2,\pm 3,\dots .
\end{align}
We define $\ln$ to be such that
\begin{align}
\ln e^w = x + i (y + 2 \pi n) .
\end{align}
Another way of saying this is, for a general complex $z$,
\begin{align}
\ln(z) = \ln|z| + i (\text{arg}(z) + 2 \pi n) .
\end{align}
One way to make sense of how to raise a complex number $z = r e^{i\theta}$ to the power of another complex number $w = x+iy$, namely $z^w$, is through the $\ln$:
\begin{align}
z^w = e^{w \ln z} = e^{(x+iy) (\ln(r) + i (\theta+2\pi n))} = e^{x \ln r - y(\theta+2\pi n)} e^{i (y \ln(r) + x (\theta+2\pi n))} .
\end{align}
This is, of course, a multi-valued function. We will have more to say about such multi-valued functions when discussing their calculus in \S \eqref{Chapter_ComplexCalculus}.
\begin{myP}
\qquad Find the inverse hyperbolic functions of eq. \eqref{Complex_HyperbolicFunctions} in terms of $\ln$. Does $\sin(z)=0$, $\cos(z)=0$ and $\tan(z)=0$ have any complex solutions? Hint: for the first question, write $e^z = w$ and $e^{-z} = 1/w$. Then solve for $w$. A similar strategy may be employed for the second question. \qed
\end{myP}
\begin{myP}
\qquad Let $\vec{\xi}$ and $\vec{\xi}'$ be vectors in a 2D Euclidean space, i.e., you may assume their Cartesian components are
\begin{align}
\vec{\xi} = (x,y) = r(\cos\phi,\sin\phi) , \qquad \qquad
\vec{\xi}' = (x',y') = r'(\cos\phi',\sin\phi') .
\end{align}
Use complex numbers, and assume that the following complex Taylor expansion of $\ln$ holds
\begin{align}
\ln(1-z) = -\sum_{\ell=1}^\infty \frac{z^\ell}{\ell} , \qquad |z| < 1 ,
\end{align}
to show that
\begin{align}
\label{Laplacian_2DGreensFunction}
\ln|\vec{\xi} - \vec{\xi}'| = \ln r_> - \sum_{\ell=1}^\infty \frac{1}{\ell} \left(\frac{r_<}{r_>}\right)^\ell \cos\Big(\ell(\phi-\phi') \Big) ,
\end{align}
where $r_>$ is the larger and $r_<$ is the smaller of the $(r,r')$, and $|\vec{\xi} - \vec{\xi}'|$ is the distance between the vectors $\vec{\xi}$ and $\vec{\xi}'$ -- not the absolute value of some complex number. Here, $\ln|\vec{\xi} - \vec{\xi}'|$ is proportional to the electric or gravitational potential generated by a point charge/mass in 2-dimensional flat space. Hint: first let $z = r e^{i\phi}$ and $z' = r' e^{i\phi'}$; then consider $\ln(z-z')$ -- how do you extract $\ln|\vec{\xi} - \vec{\xi}'|$ from it? \qed
\end{myP}

\newpage

\section{Matrix Algebra: A Review}
\label{Chapter_MatrixAlgebra}
\footnote{Much of the material here in this section were based on Chapter 1 of Cahill's {\it Physical Mathematics}.}In this section I will review some basic properties of matrices and matrix algebra, oftentimes using index notation. We will assume all matrices have complex entries unless otherwise stated. This is intended to be warmup to the next section, where I will treat Linear Algebra from a more abstract point of view.

\subsection{Basics, Matrix Operations, and Special types of matrices}

{\bf Index notation, Einstein summation, Basic Matrix Operations} \qquad Consider two matrices $M$ and $N$. The $ij$ component -- the $i$th row and $j$th column of $M$ and that of $N$ can be written as
\begin{align}
M^i_{\phantom{i}j} \qquad \text{ and } \qquad N^i_{\phantom{i}j} .
\end{align}
As an example, if $M$ is a $2 \times 2$ matrix, we have
\begin{align}
M = \left[
\begin{array}{cc}
M^{1}_{\phantom{1}1}	&	M^{1}_{\phantom{1}2} \\
M^{2}_{\phantom{2}1}	&	M^{2}_{\phantom{2}2}
\end{array}\right] .
\end{align}
I prefer to write one index up and one down, because as we shall see in the abstract formulation of linear algebra below, the row and column indices may transform differently. However, it is common to see the notation $M_{ij}$ and $M^{ij}$, etc., too. 

A vector $v$ can be written as
\begin{align}
v^i = (v^1,v^2,\dots,v^{D-1},v^D) .
\end{align}
Here, $v^5$ does not mean the fifth power of some quantity $v$, but rather the $5$th component of the vector $v$.

The matrix multiplication $M \cdot N$ can be written as
\begin{align}
\label{MatrixMultiplication}
(M \cdot N)^i_{\phantom{i}j} = \sum_{k=1}^D M^i_{\phantom{i}k} N^k_{\phantom{k}j} \equiv M^i_{\phantom{i}k} N^k_{\phantom{k}j} .
\end{align}
In words: the $ij$ component of the product $MN$, for a fixed $i$ and fixed $j$, means we are taking the $i$th row of $M$ and ``dotting" it into the $j$th column of $N$. In the second equality we have employed Einstein's summation convention, which we will continue to do so in these notes: repeated indices are summed over their relevant range -- in this case, $k \in \{1,2,\dots,D\}$. For example, if
\begin{align}
M = \left[
\begin{array}{cc}
a	&	b \\
c	&	d 
\end{array}\right], \qquad 
N = \left[
\begin{array}{cc}
1	&	2 \\
3	&	4 
\end{array}\right] ,
\end{align}
then
\begin{align}
M \cdot N &= \left[
\begin{array}{cc}
a + 3 b	& 2a + 4b \\
c + 3 d	& 2c + 4d 
\end{array}\right] .
\end{align}
{\it Note:} $M^i_{\phantom{i}k} N^k_{\phantom{k}j}$ works for multiplication of non-square matrices $M$ and $N$ too, as long as the number of columns of $M$ is equal to the number of rows of $N$, so that the sum involving $k$ makes sense.

Addition of $M$ and $N$; and multiplication of $M$ by a complex number $\lambda$ goes respectively as
\begin{align}
(M+N)^i_{\phantom{i}j} = M^i_{\phantom{i}j} + N^i_{\phantom{i}j} 
\end{align}
and
\begin{align}
(\lambda M)^i_{\phantom{i}j} = \lambda M^i_{\phantom{i}j} .
\end{align}
{\bf Associativity} \qquad The associativity of matrix multiplication means $(AB)C = A(BC) = ABC$. This can be seen using index notation
\begin{align}
A^i_{\phantom{i}k} B^k_{\phantom{k}l} C^l_{\phantom{l}j}
= (AB)^i_{\phantom{i}l} C^l_{\phantom{l}j} = A^i_{\phantom{i}k} (BC)^k_{\phantom{k}j}
= (ABC)^i_{\phantom{i}j} .
\end{align}
{\bf Tr} \qquad Tr$(A) \equiv A^i_{\phantom{i}i}$ denotes the trace of a square matrix $A$. The index notation makes it clear the trace of $AB$ is that of $BA$ because
\begin{align}
\Tr{A \cdot B} = A^l_{\phantom{i}k} B^k_{\phantom{k}l} = B^k_{\phantom{k}l} A^l_{\phantom{i}k} = \Tr{B \cdot A} .
\end{align}
This immediately implies the Tr is cyclic, in the sense that
\begin{align}
\Tr{X_1 \cdot X_2 \cdots X_N} = \Tr{X_N \cdot X_1 \cdot X_2 \cdots X_{N-1}} = \Tr{X_2 \cdot X_3 \cdots X_N \cdot X_1} .
\end{align}
\begin{myP}
\qquad Prove the linearity of the Tr, namely for $D \times D$ matrices $X$ and $Y$ and complex number $\lambda$,
\begin{align}
\Tr{X+Y} = \Tr{X} + \Tr{Y}, \qquad \Tr{\lambda X} = \lambda \Tr{X} .
\end{align} 
Comment on whether it makes sense to define Tr$(A) \equiv A^i_{\phantom{i}i}$, if $A$ is not a square matrix. \qed
\end{myP}
{\bf Identity and the Kronecker delta} \qquad The $D \times D$ identity matrix $\mathbb{I}$ has $1$ on each and every component on its diagonal and $0$ everywhere else. This is also the Kronecker delta.
\begin{align}
\mathbb{I}^i_{\phantom{i}j} = \delta^i_{\phantom{i}j} 
&= 1, \qquad i=j \nonumber\\
&= 0, \qquad i \neq j
\end{align}
The Kronecker delta is also the flat Euclidean metric in $D$ spatial dimensions; in that context we would write it with both lower indices $\delta_{ij}$ and its inverse is $\delta^{ij}$.

The Kronecker delta is also useful for representing {\it diagonal} matrices. These are matrices that have non-zero entries strictly on their diagonal, where row equals to column number. For example $A^i_{\phantom{i}j} = a_i \delta^i_j = a_j \delta^i_j$ is the diagonal matrix with $a_1, a_2, \dots, a_D$ filling its diagonal components, from the upper left to the lower right. Diagonal matrices are also often denoted, for instance, as 
\begin{align}
A = \text{diag}[a_1,\dots,a_D] .
\end{align}
Suppose we multiply $A B$, where $B$ is also diagonal ($B^i_{\phantom{i}j} = b_i \delta^i_j = b_j \delta^i_j$),
\begin{align}
(AB)^i_{\phantom{i}j} = \sum_l  a_i \delta^i_l b_j \delta^l_j .
\end{align}
If $i \neq j$ there will be no $l$ that is simultaneously equal to $i$ and $j$; therefore either one or both the Kronecker deltas are zero and the entire sum is zero. If $i = j$ then when (and only when) $l=i=j$, the Kronecker deltas are both one, and
\begin{align}
(AB)^i_{\phantom{i}j} = a_i b_j .
\end{align}
This means we have shown, using index notation, that the product of diagonal matrices yields another diagonal matrix.
\begin{align}
(AB)^i_{\phantom{i}j} = a_i b_j \delta^i_j \qquad \text{(No sum over $i,j$)}.
\end{align}
{\bf Transpose} \qquad The transpose $^T$ of {\it any} matrix $A$ is
\begin{align}
(A^T)^i_{\phantom{i}j} = A^j_{\phantom{j}i} .
\end{align}
In words: the $i$ row of $A^T$ is the $i$th column of $A$; the $j$th column of $A^T$ is the $j$th row of $A$. If $A$ is a (square) $D \times D$ matrix, you reflect it along the diagonal to obtain $A^T$.
\begin{myP}
\qquad Show using index notation that $(A \cdot B)^T = B^T A^T$. \qed
\end{myP}
{\bf Adjoint} \qquad The adjoint $^\dagger$ of {\it any} matrix is given by
\begin{align}
(A^\dagger)^i_{\phantom{i}j} = (A^j_{\phantom{j}i})^* = (A^*)^j_{\phantom{j}i} .
\end{align}
In other words, $A^\dagger = (A^T)^*$; to get $A^\dagger$, you start with $A$, take its transpose, then take its complex conjugate. An example is,
\begin{align}
A &= \left[
\begin{array}{cc}
1+i		& e^{i\theta} \\
x+iy	& \sqrt{10}
\end{array} \right] , \qquad 0 \leq \theta < 2\pi, \ x,y\in\mathbb{R} \\
A^T &=  \left[
\begin{array}{cc}
1+i			& x+iy \\
e^{i\theta}	& \sqrt{10}
\end{array} \right] , \qquad
A^\dagger = \left[
\begin{array}{cc}
1-i				& x-iy \\
e^{-i\theta}	& \sqrt{10}
\end{array} \right] .
\end{align}
{\bf Orthogonal, Unitary, Symmetric, and Hermitian} \qquad A $D \times D$ matrix $A$ is 
\begin{enumerate}
\item Orthogonal if $A^T A = A A^T = \mathbb{I}$. The set of real orthogonal matrices implement rotations in a $D$-dimensional real (vector) space.
\item Unitary if $A^\dagger A = A A^\dagger = \mathbb{I}$. Thus, a real unitary matrix is orthogonal. Moreover, unitary matrices, like their real orthogonal counterparts, implement ``rotations" in a $D$ dimensional complex (vector) space.
\item Symmetric if $A^T = A$; anti-symmetric if $A^T = -A$.
\item Hermitian if $A^\dagger = A$; anti-hermitian if $A^\dagger = -A$.
\end{enumerate}
\begin{myP}
\qquad Explain why, if $A$ is an orthogonal matrix, it obeys the equation
\begin{align}
A^i_{\phantom{i}k} A^j_{\phantom{j}l} \delta_{ij} = \delta_{kl} .
\end{align}
Now explain why, if $A$ is a unitary matrix, it obeys the equation
\begin{align}
\label{UnitaryMatricesHaveOrthogonalColumns}
(A^i_{\phantom{i}k})^* A^j_{\phantom{j}l} \delta_{ij} = \delta_{kl} .
\end{align} \qed
\end{myP}
\begin{myP}
\qquad Prove that $(AB)^T = B^T A^T$ and $(AB)^\dagger = B^\dagger A^\dagger$. This means if $A$ and $B$ are orthogonal, then $AB$ is orthogonal; and if $A$ and $B$ are unitary $AB$ is unitary. Can you explain why? \qed
\end{myP}
Simple examples of a unitary, symmetric and Hermitian matrix are, respectively (from left to right):
\begin{align}
\left[
\begin{array}{cc}
e^{i\theta}		& 0 		\\
0				& e^{i\delta}
\end{array} \right], \qquad
\left[
\begin{array}{cc}
e^{i\theta}		& X		\\
X				& e^{i\delta}
\end{array} \right], \qquad
\left[
\begin{array}{cc}
\sqrt{109}		& 1-i 		\\
1+i				& \theta^\delta
\end{array} \right], \qquad
\theta,\delta \in \mathbb{R} .
\end{align}

\subsection{Determinants, Linear (In)dependence, Inverses and Eigensystems}

{\bf Levi-Civita symbol and the Determinant} \qquad We will now define the determinant of a $D \times D$ matrix $A$ through the Levi-Civita symbol $\epsilon_{i_1 i_2 \dots i_{D-1} i_D}$:
\begin{align}
\label{DetOfMatrix}
\det A \equiv \epsilon_{i_1 i_2 \dots i_{D-1} i_D} A^{i_1}_{\phantom{i_1}1} A^{i_2}_{\phantom{i_2}2} \dots A^{i_{D-1}}_{\phantom{i_{D-1}}D-1} A^{i_D}_{\phantom{i_D}D} .
\end{align}
Every index on the Levi-Civita runs from $1$ through $D$. This definition is equivalent to the usual co-factor expansion definition. The $D$-dimensional Levi-Civita symbol is defined through the following properties.
\begin{itemize}
\item It is completely antisymmetric in its indices. This means swapping any of the indices $i_a \leftrightarrow i_b$ (for $a \neq b$) will return
\begin{align}
\epsilon_{i_1 i_2 \dots i_{a-1} i_a i_{a+1} \dots i_{b-1} i_b i_{b+1} \dots i_{D-1} i_D} 
= - \epsilon_{i_1 i_2 \dots i_{a-1} i_b i_{a+1} \dots i_{b-1} i_a i_{b+1} \dots i_{D-1} i_D} .
\end{align}
\item In matrix algebra and flat Euclidean space, $\epsilon_{1 2 3 \dots D} = \epsilon^{1 2 3 \dots D} \equiv 1$.\footnote{In Lorentzian flat spacetimes, the Levi-Civita tensor with upper indices will need to be carefully distinguished from its counterpart with lower indices.}
\end{itemize}
These are sufficient to define every component of the Levi-Civita symbol. Because $\epsilon$ is fully antisymmetric, if any of its $D$ indices are the same, say $i_a = i_b$, then the Levi-Civita symbol returns zero. (Why?) Whenever $i_1 \dots i_D$ are distinct indices, $\epsilon_{i_1 i_2 \dots i_{D-1} i_D}$ is really the sign of the permutation ($\equiv (-)^\text{nunber of swaps of index pairs}$) that brings $\{ 1,2,\dots,D-1,D \}$ to $\{ i_1,i_2,\dots,i_{D-1}, i_D \}$. Hence, $\epsilon_{i_1 i_2 \dots i_{D-1} i_D}$ is $+1$ when it takes zero/even number of swaps, and $-1$ when it takes odd.

For example, in the 2 dimensional case $\epsilon_{11}=\epsilon_{22}=0$; whereas it takes one swap to go from $12$ to $21$. Therefore,
\begin{align}
1 = \epsilon_{12} = -\epsilon_{21} .
\end{align}
In the 3 dimensional case,
\begin{align}
1 = \epsilon_{123} = -\epsilon_{213} = -\epsilon_{321} = -\epsilon_{132} = \epsilon_{231} = \epsilon_{312} .
\end{align}
Properties of the determinant include
\begin{align}
\det A^T = \det A, \qquad \det(A \cdot B) = \det A \cdot \det B, \qquad \det A^{-1} = \frac{1}{\det A},
\end{align}
for square matrices $A$ and $B$. As a simple example, let us use eq. \eqref{DetOfMatrix} to calculate the determinant of
\begin{align}
A =
\left[\begin{array}{cc}
a 	& b \\
c 	& d
\end{array}\right] .
\end{align}
Remember the only non-zero components of $\epsilon_{i_1 i_2}$ are $\epsilon_{12}=1$ and $\epsilon_{21}=-1$.
\begin{align}
\det A = \epsilon_{12} A^{1}_{\phantom{1}1} A^{2}_{\phantom{2}2} + \epsilon_{21} A^{2}_{\phantom{2}1} A^{1}_{\phantom{1}2} 
&= A^{1}_{\phantom{1}1} A^{2}_{\phantom{2}2} - A^{2}_{\phantom{2}1} A^{1}_{\phantom{1}2} \nonumber\\
&= a d - b c .
\end{align}
{\bf Linear (in)dependence} \qquad Given a set of $D$ vectors $\{v_1,\dots,v_D\}$, we say one of them is linearly dependent (say $v_i$) if we can express it in as a sum of multiples of the rest of the vectors,
\begin{align}
v_i = \sum_{j \neq i}^{D-1} \chi_j v_j \qquad \text{for some} \qquad \chi_j \in \mathbb{C} .
\end{align}
We say the $D$ vectors are linearly independent if none of the vectors are linearly dependent on the rest.

{\it Det as test of linear independence} \qquad If we view the columns or rows of a $D \times D$ matrix $A$ as vectors and if these $D$ vectors are linearly dependent, then the determinant of $A$ is zero. This is because of the antisymmetric nature of the Levi-Civita symbol. Moreover, suppose $\det A \neq 0$. Cramer's rule (cf. eq. \eqref{DetOfMatrix_CramersRule} below) tells us the inverse $A^{-1}$ exists. In fact, for finite dimensional matrix $A$, its inverse $A^{-1}$ is unique. That means the only solution to the $D$-component row (or column) vector $w$, obeying $w \cdot A =0$ (or, $A \cdot w = 0$), is $w=0$. And since $w \cdot A$ (or $A \cdot w$) describes the linear combination of the rows (or, columns) of $A$; this indicates they must be linearly independent whenever $\det A \neq 0$.
\begin{quotation}
For a square matrix $A$, $\det A=0$ iff ($\equiv$ if and only if) its columns and rows are linearly dependent. Equivalently, $\det A \neq 0$ iff its columns and rows are linearly independent.
\end{quotation}
\begin{myP}
\qquad If the columns of a square matrix $A$ are linearly dependent, use eq. \eqref{DetOfMatrix} to prove that $\det A = 0$. Hint: use the antisymmetric nature of the Levi-Civita symbol.
\end{myP}
\begin{myP}
\qquad Show that, for a $D \times D$ matrix $A$ and some complex number $\lambda$,
\begin{align}
\det(\lambda A) = \lambda^D \det A.
\end{align}
\end{myP} 
Hint: this follows almost directly from eq. \eqref{DetOfMatrix}. \qed
\begin{myP}
{\it Relation to cofactor expansion} \qquad The co-factor expansion definition of the determinant is
\begin{align}
\label{DetOfMatrix_CofactorExpansion}
\det A = \sum_{i=1}^D A^i_{\phantom{i}k} C^{i}_{\phantom{i}k} ,
\end{align}
where $k$ is an arbitrary integer from $1$ through $D$. The $C^{i}_{\phantom{i}k}$ is $(-)^{i+k}$ times the determinant of the $(D-1) \times (D-1)$ matrix formed from removing the $i$th row and $k$th column of $A$. (This definition sums over the row numbers; it is actually equally valid to define it as a sum over column numbers.)

As a $3 \times 3$ example, we have
{\allowdisplaybreaks\begin{align}
\det \left[\begin{array}{ccc}
a & b & c \\
d & e & f \\
g & h & l
\end{array}\right]
&= b (-)^{1+2} \det \left[\begin{array}{cc}
d & f \\
g & l
\end{array}\right]
+ e (-)^{2+2} \det \left[\begin{array}{ccc}
a & c \\
g & l
\end{array}\right]
+ h (-)^{3+2} \det \left[\begin{array}{ccc}
a & c \\
d & f 
\end{array}\right] .
\end{align}}
{\it Cramer's rule} \qquad Can you show the equivalence of equations \eqref{DetOfMatrix} and \eqref{DetOfMatrix_CofactorExpansion}? Can you also show that
\begin{align}
\label{DetOfMatrix_CramersRule}
\delta_{kl} \det A = \sum_{i=1}^D A^i_{\phantom{i}k} C^{i}_{\phantom{i}l} ?
\end{align}
That is, show that when $k \neq l$, the sum on the right hand side is zero. What does eq. \eqref{DetOfMatrix_CramersRule} tell us about $(A^{-1})^l_{\phantom{l}i}$?

Hint: start from the left-hand-side, namely
\begin{align}
\det A 
&= \epsilon_{j_1 \dots j_D} A^{j_1}_{\phantom{j_1}1} \dots A^{j_D}_{\phantom{j_D}D} \\
&= A^i_{\phantom{i}k} \left( \epsilon_{j_1 \dots j_{k-1} i j_{k+1} \dots j_D} A^{j_1}_{\phantom{j_1}1} \dots A^{j_{k-1}}_{\phantom{j_{k-1}}k-1} A^{j_{k+1}}_{\phantom{j_{k+1}}k+1} \dots A^{j_D}_{\phantom{j_D}D} \right) , \nonumber 
\end{align}
where $k$ is an arbitrary integer in the set $\{ 1,2,3,\dots,D-1,D \}$. Examine the term in the parenthesis. First shift the index $i$, which is located at the $k$th slot from the left, to the $i$th slot. Then argue why the result is $(-)^{i+k}$ times the determinant of $A$ with the $i$th row and $k$th column removed. \qed
\end{myP}
{\bf Pauli Matrices} \qquad The $2 \times 2$ identity together with the Pauli matrices are Hermitian matrices.
\begin{align}
\label{PauliMatrices}
\sigma^0 \equiv
\left[\begin{array}{cc}
1 & 0 \\
0 & 1
\end{array}\right], \qquad
\sigma^1 \equiv
\left[\begin{array}{cc}
0 & 1 \\
1 & 0
\end{array}\right], \qquad
\sigma^2 \equiv
\left[\begin{array}{cc}
0 	& -i \\
i 	& 0
\end{array}\right], \qquad
\sigma^3 \equiv
\left[\begin{array}{cc}
1 	& 0 	\\
0 	& -1
\end{array}\right]
\end{align}
\begin{myP}
\qquad Let $p_\mu \equiv (p_0,p_1,p_2,p_3)$ be a 4-component collection of complex numbers. Verify the following determinant, relevant for the study of Lorentz symmetry in 4-dimensional flat spacetime,
\begin{align}
\det p_\mu \sigma^\mu = \sum_{0 \leq \mu,\nu \leq 3} \eta^{\mu\nu} p_\mu p_\nu \equiv p^2 ,
\end{align}
where $p_\mu \sigma^\mu \equiv \sum_{0 \leq \mu \leq 3} p_\mu \sigma^\mu$ and
\begin{align}
\eta^{\mu\nu}
\equiv \left[\begin{array}{cccc}
1 & 0 	& 0 	& 0 \\
0 & -1 	& 0 	& 0 \\
0 & 0 	& -1 	& 0 \\
0 & 0 	& 0 	& -1 
\end{array}\right] .
\end{align}
(This is the metric in 4 dimensional flat ``Minkowski" spacetime.) Verify, for $i,j,k\in\{1,2,3\}$,
\begin{align}
\label{PauliMatrices_PropertiesI}
\det \sigma^0 = 1, \qquad \det \sigma^i = -1, \qquad \Tr{\sigma^0}=2, \qquad \Tr{\sigma^i} = 0 \\
\label{PauliMatrices_PropertiesII}
\sigma^i \sigma^j = \delta^{ij} \mathbb{I} + i \sum_{1 \leq k \leq 3} \epsilon^{ijk} \sigma^k, \qquad \sigma^2 \sigma^i \sigma^2 = -(\sigma^i)^* .
\end{align}
Also use the antisymmetric nature of the Levi-Civita symbol to aruge that
\begin{align}
\theta_i \theta_j \epsilon^{ijk} = 0 .
\end{align}
Can you use these facts to calculate
\begin{align}
\label{SU2}
U(\vec{\theta}) \equiv \exp\left[ -\frac{i}{2} \sum_{j=1}^{3} \theta_j \sigma^j \right] \equiv e^{-(i/2)\vec{\theta} \cdot \vec{\sigma}} ?
\end{align}
(Hint: Taylor expand $\exp X = \sum_{\ell=0}^{\infty} X^\ell/\ell !$, followed by applying the first relation in eq. \eqref{PauliMatrices_PropertiesII}.) For now, assume $\{\theta_i\}$ can be complex; later on you'd need to specialize to $\{\theta_i\}$ being real. Show that any $2 \times 2$ complex matrix $A$ can be built from $p_\mu \sigma^\mu$ by choosing the $p_\mu$s appropriately. Then compute $(1/2) \Tr{p_\mu \sigma^\mu \sigma^\nu}$, for $\nu=0,1,2,3$, and comment on how the trace can be used, given $A$, to solve for the $p_\mu$ in the equation
\begin{align}
p_\mu \sigma^\mu = A.
\end{align} \qed
\end{myP}
{\bf Inverse} \qquad The inverse of the $D \times D$ matrix $A$ is defined to be 
\begin{align}
A^{-1} A = A A^{-1} = \mathbb{I} .
\end{align}
The inverse $A^{-1}$ of a finite dimensional matrix $A$ is unique; moreover, the left $A^{-1}A = \mathbb{I}$ and right inverses $AA^{-1} = \mathbb{I}$ are the same object. The inverse exists if and only if ($\equiv$ iff) $\det A \neq 0$.
\begin{myP}
\qquad How does eq. \eqref{DetOfMatrix_CramersRule} allow us to write down the inverse matrix $(A^{-1})^i_{\phantom{i}k}$? \qed
\end{myP}
\begin{myP}
\qquad Why are the left and right inverses of (an invertible) matrix $A$ the same? Hint: Consider $LA=\mathbb{I}$ and $AR=\mathbb{I}$; for the first, multiply $R$ on both sides from the right. \qed
\end{myP}
\begin{myP}
\qquad Prove that $(A^{-1})^T = (A^T)^{-1}$ and $(A^{-1})^\dagger = (A^\dagger)^{-1}$. \qed
\end{myP}
{\bf Eigenvectors and Eigenvalues} \qquad If $A$ is a $D \times D$ matrix, $v$ is its ($D$-component) eigenvector with eigenvalue $\lambda$ if it obeys
\begin{align}
\label{EigenvectorEigenvalue}
A^i_{\phantom{i}j} v^j = \lambda v^i .
\end{align}
This means 
\begin{align}
\label{EigenvectorEigenvalue_NullEqn}
(A^i_{\phantom{i}j} - \lambda \delta^i_{\phantom{i}j})v^j = 0
\end{align}
has non-trivial solutions iff
\begin{align}
\label{CharacteristicEqn}
P_D(\lambda) \equiv \det \left( A - \lambda \mathbb{I} \right) = 0 .
\end{align}
Equation \eqref{CharacteristicEqn} is known as the characteristic equation. For a $D \times D$ matrix, it gives us a $D$th degree polynomial $P_D(\lambda)$ for $\lambda$, whose roots are the eigenvalues of the matrix $\lambda$ -- the set of all eigenvalues of a matrix is called its {\it spectrum}. For each solution for $\lambda$, we then proceed to solve for the $v^i$ in eq. \eqref{EigenvectorEigenvalue_NullEqn}. That there is always at least one solution -- there could be more -- for $v^i$ is because, since its determinant is zero, the columns of $A - \lambda \mathbb{I}$ are necessarily linearly dependent. As already discussed above, this amounts to the statement that there is some sum of multiples of these columns ($\equiv$ ``linear combination") that yields zero -- in fact, the components of $v^i$ are precisely the coefficients in this sum. If $\{ w_i \}$ are these columns of $A - \lambda \mathbb{I}$,
\begin{align}
A - \lambda \mathbb{I} \equiv \left[
w_1 w_2 \dots w_D
\right] \qquad \Rightarrow \qquad
(A - \lambda \mathbb{I})v = \sum_j w_j v^j = 0 .
\end{align}
(Note that, if $\sum_j w_j v^j = 0$ then $\sum_j w_j (K v^j) = 0$ too, for any complex number $K$; in other words, eigenvectors are only defined up to an overall multiplicative constant.) Every $D \times D$ matrix has $D$ eigenvalues from solving the $D$th order polynomial equation \eqref{CharacteristicEqn}; from that, you can then obtain $D$ corresponding eigenvectors. Note, however, the eigenvalues can be repeated; when this occurs, it is known as a {\it degenerate} spectrum. Moreover, not all the eigenvectors are guaranteed to be linearly independent; i.e., some eigenvectors can turn out to be sums of multiples of other eigenvectors. 

The {\it Cayley-Hamilton theorem} states that the matrix $A$ satisfies its own characteristic equation. In detail, if we express eq. \eqref{CharacteristicEqn} as $\sum_{i=0}^D q_i \lambda^i = 0$ (for appropriate complex constants $\{q_i\}$), then replace $\lambda^i \to A^i$ (namely, the $i$th power of $\lambda$ with the $i$th power of $A$), we would find
\begin{align}
P_D(A) = 0 .
\end{align}
Any $D \times D$ matrix $A$ admits a {\it Schur decomposition}. Specifically, there is some unitary matrix $U$ such that $A$ can be brought to an upper triangular form, with its eigenvalues on the diagonal:
\begin{align}
U^\dagger A U = \text{diag}(\lambda_1,\dots,\lambda_D) + N,
\end{align}
where $N$ is strictly upper triangular, with $N^i_{\phantom{i}j} = 0$ for $j \leq i$. The Schur decomposition can be proved via mathematical induction on the size of the matrix.

A special case of the Schur decomposition occurs when all the off-diagonal elements are zero. A $D \times D$ matrix $A$ can be {\it diagonalized} if there is some unitary matrix $U$ such that
\begin{align}
U^\dagger A U = \text{diag}(\lambda_1,\dots,\lambda_D),
\end{align}
where the $\{ \lambda_i \}$ are the eigenvalues of $A$. Each column of $U$ is filled with a distinct unit length eigenvector of $A$. (Unit length means $v^\dagger v = (v^i)^* v^j \delta_{ij} = 1$.) In index notation,
\begin{align}
\label{Diagonalization_v1}
A^i_{\phantom{i}j} U^j_{\phantom{j}k} 
= \lambda_k U^i_{\phantom{i}k} 
= U^i_{\phantom{j}l} \delta^l_{\phantom{l}k} \lambda_k, \qquad \text{(No sum over $k$)} .
\end{align}
In matrix notation,
\begin{align}
A U = U \text{diag}[\lambda_1,\lambda_2,\dots,\lambda_{D-1},\lambda_D] .
\end{align}
Here, $U^j_{\phantom{j}k}$ for fixed $k$, is the $k$th eigenvector, and $\lambda_k$ is the corresponding eigenvalue. By multiplying both sides with $U^\dagger$, we have
\begin{align}
\label{Diagonalization_v2}
U^\dagger A U = D, \qquad D^j_{\phantom{j}l} \equiv \lambda_l \delta^j_{\phantom{j}l} \qquad \text{(No sum over $l$)} .
\end{align}
Some jargon: the {\it null space} of a matrix $M$ is the space spanned by all vectors $\{v_i\}$ obeying $M \cdot v_i = 0$. When we solve for the eigenvector of $A$ by solving $(A-\lambda \mathbb{I})\cdot v$, we are really solving for the null space of the matrix $M \equiv A-\lambda \mathbb{I}$, because for a fixed eigenvalue $\lambda$, there could be more than one solution -- that's what we mean by degeneracy.

Real symmetric matrices can be always diagonalized via an orthogonal transformation. Complex Hermitian matrices can always be diagonalized via a unitary one. These statements can be proved readily using their Schur decomposition. For, let $A$ be Hermitian and $U$ be a unitary matrix such that
\begin{align}
U A U^\dagger = \text{diag}(\lambda_1,\dots,\lambda_D) + N ,
\end{align}
where $N$ is strictly upper triangular. Now, if $A$ is Hermitian, so is $U A U^\dagger$, because $(U A U^\dagger)^\dagger = (U^\dagger)^\dagger A^\dagger U^\dagger = U A U^\dagger$. Therefore,
\begin{align}
(U A U^\dagger)^\dagger = U A U^\dagger \qquad \Rightarrow \qquad
\text{diag}(\lambda_1^*,\dots,\lambda_D^*) + N^\dagger = \text{diag}(\lambda_1,\dots,\lambda_D) + N .
\end{align}
Because the transpose of a strictly upper triangular matrix returns a strictly lower triangular matrix, we have a strictly lower triangular matrix $N^\dagger$ plus a diagonal matrix (built out of the complex conjugate of the eigenvalues of $A$) equal to a diagonal one (built out of the eigenvalues of $A$) plus a strictly upper triangular $N$. That means $N=0$ and $\lambda_l = \lambda_l^*$. That is, any Hermitian $A$ is diagonalizable and all its eigenvalues are real.

Unitary matrices can also always be diagonalized. In fact, all its eigenvalues $\{ \lambda_i \}$ lie on the unit circle on the complex plane, i.e., $|\lambda_i|=1$. Suppose now $A$ is unitary and $U$ is another unitary matrix such that the Schur decomposition of $A$ reads
\begin{align}
U A U^\dagger = M ,
\end{align}
where $M$ is an upper triangular matrix with the eigenvalues of $A$ on its diagonal. Now, if $A$ is unitary, so is $U A U^\dagger$, because
\begin{align}
\left(U A U^\dagger\right)^\dagger (U A U^\dagger)
= U A^\dagger U^\dagger U A U^\dagger 
= U A^\dagger A U^\dagger = U U^\dagger = \mathbb{I} .
\end{align}
That means
\begin{align}
M^\dagger M = \mathbb{I} \qquad \Rightarrow \qquad 
(M^\dagger M)^k_{\phantom{k}l}
= (M^\dagger)^k_{\phantom{k}s} M^s_{\phantom{s}l} 
= \sum_s \overline{M^s_{\phantom{s}k}} M^s_{\phantom{s}l} 
= \delta_{ij} \overline{M^i_{\phantom{i}k}} M^j_{\phantom{j}l} = \delta_{kl} ,
\end{align}
where we have recalled eq. \eqref{UnitaryMatricesHaveOrthogonalColumns} in the last equality. If $w_i$ denotes the $i$th column of $M$, the unitary nature of $M$ implies all its columns are orthogonal to each other and each column has length one. Since $M$ is upper triangular, we see that the only non-zero component of the first column is its first row, i.e., $w_1^i = M^i_{\phantom{i}1} = \lambda_1 \delta_1^i$. Unit length means $w_1^\dagger w_1 = 1 \Rightarrow |\lambda_1|^2=1$. That $w_1$ is orthogonal to every other column $w_{i>1}$ means the latter have their first rows equal to zero; $\overline{M^1_{\phantom{1}1}} M^1_{\phantom{1}l} = \overline{\lambda_1} M^1_{\phantom{1}l} = 0 \Rightarrow M^1_{\phantom{1}l} = 0$ for $l \neq 1$ -- remember $\overline{M^1_{\phantom{1}1}} = \overline{\lambda_1}$ itself cannot be zero because it lies on the unit circle on the complex plane. Now, since its first component is necessarily zero, the only non-zero component of the second column is its second row, i.e., $w_2^i = M^i_{\phantom{i}2} = \lambda_2 \delta_2^i$. Unit length again means $|\lambda_2|^2=1$. And, by demanding that $w_2$ be orthogonal to every other column means their second components are zero: $\overline{M^2_{\phantom{2}2}} M^2_{\phantom{2}l} = \overline{\lambda_2} M^2_{\phantom{2}l} = 0 \Rightarrow M^2_{\phantom{2}l} = 0$ for $l > 2$ -- where, again,  $\overline{M^2_{\phantom{2}2}} = \overline{\lambda_2}$ cannot be zero because it lies on the complex plane unit circle. By induction on the column number, we see that the only non-zero component of the $i$th column is the $i$th row. That is, any unitary $A$ is diagonalizable and all its eigenvalues lie on the circle: $|\lambda_{1 \leq i \leq D}|=1$.

{\it Diagonalization example} \qquad As an example, let's diagonalize $\sigma^2$ from eq. \eqref{PauliMatrices}.
\begin{align}
P_2(\lambda) = \det 
\left[\begin{array}{cc}
-\lambda 	& -i \\
i 			& -\lambda
\end{array}\right] = \lambda^2 - 1 = 0
\end{align}
(We can even check Caley-Hamilton here: $P_2(\sigma^2) = (\sigma^2)^2 - \mathbb{I} = \mathbb{I}-\mathbb{I} = 0$; see eq. \eqref{PauliMatrices_PropertiesII}.) The solutions are $\lambda = \pm 1$ and
\begin{align}
\left[\begin{array}{cc}
\mp 1 	& -i \\
i 		& \mp 1
\end{array}\right] \left[\begin{array}{c} v^1 \\ v^2 \end{array}\right] 
= \left[\begin{array}{c} 0 \\ 0 \end{array}\right] \qquad \Rightarrow \qquad v^1_\pm = \mp i v^2_\pm .
\end{align}
The subscripts on $v$ refer to their eigenvalues, namely
\begin{align}
\label{DiagonalizationExample}
\sigma^2 v_\pm = \pm v_\pm .
\end{align}
By choosing $v^2 = 1/\sqrt{2}$, we can check $(v_\pm^i)^* v_\pm^j \delta_{ij} = 1$ and therefore the normalized eigenvectors are
\begin{align}
v_\pm = \frac{1}{\sqrt{2}}\left[\begin{array}{c} \mp i \\ 1 \end{array}\right] .
\end{align}
Furthermore you can check directly that eq. \eqref{DiagonalizationExample} is satisfied. We therefore have
\begin{align}
\underbrace{\left(\frac{1}{\sqrt{2}} \left[\begin{array}{cc}
i 	& 1 \\
-i	& 1
\end{array}\right]\right)}_{\equiv U^\dagger}
\sigma^2 \underbrace{\left(\frac{1}{\sqrt{2}} \left[\begin{array}{cc}
-i 	& i \\
1	& 1
\end{array}\right]\right)}_{\equiv U}
= \left[\begin{array}{cc}
1 		& 0 \\
0 		& -1
\end{array}\right] .
\end{align}
An example of a matrix that cannot be diagonalized is
\begin{align}A \equiv \left[
\begin{array}{cc}
0 & 0 \\
1 & 0
\end{array} \right] .
\end{align}
The characteristic equation is $\lambda^2=0$, so both eigenvalues are zero. Therefore $A-\lambda\mathbb{I}=A$, and
\begin{align}
\left[
\begin{array}{cc}
0 & 0 \\
1 & 0
\end{array} \right] \left[
\begin{array}{cc}
v^1 \\
v^2 
\end{array} \right] = 
\left[
\begin{array}{cc}
0 \\
0 
\end{array} \right] \qquad \Rightarrow \qquad v^1 = 0, \ v^2 \text{ arbitrary.}
\end{align}
There is a repeated eigenvalue of $0$, but there is only one linearly independent eigenvector $(0,1)$. It is not possible to build a unitary $2 \times 2$ matrix $U$ whose columns are distinct unit length eigenvectors of $\sigma^2$.
\begin{myP}
\qquad Show how to go from eq. \eqref{Diagonalization_v1} to eq. \eqref{Diagonalization_v2} using index notation. \qed
\end{myP}
\begin{myP}
\qquad Use the Schur decomposition to explain why, for any matrix $A$, $\Tr{A}$ is equal to the sum of its eigenvalues and $\det A$ is equal to their product:
\begin{align}
\Tr{A} = \sum_{l=1}^{D} \lambda_l, \qquad \det A = \prod_{l=1}^{D} \lambda_l.
\end{align}
Hint: for $\det A$, the key question is how to take the determinant of an upper triangular matrix. \qed
\end{myP}
\begin{myP}
\qquad For a {\it strictly} upper triangular matrix $N$, prove that $N$ multiplied to itself any number of times still returns a strictly upper triangular matrix. Can a strictly upper triangular matrix be diagonalized? (Explain.) \qed
\end{myP}
\begin{myP}
\qquad Suppose $A = U X U^\dagger$, where $U$ is a unitary matrix. If $f(z)$ is a function of $z$ that can be Taylor expanded about some point $z_0$, explain why $f(A) = U f(X) U^\dagger$. Hint: Can you explain why $(U B U^\dagger)^\ell = U B^\ell U^\dagger$, for $B$ some arbitrary matrix, $U$ unitary, and $\ell = 1,2,3,\dots$? \qed
\end{myP}
\begin{myP}
\qquad Can you provide a simple explanation to why the eigenvalues $\{\lambda_l\}$ of a unitary matrix are always of unit absolute magnitude; i.e. why are the $|\lambda_l|=1$? \qed
\end{myP}
\begin{myP}
{\it Simplified example of neutrino oscillations.} \qquad We begin with the observation that the solution to the first order equation
\begin{align}
i \partial_t \psi(t) = E \psi(t),
\end{align}
for $E$ some real constant, is 
\begin{align}
\psi(t) = e^{-iEt} \psi_0 .
\end{align}
The $\psi_0$ is some arbitrary (possibly complex) constant, corresponding to the initial condition $\psi(t=0)$. Now solve the matrix differential equation
\begin{align}
i \partial_t N(t) = H N(t), \qquad N(t) \equiv \left[ \begin{array}{c} \nu_1(t) \\ \nu_2(t) \end{array} \right],
\end{align}
with the initial condition -- describing the production of $\nu_1$-type of neutrino, say --
\begin{align}
\left[ \begin{array}{c} \nu_1(t=0) \\ \nu_2(t=0) \end{array}\right] = \left[ \begin{array}{c} 1 \\ 0 \end{array}\right],
\end{align}
where the Hamiltonian $H$ is
\begin{align}
H &\equiv \left[
\begin{array}{cc}
	p & 0 \\
	0 & p 
\end{array}
\right] + \frac{1}{4p} M , \\
M &\equiv \left[
\begin{array}{cc}
m_1^2+m_2^2 + (m_1^2-m_2^2) \cos (2 \theta ) 	& (m_1^2-m_2^2) \sin (2 \theta ) \\
(m_1^2-m_2^2) \sin (2 \theta) 					& m_1^2+m_2^2+\left(m_2^2-m_1^2\right) \cos (2 \theta )
\end{array}
\right] .
\end{align}
The $p$ is the magnitude of the momentum, $m_{1,2}$ are masses, and $\theta$ is the ``mixing angle". Then calculate
\begin{align}
P_{1 \to 1} \equiv \left\vert N(t)^\dagger \left[ \begin{array}{c} 1 \\ 0 \end{array} \right] \right\vert^2
\qquad \text{   and   } \qquad
P_{1 \to 2} \equiv \left\vert N(t)^\dagger \left[ \begin{array}{c} 0 \\ 1 \end{array} \right] \right\vert^2 .
\end{align}
Express $P_{1 \to 1}$ and $P_{1 \to 2}$ in terms of $\Delta m^2 \equiv m_1^2 - m_2^2$. (In quantum mechanics, they respectively correspond to the probability of observing the neutrinos $\nu_1$ and $\nu_2$ at time $t>0$, given $\nu_1$ was produced at $t=0$.) Hint: Start by diagonalizing $M = U^T A U$ where
\begin{align}
U \equiv \left[
\begin{array}{cc}
\cos\theta 	& \sin\theta \\
-\sin\theta & \cos\theta \\
\end{array}
\right] .
\end{align}
The $U N(t)$ is known as the ``mass-eigenstate" basis. Can you comment on why? Note that, in the highly relativistic limit, the energy $E$ of a particle of mass $m$ is
\begin{align}
E = \sqrt{p^2 + m^2} \to p + \frac{m^2}{2p} + \mathcal{O}(1/p^2) .
\end{align}
Note: In this problem, we have implicitly set $\hbar = c = 1$, where $\hbar$ is the reduced Planck's constant and $c$ is the speed of light in vacuum. \qed
\end{myP}

\subsection{Special Topic 1: 2D real orthogonal matrices}

In this subsection we will illustrate what a real orthogonal matrix is by studying the 2D case in some detail. Let $A$ be such a $2 \times 2$ real orthogonal matrix. We will begin by writing its components as follows
\begin{align}
A \equiv \left[
\begin{array}{cc}
v^1 	& v^2 \\
w^1		& w^2
\end{array}
\right] .
\end{align}
(As we will see, it is useful to think of $v^{1,2}$ and $w^{1,2}$ as components of 2D vectors.) That $A$ is orthogonal means $A A^T = \mathbb{I}$.
\begin{align}
\left[
\begin{array}{cc}
v^1 	& v^2 \\
w^1		& w^2
\end{array}
\right] \cdot
\left[
\begin{array}{cc}
v^1 	& w^1 \\
v^2		& w^2
\end{array}
\right] 
=
\left[
\begin{array}{cc}
\vec{v} \cdot \vec{v}	& \vec{v} \cdot \vec{w} \\
\vec{w} \cdot \vec{v}	& \vec{w} \cdot \vec{w}
\end{array}
\right] = 
\left[
\begin{array}{cc}
1	& 0 \\
0	& 1
\end{array}
\right] .
\end{align}
This translates to: $\vec{w}^2 \equiv \vec{w}\cdot\vec{w} = 1$, $\vec{v}^2 \equiv \vec{v}\cdot\vec{v} = 1$ (length of both the 2D vectors are one); and $\vec{w} \cdot \vec{v} = 0$ (the two vectors are perpendicular). In 2D any vector can be expressed in polar coordinates; for example, the Cartesian components of $\vec{v}$ are
\begin{align}
v^i = r(\cos\phi,\sin\phi), \qquad r \geq 0, \ \phi \in [0,2\pi) .
\end{align}
But $\vec{v}^2 = 1$ means $r=1$. Similarly, 
\begin{align}
w^i = (\cos\phi',\sin\phi'), \qquad \phi' \in [0,2\pi).
\end{align}
Because $\vec{v}$ and $\vec{w}$ are perpendicular,
\begin{align}
\vec{v}\cdot\vec{w} = \cos\phi \cdot \cos\phi' + \sin\phi \cdot \sin\phi' = \cos(\phi-\phi') = 0 .
\end{align}
This means $\phi' = \phi \pm \pi/2$. (Why?) Furthermore
\begin{align}
w^i 
= (\cos(\phi \pm \pi/2), \sin(\phi \pm \pi/2))
= (\mp \sin(\phi), \pm \cos(\phi) ) .
\end{align}
What we have figured out is that, any real orthogonal matrix can be parametrized by an angle $0 \leq \phi < 2\pi$; and for each $\phi$ there are two distinct solutions.
\begin{align}
\label{O2Solutions}
R_1(\phi) = \left[
\begin{array}{cc}
\cos\phi	& \sin\phi \\
-\sin\phi	& \cos\phi
\end{array}
\right] ,
\qquad\qquad
R_2(\phi) = \left[
\begin{array}{cc}
\cos\phi	& \sin\phi \\
\sin\phi	& -\cos\phi
\end{array}
\right] .
\end{align}
By a direct calculation you can check that $R_1(\phi > 0)$ rotates an arbitrary 2D vector {\it clockwise} by $\phi$. Whereas, $R_2(\phi > 0)$ rotates the vector, followed by flipping the sign of its $y$-component; this is because
\begin{align}
R_2(\phi) = \left[
\begin{array}{cc}
1	& 0 \\
0	& -1
\end{array}
\right] \cdot R_1(\phi) .
\end{align}
In other words, the $R_2(\phi=0)$ in eq. \eqref{O2Solutions} corresponds to a ``parity flip" where the vector is reflected about the $x$-axis. 
\begin{myP}
\qquad What about the matrix that reflects 2D vectors about the $y$-axis? What value of $\theta$ in $R_2(\theta)$ would it correspond to?

Find the determinants of $R_1(\phi)$ and $R_2(\phi)$. You should be able to use that to argue, there is no $\theta_0$ such that $R_1(\theta_0) = R_2(\theta_0)$. Also verify that
\begin{align}
\label{SO2RotationLaw}
R_1(\phi) R_1(\phi') = R_1(\phi+\phi') .
\end{align} 
This makes geometric sense: rotating a vector clockwise by $\phi$ then by $\phi'$ should be the same as rotation by $\phi+\phi'$. Mathematically speaking, this composition law in eq. \eqref{SO2RotationLaw} tells us rotations form the SO$_2$ {\it group}. The set of $D \times D$ real orthogonal matrices obeying $R^T R = \mathbb{I}$, including both rotations and reflections, forms the group O$_D$. The group involving only rotations is known as SO$_D$; where the `S' stands for ``special" ($\equiv$ determinant equals one). \qed
\end{myP}
\begin{myP}
{\it $2 \times 2$ Unitary Matrices.} \qquad Can you construct the most general $2 \times 2$ unitary matrix? First argue that the most general complex 2D vector $\vec{v}$ that satisfies $\vec{v}^\dagger \vec{v} = 1$ is
\begin{align}
v^i = e^{i\phi_1} (\cos\theta, e^{i\phi_2} \sin\theta), \qquad \phi_{1,2}, \theta \in [0,2\pi).
\end{align}
Then consider $\vec{v}^\dagger \vec{w} = 0$, where
\begin{align}
w^i = e^{i\phi'_1} (\cos\theta', e^{i\phi'_2} \sin\theta'), \qquad \phi'_{1,2}, \theta' \in [0,2\pi).
\end{align}
You should arrive at
\begin{align}
\sin(\theta) \sin(\theta') e^{i(\phi'_2 - \phi_2)} + \cos(\theta) \cos(\theta') = 0.
\end{align}
By taking the real and imaginary parts of this equation, argue that
\begin{align}
\phi'_2 = \phi_2, \qquad
\theta = \theta' \pm \frac{\pi}{2} .
\end{align}
or
\begin{align}
\phi'_2 = \phi_2 + \pi, \qquad 
\theta = -\theta' \pm \frac{\pi}{2} .
\end{align} 
From these, deduce that the most general $2 \times 2$ unitary matrix $U$ can be built from the most general real orthogonal one $O(\theta)$ via
\begin{align}
\label{U2}
U = \left[
\begin{array}{cc}
e^{i\phi_1}	& 0 \\
0			& e^{i\phi_2}
\end{array}
\right] \cdot O(\theta) \cdot 
\left[ \begin{array}{cc}
1	& 0 \\
0	& e^{i\phi_3}
\end{array}
\right] .
\end{align}
As a simple check: note that $\vec{v}^\dagger \vec{v} = \vec{w}^\dagger \vec{w} = 1$ together with $\vec{v}^\dagger \vec{w} = 0$ provides 4 constraints for 8 parameters -- 4 complex entries of a $2 \times 2$ matrix -- and therefore we should have 4 free parameters left. 

{\it Bonus problem:} \qquad By imposing $\det U = 1$, can you connect eq. \eqref{U2} to eq. \eqref{SU2}? \qed
\end{myP}

\newpage
\section{Linear Algebra}
\label{Chapter_LinearAlgebra}
\subsection{Definition}

Loosely speaking, the notion of a vector space -- as the name suggests -- amounts to abstracting the algebraic properties -- addition of vectors, multiplication of a vector by a number, etc. -- obeyed by the familiar $D \in \{1,2,3,\dots\}$ dimensional Euclidean space $\mathbb{R}^D$. We will discuss the linear algebra of vector spaces using Paul Dirac's bra-ket notation. This will not only help you understand the logical foundations of linear algebra and the matrix algebra you encountered earlier, it will also prepare you for the study of quantum theory, which is built entirely on the theory of both finite and infinite dimensional vector spaces.\footnote{The material in this section of our notes was drawn heavily from the contents and problems provided in Chapter 1 of Sakurai's {\it Modern Quantum Mechanics}.}

We will consider a vector space over complex numbers. A member of the vector space will be denoted as $\ket{\alpha}$; we will use the words ``ket", ``vector" and ``state" interchangeably in what follows. We will allude to aspects of quantum theory, but point out everything we state here holds in a more general context; i.e., quantum theory is not necessary but merely an application -- albeit a very important one for physics. For now $\alpha$ is just some arbitrary label, but later on it will often correspond to the eigenvalue of some linear operator. We may also use $\alpha$ as an enumeration label, where $\ket{\alpha}$ is the $\alpha$th element in the collection of vectors. In quantum mechanics, a physical system is postulated to be completely described by some $\ket{\alpha}$ in a vector space, whose time evolution is governed by some Hamiltonian. (The latter is what Schr\"{o}dinger's equation is about.)

Here is what defines a ``vector space over complex numbers":

\begin{enumerate}

\item {\bf Addition} \qquad Any two vectors can be added to yield another vector
\begin{align}
\ket{\alpha} + \ket{\beta} = \ket{\gamma} .
\end{align}
Addition is commutative and associative:
\begin{align}
\ket{\alpha} + \ket{\beta} &= \ket{\beta} + \ket{\alpha} \\
\ket{\alpha} + (\ket{\beta} + \ket{\gamma}) &= (\ket{\alpha} + \ket{\beta}) + \ket{\gamma} .
\end{align}
\item {\bf Additive identity (zero vector) and existence of inverse} \qquad There is a zero vector $\ket{\text{zero}}$ -- which can be gotten by multiplying any vector by $0$, i.e.,
\begin{align}
0 \ket{\alpha} = \ket{\text{zero}} 
\end{align} 
-- that acts as an additive identity.\footnote{In this section we will be careful and denote the zero vector as $\ket{\text{zero}}$. For the rest of the notes, whenever the context is clear, we will often use $0$ to denote the zero vector.} Namely, adding $\ket{\text{zero}}$ to any vector returns the vector itself:
\begin{align}
\ket{\text{zero}} + \ket{\beta} = \ket{\beta} .
\end{align}
For any vector $\ket{\alpha}$ there exists an additive inverse; if $+$ is the usual addition, then the inverse of $\ket{\alpha}$ is just $(-1)\ket{\alpha}$.
\begin{align}
\ket{\alpha} + (-\ket{\alpha}) = \ket{\text{zero}} .
\end{align}
\item {\bf Multiplication by scalar} \qquad Any ket can be multiplied by an arbitrary complex number $c$ to yield another vector
\begin{align}
c \ket{\alpha} = \ket{\gamma} .
\end{align}
(In quantum theory, $\ket{\alpha}$ and $c \ket{\alpha}$ are postulated to describe the same system.) This multiplication is distributive with respect to both vector and scalar addition; if $a$ and $b$ are arbitrary complex numbers,
\begin{align}
a(\ket{\alpha} + \ket{\beta}) 	&= a\ket{\alpha} + a \ket{\beta} \\
(a + b) \ket{\alpha} 			&= a \ket{\alpha} + b \ket{\alpha} .
\end{align}

\end{enumerate}
{\it Note: }If you define a ``vector space over scalars," where the scalars can be more general objects than complex numbers, then in addition to the above axioms, we have to add: (I) Associativity of scalar multiplication, where $a(b\ket{\alpha}) = (ab)\ket{\alpha}$ for any scalars $a$, $b$ and vector $\ket{\alpha}$; (II) Existence of a scalar identity $1$, where $1 \ket{\alpha} = \ket{\alpha}$.

{\bf Examples} \qquad The Euclidean space $\mathbb{R}^D$ itself, the space of $D$-tuples of real numbers
\begin{align}
\ket{\vec{a}} \equiv (a^1,a^2,\dots,a^D),
\end{align}
with $+$ being the usual addition operation is, of course, {\it the} example of a vector space. We shall check explicitly that $\mathbb{R}^D$ does in fact satisfy all the above axioms. To begin, let 
\begin{align}
\ket{\vec{v}} &= (v^1,v^2,\dots,v^D) , \nonumber\\
\ket{\vec{w}} &= (w^1,w^2,\dots,w^D) \qquad \text{ and } \\
\ket{\vec{x}} &= (x^1,x^2,\dots,x^D)
\end{align}
be vectors in $\mathbb{R}^D$. 
\begin{enumerate}
	\item {\bf Addition} \qquad Any two vectors can be added to yield another vector
	\begin{align}
	\ket{\vec{v}} + \ket{\vec{w}} = (v^1+w^1, \dots, v^D + w^D) \equiv \ket{\vec{v}+\vec{w}} .
	\end{align}
	Addition is commutative and associative because we are adding/subtracting the vectors component-by-component:
	{\allowdisplaybreaks\begin{align}
	\ket{\vec{v}} + \ket{\vec{w}} 	&= \ket{\vec{v}+\vec{w}} = (v^1+w^1, \dots, v^D + w^D) \nonumber\\ 
									&= (w^1 + v^1, \dots, w^D + v^D) \nonumber\\
									&= \ket{\vec{w}} + \ket{\vec{v}} = \ket{\vec{w}+\vec{v}} , \\
	\ket{\vec{v}} + \ket{\vec{w}} + \ket{\vec{x}} 
				&= (v^1+w^1+x^1, \dots, v^D + w^D + x^D) \nonumber\\
				&= (v^1+(w^1+x^1), \dots, v^D + (w^D + x^D)) \nonumber\\
				&= ((v^1+w^1)+x^1, \dots, (v^D + w^D) + x^D) \nonumber\\
				&= \ket{\vec{v}} + (\ket{\vec{w}} + \ket{\vec{x}}) = (\ket{\vec{v}} + \ket{\vec{w}}) + \ket{\vec{x}} = \ket{\vec{v} + \vec{w} + \vec{x}} .
	\end{align}}
	\item {\bf Additive identity (zero vector) and existence of inverse} \qquad There is a zero vector $\ket{\text{zero}}$ -- which can be gotten by multiplying any vector by $0$, i.e.,
	\begin{align}
	0 \ket{\vec{v}} = 0 (v^1,\dots,v^D) = (0,\dots,0) = \ket{\text{zero}} 
	\end{align} 
	-- that acts as an additive identity. Namely, adding $\ket{\text{zero}}$ to any vector returns the vector itself:
	\begin{align}
	\ket{\text{zero}} + \ket{\vec{w}} = (0,\dots,0) + (w^1,\dots,w^D) = \ket{\vec{w}} .
	\end{align}
	For any vector $\ket{\vec{x}}$ there exists an additive inverse; in fact, the inverse of $\ket{\vec{x}}$ is just $(-1)\ket{\vec{x}} = \ket{-\vec{x}}$.
	\begin{align}
	\ket{\vec{x}} + (-\ket{\vec{x}}) = (x^1,\dots,x^D) - (x^1,\dots,x^D) = \ket{\text{zero}} .
	\end{align}
	\item {\bf Multiplication by scalar} \qquad Any ket can be multiplied by an arbitrary real number $c$ to yield another vector
	\begin{align}
	c \ket{\vec{v}} = c(v^1,\dots,v^D) = (cv^1,\dots,cv^D) \equiv \ket{c\vec{v}} .
	\end{align}
	This multiplication is distributive with respect to both vector and scalar addition; if $a$ and $b$ are arbitrary real numbers,
	\begin{align}
	a(\ket{\vec{v}} + \ket{\vec{w}}) 	&= (av^1+aw^1,av^2+aw^2,\dots,av^D+aw^D) \nonumber\\
										&= \ket{a\vec{v}} + \ket{a\vec{w}} = a\ket{\vec{v}} + a\ket{\vec{w}} , \\
	(a + b) \ket{\vec{x}} 	&= (ax^1 + bx^1,\dots,ax^D + bx^D) \nonumber\\
							&= \ket{a\vec{x}} + \ket{b\vec{x}} = a \ket{\vec{x}} + b \ket{\vec{x}} . 
	\end{align} \qed
\end{enumerate}

The following are some further examples of vector spaces. 

\begin{enumerate}
	
\item The space of polynomials with complex coefficients.

\item The space of square integrable functions on $\mathbb{R}^D$ (where $D$ is an arbitrary integer greater or equal to 1); i.e., all functions $f(\vec{x})$ such that $\int_{\mathbb{R}^D} \dd^D \vec{x} |f(\vec{x})|^2 < \infty$.

\item The space of all homogeneous solutions to a linear (ordinary or partial) differential equation.

\item The space of $M \times N$ matrices of complex numbers, where $M$ and $N$ are arbitrary integers greater or equal to 1.

\end{enumerate}
\begin{myP}
\qquad Prove that the examples in (1), (3), and (4) are indeed vector spaces, by running through the above axioms. \qed
\end{myP}
{\bf Linear (in)dependence, Basis, Dimension} \qquad Suppose we pick $N$ vectors from a vector space, and find that one of them can be expressed as a linear combination of the rest,
\begin{align}
\ket{N} = \sum_{i=1}^{N-1} c^i \ket{i} ,
\end{align}
where the $\left\{c^i\right\}$ are complex numbers. Then we say that this set of $N$ vectors are linearly dependent. Suppose we have picked $M$ vectors $\{ \ket{1}, \ket{2}, \ket{3}, \dots, \ket{M}\}$ such that they are linearly independent, i.e., no vector is a linear combination of any others, and suppose further that any arbitrary vector $\ket{\alpha}$ from the vector space can now be expressed as a linear combination (aka superposition) of these vectors
\begin{align}
\ket{\alpha} = \sum_{i=1}^D \chi^i \ket{i}, \qquad \{ \chi^i \in \mathbb{C} \} .
\end{align}
In other words, we now have a maximal number of linearly independent vectors -- then, $M$ is called the {\it dimension} of the vector space. The $\{\ket{i} \vert i = 1,2,\dots,M\}$ is a complete set of {\it basis vectors}; and such a set of (basis) vectors is said to {\it span} the vector space.

For instance, for the $D$-tuple $\ket{\vec{a}} \equiv (a^1,\dots,a^D)$ from the real vector space of $\mathbb{R}^D$, we may choose
\begin{align}
\ket{1} = (1,0,0,\dots), \qquad 
\ket{2} = (0,1,0,0,\dots), \qquad \nonumber\\
\ket{3} = (0,0,1,0,0,\dots), \qquad \dots \qquad
\ket{D} = (0,0,\dots,0,0,1) .
\end{align}
Then, any arbitrary $\ket{\vec{a}}$ can be written as
\begin{align}
\ket{\vec{a}} = (a^1,\dots,a^D) = \sum_{i=1}^D a^i \ket{i} .
\end{align}
The basis vectors are the $\{\ket{i}\}$ and the dimension is $D$.
\begin{myP}
\qquad Is the space of polynomials of complex coefficients of degree less than or equal to $(n \geq 1)$ a vector space? (Namely, this is the set of polynomials of the form $P_n(x) = c_0 + c_1 x + \dots + c_n x^n$, where the $\{ c_i \vert i = 1,2,\dots,n \}$ are complex numbers.) If so, write down a set of basis vectors. What is its dimension? Answer the same questions for the space of $D \times D$ matrices of complex numbers. \qed
\end{myP}

\subsection{Inner Products}

In Euclidean $D$-space $\mathbb{R}^D$ the ordinary dot product, between the real vectors $\ket{\vec{a}} \equiv (a^1,\dots,a^D)$ and $\vert \vec{b} \rangle \equiv (b^1,\dots,b^D)$, is defined as
\begin{align}
\vec{a}\cdot\vec{b} \equiv \sum_{i=1}^D a^i b^i = \delta_{ij} a^i b^j .
\end{align}
The inner product of linear algebra is again an abstraction of this notion of the dot product, where the analog of $\vec{a} \cdot \vec{b}$ will be denoted as $\langle \vec{a} \vert \vec{b} \rangle$. Like the dot product for Euclidean space, the inner product will allow us to define a notion of the length of vectors and angles between different vectors. 

{\bf Dual/``bra" space} \qquad Given a vector space, an inner product is defined by first introducing a {\it dual space} (aka {\it bra space}) to this vector space. Specifically, given a vector $\ket{\alpha}$ we write its dual as $\bra{\alpha}$. We also introduce the notation
\begin{align}
\ket{\alpha}^\dagger \equiv \bra{\alpha} .
\end{align}
Importantly, for some complex number $c$, the dual of $c \ket{\alpha}$ is
\begin{align}
(c \ket{\alpha})^\dagger \equiv c^* \bra{\alpha} .
\end{align}
Moreover, for complex numbers $a$ and $b$,
\begin{align}
\left(a \ket{\alpha} + b \ket{\beta}\right)^\dagger \equiv a^* \bra{\alpha} + b^* \bra{\beta} .
\end{align}
Since there is a one-to-one correspondence between the vector space and its dual, it is not difficult to see this dual space is indeed a vector space. 

Now, the primary purpose of these dual vectors is that they act on vectors of the original vector space to return a complex number:
\begin{align}
\braket{\alpha}{\beta} \in \mathbb{C} .
\end{align}
{\bf Definition.} \qquad The inner product is now defined by the following properties. For an arbitrary complex number $c$,
{\allowdisplaybreaks\begin{align}
\bra{\alpha}\left(\ket{\beta} + \ket{\gamma}\right) &= \braket{\alpha}{\beta} + \braket{\alpha}{\gamma}  \\
\bra{\alpha}(c \ket{\beta}) &= c \braket{\alpha}{\beta} \\
\braket{\alpha}{\beta}^* = \overline{\braket{\alpha}{\beta}} &= \braket{\beta}{\alpha} \\
\braket{\alpha}{\alpha} &\geq 0
\end{align}}
and 
\begin{align}
\braket{\alpha}{\alpha} = 0
\end{align}
if and only if $\ket{\alpha}$ is the zero vector.

Some words on notation here. Especially in the math literature, the bra-ket notation is not used. There, the inner product is often denoted by $(\alpha,\beta)$, where $\alpha$ and $\beta$ are vectors. Then the defining properties of the inner product would read instead
{\allowdisplaybreaks\begin{align}
(\alpha,\beta+\gamma) &= (\alpha,\beta) + (\alpha,\gamma)  \\
(\alpha,\beta)^* = \overline{(\alpha,\beta)} &= (\beta,\alpha) \\
(\alpha,\alpha) &\geq 0
\end{align}}
and
\begin{align}
(\alpha,\alpha) = 0
\end{align}
if and only if $\alpha$ is the zero vector.
\begin{myP}
\qquad Prove that $\braket{\alpha}{\alpha}$ is a real number. \qed
\end{myP}
The following are examples of inner products.
\begin{itemize}
\item Take the $D$-tuple of complex numbers $\ket{\alpha} \equiv (\alpha^1,\dots,\alpha^D)$ and $\ket{\beta} \equiv (\beta^1,\dots,\beta^D)$; and define the inner product to be
\begin{align}
\label{InnerProduct_ExampleI}
\braket{\alpha}{\beta} \equiv \sum_{i=1}^D (\alpha^i)^* \beta^i = \delta_{ij} (\alpha^i)^* \beta^j = \alpha^\dagger \beta .
\end{align}
\item Consider the space of $D \times D$ complex matrices. Consider two such matrices $X$ and $Y$ and define their inner product to be
\begin{align}
\braket{X}{Y} \equiv \Tr{X^\dagger Y} .
\end{align}
Here, Tr means the matrix trace and $X^\dagger$ is the adjoint of the matrix $X$.
\item Consider the space of polynomials. Suppose $\ket{f}$ and $\ket{g}$ are two such polynomials of the vector space. Then 
\begin{align}
\braket{f}{g} \equiv \int_{-1}^{1} \dd x f(x)^* g(x) 
\end{align}
defines an inner product. Here, $f(x)$ and $g(x)$ indicates the polynomials are expressed in terms of the variable $x$.
\end{itemize}
\begin{myP}
\qquad Prove the above examples are indeed inner products. \qed
\end{myP}
\begin{myP}
\qquad Prove the Schwarz inequality:
\begin{align}
\braket{\alpha}{\alpha} \braket{\beta}{\beta} \geq \left\vert \braket{\alpha}{\beta} \right\vert^2 .
\end{align}
The analogy in Euclidean space is $|\vec{x}|^2 |\vec{y}|^2 \geq |\vec{x} \cdot \vec{y}|^2$. Hint: Start with
\begin{align}
\left(\bra{\alpha} + c^* \bra{\beta}\right) \left(\ket{\alpha} + c \ket{\beta}\right) \geq 0 .
\end{align}
for any complex number $c$. (Why is this true?) Now choose an appropriate $c$ to prove the Schwarz inequality. \qed
\end{myP}
{\bf Orthogonality} \qquad Just as we would say two real vectors in $\mathbb{R}^D$ are perpendicular (aka orthogonal) when their dot product is zero, we may now define two vectors $\ket{\alpha}$ and $\ket{\beta}$ in a vector space to be orthogonal when their inner product is zero:
\begin{align}
\braket{\alpha}{\beta} = 0 = \braket{\beta}{\alpha}.
\end{align}
We also call $\sqrt{\braket{\alpha}{\alpha}}$ the norm of the vector $\ket{\alpha}$; recall, in Euclidean space, the analogous $|\vec{x}| = \sqrt{\vec{x} \cdot \vec{x}}$. Given any vector $\ket{\alpha}$ that is not the zero vector, we can always construct a vector from it that is of unit length,
\begin{align}
\ket{\tilde{\alpha}} \equiv \frac{\ket{\alpha}}{\sqrt{\braket{\alpha}{\alpha}}} \qquad \Rightarrow \qquad \braket{\tilde{\alpha}}{\tilde{\alpha}} = 1.
\end{align}
Suppose we are given a set of basis vectors $\{\ket{i'}\}$ of a vector space. Through what is known as the Gram-Schmidt process, one can always build from them a set of orthonormal basis vectors $\{\ket{i}\}$; where every basis vector has unit norm and is orthogonal to every other basis vector,
\begin{align}
\braket{i}{j} = \delta^i_j .
\end{align}
As you will see, just as vector calculus problems are often easier to analyze when you choose an orthogonal coordinate system, linear algebra problems are often easier to study when you use an orthonormal basis to describe your vector space.
\begin{myP}
\qquad Suppose $\ket{\alpha}$ and $\ket{\beta}$ are linearly dependent -- they are scalar multiples of each other. However, their inner product is zero. What are $\ket{\alpha}$ and $\ket{\beta}$? \qed
\end{myP}
\begin{myP}
\qquad Let $\{\ket{1},\ket{2},\dots,\ket{N}\}$ be a set of $N$ orthonormal vectors. Let $\ket{\alpha}$ be an arbitrary vector lying in the same vector space. Show that the following vector constructed from $\ket{\alpha}$ is orthogonal to all the $\{\ket{i}\}$.
\begin{align}
\label{OrthogonalProjection}
\ket{\widetilde{\alpha}} \equiv \ket{\alpha} - \sum_{j=1}^N \ket{j} \braket{j}{\alpha} .
\end{align} 
This result is key to the following Gram-Schmidt process. \qed
\end{myP}
{\bf Gram-Schmidt} \qquad Let $\{ \ket{\alpha_1}, \ket{\alpha_2}, \dots, \ket{\alpha_D} \}$ be a set of $D$ linearly independent vectors that spans some vector space. The Gram-Schmidt process is an iterative algorithm, based on the observation in eq. \eqref{OrthogonalProjection}, to generate from it a set of orthonormal set of basis vectors. 
\begin{enumerate}
\item Take the first vector $\ket{\alpha_1}$ and normalize it to unit length:
\begin{align}
\ket{\widetilde{\alpha}_1} = \frac{\ket{\alpha_1}}{\sqrt{\braket{\alpha_1}{\alpha_1}}}.
\end{align}
\item Take the second vector $\ket{\alpha_2}$ and project out $\ket{\widetilde{\alpha}_1}$:
\begin{align}
\ket{\alpha'_2} \equiv \ket{\alpha_2} - \ket{\widetilde{\alpha}_1}\braket{\widetilde{\alpha}_1}{\alpha_2},
\end{align}
and normalize it to unit length
\begin{align}
\ket{\widetilde{\alpha}_2} \equiv \frac{\ket{\alpha'_2}}{\sqrt{\braket{\alpha'_2}{\alpha'_2}}} .
\end{align}
\item Take the third vector $\ket{\alpha_3}$ and project out $\ket{\widetilde{\alpha}_1}$ and $\ket{\widetilde{\alpha}_2}$:
\begin{align}
\ket{\alpha'_3} \equiv \ket{\alpha_3} - \ket{\widetilde{\alpha}_1}\braket{\widetilde{\alpha}_1}{\alpha_3} - \ket{\widetilde{\alpha}_2}\braket{\widetilde{\alpha}_2}{\alpha_3},
\end{align}
then normalize it to unit length
\begin{align}
\ket{\widetilde{\alpha}_3} \equiv \frac{\ket{\alpha'_3}}{\sqrt{\braket{\alpha'_3}{\alpha'_3}}} .
\end{align}
\item Repeat \dots Take the $i$th vector $\ket{\alpha_i}$ and project out $\ket{\widetilde{\alpha}_1}$ through $\ket{\widetilde{\alpha}_{i-1}}$:
\begin{align}
\ket{\alpha'_i} \equiv \ket{\alpha_i} - \sum_{j=1}^{i-1} \ket{\widetilde{\alpha}_j}\braket{\widetilde{\alpha}_j}{\alpha_i} ,
\end{align}
then normalize it to unit length
\begin{align}
\ket{\widetilde{\alpha}_i} \equiv \frac{\ket{\alpha'_i}}{\sqrt{\braket{\alpha'_i}{\alpha'_i}}} .
\end{align}
\end{enumerate}
By construction, $\ket{\widetilde{\alpha}_i}$ will be orthogonal to $\ket{\widetilde{\alpha}_1}$ through $\ket{\widetilde{\alpha}_{i-1}}$. Therefore, at the end of the process, we will have $D$ mutually orthogonal and unit norm vectors. Because they are orthogonal they are linearly independent -- hence, we have succeeded in constructing an orthonormal set of basis vectors.

{\it Example} \qquad Here is a simple example in 3D Euclidean space endowed with the usual dot product. Let us have
\begin{align}
\ket{\alpha_1} \dot{=} (2,0,0), \qquad 
\ket{\alpha_2} \dot{=} (1,1,1), \qquad 
\ket{\alpha_3} \dot{=} (1,0,1) .
\end{align}
You can check that these vectors are linearly independent by taking the determinant of the $3 \times 3$ matrix formed from them. Alternatively, the fact that they generate a set of basis vectors from the Gram-Schmidt process also implies they are linearly independent.

Normalizing $\ket{\alpha_1}$ to unity,
\begin{align}
\ket{\widetilde{\alpha}_1} 
= \frac{\ket{\alpha_1}}{\sqrt{\braket{\alpha_1}{\alpha_1}}} 
= \frac{(2,0,0)}{2} = (1,0,0) .
\end{align}
Next we project out $\ket{\widetilde{\alpha}_1}$ from $\ket{\alpha_2}$.
\begin{align}
\ket{\alpha'_2} = \ket{\alpha_2} - \ket{\widetilde{\alpha}_1} \braket{\widetilde{\alpha}_1}{\alpha_2} 
= (1,1,1) - (1,0,0)  (1+0+0) = (0,1,1) .
\end{align}
Then we normalize it to unit length.
\begin{align}
\ket{\widetilde{\alpha}_2} = \frac{\ket{\alpha'_2}}{\sqrt{\braket{\alpha'_2}{\alpha'_2}}} = \frac{(0,1,1)}{\sqrt{2}} .
\end{align}
Next we project out $\ket{\widetilde{\alpha}_1}$ and $\ket{\widetilde{\alpha}_2}$ from $\ket{\alpha_3}$.
\begin{align}
\ket{\alpha'_3} 
&= \ket{\alpha_3} 
		- \ket{\widetilde{\alpha}_1} \braket{\widetilde{\alpha}_1}{\alpha_3} 
		- \ket{\widetilde{\alpha}_2} \braket{\widetilde{\alpha}_2}{\alpha_3} \nonumber\\
&= (1,0,1) - (1,0,0)  (1+0+0) - \frac{(0,1,1)}{\sqrt{2}} \frac{0+0+1}{\sqrt{2}} \nonumber\\
&= (1,0,1) - (1,0,0) - \frac{(0,1,1)}{2} = \left(0,-\frac{1}{2},\frac{1}{2}\right).
\end{align}
Then we normalize it to unit length.
\begin{align}
\ket{\widetilde{\alpha}_3}
= \frac{\ket{\alpha'_3}}{\sqrt{\braket{\alpha'_3}{\alpha'_3}}} 
= \frac{\left(0,-1,1\right)}{\sqrt{2}} .
\end{align}
You can check that
\begin{align}
\ket{\widetilde{\alpha}_1} = (1,0,0), \qquad
\ket{\widetilde{\alpha}_2} = \frac{(0,1,1)}{\sqrt{2}}, \qquad
\ket{\widetilde{\alpha}_3} = \frac{\left(0,-1,1\right)}{\sqrt{2}} ,
\end{align}
are mutually perpendicular and of unit length.
\begin{myP}
\qquad Consider the space of polynomials with complex coefficients. Let the inner product be
\begin{align}
\braket{f}{g} \equiv \int_{-1}^{+1} \dd x f(x)^* g(x) .
\end{align}
Starting from the set $\{ \ket{0} = 1, \ket{1} = x, \ket{2} = x^2 \}$, construct from them a set of orthonormal basis vectors spanning the subspace of polynomials of degree equal to or less than $2$. Compare your results with the Legendre polynomials
\begin{align}
P_\ell(x) \equiv \frac{1}{2^\ell \ell !} \frac{\dd^\ell}{\dd x^\ell} \left(x^2 - 1\right)^\ell, \qquad \ell = 0,1,2.
\end{align} \qed
\end{myP}
{\bf Orthogonality and Linear independence.} \qquad We close this subsection with an observation. If a set of non-zero kets $\{\ket{i} \vert i = 1,2,\dots,N-1,N\}$ are orthogonal, then they are necessarily linearly independent. This can be proved readily by contradiction. Suppose these kets were linearly dependent. Then it must be possible to find non-zero complex numbers $\{C^i\}$ such that
\begin{align}
\sum_{i=1}^{N} C^i \ket{i} = 0 .
\end{align}
If we now act $\bra{j}$ on this equation, for any $j \in \{ 1,2,3,\dots,N \}$,
\begin{align}
\sum_{i=1}^{N} C^i \braket{j}{i} = \sum_{i=1}^{N} C^i \delta_{ij} \braket{j}{j} = C^j \braket{j}{j} = 0 .
\end{align} 
That means all the $\{C^j \vert j=1,2,\dots,N\}$ are in fact zero. 

A simple application of this observation is, if you have found $D$ mutually orthogonal kets $\{\ket{i}\}$ in a $D$ dimensional vector space, then these kets form a basis. By normalizing them to unit length, you'd have obtained an orthonormal basis. Such an example is that of the Pauli matrices $\{\sigma^\mu \vert \mu = 0,1,2,3 \}$ in eq. \eqref{PauliMatrices}. The vector space of $2 \times 2$ complex matrices is 4-dimensional, since there are 4 independent components. Moreover, we have already seen that the trace $\Tr{X^\dagger Y}$ is one way to define an inner product of matrices $X$ and $Y$. Since
\begin{align}
\frac{1}{2}\Tr{\left(\sigma^\mu\right)^\dagger \sigma^\nu}
= \frac{1}{2}\Tr{\sigma^\mu \sigma^\nu}
= \delta^{\mu\nu}, \qquad \mu,\nu \in \{ 0,1,2,3 \} ,
\end{align}
that means, by the argument just given, the 4 Pauli matrices $\{\sigma^\mu\}$ form an orthogonal set of basis vectors for the vector space of complex $2 \times 2$ matrices. That means it must be possible to choose $\{p_\mu\}$ such that the superposition $p_\mu \sigma^\mu$ is equal to any given $2 \times 2$ complex matrix $A$. In fact,
\begin{align}
p_\mu \sigma^\mu = A, \qquad \Leftrightarrow \qquad p_\mu = \frac{1}{2} \Tr{\sigma^\mu A} .
\end{align}

\subsection{Linear Operators}

\subsubsection{Definitions and Fundamental Concepts}

In quantum theory, a physical observable is associated with a (Hermitian) linear operator acting on the vector space. What defines a linear operator? Let $A$ be one. Firstly, when it acts from the left on a vector, it returns another vector
\begin{align}
A \ket{\alpha} = \ket{\alpha'} .
\end{align}
In other words, if you can tell me what you want the ``output" $\ket{\alpha'}$ to be, after $A$ acts on any vector of the vector space $\ket{\alpha}$ -- you'd have defined $A$ itself. But that's not all -- linearity also means, for otherwise arbitrary operators $A$ and $B$ and complex numbers $c$ and $d$,
\begin{align}
\label{LinearOperators_Properties}
(A+B) \ket{\alpha} &= A\ket{\alpha} + B\ket{\alpha} \\
A (c \ket{\alpha} + d \ket{\beta}) &= c \ A\ket{\alpha} + d \ A\ket{\beta} . \nonumber
\end{align}
An operator always acts on a bra from the right, and returns another bra,
\begin{align}
\bra{\alpha} A = \bra{\alpha'} .
\end{align}
{\bf Adjoint} \qquad We denote the adjoint of the linear operator $X$, by taking the $^\dagger$ of the ket $X \ket{\alpha}$ in the following way:
\begin{align}
(X \ket{\alpha})^\dagger = \bra{\alpha} X^\dagger. 
\end{align}
{\bf Multiplication} \qquad If $X$ and $Y$ are both linear operators, since $Y \ket{\alpha}$ is a vector, we can apply $X$ to it to obtain another vector, $X(Y \ket{\alpha})$. This means we ought to be able to multiply operators, for e.g., $XY$. We will assume this multiplication is associative, namely
\begin{align}
XYZ = (XY)Z = X(YZ).
\end{align}
\begin{myP}
\qquad By considering the adjoint of $XY \ket{\alpha}$, where $X$ and $Y$ are arbitrary linear operators and $\ket{\alpha}$ is an arbitrary vector, prove that
\begin{align}
(XY)^\dagger = Y^\dagger X^\dagger .
\end{align} 
Hint: take the adjoint of $(XY) \ket{\alpha}$ and $X(Y \ket{\alpha})$. \qed
\end{myP}
{\bf Eigenvectors and eigenvalues} \qquad An eigenvector of some linear operator $A$ is a vector that, when acted upon by $A$, returns the vector itself multiplied by a complex number $a$:
\begin{align}
X \ket{a} = a \ket{a} .
\end{align}
This number $a$ is called the eigenvalue of $A$.

{\bf Ket-bra operator} \qquad Notice that the product $\ket{\alpha} \bra{\beta}$ can be considered a linear operator. To see this, we apply it on some arbitrary vector $\ket{\gamma}$ and observe it returns the vector $\ket{\alpha}$ multiplied by a complex number describing the projection of $\ket{\gamma}$ on $\ket{\beta}$,
\begin{align}
\left(\ket{\alpha} \bra{\beta}\right) \ket{\gamma} = \ket{\alpha} (\braket{\beta}{\gamma}) = (\braket{\beta}{\gamma}) \cdot \ket{\alpha} ,
\end{align}
as long as we assume these products are associative. It obeys the following ``linearity" rules. If $\ket{\alpha} \bra{\beta}$ and $\ket{\alpha'} \bra{\beta'}$ are two different ket-bra operators,
\begin{align}
(\ket{\alpha} \bra{\beta}+\ket{\alpha'} \bra{\beta'}) \ket{\gamma} &= \ket{\alpha} \braket{\beta}{\gamma} + \ket{\alpha'} \braket{\beta'}{\gamma} ;
\end{align}
and for complex numbers $c$ and $d$,
\begin{align}
\ket{\alpha} \bra{\beta} \left( c \ket{\gamma} + d \ket{\gamma'} \right) 
&= c \ \ket{\alpha} \braket{\beta}{\gamma} + d \ \ket{\alpha} \braket{\beta}{\gamma'} .
\end{align}
\begin{myP}
\qquad Show that
\begin{align}
\left( \ket{\alpha} \bra{\beta}\right)^\dagger = \ket{\beta} \bra{\alpha} .
\end{align}
Hint: Act $\ket{\alpha} \bra{\beta}$ on an arbitrary vector, and then take its adjoint. \qed
\end{myP}
{\bf Projection operator} \qquad The special case $\ket{\alpha} \bra{\alpha}$ acting on any vector $\ket{\gamma}$ will return $\ket{\alpha} \braket{\alpha}{\gamma}$. Thus, we can view it as a projection operator -- it takes an arbitrary vector and extracts the portion of it ``parallel" to $\ket{\alpha}$.

{\bf Identity} \qquad The identity operator obeys
\begin{align}
\mathbb{I} \ket{\gamma} = \ket{\gamma} .
\end{align}
{\bf Inverse} \qquad The inverse of the operator $X$ is still defined as one that obeys
\begin{align}
X^{-1} X = X X^{-1} = \mathbb{I} .
\end{align}
Strictly speaking, we need to distinguish between the left and right inverse, but in finite dimensional vector spaces, they are the same object.

{\bf Superposition, the identity operator, and vector components} \qquad We will now see that (square) matrices can be viewed as representations of linear operators on a vector space. Let $\{\ket{i}\}$ denote the basis orthonormal vectors of the vector space,
\begin{align}
\braket{i}{j} = \delta^i_j .
\end{align}
Then we may consider acting an linear operator $X$ on some arbitrary vector $\ket{\gamma}$, which we will express as a linear combination of the $\{\ket{i}\}$:
\begin{align}
\ket{\gamma} = \sum_i \widehat{\gamma}^i \ket{i}, \qquad\qquad \{ \widehat{\gamma}^i \in \mathbb{C} \} .
\end{align}
By acting both sides with respect to $\bra{j}$, we have
\begin{align}
\braket{j}{\gamma} = \widehat{\gamma}^j .
\end{align}
In other words,
\begin{align}
\label{VectorComponents}
\ket{\gamma} = \sum_i \ket{i} \braket{i}{\gamma} .
\end{align}
Since $\ket{\gamma}$ was arbitrary, we have identified the identity operator as
\begin{align}
\label{CompletenessRelation}
\mathbb{I} = \sum_i \ket{i} \bra{i} .
\end{align}
This is also often known as a completeness relation: summing over the ket-bra operators built out of the orthonormal basis vectors of a vector space returns the unit (aka identity) operator. $\mathbb{I}$ acting on any vector yields the same vector.

Once a set of orthonormal basis vectors are chosen, notice from the expansion in eq. \eqref{VectorComponents}, that to specify a vector $\ket{\gamma}$ all we need to do is to specify the complex numbers $\{ \braket{i}{\gamma} \}$. These can be arranged as a column vector; if the dimension of the vector space is $D$, then
\begin{align}
\ket{\gamma} \dot{=} 
\left[\begin{array}{c}
\braket{1}{\gamma} \\ \braket{2}{\gamma} \\ \braket{3}{\gamma} \\ \dots \\ \braket{D}{\gamma} 
\end{array}\right] .
\end{align}
The $\dot{=}$ is not quite an equality; rather it means ``represented by," in that this column vector contains as much information as eq. \eqref{VectorComponents}, provided the orthonormal basis vectors are known.

We may also express an arbitrary bra through a superposition of the basis bras $\{\bra{i}\}$, using eq. \eqref{CompletenessRelation}.
\begin{align}
\bra{\alpha} = \sum_i \braket{\alpha}{i} \bra{i} .
\end{align}
{\bf Matrix elements} \qquad Consider now some operator $X$ acting on an arbitrary vector $\ket{\gamma}$, expressed through the orthonormal basis vectors $\{\ket{i}\}$.
\begin{align}
X \ket{\gamma} = \sum_i X \ket{i} \braket{i}{\gamma} .
\end{align}
We can insert an identity operator from the left,
\begin{align}
\label{LinearOperator_ModeExpansion}
X \ket{\gamma} = \sum_{i,j} \ket{j} \braOket{j}{X}{i} \braket{i}{\gamma} .
\end{align}
We can also apply the $l$th basis bra $\bra{l}$ from the left on both sides and obtain
\begin{align}
\bra{l} X \ket{\gamma} = \sum_i \braOket{l}{X}{i} \braket{i}{\gamma} .
\end{align}
Just as we read off the components of the vector in eq. \eqref{VectorComponents} as a column vector, we can do the same here. Again supposing a $D$ dimensional vector space (for notational convenience),
\begin{align}
\label{LinearOperator_MatrixRepresentation}
X \ket{\gamma} \dot{=} 
\left[
\begin{array}{ccccc}
\braOket{1}{X}{1}	& \braOket{1}{X}{2}		& \dots		& \braOket{1}{X}{D}   	\\
\braOket{2}{X}{1}	& \braOket{2}{X}{2}		& \dots		& \braOket{2}{X}{D}   	\\
\dots				& \dots					& \dots		& \dots					\\
\braOket{D}{X}{1}	& \braOket{D}{X}{2}		& \dots		& \braOket{D}{X}{D}
\end{array}
\right]
\left[\begin{array}{c}
\braket{1}{\gamma} \\ \braket{2}{\gamma} \\ \braket{3}{\gamma} \\ \dots \\ \braket{D}{\gamma} 
\end{array}\right] .
\end{align}
In words: $X$ acting on some vector $\ket{\gamma}$ can be represented by the column vector gotten from acting the matrix $\braOket{j}{X}{i}$, with row number $j$ and column number $i$, acting on the column vector $\braket{i}{\gamma}$. In index notation, with\footnote{In this chapter on the abstract formulation of Linear Algebra, I use a $\widehat{\cdot}$ to denote a matrix (representation), in order to distinguish it from the linear operator itself.}
\begin{align}
\widehat{X}^i_{\phantom{i}j} \equiv \braOket{i}{X}{j} \qquad \text{and} \qquad \gamma^j \equiv \braket{j}{\gamma},
\end{align}
we have
\begin{align}
\bra{i} X \ket{\gamma} = \widehat{X}^i_{\phantom{i}j} \gamma^j .
\end{align}
Since $\ket{\gamma}$ in eq. \eqref{LinearOperator_ModeExpansion} was arbitrary, we may record that any linear operator $X$ admits an ket-bra operator expansion:
\begin{align}
\label{LinearOperator_KetBraExpansion}
X = \sum_{i,j} \ket{j} \braOket{j}{X}{i} \bra{i} = \sum_{i,j} \ket{j} \widehat{X}^j_{\phantom{j}i} \bra{i} .
\end{align}
Equivalently, this result follows from inserting the completeness relation in eq. \eqref{CompletenessRelation} on the left and right of $X$. We see that specifying the matrix $\widehat{X}^j_{\phantom{j}i}$ amounts to defining the linear operator $X$ itself. 

{\it Vector Space of Linear Operators} \qquad You may step through the axioms of Linear Algebra to verify that the space of Linear operators is, in fact, a vector space itself. Given an orthonormal basis $\{ \ket{i} \}$ for the original vector space upon which these linear operators are acting, we see that the expansion in eq. \eqref{LinearOperator_KetBraExpansion} -- which holds for an arbitrary linear operator $X$ -- teaches us the set of ket-bra operators
\begin{align}
\left\{ \ketbra{j}{i} \left\vert j,i = 1,2,3,\dots,D \right.\right\}
\end{align}
form the basis of the space of linear operators. The matrix elements $\braOket{j}{X}{i} = \widehat{X}^j_{\phantom{j}i}$ are the expansion coefficients. 

{\it Example} \qquad What is the matrix representation of $\ket{\beta} \bra{\alpha}$? We apply $\bra{i}$ from the left and $\ket{j}$ from the right to obtain the $ij$ component
\begin{align}
\bra{i}\left( \ketbra{\alpha}{\beta} \right)\ket{j} = \braket{i}{\alpha} \braket{\beta}{j} \dot{=} \alpha^i \overline{\beta^j} .
\end{align}
{\it Products of operators} \qquad We can consider $YX$, where $X$ and $Y$ are linear operators. By inserting the completeness relation in eq. \eqref{CompletenessRelation},
\begin{align}
Y X \ket{\gamma} 
&= \sum_{i,j,k} \ket{k} \bra{k} Y \ket{j} \braOket{j}{X}{i} \braket{i}{\gamma} \nonumber\\
&= \sum_{k} \ket{k} \widehat{Y}^k_{\phantom{k}j} \widehat{X}^j_{\phantom{j}i} \gamma^i .
\end{align}
The product $YX$ can therefore be represented as
\begin{align}
\label{LinearOperator_MatrixRepresentation_YX}
YX \dot{=} \left[
\begin{array}{ccccc}
\braOket{1}{Y}{1}	& \braOket{1}{Y}{2}		& \dots		& \braOket{1}{Y}{D}   	\\
\braOket{2}{Y}{1}	& \braOket{2}{Y}{2}		& \dots		& \braOket{2}{Y}{D}   	\\
\dots				& \dots					& \dots		& \dots					\\
\braOket{D}{Y}{1}	& \braOket{D}{Y}{2}		& \dots		& \braOket{D}{Y}{D}
\end{array}
\right]
\left[
\begin{array}{ccccc}
\braOket{1}{X}{1}	& \braOket{1}{X}{2}		& \dots		& \braOket{1}{X}{D}   	\\
\braOket{2}{X}{1}	& \braOket{2}{X}{2}		& \dots		& \braOket{2}{X}{D}   	\\
\dots				& \dots					& \dots		& \dots					\\
\braOket{D}{X}{1}	& \braOket{D}{X}{2}		& \dots		& \braOket{D}{X}{D}
\end{array}
\right] .
\end{align}
Notice how the rules of matrix multiplication emerges from this abstract formulation of linear operators acting on a vector space.

{\it Inner product of two kets} \qquad In an orthonormal basis, the inner product of $\ket{\alpha}$ and $\ket{\beta}$ can be written as a complex ``dot product" because we may insert the completeness relation in eq. \eqref{CompletenessRelation},
\begin{align}
\braket{\alpha}{\beta} = \braOket{\alpha}{\mathbb{I}}{\beta} 
= \sum_i \braket{\alpha}{i} \braket{i}{\beta} \dot{=} \delta_{ij} \overline{\alpha^i} \beta^j = \alpha^\dagger \beta .
\end{align}
This means if $\braket{i}{\beta}$ is the column vector representing $\ket{\beta}$ in a given orthonormal basis; then $\braket{\alpha}{i}$, the adjoint of the column $\braket{i}{\alpha}$ representing $\ket{\alpha}$, should be viewed as a row vector.

Furthermore, if $\ket{\gamma}$ has unit norm, then
\begin{align}
1 = \braket{\gamma}{\gamma} 
= \sum_i \braket{\gamma}{i} \braket{i}{\gamma} 
= \sum_i \left\vert \braket{i}{\gamma} \right\vert^2 
\dot{=} \delta_{ij} \overline{\gamma^i} \gamma^j = \gamma^\dagger \gamma .
\end{align}
{\it Adjoint} \qquad Through the associativity of products, we also see that, for any states $\ket{\alpha}$ and $\ket{\beta}$; and for any linear operator $X$,
\begin{align}
\label{Adjoint_MatrixRep}
\overline{\bra{\alpha} X \ket{\beta}} = \overline{\bra{\alpha} (X \ket{\beta})}
= ((X \ket{\beta}))^\dagger\ket{\alpha}
= (\bra{\beta} X^\dagger) \ket{\alpha} = \bra{\beta} X^\dagger \ket{\alpha}
\end{align}
If we take matrix elements of $X$ with respect to an orthonormal basis $\{ \ket{i} \}$, we recover our previous (matrix algebra) definition of the adjoint:
\begin{align}
\braOket{j}{X^\dagger}{i} = \braOket{i}{X}{j}^* .
\end{align}
{\bf Mapping finite dimensional vector spaces to $\mathbb{C}^D$} \qquad We summarize our preceding discussion. Even though it is possible to discuss finite dimensional vector spaces in the abstract, it is always possible to translate the setup at hand to one of the $D$-tuple of complex numbers, where $D$ is the dimensionality. First choose a set of orthonormal basis vectors $\{\ket{1},\dots,\ket{D}\}$. Then, every vector $\ket{\alpha}$ can be represented as a column vector; the $i$th component is the result of projecting the abstract vector on the $i$th basis vector $\braket{i}{\alpha}$; conversely, writing a column of complex numbers can be interpreted to define a vector in this orthonormal basis. The inner product between two vectors $\braket{\alpha}{\beta} = \sum_i \braket{\alpha}{i} \braket{i}{\beta}$ boils down to the complex conjugate of the $\braket{i}{\alpha}$ column vector dotted into the $\braket{i}{\beta}$ vector. Moreover, every linear operator $O$ can be represented as a matrix with the element on the $i$th row and $j$th column given by $\braOket{i}{O}{j}$; and conversely, writing any square matrix $\widehat{O}^i_{\phantom{i}j}$ can be interpreted to define a linear operator, on this vector space, with matrix elements $\braOket{i}{O}{j}$. Product of linear operators becomes products of matrices, with the usual rules of matrix multiplication.
\begin{center}
\begin{tabular}{| l | l |}
\hline
Object 																					& Representation \\ \hline
Vector/Ket:	$\ket{\alpha} = \sum_i \ket{i} \braket{i}{\alpha}$							
													& $\alpha^i = (\braket{1}{\alpha},\dots,\braket{D}{\alpha})^T$ \\
Dual Vector/Bra:	$\bra{\alpha} = \sum_i \braket{\alpha}{i} \bra{i}$							
													& $(\alpha^\dagger)^i = (\braket{\alpha}{1},\dots,\braket{\alpha}{D})$ \\
Inner product: $\braket{\alpha}{\beta} = \sum_i \braket{\alpha}{i} \braket{i}{\beta}$
													& $\alpha^\dagger \beta = \delta_{ij} \overline{\alpha^i} \beta^j$ \\
Linear operator (LO): $X = \sum_{i,j} \ket{i} \braOket{i}{X}{j} \bra{j}$		
													& $\widehat{X}^i_{\phantom{i}j} = \braOket{i}{X}{j}$ \\
LO acting on ket: $X \ket{\gamma} = \sum_{i,j} \ket{i} \braOket{i}{X}{j} \braket{j}{\gamma}$		
													& $(\widehat{X}\gamma)^i = \widehat{X}^i_{\phantom{i}j} \gamma^j$\\
Products of LOs: $XY = \sum_{i,j,k} \ket{i} \braOket{i}{X}{j} \braOket{j}{Y}{k} \bra{k}$		
													& $(\widehat{XY})^i_{\phantom{i}k} = \widehat{X}^i_{\phantom{i}j} \widehat{Y}^j_{\phantom{j}k}$ \\
Adjoint of LO: $X^\dagger = \sum_{i,j} \ket{j} \overline{\braOket{i}{X}{j}} \bra{i}$
													& $(\widehat{X}^\dagger)^j_{\phantom{j}i} = \overline{\braOket{i}{X}{j}} = \overline{(\widehat{X}^T)^j_{\phantom{j}i}}$ \\
\hline
\end{tabular}
\end{center}
Next we highlight two special types of linear operators.

\subsubsection{Hermitian Operators}

A hermitian linear operator $X$ is one that is equal to its own adjoint, namely
\begin{align}
X^\dagger = X. 
\end{align}
From eq. \eqref{Adjoint_MatrixRep}, we see that a linear operator $X$ is hermitian if and only if
\begin{align}
\bra{\alpha} X \ket{\beta} = \bra{\beta} X \ket{\alpha}^* 
\end{align}
for arbitrary vectors $\ket{\alpha}$ and $\ket{\beta}$. In particular, if $\{ \ket{i} \vert i = 1,2,3,\dots,D \}$ form an orthonormal basis, we recover the definition of a Hermitian matrix,
\begin{align}
\bra{j} X \ket{i} = \bra{i} X \ket{j}^* .
\end{align}
We now turn to the following important facts about Hermitian operators.
\begin{quotation}
{\bf Hermitian Operators Have Real Spectra:} \qquad If $X$ is a Hermitian operator, all its eigenvalues are real and eigenvectors corresponding to different eigenvalues are orthogonal.
\end{quotation}
{\it Proof} \qquad Let $\ket{a}$ and $\ket{a'}$ be eigenvectors of $X$, i.e.,
\begin{align}
\label{HermitianOperator_RealSpectrun_I}
X \ket{a} = a \ket{a}
\end{align} 
Taking the adjoint of the analogous equation for $\ket{a'}$, and using $X = X^\dagger$,
\begin{align}
\label{HermitianOperator_RealSpectrun_II}
\bra{a'} X = a'^* \bra{a'} .
\end{align} 
We can multiply $\bra{a'}$ from the left on both sides of eq. \eqref{HermitianOperator_RealSpectrun_I}; and multiply $\ket{a}$ from the right on both sides of eq. \eqref{HermitianOperator_RealSpectrun_II}.
\begin{align}
\bra{a'} X \ket{a} = a \braket{a'}{a}, \qquad \bra{a'} X \ket{a} = a'^* \braket{a'}{a} 
\end{align}
Subtracting these two equations,
\begin{align}
0 = (a-a'^*) \braket{a'}{a} .
\end{align}
Suppose the eigenvalues are the same, $a = a'$. Then $0 = (a-a^*) \braket{a}{a}$; because $\ket{a}$ is not a null vector, this means $a = a^*$; eigenvalues of Hermitian operators are real. Suppose instead the eigenvalues are distinct, $a \neq a'$. Because we have just proven that $a'$ can be assumed to be real, we have $0=(a-a')\braket{a'}{a}$. By assumption the factor $a-a'$ is not zero. Therefore $\braket{a'}{a} = 0$, namely, eigenvectors corresponding to different eigenvalues of a Hermitian operator are orthogonal.
\begin{quotation}
{\bf Completeness of Hermitian Eigensystem:} \qquad The eigenkets $\{ \ket{\lambda_k} \vert k=1,2,\dots,D \}$ of a Hermitian operator span the vector space upon which it is acting. The full set of eigenvalues $\{ \lambda_k \vert k=1,2,\dots,D \}$ of some Hermitian operator is called its {\it spectrum}; and from eq. \eqref{CompletenessRelation}, completeness of its eigenvectors reads
\begin{align}
\label{HermitianOperator_Completeness}
\mathbb{I} = \sum_{k=1}^{D} \ketbra{\lambda_k}{\lambda_k} .
\end{align}
In the language of matrix algebra, we'd say that a Hermitian matrix is always diagonalizable via a unitary transformation.
\end{quotation}
In quantum theory, we postulate that observables such as spin, position, momentum, etc., correspond to Hermitian operators; their eigenvalues are then the possible outcomes of the measurements of these observables. This is because their spectrum are real, which guarantees we get a real number from performing a measurement on the system at hand.

{\bf Degeneracy} \qquad If more than one eigenket of $A$ has the same eigenvalue, we say $A$'s spectrum is degenerate. The simplest example is the identity operator itself: every basis vector is an eigenvector with eigenvalue $1$. 

When an operator is degenerate, the labeling of eigenkets using their eigenvalues become ambiguous -- which eigenket does $\ket{\lambda}$ correspond to, if this subspace is 5 dimensional, say? What often happens is that one can find a different observable $B$ to distinguish between the eigenkets of the same $\lambda$. For example, we will see below that the negative Laplacian on the $2$-sphere -- known as the ``square of total angular momentum," when applied to quantum mechanics -- will have eigenvalues $\ell(\ell+1)$, where $\ell \in \{ 0,1,2,3,\dots\}$. It will also turn out to be $(2\ell+1)$-fold degenerate, but this degeneracy can be labeled by an integer $m$, corresponding to the eigenvalues of the generator-of-rotation about the North pole $J(\phi)$ (where $\phi$ is the azimuthal angle). A closely related fact is that $[-\vec{\nabla}_{\mathbb{S}^2}^2, J(\phi)] = 0$, where $[X,Y] \equiv XY-YX$.
\begin{align}
&-\vec{\nabla}_{\mathbb{S}^2}^2 \ket{\ell,m} = \ell(\ell+1) \ket{\ell,m}, \\
\ell &\in \{ 0,1,2,\dots \}, \qquad m \in \{ -\ell,-\ell+1,\dots,-1,0,1,\dots,\ell-1,\ell \} . \nonumber
\end{align}
It's worthwhile to mention, in the context of quantum theory -- degeneracy in the spectrum is often associated with the presence of symmetry. For example, the Stark and Zeeman effects can be respectively thought of as the breaking of rotational symmetry of an atomic system by a non-zero magnetic and electric field. Previously degenerate spectral lines become split into distinct ones, due to these $\vec{E}$ and $\vec{B}$ fields.\footnote{See Wikipedia articles on the \href{https://en.wikipedia.org/wiki/Stark_effect}{Stark} and \href{https://en.wikipedia.org/wiki/Zeeman_effect}{Zeeman} effects for plots of the energy levels vs. electric/magnetic field strengths.} In the context of classical field theory, we will witness in the section on continuous vector spaces below, how the translation invariance of space leads to a degenerate spectrum of the Laplacian.
\begin{myP}
\qquad Let $X$ be a linear operator with eigenvalues $\{\lambda_i \vert i = 1,2,3,\dots,D\}$ and orthonormal eigenvectors $\{ \ket{\lambda_i} \vert i = 1,2,3,\dots,D \}$ that span the given vector space. Show that $X$ can be expressed as
\begin{align}
\label{LinearOperator_DiagonalModeExpansion}
X = \sum_i \lambda_i \ket{\lambda_i} \bra{\lambda_i} .
\end{align}
(Assume a non-degenerate spectra for now.) Verify that the right hand side is represented by a diagonal matrix in this basis $\{ \ket{\lambda_i} \}$. Of course, a Hermitian linear operator is a special case of eq. \eqref{LinearOperator_DiagonalModeExpansion}, where all the $\{\lambda_i\}$ are real. Hint: Given that the eigenkets of $X$ span the vector space, all you need to verify is that all possible matrix elements of $X$ return what you expect. \qed
\end{myP}
{\bf How to diagonalize a Hermitian operator?} \qquad Suppose you are given a Hermitian operator $H$ in some orthonormal basis $\{\ket{i}\}$, namely  
\begin{align}
H = \sum_{i,j} \ket{i} \widehat{H}^i_{\phantom{i}j} \bra{j} .
\end{align}
How does one go about diagonalizing it? Here is where the matrix algebra you are familiar with comes in. By treating $\widehat{H}^i_{\phantom{i}j}$ as a matrix, you can find its eigenvectors and eigenvalues $\{ \lambda_k \}$. Specifically, what you are solving for is the unitary matrix $\widehat{U}^j_{\phantom{j}k}$, whose $k$th column is the $k$th unit length eigenvector of $\widehat{H}^i_{\phantom{i}j}$, with eigenvalue $\lambda_k$:
\begin{align}
\label{HermitianOperator_EigenvalueEquation}
\widehat{H}^i_{\phantom{i}j} \widehat{U}^j_{\phantom{j}k} = \lambda_k \widehat{U}^j_{\phantom{j}k} \qquad \Leftrightarrow \qquad
\sum_j \braOket{i}{H}{j} \braket{j}{\lambda_k} = \lambda_k \braket{i}{\lambda_k} ,
\end{align}
with
\begin{align}
\braOket{i}{H}{j} \equiv \widehat{H}^i_{\phantom{i}j} 
\qquad\qquad
\text{and} 
\qquad\qquad
\braket{j}{\lambda_k} \equiv \widehat{U}^j_{\phantom{j}k} .
\end{align}
Once you have the explicit solutions for $(\braket{1}{\lambda_k}, \braket{2}{\lambda_k}, \dots, \braket{D}{\lambda_k})^T$, you can then write the eigenket itself as
\begin{align}
\label{LinearOperator_EigenketExpansion}
\ket{\lambda_k} = \sum_i \ket{i} \braket{i}{\lambda_k} = \sum_{i} \ket{i} \widehat{U}^i_{\phantom{j}k} .
\end{align}
The operator $H$ has now been diagonalized as
\begin{align}
H = \sum_{k} \lambda_k \ketbra{\lambda_k}{\lambda_k}
\end{align}
because eq. \eqref{LinearOperator_EigenketExpansion} says
\begin{align}
H = \sum_{k} \lambda_k \ketbra{\lambda_k}{\lambda_k}
&= \sum_{k} \lambda_k \sum_{i,j} \ket{i} \braket{i}{\lambda_k} \overline{\braket{j}{\lambda_k}} \bra{j} \\
&= \sum_{i,j} \ket{i} \left(\sum_{k} \lambda_k \braket{i}{\lambda_k} \overline{\braket{j}{\lambda_k}} \right) \bra{j} .
\end{align}
We may multiply $\widehat{U}^\dagger$ on both sides of eq. \eqref{HermitianOperator_EigenvalueEquation} and remember $\braket{j}{\lambda_k} \equiv \widehat{U}^j_{\phantom{j}k}$ and $(\widehat{U}^\dagger)^{i}_{\phantom{i}j} = \overline{\widehat{U}^{j}_{\phantom{j}i}}$, to write the Hermitian matrix diagonalization problem as
\begin{align}
\widehat{H}^{i}_{\phantom{i}j} 
= \sum_k \widehat{U}^{i}_{\phantom{i}k} \cdot \lambda_k \cdot \overline{\widehat{U}^j_{\phantom{j}k}} 
= \sum_{k} \lambda_k \braket{i}{\lambda_k} \overline{\braket{j}{\lambda_k}} .
\end{align}
In summary,
\begin{align}
H
&= \sum_{i,j} \ket{i} \widehat{H}^i_{\phantom{i}j} \bra{j} 
= \sum_k \lambda_k \ketbra{\lambda_k}{\lambda_k} \\
&= \sum_{i,j} \ket{i} \left(\widehat{U} \cdot \text{diag}\left[\lambda_1,\dots,\lambda_D\right] \cdot \widehat{U}^\dagger \right)^i_{\phantom{i}j} \bra{j} .
\end{align}
{\bf Compatible observables} \qquad Let $X$ and $Y$ be observables -- aka Hermitian operators. We shall define compatible observables to be ones where the operators commute,
\begin{align}
[A,B] \equiv AB-BA = 0 .
\end{align}
They are incompatible when $[A,B] \neq 0$. Finding the maximal set of mutually compatible set of observables in a given physical system will tell us the range of eigenvalues that fully capture the quantum state of the system. To understand this we need the following result. %\footnote{To have a full understanding we also need to invoke the measurement postulate in quantum theory. However, since this is a math-for-physics course I will not discuss that here.}

\begin{quotation}
	{\bf Theorem} \qquad Suppose $X$ and $Y$ are observables -- they are Hermitian operators. Then $X$ and $Y$ are compatible (i.e., commute with each other) if and only if they are simultaneously diagonalizable.
\end{quotation}
{\it Proof} \qquad We will provide the proof for the case where the spectrum of $X$ is non-degenerate. We have already stated earlier that if $X$ is Hermitian we can expand it in its basis eigenkets.
\begin{align}
X = \sum_{a} a \ket{a} \bra{a} 
\end{align}
In this basis $X$ is already diagonal. But what about $Y$? Suppose $[X,Y]=0$. We consider, for distinct eigenvalues $a$ and $a'$ of $X$,
\begin{align}
\braOket{a'}{[X,Y]}{a} = \braOket{a'}{XY-YX}{a} = (a'-a)\braOket{a'}{Y}{a} = 0 .
\end{align}
Since $a-a' \neq 0$ by assumption, we must have $\braOket{a'}{Y}{a}=0$. That means the only non-zero matrix elements are the diagonal ones $\braOket{a}{Y}{a}$.\footnote{If the spectrum of $X$ were $N$-fold degenerate, $\{\ket{a;i} \vert i = 1,2,\dots,N\}$ with $X \ket{a;i} = a \ket{a;i}$, to extend the proof to this case, all we have to do is to diagonalize the $N \times N$ matrix $\braOket{a;i}{Y}{a;j}$. That this is always possible is because $Y$ is Hermitian. Within the subspace spanned by these $\{\ket{a;i}\}$, $X = \sum_i a \ketbra{a;i}{a;i} + \dots$ acts like $a$ times the identity operator, and will therefore definitely commute with $Y$.}

We have thus shown $[X,Y] = 0 \Rightarrow$ $X$ and $Y$ are simultaneously diagonalizable. We now turn to proving, if $X$ and $Y$ are simultaneously diagonalizable, then $[X,Y]=0$. That is, suppose
\begin{align}
X = \sum_{a,b} a \ket{a,b} \bra{a,b} 
\qquad \text{ and } \qquad
Y = \sum_{a,b} b \ket{a,b} \bra{a,b} ,
\end{align}
let's compute the commutator
\begin{align}
[X,Y]
&= \sum_{a,b,a',b'} a b' \left(\ket{a,b} \braket{a,b}{a',b'} \bra{a',b'} - \ket{a',b'} \bra{a,b}\right) . 
\end{align}
Remember that eigenvectors corresponding to distinct eigenvalues are orthogonal, namely $\braket{a,b}{a',b'}$ is unity only when $a=a'$ and $b=b'$ simultaneously. This means we may discard the summation over $(a',b')$ and set $a=a'$ and $b=b'$ within the summand.
\begin{align}
[X,Y] = \sum_{a,b} a b \left(\ket{a,b} \bra{a,b} - \ket{a,b} \braket{a,b}{a,b} \bra{a,b}\right) = 0 .
\end{align}
%In particular, in the orthonormal eigenket basis the matrix representations $\widehat{X}$ and $\widehat{Y}$ are both diagonal (cf. eq. \eqref{LinearOperator_DiagonalModeExpansion}). Since any two diagonal matrices commute, we see that the operators must too: $[X,Y]=0$.\footnote{This is, of course, an illustration of the one-to-one correspondence between linear operators and their representations. If the matrix representations of $X$ and $Y$ commute, so must the linear operators themselves.} \qed
%We see that incompatible observables definitely cannot be simultaneously diagonalized. Notice this means if $X$ and $Y$ describe some physical system, then the kets of the system can be labeled at least by two numbers, for example $\ket{a,b}$, where $a$ and $b$ are respectively the eigenvalues of $X$ and of $Y$. We have already seen an example: the negative Laplacian on the $2$-sphere $-\vec{\nabla}_{\mathbb{S}^2}^2$ and the rotation operator $-i\partial_\phi$ commutes, and a state on the $2$-sphere can be labeled by their eigenvalues.
%\begin{myP}
%\qquad Complete the proof for the case where there is degeneracy in the spectrum of $X$. Hint: you only need to study the matrix elements of $Y$ in the degenerate subspace.
%\end{myP}
\begin{myP}
\qquad Assuming the spectrum of $X$ is non-degenerate, show that the $Y$ in the preceding theorem can be expanded in terms of the eigenkets of $X$ as
\begin{align}
Y = \sum_a \ket{a} \braOket{a}{Y}{a} \bra{a} .
\end{align}
Read off the eigenvalues.
\end{myP}
{\bf Probabilities and Expectation value} \qquad In the context of quantum theory, given a state $\ket{\alpha}$ and an observable $O$, we may expand the former in terms of the orthonormal eigenkets $\{ \ket{\lambda_i} \}$ of the latter,
\begin{align}
\ket{\alpha} = \sum_i \ket{\lambda_i} \braket{\lambda_i}{\alpha}, \qquad O \ket{\lambda_i} = \lambda_i \ket{\lambda_i} .
\end{align}
It is a postulate of quantum theory that the probability of obtaining a specific $\lambda_j$ in an experiment designed to observe $O$ (which can be energy, spin, etc.) is given by $|\braket{\lambda_j}{\alpha}|^2 = \braket{\alpha}{\lambda_i} \braket{\lambda_i}{\alpha}$; if the spectrum is degenerate, so that there are $N$ eigenkets $\{ \ket{\lambda_i;j} \vert j =1,2,3,\dots,N \}$ corresponding to $\lambda_i$, then the probability will be
\begin{align}
P(\lambda_i) = \sum_j \braket{\alpha}{\lambda_i;j} \braket{\lambda_i;j}{\alpha} .
\end{align}
This is known as the Born rule.

The expectation value of some operator $O$ with respect to some state $\ket{\alpha}$ is defined to be
\begin{align}
\braOket{\alpha}{O}{\alpha} .
\end{align}
If $O$ is Hermitian, then the expectation value is real, since
\begin{align}
\braOket{\alpha}{O}{\alpha}^* = \braOket{\alpha}{O^\dagger}{\alpha} = \braOket{\alpha}{O}{\alpha} .
\end{align}
In the quantum context, because we may interpret $O$ to be an observable, its expectation value with respect to some state can be viewed as the average value of the observable. This can be seen by expanding $\ket{\alpha}$ in terms of the eigenstates of $O$.
\begin{align}
\braOket{\alpha}{O}{\alpha}
&= \sum_{i,j} \braket{\alpha}{\lambda_i} \braOket{\lambda_i}{O}{\lambda_j} \braket{\lambda_j}{\alpha} \nonumber\\
&= \sum_{i,j} \braket{\alpha}{\lambda_i} \lambda_i \braket{\lambda_i}{\lambda_j} \braket{\lambda_j}{\alpha} \nonumber\\
&= \sum_{i} |\braket{\alpha}{\lambda_i}|^2 \lambda_i .
\end{align}
The probability of finding $\lambda_i$ is $|\braket{\alpha}{\lambda_i}|^2$, therefore the expectation value is an average. (In the sum here, we assume a non-degenerate spectrum for simplicity.)

Suppose instead $O$ is anti-Hermitian, $O^\dagger = -O$. Then we see its expectation value with respect to some state $\ket{\alpha}$ is purely imaginary.
\begin{align}
\braOket{\alpha}{O}{\alpha}^* = \braOket{\alpha}{O^\dagger}{\alpha} = -\braOket{\alpha}{O}{\alpha} 
\end{align}
%\begin{myP}
%\qquad Compute the adjoint of the commutator $[X,Y]$ and express the answer as a commutator. What happens when $X$ and $Y$ are Hermitian?
%\end{myP}
{\bf Pauli matrices from their algebra.} \qquad Before moving on to unitary operators, let us now try to construct (up to a phase) the Pauli matrices in eq. \eqref{PauliMatrices}. We assume the following.
\begin{itemize}
	\item The $\{ \sigma^i \vert i = 1,2,3 \}$ are Hermitian linear operators acting on a 2 dimensional vector space. 
	\item They obey the algebra
	\begin{align}
	\label{PauliAlgebra}
	\sigma^i \sigma^j = \delta^{ij} \mathbb{I} + i \sum_k \epsilon^{ijk} \sigma^k .
	\end{align}
	That this is consistent with the Hermitian nature of the $\{ \sigma^i \}$ can be checked by taking $^\dagger$ on both sides. We have $(\sigma^i \sigma^j)^\dagger = \sigma^j \sigma^i$ on the left-hand-side; whereas on the right-hand-side $(\delta^{ij} \mathbb{I} + i \sum_k \epsilon^{ijk} \sigma^k)^\dagger = \delta^{ij} \mathbb{I} - i \epsilon^{ijk} \sigma^k = \delta^{ij} \mathbb{I} + i \epsilon^{jik} \sigma^k = \sigma^j \sigma^i$.
\end{itemize}
We begin by noting
\begin{align}
[\sigma^i,\sigma^j] 
= (\delta^{ij}-\delta^{ji}) \mathbb{I} + \sum_k i (\epsilon^{ijk} - \epsilon^{jik}) \sigma^k
= 2 i \sum_k \epsilon^{ijk} \sigma^k .
\end{align}
We then define the operators
\begin{align}
\sigma^\pm \equiv \sigma^1 \pm i \sigma^2 \qquad \Rightarrow \qquad 
(\sigma^\pm)^\dagger = \sigma^\mp.
\end{align}
and calculate\footnote{The commutator is linear in that $[X,Y+Z] = X(Y+Z) - (Y+Z)X = (XY-YX) + (XZ - ZX) = [X,Y] + [X,Z]$.}
\begin{align}
[\sigma^3, \sigma^\pm] 
= [\sigma^3, \sigma^1] \pm i [\sigma^3, \sigma^2] 
&= 2 i \epsilon^{312} \sigma^2 \pm 2 i^2 \epsilon^{321} \sigma^1 \\
&= 2 i \sigma^2 \pm 2 \sigma^1 
= \pm 2 (\sigma^1 \pm i \sigma^2) , \nonumber\\
\Rightarrow \qquad 
[\sigma^3, \sigma^\pm] &= \pm 2 \sigma^\pm .
\end{align}
Also,
{\allowdisplaybreaks\begin{align}
\sigma^\mp \sigma^\pm 
&= (\sigma^1 \mp i \sigma^2) (\sigma^1 \pm i \sigma^2) \nonumber\\
&= (\sigma^1)^2 + (\mp i)(\pm i) (\sigma^2)^2 \mp i \sigma^2 \sigma^1 \pm i \sigma^1 \sigma^2 \nonumber\\
&= 2 \mathbb{I} \pm i (\sigma^1 \sigma^2 - \sigma^2 \sigma^1) 
= 2 \mathbb{I} \pm i [\sigma^1,\sigma^2] = 2 \mathbb{I} \pm 2 i^2 \epsilon^{123} \sigma^3 \nonumber\\
\Rightarrow \qquad
\sigma^\mp \sigma^\pm &= 2 (\mathbb{I} \mp \sigma^3) .
\end{align}}
{\it $\sigma^3$ and its Matrix representation.} \qquad Suppose $\ket{\lambda}$ is a unit norm eigenket of $\sigma^3$. Using $\sigma^3 \ket{\lambda} = \lambda \ket{\lambda}$ and $(\sigma^3)^2 = \mathbb{I}$,
\begin{align}
1 = \braket{\lambda}{\lambda} = \braOket{\lambda}{\sigma^3 \sigma^3}{\lambda} 
= \left(\sigma^3 \ket{\lambda}\right)^\dagger\left(\sigma^3 \ket{\lambda}\right) 
= \lambda^2 \braket{\lambda}{\lambda} = \lambda^2 .
\end{align}
We see immediately that the spectrum is at most $\lambda_\pm = \pm 1$. (We will prove below that the vector space is indeed spanned by both $\ket{\pm}$.) Since the vector space is 2 dimensional, and since the eigenvectors of a Hermitian operator with distinct eigenvalues are necessarily orthogonal, we see that $\ket{\pm}$ span the space at hand. We may thus say
\begin{align}
\sigma^3 = \ketbra{+}{+} - \ketbra{-}{-} ,
\end{align}
which immediately allows us to read off its matrix representation in this basis $\{ \ket{\pm} \}$, with $\braOket{+}{\sigma^3}{+}$ being the top left hand corner entry:
\begin{align}
\braOket{j}{\sigma^3}{i} 
= \left[
\begin{array}{cc}
1	& 0 \\
0	& -1
\end{array}
\right] .
\end{align}
Observe that we could have considered $\braOket{\lambda}{\sigma^i \sigma^i}{\lambda}$ for any $i \in \{ 1,2,3 \}$; we are just picking $i=3$ for concreteness. In particular, we see from their algebraic properties that {\it all three Pauli operators $\sigma^{1,2,3}$ have the same spectrum $\{+1,-1\}$}. Moreover, since the $\sigma^i$s do not commute, we already know they cannot be simultaneously diagonalized.

{\it Raising and lowering (aka Ladder) operators $\sigma^\pm$, and $\sigma^{1,2}$.} \qquad Let us now consider
{\allowdisplaybreaks\begin{align}
\sigma^3 \sigma^\pm \ket{\lambda} 
&= (\sigma^3 \sigma^\pm - \sigma^\pm \sigma^3 + \sigma^\pm \sigma^3) \ket{\lambda} \nonumber\\
&= ([\sigma^3, \sigma^\pm] + \sigma^\pm \sigma^3) \ket{\lambda} = (\pm 2 \sigma^\pm + \lambda \sigma^\pm) \ket{\lambda} \nonumber\\
&= (\lambda \pm 2) \sigma^\pm \ket{\lambda} \qquad \Rightarrow \qquad
\sigma^\pm \ket{\lambda} = K_\lambda^\pm \ket{\lambda \pm 2} , \qquad K_\lambda^\pm \in\mathbb{C} .
\end{align}}
This is why the $\sigma^\pm$ are often called raising/lowering operators: when applied to the eigenket $\ket{\lambda}$ of $\sigma^3$ it returns an eigenket with eigenvalue raised/lowered by $2$ relative to $\lambda$. This sort of algebraic reasoning is important for the study of group representations; solving the energy levels of the quantum harmonic oscillator and the Hydrogen atom\footnote{For the H atom, the algebraic derivation of its energy levels involve the quantum analog of the classical Laplace-Runge-Lenz vector.}; and even the notion of particles in quantum field theory. 

What is the norm of $\sigma^\pm \ket{\lambda}$?
\begin{align}
\braOket{\lambda}{\sigma^\mp \sigma^\pm}{\lambda} &= |K_\lambda^\pm|^2 \braket{\lambda \pm 2}{\lambda \pm 2} \nonumber\\
\braOket{\lambda}{2(\mathbb{I} \mp \sigma^3)}{\lambda} &= |K_\lambda^\pm|^2 \nonumber\\
2 (1 \mp \lambda) &= |K_\lambda^\pm|^2 .
\end{align}
This means we can solve $K_\lambda^\pm$ up to a phase
\begin{align}
K_\lambda^\pm = e^{i\delta_\pm^{(\lambda)}} \sqrt{2(1 \mp \lambda)}, \qquad \lambda\in\{-1,+1\} .
\end{align}
Note that $K_+^+ = e^{i\delta_+^{(+)}} \sqrt{2(1 - (+1))} = 0$, and $K_-^- = e^{i\delta_-^{(-)}} \sqrt{2(1 + (-1))} = 0$, which means
\begin{align}
\sigma^+ \ket{+} = 0, \qquad \sigma^- \ket{-} = 0.
\end{align}
We can interpret this as saying, there are no larger eigenvalues than $+1$ and no smaller than $-1$ -- this is consistent with our assumption that we have a 2-dimensional vector space. Moreover, $K_+^- = e^{i\delta_-^{(+)}} \sqrt{2(1 + (+1))} = 2 e^{i\delta_-^{(+)}}$ and $K_-^+ = e^{i\delta_+^{(-)}} \sqrt{2(1 - (-1))} = 2 e^{i\delta_-^{(+)}}$.
\begin{align}
\sigma^+ \ket{-} = 2 e^{i\delta_+^{(-)}} \ket{+}, \qquad 
\sigma^- \ket{+} = 2 e^{i\delta_-^{(+)}} \ket{-} .
\end{align}
At this point, we have proved that the spectrum of $\sigma^3$ has to include both $\ket{\pm}$, because we can get from one to the other by applying $\sigma^\pm$ appropriately. In other words, if $\ket{+}$ exists, so does $\ket{-} \propto \sigma^- \ket{+}$; and if $\ket{-}$ exists, so does $\ket{+} \propto \sigma^+ \ket{-}$. 

Also notice we have figured out how $\sigma^\pm$ acts on the basis kets (up to phases), just from their algebraic properties. We may now turn this around to write them in terms of the basis bras/kets:
\begin{align}
\sigma^+ = 2 e^{i\delta_+^{(-)}} \ketbra{+}{-}, \qquad \sigma^- = 2 e^{i\delta_-^{(+)}} \ketbra{-}{+} .
\end{align}
Since $(\sigma^+)^\dagger = \sigma^-$, we must have $\delta_+^{(-)} = -\delta_-^{(+)} \equiv \delta$.
\begin{align}
\sigma^+ = 2 e^{i\delta} \ketbra{+}{-}, \qquad \sigma^- = 2 e^{-i\delta} \ketbra{-}{+} .
\end{align}
with the corresponding matrix representations, with $\braOket{+}{\sigma^\pm}{+}$ being the top left hand corner entry: 
\begin{align}
\braOket{j}{\sigma^+}{i} 
= \left[
\begin{array}{cc}
0	& 2 e^{i\delta} \\
0	& 0
\end{array}
\right]
, \qquad 
\braOket{j}{\sigma^-}{i} = 
\left[
\begin{array}{cc}
0				& 0 \\
2 e^{-i\delta}	& 0
\end{array}
\right] .
\end{align}
Now, we have $\sigma^\pm = \sigma^1 \pm i \sigma^2$, which means we can solve for
\begin{align}
2 \sigma^1 = \sigma^+ + \sigma^-, \qquad 2 i \sigma^2 = \sigma^+ - \sigma^- .
\end{align}
We have
\begin{align}
\sigma^1 &= e^{i\delta} \ketbra{+}{-} + e^{-i\delta} \ketbra{-}{+}, \\
\sigma^2 &= -i e^{i\delta} \ketbra{+}{-} + i e^{-i\delta} \ketbra{-}{+}, \qquad \delta \in \mathbb{R} ,
\end{align}
with matrix representations
\begin{align}
\braOket{j}{\sigma^1}{i} 
= \left[
\begin{array}{cc}
0				& e^{i\delta} \\
e^{-i\delta}	& 0
\end{array}
\right]
, \qquad 
\braOket{j}{\sigma^2}{i} = 
\left[
\begin{array}{cc}
0				& -i e^{i\delta} \\
i e^{-i\delta}	& 0
\end{array}
\right] .
\end{align}
You can check explicitly that the algebra in eq. \eqref{PauliAlgebra} holds for any $\delta$. However, we can also use the fact that unit normal eigenkets can be re-scaled by a phase and still remain unit norm eigenkets. 
\begin{align}
\sigma^3 \left(e^{i\theta} \ket{\pm} \right) = \pm \left( e^{i\theta} \ket{\pm} \right), \qquad 
\left(e^{i\theta} \ket{\pm} \right)^\dagger \left(e^{i\theta} \ket{\pm} \right) = 1 ,\qquad \theta\in\mathbb{R}.
\end{align}
We re-group the phases occurring within our $\sigma^3$ and $\sigma^\pm$ as follows.
\begin{align}
\sigma^3 = (e^{i\delta/2} \ket{+})(e^{i\delta/2} \ket{+})^\dagger - (e^{-i\delta/2} \ket{-})(e^{-i\delta/2} \ket{-})^\dagger, \\
\sigma^+ = 2 (e^{i\delta/2} \ket{+})(e^{-i\delta/2} \ket{-})^\dagger, \qquad 
\sigma^- = 2 (e^{-i\delta/2} \ket{-})(e^{i\delta/2} \ket{+})^\dagger .
\end{align}
That is, if we re-define $\ket{\pm '} \equiv e^{\pm i\delta/2} \ket{\pm}$, followed by dropping the primes, we would have
\begin{align}
\sigma^3 = \ketbra{+}{+} - \ketbra{-}{-}, \\
\sigma^+ = 2 \ketbra{+}{-}, \qquad 
\sigma^- = 2 \ketbra{-}{+} ,
\end{align}
and again using $\sigma^1 = (\sigma^1 + \sigma^2)/2$ and $\sigma^2 = -i(\sigma^1 - \sigma^2)/2$,
\begin{align}
\sigma^1 &= \ketbra{+}{-} + \ketbra{-}{+}, \\
\sigma^2 &= -i \ketbra{+}{-} + i \ketbra{-}{+}, \qquad \delta \in \mathbb{R} .
\end{align}
We see that the Pauli matrices in eq. \eqref{PauliMatrices} correspond to the matrix representations of $\sigma^i$ in the basis built out of the unit norm eigenkets of $\sigma^3$, with an appropriate choice of phase.

Note that there is nothing special about choosing our basis as the eigenkets of $\sigma^3$ -- we could have chosen the eigenkets of $\sigma^1$ or $\sigma^2$ as well. The analogous raising and lower operators can then be constructed from the remaining $\sigma^i$s.

Finally, for $\widehat{U}$ unitary we have already noted that $\det(\widehat{U} \widehat{\sigma}^i \widehat{U}^\dagger) = \det \widehat{\sigma}^i$ and $\Tr{\widehat{U} \widehat{\sigma}^i \widehat{U}^\dagger} = \Tr{\widehat{\sigma}^i}$. Therefore, if we choose $\widehat{U}$ such that $\widehat{U} \widehat{\sigma}^i \widehat{U}^\dagger = \text{diag}(1,-1)$ -- since we now know the eigenvalues of each $\widehat{\sigma}^i$ are $\pm 1$ -- we readily deduce that
\begin{align}
\det \widehat{\sigma}^i = -1, \qquad \Tr{\widehat{\sigma}^i} = 0 .
\end{align}
(However, $\widehat{\sigma}^2 \widehat{\sigma}^i \widehat{\sigma}^2 = -(\widehat{\sigma}^i)^*$ does not hold unless $\delta=0$.)

\subsubsection{Unitary Operation as Change of Orthonormal Basis}

A unitary operator $U$ is one whose inverse is its adjoint, i.e.,
\begin{align}
\label{UnitaryOperator_Definition}
U^\dagger U = U U^\dagger = \mathbb{I} .
\end{align}
Like their Hermitian counterparts, unitary operators play a special role in quantum theory. At a somewhat mundane level, they describe the change from one set of basis vectors to another. The analog in Euclidean space is the rotation matrix. But when the quantum dynamics is invariant under a particular change of basis -- i.e., there is a symmetry enjoyed by the system at hand -- then the eigenvectors of these unitary operators play a special role in classifying the dynamics itself. Also, in order to conserve probabilities, the time evolution operator, which takes an initial wave function(nal) of the quantum system and evolves it forward in time, is in fact a unitary operator itself.

Let us begin by understanding the action of a unitary operator as a change of basis vectors. Up till now we have assumed we can always find an orthonormal set of basis vectors $\{\ket{i}|i=1,2,\dots,D\}$, for a $D$ dimensional vector space. But just as in Euclidean space, this choice of basis vectors is not unique -- in 3-space, for instance, we can rotate $\{\widehat{x},\widehat{y},\widehat{z}\}$ to some other $\{\widehat{x}',\widehat{y}',\widehat{z}'\}$ (i.e., redefine what we mean by the $x$, $y$ and $z$ axes). Hence, let us suppose we have found two such sets of orthonormal basis vectors
\begin{align}
\left\{ \ket{1}, \dots, \ket{D} \right\} \qquad \text{ and } \qquad \left\{ \ket{1'}, \dots, \ket{D'} \right\} .
\end{align}
(For concreteness the dimension of the vector space is $D$.) Remember a linear operator is defined by its action on every element of the vector space; equivalently, by linearity and completeness, it is defined by how it acts on each basis vector. We may thus define our unitary operator $U$ via
\begin{align}
\label{UnitaryOperatorAsChangeOfBasis}
U \ket{i} = \ket{i'}, \qquad i \in \{1,2,\dots,D\} .
\end{align}
Its matrix representation in the unprimed basis $\{\ket{i}\}$ is gotten by projecting both sides along $\ket{j}$.
\begin{align}
\label{UnitaryOperatorAsChangeOfBasis_MatrixElement}
\braOket{j}{U}{i} = \braket{j}{i'}, \qquad i,j \in \{1,2,\dots,D\} .
\end{align}
Is $U$ really unitary? One way to verify this is through its matrix representation. We have
\begin{align}
\label{UnitaryOperatorAsChangeOfBasis_Udagger_MatrixElement}
\bra{j} U^\dagger \ket{i} = \bra{i} U \ket{j}^* = \braket{j'}{i} .
\end{align}
Whereas $U^\dagger U$ in matrix form is
\begin{align}
\sum_k \bra{j} U^\dagger \ket{k} \bra{k} U \ket{i} 
= \sum_k \bra{k} U \ket{j}^* \bra{k} U \ket{i} 
= \sum_k \braket{k}{i'} \braket{k}{j'}^* 
= \sum_k \braket{j'}{k} \braket{k}{i'} .
\end{align}
Because both $\{\ket{k}\}$ and $\{\ket{k'}\}$ form an orthonormal basis, we may invoke the completeness relation eq. \eqref{CompletenessRelation} to deduce
\begin{align}
\sum_k \bra{j} U^\dagger \ket{k} \bra{k} U \ket{i} = \braket{j'}{i'} = \delta^j_i .
\end{align}
That is, we recover the unit matrix when we multiply the matrix representation of $U^\dagger$ to that of $U$.\footnote{Strictly speaking we have only verified that the left inverse of $U$ is $U^\dagger$, but for finite dimensional matrices, the left inverse is also the right inverse.} Since we have not made any additional assumptions about the two arbitrary sets of orthonormal basis vectors, this verification of the unitary nature of $U$ is itself independent of the choice of basis.

Alternatively, let us observe that the $U$ defined in eq. \eqref{UnitaryOperatorAsChangeOfBasis} can be expressed as
\begin{align}
U = \sum_j \ket{j'} \bra{j} .
\end{align}
All we have to verify is $U \ket{i} = \ket{i'}$ for any $i \in \{ 1,2,3,\dots,D\}$.
\begin{align}
U \ket{i} = \sum_j \ket{j'} \braket{j}{i} = \sum_j \ket{j'} \delta^j_i = \ket{i'} .
\end{align}
The unitary nature of $U$ can also be checked explicitly. Remember $(\ketbra{\alpha}{\beta})^\dagger = \ketbra{\beta}{\alpha}$.
\begin{align}
U^\dagger U 
&= \sum_j \ket{j} \bra{j'} \sum_k \ket{k'} \bra{k} 
= \sum_{j,k} \ket{j} \braket{j'}{k'} \bra{k} \nonumber\\
&= \sum_{j,k} \ket{j} \delta^j_k \bra{k} = \sum_{j} \ket{j} \bra{j} = \mathbb{I} .
\end{align}
The very last equality is just the completeness relation in eq. \eqref{CompletenessRelation}.

Starting from $U$ defined in eq. \eqref{UnitaryOperatorAsChangeOfBasis} as a change-of-basis operator, we have shown $U$ is unitary whenever the old $\{ \ket{i} \}$ and new $\{ \ket{i'} \}$ basis are given. Turning this around -- suppose $U$ is some arbitrary unitary linear operator, given some orthonormal basis $\{ \ket{i} \}$ we can construct a new orthonormal basis $\{ \ket{j'} \}$ by {\it defining}
\begin{align}
\ket{i'} \equiv U \ket{i} .
\end{align}
All we have to show is that $\{ \ket{i'} \}$ form an orthonormal set.
\begin{align}
\braket{j'}{i'} = \left( U \ket{j}\right)^\dagger \left( U \ket{i} \right) = \braOket{j}{U^\dagger U}{i} = \braket{j}{i} = \delta^j_i .
\end{align}
We may therefore pause to summarize our findings as follows.
\begin{quote}
	A linear operator $U$ implements a change-of-basis from the orthonormal set $\{ \ket{i} \}$ to some other (appropriately defined) orthonormal set $\{ \ket{i'} \}$ if and only if $U$ is unitary.
\end{quote}
{\bf Change-of-basis of $\braket{\alpha}{i}$} \qquad Given a bra $\bra{\alpha}$, we may expand it either in the new $\{ \bra{i'} \}$ or old $\{ \bra{i} \}$ basis bras,
\begin{align}
\bra{\alpha} = \sum_i \braket{\alpha}{i} \bra{i} = \sum_i \braket{\alpha}{i'} \bra{i'} .
\end{align}
We can relate the components of expansions using $\braOket{i}{U}{k} = \braket{i}{k'}$ (cf. eq. \eqref{UnitaryOperatorAsChangeOfBasis_MatrixElement}),
\begin{align}
\sum_{k} \braket{\alpha}{k'} \bra{k'} 
&= \sum_i \braket{\alpha}{i} \bra{i}	\nonumber\\
&= \sum_{i,k} \braket{\alpha}{i} \braket{i}{k'} \bra{k'} = \sum_k \left(\sum_i \braket{\alpha}{i} \braOket{i}{U}{k}\right) \bra{k'} .
\end{align}
Equating the coefficients of $\bra{k'}$ on the left and (far-most) right hand sides, we see the components of the bra in the new basis can be gotten from that in the old basis using $\widehat{U}$,
\begin{align}
\braket{\alpha}{k'} = \sum_i \braket{\alpha}{i} \braOket{i}{U}{k} .
\end{align}
In words: the $\bra{\alpha}$ row vector in the basis $\{\bra{i'}\}$ is equal to $U$, written in the basis $\{\braOket{j}{U}{i}\}$, acting (from the right) on the $\braket{\alpha}{i}$ row vector, the $\bra{\alpha}$ in the basis $\{\bra{i}\}$. Moreover, in index notation,
\begin{align}
\alpha_{k'} = \alpha_{i} \widehat{U}^i_{\phantom{i}k} .
\end{align}
\begin{myP}
\qquad Given a vector $\ket{\alpha}$, and the orthonormal basis vectors $\{\ket{i}\}$, we can represent it as a column vector, where the $i$th component is $\braket{i}{\alpha}$. What does this column vector look like in the basis $\{ \ket{i'} \}$? Show that it is given by the matrix multiplication 
\begin{align}
\braket{i'}{\alpha} = \sum_k \braOket{i}{U^\dagger}{k} \braket{k}{\alpha}, \qquad U \ket{i} = \ket{i'} .
\end{align}
In words: the $\ket{\alpha}$ column vector in the basis $\{\ket{i'}\}$ is equal to $U^\dagger$, written in the basis $\{\braOket{j}{U^\dagger}{i}\}$, acting (from the left) on the $\braket{i}{\alpha}$ column vector, the $\ket{\alpha}$ in the basis $\{\ket{i}\}$. 

Furthermore, in index notation,
\begin{align}
\alpha^{i'} = (\widehat{U}^\dagger)^{i}_{\phantom{i}k} \alpha^{k} .
\end{align}
From the discussion on how components of bra(s) transform under a change-of-basis, together the analogous discussion of linear operators below, you will begin to see why in index notation, there is a need to distinguish between upper and lower indices -- {\it they transform oppositely from each other}. \qed
\end{myP}
\begin{myP}
{\it 2D rotation in 3D.} \qquad Let's rotate the basis vectors of the 2D plane, spanned by the $x$- and $z$-axis, by an angle $\theta$. If $\ket{1}$, $\ket{2}$, and $\ket{3}$ respectively denote the unit vectors along the $x$, $y$, and $z$ axes, how should the operator $U(\theta)$ act to rotate them? For example, since we are rotating the $13$-plane, $U \ket{2} = \ket{2}$. (Drawing a picture may help.) Can you then write down the matrix representation $\braOket{j}{U(\theta)}{i}$? \qed
\end{myP}
{\bf Change-of-basis of $\braOket{i}{X}{j}$} \qquad Now we shall proceed to ask, how do we use $U$ to change the matrix representation of some linear operator $X$ written in the basis $\{\ket{i}\}$ to one in the basis $\{\ket{i}'\}$? Starting from $\braOket{i'}{X}{j'}$ we insert the completeness relation eq. \eqref{CompletenessRelation} in the basis $\{\ket{i}\}$, on both the left and the right,
\begin{align}
\label{ChangeOfBasisX}
\braOket{i'}{X}{j'} 
&= \sum_{k,l} \braket{i'}{k} \braOket{k}{X}{l} \braket{l}{j'} \nonumber\\
&= \sum_{k,l} \braOket{i}{U^\dagger}{k} \braOket{k}{X}{l} \braOket{l}{U}{j} 
= \braOket{i}{U^\dagger X U}{j},
\end{align}
where we have recognized (from equations \eqref{UnitaryOperatorAsChangeOfBasis_MatrixElement} and \eqref{UnitaryOperatorAsChangeOfBasis_Udagger_MatrixElement}) $\braket{i'}{k} = \braOket{i}{U^\dagger}{k}$ and $\braket{l}{j'} = \braOket{l}{U}{j}$. If we denote $\widehat{X}'$ as the matrix representation of $X$ with respect to the primed basis; and $\widehat{X}$ and $\widehat{U}$ as their corresponding operators with respect to the unprimed basis, we recover the similarity transformation
\begin{align}
\widehat{X}' = \widehat{U}^\dagger \widehat{X} \widehat{U} .
\end{align}
In index notation, with primes on the indices reminding us that the matrix is written in the primed basis $\{ \ket{i'} \}$ and the unprimed indices in the unprimed basis $\{ \ket{i} \}$,
\begin{align}
\widehat{X}^{i'}_{\phantom{i'}j'} = (\widehat{U}^\dagger)^i_{\phantom{i}k} \widehat{X}^{k}_{\phantom{k}l} \widehat{U}^l_{\phantom{l}j} .
\end{align}
As already alluded to, we see here the $i$ and $j$ indices transform ``oppositely" from each other -- so that, even in matrix algebra, if we view square matrices as (representations of) linear operators acting on some vector space, then the row index $i$ should have a different position from the column index $j$ so as to distinguish their transformation properties. This will allow us to readily implement that fact, when upper and lower indices are repeated, the pair transform as a scalar -- for example, $X^{i'}_{\phantom{i'}i'} = X^{i}_{\phantom{i}i}$.\footnote{This issue of upper versus lower indices will also appear in differential geometry. Given a pair of indices that transform oppositely from each other, we want them to be placed differently (upper vs. lower), so that when we set their labels equal -- with Einstein summation in force -- they automatically transforms as a scalar, since the pair of transformations will undo each other.}

On the other hand, from the last equality of eq. \eqref{ChangeOfBasisX}, we may also view $\widehat{X}'$ as the matrix representation of the operator
\begin{align}
X' \equiv U^\dagger X U
\end{align}
written in the old basis $\{ \ket{i} \}$. To reiterate,
\begin{align}
\braOket{i'}{X}{j'} = \braOket{i}{U^\dagger X U}{j} .
\end{align}
The next two theorems can be interpreted as telling us that the Hermitian/unitary nature of operators and their spectra are really basis-independent constructs.
\begin{quotation}
{\bf Theorem} \qquad Let $X' \equiv U^\dagger X U$. If $U$ is a unitary operator, $X$ and $X'$ shares the same spectrum.
\end{quotation}
{\it Proof} \qquad Let $\ket{\lambda}$ be the eigenvector and $\lambda$ be the corresponding eigenvalue of $X$.
\begin{align}
X \ket{\lambda} = \lambda \ket{\lambda} 
\end{align}
By inserting a $\mathbb{I} = U^\dagger U$ and multiplying both sides on the left by $U^\dagger$,
\begin{align}
U^\dagger X U U^\dagger \ket{\lambda} &= \lambda U^\dagger \ket{\lambda} , \\
X' (U^\dagger \ket{\lambda}) &= \lambda (U^\dagger \ket{\lambda}) .
\end{align}
That is, given the eigenvector $\ket{\lambda}$ of $X$ with eigenvalue $\lambda$, the corresponding eigenvector of $X'$ is $U^\dagger \ket{\lambda}$ with precisely the same eigenvalue $\lambda$.
\begin{quotation}
	{\bf Theorem.} \qquad Let $X' \equiv U^\dagger X U$. If $X$ is Hermitian, so is $X'$. If $X$ is unitary, so is $X'$.
\end{quotation}
{\it Proof} \qquad If $X$ is Hermitian, we consider $X'^\dagger$.
\begin{align}
X'^\dagger = \left(U^\dagger X U\right)^\dagger = U^\dagger X^\dagger (U^\dagger)^\dagger = U^\dagger X U = X'.
\end{align}
If $X$ is unitary we consider $X'^\dagger X'$.
\begin{align}
X'^\dagger X' 
= \left(U^\dagger X U\right)^\dagger (U^\dagger X U) 
= U^\dagger X^\dagger U U^\dagger X U = U^\dagger X^\dagger X U 
= U^\dagger U = \mathbb{I} .
\end{align}
{\it Remark} \qquad We won't prove it here, but it is possible to find a unitary operator $U$, related to rotation in $\mathbb{R}^3$, that relates any one of the Pauli operators to the other
\begin{align}
U^\dagger \sigma^i U = \sigma^j , \qquad i \neq j .
\end{align}
This is consistent with what we have already seen earlier, that all the $\{ \sigma^k \}$ have the same spectrum $\{ -1, +1 \}$.

{\it Physical Significance} \qquad To put the significance of these statements in a physical context, recall the eigenvalues of an observable are possible outcomes of a physical experiment, while $U$ describes a change of basis. Just as classical observables such as lengths, velocity, etc. should not depend on the coordinate system we use to compute the predictions of the underlying theory -- in the discussion of curved space(time)s we will see the analogy there is called {\it general covariance} -- we see here that the possible experimental outcomes from a quantum system is independent of the choice of basis vectors we use to predict them. Also notice the very Hermitian and Unitary nature of a linear operator is invariant under a change of basis.

%{\it Schur Decomposition.} \qquad We have just witnessed that the spectrum of {\it any} linear operator is invariant under the change of basis. We may now in fact exploit this to prove the Schur decomposition you've encountered in our discussion of matrix algebra. Since {\it any} basis would do the job, we just have to find one basis where our linear operator $A$ admits an upper triangular form, with its eigenvalues filling its diagonal elements. Let us begin with some complex $D \times D$ matrix $\widehat{A}$, which we shall view as some representation of $A$ which is not necessarily upper triangular for now. After you have found its $D$ eigenvalues, start with the first $\lambda_1$. If we now consider the matrix where the column has  has $\lambda_1$ as its first element, namely $v_1^i = (\lambda_1,\vec{v}_\perp)$, its eigenvector is .

{\it Diagonalization of observable} \qquad Diagonalization of a matrix is nothing but the change-of-basis, expressing a linear operator $X$ in some orthonormal basis $\{ \ket{i} \}$ to one where it becomes a diagonal matrix with respect to the orthonormal eigenket basis $\{\ket{\lambda}\}$. That is, suppose you started with
\begin{align}
X = \sum_k \lambda_k \ketbra{\lambda_k}{\lambda_k} 
\end{align}
and defined the unitary operator
\begin{align}
U \ket{k} = \ket{\lambda_k}  \qquad \Leftrightarrow \qquad \braOket{i}{U}{k} = \braket{i}{\lambda_k} .
\end{align}
Notice the $k$th column of $\widehat{U}^i_{\phantom{i}k} \equiv \braOket{i}{U}{k}$ are the components of the $k$th unit norm eigenvector $\ket{\lambda_k}$ written in the $\{ \ket{i} \}$ basis. This implies, via two insertions of the completeness relation in eq. \eqref{CompletenessRelation},
\begin{align}
X &= \sum_{i,j,k} \lambda_k \ket{i} \braket{i}{\lambda_k} \braket{\lambda_k}{j} \bra{j} .
\end{align}
Taking matrix elements,
\begin{align}
\braOket{i}{X}{j} = \widehat{X}^i_{\phantom{i}j} 
= \sum_{k,l} \braket{i}{\lambda_k} \lambda_k \delta^k_l \braket{\lambda_l}{j} 
= \sum_{k,l} \widehat{U}^i_{\phantom{i}k} \lambda_k \delta^k_l (\widehat{U}^\dagger)^l_{\phantom{l}j} .
\end{align}
Multiplying both sides by $\widehat{U}^\dagger$ on the left and $\widehat{U}$ on the right, we have
\begin{align}
\widehat{U}^\dagger \widehat{X} \widehat{U} &= \text{diag}(\lambda_1, \lambda_2, \dots, \lambda_D) .
\end{align}
{\it Schur decomposition.} \qquad Not all linear operators are diagonalizable. However, we already know that any square matrix $\widehat{X}$ can be brought to an upper triangular form
\begin{align}
\widehat{U}^\dagger \widehat{X} \widehat{U} = \widehat{\Gamma} + \widehat{N} , \qquad 
\widehat{\Gamma} \equiv \text{diag}\left(\lambda_1, \dots, \lambda_D\right),
\end{align}
where the $\{\lambda_i\}$ are the eigenvalues of $X$ and $\widehat{N}$ is strictly upper triangular. We may now phrase the Schur decomposition as a change-of-basis from $\widehat{X}$ to its upper triangular form.
\begin{quote}
Given a linear operator $X$, it is always possible to find an orthonormal basis such that its matrix representation is upper triangular, with its eigenvalues forming its diagonal elements.
\end{quote}
{\bf Trace} \qquad Define the trace of a linear operator $X$ as
\begin{align}
\Tr{X} = \sum_i \braOket{i}{X}{i}, \qquad \braket{i}{j} = \delta^i_j .
\end{align}
The Trace yields a complex number. Let us see that this definition is independent of the orthonormal basis $\{ \ket{i} \}$. Suppose we found a different set of orthonormal basis $\{ \ket{i'} \}$, with $\braket{i'}{j'} = \delta^i_j$. Now consider
\begin{align}
\sum_i \braOket{i'}{X}{i'} 
&= \sum_{i,j,k} \braket{i'}{j} \braOket{j}{X}{k} \braket{k}{i'}
= \sum_{i,j,k} \braket{k}{i'} \braket{i'}{j} \braOket{j}{X}{k} \nonumber\\
&= \sum_{j,k} \braket{k}{j} \braOket{j}{X}{k} = \sum_{k} \braOket{k}{X}{k} . 
\end{align}
Because Tr is invariant under a change of basis, we can view the trace operation that turns an operator into a genuine scalar. This notion of a scalar is analogous to the quantities (pressure of a gas, temperature, etc.) that do not change no matter what coordinates one uses to compute/measure them.
\begin{myP}
\qquad Prove the following statements. For linear operators $X$ and $Y$,
\begin{align}
\Tr{XY} &= \Tr{YX} \\
\Tr{U^\dagger X U} &= \Tr{X} 
\end{align} \qed
\end{myP}
%\begin{myP}
%\qquad If $\{\ket{i}|i=1,2,3\dots,D\}$ is a set of orthonormal basis vectors, what is $\Tr{\ket{j} \bra{k}}$, where $j,k \in\{1,2,\dots,D\}$? \qed
%\end{myP}
%\begin{myP}
%\qquad Verify the following Jacobi identity. For linear operators $X$, $Y$ and $Z$,
%\begin{align}
%[X,[Y,Z]] + [Y,[Z,X]] + [Z,[X,Y]] = 0.
%\end{align}
%Furthermore, verify that
%\begin{align}
%[X,Y]=-[Y,X], \qquad [X,Y+Z] = [X,Y]+[X,Z], \qquad [X,YZ] = [X,Y]Z + Y[X,Z] .
%\end{align}
%\end{myP}
\begin{myP}
\qquad Find the unit norm eigenvectors that can be expressed as a linear combination of $\ket{1}$ and $\ket{2}$, and their corresponding eigenvalues, of the operator
\begin{align}
\label{DiagonalizationProblem1_1of2}
X \equiv a \left(\ket{1}\bra{1} - \ket{2}\bra{2} + \ket{1}\bra{2} + \ket{2}\bra{1}\right) .
\end{align}
Assume that $\ket{1}$ and $\ket{2}$ are orthogonal and of unit norm. (Hint: First calculate the matrix $\braOket{j}{X}{i}$.)

Now consider the operators built out of the orthonormal basis vectors $\{ \ket{i} | i=1,2,3 \}$.
\begin{align}
\label{DiagonalizationProblem1_2of2}
Y &\equiv a \left( \ketbra{1}{1} - \ketbra{2}{2} - \ketbra{3}{3} \right) , \\
Z &\equiv b \ketbra{1}{1} -ib \ketbra{2}{3} +ib \ketbra{3}{2} . \nonumber
\end{align}
(In equations \eqref{DiagonalizationProblem1_1of2} and \eqref{DiagonalizationProblem1_2of2}, $a$ and $b$ are real numbers.) Are $Y$ and $Z$ hermitian? Write down their matrix representations. Verify $[Y,Z]=0$ and proceed to simultaneously diagonalize $Y$ and $Z$. \qed
\end{myP}
\begin{myP}
{\it Pauli matrices re-visited.} \qquad Refer to the Pauli matrices $\{\sigma^\mu\}$ defined in eq. \eqref{PauliMatrices}. Let $p_\mu$ be a 4-component collection of real numbers. We may then view $p_\mu \sigma^\mu$ (where $\mu$ sums over $0$ through $3$) as a Hermitian operator acting on a 2 dimensional vector space.
\begin{enumerate}
\item Find the eigenvalues $\lambda_\pm$ and corresponding unit norm eigenvectors $\xi^\pm$ of $p_i \sigma^i$ (where $i$ sums over $1$ through $3$). These are called the {\it helicity eigenstates}. Are they also eigenstates of $p_\mu \sigma^\mu$? (Hint: consider $[p_i \sigma^i, p_\mu \sigma^\mu]$.)
\item Explain why
\begin{align}
p_i \widehat{\sigma}^i = \lambda_+ \xi^+ (\xi^+)^\dagger + \lambda_- \xi^- (\xi^-)^\dagger .
\end{align}
Can you write down the analogous expansion for $p_\mu \widehat{\sigma}^\mu$?
\item If we define the square root of an operator or matrix $\sqrt{A}$ as the solution to $\sqrt{A} \sqrt{A} = A$, write down the expansion for $\sqrt{p_\mu \widehat{\sigma}^\mu}$.
\item These 2 component spinors $\xi^\pm$ play a key role in the study of Lorentz symmetry in 4 spacetime dimensions. Consider applying an invertible transformation $L_\text{A}^{\phantom{A}B}$ on these spinors, i.e., replace
\begin{align}
(\xi^\pm)_\text{A} \to L_\text{A}^{\phantom{A}B} (\xi^\pm)_\text{B} .
\end{align}
(The A and B indices run from $1$ to $2$, the components of $\xi^\pm$.) How does $p_\mu \widehat{\sigma}^\mu$ change under such a transformation? And, how does its determinant change?
\end{enumerate} \qed
\end{myP}
\begin{myP}
{\bf Schr\"{o}dinger's equation} \qquad The primary equation in quantum mechanics (and quantum field theory), governing how states evolve in time, is
\begin{align}
\label{SchrodingerEquation}
i\hbar \partial_t \ket{\psi(t)} = H \ket{\psi(t)} ,
\end{align}
where $\hbar \approx 1.054572 \times 10^{-34}$ J s is the reduced Planck's constant, and $H$ is the Hamiltonian ($\equiv$ Hermitian total energy linear operator) of the system. The physics of a particular system is encoded within $H$. 
	
Suppose $H$ is independent of time, and suppose its orthonormal eigenkets $\{ \ket{E_i;n_j} \}$ are known ($n_j$ being the degeneracy label, running over all eigenkets with the same energy $E_j$), with $H \ket{E_i;n_i} = E_i \ket{E_i;n_i}$ and $\{ E_i \in\mathbb{R} \}$, where we will assume the energies are discrete. Show that the solution to Schr\"{o}dinger's equation in \eqref{SchrodingerEquation} is
\begin{align}
\ket{\psi(t)} = \sum_{j,n_j} e^{-(i/\hbar) E_j t} \ket{E_j;n_j} \braket{E_j;n_j}{\psi(t=0)} ,
\end{align}
where $\ket{\psi(t=0)}$ is the initial condition, i.e., the state $\ket{\psi(t)}$ at $t=0$. (Hint: Check that eq. \eqref{SchrodingerEquation} and the initial condition are satisfied.) Since the initial state was arbitrary, what you have verified is that the operator
\begin{align}
U(t,t') \equiv \sum_{j,n_j} e^{-(i/\hbar) E_j (t-t')} \ketbra{E_j;n_j}{E_j;n_j}
\end{align}
obeys Schr\"{o}dinger's equation,
\begin{align}
i\hbar \partial_t U(t,t') = H U(t,t') .
\end{align} 
Is $U(t,t')$ unitary? Explain what is the operator $U(t=t')$?

Express the expectation value $\braOket{\psi(t)}{H}{\psi(t)}$ in terms of the energy eigenkets and eigenvalues. Compare it with the expectation value $\braOket{\psi(t=0)}{H}{\psi(t=0)}$.

What if the Hamiltonian in Schr\"{o}dinger's equation depends on time -- what is the corresponding $U$? Consider the following (somewhat formal) solution for $U$.
\begin{align}
U(t,t') &\equiv \mathbb{I} 
		- \frac{i}{\hbar} \int_{t'}^{t} \dd \tau_1 H(\tau_1) 
		+ \left(-\frac{i}{\hbar}\right)^2 \int_{t'}^{t} \dd \tau_2 \int_{t'}^{\tau_2} \dd \tau_1 H(\tau_2) H(\tau_1) + \dots \\
&= \mathbb{I} + \sum_{\ell=1}^\infty \mathcal{I}_\ell(t,t') ,
\end{align}
where the $\ell$-nested integral $\mathcal{I}_\ell(t,t')$ is
\begin{align}
\mathcal{I}_\ell(t,t') \equiv \left(-\frac{i}{\hbar}\right)^\ell 
	\int_{t'}^{t} \dd \tau_\ell \int_{t'}^{\tau_\ell} \dd \tau_{\ell-1} \dots \int_{t'}^{\tau_3} \dd \tau_{2} \int_{t'}^{\tau_2} \dd \tau_{1}
	H(\tau_\ell) H(\tau_{\ell-1}) \dots H(\tau_2) H(\tau_1) .
\end{align}
(Be aware that, if the Hamiltonian $H(t)$ depends on time, it may not commute with itself at different times, namely one {\it cannot} assume $[H(\tau_1),H(\tau_2)] = 0$ if $\tau_1 \neq \tau_2$.) Verify that, for $t > t'$,
\begin{align}
i \hbar \partial_t U(t,t') = H(t) U(t,t') .
\end{align}
What is $U(t=t')$? You should be able to conclude that $\ket{\psi(t)} = U(t,t') \ket{\psi(t')}$. Hint: Start with $i \hbar \partial_t \mathcal{I}_\ell(t,t')$ and employ Leibniz's rule:
\begin{align}
\frac{\dd}{\dd t} \left(\int_{\alpha(t)}^{\beta(t)} F(t,z) \dd z\right)
= \int_{\alpha(t)}^{\beta(t)} \frac{\partial F(t,z)}{\partial t} \dd z 
+ F\left(t,\beta(t)\right) \beta'(t) - F\left(t,\alpha(t)\right) \alpha'(t) . 
\end{align}
{\it Bonus:} Can you prove Leibniz's rule, by say, using the limit definition of the derivative? \qed

\end{myP}

\subsection{Tensor Products of Vector Spaces}

In this section we will introduce the concept of a tensor product. It is a way to ``multiply" vector spaces, through the product $\otimes$, to form a larger vector space. Tensor products not only arise in quantum theory but is present even in classical electrodynamics, gravitation and field theories of non-Abelian gauge fields interacting with spin$-1/2$ matter. In particular, tensor products arise in quantum theory when you need to, for example, describe both the spatial wave-function and the spin of a particle.

{\bf Definition} \qquad To set our notation, let us consider multiplying $N \geq 2$ distinct vector spaces, i.e., $V_1 \otimes V_2 \otimes \dots \otimes V_N$ to form a $V_\text{L}$. We write the tensor product of a vector $\ket{\alpha_1;1}$ from $V_1$, $\ket{\alpha_2;2}$ from $V_2$ and so on through $\ket{\alpha_N;N}$ from $V_N$ as
\begin{align}
\label{TensorProduct}
\ket{\mathfrak{A};L} \equiv \ket{\alpha_1; 1} \otimes \ket{\alpha_2; 2} \otimes \dots \otimes \ket{\alpha_N; N} ,
\end{align}
where it is understood the vector $\ket{\alpha_i; i}$ in the $i$th slot (from the left) is an element of the $i$th vector space $V_i$. As we now see, the tensor product is multi-linear because it obeys the following algebraic rules.
\begin{enumerate}
\item The tensor product is distributive over addition. For example,
\begin{align}
\label{TensorProduct_x1}
\ket{\alpha} \otimes \left(\ket{\alpha'} + \ket{\beta'}\right) \otimes \ket{\alpha''}
= \ket{\alpha} \otimes \ket{\alpha'} \otimes \ket{\alpha''}
+ \ket{\alpha} \otimes \ket{\beta'} \otimes \ket{\alpha''} .
\end{align}
\item Scalar multiplication can be factored out. For example,
\begin{align}
\label{TensorProduct_x2}
c \left( \ket{\alpha} \otimes \ket{\alpha'} \right)
= (c \ket{\alpha}) \otimes \ket{\alpha'}
= \ket{\alpha} \otimes (c\ket{\alpha'}) .
\end{align}
\end{enumerate}
Our larger vector space $V_\text{L}$ is spanned by all vectors of the form in eq. \eqref{TensorProduct}, meaning every vector in $V_\text{L}$ can be expressed as a linear combination: 
\begin{align}
\ket{\mathfrak{A}';L} \equiv \sum_{\alpha_1,\dots,\alpha_N} C^{\alpha_1,\dots,\alpha_N} 
\ket{\alpha_1;1} \otimes \ket{\alpha_2;2} \otimes \dots \otimes \ket{\alpha_N;N} \in V_\text{L} .
\end{align}
(The $C^{\alpha_1,\dots,\alpha_N}$ is just a collection complex numbers.) In fact, if we let $\{ \ket{i;j} | i=1,2,\dots,D_j \}$ be the basis vectors of the $j$th vector space $V_j$, 
\begin{align}
\ket{\mathfrak{A}';L} &= \sum_{\alpha_1,\dots,\alpha_N} \sum_{i_1,\dots,i_N} 
C^{\alpha_1,\dots,\alpha_N} 
\braket{i_1;1}{\alpha_1} \braket{i_2;2}{\alpha_2} \dots \braket{i_N;N}{\alpha_N} \nonumber\\
&\qquad\qquad \times \ket{i_1;1} \otimes \ket{i_2;2} \otimes \dots \otimes \ket{i_N;N} .
\end{align}
In other words, the basis vectors of this tensor product space $V_\text{L}$ are formed from products of the basis vectors from each and every vector space $\{V_i\}$.

{\bf Dimension} \qquad If the $i$th vector space $V_i$ has dimension $D_i$, then the dimension of $V_L$ itself is $D_1 D_2 \dots D_{N-1} D_N$. The reason is, for a given tensor product $\ket{i_1;1} \otimes \ket{i_2;2} \otimes \dots \otimes \ket{i_N;N}$, there are $D_1$ choices for $\ket{i_1;1}$, $D_2$ choices for $\ket{i_2;2}$, and so on.

{\it Example} \qquad Suppose we tensor two copies of the 2-dimensional vector space that the Pauli operators $\{ \sigma^i \}$ act on. Each space is spanned by $\ket{\pm}$. The tensor product space is then spanned by the following 4 vectors
\begin{align}
\ket{1;L} = \ket{+} \otimes \ket{+}, \qquad \ket{2;L} = \ket{+} \otimes \ket{-}, \\
\ket{3;L} = \ket{-} \otimes \ket{+}, \qquad \ket{4;L} = \ket{-} \otimes \ket{-} .
\end{align}
(Note that this ordering of the vectors is of course {\it not} unique.)

{\bf Adjoint and Inner Product} \qquad Just as we can form tensor products of kets, we can do so for bras. We have
\begin{align}
\left(\ket{\alpha_1} \otimes \ket{\alpha_2} \otimes \dots \otimes \ket{\alpha_N}\right)^\dagger
= \bra{\alpha_1} \otimes \bra{\alpha_2} \otimes \dots \otimes \bra{\alpha_N} ,
\end{align}
where the $i$th slot from the left is a bra from the $i$th vector space $V_i$. We also have the inner product
\begin{align}
&\left(\bra{\alpha_1} \otimes \bra{\alpha_2} \otimes \dots \otimes \bra{\alpha_N}\right)
\left( c \ket{\beta_1} \otimes \ket{\beta_2} \otimes \dots \otimes \ket{\beta_N} + d \ket{\gamma_1} \otimes \ket{\gamma_2} \otimes \dots \otimes \ket{\gamma_N} \right) \nonumber\\
&= c \braket{\alpha_1}{\beta_1} \braket{\alpha_2}{\beta_2} \dots \braket{\alpha_N}{\beta_N} 
+ d \braket{\alpha_1}{\gamma_1} \braket{\alpha_2}{\gamma_2} \dots \braket{\alpha_N}{\gamma_N} ,
\end{align}
where $c$ and $d$ are complex numbers. For example, the orthonormal nature of the $\{ \ket{i_1;1} \otimes \dots \otimes \ket{i_N;N} \}$ follow from
\begin{align}
\left(\bra{j_1;1} \otimes \dots \otimes \bra{j_N;N}\right)\left(\ket{i_1;1} \otimes \dots \otimes \ket{i_N;N}\right)
&= \braket{j_1;1}{i_1;1} \braket{j_2;2}{i_2;2} \dots \braket{j_N;N}{i_N;N}  \nonumber\\
&= \delta^{j_1}_{i_1} \dots \delta^{j_N}_{i_N} .
\end{align}
{\bf Linear Operators} \qquad If $X_i$ is a linear operator acting on the $i$th vector space $V_i$, we can form a tensor product of them. Their operation is defined as
\begin{align}
&\left(X_1 \otimes X_2 \otimes \dots \otimes X_N\right)\left( c \ket{\beta_1} \otimes \ket{\beta_2} \otimes \dots \otimes \ket{\beta_N} + d \ket{\gamma_1} \otimes \ket{\gamma_2} \otimes \dots \otimes \ket{\gamma_N} \right) \\
&= c (X_1\ket{\beta_1}) \otimes (X_2\ket{\beta_2}) \otimes \dots \otimes (X_N\ket{\beta_N})
+ d (X_1\ket{\gamma_1}) \otimes (X_2\ket{\gamma_2}) \otimes \dots \otimes (X_N\ket{\gamma_N}) , \nonumber
\end{align}
where $c$ and $d$ are complex numbers.

The most general linear operator $Y$ acting on our tensor product space $V_\text{L}$ can be built out of the basis ket-bra operators.
\begin{align}
Y = \sum_{\substack{i_1, \dots, i_N \\ j_1, \dots, j_N}}
\left( \ket{i_1;1} \otimes \dots \otimes \ket{i_N;N} \right)
&\widehat{Y}^{i_1 \dots i_N}_{\phantom{i_1 \dots i_N} j_1 \dots j_N}
\left( \bra{j_1;1} \otimes \dots \otimes \bra{j_N;N} \right) , \\
&\widehat{Y}^{i_1 \dots i_N}_{\phantom{i_1 \dots i_N} j_1 \dots j_N} \in \mathbb{C} .
\end{align}
\begin{myP}
{\it Tensor transformations.} \qquad Consider the state
\begin{align}
\ket{\mathfrak{A}';L} &= \sum_{1 \leq i_1 \leq D_1} \sum_{1 \leq i_2 \leq D_2} \dots \sum_{1 \leq i_N \leq D_N} 
T^{i_1 i_2 \dots i_{N-1} i_N} \ket{i_1;1} \otimes \ket{i_2;2} \otimes \dots \otimes \ket{i_N;N} ,
\end{align}
where $\{\ket{i_j;j}\}$ are the $D_j$ orthonormal basis vectors spanning the $j$th vector space $V_j$, and $T^{i_1 i_2 \dots i_{N-1} i_N}$ are complex numbers. Consider a change of basis for each vector space, i.e., $\ket{i;j} \to \ket{i';j}$. By defining the unitary operator that implements this change-of-basis
\begin{align}
U &\equiv \,_{(1)}U \otimes \,_{(2)}U \otimes \dots \otimes \,_{(N)}U , \\
\,_{(i)}U &\equiv \sum_{1 \leq j \leq D_i} \ketbra{j';i}{j;i} ,
\end{align}
expand $\ket{\mathfrak{A}';L}$ in the new basis $\{\ket{j'_1;1} \otimes \dots \otimes \ket{j'_N;N}\}$; this will necessarily involve the $U^\dagger$'s. Define the coefficients of this new basis via
\begin{align}
\ket{\mathfrak{A}';L} &= \sum_{1 \leq i'_1 \leq D_1} \sum_{1 \leq i'_2 \leq D_2} \dots \sum_{1 \leq i'_N \leq D_N} 
T'^{i'_1 i'_2 \dots i'_{N-1} i'_N} \ket{i'_1;1} \otimes \ket{i'_2;2} \otimes \dots \otimes \ket{i'_N;N} .
\end{align}
Now relate $T'^{i'_1 i'_2 \dots i'_{N-1} i'_N}$ to the coefficients in the old basis $T^{i_1 i_2 \dots i_{N-1} i_N}$ using the matrix elements
\begin{align}
\left( \,_{(i)}\widehat{U}^\dagger \right)^{j}_{\phantom{j}k} \equiv \braOket{j;i}{\left( \,_{(i)}U \right)^\dagger}{k;i} .
\end{align}
Can you perform a similar change-of-basis for the following dual vector?
\begin{align}
\bra{\mathfrak{A}';L} &= \sum_{1 \leq i_1 \leq D_1} \sum_{1 \leq i_2 \leq D_2} \dots \sum_{1 \leq i_N \leq D_N} 
T_{i_1 i_2 \dots i_{N-1} i_N} \bra{i_1;1} \otimes \bra{i_2;2} \otimes \dots \otimes \bra{i_N;N} 
\end{align}
In differential geometry, tensors will transform in analogous ways. \qed
\end{myP}

\subsection{Continuous Spaces and Infinite $D-$Space}

For the final section we will deal with vector spaces with continuous spectra, with infinite dimensionality. To make this topic rigorous is beyond the scope of these notes; but the interested reader should consult the functional analysis portion of the math literature. Our goal here is a practical one: we want to be comfortable enough with continuous spaces to solve problems in quantum mechanics and (quantum and classical) field theory.

\subsubsection{Preliminaries: Dirac's $\delta$ and eigenket integrals}

{\bf Dirac's $\delta$-``function"} \qquad We will see that transitioning from discrete, finite dimensional vector spaces to continuous ones means summations become integrals; while Kronecker-$\delta$s will be replaced with Dirac-$\delta$ functions. In case the latter is not familiar, the Dirac-$\delta$ function of one variable is to be viewed as an object that occurs within an integral, and is defined via
\begin{align}
\int_{a}^{b} f(x') \delta(x'-x) \dd x' = f(x) ,
\end{align}
for all $a$ less than $x$ and all $b$ greater than $x$, i.e., $a < x < b$. This indicates $\delta(x'-x)$ has to be sharply peaked at $x'=x$ and zero everywhere, since the result of integral picks out the value of $f$ solely at $x$.

The Dirac $\delta$-function is often loosely viewed as $\delta(x) = 0$ when $x \neq 0$ and $\delta(x) = \infty$ when $x=0$. An alternate approach is to define $\delta(x)$ as a sequence of functions more and more sharply peaked at $x=0$, whose integral over the real line is unity. Three examples are
{\allowdisplaybreaks\begin{align}
\delta(x) 
&= \lim_{\epsilon \to 0^+} \Theta\left( \frac{\epsilon}{2}-|x| \right) \frac{1}{\epsilon} \\
&= \lim_{\epsilon \to 0^+} \frac{e^{-\frac{|x|}{\epsilon}}}{2 \epsilon} \\
&= \lim_{\epsilon \to 0^+} \frac{1}{\pi} \frac{\epsilon}{x^2 + \epsilon^2}
\end{align}}
For the first equality, $\Theta(z)$ is the step function, defined to be 
\begin{align}
\Theta(z) 
&= 1, \qquad \text{for $z>0$} \nonumber\\
&= 0, \qquad \text{for $z<0$} .		
\end{align}
\begin{myP}
\qquad Justify these three definitions of $\delta(x)$. What happens, for finite $x \neq 0$, when $\epsilon \to 0^+$? Then, by holding $\epsilon$ fixed, integrate them over the real line, before proceeding to set $\epsilon \to 0^+$. \qed
\end{myP}
For later use, we record the following integral representation of the Dirac $\delta$-function.
\begin{align}
\label{DiracDelta_IntegralRep}
\int_{-\infty}^{+\infty} \frac{\dd\omega}{2\pi} e^{i\omega(z-z')} = \delta(z-z') 
\end{align}
\begin{myP}
\qquad Can you justify the following?
\begin{align}
\Theta(z-z') = \int_{z_0}^{z} \dd z'' \delta(z''-z'), \qquad z' > z_0.
\end{align}
We may therefore assert the derivative of the step function is the $\delta$-function,
\begin{align}
\label{DiracDelta_ThetaPrime}
\Theta'(z-z') = \delta(z-z') .
\end{align} \qed
\end{myP}
A few properties of the $\delta$-function are worth highlighting.
\begin{itemize}
\item From eq. \eqref{DiracDelta_ThetaPrime} -- that a $\delta(z-z')$ follows from taking the derivative of a discontinuous function -- in this case, $\Theta(z-z')$ -- will be important for the study of Green's functions. 
\item If the argument of the $\delta$-function is a function $f$ of some variable $z$, then as long as $f'(z) \neq 0$ whenever $f(z)=0$, it may be re-written as
\begin{align}
\label{DiracDelta_Jacobian_1D}
\delta\left( f(z) \right) = \sum_{z_i \equiv i\text{th zero of $f(z)$}} \frac{\delta(z-z_i)}{|f'(z_i)|} .
\end{align}
To justify this we recall the fact that, the $\delta$-function itself is non-zero only when its argument is zero. This explains why we sum over the zeros of $f(z)$. Now we need to fix the coefficient of the $\delta$-function near each zero. That is, what are the $\varphi_i$'s in
\begin{align}
\delta\left( f(z) \right) = \sum_{z_i \equiv \text{$i$th zero of $f(z)$}} \frac{\delta(z-z_i)}{\varphi_i} ?
\end{align}
We now use the fact that integrating a $\delta$-function around the small neighborhood of the $i$th zero of $f(z)$ {\it with respect to $f$} has to yield unity. It makes sense to treat $f$ as an integration variable near its zero because we have assumed its slope is non-zero, and therefore near its $i$th zero,
\begin{align}
f(z) &= f'(z_i) (z-z_i) + \mathcal{O}((z-z_i)^2) , \\
\Rightarrow \qquad \dd f &= f'(z_i) \dd z + \mathcal{O}((z-z_i)^1) \dd z .
\end{align}
The integration around the $i$th zero reads, for $0 < \epsilon \ll 1$,
\begin{align}
1 
= \int_{z = z_i - \epsilon}^{z = z_i + \epsilon} \dd f \delta\left( f \right)
&= \int_{z = z_i - \epsilon}^{z = z_i + \epsilon} \dd z \left\vert \left( f'(z_i) + \mathcal{O}((z-z_i)^1) \right) \right\vert 
\frac{\delta\left( z-z_i \right)}{\varphi_i} \\
&\stackrel{\epsilon \to 0}{\to} \frac{\left\vert f'(z_i) \right\vert}{\varphi_i} .
\end{align}
(When you change variables within an integral, remember to include the absolute value of the Jacobian, which is essentially $|f'(z_i)|$ in this case.)  The $\mathcal{O}(z^p)$ means ``the next term in the series has a dependence on the variable $z$ that goes as $z^p$"; this first correction can be multiplied by other stuff, but has to be proportional to $z^p$. 
	
A simple application of eq. \eqref{DiracDelta_Jacobian_1D} is, for $a \in \mathbb{R}$,
\begin{align}
\delta(az) = \frac{\delta(z)}{|a|} .
\end{align}
\item Since $\delta(z)$ is non-zero only when $z=0$, it must be that $\delta(-z)=\delta(z)$ and more generally
\begin{align}
\delta(z-z') = \delta(z'-z) .
\end{align}
\item We may also take the derivative of a $\delta$-function. Under an integral sign, we may apply integration-by-parts as follows:
\begin{align}
\int_{a}^{b} \delta'(x-x') f(x) \dd x = [\delta(x-x') f(x)]_{x=a}^{x=b} - \int_{a}^{b} \delta(x-x') f'(x) \dd x = -f'(x')
\end{align}
as long as $x'$ lies strictly between $a$ and $b$, $a < x' < b$, where $a$ and $b$ are both real.
\item {\it Dimension} \qquad What is the dimension of the $\delta$-function? Turns out $\delta(\xi)$ has dimensions of $1/[\xi]$, i.e., the reciprocal of the dimension of its argument. The reason is
\begin{align}
\int \dd \xi \delta(\xi) = 1 \qquad \Rightarrow \qquad [\xi] \left[ \delta(\xi) \right] = 1 .
\end{align}
\end{itemize}
{\bf Continuous spectrum} \qquad Let $\Omega$ be a Hermitian operator whose spectrum is continuous; i.e., $\Omega \ket{\omega} = \omega \ket{\omega}$ with $\omega$ being a continuous parameter. If $\ket{\omega}$ and $\ket{\omega'}$ are both ``unit norm" eigenvectors of different eigenvalues $\omega$ and $\omega'$, we have
\begin{align}
\braket{\omega}{\omega'} = \delta(\omega-\omega') .
\end{align}
(This assumes a ``translation symmetry" in this $\omega$-space; we will see later how to modify this inner product when the translation symmetry is lost.) The completeness relation in eq. \eqref{CompletenessRelation} is given by
\begin{align}
\int \dd\omega \ket{\omega} \bra{\omega} = \mathbb{I} .
\end{align}
An arbitrary vector $\ket{\alpha}$ can be expressed as
\begin{align}
\ket{\alpha} = \int \dd\omega \ket{\omega} \braket{\omega}{\alpha} .
\end{align}
When the state is normalized to unity, we say
\begin{align}
\braket{\alpha}{\alpha} = \int \dd\omega \braket{\alpha}{\omega} \braket{\omega}{\alpha} = \int \dd\omega |\braket{\omega}{\alpha}|^2 = 1.
\end{align}
The inner product between arbitrary vectors $\ket{\alpha}$ and $\ket{\beta}$ now reads
\begin{align}
\braket{\alpha}{\beta} = \int \dd\omega \braket{\alpha}{\omega} \braket{\omega}{\beta} .
\end{align}
Since by assumption $\Omega$ is diagonal, i.e.,
\begin{align}
\Omega = \int \dd\omega \omega \ketbra{\omega}{\omega} ,
\end{align}
the matrix elements of $\Omega$ are
\begin{align}
\braOket{\omega}{\Omega}{\omega'} = \omega \delta(\omega-\omega') = \omega' \delta(\omega-\omega') .
\end{align}
Because of the $\delta$-function, it does not matter if we write $\omega$ or $\omega'$ on the right hand side.

\subsubsection{Continuous Operators, Translations, and the Fourier transform}

An important example that we will deal in detail here, is that of the eigenket of the position operator $\vec{X}$, where we assume there is some underlying infinite $D$-space $\mathbb{R}^D$. The arrow indicates the position operator itself has $D$ components, each one corresponding to a distinct axis of the $D$-dimensional Euclidean space. $\ket{\vec{x}}$ would describe the state that is (infinitely) sharply localized at the position $\vec{x}$; namely, it obeys the $D$-component equation
\begin{align}
\vec{X} \ket{\vec{x}} = \vec{x} \ket{\vec{x}} .
\end{align}
Or, in index notation,
\begin{align}
X^k \ket{\vec{x}} = x^k \ket{\vec{x}}, \qquad k \in \{ 1,2,\dots,D \} .
\end{align}
The position eigenkets are normalized as, in Cartesian coordinates,
\begin{align}
\braket{\vec{x}}{\vec{x}'} = \delta^{(D)}(\vec{x}-\vec{x}') 
\equiv \prod_{i=1}^D \delta(x^i - x'^i)
= \delta(x^1-x'^1) \delta(x^2-x'^2) \dots \delta(x^D-x'^D) .
\end{align}
\footnote{As an important aside, the generalization of the 1D transformation law in eq. \eqref{DiracDelta_Jacobian_1D} involving the $\delta$-function has the following higher dimensional generalization. If we are given a transformation $\vec{x} \equiv \vec{x}(\vec{y})$ and $\vec{x}' \equiv \vec{x}'(\vec{y}')$, then
\begin{align}
\label{DiracDelta_Jacobian_HigherD}
\delta^{(D)}\left( \vec{x}-\vec{x}' \right) 
= \frac{\delta^{(D)}(\vec{y}-\vec{y}')}{ \left\vert \det \partial x^a(\vec{y})/\partial y^b \right\vert } 
= \frac{\delta^{(D)}(\vec{y}-\vec{y}')}{ \left\vert \det \partial x'^a(\vec{y}')/\partial y'^b \right\vert } ,
\end{align}
where $\delta^{(D)}(\vec{x}-\vec{x}') \equiv \prod_{i=1}^D \delta(x^i-x'^i)$, $\delta^{(D)}(\vec{y}-\vec{y}') \equiv \prod_{i=1}^D \delta(y^i-y'^i)$, and the Jacobian inside the absolute value occurring in the denominator on the right hand side is the usual determinant of the matrix whose $a$th row and $b$th column is given by $\partial x^a(\vec{y})/\partial y^b$. (The second and third equalities follow from each other because the $\delta$-functions allow us to assume $\vec{y}=\vec{y}'$.) Equation \eqref{DiracDelta_Jacobian_HigherD} can be justified by demanding that its integral around the point $\vec{x}=\vec{x}'$ gives one. For $0 < \epsilon \ll 1$, and denoting $\delta^{(D)}(\vec{x}-\vec{x}') = \delta^{(D)}(\vec{y}-\vec{y}')/\varphi(\vec{y}')$,
\begin{align}
1 
= \int_{|\vec{x}-\vec{x}'| \leq \epsilon} \dd^D \vec{x} \delta^{(D)}(\vec{x}-\vec{x}')
= \int_{|\vec{x}-\vec{x}'| \leq \epsilon} \dd^D \vec{y} \left\vert \det \frac{\partial x^a(\vec{y})}{\partial y^b} \right\vert \frac{\delta^{(D)}(\vec{y}-\vec{y}')}{\varphi(\vec{y}')} 
= \frac{\left\vert \det \frac{\partial x'^a(\vec{y}')}{\partial y'^b} \right\vert}{\varphi(\vec{y}')} .
\end{align}}Any other vector $\ket{\alpha}$ in the Hilbert space can be expanded in terms of the position eigenkets.
\begin{align}
\ket{\alpha} = \int_{\mathbb{R}^D} \dd^D \vec{x} \ket{\vec{x}} \braket{\vec{x}}{\alpha} .
\end{align}
Notice $\braket{\vec{x}}{\alpha}$ is an ordinary (possibly complex) function of the spatial coordinates $\vec{x}$. We see that the space of functions emerges from the vector space spanned by the position eigenkets. Just as we can view $\braket{i}{\alpha}$ in $\ket{\alpha} = \sum_i \ket{i} \braket{i}{\alpha}$ as a column vector, the function $f(\vec{x}) \equiv \braket{\vec{x}}{f}$ is in some sense a continuous (infinite dimensional) ``vector" in this position representation.

In the context of {\it quantum mechanics} $\braket{\vec{x}}{\alpha}$ would be identified as a wave function, more commonly denoted as $\psi(\vec{x})$; in particular, $|\braket{\vec{x}}{\alpha}|^2$ is interpreted as the probability density that the system is localized around $\vec{x}$ when its position is measured. This is in turn related to the demand that the wave function obey $\int \dd^D \vec{x} |\braket{\vec{x}}{\alpha}|^2 = 1$. However, it is worth highlighting here that our discussion regarding the Hilbert spaces spanned by the position eigenkets $\{\ket{\vec{x}}\}$ (and later below, by their momentum counterparts $\{\vert \vec{k} \rangle\}$) does not necessarily have to involve quantum theory.\footnote{This is especially pertinent for those whose first contact with continuous Hilbert spaces was in the context of a quantum mechanics course.} We will provide concrete examples below, such as how the concept of Fourier transform emerges and how classical field theory problems -- the derivation of the Green's function of the Laplacian in eq. \eqref{GreensFunctionLaplacian}, for instance -- can be tackled using the methods/formalism delineated here.

{\bf Matrix elements} \qquad Suppose we wish to calculate the matrix element $\braOket{\alpha}{Y}{\beta}$ in the position representation. It is
\begin{align}
\braOket{\alpha}{Y}{\beta} 
&= \int \dd^D \vec{x} \int \dd^D \vec{x}' \braket{\alpha}{\vec{x}} \braOket{\vec{x}}{Y}{\vec{x}'} \braket{\vec{x}'}{\beta} \nonumber\\
&= \int \dd^D \vec{x} \int \dd^D \vec{x}' \braket{\vec{x}}{\alpha}^* \braOket{\vec{x}}{Y}{\vec{x}'} \braket{\vec{x}'}{\beta} .
\end{align}
If the operator $Y(\vec{X})$ were built solely from the position operator $\vec{X}$, then 
\begin{align}
\braOket{\vec{x}}{Y(\vec{X})}{\vec{x}'} 
= Y(\vec{x}) \delta^{(D)}(\vec{x}-\vec{x}') 
= Y(\vec{x}') \delta^{(D)}(\vec{x}-\vec{x}') ;
\end{align}
and the double integral collapses into one,
\begin{align}
\braOket{\alpha}{Y(\vec{X})}{\beta} &= \int \dd^D \vec{x} \braket{\vec{x}}{\alpha}^* \braket{\vec{x}'}{\beta} Y(\vec{x}) .
\end{align}
\begin{myP}
\qquad Show that if $U$ is a unitary operator and $\ket{\alpha}$ is an arbitrary vector, then $\ket{\alpha}$, $U \ket{\alpha}$ and $U^\dagger \ket{\alpha}$ have the same norm. \qed
\end{myP}

{\bf Continuous unitary operators} \qquad Translations and rotation are examples of operations that involve continuous parameter(s) -- for translation it involves a displacement vector; for rotation we have to specify the axis of rotation as well as the angle of rotation itself. 
\begin{quotation}
	{\it Exp(anti-Hermitian operator) is unitary} \qquad It will often be the case that, when realized as linear operators on some Hilbert space, these continuous operators will be unitary. Suppose further, when their parameters $\vec{\xi}$ -- which we will assume to be real -- are tuned to zero $\vec{0}$, the identity operator is recovered. When these conditions are satisfied, the continuous unitary operator $U$ can (in most cases of interest) be expressed as the exponential of the anti-Hermitian operator $-i \vec{\xi} \cdot \vec{K}$, namely
\begin{align}
\label{ExpoentialMap}
U(\vec{\xi}) = \exp\left( -i \vec{\xi} \cdot \vec{K} \right).
\end{align}
\end{quotation}
To be clear, we are allowing for $N \geq 1$ continuous real parameter(s), which we collectively denote as $\vec{\xi}$. The Hermitian operator will also have $N$ distinct components, so $\vec{\xi} \cdot \vec{K} \equiv \sum_i \xi^i K_i$. 

To check the unitary nature of eq. \eqref{ExpoentialMap}, we first record that the exponential of an operator $X$ is defined through the Taylor series
\begin{align}
e^X \equiv \mathbb{I} + X + \frac{X^2}{2!} + \frac{X^3}{3!} + \dots = \sum_{\ell=0}^\infty \frac{X^\ell}{\ell !} .
\end{align}
For $X =  -i \vec{\xi} \cdot \vec{K}$, where $\vec{\xi}$ is real and $\vec{K}$ is Hermitian, we have
\begin{align}
X^\dagger = \left(-i \vec{\xi} \cdot \vec{K}\right)^\dagger = i \vec{\xi} \cdot \vec{K}^\dagger = i \vec{\xi} \cdot \vec{K} = -X .
\end{align} 
That is, $X$ is anti-Hermitian. Now, for $\ell$ integer, $(X^\ell)^\dagger = (X^\dagger)^\ell = (-X)^\ell$. Thus,
\begin{align}
(e^X)^\dagger 
= \sum_{\ell=0}^\infty \frac{(X^\ell)^\dagger}{\ell !} 
= \sum_{\ell=0}^\infty \frac{(-X)^\ell}{\ell !} = e^{-X} .
\end{align}
Generically, operators do not commute $AB \neq BA$. (Such a non-commuting example is rotation.) In that case, $\exp(A+B) \neq \exp(A) \exp(B)$. However, because $X$ and $-X$ do commute, we can check the unitary nature of $\exp X$ by Taylor expanding each of the exponentials in $\exp(X) (\exp(X))^\dagger = \exp(X) \exp(-X)$ and finding that the series can be re-arranged to that of $\exp(X-X)=\mathbb{I}$. Specifically, whenever $A$ and $B$ {\it do} commute
\begin{align}
\exp(A) \exp(B) 
&= \sum_{\ell_1,\ell_2 = 0}^\infty \frac{A^{\ell_1} B^{\ell_2}}{\ell_1! \ell_2!} 
= \sum_{\ell=0}^\infty \frac{1}{\ell !} \sum_{s=0}^{\ell} \binom{\ell}{s} A^s B^{\ell-s} \\
&= \sum_{\ell=0}^\infty \frac{(A+B)^\ell}{\ell !}
= \exp(A+B) .
\end{align}
%\begin{myP}
%\qquad Suppose $Y$ is some hermitian operator with known eigenvalues. Let $f$ be some smooth function that admits a Taylor series. What does $\exp[if(Y)]$ mean? Consider both the case where $Y$ has a discrete spectrum and when it has a continuous one. \qed
%\end{myP}
{\bf Translation in $\mathbb{R}^D$} \qquad To make these ideas regarding continuous operators more concrete, we will now study the case of translation in some detail, realized on a Hilbert space spanned by the position eigenkets $\{\ket{\vec{x}}\}$. To be specific, let $\mathcal{T}(\vec{d})$ denote the translation operator parameterized by the displacement vector $\vec{d}$. We shall work in $D$ space dimensions. We define the translation operator by its action
\begin{align}
\label{TranslationOperator_Property1}
\mathcal{T}(\vec{d}) \ket{\vec{x}} = \ket{\vec{x}+\vec{d}} .
\end{align}
Since $\ket{\vec{x}}$ and $|\vec{x}+\vec{d}\rangle$ can be viewed as distinct elements of the set of basis vectors, we shall see that the translation operator can be viewed as a unitary operator, changing basis from $\{\ket{\vec{x}} | \vec{x} \in \mathbb{R}^D \}$ to $\{ \vert\vec{x}+\vec{d}\rangle | \vec{x} \in \mathbb{R}^D \}$. The inverse transformation is
\begin{align}
\label{TranslationOperator_Property2}
\mathcal{T}(\vec{d})^\dagger \ket{\vec{x}} = \ket{\vec{x}-\vec{d}} .
\end{align}
Of course we have the identity operator $\mathbb{I}$ when $\vec{d}=\vec{0}$,
\begin{align}
\label{TranslationOperator_Property3}
\mathcal{T}(\vec{0}) \ket{\vec{x}} = \ket{\vec{x}}  \qquad \Rightarrow \qquad \mathcal{T}(\vec{0}) = \mathbb{I}.
\end{align}
The following composition law has to hold
\begin{align}
\label{TranslationOperator_Property4}
\mathcal{T}(\vec{d}_1) \mathcal{T}(\vec{d}_2) = \mathcal{T}(\vec{d}_1+\vec{d}_2),
\end{align}
because translation is commutative
\begin{align}
\mathcal{T}(\vec{d}_1) \mathcal{T}(\vec{d}_2) \ket{\vec{x}} = \mathcal{T}(\vec{d}_1) \ket{\vec{x}+\vec{d}_2} 
= \ket{\vec{x}+\vec{d}_2+\vec{d}_1} = \ket{\vec{x}+\vec{d}_1+\vec{d}_2} 
= \mathcal{T}(\vec{d}_1+\vec{d}_2) \ket{\vec{x}}.
\end{align}
\begin{myP}
{\it Translation operator is unitary.} \qquad Show that 
\begin{align}
\label{TranslationOperator_EuclideanSpace}
\mathcal{T}(\vec{d}) = \int_{\mathbb{R}^D} \dd^D \vec{x}' \ketbra{\vec{d} + \vec{x}'}{\vec{x}'}
\end{align}
satisfies eq. \eqref{TranslationOperator_Property1} and therefore is the correct ket-bra operator representation of the translation operator. Check explicitly that $\mathcal{T}(\vec{d})$ is unitary.	\qed
\end{myP}
{\it Momentum operator} \qquad Since eq. \eqref{TranslationOperator_EuclideanSpace} tells us the translation operator is unitary, we may now invoke the form of the continuous unitary operator in eq. \eqref{ExpoentialMap} to deduce
\begin{align}
\label{TranslationOperator}
\mathcal{T}(\vec{\xi}) 
= \exp\left( -i \vec{\xi} \cdot \vec{P} \right) 
= \exp\left( -i \xi^k P_k \right) 
\end{align}
We will call the Hermitian operator $\vec{P}$ the momentum operator.\footnote{Strictly speaking $P_j$ here has dimensions of $1/$[length], whereas the momentum you might be familiar with has units of [mass $\times$ length/time$^2$].} In this exp form, eq. \eqref{TranslationOperator_Property4} reads
\begin{align}
\exp\left( -i \vec{d}_1 \cdot \vec{P} \right) \exp\left( -i \vec{d}_2 \cdot \vec{P} \right) = \exp\left( -i (\vec{d}_1 + \vec{d}_2) \cdot \vec{P} \right) .
\end{align}
{\it Translation invariance} \qquad Infinite (flat) $D$-space $\mathbb{R}^D$ is the same everywhere and in every direction. This intuitive fact is intimately tied to the property that $\mathcal{T}(\vec{d})$ is a unitary operator: it just changes one orthonormal basis to another, and physically speaking, there is no privileged set of basis vectors. For instance, the norm of vectors is position independent:
\begin{align}
\braket{\vec{x}+\vec{d}}{\vec{x}'+\vec{d}} = \delta^{(D)}\left(\vec{x}-\vec{x}'\right) = \braket{\vec{x}}{\vec{x}'} .
\end{align}
As we will see below, if we confine our attention to some finite domain in $\mathbb{R}^D$ or if space is no longer flat, then (global) translation symmetry is lost and the translation operator still exists but is no longer unitary.\footnote{When we restrict the domain to a finite one embedded within flat $\mathbb{R}^D$, there is still local translation symmetry in that, performing the same experiment at $\vec{x}$ and at $\vec{x}'$ should not lead to any physical differences {\it as long as} both $\vec{x}$ and $\vec{x}'$ lie within the said domain. But global translation symmetry is ``broken" because, the domain is ``here" and not ``there"; as illustrated by the 1D example, translating too much in one direction would bring you out of the domain.} In particular, when the domain is finite eq. \eqref{TranslationOperator_EuclideanSpace} may no longer make sense; a 1D example would be to consider $\{ \ket{z} \vert 0 \leq z \leq L \}$,
\begin{align}
\label{TranslationOperator_FinitBox}
\mathcal{T}(d > 0) \stackrel{?}{=} \int_0^L \dd z' \ketbra{z'+d}{z'} .
\end{align}
When $z' = L$, say, the bra in the integrand is $\bra{L}$ but the ket $\ket{L+d}$ would make no sense because $L+d$ lies outside the domain.

{\it Commutation relations between $X^i$ and $P_j$} \qquad We have seen, just from postulating a Hermitian position operator $X^i$, and considering the translation operator acting on the space spanned by its eigenkets $\{ \ket{\vec{x}} \}$, that there exists a Hermitian momentum operator $P_j$ that occurs in the exponent of said translation operator. This implies the continuous space at hand can be spanned by either the position eigenkets $\{ \ket{\vec{x}} \}$ or the momentum eigenkets, which obey
\begin{align}
P_j \ketk = k_j \ketk .
\end{align}
Are the position and momentum operators simultaneously diagonalizable? Can we label a state with both position and momentum? The answer is no.

To see this, we now consider an infinitesimal displacement operator $\mathcal{T}(\dd\vec{\xi})$.
\begin{align}
\vec{X} \mathcal{T}(\dd\vec{\xi}) \ket{\vec{x}} = \vec{X} \ket{\vec{x}+\dd\vec{\xi}} = (\vec{x} + \dd\vec{\xi}) \ket{\vec{x}+\dd\vec{\xi}} ,
\end{align}
and
\begin{align}
\mathcal{T}(\dd\vec{\xi}) \vec{X} \ket{\vec{x}} = \vec{x} \ket{\vec{x}+\dd\vec{\xi}} .
\end{align}
Since $\ket{\vec{x}}$ was an arbitrary vector, we may subtract the two equations 
\begin{align}
\left[ \vec{X}, \mathcal{T}(\dd\vec{\xi}) \right] \ket{\vec{x}} 
= \dd\vec{\xi} \ket{\vec{x}+\dd\vec{\xi}} 
= \dd\vec{\xi} \ket{\vec{x}} + \mathcal{O}\left( \dd\vec{\xi}^2 \right) .
\end{align}
At first order in $\dd\vec{\xi}$, we have the operator identity
\begin{align}
\label{Commutation_X_and_Translation}
\left[ \vec{X}, \mathcal{T}(\dd\vec{\xi}) \right] = \dd\vec{\xi} .
\end{align}
The left hand side involves operators, but the right hand side only real numbers. At this point we invoke eq. \eqref{TranslationOperator}, and deduce, for infinitesimal displacements,
\begin{align}
\mathcal{T}(\dd\vec{\xi}) = 1 - i \dd\vec{\xi} \cdot \vec{P} + \mathcal{O}(\dd\vec{\xi}^2) 
\end{align}
which in turn means eq. \eqref{Commutation_X_and_Translation} now reads, as $\dd\vec{\xi} \to \vec{0}$,
\begin{align}
\left[ \vec{X}, -i \dd\vec{\xi} \cdot \vec{P} \right] &= \dd\vec{\xi} \\
\left[ X^l, P_j \right] \dd \xi^j &= i \delta^l_j \dd\xi^j \qquad \text{(the $l$th component)}
\end{align}
Since the $\{\dd\xi^j\}$ are independent, the coefficient of $\dd\xi^j$ on both sides must be equal. This leads us to the fundamental commutation relation between $k$th component of the position operator with the $j$ component of the momentum operator:
\begin{align}
\label{Commutation_X_and_P}
\left[ X^k, P_j \right] &= i \delta^k_j, \qquad j,k \in \{1,2,\dots,D\} .
\end{align}
To sum: although $X^k$ and $P_j$ are both Hermitian operators in infinite flat $\mathbb{R}^D$, we see they are incompatible and thus, to span the continuous vector space at hand we can use either the eigenkets of $X^i$ or that of $P_j$ but not both. We will, in fact, witness below how changing from the position to momentum eigenket basis gives rise to the Fourier transform and its inverse.
\begin{align}
\ket{f} &= \int_{\mathbb{R}^D} \dd^D \vec{x}' \ket{\vec{x}'} \braket{\vec{x}'}{f}, \qquad X^i \ket{\vec{x}'} = x'^i \ket{\vec{x}'} \\
\ket{f} &= \int_{\mathbb{R}^D} \dd^D \vec{k}' \ket{\vec{k}'} \braket{\vec{k}'}{f}, \qquad P_j \ket{\vec{k}'} = k'_j \ket{\vec{k}'} .
\end{align}
For those already familiar with quantum theory, notice there is no $\hbar$ on the right hand side; nor will there be any throughout this section. This is not because we have ``set $\hbar=1$" as is commonly done in theoretical physics literature. Rather, it is because we wish to reiterate that the linear algebra of continuous operators, just like its discrete finite dimension counterparts, is really an independent structure on its own. Quantum theory is merely one of its application, albeit a very important one.
\begin{myP}
\qquad Because translation is commutative, $\vec{d}_1 + \vec{d}_2 = \vec{d}_2 + \vec{d}_1$, argue that the translation operators commute:
\begin{align}
\left[ \mathcal{T}(\vec{d}_1), \mathcal{T}(\vec{d}_2) \right] = 0.
\end{align}
By considering infinitesimal displacements $\vec{d}_1 = \dd\vec{\xi}_1$ and $\vec{d}_2 = \dd\vec{\xi}_2$, show that eq. \eqref{TranslationOperator} leads to us to conclude that momentum operators commute among themselves,
\begin{align}
[P_i,P_j] = 0, \qquad i,j \in\{1,2,3,\dots,D\} .
\end{align} \qed
\end{myP}
\begin{myP}
\qquad Let $\ketk$ be an eigenket of the momentum operator $\vec{P}$. Is $\ketk$ an eigenvector of $\mathcal{T}(\vec{d})$? If so, what is the corresponding eigenvalue? \qed
\end{myP}
\begin{myP}
\qquad Derive the momentum operator $\vec{P}$ in the position eigenket basis,
\begin{align}
\label{MomentumOperator_PositionRep}
\braOket{\vec{x}}{\vec{P}}{\alpha} = -i \frac{\partial}{\partial \vec{x}} \braket{\vec{x}}{\alpha} ,
\end{align}
for an arbitrary state $\ket{\alpha}$. Hint: begin with
\begin{align}
\braOket{\vec{x}}{\mathcal{T}(\dd\vec{\xi})}{\alpha} = \braket{\vec{x}-\dd\vec{\xi}}{\alpha} .
\end{align}
Taylor expand both the operator $\mathcal{T}(\dd\vec{\xi})$ as well as the function $\braket{\vec{x}-\dd\vec{\xi}}{\alpha}$, and take the $\dd\vec{\xi} \to \vec{0}$ limit, keeping only the $\mathcal{O}(\dd\vec{\xi})$ terms. 

Next, check that this representation of $\vec{P}$ is consistent with eq. \eqref{Commutation_X_and_P} by considering
\begin{align}
\braOket{\vec{x}}{\left[ X^k, P_j \right]}{\alpha} &= i \delta^k_j \braket{\vec{x}}{\alpha} .
\end{align}
Start by expanding the commutator on the left hand side, and show that you can recover eq. \eqref{MomentumOperator_PositionRep}. \qed
\end{myP}
\begin{myP}
\qquad Express the following matrix element in the position space representation
\begin{align}
\braOket{\alpha}{\vec{P}}{\beta} = \int \dd^D \vec{x} \Big( \qquad ? \qquad \Big) .
\end{align} \qed
\end{myP}
\begin{myP}
\qquad Show that the negative of the Laplacian, namely 
\begin{align}
-\vec{\nabla}^2 \equiv -\sum_i \frac{\partial}{\partial x^i} \frac{\partial}{\partial x^i} \qquad \text{(in Cartesian coordinates $\{ x^i \}$)},
\end{align}
is the square of the momentum operator. That is, for an arbitrary state $\ket{\alpha}$, show that
\begin{align}
\label{MomentumSquaredIsNegativeLaplacian}
\braOket{\vec{x}}{\vec{P}^2}{\alpha} = - \delta^{ij} \frac{\partial}{\partial x^i} \frac{\partial}{\partial x^j} \braket{\vec{x}}{\alpha} 
\equiv -\vec{\nabla}^2 \braket{\vec{x}}{\alpha} .
\end{align} \qed
\end{myP}
\begin{myP}
{\it Translation as Taylor series.} \qquad Use equations \eqref{TranslationOperator} and \eqref{MomentumOperator_PositionRep} to infer, for an arbitrary state $\ket{f}$,
\begin{align}
\braket{\vec{x} + \vec{\xi}}{f} = \exp\left( \vec{\xi} \cdot \frac{\partial}{\partial \vec{x}}\right) \braket{\vec{x}}{f} .
\end{align}
Compare the right hand side with the Taylor expansion of the function $f(\vec{x}+\vec{\xi})$ about $\vec{x}$. \qed
\end{myP}
%
%\underline{\it Problem:} \qquad Show that, if $G$ and $F$ admit a Taylor series,
%\begin{align}
%\left[ \vec{X}, G(\vec{P}) \right] = i \frac{\partial G}{\partial \vec{P}}, \qquad
%\left[ \vec{P}, F(\vec{X}) \right] = -i \frac{\partial F}{\partial \vec{X}} .
%\end{align}
%The $\vec{X}$ and $\vec{P}$ are the position and momentum operators.
\begin{myP}
\qquad Prove the Campbell-Baker-Hausdorff lemma. For linear operators $A$ and $B$, and complex number $\alpha$,
\begin{align}
e^{i \alpha A} B e^{-i \alpha A} = \sum_{\ell=0}^\infty \frac{(i\alpha)^\ell}{\ell !} \underbrace{[A,[A,\dots[A}_{\text{$\ell$ of these}},B]]] ,
\end{align}
where the $\ell=0$ term is understood to be just $B$. (Hint: Taylor expand the left-hand-side and use mathematical induction.) 

Next, consider the expectation values of the position $\vec{X}$ and momentum $\vec{P}$ operator with respect to a general state $\ket{\psi}$:
\begin{align}
\braOket{\psi}{\vec{X}}{\psi} \qquad \text{ and } \qquad \braOket{\psi}{\vec{P}}{\psi} .
\end{align}
What happens to these expectation values when we replace $\ket{\psi} \to \mathcal{T}(\vec{d}) \ket{\psi}$? \qed
\end{myP}
{\bf (Lie) Group theory} \qquad Our discussion here on the unitary operator $\mathcal{T}$ that implements translations on the Hilbert space spanned by the position eigenkets $\{\ket{\vec{x}}\}$, is really an informal introduction to the theory of continuous groups. The collection of continuous unitary translation operators $\{\mathcal{T}(\vec{d})\}$ forms a group, which like a vector space is defined by a set of axioms. Continuous unitary group elements that can be brought to the identity operator, by setting the continuous real parameters to zero, can always be expressed in the exponential form in eq. \eqref{ExpoentialMap}. The Hermitian operators $\vec{K}$, that is said to ``generate" the group elements, may obey non-trivial commutation relations (aka Lie algebra). For instance, because rotation operations in Euclidean space do not commute -- rotating about the $z$-axis followed by rotation about the $x$-axis, is not the same as rotation about the $x$-axis followed by about the $z$-axis -- their corresponding unitary operators acting on the Hilbert space spanned by $\{\ket{\vec{x}}\}$ will give rise to, in 3 dimensional space, $[K_i, K_j] = i \epsilon_{ijl} K_l$, for $i,j,l \in \{1,2,3\}$. 

{\bf Fourier analysis} \qquad We will now show how the concept of a Fourier transform readily arises from the formalism we have developed so far. To initiate the discussion we start with eq. \eqref{MomentumOperator_PositionRep}, with $\ket{\alpha}$ replaced with a momentum eigenket $\ketk$. This yields the eigenvalue/vector equation for the momentum operator in the position representation.
\begin{align}
\label{xk_PDE}
\braOket{\vec{x}}{\vec{P}}{\vec{k}} = \vec{k} \langle \vec{x} | \vec{k} \rangle 
= -i \frac{\partial}{\partial \vec{x}} \langle \vec{x} | \vec{k} \rangle , \qquad \Leftrightarrow \qquad
k_j \langle \vec{x} | \vec{k} \rangle = -i \frac{\partial \langle \vec{x} | \vec{k} \rangle}{\partial x^j} .
\end{align}
In $D$-space, this is a set of $D$ first order differential equations for the function $\langle \vec{x} | \vec{k} \rangle$. Via a direct calculation you can verify that the solution to eq. \eqref{xk_PDE} is simply the plane wave
\begin{align}
\label{xk_PDE_Soln}
\langle \vec{x} | \vec{k} \rangle = \chi \exp\left( i \vec{k} \cdot \vec{x} \right) .
\end{align}
where $\chi$ is complex constant to be fixed in the following way. We want
\begin{align}
\int_{\mathbb{R}^D} \dd^D k \langle \vec{x} | \vec{k} \rangle \langle \vec{k} | \vec{x}' \rangle 	
&= \braket{\vec{x}}{\vec{x}'} = \delta^{(D)}(\vec{x}-\vec{x}') .
\end{align}
Using the plane wave solution,
\begin{align}
(2\pi)^D |\chi|^2 \int \frac{\dd^D k}{(2\pi)^D} e^{i\vec{k}\cdot(\vec{x}-\vec{x}')} &= \delta^{(D)}(\vec{x}-\vec{x}') .
\end{align}
Now, recall the representation of the $D$-dimensional $\delta$-function
\begin{align}
\label{FourierExpansion_DiracDelta}
\int_{\mathbb{R}^D} \frac{\dd^D k}{(2\pi)^D} e^{i \vec{k} \cdot (\vec{x}-\vec{x}')} = \delta^{(D)}(\vec{x}-\vec{x}') .
\end{align}
Therefore, up to an overall multiplicative phase $e^{i\delta}$, which we will choose to be unity, $\chi = 1/(2\pi)^{D/2}$ and eq. \eqref{xk_PDE_Soln} becomes
\begin{align}
\label{PlaneWave_UnitaryOperator}
\langle \vec{x} | \vec{k} \rangle = (2\pi)^{-D/2} \exp\left( i \vec{k} \cdot \vec{x} \right) .
\end{align}
By comparing eq. \eqref{PlaneWave_UnitaryOperator} with eq. \eqref{UnitaryOperatorAsChangeOfBasis_MatrixElement}, we see that the plane wave in eq. \eqref{PlaneWave_UnitaryOperator} can be viewed as the matrix element of the unitary operator implementing the change-of-basis from position to momentum space, and vice versa.

We may now examine how the position representation of an arbitrary state $\braket{\vec{x}}{f}$ can be expanded in the momentum eigenbasis.
\begin{align}
\label{FourierExpansion_I}
\braket{\vec{x}}{f} 
= \int_{\mathbb{R}^D} \dd^D \vec{k} \langle \vec{x} | \vec{k} \rangle \braket{\vec{k}}{f} 
= \int_{\mathbb{R}^D} \frac{\dd^D \vec{k}}{(2\pi)^{D/2}} e^{i \vec{k} \cdot \vec{x}} \braket{\vec{k}}{f}
\end{align}
Similarly, we may expand the momentum representation of an arbitrary state $\braket{\vec{k}}{f}$ in the position eigenbasis.
\begin{align}
\label{FourierExpansion_II}
\braket{\vec{k}}{f} 
= \int_{\mathbb{R}^D} \dd^D \vec{x} \braket{\vec{k}}{\vec{x}} \braket{\vec{x}}{f} 
= \int_{\mathbb{R}^D} \frac{\dd^D \vec{x}}{(2\pi)^{D/2}} e^{-i \vec{k} \cdot \vec{x}} \braket{\vec{x}}{f}
\end{align}
Equations \eqref{FourierExpansion_I} and \eqref{FourierExpansion_II} are nothing but the Fourier expansion of some function $f(\vec{x})$ and its inverse transform.\footnote{A warning on conventions: everywhere else in these notes, our Fourier transform conventions will be $\int \dd^D k/(2\pi)^D$ for the momentum integrals and $\int \dd^D x$ for the position space integrals. This is just a matter of where the $(2\pi)$s are allocated, and no math/physics content is altered.} 

{\it Plane waves as orthonormal basis vectors} \qquad For practical calculations, it is of course cumbersome to carry around the position $\{ \ket{\vec{x}} \}$ or momentum eigenkets $\{ \vert \vec{k} \rangle \}$. As far as the space of functions in $\mathbb{R}^D$ is concerned, i.e., if one works solely in terms of the components $f(\vec{x}) \equiv \braket{\vec{x}}{f}$, as opposed to the space spanned by $\ket{\vec{x}}$, then one can view the plane waves $\{ \exp(i\vec{k}\cdot\vec{x})/(2\pi)^{D/2} \}$ in the Fourier expansion of eq. \eqref{FourierExpansion_I} as the orthonormal basis vectors. The coefficients of the expansion are then the $\widetilde{f}(\vec{k}) \equiv \langle \vec{k} \vert f \rangle$.
\begin{align}
f(\vec{x}) &= \int_{\mathbb{R}^D} \frac{\dd^D \vec{k}}{(2\pi)^{D/2}} e^{i \vec{k} \cdot \vec{x}} \widetilde{f}(\vec{k})
\end{align}
By multiplying both sides by $\exp(-i\vec{k}'\cdot\vec{x})/(2\pi)^{D/2}$, integrating over all space, using the integral representation of the $\delta$-function in eq. \eqref{DiracDelta_IntegralRep}, and finally replacing $\vec{k}' \to \vec{k}$,
\begin{align}
\widetilde{f}(\vec{k}) = \int_{\mathbb{R}^D} \frac{\dd^D \vec{x}}{(2\pi)^{D/2}} e^{-i \vec{k} \cdot \vec{x}} f(\vec{x}) .
\end{align}
\begin{myP}
\qquad Prove that, for the eigenstate of momentum $\ketk$, arbitrary states $\ket{\alpha}$ and $\ket{\beta}$,
\begin{align}
\braOket{\vec{k}}{\vec{X}}{\alpha} 	&= i \frac{\partial}{\partial \vec{k}} \braket{\vec{k}}{\alpha} \\
\braOket{\beta}{\vec{X}}{\alpha} 	&= \int \dd^D \vec{k} \braket{\vec{k}}{\beta}^* i \frac{\partial}{\partial \vec{k}} \braket{\vec{k}}{\alpha} .
\end{align}
The $\vec{X}$ is the position operator. \qed
\end{myP}
\begin{myP}
\qquad Consider the function, with $d>0$,
\begin{align}
\braket{\vec{x}}{\psi} = \left(\sqrt{\pi} d\right)^{-D/2} e^{i\vec{k}\cdot\vec{x}} \exp\left( -\frac{\vec{x}^2}{2 d^2} \right) .
\end{align}
Compute $\braket{\vec{k}'}{\psi}$, the state $\ket{\psi}$ in the momentum eigenbasis. Let $\vec{X}$ and $\vec{P}$ denote the position and momentum operators. Calculate the following expectation values:
\begin{align}
\braOket{\psi}{\vec{X}}{\psi}, \qquad \braOket{\psi}{\vec{X}^2}{\psi}, \qquad \braOket{\psi}{\vec{P}}{\psi}, \qquad \braOket{\psi}{\vec{P}^2}{\psi} .
\end{align}
What is the value of
\begin{align}
\left( \braOket{\psi}{\vec{X}^2}{\psi} - \braOket{\psi}{\vec{X}}{\psi}^2 \right)
\left( \braOket{\psi}{\vec{P}^2}{\psi} - \braOket{\psi}{\vec{P}}{\psi}^2 \right)?
\end{align} 
Hint: In this problem you will need the following results
\begin{align}
\int_{-\infty}^{+\infty} \dd x e^{-a (x + i y)^2} = \int_{-\infty}^{+\infty} \dd x e^{-a x^2} = \sqrt{\frac{\pi}{a}}, \qquad a > 0, y \in \mathbb{R} .
\end{align} 
If you encounter an integral of the form
\begin{align}
\int_{\mathbb{R}^D} \dd^D \vec{x}' e^{-\alpha \vec{x}^2} e^{i \vec{x} \cdot (\vec{q} - \vec{q}')} , \qquad \alpha > 0,
\end{align}
you should try to combine the exponents and ``complete the square". \qed
\end{myP}
{\bf Translation in momentum space} \qquad We have discussed how to implement translation in position space using the momentum operator $\vec{P}$, namely $\mathcal{T}(\vec{d}) = \exp(-i \vec{d} \cdot \vec{P})$. What would be the corresponding translation operator in momentum space?\footnote{This question was suggested by Jake Leistico, who also correctly guessed the essential form of eq. \eqref{TranslationInMomentumSpace}.} That is, what is $\widetilde{\mathcal{T}}$ such that
\begin{align}
\widetilde{\mathcal{T}}(\vec{d}) \ketk = \ket{\vec{k} + \vec{d}}, \qquad P_j \ketk = k_j \ketk ?
\end{align}
Of course, one representation would be the analog of eq. \eqref{TranslationOperator_EuclideanSpace}.
\begin{align}
\widetilde{\mathcal{T}}(\vec{d}) = \int_{\mathbb{R}^D} \dd^D \vec{k}' \ketbra{\vec{k}'+\vec{d}}{\vec{k}'}
\end{align}
But is there an exponential form, like there is one for the translation in position space (eq. \eqref{TranslationOperator})? We start with the observation that the momentum eigenstate $\ketk$ can be written as a superposition of the position eigenkets using eq. \eqref{PlaneWave_UnitaryOperator},
\begin{align}
\ketk 
= \int_{\mathbb{R}^D} \dd^D \vec{x}' \ket{\vec{x}'} \left\langle \vec{x}' \left\vert \vec{k} \right\rangle \right. 
= \int_{\mathbb{R}^D} \frac{\dd^D \vec{x}'}{(2\pi)^{D/2}} e^{i \vec{k} \cdot \vec{x}'} \ket{\vec{x}'} .
\end{align}
Now consider
\begin{align}
\exp(+i \vec{d} \cdot \vec{X}) \ketk 
&= \int_{\mathbb{R}^D} \frac{\dd^D \vec{x}'}{(2\pi)^{D/2}} e^{i \vec{k} \cdot \vec{x}'} e^{i \vec{d} \cdot \vec{x}'} \ket{\vec{x}'} \nonumber\\
&= \int_{\mathbb{R}^D} \frac{\dd^D \vec{x}'}{(2\pi)^{D/2}} e^{i (\vec{k} + \vec{d})\cdot \vec{x}'} \ket{\vec{x}'}
= \ket{\vec{k}+\vec{d}} .
\end{align}
That means
\begin{align}
\label{TranslationInMomentumSpace}
\widetilde{\mathcal{T}}(\vec{d}) = \exp\left( i \vec{d} \cdot \vec{X} \right) .
\end{align}
{\bf Spectra of $\vec{P}$ and $\vec{P}^2$ in infinite $\mathbb{R}^D$} \qquad We conclude this section by summarizing the several interpretations of the plane waves $\{ \langle \vec{x} \vert \vec{k} \rangle \equiv \exp(i\vec{k} \cdot \vec{x})/(2\pi)^{D/2} \}$. 
\begin{enumerate}
\item They can be viewed as the orthonormal basis vectors (in the $\delta$-function sense) spanning the space of complex functions on $\mathbb{R}^D$.
\item They can be viewed as the matrix element of the unitary operator $U$ that performs a change-of-basis between the position and momentum eigenbasis, namely $U \ket{\vec{x}} = \vert \vec{k} \rangle$.
\item They are simultaneous eigenstates of the momentum operators $\{ -i \partial_j \equiv -i \partial/\partial x^j \vert j = 1,2,\dots,D \}$ and the negative Laplacian $-\vec{\nabla}^2$ in the position representation.
\begin{align}
-\vec{\nabla}_{\vec{x}}^2 \langle \vec{x} \vert \vec{k} \rangle = \vec{k}^2 \langle \vec{x} \vert \vec{k} \rangle, \qquad
- i \partial_j \langle \vec{x} \vert \vec{k} \rangle = k_j \langle \vec{x} \vert \vec{k} \rangle, \qquad \vec{k}^2 \equiv \delta^{ij} k_i k_j .
\end{align}
The eigenvector/value equation for the momentum operators had been solved previously in equations \eqref{xk_PDE} and \eqref{xk_PDE_Soln}. For the negative Laplacian, we may check
\begin{align}
-\vec{\nabla}_{\vec{x}}^2 \langle \vec{x} \vert \vec{k} \rangle
= \braOket{\vec{x}}{\vec{P}^2}{\vec{k}} = \vec{k}^2 \langle \vec{x} \vert \vec{k} \rangle .
\end{align}
That the plane waves are simultaneous eigenvectors of $P_j$ and $\vec{P}^2 = -\vec{\nabla}^2$ is because these operators commute amongst themselves: $[P_j,\vec{P}^2] = [P_i,P_j] = 0$. This is therefore an example of {\it degeneracy}. For a fixed eigenvalue $k^2$ of the negative Laplacian, there is a continuous infinity of eigenvalues of the momentum operators, only constrained by
\begin{align}
\sum_{j=1}^D (k_j)^2 = k^2, \qquad \qquad \vec{P}^2 \ket{k^2;k_1 \dots k_D} = k^2 \ket{k^2;k_1 \dots k_D} .
\end{align}
Physically speaking we may associate this degeneracy with the presence of translation symmetry of the underlying infinite flat $\mathbb{R}^D$.

\end{enumerate}

\subsubsection{Boundary Conditions, Finite Box, Periodic functions and the Fourier Series}

Up to now we have not been terribly precise about the boundary conditions obeyed by our states $\braket{\vec{x}}{f}$, except to say they are functions residing in an infinite space $\mathbb{R}^D$. Let us now rectify this glaring omission -- drop the assumption of infinite space $\mathbb{R}^D$ -- and study how, in particular, the Hermitian nature of the $\vec{P}^2 \equiv -\vec{\nabla}^2$ operator now depends crucially on the boundary conditions obeyed by its eigenstates. If $\vec{P}^2$ is Hermitian,
\begin{align}
\braOket{\psi_1}{\vec{P}^2}{\psi_2} = \braOket{\psi_1}{\left(\vec{P}^2\right)^\dagger}{\psi_2} = \braOket{\psi_2}{\vec{P}^2}{\psi_1}^* , 
\end{align}
for any states $\ket{\psi_{1,2}}$. Inserting a complete set of position eigenkets, and using
\begin{align}
\braOket{\vec{x}}{\vec{P}^2}{\psi_{1,2}} &= -\vec{\nabla}^2_{\vec{x}} \braket{\vec{x}}{\psi_{1,2}} ,
\end{align}
we arrive at the condition that, if $\vec{P}^2$ is Hermitian then the negative Laplacian can be ``integrated-by-parts" to act on either $\psi_1$ or $\psi_2$.
\begin{align}
\int_{\mathfrak{D}} \dd^D x \braket{\psi_1}{\vec{x}} \braOket{\vec{x}}{\vec{P}^2}{\psi_2} 
&\stackrel{?}{=} \int_{\mathfrak{D}} \dd^D x \braket{\psi_2}{\vec{x}}^* \braOket{\vec{x}}{\vec{P}^2}{\psi_1}^* , \nonumber\\
\int_{\mathfrak{D}} \dd^D x \psi_1(\vec{x})^* \left(-\vec{\nabla}^2_{\vec{x}} \psi_2(\vec{x})\right)
&\stackrel{?}{=} \int_{\mathfrak{D}} \dd^D x \left(-\vec{\nabla}^2_{\vec{x}} \psi_1(\vec{x})^* \right) \psi_2(\vec{x}), \qquad 
\psi_{1,2}(\vec{x}) \equiv \braket{\vec{x}}{\psi_{1,2}} .
\end{align}
Notice we have to specify a domain $\mathfrak{D}$ to perform the integral. If we now proceed to work from the left hand side, and use Gauss' theorem from vector calculus,
{\allowdisplaybreaks\begin{align}
\int_{\mathfrak{D}} \dd^D x \psi_1(\vec{x})^* \left(-\vec{\nabla}^2_{\vec{x}} \psi_2(\vec{x})\right)
&= \int_{\partial\mathfrak{D}} \dd^{D-1} \vec{\Sigma} \cdot \left(-\vec{\nabla} \psi_1(\vec{x})^* \right) \psi_2(\vec{x})
+ \int_{\mathfrak{D}} \dd^D x \vec{\nabla} \psi_1(\vec{x})^* \cdot \vec{\nabla} \psi_2(\vec{x}) \nonumber\\
&=
\int_{\partial\mathfrak{D}} \dd^{D-1} \vec{\Sigma} \cdot \left\{ \left(-\vec{\nabla} \psi_1(\vec{x})^* \right) \psi_2(\vec{x}) + \psi_1(\vec{x})^* \vec{\nabla} \psi_2(\vec{x}) \right\} \nonumber\\
&\qquad+ \int_{\mathfrak{D}} \dd^D x \psi_1(\vec{x})^* \left(- \vec{\nabla}^2 \psi_2(\vec{x})\right)
\end{align}}
Here, $\dd^{D-1} \vec{\Sigma}$ is the $(D-1)$-dimensional analog of the 2D infinitesimal area element $\dd\vec{A}$ in vector calculus, and is proportional to the unit (outward) normal $\vec{n}$ to the boundary of the domain $\partial\mathfrak{D}$. We see that integrating-by-parts the $\vec{P}^2$ from $\psi_1$ onto $\psi_2$ can be done, but would incur the two surface integrals. To get rid of them, we may demand the eigenfunctions $\{\psi_\lambda\}$ of $\vec{P}^2$ or their normal derivatives $\{\vec{n} \cdot \vec{\nabla} \psi_\lambda\}$ to be zero:
\begin{align}
\psi_\lambda(\partial\mathfrak{D}) = 0 \ \text{(Dirichlet)} 
\qquad \text{    or    } \qquad
\vec{n} \cdot \vec{\nabla} \psi_\lambda(\partial\mathfrak{D}) &= 0 \ \text{(Neumann)}.
\end{align}
\footnote{Actually we may also allow the eigenfunctions to obey a mixed boundary condition, but we will stick to either Dirichlet or Neumann for simplicity.}{\bf No boundaries} \qquad The exception to the requirement for boundary conditions, is when the domain $\mathfrak{D}$ itself has {\it no boundaries} -- there will then be no ``surface terms" to speak of, and the Laplacian is hence automatically Hermitian. In this case, the eigenfunctions often obey periodic boundary conditions; we will see examples below.

\noindent{\bf Summary} \qquad The abstract bra-ket notation $\langle \psi_1 \vert \vec{P}^2 \vert \psi_2 \rangle$ obscures the fact that boundary conditions are required to ensure the Hermitian nature of $\vec{P}^2$. By going to the position basis, we see not only do we have to specify what the domain $\mathfrak{D}$ of the underlying space is, we have to either demand the eigenfunctions or their normal derivatives vanish on the boundary $\partial \mathfrak{D}$. In the discussion of partial differential equations below, we will generalize this analysis to curved spaces.

{\bf Example: Finite box} \qquad The first illustrative example is as follows. Suppose our system is defined only in a finite box. For the $i$th Cartesian axis, the box is of length $L^i$. If we demand that the eigenfunctions of $-\vec{\nabla}^2$ vanish at the boundary of the box, we find the eigensystem
\begin{align}
\label{FourierSeries_FiniteBox_I}
-\vec{\nabla}^2_{\vec{x}} \braket{\vec{x}}{\vec{n}} = \lambda(\vec{n}) \braket{\vec{x}}{\vec{n}},
\qquad 	& \braket{\vec{x};x^i = 0}{\vec{n}} = \braket{\vec{x};x^i = L^i}{\vec{n}} = 0, \\
		& i=1,2,3,\dots,D, 
\end{align}
admits the solution
\begin{align}
\label{FourierSeries_FiniteBox_II}
\braket{\vec{x}}{\vec{n}} \propto \prod_{i=1}^{D} \sin\left( \frac{\pi n^i}{L^i} x^i \right) ,
\qquad \lambda(\vec{n}) = \sum_{i=1}^{D} \left(\frac{\pi n^i}{L^i}\right)^2 .
\end{align}
These $\{n^i\}$ runs over the positive integers only; because sine is an odd function, the negative integers do not yield new solutions.

{\it Remark} \qquad Notice that, even though $\vec{P}^2$ is Hermitian in this finite box, the translation operator $\mathcal{T}(\xi) \equiv e^{-i \vec{\xi} \cdot \vec{P}}$ is no longer unitary and the momentum operator $P_j$ no longer Hermitian, because $\mathcal{T}(\xi)$ may move $\ket{x}$ out of the box if the translation distance is larger than $L$, and thus can no longer be viewed as a change of basis. (Recall the discussion around eq. \eqref{TranslationOperator_FinitBox}.) More explicitly, the eigenvectors of $\vec{P}$ in eq. \eqref{PlaneWave_UnitaryOperator} do not vanish on the walls of the box -- for e.g., in 1D, $\exp(ikx) \to 1$ when $x=0$ and $\exp(ikx) \to \exp(ikL) \neq 0$ when $x=L$ -- and therefore do not even lie in the vector space spanned by the eigenfunctions of $\vec{P}^2$. (Of course, you can superpose the momentum eigenstates of {\it different} eigenvalues to obtain the states in eq. \eqref{FourierSeries_FiniteBox_II}, but they will no longer be eigenstates of $P_j$.) Furthermore, if we had instead demanded the vanishing of the normal derivative, $\partial_x\exp(ikx) = ik \exp(ikx) \to ik \neq 0$ either, unless $k=0$.
\begin{myP}
\qquad Verify that the basis eigenkets in eq. \eqref{FourierSeries_FiniteBox_II} do solve eq. \eqref{FourierSeries_FiniteBox_I}. What is the correct normalization for $\braket{\vec{x}}{\vec{n}}$? Also verify that the basis plane waves in eq. \eqref{FourierSeries_Basis} satisfy the normalization condition in eq. \eqref{FourierSeries_Normalization}. \qed
\end{myP}
{\bf Periodic B.C.'s: the Fourier Series.} \qquad If we stayed within the infinite space, but now imposed periodic boundary conditions,
\begin{align}
\braket{\vec{x};x^i \to x^i+L^i}{f} &= \braket{\vec{x};x^i}{f} , \\
\label{FourierSeries_BC}
f(x^1,\dots,x^i+L^i,\dots,x^D) 		&= f(x^1,\dots,x^i,\dots,x^D) = f(\vec{x}) ,
\end{align}
this would mean, not all the basis plane waves from eq. \eqref{PlaneWave_UnitaryOperator} remains in the Hilbert space. Instead, periodicity means
\begin{align}
\langle \vec{x} ; x^j = x^j + L^j | \vec{k} \rangle &= \langle \vec{x} ; x^j = x^j | \vec{k} \rangle \nonumber\\
e^{i k_j (x^j + L^j)} 								&= e^{i k_j x^j} , \qquad \qquad \text{(No sum over $j$.)}
\end{align}
(The rest of the plane waves, $e^{i k_l x^l}$ for $l \neq j$, cancel out of the equation.) This further implies the eigenvalue $k_j$ becomes discrete:
\begin{align}
e^{i k_j L^j} 	= 1 \ \text{(No sum over $j$.)}\qquad \Rightarrow \qquad 
k_j L^j 		= 2\pi n \qquad \Rightarrow \qquad k_j = \frac{2\pi n^j}{L^j}, \nonumber\\
\qquad n^j=0,\pm 1,\pm 2, \pm 3,\dots.
\end{align}
We need to re-normalize our basis plane waves. In particular, since space is now periodic, we ought to only need to integrate over one typical volume.
\begin{align}
\label{FourierSeries_Normalization}
\int_{\{0 \leq x^i \leq L^i | i=1,2,\dots,D\}} \dd^D \vec{x} \braket{\vec{n}'}{\vec{x}} \braket{\vec{x}}{\vec{n}} = \delta^{\vec{n}}_{\vec{n}'}
\equiv \prod_{i=1}^{D} \delta^{n'^i}_{n^i} .
\end{align}
Because we have a set of orthonormal eigenvectors of the negative Laplacian,
\begin{align}
\label{FourierSeries_Basis}
\left\langle \vec{x} \left\vert \vec{n} \right\rangle\right. \equiv \prod_{j=1}^D \frac{\exp\left( i \frac{2\pi n^j}{L^j} x^j \right)}{\sqrt{L^j}} , \\
-\vec{\nabla}^2 \left\langle \vec{x} \left\vert \vec{n} \right\rangle\right.
= \lambda(\vec{n}) \left\langle \vec{x} \left\vert \vec{n} \right\rangle\right.,
\qquad \lambda(\vec{n}) = \sum_i \left(\frac{2\pi n^i}{L^i}\right)^2 ;
\end{align}
they obey the completeness relation
\begin{align}
\braket{\vec{x}}{\vec{x}'} 
= \delta^{(D)}(\vec{x}-\vec{x}')
= \sum_{n^1 = -\infty}^\infty \dots \sum_{n^D=-\infty}^\infty \braket{\vec{x}}{\vec{n}} \braket{\vec{n}}{\vec{x}'} .
\end{align}
To sum: any periodic function $f$, subject to eq. \eqref{FourierSeries_BC}, can be expanded as a superposition of periodic plane waves in eq. \eqref{FourierSeries_Basis},
\begin{align}
\label{FourierSeries_I}
f(\vec{x}) 
&= \sum_{n^1 = -\infty}^\infty \dots \sum_{n^D=-\infty}^\infty \widetilde{f}(n^1,\dots,n^D) \prod_{j=1}^D (L^j)^{-1/2} \exp\left( i \frac{2\pi n^j}{L^j} x^j \right) .
\end{align}
This is known as the {\it Fourier series}. By using the inner product in eq. \eqref{FourierSeries_Normalization}, or equivalently, multiplying both sides of eq. \eqref{FourierSeries_I} by $\prod_j (L^j)^{-1/2} \exp( -i (2\pi n'^j/L^j) x^j )$ and integrating over a typical volume, we obtain the coefficients of the Fourier series expansion
\begin{align}
\label{FourierSeries_II}
\widetilde{f}(n^1,n^2,\dots,n^D) = \int_{0 \leq x^j \leq L^j} \dd^D \vec{x} f(\vec{x}) \prod_{j=1}^D (L^j)^{-1/2} \exp\left( -i \frac{2\pi n^j}{L^j} x^j \right) .
\end{align}
\noindent{\it Remark I} \qquad The $\exp$ in eq. \eqref{FourierSeries_Basis} are not a unique set of basis vectors, of course. One could use sines and cosines instead, for example.

%If $f(\vec{x};x^i=x^i+L^i)=f(\vec{x})$ denotes a member of the set of all periodic functions, where the $i$th Cartesian axis has a period of length $L^i$, then it admits the Fourier series expansion:
%\begin{align}
%f(\vec{x}) \equiv \braket{\vec{x}}{f} &= \sum_{\vec{n}} \braket{\vec{x}}{\vec{n}} \braket{\vec{n}}{f} 
%= \frac{1}{L^D} \sum_{\vec{n}} \widetilde{f}(\vec{n}) \exp\left( i \vec{k}(\vec{n}) \cdot \vec{x} \right), \nonumber\\
%& k^i(\vec{n}) \equiv \frac{2\pi}{L^i} n^i; \ n^i = 0, \pm 1, \pm 2,\dots, \qquad (\text{No sum over $i$.})
%\end{align}
%The inverse Fourier expansion is
%\begin{align}
%\widetilde{f}(\vec{n}) \equiv \braket{\vec{n}}{f} 
%&= \int_{\{0 \leq x^i \leq L^i | i=1,2,\dots,D\}} \dd^D \vec{x} \braket{\vec{n}}{\vec{x}} \braket{\vec{x}}{f} \nonumber\\
%&= \frac{1}{L^D} \int_{0}^{L^1} \dd x^1 \dots \int_{0}^{L^D} \dd x^D f(\vec{x}) \exp\left( -i \vec{k}(\vec{n}) \cdot \vec{x} \right) .
%\end{align}

\noindent{\it Remark II} \qquad Even though we are explicitly integrating the $i$th Cartesian coordinate from $0$ to $L^i$ in eq. \eqref{FourierSeries_II}, since the function is periodic, we really just need only to integrate over a complete period, from $\kappa$ to $\kappa + L^i$ (for $\kappa$ real), to achieve the same result. For example, in 1D, and whenever $f(x)$ is periodic (with a period of $L$),
\begin{align}
\int_{0}^{L} \dd x f(x) = \int_{\kappa}^{\kappa + L} \dd x f(x) .
\end{align}
(Drawing a plot here may help to understand this statement.)

\newpage

\section{Calculus on the Complex Plane}
\label{Chapter_ComplexCalculus}
\subsection{Differentiation}

\footnote{Much of the material here on complex analysis is based on Arfken et al's {\it Mathematical Methods for Physicists}.}The derivative of a complex function $f(z)$ is defined in a similar way as its real counterpart:
\begin{align}
f'(z) \equiv \frac{\dd f(z)}{\dd z} \equiv \lim_{\Delta z \to 0} \frac{f(z+\Delta z) - f(z)}{\Delta z} .
\end{align}
However, the meaning is more subtle because $\Delta z$ (just like $z$ itself) is now complex. What this means is that, in taking this limit, it has to yield the same answer no matter what direction you approach $z$ on the complex plane. For example, if $z=x+iy$, taking the derivative along the real direction must be equal to that along the imaginary one,
\begin{align}
f'(z) 
&= \lim_{\Delta x \to 0} \frac{f(x+\Delta x+iy) - f(x+iy)}{\Delta x} = \partial_x f(z) 	\nonumber\\
&= \lim_{\Delta y \to 0} \frac{f(x+i(y+\Delta y)) - f(x+iy)}{i \Delta y} = \frac{\partial f(z)}{\partial (iy)} = \frac{1}{i} \partial_y f(z) , 
\end{align}
where $x$, $y$, $\Delta x$ and $\Delta y$ are real. This direction independence imposes very strong constraints on complex differentiable functions: they will turn out to be extremely smooth, in that if you can differentiate them at a given point $z$, you are guaranteed they are differentiable infinite number of times there. (This is not true of real functions.) If $f(z)$ is differentiable in some region on the complex plane, we say $f(z)$ is analytic there. 

If the first derivatives of $f(z)$ are continuous, the criteria for determining whether $f(z)$ is differentiable comes in the following pair of partial differential equations.
\begin{quotation}
	{\bf Cauchy-Riemann conditions for analyticity} \qquad Let $z=x+iy$ and $f(z) = u(x,y) + i v(x,y)$, where $x$, $y$, $u$ and $v$ are real. Let $u$ and $v$ have continuous first partial derivatives in $x$ and $y$. Then $f(z)$ is an analytic function in the neighborhood of $z$ if and only if the following (Cauchy-Riemann) equations are satisfied by the real and imaginary parts of $f$:
\begin{align}
\label{Complex_CauchyRiemann}
\partial_x u = \partial_y v, \qquad \partial_y u = -\partial_x v .
\end{align}
\end{quotation}
To understand why this is true, we first consider differentiating along the (real) $x$ direction, we'd have
\begin{align}
\frac{\dd f(z)}{\dd z} = \partial_x u + i \partial_x v .
\end{align}
If we differentiate along the (imaginary) $iy$ direction instead, we'd have
\begin{align}
\frac{\dd f(z)}{\dd z} = \frac{1}{i} \partial_y u + \partial_y v = \partial_y v - i \partial_y u .
\end{align}
Since these two results must be the same, we may equate their real and imaginary parts to obtain eq. \eqref{Complex_CauchyRiemann}. (It is at this point, if we did not assume $u$ and $v$ have continuous first derivatives, that we see the Cauchy-Riemann conditions in eq. \eqref{Complex_CauchyRiemann} are necessary but not necessarily sufficient ones for analyticity.)

Conversely, if eq. \eqref{Complex_CauchyRiemann} are satisfied and if we do assume $u$ and $v$ have continuous first derivatives, we may consider an arbitrary variation of the function $f$ along the direction $\dd z = \dd x + i \dd y$ via
{\allowdisplaybreaks\begin{align}
\dd f(z) = \partial_x f(z) \dd x + \partial_y f(z) \dd y
&= (\partial_x u + i \partial_x v)\dd x + (\partial_y u + i \partial_y v) \dd y \nonumber\\
&\text{(Use eq. \eqref{Complex_CauchyRiemann} on the $\dd y$ terms.)} \nonumber\\
&= (\partial_x u + i \partial_x v)\dd x + ( - \partial_x v + i \partial_x u) \dd y \nonumber\\
&= (\partial_x u + i \partial_x v)\dd x + (\partial_x u + i \partial_x v) i \dd y \nonumber\\
&= (\partial_x u + i \partial_x v) \dd z .
\end{align}}
\footnote{In case the assumption of continuous first derivatives is not clear -- note that, if $\partial_x f$ and $\partial_y f$ were not continuous, then $\dd f$ (the variation of $f$) in the direction across the discontinuity cannot be computed in terms of the first derivatives. Drawing a plot for a real function $F(x)$ with a discontinuous first derivative (i.e., a ``kink") would help.}Therefore, the complex derivative $\dd f/\dd z$ yields the same answer regardless of the direction of variation $\dd z$, and is given by
\begin{align}
\frac{\dd f(z)}{\dd z} = \partial_x u + i \partial_x v .
\end{align}
\noindent{\it Polar coordinates} \qquad It is also useful to express the Cauchy-Riemann conditions in polar coordinates $(x,y) = r(\cos\theta,\sin\theta)$. We have
\begin{align}
\partial_r 		
		&= \frac{\partial x}{\partial r} \partial_x + \frac{\partial y}{\partial r} \partial_y 
		= \cos\theta \partial_x + \sin\theta \partial_y \\
\partial_\theta 
		&= \frac{\partial x}{\partial \theta} \partial_x + \frac{\partial y}{\partial \theta} \partial_y 
		= -r\sin\theta \partial_x + r \cos\theta \partial_y .
\end{align}
By viewing this as a matrix equation $(\partial_r, \partial_\theta)^T = M (\partial_x, \partial_y)^T$, we may multiply $M^{-1}$ on both sides and obtain the $(\partial_x,\partial_y)$ in terms of the $(\partial_r, \partial_\theta)$.
\begin{align}
\partial_x	&= \cos\theta \partial_r - \frac{\sin\theta}{r} \partial_\theta \\
\partial_y 	&= \sin\theta \partial_r + \frac{\cos\theta}{r} \partial_\theta .
\end{align}
The Cauchy-Riemann conditions in eq. \eqref{Complex_CauchyRiemann} can now be manipulated by replacing the $\partial_x$ and $\partial_y$ with the right hand sides above. Denoting $c \equiv \cos\theta$ and $s \equiv \sin\theta$,
\begin{align}
\left(cs \partial_r - \frac{s^2}{r} \partial_\theta\right) u &= \left(s^2 \partial_r + \frac{cs}{r} \partial_\theta\right) v, \\
\left(sc \partial_r + \frac{c^2}{r} \partial_\theta\right) u &= -\left(c^2 \partial_r - \frac{sc}{r} \partial_\theta\right) v,
\end{align}
and
\begin{align}
\left(c^2 \partial_r - \frac{sc}{r} \partial_\theta\right) u &= \left(sc\partial_r 		+ \frac{c^2}{r} \partial_\theta\right) v, \\
\left(s^2 \partial_r + \frac{sc}{r} \partial_\theta\right) u &= -\left(cs \partial_r 	- \frac{s^2}{r} \partial_\theta\right) v .
\end{align}
(We have multiplied both sides of eq. \eqref{Complex_CauchyRiemann} with appropriate factors of sines and cosines.) Subtracting the first pair and adding the second pair of equations, we arrive at the polar coordinates version of Cauchy-Riemann:
\begin{align}
\label{Complex_CauchyRiemann_Polar}
\frac{1}{r} \partial_\theta u 	= -\partial_r v , \qquad \qquad
\partial_r u 					= \frac{1}{r} \partial_\theta v .
\end{align}
{\it Examples} \qquad Complex differentiability is much more restrictive than the real case. An example is $f(z) = |z|$. If $z$ is real, then at least for $z \neq 0$, we may differentiate $f(z)$ -- the result is $f'(z) = 1$ for $z>0$ and $f'(z) = -1$ for $z<0$. But in the complex case we would identify, with $z=x+iy$,
\begin{align}
f(z) = |z| = \sqrt{x^2+y^2} = u(x,y) + i v(x,y) \qquad \Rightarrow \qquad v(x,y) = 0 .
\end{align}
It's not hard to see that the Cauchy-Riemann conditions in eq. \eqref{Complex_CauchyRiemann} cannot be satisfied since $v$ is zero while $u$ is non-zero. In fact, any $f(z)$ that remains strictly real across the complex $z$ plane is not differentiable unless $f(z)$ is constant.
\begin{align}
f(z) = u(x,y) \qquad \Rightarrow \qquad \partial_x u = \partial_y v = 0, \quad \partial_y u = -\partial_x v = 0 .
\end{align}
Similarly, if $f(z)$ were purely imaginary across the complex $z$ plane, it is not differentiable unless $f(z)$ is constant.
\begin{align}
f(z) = i v(x,y) \qquad \Rightarrow \qquad 0 = \partial_x u = \partial_y v, \quad 0 = -\partial_y u = \partial_x v .
\end{align}
%If its real part only depends on $x$ and the imaginary part on $y$, then $f(z) = u(x) + i v(y)$ not differentiable as well. The $\partial_y u = 0 = -\partial_x v$ but
%\begin{align}
%\partial_x u(x) = \partial_y u(y)
%\end{align}
%make no sense because the left-hand-side only depends on $x$ and the right-hand-side only on $y$.

{\bf Differentiation rules} \qquad If you know how to differentiate a function $f(z)$ when $z$ is real, then as long as you can show that $f'(z)$ exists, the differentiation formula for the complex case would carry over from the real case. That is, suppose $f'(z) = g(z)$ when $f$, $g$ and $z$ are real; then this {\it form} has to hold for complex $z$. For example, powers are differentiated the same way
\begin{align}
\frac{\dd}{\dd z} z^\alpha = \alpha z^{\alpha-1}, \qquad \alpha \in\mathbb{R} ,
\end{align}
and
\begin{align}
\frac{\dd \sin(z)}{\dd z} = \cos z, \qquad 
\frac{\dd a^z}{\dd z} = \frac{\dd e^{z \ln a}}{\dd z} = a^z \ln a .
\end{align}
It is not difficult to check the first derivatives of $z^\alpha$, $\sin(z)$ and $a^z$ are continuous; and the Cauchy-Riemann conditions are satisfied. For instance, $z^\alpha = r^\alpha e^{i\alpha \theta} = r^\alpha \cos(\alpha\theta) + i r^\alpha \sin(\alpha\theta)$ and eq. \eqref{Complex_CauchyRiemann_Polar} can be verified.
\begin{align}
r^{\alpha-1} \partial_\theta \cos(\alpha\theta) 
		= -\alpha r^{\alpha-1} \sin(\alpha\theta) \stackrel{?}{=} -\sin(\alpha\theta) \partial_r r^\alpha = -\alpha r^{\alpha-1} \sin(\alpha\theta), \\
\cos(\alpha\theta) \partial_r r^\alpha = \alpha r^{\alpha-1} \cos(\alpha\theta) \stackrel{?}{=} r^{\alpha-1} \partial_\theta \sin(\alpha\theta) = \alpha r^{\alpha-1} \cos(\alpha\theta) .
\end{align}
(This proof that $z^\alpha$ is analytic fails at $r=0$; in fact, for $\alpha<1$, we see that $z^\alpha$ is not analytic there.) In particular, differentiability is particularly easy to see if $f(z)$ can be {\it defined} through its power series.

{\it Product and chain rules} \qquad The product and chain rules apply too. For instance,
\begin{align}
(fg)' = f' g + f g' .
\end{align}
because
{\allowdisplaybreaks\begin{align}
(fg)' 
&= \lim_{\Delta z \to 0} \frac{f(z+\Delta z) g(z+\Delta z) - f(z) g(z)}{\Delta z} \nonumber\\
&= \lim_{\Delta z \to 0} \frac{(f(z) + f' \cdot \Delta z)(g(z) + g' \Delta z) - f(z) g(z)}{\Delta z} \nonumber\\
&= \lim_{\Delta z \to 0} \frac{fg + f g' \Delta z + f' g \Delta z + \mathcal{O}((\Delta z)^2) - fg}{\Delta z} 
= f' g + f g' .
\end{align}}
We will have more to say later about carrying over properties of real differentiable functions to their complex counterparts.
\begin{myP}
{\it Conformal transformations} \qquad Complex functions can be thought of as a map from one 2D plane to another. In this problem, we will see how they define angle preserving transformations. Consider two paths on a complex plane $z=x+iy$ that intersects at some point $z_0$. Let the angle between the two lines at $z_0$ be $\theta$. Given some complex function $f(z) = u(x,y) + i v(x,y)$, this allows us to map the two lines on the $(x,y)$ plane into two lines on the $(u,v)$ plane. Show that, as long as $\dd f(z)/\dd z \neq 0$, the angle between these two lines on the $(u,v)$ plane at $f(z_0)$, is still $\theta$. Hint: imagine parametrizing the two lines with $\lambda$, where the first line is $\xi_1(\lambda) = x_1(\lambda)+iy_1(\lambda)$ while the second line is $\xi_2(\lambda) = x_2(\lambda)+iy_2(\lambda)$. Let their intersection point be $\xi_1(\lambda_0) = \xi_2(\lambda_0)$. Now also consider the two lines on the $(u,v)$ plane: $f(\xi_1(\lambda)) = u(\xi_1(\lambda)) + i v(\xi_1(\lambda))$ and $f(\xi_2(\lambda)) = u(\xi_2(\lambda)) + i v(\xi_2(\lambda))$. On the $(x,y)$-plane, consider $\arg [(\dd 
\xi_1/\dd\lambda)/(\dd 
\xi_2/\dd\lambda)]$; whereas on the $(u,v)$-plane consider $\arg [(\dd 
f(\xi_1)/\dd\lambda)/(\dd 
f(\xi_2)/\dd\lambda)]$.
\end{myP}
\begin{quotation}
{\bf 2D Laplace's equation} \qquad Suppose $f(z) = u(x,y) + i v(x,y)$, where $z=x+iy$ and $x$, $y$, $u$ and $v$ are real. If $f(z)$ is complex-differentiable then the Cauchy-Riemann relations in eq. \eqref{Complex_CauchyRiemann} imply that both the real and imaginary parts of a complex function obey Laplace's equation, namely
\begin{align}
\label{Complex_2DLaplace}
(\partial_x^2 + \partial_y^2) u(x,y) = (\partial_x^2 + \partial_y^2) v(x,y) = 0.
\end{align}
\end{quotation}
To see this we differentiate eq. \eqref{Complex_CauchyRiemann} appropriately,
\begin{align}
\label{Complex_CauchyRiemann_MixedDs}
\partial_x \partial_y u = \partial_y^2 v, \qquad \partial_x \partial_y u = -\partial_x^2 v \\
\partial_x^2 u = \partial_x \partial_y v, \qquad -\partial_y^2 u = \partial_x \partial_y v .
\end{align}
We now can equate the right hand sides of the first line; and the left hand sides of the second line. This leads to \eqref{Complex_2DLaplace}.

Because of eq. \eqref{Complex_2DLaplace}, complex analysis can be very useful for 2D electrostatic problems.

Moreover, $u$ and $v$ cannot admit local minimum or maximums, as long as $\partial_x^2 u$ and $\partial_x^2 v$ are non-zero. In particular, the determinants of the $2 \times 2$ Hessian matrices $\partial^2 u/\partial(x,y)^i \partial(x,y)^j$ and $\partial^2 v/\partial(x,y)^i \partial(x,y)^j$ -- and hence the product of their eigenvalues -- are negative. For,
\begin{align}
\det \frac{\partial^2 u}{\partial(x,y)^i \partial(x,y)^j}
&= \det\left[\begin{array}{cc}
\partial_x^2 u 			& \partial_x \partial_y u 	\\
\partial_x \partial_y u & \partial_y^2 u 
\end{array}\right] \nonumber\\
		&= \partial_x^2 u \partial_y^2 u - (\partial_x \partial_y u)^2 = - (\partial_y^2 u)^2 - (\partial_y^2 v)^2 \leq 0 , \\
\det \frac{\partial^2 v}{\partial(x,y)^i \partial(x,y)^j}
&= \det\left[\begin{array}{cc}
\partial_x^2 v 			& \partial_x \partial_y v 	\\
\partial_x \partial_y v & \partial_y^2 v 
\end{array}\right] \nonumber\\
		&= \partial_x^2 v \partial_y^2 v - (\partial_x \partial_y v)^2 = - (\partial_y^2 v)^2 - (\partial_y^2 u)^2 \leq 0 ,
\end{align}
where both equations \eqref{Complex_2DLaplace} and \eqref{Complex_CauchyRiemann_MixedDs} were employed.

\subsection{Cauchy's integral theorems, Laurent Series, Analytic Continuation}

Complex integration is really a line integral $\int \vec{\xi} \cdot (\dd x,\dd y)$ on the 2D complex plane. Given some path (aka ``contour") $C$, defined by $z(\lambda_1 \leq \lambda \leq \lambda_2) = x(\lambda) + i y(\lambda)$, with $z(\lambda_1) = z_1$ and $z(\lambda_2) = z_2$,
{\allowdisplaybreaks\begin{align}
\label{Complex_ContourIntegral}
\int_C \dd z f(z)
&= \int_{z(\lambda_1 \leq \lambda \leq \lambda_2)} (\dd x + i \dd y) \left( u(x,y) + i v(x,y) \right) \nonumber\\
&= \int_{z(\lambda_1 \leq \lambda \leq \lambda_2)} \left( u \dd x - v \dd y \right) + i \int_{z(\lambda_1 \leq \lambda \leq \lambda_2)} \left( v \dd x + u \dd y \right) \nonumber\\
&= \int_{\lambda_1}^{\lambda_2} \dd\lambda \left( u \frac{\dd x(\lambda)}{\dd \lambda} - v \frac{\dd y(\lambda)}{\dd \lambda} \right)
+ i \int_{\lambda_1}^{\lambda_2} \dd\lambda \left( v \frac{\dd x(\lambda)}{\dd \lambda} + u \frac{\dd y(\lambda)}{\dd \lambda} \right) .
\end{align}}
The real part of the line integral involves Re$\vec{\xi} = (u,-v)$ and its imaginary part Im$\vec{\xi} = (v,u)$.

\noindent{\it Remark I} \qquad Because complex integration is a line integral, reversing the direction of contour $C$ (which we denote as $-C$) would yield return negative of the original integral.
\begin{align}
\int_{-C} \dd z f(z) = -\int_C \dd z f(z) 
\end{align}
\noindent{\it Remark II} \qquad The complex version of the fundamental theorem of calculus has to hold, in that
\begin{align}
\int_C \dd z f'(z) = \int_C \dd f &= f(\text{``upper" end point of $C$}) - f(\text{``lower" end point of $C$}) \nonumber\\
= \int_{z_1}^{z_2} \dd z f'(z) &= f(z_2) - f(z_1) .
\end{align}
\begin{quotation}
{\bf Cauchy's integral theorem} \qquad In introducing the contour integral in eq. \eqref{Complex_ContourIntegral}, we are not assuming any properties about the integrand $f(z)$. However, if the complex function $f(z)$ is analytic throughout some simply connected region\footnote{A simply connected region is one where every closed loop in it can be shrunk to a point.} containing the contour $C$, then we are lead to one of the key results of complex integration theory: the integral of $f(z)$ within any {\it closed} path $C$ there is zero.
\begin{align}
\label{Complex_CauchyIntegralTheorem}
\oint_C f(z) \dd z = 0 
\end{align}
\end{quotation}
Unfortunately the detailed proof will take up too much time and effort, but the mathematically minded can consult, for example, Brown and Churchill's {\it Complex Variables and Applications}. 
\begin{myP}
\qquad If the first derivatives of $f(z)$ are assumed to be continuous, then a proof of this modified Cauchy's theorem can be carried out by starting with the view that $\oint_C f(z) \dd z$ is a (complex) line integral around a closed loop. Then apply Stokes' theorem followed by the Cauchy-Riemann conditions in eq. \eqref{Complex_CauchyRiemann}. Can you fill in the details? \qed
\end{myP}
\noindent{\it Important Remarks} \qquad Cauchy's theorem has an important implication. Suppose we have a contour integral $\int_C g(z) \dd z$, where $C$ is some arbitrary (not necessarily closed) contour. Suppose we have another contour $C'$ whose end points coincide with those of $C$. If the function $g(z)$ is analytic inside the region bounded by $C$ and $C'$, then it has to be that 
\begin{align}
\int_C g(z) \dd z = \int_{C'} g(z) \dd z .
\end{align}
The reason is that, by subtracting these two integrals, say $(\int_C - \int_{C'}) g(z) \dd z$, the $-$ sign can be absorbed by reversing the direction of the $C'$ integral. We then have a closed contour integral $(\int_C - \int_{C'}) g(z) \dd z = \oint g(z) \dd z$ and Cauchy's theorem in eq. \eqref{Complex_CauchyIntegralTheorem} applies. 

This is a very useful observation because it means, for a given contour integral, you can deform the contour itself to a shape that would make the integral easier to evaluate. Below, we will generalize this and show that, even if there are isolated points where the function is not analytic, you can still pass the contour over these points, but at the cost of incurring additional terms resulting from taking the residues there. Another possible type of singularity is known as a branch point, which will then require us to introduce a branch cut.

Note that the simply connected requirement can often be circumvented by considering an appropriate cut line. For example, suppose $C_1$ and $C_2$ were both counterclockwise (or both clockwise) contours around an annulus region, within which $f(z)$ is analytic. Then 
\begin{align}
\oint_{C_1} f(z) \dd z = \oint_{C_2} f(z) \dd z .
\end{align}
{\it Example I} \qquad A simple but important example is the following integral, where the contour $C$ is an arbitrary counterclockwise closed loop that encloses the point $z=0$. 
\begin{align}
I \equiv \oint_C \frac{\dd z}{z} 
\end{align}
Cauchy's integral theorem does not apply directly because $1/z$ is not analytic at $z=0$. By considering a counterclockwise circle $C'$ of radius $R > 0$, however, we may argue 
\begin{align}
\oint_C \frac{\dd z}{z} = \oint_{C'} \frac{\dd z}{z} .
\end{align}
\footnote{This is where drawing a picture would help: for simplicity, if $C'$ lies entirely within $C$, the first portion of the cut lines would begin anywhere from $C'$ to anywhere to $C$, followed by the reverse trajectory from $C$ to $C'$ that runs infinitesimally close to the first portion. Because they are infinitesimally close, the contributions of these two portions cancel; but we now have a simply connected closed contour integral that amounts to $0 = (\int_C - \int_{C'})\dd z/z$.}We may then employ polar coordinates, so that the path $C'$ could be described as $z = R e^{i\theta}$, where $\theta$ would run from $0$ to $2\pi$.
\begin{align}
\oint_C \frac{\dd z}{z} = \int_0^{2\pi} \frac{\dd (R e^{i\theta})}{R e^{i\theta}}
= \int_0^{2\pi} i \dd\theta = 2\pi i. 
\end{align}
\noindent{\it Example II} \qquad Let's evaluate $\oint_C z \dd z$ and $\oint_C \dd z$ directly and by using Cauchy's integral theorem. Here, $C$ is some closed contour on the complex plane. Directly:
\begin{align}
\oint_C z \dd z = \left.\frac{z^2}{2}\right\vert_{z=z_0}^{z=z_0} = 0,\qquad
\oint_C \dd z = \left.z\right\vert_{z=z_0}^{z=z_0} = 0 .
\end{align}
Using Cauchy's integral theorem -- we first note that $z$ and $1$ are analytic, since they are powers of $z$; we thus conclude the integrals are zero.
\begin{myP}
\qquad For some contour $C$, let $M$ be the maximum of $|f(z)|$ along it and $L \equiv \int_C \sqrt{\dd x^2 + \dd y^2}$ be the length of the contour itself, where $z=x+iy$ (for $x$ and $y$ real). Argue that
\begin{align}
\left\vert \int_C f(z) \dd z \right\vert \leq \int_C |f(z)| |\dd z| \leq M \cdot L .
\end{align}
Note: $|\dd z| = \sqrt{\dd x^2 + \dd y^2}$. (Why?) Hints: Can you first argue for the triangle inequality, $|z_1 + z_2| \leq |z_1|+|z_2|$, for any two complex numbers $z_{1,2}$? What about $|z_1+z_2+\dots+z_N| \leq |z_1|+|z_2|+ \dots +|z_N|$? Then view the integral as a discrete sum, and apply this generalized triangle inequality to it. \qed
\end{myP}
\begin{myP}
Evaluate
\begin{align}
\oint_C \frac{\dd z}{z(z+1)} ,
\end{align}
where $C$ is an arbitrary contour enclosing the points $z=0$ and $z=-1$. Note that Cauchy's integral theorem is not directly applicable here. Hint: Apply a partial fractions decomposition of the integrand, then for each term, convert this arbitrary contour to an appropriate circle. \qed
\end{myP}
The next major result allows us to deduce $f(z)$, for $z$ lying within some contour $C$, by knowing its values on $C$. 
\begin{quotation}
{\bf Cauchy's integral formula} \qquad If $f(z)$ is analytic on and within some closed counterclockwise contour $C$, then
\begin{align}
\label{Complex_CauchyIntegral}
\oint_C \frac{\dd z'}{2\pi i} \frac{f(z')}{z'-z} 
&= f(z) \qquad \text{if $z$ lies inside $C$} \nonumber\\
&= 0 \qquad \text{if $z$ lies outside $C$} .
\end{align}
\end{quotation}
{\it Proof} \qquad If $z$ lies outside $C$ then the integrand is analytic within its interior and therefore Cauchy's integral theorem applies. If $z$ lies within $C$ we may then deform the contour such that it becomes an infinitesimal counterclockwise circle around $z' \approx z$,
\begin{align}
z' \equiv z + \epsilon e^{i\theta}, \qquad 0 < \epsilon \ll 1 .
\end{align}
We then have
\begin{align}
\oint_C \frac{\dd z'}{2\pi i} \frac{f(z')}{z'-z} 
&= \frac{1}{2\pi i} \int_{0}^{2\pi} \epsilon e^{i\theta} i \dd\theta \frac{f(z + \epsilon e^{i\theta})}{\epsilon e^{i\theta}} \nonumber\\
&= \int_{0}^{2\pi} \frac{\dd\theta}{2\pi} f(z + \epsilon e^{i\theta}) .
\end{align}
By taking the limit $\epsilon \to 0^+$, we get $f(z)$, since $f(z')$ is analytic and thus continuous at $z'=z$.
\begin{quotation}
{\bf Cauchy's integral formula for derivatives} \qquad By applying the limit definition of the derivative, we may obtain an analogous definition for the $n$th derivative of $f(z)$. For some closed counterclockwise contour $C$,
\begin{align}
\label{Complex_CauchyIntegral_Derivative}
\oint_C \frac{\dd z'}{2\pi i} \frac{f(z')}{(z'-z)^{n+1}} 
&= \frac{f^{(n)}(z)}{n!} \qquad \text{if $z$ lies inside $C$} \nonumber\\
&= 0 \qquad \text{if $z$ lies outside $C$} .
\end{align}
\end{quotation}
This implies -- as already advertised earlier -- once $f'(z)$ exists, $f^{(n)}(z)$ also exists for any $n$. Complex-differentiable functions are infinitely smooth.

The converse of Cauchy's integral formula is known as Morera's theorem, which we will simply state without proof.
\begin{quotation}
{\bf Morera's theorem} \qquad If $f(z)$ is continuous in a simply connected region and $\oint_C f(z) \dd z = 0$ for {\it any} closed contour $C$ within it, then $f(z)$ is analytic throughout this region.
\end{quotation}
Now, even though $f^{(n > 1)}(z)$ exists once $f'(z)$ exists (cf. \eqref{Complex_CauchyIntegral_Derivative}), $f(z)$ cannot be infinitely smooth everywhere on the complex $z-$plane..
\begin{quotation}
{\bf Liouville's theorem} \qquad If $f(z)$ is analytic and bounded -- i.e., $|f(z)|$ is less than some positive constant $M$ -- for all complex $z$, then $f(z)$ must in fact be a constant. Apart from the constant function, analytic functions must blow up somewhere on the complex plane.
\end{quotation}
{\it Proof} \qquad To prove this result we employ eq. \eqref{Complex_CauchyIntegral_Derivative}. Choose a counterclockwise circular contour $C$ that encloses some arbitrary point $z$,
\begin{align}
|f^{(n)}(z)| 
&\leq n! \oint_C \frac{|\dd z'|}{2\pi} \frac{|f(z')|}{|(z'-z)^{n+1}|} \\
&\leq n! \frac{M}{2 \pi r^{n+1}} \oint_C |\dd z'| = n! \frac{M}{r^{n}} .
\end{align}
Here, $r$ is the radius from $z$ to $C$. But by Cauchy's theorem, the circle can be made arbitrarily large. By sending $r\to \infty$, we see that $|f^{(n)}(z)|=0$, the $n$th derivative of the analytic function at an arbitrary point $z$ is zero for any integer $n \geq 1$. This proves the theorem.

{\it Examples} \qquad The exponential $e^z$ while differentiable everywhere on the complex plane, does in fact blow up at Re $z \to \infty$. Sines and cosines are oscillatory and bounded on the real line; and are differentiable everywhere on the complex plane. However, they blow up as one move towards positive or negative imaginary infinity. Remember $\sin(z) = (e^{iz} - e^{-iz})/(2i)$ and $\cos(z) = (e^{iz} + e^{-iz})/2$. Then, for $R \in \mathbb{R}$,
\begin{align}
\sin(iR) = \frac{e^{-R} - e^{R}}{2i}, \qquad \cos(iR) = \frac{e^{-R} + e^{R}}{2} .
\end{align}
Both $\sin(iR)$ and $\cos(iR)$ blow up as $R \to \pm \infty$.
\begin{myP}
{\it Fundamental theorem of algebra.} \qquad Let $P(z) = p_0 + p_1 z + \dots p_n z^n$ be an $n$th degree polynomial, where $n$ is an integer greater or equal to $1$. By considering $f(z) = 1/P(z)$, show that $P(z)$ has at least one root. (Once a root has been found, we can divide it out from $P(z)$ and repeat the argument for the remaining $(n-1)$-degree polynomial. By induction, this implies an $n$th degree polynomial has exactly $n$ roots -- this is the fundamental theorem of algebra.) \qed
\end{myP}
{\bf Taylor series} \qquad The generalization of the Taylor series of a real differentiable function to the complex case is known as the Laurent series. If the function is completely smooth in some region on the complex plane, then we shall see that it can in fact be Taylor expanded the usual way, except the expressions are now complex. If there are isolated points where the function blows up, then it can be (Laurent) expanded about those points, in powers of the complex variable -- except the series begins at some negative integer power, as opposed to the zeroth power in the usual Taylor series.

To begin, let us show that the geometric series still works in the complex case.
\begin{myP}
\qquad By starting with the $N$th partial sum,
\begin{align}
S_N \equiv \sum_{\ell=0}^{N} t^\ell ,
\end{align}
prove that, as long as $|t| < 1$,
\begin{align}
\label{Complex_GeometricSeries}
\frac{1}{1-t} = \sum_{\ell=0}^\infty t^\ell .
\end{align} \qed
\end{myP}
Now pick a point $z_0$ on the complex plane and identify the nearest point, say $z_1$, where $f$ is no longer analytic. Consider some closed counterclockwise contour $C$ that lies within the circular region $|z-z_0| < |z_1-z_0|$. Then we may apply Cauchy's integral formula eq. \eqref{Complex_CauchyIntegral}, and deduce a series expansion about $z_0$:
\begin{align}
f(z)
&= \oint_C \frac{\dd z'}{2\pi i} \frac{f(z')}{z'-z} \nonumber\\
&= \oint_C \frac{\dd z'}{2\pi i} \frac{f(z')}{(z'-z_0) - (z-z_0)} = \oint_C \frac{\dd z'}{2\pi i} \frac{f(z')}{(z'-z_0)(1 - (z-z_0)/(z'-z_0))} \nonumber\\
&= \sum_{\ell=0}^\infty \oint_C \frac{\dd z'}{2\pi i} \frac{f(z')}{(z'-z_0)^{\ell+1}} \left(z-z_0\right)^\ell .
\end{align}
We have used the geometric series in eq. \eqref{Complex_GeometricSeries} and the fact that it converges uniformly to interchange the order of integration and summation. At this point, if we now recall Cauchy's integral formula for the $n$th derivative of an analytic function, eq. \eqref{Complex_CauchyIntegral_Derivative}, we have arrived at its Taylor series.
\begin{quotation}
{\it Taylor series} \qquad For $f(z)$ complex analytic within the circular region $|z-z_0| < |z_1-z_0|$, where $z_1$ is the nearest point to $z_0$ where $f$ is no longer differentiable,
\begin{align}
\label{Complex_TaylorSeries}
f(z) = \sum_{\ell=0}^\infty (z-z_0)^\ell \frac{f^{(\ell)}(z_0)}{\ell !} ,
\end{align}
where $f^{(\ell)}(z)/\ell !$ is given by eq. \eqref{Complex_CauchyIntegral_Derivative}.
\end{quotation}
\begin{myP}
\qquad Complex binomial theorem. For $p$ any real number and $z$ any complex number obeying $|z|<1$, prove the complex binomial theorem using eq. \eqref{Complex_TaylorSeries},
\begin{align}
\label{Complex_Binomial}
(1+z)^p = \sum_{\ell=0}^\infty \binom{p}{\ell} z^\ell, \qquad 
\binom{p}{0} \equiv 1, \qquad \binom{p}{\ell} = \frac{p(p-1) \dots (p-(\ell-1))}{\ell !} .
\end{align}
\end{myP}
{\bf Laurent series} \qquad We are now ready to derive the Laurent expansion of a function $f(z)$ that is analytic within an annulus, say bounded by the circles $|z-z_0| = r_1$ and $|z-z_0| = r_2 > r_1$. That is, the center of the annulus region is $z_0$ and the smaller circle has radius $r_1$ and larger one $r_2$. To start, we let $C_1$ be a clockwise circular contour with radius $r_2 > r'_1 > r_1$ and let $C_2$ be a counterclockwise circular contour with radius $r_2 > r'_2 > r'_1 > r_1$. As long as $z$ lies between these two circular contours, we have
\begin{align}
f(z) = \left(\int_{C_1} + \int_{C_2}\right) \frac{\dd z'}{2\pi i} \frac{f(z')}{z'-z}  .
\end{align}
Strictly speaking, we need to integrate along a cut line joining the $C_1$ and $C_2$ -- and another one infinitesimally close to it, in the opposite direction -- so that we can form a closed contour. But by assumption $f(z)$ is analytic and therefore continuous; the integrals along these pair of cut lines must cancel. For the $C_1$ integral, we may write $z'-z = -(z-z_0)(1 - (z'-z_0)/(z-z_0))$ and apply the geometric series in eq. \eqref{Complex_GeometricSeries} because $|(z'-z_0)/(z-z_0)| < 1$. Similarly, for the $C_2$ integral, we may write $z'-z = (z'-z_0)(1 - (z-z_0)/(z'-z_0))$ and geometric series expand the right factor because $|(z-z_0)/(z'-z_0)| < 1$. These lead us to
\begin{align}
f(z) 
&= \sum_{\ell=0}^\infty (z-z_0)^\ell \int_{C_2} \frac{\dd z'}{2\pi i} \frac{f(z')}{(z'-z_0)^{\ell+1}} 
- \sum_{\ell=0}^{\infty} \frac{1}{(z-z_0)^{\ell+1}}\int_{C_1} \frac{\dd z'}{2\pi i} (z'-z_0)^\ell f(z') .
\end{align}
Remember complex integration can be thought of as a line integral, which reverses sign if we reverse the direction of the line integration. Therefore we may absorb the $-$ sign in front of the $C_1$ integral(s) by turning $C_1$ from a clockwise circle into $C'_1 = -C_1$, a counterclockwise one. Moreover, note that we may now deform the contour $C'_1$ into $C_2$,
\begin{align}
\int_{C'_1} \frac{\dd z'}{2\pi i} (z'-z_0)^\ell f(z') = \int_{C_2} \frac{\dd z'}{2\pi i} (z'-z_0)^\ell f(z') ,
\end{align}
because for positive $\ell$ the integrand $(z'-z_0)^\ell f(z')$ is analytic in the region lying between the circles $C'_1$ and $C_2$. At this point we have 
\begin{align}
f(z) 
&= \sum_{\ell=0}^\infty \int_{C_2} \frac{\dd z'}{2\pi i} \left( (z-z_0)^\ell \frac{f(z')}{(z'-z_0)^{\ell+1}} + \frac{1}{(z-z_0)^{\ell+1}} (z'-z_0)^\ell f(z') \right) .
\end{align}
Proceeding to re-label the second series by replacing $\ell+1 \to -\ell'$, so that the summation then runs from $-1$ through $-\infty$, the Laurent series emerges. 
\begin{quotation}
{\it Laurent series} \qquad Let $f(z)$ be analytic within the annulus $r_1 < |z-z_0| < r_2 < |z_1-z_0|$, where $z_0$ is some complex number such that $f(z)$ may not be analytic within $|z-z_0| < r_1$; $z_1$ is the nearest point outside of $|z-z_0| \geq r_1$ where $f(z)$ fails to be differentiable; and the radii $r_2 > r_1 > 0$ are real positive numbers. The Laurent expansion of $f(z)$ about $z_0$, valid throughout the entire annulus, reads
\begin{align}
\label{Complex_LaurentSeries}
f(z) 		&= \sum_{\ell=-\infty}^\infty L_\ell(z_0) \cdot (z-z_0)^\ell , \\
L_\ell(z_0) &\equiv \int_C \frac{\dd z'}{2\pi i} \frac{f(z')}{(z'-z_0)^{\ell+1}} .
\end{align}
The $C$ is any counterclockwise closed contour containing both $z$ and the inner circle $|z-z_0| = r_1$.
\end{quotation}
{\it Uniqueness} \qquad It is worth asserting that the Laurent expansion of a function, in the region where it is analytic, is unique. That means it is not always necessary to perform the integrals in eq. \eqref{Complex_LaurentSeries} to obtain the expansion coefficients $L_\ell$.
\begin{myP}
\qquad For complex $z$, $a$ and $b$, obtain the Laurent expansion of 
\begin{align}
f(z) \equiv \frac{1}{(z-a)(z-b)} , \qquad a \neq b,
\end{align}
about $z=a$, in the region $0 < |z-a| < |a-b|$ using eq. \eqref{Complex_LaurentSeries}. Check your result either by writing 
\begin{align}
\frac{1}{z-b} = -\frac{1}{1 - (z-a)/(b-a)} \frac{1}{b-a} .
\end{align}
and employing the geometric series in eq. \eqref{Complex_GeometricSeries}, or directly performing a Taylor expansion of $1/(z-b)$ about $z=a$.
\end{myP}
\begin{myP}
{\it Schwarz reflection principle.} \qquad Proof the following statement using Laurent expansion. If a function $f(z=x+iy) = u(x,y) + iv(x,y)$ can be Laurent expanded (for $x$, $y$, $u$, and $v$ real) about some point on the real line, and if $f(z)$ is real whenever $z$ is real, then
\begin{align}
(f(z))^* = u(x,y) - iv(x,y) = f(z^*) = u(x,-y) + iv(x,-y) .
\end{align}
Comment on why this is called the ``reflection principle". \qed
\end{myP}
We now turn to an important result that allows us to extend the definitions of complex differentiable functions beyond their original range of validity.
\begin{quotation}
{\bf Analytic continuation} \qquad An analytic function $f(z)$ is fixed uniquely throughout a given region $\Sigma$ on the complex plane, once its value is specified on a line segment lying within $\Sigma$. 
\end{quotation}
This in turn means, suppose we have an analytic function $f_1(z)$ defined in a region $\Sigma_1$ on the complex plane, and suppose we found another analytic function $f_2(z)$ defined in some region $\Sigma_2$ such that $f_2(z)$ agrees with $f_1(z)$ in their common region of intersection. (It is important that $\Sigma_2$ does have some overlap with $\Sigma_1$.) Then we may view $f_2(z)$ as an analytic continuation of $f_1(z)$, because this extension is unique -- it is not possible to find a $f_3(z)$ that agrees with $f_1(z)$ in the common intersection between $\Sigma_1$ and $\Sigma_2$, yet behave different in the rest of $\Sigma_2$.

These results inform us, any real differentiable function we are familiar with can be extended to the complex plane, simply by knowing its Taylor expansion. For example, $e^x$ is infinitely differentiable on the real line, and its definition can be readily extended into the complex plane via its Taylor expansion.

An example of analytic continuation is that of the geometric series. If we define
\begin{align}
f_1(z) \equiv \sum_{\ell=0}^{\infty} z^\ell, \qquad |z| < 1,
\end{align}
and
\begin{align}
f_2(z) &\equiv \frac{1}{1-z} ,
\end{align}
then we know they agree in the region $|z| < 1$ and therefore any line segment within it. But while $f_1(z)$ is defined only in this region, $f_2(z)$ is valid for any $z \neq 1$. Therefore, we may view $1/(1-z)$ as the analytic continuation of $f_1(z)$ for the region $|z|>1$. Also observe that we can now understand why the series is valid only for $|z| < 1$: the series of $f_1(z)$ is really the Taylor expansion of $f_2(z)$ about $z=0$, and since the nearest singularity is at $z=1$, the circular region of validity employed in our (constructive) Taylor series proof is in fact $|z| < 1$.
\begin{myP} \qquad
One key application of analytic continuation is that, some special functions in mathematical physics admit a power series expansion that has a finite radius of convergence. This can occur if the differential equations they solve have singular points. Many of these special functions also admit an integral representation, whose range of validity lies beyond that of the power series. This allows the domain of these special functions to be extended.

The hypergeometric function $\,_2F_1(\alpha,\beta;\gamma;z)$ is such an example. For $|z| < 1$ it has a power series expansion
\begin{align}
\label{Complex_2F1_Series}
\,_2F_1(\alpha,\beta;\gamma;z) &= \sum_{\ell=0}^{\infty} C_\ell(\alpha,\beta;\gamma) \frac{z^\ell}{\ell !} , \nonumber\\
C_0(\alpha,\beta;\gamma) &\equiv 1 , \nonumber\\
C_{\ell \geq 1}(\alpha,\beta;\gamma) &\equiv \frac{\alpha (\alpha+1) \dots (\alpha+(\ell-1)) \cdot \beta (\beta+1) \dots (\beta+(\ell-1))}{\gamma (\gamma+1) \dots (\gamma+(\ell-1))} .
\end{align}
On the other hand, it also has the following integral representation,
\begin{align}
\label{Complex_2F1_IntegralRep}
\,_2F_1(\alpha,\beta;\gamma;z) &= \frac{\Gamma(\gamma)}{\Gamma(\gamma-\beta) \Gamma(\beta)}
\int_{0}^{1} t^{\beta-1} (1-t)^{\gamma-\beta-1} (1-tz)^{-\alpha} \dd t, \qquad \text{Re$(\gamma) > $Re$(\beta) > 0$.}
\end{align}
(Here, $\Gamma(z)$ is known as the Gamma function; see http://dlmf.nist.gov/5.) Show that eq. \eqref{Complex_2F1_IntegralRep} does in fact agree with eq. \eqref{Complex_2F1_Series} for $|z| < 1$. You can apply the binomial expansion in eq. \eqref{Complex_Binomial} to $(1-tz)^{-\alpha}$, followed by result
\begin{align}
\int_{0}^{1} \dd t (1-t)^{\alpha-1} t^{\beta-1} = \frac{\Gamma(\alpha) \Gamma(\beta)}{\Gamma(\alpha+\beta)}, \qquad \text{Re$(\alpha)$, Re$(\beta) > 0$} .
\end{align}
You may also need the property
\begin{align}
z \Gamma(z) = \Gamma(z+1) .
\end{align}
Therefore eq. \eqref{Complex_2F1_IntegralRep} extends eq. \eqref{Complex_2F1_Series} into the region $|z|>1$. \qquad \qed
\end{myP}

\subsection{Poles and Residues}

In this section we will consider the closed counterclockwise contour integral
\begin{align}
\oint_C \frac{\dd z}{2\pi i} f(z) ,
\end{align}
where $f(z)$ is analytic everywhere on and within $C$ except at isolated singular points of $f(z)$ -- which we will denote as $\{z_1,\dots,z_n\}$, for $(n \geq 1)$-integer. That is, we will assume there is no other type of singularities. We will show that the result is the sum of the residues of $f(z)$ at these points. This case will turn out to have a diverse range of physical applications, including the study of the vibrations of black holes.

We begin with some jargon.

{\bf Nomenclature} \qquad If a function $f(z)$ admits a Laurent expansion about $z=z_0$ starting from $1/(z-z_0)^m$, for $m$ some positive integer,
\begin{align}
f(z) = \sum_{\ell=-m}^{\infty} L_\ell \cdot (z-z_0)^\ell ,
\end{align}
we say the function has a pole of order $m$ at $z=z_0$. If $m=\infty$ we say the function has an essential singularity. The residue of a function $f$ at some location $z_0$ is simply the coefficient $L_{-1}$ of the negative one power ($\ell=-1$ term) of the Laurent series expansion about $z=z_0$.

The key to the result already advertised is the following.
\begin{myP}
\qquad If $n$ is an arbitrary integer, show that
\begin{align}
\label{Complex_IntegralPowerLaw}
\oint_C (z'-z)^n \frac{\dd z'}{2\pi i} 
&= 1 ,\qquad \text{when $n=-1$}, \nonumber \\
&= 0 ,\qquad \text{when $n \neq -1$},
\end{align}
where $C$ is any contour (whose interior defines a simply connected domain) that encloses the point $z' = z$. \qed
\end{myP}
By assumption, we may deform our contour $C$ so that they become the collection of closed counterclockwise contours $\{C'_i|i=1,2,\dots,n\}$ around each and every isolated point. This means
\begin{align}
\oint_{C} f(z') \frac{\dd z'}{2\pi i} = \sum_i \oint_{C'_i} f(z') \frac{\dd z'}{2\pi i} .
\end{align}
Strictly speaking, to preserve the full closed contour structure of the original $C$, we need to join these new contours -- say $C'_i$ to $C'_{i+1}$, $C'_{i+1}$ to $C'_{i+2}$, and so on -- by a pair of contour lines placed infinitesimally apart, for e.g., one from $C'_i \to C'_{i+1}$ and the other $C'_{i+1} \to C'_{i}$. But by assumption $f(z)$ is analytic and therefore continuous there, and thus the contribution from these pairs will surely cancel. Let us perform a Laurent expansion of $f(z)$ about $z_i$, the $i$th singular point, and then proceed to integrate the series term-by-term using eq. \eqref{Complex_IntegralPowerLaw}.
\begin{align}
\oint_{C'_i} f(z') \frac{\dd z'}{2\pi i} 
= \int_{C'_i} \sum_{\ell=-m_i}^{\infty} L_\ell^{(i)} \cdot (z'-z_i)^\ell \frac{\dd z'}{2\pi i} 
= L_{-1}^{(i)} .
\end{align}
\begin{quotation}
{\bf Residue theorem} \qquad As advertised, the closed counterclockwise contour integral of a function that is analytic everywhere on and within the contour, except at isolated points $\{ z_i \}$, yields the sum of the residues at each of these points. In equation form,
\begin{align}
\oint_{C} f(z') \frac{\dd z'}{2\pi i} = \sum_i L_{-1}^{(i)} ,
\end{align}
where $L_{-1}^{(i)}$ is the residue at the $i$th singular point $z_i$.
\end{quotation}
{\it Example I} \qquad Let us start with a simple application of this result. Let $C$ be some closed counterclockwise contour containing the points $z=0,a,b$.
\begin{align}
I = \oint_C \frac{\dd z}{2\pi i} \frac{1}{z(z-a)(z-b)} .
\end{align}
One way to do this is to perform a partial fractions expansion first.
\begin{align}
I = \oint_C \frac{\dd z}{2\pi i} \left( \frac{1}{abz} + \frac{1}{a(a-b)(z-a)} + \frac{1}{b(b-a)(z-b)} \right) .
\end{align}
In this form, the residues are apparent, because we can view the first term as some Laurent expansion about $z=0$ with only the negative one power; the second term as some Laurent expansion about $z=a$; the third about $z=b$. Therefore, the sum of the residues yield
\begin{align}
I = \frac{1}{ab} + \frac{1}{a(a-b)} + \frac{1}{b(b-a)} = \frac{(a-b) + b - a}{ab(a-b)}= 0 .
\end{align}
If you don't do a partial fractions decomposition, you may instead recognize, as long as the 3 points $z=0,a,b$ are distinct, then near $z=0$ the factor $1/((z-a)(z-b))$ is analytic and admits an ordinary Taylor series that begins at the zeroth order in $z$, i.e.,
\begin{align}
\frac{1}{z(z-a)(z-b)} = \frac{1}{z}\left( \frac{1}{ab} + \mathcal{O}(z) \right) .
\end{align}
Because the higher positive powers of the Taylor series cannot contribute to the $1/z$ term of the Laurent expansion, to extract the negative one power of $z$ in the Laurent expansion of the integrand, we simply evaluate this factor at $z=0$. Likewise, near $z=a$, the factor $1/(z(z-b))$ is analytic and can be Taylor expanded in zero and positive powers of $(z-a)$. To understand the residue of the integrand at $z=a$ we simply evaluate $1/(z(z-b))$ at $z=a$. Ditto for the $z=b$ singularity.
\begin{align}
\oint_C \frac{\dd z}{2\pi i} \frac{1}{z(z-a)(z-b)}
&= \sum_{z_i = 0,a,b} \left(\text{Residue of } \frac{1}{z(z-a)(z-b)} \text{ at $z_i$}\right) \nonumber\\
&= \frac{1}{ab} + \frac{1}{a(a-b)} +  \frac{1}{b(b-a)} = 0.
\end{align}
The reason why the result is zero can actually be understood via contour integration as well. If you now consider a closed clockwise contour $C_\infty$ at infinity and view the integral $(\int_C + \int_{C_\infty}) f(z) \dd z$, you will be able to convert it into a closed contour integral by linking $C$ and $C_\infty$ via two infinitesimally close radial lines which would not actually contribute to the answer. But $(\int_C + \int_{C_\infty}) f(z) \dd z = \int_{C_\infty} f(z) \dd z$ because $C_\infty$ does not contribute either -- why? Therefore, since there are no poles in the region enclosed by $C_\infty$ and $C$, the answer has to be zero.

{\it Example II} \qquad Let $C$ be a closed counterclockwise contour around the origin $z=0$. Let us do
\begin{align}
I \equiv \oint_C \exp(1/z^2) \dd z .
\end{align} 
We Taylor expand the exp, and notice there is no term that goes as $1/z$. Hence,
\begin{align}
I = \sum_{\ell=0}^\infty \frac{1}{\ell !} \oint_C \frac{\dd z}{z^{2\ell}} = 0 .
\end{align} 
A major application of contour integration is to that of integrals involving real variables. 

{\bf Application I: Trigonometric integrals} \qquad If we have an integral of the form
\begin{align}
\int_0^{2\pi} \dd \theta f(\cos\theta,\sin\theta) ,
\end{align} 
then it may help to change from $\theta$ to
\begin{align}
z \equiv e^{i\theta} \qquad \Rightarrow \qquad \dd z = i \dd \theta \cdot e^{i\theta} = i \dd\theta \cdot z, 
\end{align}
and
\begin{align}
\sin\theta = \frac{z - 1/z}{2i}, \qquad \cos\theta = \frac{z + 1/z}{2} .
\end{align}
The integral is converted into a sum over residues:
\begin{align}
\label{Complex_TrigIntegral}
\int_0^{2\pi} \dd \theta f(\cos\theta,\sin\theta) 
&= 2\pi \oint_{|z| = 1} \frac{\dd z}{2\pi i z} f\left( \frac{z + 1/z}{2}, \frac{z - 1/z}{2i}\right) \nonumber\\
&= 2\pi \sum_j \left(\text{$j$th residue of } \frac{f\left( \frac{z + 1/z}{2}, \frac{z - 1/z}{2i}\right)}{z} \text{ for $|z| < 1$}\right) .
\end{align} 
{\it Example} \qquad For $a \in \mathbb{R}$,
{\allowdisplaybreaks\begin{align}
I = \int_{0}^{2\pi} \frac{\dd \theta}{a + \cos\theta} 
&= \oint_{|z| = 1} \frac{\dd z}{i z} \frac{1}{a + (1/2)(z+1/z)} = \oint_{|z| = 1} \frac{\dd z}{i} \frac{1}{az+ (1/2)(z^2+1)} \nonumber\\
&= 4\pi \oint_{|z| = 1} \frac{\dd z}{2\pi i} \frac{1}{(z- z_+)(z-z_-)}, \qquad z_\pm \equiv -a \pm \sqrt{a^2-1} .
\end{align}}
Assume, for the moment, that $|a| < 1$. Then $|-a \pm \sqrt{a^2-1}|^2 = |-a \pm i \sqrt{1-a^2}|^2 = |a^2 + (1-a^2)|^2 = 1$. Both $z_\pm$ lie on the unit circle, and the contour integral does not make much sense as it stands because the contour $C$ passes through both $z_\pm$. So let us assume that $a$ is real but $|a| > 1$. When $a$ runs from 1 to infinity, $-a-\sqrt{a^2-1}$ runs from $-1$ to $-\infty$; while $-a+\sqrt{a^2-1} = -(a-\sqrt{a^2-1})$ runs from $-1$ to $0$ because $a > \sqrt{a^2-1}$. When $-a$ runs from $1$ to $\infty$, on the other hand, $-a-\sqrt{a^2-1}$ runs from $1$ to $0$; while $-a+\sqrt{a^2-1}$ runs from $1$ to $\infty$. In other words, for $a>1$, $z_+ = -a+\sqrt{a^2-1}$ lies within the unit circle and the relevant residue is $1/(z_+-z_-) = 1/(2\sqrt{a^2-1}) = \text{sgn}(a)/(2\sqrt{a^2-1})$. For $a<-1$ it is $z_- = -a-\sqrt{a^2-1}$ that lies within the unit circle and the relevant residue is $1/(z_--z_+) = - 1/(2\sqrt{a^2-1}) = \text{sgn}(a)/(2\sqrt{a^2-1})$. Therefore,
\begin{align}
\int_{0}^{2\pi} \frac{\dd \theta}{a + \cos\theta} = \frac{2\pi \text{sgn}(a)}{\sqrt{a^2-1}}, \qquad a\in\mathbb{R}, \ |a|>1 .
\end{align}
{\bf Application II: Integrals along the real line} \qquad If you need to do $\int_{-\infty}^{+\infty}f(z)\dd z$, it may help to view it as a complex integral and ``close the contour" either in the upper or lower half of the complex plane -- thereby converting the integral along the real line into one involving the sum of residues in the upper or lower plane.

An example is the following
\begin{align}
I \equiv \int_{-\infty}^{\infty} \frac{\dd z}{z^2+z+1} .
\end{align}
Let us complexify the integrand and consider its behavior in the limit $z = \lim_{\rho \to \infty} \rho e^{i\theta}$, either for $0 \leq \theta \leq \pi$ (large semi-circle in the upper half plane) or $\pi \leq \theta \leq 2\pi$ (large semi-circle in the lower half plane).
\begin{align}
\lim_{\rho\to\infty} \left\vert \frac{i\dd\theta \cdot \rho e^{i\theta}}{\rho^2 e^{i2\theta}+\rho e^{i\theta}+1} \right\vert
\to \lim_{\rho\to\infty} \frac{\dd\theta}{\rho} = 0 .
\end{align}
This is saying the integral along this large semi-circle either in the upper or lower half complex plane is zero. Therefore $I$ is equal to the integral along the real axis plus the contour integral along the semi-circle, since the latter contributes nothing. But the advantage of this view is that we now have a closed contour integral. Because the roots of the polynomial in the denominator of the integrand are $e^{-i2\pi/3}$ and $e^{i2\pi/3}$, so we may write
\begin{align}
I = 2\pi i \oint_C \frac{\dd z}{2\pi i}\frac{1}{(z-e^{-i2\pi/3})(z-e^{i2\pi/3})} .
\end{align}
Closing the contour in the upper half plane yields a counterclockwise path, which yields
\begin{align}
I = \frac{2\pi i}{e^{i2\pi/3}-e^{-i2\pi/3}} = \frac{\pi}{\sin(2\pi/3)}  .
\end{align}
Closing the contour in the lower half plane yields a clockwise path, which yields
\begin{align}
I = \frac{-2\pi i}{e^{-i2\pi/3}-e^{i2\pi/3}} = \frac{\pi}{\sin(2\pi/3)} .
\end{align}
Of course, the two answers have to match.

{\it Example: Fourier transform} \qquad The Fourier transform is in fact a special case of the integral on the real line that can often be converted to a closed contour integral.
\begin{align}
\label{ComplexAnalysis_FT}
f(t) = \int_{-\infty}^{\infty} \widetilde{f}(\omega) e^{i\omega t} \frac{\dd \omega}{2\pi}, \qquad t \in\mathbb{R} .
\end{align}
We will assume $t$ is real and $\widetilde{f}$ has only isolated singularities.\footnote{In physical applications $\widetilde{f}$ may have branch cuts; this will be dealt with in the next section.} Let $C$ be a large semi-circular path, either in the upper or lower complex plane; consider the following integral along $C$.
\begin{align}
I' \equiv \int_C \widetilde{f}(\omega) e^{i\omega t} \frac{\dd \omega}{2\pi}
&= \lim_{\rho \to \infty} \int \widetilde{f}\left( \rho e^{i\theta} \right) e^{i\rho (\cos\theta) t} e^{- \rho (\sin\theta) t} \frac{i \dd\theta \cdot \rho e^{i\theta}}{2\pi}
\end{align}
At this point we see that, for $t<0$, unless $\widetilde{f}$ goes to zero much faster than the $e^{- \rho (\sin\theta) t}$ for large $\rho$, the integral blows up in the upper half plane where $(\sin\theta)>0$. For $t>0$, unless $f$ goes to zero much faster than the $e^{- \rho (\sin\theta) t}$ for large $\rho$, the integral blows up in the lower half plane where $(\sin\theta)<0$. In other words, the sign of $t$ will determine how you should ``close the contour" -- in the upper or lower half plane.

Let us suppose $|\widetilde{f}| \leq M$ on the semi-circle and consider the magnitude of this integral,
\begin{align}
|I'| \leq \lim_{\rho \to \infty} \left(\rho M \int e^{- \rho (\sin\theta) t} \frac{\dd\theta}{2\pi} \right),
\end{align}
Remember if $t>0$ we integrate over $\theta \in [0,\pi]$, and if $t<0$ we do $\theta \in [-\pi,0]$. Either case reduces to
\begin{align}
|I'| \leq \lim_{\rho \to \infty} \left(2 \rho M \int_0^{\pi/2} e^{- \rho (\sin\theta) |t|} \frac{\dd\theta}{2\pi} \right),
\end{align}
because 
\begin{align}
\int_0^\pi F(\sin(\theta)) \dd\theta = 2 \int_0^{\pi/2} F(\sin(\theta)) \dd\theta 
\end{align}
for any function $F$. The next observation is that, over the range $\theta \in [0,\pi/2]$,
\begin{align}
\frac{2\theta}{\pi} \leq \sin\theta ,
\end{align}
because $y = 2\theta/\pi$ is a straight line joining the origin to the maximum of $y=\sin\theta$ at $\theta=\pi/2$. (Making a plot here helps.) This in turn means we can replace $\sin\theta$ with $2\theta/\pi$ in the exponent, i.e., exploit the inequality $e^{-X} < e^{-Y}$ if $X > Y > 0$, and deduce
\begin{align}
|I'| &\leq \lim_{\rho \to \infty} \left(2 \rho M \int_0^{\pi/2} e^{- 2\rho \theta |t|/\pi} \frac{\dd\theta}{2\pi} \right) \\
&= \lim_{\rho \to \infty} \left(\frac{\rho M}{\pi} \pi \frac{e^{- \rho \pi |t|/\pi}-1}{-2\rho |t|} \right) 
= \frac{1}{2|t|} \lim_{\rho \to \infty} M 
\end{align}
As long as $|\widetilde{f}(\omega)|$ goes to zero as $\rho \to \infty$, we see that $I'$ (which is really $0$) can be added to the Fourier integral $f(t)$ along the real line, converting $f(t)$ to a closed contour integral. If $\widetilde{f}(\omega)$ is analytic except at isolated points, then $I$ can be evaluated through the sum of residues at these points.

To summarize, when faced with the frequency-transform type integral in eq. \eqref{ComplexAnalysis_FT},
\begin{itemize}
\item If $t>0$ and if $|\widetilde{f}(\omega)|$ goes to zero as $|\omega| \to \infty$ on the large semi-circle path of radius $|\omega|$ on the upper half complex plane, then we close the contour there and convert the integral $f(t) = \int_{-\infty}^{\infty} \widetilde{f}(\omega) e^{i\omega t} \frac{\dd \omega}{2\pi}$ to $i$ times the sum of the residues of $\widetilde{f}(\omega) e^{i\omega t}$ for Im$(\omega)>0$ -- provided the function $\widetilde{f}(\omega)$ is analytic except at isolated points there.

\item If $t<0$ and if $|\widetilde{f}(\omega)|$ goes to zero as $|\omega| \to \infty$ on the large semi-circle path of radius $|\omega|$ on the lower half complex plane, then we close the contour there and convert the integral $f(t) = \int_{-\infty}^{\infty} \widetilde{f}(\omega) e^{i\omega t} \frac{\dd \omega}{2\pi}$ to $-i$ times the sum of the residues of $\widetilde{f}(\omega) e^{i\omega t}$ for Im$(\omega)<0$ -- provided the function $\widetilde{f}(\omega)$ is analytic except at isolated points there.

\item A quick guide to how to close the contour is to evaluate the exponential on the imaginary $\omega$ axis, and take the infinite radius limit of $|\omega|$, namely $\lim_{|\omega| \to \infty} e^{it(\pm i|\omega|)} = \lim_{|\omega| \to \infty} e^{\mp t|\omega|}$, where the upper sign is for the positive infinity on the imaginary axis and the lower sign for negative infinity. We want the exponential to go to zero, so we have to choose the upper/lower sign based on the sign of $t$.

\end{itemize}

If $\widetilde{f}(\omega)$ requires branch cut(s) in either the lower or upper half complex planes -- branch cuts will be discussed shortly -- we may still use this closing of the contour to tackle the Fourier integral $f(t)$. In such a situation, there will often be additional contributions from the part of the contour hugging the branch cut itself.

An example is the following integral
\begin{align}
I(t) \equiv \int_{-\infty}^{+\infty} \frac{\dd\omega}{2\pi} \frac{e^{i\omega t}}{(\omega+i)^2 (\omega-2i)} , \qquad t\in\mathbb{R} .
\end{align}
The denominator $(\omega+i)^2 (\omega-2i)$ has a double root at $\omega=-i$ (in the lower half complex plane) and a single root at $\omega=2i$ (in the upper half complex plane). You can check readily that $1/((\omega+i)^2 (\omega-2i))$ does go to zero as $|\omega| \to \infty$. If $t>0$ we close the integral on the upper half complex plane. Since $e^{i\omega t}/(\omega+i)^2$ is analytic there, we simply apply Cauchy's integral formula in eq. \eqref{Complex_CauchyIntegral}.
\begin{align}
I(t>0) = i \frac{e^{i(2i) t}}{(2i+i)^2} = -i \frac{e^{-2t}}{9}  .
\end{align}
If $t<0$ we then need form a closed {\it clockwise} contour $C$ by closing the integral along the real line in the lower half plane. Here, $e^{i\omega t}/(\omega-2i)$ is analytic, and we can invoke eq. \eqref{Complex_CauchyIntegral_Derivative},
\begin{align}
I(t<0) 
= i \oint_C \frac{\dd\omega}{2\pi i} \frac{e^{i \omega t}}{(\omega+i)^2 (\omega-2i)}
&= -i \frac{\dd}{\dd \omega} \left( \frac{e^{i\omega t}}{\omega - 2i} \right)_{\omega = -i} \nonumber\\
&= -i e^{t} \frac{1-3t}{9}
\end{align}
To summarize,
\begin{align}
\int_{-\infty}^{+\infty} \frac{\dd\omega}{2\pi} \frac{e^{i\omega t}}{(\omega+i)^2 (\omega-2i)}
= -i \frac{e^{-2t}}{9} \Theta(t) - i e^{t} \frac{1-3t}{9} \Theta(-t) ,
\end{align}
where $\Theta(t)$ is the step function. 

We can check this result as follows. Since $I(t=0) = -i/9$ can be evaluated independently, this indicates we should expect the $I(t)$ to be continuous there: $I(t=0^+) = I(t=-0^+) = -i/9$. Also notice, if we apply a $t$-derivative on $I(t)$ and interchange the integration and derivative operation, each $\dd/\dd t$ amounts to a $i \omega$. Therefore, we can check the following differential equations obeyed by $I(t)$:
{\allowdisplaybreaks\begin{align}
\left(\frac{1}{i} \frac{\dd}{\dd t} + i \right)^2 \left(\frac{1}{i} \frac{\dd}{\dd t} - 2i \right) I(t) &= \delta(t) , \\
\left(\frac{1}{i} \frac{\dd}{\dd t} + i \right)^2 I(t) 
			&= \int_{-\infty}^{+\infty} \frac{\dd\omega}{2\pi} \frac{e^{i\omega t}}{\omega-2i} = i \Theta(t) e^{-2t} , \\
\left(\frac{1}{i} \frac{\dd}{\dd t} - 2i \right) I(t) 
			&= \int_{-\infty}^{+\infty} \frac{\dd\omega}{2\pi} \frac{e^{i\omega t}}{(\omega+i)^2} = -i \Theta(-t) i t e^{t} = \Theta(-t) t e^{t} .
\end{align}}
\begin{myP}
\qquad Evaluate
\begin{align}
\int_{-\infty}^{\infty} \frac{\dd z}{z^3 + i} .
\end{align} \qed
\end{myP}
\begin{myP}
\qquad Show that the integral representation of the step function $\Theta(t)$ is
\begin{align}
\Theta(t) = \int_{-\infty}^{+\infty} \frac{\dd \omega}{2\pi i} \frac{e^{i\omega t}}{\omega - i0^+} .
\end{align}
The $\omega - i0^+$ means the purely imaginary root lies very slightly above $0$; alternatively one would view it as an instruction to deform the contour by making an infinitesimally small counterclockwise semi-circle going slightly below the real axis around the origin.

Next, let $a$ and $b$ be non-zero real numbers. Evaluate
\begin{align}
I(a,b) \equiv \int_{-\infty}^{+\infty} \frac{\dd \omega}{2\pi i} \frac{e^{i\omega a}}{\omega + i b} .
\end{align} \qed
\end{myP}
\begin{myP}
\qquad (From Arfken et al.) Sometimes this ``closing-the-contour" trick need not involve closing the contour at infinity. Show by contour integration that
\begin{align}
\label{BranchCutExample}
I \equiv \int_{0}^{\infty} \frac{(\ln x)^2}{1+x^2} \dd x = \frac{\pi^3}{8} .
\end{align}
Hint: Put $x = z \equiv e^t$ and try to evaluate the integral now along the contour that runs along the real line from $t=-R$ to $t=R$ -- for $R \gg 1$ -- then along a vertical line from $t=R$ to $t=R+i\pi$, then along the horizontal line from $t=R+i\pi$ to $t=-R+i\pi$, then along the vertical line back to $t=-R$; then take the $R \to +\infty$ limit. \qed
\end{myP}
\begin{myP}
\qquad Evaluate 
\begin{align}
I(a) \equiv \int_{-\infty}^{\infty} \frac{\sin(a x)}{x} \dd x, \qquad a \in \mathbb{R} .
\end{align}
Hint(s): First convert the sine into exponentials and deform the contour along the real line into one that makes a infinitesimally small semi-circular detour around the origin $z=0$. The semi-circle can be clockwise, passing above $z=0$ or counterclockwise, going below $z=0$. Make sure you justify why making such a small deformation does not affect the answer. \qed
\end{myP}
\begin{myP}
\qquad Evaluate
\begin{align}
I(t) \equiv \int_{-\infty}^{+\infty} \frac{\dd\omega}{2\pi} \frac{e^{-i\omega t}}{(\omega-ia)^2 (\omega+ib)^2} , \qquad t\in\mathbb{R}; \ a,b > 0 .
\end{align} \qed
\end{myP}

\subsection{Branch Points, Branch Cuts}

{\bf Branch points and Riemann sheets} \qquad A branch point of a function $f(z)$ is a point $z_0$ on the complex plane such that going around $z_0$ in an infinitesimally small circle does not give you back the same function value. That is,
\begin{align}
f\left( z_0 + \epsilon \cdot e^{i\theta} \right) \neq f\left( z_0 + \epsilon \cdot e^{i(\theta+2\pi)} \right), 
\qquad\qquad
 0 < \epsilon \ll 1 .
\end{align}
{\it Example I} \qquad One example is the power $z^\alpha$, for $\alpha$ non-integer. Zero is a branch point because, for $0 < \epsilon \ll 1$, we may considering circling it $n \in \mathbb{Z}^+$ times.
\begin{align}
(\epsilon e^{2\pi n i})^\alpha = \epsilon^\alpha e^{2\pi n \alpha i} \neq \epsilon^\alpha .
\end{align} 
If $\alpha = 1/2$, then circling zero twice would bring us back to the same function value. If $\alpha = 1/m$, where $m$ is a positive integer, we would need to circle zero $m$ times to get back to the same function value. What this is teaching us is that, to define the function $f(z) = z^{1/m}$ properly, we need $m$ ``Riemann sheets" of the complex plane. To see this, we first define a cut line along the positive real line and proceed to explore the function $f$ by sampling its values along a continuous line. If we start from a point slightly above the real axis, $z^{1/m}$ there is defined as $|z|^{1/m}$, where the positive root is assumed here. As we move around the complex plane, let us use polar coordinates to write $z = \rho e^{i\theta}$; once $\theta$ runs beyond $2\pi$, i.e., once the contour circles around the origin more than one revolution, we exit the first complex plane and enter the second. For example, when $z$ is slightly above the real axis on the second sheet, we define $z=|z|^{1/m} e^{i2\pi/m}$; and anywhere else on the second sheet we have $z=|z|^{1/m} e^{i(2\pi/m) + i \theta}$, where $\theta$ is still measured with respect to the real axis. We can continue this process, circling the origin, with each increasing counterclockwise revolution taking us from one sheet to the next. On the $n$th sheet our function reads $z=|z|^{1/m} e^{i(2\pi n/m) + i \theta}$. It is the $m$th sheet that needs to be joined with the very first sheet, because by the $m$th sheet we have covered all the $m$ solutions of what we mean by taking the $m$th root of a complex number. (If we had explored the function using a clockwise path instead, we'd migrated from the first sheet to the $m$th sheet, then to the $(m-1)$th sheet and so on.) Finally, if $\alpha$ were not rational -- it is not the ratio of two integers -- we would need an infinite number of Riemann sheets to fully describe $z^\alpha$ as a complex differentiable function of $z$.

The presence of the branch cut(s) is necessary because we need to join one Riemann sheet to the next, so as to construct an analytic function mapping the full domain back to the complex plane. However, as long as one Riemann sheet is joined to the next so that the function is analytic across this boundary, and as long as the full domain is mapped properly onto the complex plane, the location of the branch cut(s) is arbitrary. For example, for the $f(z) = z^\alpha$ case above, as opposed to the real line, we can define our branch cut to run along the radial line $\{\rho e^{i\theta_0} | \rho \geq 0\}$ for any $0 < \theta_0 \leq 2\pi$. All we are doing is re-defining where to join one sheet to another, with the $n$th sheet mapping one copy of the complex plane $\{\rho e^{i(\theta_0+\varphi)} | \rho \geq 0, 0 \leq \varphi < 2\pi \}$ to $\{ |z|^\alpha e^{i\alpha(\theta_0 +\varphi)} | \rho \geq 0, 0 \leq \varphi < 2\pi \}$. Of course, in this new definition, the $2\pi-\theta_0 \leq \varphi < 2\pi$ portion of the $n$th sheet would have belonged to the $(n+1)$th sheet in the old definition -- but, taken as a whole, the collection of all relevant Riemann sheets still cover the same domain as before.

{\it Example II} \qquad $\ln$ is another example. You already know the answer but let us work out the complex derivative of $\ln z$. Because $e^{\ln z} = z$, we have
\begin{align}
(e^{\ln z})' = e^{\ln z} \cdot (\ln z)' = z \cdot (\ln z)' = 1 .
\end{align}
This implies,
\begin{align}
\frac{\dd \ln z}{\dd z} = \frac{1}{z}, \qquad z \neq 0,
\end{align}
which in turn says $\ln z$ is analytic away from the origin. We may now consider making $m$ infinitesimal circular trips around $z=0$.
\begin{align}
\ln (\epsilon e^{i 2\pi m}) = \ln (\epsilon e^{i 2\pi m}) = \ln \epsilon + i 2\pi m \neq \ln \epsilon .
\end{align}
Just as for $f(z)=z^\alpha$ when $\alpha$ is irrational, it is in fact not possible to return to the same function value -- the more revolutions you take, the further you move in the imaginary direction. $\ln(z)$ for $z=x+iy$ actually maps the $m$th Riemann sheet to a horizontal band on the complex plane, lying between $2\pi(m-1) \leq \text{Im} \ln(z) \leq 2\pi m$.

{\it Breakdown of Laurent series} \qquad To understand the need for multiple Riemann sheets further, it is instructive to go back to our discussion of the Laurent series using an annulus around the isolated singular point, which lead up to eq. \eqref{Complex_LaurentSeries}. For both $f(z)=z^\alpha$ and $f(z)=\ln(z)$, the branch point is at $z=0$. If we had used a single complex plane, with say a branch cut along the positive real line, $f(z)$ would not even be continuous -- let alone analytic -- across the $z=x>0$ line: $f(z=x + i0^+) = x^\alpha \neq f(z=x - i0^+) = x^\alpha e^{i 2\pi \alpha}$, for instance. Therefore the derivation there would not go through, and a Laurent series for either $z^\alpha$ or $\ln z$ about $z=0$ cannot be justified. But as far as integration is concerned, provided we keep track of how many times the contour wraps around the origin -- and therefore how many Riemann sheets have been transversed -- both $z^\alpha$ and $\ln z$ are analytic once all relevant Riemann sheets have been taken into account. For example, let us do $\oint_C \ln(z) \dd z$, where $C$ begins from the point $z_1 \equiv r_1 e^{i\theta_1}$ and loops around the origin $n$ times and ends on the point $z_2 \equiv r_2 e^{i\theta_2 + i 2\pi n}$ for ($n \geq 1$)-integer. Across these $n$ sheets and away from $z=0$, $\ln(z)$ is analytic. We may therefore invoke Cauchy's theorem in eq. \eqref{Complex_CauchyIntegralTheorem} to deduce the result depends on the path only through its `winding number' $n$. Because $(z \ln(z) - z)' = \ln z$,
\begin{align}
\int_{z_1}^{z_2} \ln(z) \dd z 
=  r_2 e^{i\theta_2} \left( \ln r_2 + i (\theta_2 + 2\pi n) - 1 \right) 
- r_1 e^{i\theta_1} \left( \ln r_1 + i \theta_1 - 1 \right) .
\end{align}
Likewise, for the same integration contour $C$,
\begin{align}
\int_{z_1}^{z_2} z^\alpha \dd z 
= \frac{r_2^{\alpha+1}}{\alpha+1} e^{i(\alpha+1)(\theta_2 + 2\pi n)} - \frac{r_1^{\alpha+1}}{\alpha+1} e^{i(\alpha+1)\theta_1} .
\end{align}
{\bf Branches} \qquad On the other hand, the purpose of defining a branch cut, is that it allows us to define a single-valued function on a {\it single} complex plane -- a branch of a multivalued function -- as long as we agree never to cross over this cut when moving about on the complex plane. For example, a branch cut along the negative real line means $\sqrt{z} = \sqrt{r} e^{i\theta}$ with $-\pi < \theta < \pi$; you don't pass over the cut line along $z<0$ when you move around on the complex plane. 

Another common example is given by the following branch of $\sqrt{z^2-1}$:
\begin{align}
\sqrt{z+1}\sqrt{z-1} = \sqrt{r_1 r_2} e^{i(\theta_1 + \theta_2)/2},
\end{align}
where $z+1 \equiv r_1 e^{i\theta_1}$ and $z-1 \equiv r_2 e^{i\theta_2}$; and $\sqrt{r_1 r_2}$ is the positive square root of $r_1 r_2 > 0$. By circling the branch point you can see the function is well defined if we cut along $-1 < z < +1$, because $(\theta_1 + \theta_2)/2$ goes from $0$ to $(\theta_1 + \theta_2)/2 = 2\pi$.\footnote{Arfken et al. goes through various points along this circling-the-($z= \pm 1$) process, but the main point is that there is no jump after a complete circle, unlike what you'd get circling the branch point of, say $z^{1/3}$. On the other hand, you may want to use the $z+1 \equiv r_1 e^{i\theta_1}$ and $z-1 \equiv r_2 e^{i\theta_2}$ parametrization here and understand how many Riemann sheets it would take define the whole $\sqrt{z^2-1}$.} Otherwise, if the cut is defined as $z<-1$ (on the negative real line) together with $z>1$ (on the positive real line), the branch points at $z=\pm 1$ cannot be circled and the function is still well defined and single-valued.

Yet another example is given by the Legendre function
\begin{align}
Q_0(z) = \ln \left[ \frac{z+1}{z-1} \right] .
\end{align}
The branch points, where the argument of the $\ln$ goes to zero, is at $z=\pm 1$. $Q_\nu(z)$ is usually defined with a cut line along $-1 < z <+1$ on the real line. Let's circle the branch points counterclockwise, with 
\begin{align}
z+1 \equiv r_1 e^{i\theta_1} \qquad \text{and} \qquad z-1 \equiv r_2 e^{i\theta_2}
\end{align} 
as before. Then,
\begin{align}
Q_0(z) = \ln \left[ \frac{z+1}{z-1} \right] = \ln\frac{r_1}{r_2} + i \left( \theta_1 - \theta_2 \right) .
\end{align}
After one closed loop, we go from $\theta_1 - \theta_2 = 0-0 = 0$ to $\theta_1 - \theta_2 = 2\pi-2\pi = 0$; there is no jump. When $x$ lies on the real line between $-1$ and $1$, $Q_0(x)$ is then defined as
\begin{align}
Q_0(x) = \frac{1}{2} Q_0(x+i0^+) + \frac{1}{2} Q_0(x-i0^+) ,
\end{align}
where the $i0^+$ in the first term on the right means the real line is approached from the upper half plane and the second term means it is approached from the lower half plane. What does that give us? Approaching from above means $\theta_1 = 0$ and $\theta_2 = \pi$; so $\ln (z+i0^++1)/(z+i0^+-1) = \ln |(z+1)/(z-1)| - i\pi$. Approaching from below means $\theta_1=2\pi$ and $\theta_2=\pi$; therefore $\ln (z-i0^++1)/(z-i0^+-1) = \ln |(z+1)/(z-1)| + i\pi$. Hence the average of the two yields
\begin{align}
Q_0(x) = \ln \left[ \frac{1+x}{1-x} \right], \qquad -1 < x < +1.
\end{align}
because the imaginary parts cancel while $|z+1|=x+1$ and $|z-1| = 1-x$ in this region. 

{\it Example} \qquad Let us exploit the following branch of natural log
\begin{align}
\ln z = \ln r + i \theta, \qquad z = r e^{i\theta} , \qquad 0 \leq \theta < 2\pi 
\end{align}
to evaluate the integral encountered in eq. \eqref{BranchCutExample}. 
\begin{align}
\label{BranchCutExample_I}
I \equiv \int_{0}^{\infty} \frac{(\ln x)^2}{1+x^2} \dd x = \frac{\pi^3}{8} .
\end{align}
To begin we will actually consider 
\begin{align}
\label{BranchCutExample_Iprime}
I' \equiv \lim_{\stackrel{R \to \infty}{\epsilon \to 0}} \oint_{C_1+C_2+C_3+C_4} \frac{(\ln z)^2}{1+z^2} \dd z ,
\end{align}
where $C_1$ runs over $z \in (-\infty,-\epsilon]$ (for $0 \leq \epsilon \ll 1$), $C_2$ over the infinitesimal semi-circle $z = \epsilon e^{i\theta}$ (for $\theta \in [\pi,0]$), $C_3$ over $z \in [\epsilon,+\infty)$ and $C_4$ over the (infinite) semi-circle $R e^{i\theta}$ (for $R \to +\infty$ and $\theta \in [0,\pi]$). 

First, we show that the contribution from $C_2$ and $C_4$ are zero once the limits $R \to \infty$ and $\epsilon \to 0$ are taken.
\begin{align}
\left\vert \lim_{\epsilon \to 0} \int_{C_2} \frac{(\ln z)^2}{1+z^2} \dd z \right\vert
&= \left\vert \lim_{\epsilon \to 0} \int_\pi^0 i \dd\theta \epsilon e^{i\theta} \frac{(\ln \epsilon + i \theta)^2}{1+\epsilon^2 e^{2i\theta}} \right\vert \nonumber\\
&\leq \lim_{\epsilon \to 0} \int_0^\pi \dd\theta \epsilon |\ln \epsilon + i \theta|^2 = 0  .
\end{align}
and
\begin{align}
\left\vert \lim_{R \to \infty} \int_{C_4} \frac{(\ln z)^2}{1+z^2} \dd z \right\vert
&= \left\vert \lim_{R \to \infty} \int_0^\pi i \dd\theta R e^{i\theta} \frac{(\ln R + i \theta)^2}{1+R^2 e^{2i\theta}} \right\vert \nonumber\\
&\leq \lim_{R \to \infty} \int_0^\pi \dd\theta |\ln R + i \theta|^2/R = 0  .
\end{align}
Moreover, $I'$ can be evaluated via the residue theorem; within the closed contour, the integrand blows up at $z=i$.
\begin{align}
I' 
&\equiv 2\pi i \lim_{\stackrel{R \to \infty}{\epsilon \to 0}} \oint_{C_1+C_2+C_3+C_4} \frac{(\ln z)^2}{(z+i)(z-i)} \frac{\dd z}{2\pi i}  \nonumber\\
&= 2\pi i \frac{(\ln i)^2}{2i} = \pi ( \ln(1) + i (\pi/2) )^2 = - \frac{\pi^3}{4} .
\end{align}
This means the sum of the integral along $C_1$ and $C_3$ yields $-\pi^3/4$. If we use polar coordinates along both $C_1$ and $C_2$, namely $z = r e^{i\theta}$,
\begin{align}
\int_{\infty}^{0} \dd r e^{i\pi} \frac{(\ln r + i \pi)^2}{1+r^2 e^{i2\pi}} + \int_{0}^{\infty}\frac{(\ln r)^2}{1+r^2} \dd r 
&= -\frac{\pi^3}{4} \\
\int_{0}^{\infty} \dd r \frac{2 (\ln r)^2  + i 2 \pi \ln r - \pi^2}{1+r^2} &= -\frac{\pi^3}{4} 
\end{align}
We may equate the real and imaginary parts of both sides. The imaginary one, in particular, says
\begin{align}
\int_{0}^{\infty} \dd r \frac{\ln r}{1+r^2} &= 0 ,
\end{align}
while the real part now hands us
\begin{align}
2 I &= \pi^2 \int_{0}^{\infty} \frac{\dd r}{1+r^2} - \frac{\pi^3}{4} \nonumber\\
&= \pi^2 \left[\arctan(r)\right]_{r=0}^{r=\infty} - \frac{\pi^3}{4} = \frac{\pi^3(2-1)}{4} = \frac{\pi^3}{4} 
\end{align}
We have managed to solve for the integral $I$
\begin{myP}
\qquad If $x$ is a non-zero real number, justify the identity
\begin{align}
\ln(x + i0^+) = \ln|x| + i\pi \Theta(-x) ,
\end{align}
where $\Theta$ is the step function. \qed
\end{myP}
\begin{myP}
\qquad (From Arfken et al.) For $-1 < a <1$, show that
\begin{align}
\int_{0}^{\infty} \dd x \frac{x^a}{(x+1)^2} = \frac{\pi a}{\sin(\pi a)} .
\end{align}
Hint: Complexify the integrand, then define a branch cut along the positive real line. Consider the closed counterclockwise contour that starts at the origin $z=0$, goes along the positive real line, sweeps out an infinite counterclockwise circle which returns to the positive infinity end of the real line, then runs along the positive real axis back to $z=0$. \qed
\end{myP}

\subsection{Fourier Transforms}

We have seen how the Fourier transform pairs arise within the linear algebra of states represented in some position basis corresponding to some $D$ dimensional infinite flat space. Denoting the state/function as $f$, and using Cartesian coordinates, the pairs read
\begin{align}
\label{FT_kInt}
f(\vec{x}) 				&= \int_{\mathbb{R}^D} \frac{\dd^D \vec{k}}{(2\pi)^D} \widetilde{f}(\vec{k}) e^{i \vec{k} \cdot \vec{x}} \\
\label{FT_xInt}
\widetilde{f}(\vec{k}) 	&= \int_{\mathbb{R}^D} \dd^D\vec{x} f(\vec{x}) e^{-i \vec{k} \cdot \vec{x}} 
\end{align}
Note that we have normalized our integrals differently from the linear algebra discussion. There, we had a $1/(2\pi)^{D/2}$ in both integrals, but here we have a $1/(2\pi)^D$ in the momentum space integrals and no $(2\pi)$s in the position space ones. Always check the Fourier conventions of the literature you are reading. By inserting eq. \eqref{FT_xInt} into eq. \eqref{FT_kInt} we may obtain the integral representation of the $\delta$-function
\begin{align}
\delta^{(D)}(\vec{x}-\vec{x}')  = \int_{\mathbb{R}^D} \frac{\dd^D k}{(2\pi)^D} e^{i\vec{k} \cdot (\vec{x}-\vec{x}')} .
\end{align}
In physical applications, almost any function residing in infinite space can be Fourier transformed. The meaning of the Fourier expansion in eq. \eqref{FT_kInt} is that of resolving a given profile $f(\vec{x})$ -- which can be a wave function of an elementary particle, or a component of an electromagnetic signal -- into its basis wave vectors. Remember the magnitude of the wave vector is the reciprocal of the wave length, $|\vec{k}| \sim 1/\lambda$. Heuristically, this indicates the coarser features in the profile -- those you'd notice at first glance -- come from the modes with longer wavelengths, small $|\vec{k}|$ values. The finer features requires us to know accurately the Fourier coefficients of the waves with very large $|\vec{k}|$, i.e., short wavelengths. 

In many physical problems we only need to understand the coarser features, the Fourier modes up to some inverse wavelength $|\vec{k}| \sim \Lambda_\text{UV}$. (This in turn means $\Lambda_\text{UV}$ lets us {\it define} what we mean by coarse ($\equiv |\vec{k}| < \Lambda_\text{UV}$) and fine $(\equiv |\vec{k}| > \Lambda_{\text{UV}})$ features.) In fact, it is often {\it not possible} to experimentally probe the Fourier modes of very small wavelengths, or equivalently, phenomenon at very short distances, because it would expend too much resources to do so. For instance, it much easier to study the overall appearance of the desk you are sitting at -- its physical size, color of its surface, etc. -- than the atoms that make it up. This is also the essence of why it is very difficult to probe quantum aspects of gravity: humanity does not currently have the resources to construct a powerful enough accelerator to understand elementary particle interactions at the energy scales where quantum gravity plays a significant role.
\begin{myP}
\qquad A simple example illustrating how Fourier transforms help us understand the coarse ($\equiv$ long wavelength) versus fine ($\equiv$ short wavelength) features of some profile is to consider a Gaussian of width $\sigma$, but with some small oscillations added on top of it.
\begin{align}
f(x) = \exp\left(-\frac{1}{2}\left(\frac{x-x_0}{\sigma}\right)^2\right) \left( 1 + \epsilon \sin(\omega x)\right), \qquad |\epsilon| \ll 1 .
\end{align}
Assume that the wavelength of the oscillations is much shorter than the width of the Gaussian, $1/\omega \ll \sigma$. Find the Fourier transform $\widetilde{f}(k)$ of $f(x)$ and comment on how, discarding the short wavelength coefficients of the Fourier expansion of $f(x)$ still reproduces its gross features, namely the overall shape of the Gaussian itself. Notice, however, if $\epsilon$ is not small, then the oscillations -- and hence the higher $|\vec{k}|$ modes -- cannot be ignored. \qed
\end{myP}
\begin{myP}
\qquad Find the inverse Fourier transform of the ``top hat" in 3 dimensions:
\begin{align}
\widetilde{f}(\vec{k}) &\equiv \Theta\left(\Lambda-|\vec{k}|\right) \\
f(\vec{x}) &= ? 
\end{align}
{\it Bonus problem:} Can you do it for arbitrary $D$ dimensions? Hint: You may need to know how to write down spherical coordinates in $D$ dimensions. Then examine eq. 10.9.4 of the NIST page \href{http://dlmf.nist.gov/10.9}{here}. \qed
\end{myP}
\begin{myP}
\qquad What is the Fourier transform of a multidimensional Gaussian
\begin{align}
f(\vec{x}) = \exp\left(-x^i M_{ij} x^j\right),
\end{align}
where $M_{ij}$ is a real symmetric matrix? (You may assume all its eigenvalues are strictly positive.) Hint: You need to diagonalize $M_{ij}$. The Fourier transform result would involve both its inverse and determinant. Furthermore, your result should justify the statement: ``The Fourier transform of a Gaussian is another Gaussian". \qed
\end{myP}
\begin{myP}
\qquad If $f(\vec{x})$ is real, show that $\widetilde{f}(\vec{k})^* = \widetilde{f}(-\vec{k})$. Similarly, if $f(\vec{x})$ is a real periodic function in $D$-space, show that the Fourier series coefficients in eq. \eqref{FourierSeries_I} and \eqref{FourierSeries_II} obey $\widetilde{f}(n^1,\dots,n^D)^* = \widetilde{f}(-n^1,\dots,-n^D)$. 

Suppose we restrict the space of functions on infinite $\mathbb{R}^D$ to those that are even under parity, $f(\vec{x}) = f(-\vec{x})$. Show that 
\begin{align}
f(\vec{x}) = \int_{\mathbb{R}^D} \frac{\dd^D \vec{k}}{(2\pi)^D} \cos\left( \vec{k} \cdot \vec{x} \right) \widetilde{f}(\vec{k}) .
\end{align}
What's the inverse Fourier transform? If instead we restrict to the space of odd parity functions, $f(-\vec{x}) = -f(\vec{x})$, show that
\begin{align}
f(\vec{x}) = i \int_{\mathbb{R}^D} \frac{\dd^D \vec{k}}{(2\pi)^D} \sin\left( \vec{k} \cdot \vec{x} \right) \widetilde{f}(\vec{k}) .
\end{align}
Again, write down the inverse Fourier transform. Can you write down the analogous Fourier/inverse Fourier series for even and odd parity periodic functions on $\mathbb{R}^D$? \qed
\end{myP}
\begin{myP}
\qquad For a complex $f(\vec{x})$, show that
\begin{align}
\int_{\mathbb{R}^D} \dd^D x |f(\vec{x})|^2 &= \int_{\mathbb{R}^D} \frac{\dd^D k}{(2\pi)^D} |\widetilde{f}(\vec{k})|^2 , \\
\int_{\mathbb{R}^D} \dd^D x M^{ij} \partial_i f(\vec{x})^* \partial_j f(\vec{x}) &= \int_{\mathbb{R}^D} \frac{\dd^D k}{(2\pi)^D} M^{ij} k_i k_j |\widetilde{f}(\vec{k})|^2 ,
\end{align}
where you should assume the matrix $M^{ij}$ does not depend on position $\vec{x}$.

Next, prove the {\it convolution theorem}: the Fourier transform of the convolution of two functions $F$ and $G$ 
\begin{align}
\label{ConvolutionTheorem_I}
f(\vec{x}) \equiv \int_{\mathbb{R}^D} \dd^D y F(\vec{x}-\vec{y}) G(\vec{y}) 
\end{align} 
is the product of their Fourier transforms
\begin{align}
\widetilde{f}(\vec{k}) = \widetilde{F}(\vec{k}) \widetilde{G}(\vec{k}) . 
\end{align}
You may need to employ the integral representation of the $\delta$-function; or invoke linear algebraic arguments. \qed
\end{myP}

\subsubsection{Application: Damped Driven Simple Harmonic Oscillator}

Many physical problems -- from RLC circuits to perturbative Quantum Field Theory (pQFT) -- reduces to some variant of the driven damped harmonic oscillator.\footnote{In pQFT the different Fourier modes of (possibly multiple) fields are the harmonic oscillators. If the equations are nonlinear, that means modes of different momenta drive/excite each other. Similar remarks apply for different fields that appear together in their differential equations. If you study fields residing in an expanding universe like ours, you'll find that the expansion of the universe provides friction and hence each Fourier mode behaves as a damped oscillator. The quantum aspects include the perspective that the Fourier modes themselves are both waves propagating in spacetime as well as particles that can be localized, say by the silicon wafers of the detectors at the Large Hadron Collider (LHC) in Geneva. These particles -- the Fourier modes -- can also be created from and absorbed by the vacuum.} We will study it in the form of the 2nd order ordinary differential equation (ODE)
\begin{align}
m \ \ddot{x}(t) + f \ \dot{x}(t) + k \ x(t) = F(t) , \qquad f,k > 0,
\end{align}
where each dot represents a time derivative; for e.g., $\ddot{x} \equiv \dd^2 x/\dd t^2$. You can interpret this equation as Newton's second law (in 1D) for a particle with trajectory $x(t)$ of mass $m$. The $f$ term corresponds to some frictional force that is proportional to the velocity of the particle itself; the $k>0$ refers to the spring constant, if the particle is in some locally-parabolic potential; and $F(t)$ is some other time-dependent external force. For convenience we will divide both sides by $m$ and re-scale the constants and $F(t)$ so that our ODE now becomes
\begin{align}
\label{ODE_DampedSHO}
\ddot{x}(t) + 2 \gamma \dot{x}(t) + \Omega^2 x(t) = F(t), \qquad \Omega \geq \gamma > 0 .
\end{align}
(For technical convenience, we have further restricted $\Omega$ to be greater or equal to $\gamma$.) We will perform a Fourier analysis of this problem by transforming both the trajectory and the external force,
\begin{align}
\label{ODE_DampedSHO_FT}
x(t) = \int_{-\infty}^{+\infty} \widetilde{x}(\omega) e^{i\omega t} \frac{\dd \omega}{2\pi}, \qquad
F(t) = \int_{-\infty}^{+\infty} \widetilde{F}(\omega) e^{i\omega t} \frac{\dd \omega}{2\pi} .
\end{align}
I will first find the particular solution $x_p(t)$ for the trajectory due to the presence of the external force $F(t)$, through the Green's function $G(t-t')$ of the differential operator $(\dd/\dd t)^2 + 2 \gamma (\dd/\dd t) + \Omega^2$. I will then show the fundamental importance of the Green's function by showing how you can obtain the homogeneous solution to the damped simple harmonic oscillator equation, once you have specified the position $x(t')$ and velocity $\dot{x}(t')$ at some initial time $t'$. (This is, of course, to be expected, since we have a 2nd order ODE.)

First, we begin by taking the Fourier transform of the ODE itself.
\begin{myP}
\qquad Show that, in frequency space, eq. \eqref{ODE_DampedSHO} is
\begin{align}
\label{ODE_DampedSHO_Fourier}
\left(-\omega^2 + 2 i \omega \gamma + \Omega^2\right) \widetilde{x}(\omega) = \widetilde{F}(\omega) .
\end{align}
In effect, each time derivative $\dd/\dd t$ is replaced with $i\omega$. We see that the differential equation in eq. \eqref{ODE_DampedSHO} is converted into an algebraic one in eq. \eqref{ODE_DampedSHO_Fourier}. \qed
\end{myP}
\noindent{\bf Inhomogeneous (particular) solution} \qquad For $F \neq 0$, we may infer from eq. \eqref{ODE_DampedSHO_Fourier} that the particular solution -- the part of $\widetilde{x}(\omega)$ that is due to $\widetilde{F}(\omega)$ -- is
\begin{align}
\widetilde{x}_p(\omega) = \frac{\widetilde{F}(\omega)}{-\omega^2 + 2 i \omega \gamma + \Omega^2} ,
\end{align}
which in turn implies
\begin{align}
x_p(t) 
&= \int_{-\infty}^{+\infty} \frac{\dd\omega}{2\pi} e^{i\omega t} \frac{\widetilde{F}(\omega)}{-\omega^2 + 2 i \omega \gamma + \Omega^2} \nonumber\\
\label{ODE_DampedSHO_SourceIntegral}
&= \int_{-\infty}^{+\infty} \dd t' F(t') G(t-t') 
\end{align}
where
\begin{align}
\label{ODE_DampedSHO_GreensFunction_IntegralRep}
G(t-t') &= \int_{-\infty}^{+\infty} \frac{\dd\omega}{2\pi} \frac{e^{i\omega (t-t')}}{-\omega^2 + 2 i \omega \gamma + \Omega^2} .
\end{align}
To get to eq. \eqref{ODE_DampedSHO_SourceIntegral} we have inserted the inverse Fourier transform
\begin{align}
\widetilde{F}(\omega) = \int_{-\infty}^{+\infty} \dd t' F(t') e^{-i\omega t'} .
\end{align}
\begin{myP}
\qquad Show that the Green's function in eq. \eqref{ODE_DampedSHO_GreensFunction_IntegralRep} obeys the damped harmonic oscillator equation eq. \eqref{ODE_DampedSHO}, but driven by a impulsive force (``point-source-at-time $t'$")
\begin{align}
\label{ODE_DampedSHO_GreensFunction_ODE}
\left( \frac{\dd^2}{\dd t^2} + 2\gamma \frac{\dd}{\dd t} + \Omega^2 \right) G(t-t') 
= \left( \frac{\dd^2}{\dd t'^2} - 2\gamma \frac{\dd}{\dd t'} + \Omega^2 \right) G(t-t') 
= \delta(t-t') ,
\end{align}
so that eq. \eqref{ODE_DampedSHO_SourceIntegral} can be interpreted as the $x_p(t)$ sourced/driven by the superposition of impulsive forces over all times, weighted by $F(t')$. Explain why the differential equation with respect to $t'$ has a different sign in front of the $2 \gamma$ term. By ``closing the contour" appropriately, verify that eq. \eqref{ODE_DampedSHO_GreensFunction_IntegralRep} yields
\begin{align}
\label{ODE_DampedSHO_GreensFunction_Result}
G(t-t') = \Theta(t-t') e^{-\gamma (t-t')} \frac{\sin\left( \sqrt{\Omega^2-\gamma^2} (t-t')\right)}{\sqrt{\Omega^2-\gamma^2}} .
\end{align} \qed
\end{myP}
Notice the Green's function obeys causality. Any force $F(t')$ from the future of $t$, i.e., $t'>t$, does not contribute to the trajectory in eq. \eqref{ODE_DampedSHO_SourceIntegral} due to the step function $\Theta(t-t')$ in eq. \eqref{ODE_DampedSHO_GreensFunction_Result}. That is,
\begin{align}
x_p(t) &= \int_{-\infty}^{t} \dd t' F(t') G(t-t') .
\end{align}
\noindent{\bf Initial value formulation and homogeneous solutions} \qquad With the Green's function $G(t-t')$ at hand and the particular solution sourced by $F(t)$ understood -- let us now move on to use $G(t-t')$ to obtain the homogeneous solution of the damped simple harmonic oscillator. Let $x_h(t)$ be the homogeneous solution satisfying
\begin{align}
\label{ODE_DampedSHO_Homo}
\left( \frac{\dd^2}{\dd t^2} + 2\gamma \frac{\dd}{\dd t} + \Omega^2 \right) x_h(t) = 0.
\end{align}
We then start by examining the following integral
\begin{align}
\label{ODE_DampedSHO_Homo_Step0}
I(t,t') &\equiv \int_{t'}^{\infty} \dd t''\Big\{
x_h(t'') \left( \frac{\dd^2}{\dd t''^2} - 2\gamma \frac{\dd}{\dd t''} + \Omega^2 \right) G(t-t'') \nonumber\\
&\qquad\qquad
- G(t-t'') \left( \frac{\dd^2}{\dd t''^2} + 2\gamma \frac{\dd}{\dd t''} + \Omega^2 \right) x_h(t'') 
\Big\}.
\end{align}
Using the equations \eqref{ODE_DampedSHO_GreensFunction_ODE} and \eqref{ODE_DampedSHO_Homo} obeyed by $G(t-t')$ and $x_h(t)$, we may immediately infer that
\begin{align}
\label{ODE_DampedSHO_Homo_Step1}
I(t,t') = \int_{t'}^{\infty} \dd t' x_h(t'') \delta(t-t'') = \Theta(t-t') x_h(t) .
\end{align}
(The step function arises because, if $t$ lies outside of $[t',\infty)$, and is therefore less than $t'$, the integral will not pick up the $\delta$-function contribution and the result would be zero.) On the other hand, we may in eq. \eqref{ODE_DampedSHO_Homo_Step0} cancel the $\Omega^2$ terms, and then integrate-by-parts one of the derivatives from the $\ddot{G}$, $\dot{G}$, and $\ddot{x}_h$ terms.
\begin{align}
I(t,t') &= \left[ x_h(t'') \left( \frac{\dd}{\dd t''} - 2\gamma \right) G(t-t'') 
		- G(t-t'') \frac{\dd x_h(t'')}{\dd t''} \right]_{t''=t'}^{t''=\infty} \\
&+ \int_{t'}^{\infty} \dd t''\Big( - \frac{\dd x_h(t'')}{\dd t''} \frac{\dd G(t-t'')}{\dd t''} + 2\gamma \frac{\dd x_h(t'')}{\dd t''} G(t-t'') \nonumber\\
&\qquad\qquad+ \frac{\dd G(t-t'')}{\dd t''} \frac{\dd x_h(t'')}{\dd t''} - 2\gamma G(t-t'') \frac{\dd x_h(t'')}{\dd t''} \Big) . \nonumber
\end{align}
Observe that the integral on the second and third lines is zero because the integrands cancel. Moreover, because of the $\Theta(t-t')$ (namely, causality), we may assert $\lim_{t' \to \infty} G(t-t') = G(t'>t) = 0$. Recalling eq. \eqref{ODE_DampedSHO_Homo_Step1}, we have arrived at
\begin{align}
\label{ODE_DampedSHO_Homo_InitialValue}
\Theta(t-t') x_h(t) &= G(t-t') \frac{\dd x_h(t')}{\dd t'} + \left( 2\gamma G(t-t') + \frac{\dd G(t-t')}{\dd t} \right) x_h(t')  .
\end{align}
Because we have not made any assumptions about our trajectory -- except it satisfies the homogeneous equation in eq. \eqref{ODE_DampedSHO_Homo} -- we have shown that, for an arbitrary initial position $x_h(t')$ and velocity $\dot{x}_h(t')$, the Green's function $G(t-t')$ can in fact also be used to obtain the homogeneous solution for $t>t'$, where $\Theta(t-t')=1$. In particular, since $x_h(t')$ and $\dot{x}_h(t')$ are freely specifiable, they must be completely independent of each other. Furthermore, the right hand side of eq. \eqref{ODE_DampedSHO_Homo_InitialValue} must span the 2-dimensional space of solutions to eq. \eqref{ODE_DampedSHO_Homo}. Therefore, the coefficients of $x_h(t')$ and $\dot{x}_h(t')$ must in fact be the two linearly independent homogeneous solutions to $x_h(t)$,
\begin{align}
\label{ODE_DampedSHO_Homo_S1}
x_h^{(1)}(t) &= G(t>t') = e^{-\gamma (t-t')} \frac{\sin\left( \sqrt{\Omega^2-\gamma^2} (t-t')\right)}{\sqrt{\Omega^2-\gamma^2}}, \\
\label{ODE_DampedSHO_Homo_S2}
x_h^{(2)}(t) &= 2\gamma G(t>t') + \partial_t G(t>t') \nonumber\\
&= e^{-\gamma (t-t')} \left( \frac{\gamma \cdot \sin\left( \sqrt{\Omega^2-\gamma^2} (t-t')\right)}{\sqrt{\Omega^2-\gamma^2}} + \cos\left( \sqrt{\Omega^2-\gamma^2} (t-t')\right) \right).
\end{align}
\footnote{Note that
\begin{align}
\frac{\dd G(t-t')}{\dd t} 
= \Theta(t-t') \frac{\dd}{\dd t} \left( e^{-\gamma (t-t')} \frac{\sin\left( \sqrt{\Omega^2-\gamma^2} (t-t')\right)}{\sqrt{\Omega^2-\gamma^2}} \right) .
\end{align} 
Although differentiating $\Theta(t-t')$ gives $\delta(t-t')$, its coefficient is proportional to $\sin( \sqrt{\Omega^2-\gamma^2} (t-t'))/\sqrt{\Omega^2-\gamma^2}$, which is zero when $t=t'$, even if $\Omega=\gamma$.}That $x_h^{(1,2)}$ must be independent for any $\gamma > 0$ and $\Omega^2$ is worth reiterating, because this is a potential issue for the damped harmonic oscillator equation when $\gamma = \Omega$. We can check directly that, in this limit, $x_h^{(1,2)}$ remain linearly independent. On the other hand, if we had solved the homogeneous equation by taking the real (or imaginary part) of an exponential; namely, try
\begin{align}
\label{ODE_DampedSHO_Homo_Trial}
x_h(t) = \text{Re} \ e^{i \omega t} ,
\end{align}
we would find, upon inserting eq. \eqref{ODE_DampedSHO_Homo_Trial} into eq. \eqref{ODE_DampedSHO_Homo}, that
\begin{align}
\label{ODE_DampedSHO_Homo_Roots}
\omega = \omega_\pm \equiv i \gamma \pm \sqrt{\Omega^2 - \gamma^2} .
\end{align}
This means, when $\Omega=\gamma$, we obtain repeated roots and the otherwise linearly independent solutions 
\begin{align}
\label{ODE_DampedSHO_SolutionsFromExp}
x_h^{(\pm)}(t) = \text{Re} \ e^{-\gamma t \pm i \sqrt{\Omega^2-\gamma^2} t} 
\end{align}
become linearly dependent there -- both $x_h^{(\pm)}(t) = e^{-\gamma t}$.
\begin{myP}
\qquad Explain why the real or imaginary part of a complex solution to a homogeneous real linear differential equation is also a solution. Now, start from eq. \eqref{ODE_DampedSHO_Homo_Trial} and verify that eq. \eqref{ODE_DampedSHO_SolutionsFromExp} are indeed solutions to eq. \eqref{ODE_DampedSHO_Homo} for $\Omega \neq \gamma$. Comment on why the presence of $t'$ in equations \eqref{ODE_DampedSHO_Homo_S1} and \eqref{ODE_DampedSHO_Homo_S2} amount to arbitrary constants multiplying the homogeneous solutions in eq. \eqref{ODE_DampedSHO_SolutionsFromExp}. \qquad\qed
\end{myP}
\begin{myP}
\qquad Suppose for some initial time $t_0$, $x_h(t_0) = 0$ and $\dot{x}_h(t_0) = V_0$. There is an external force given by
\begin{align}
F(t) = \text{Im} \left( e^{-(t/\tau)^2} e^{i \mu t} \right), \qquad \text{for $-2\pi n/\mu \leq t \leq 2\pi n/\mu$}, \qquad \mu > 0, \qquad  .
\end{align}
and $F(t)=0$ otherwise. ($n$ is an integer greater than 1.) Solve for the motion $x(t>t_0)$ of the damped simple harmonic oscillator, in terms of $t_0$, $V_0$, $\tau$, $\mu$ and $n$. \qquad\qed
\end{myP}

\subsection{Fourier Series}

Consider a periodic function $f(x)$ with period $L$, meaning
\begin{align}
f(x+L) = f(x) .
\end{align}
Then its Fourier series representation is given by
\begin{align}
\label{FourierSeries}
f(x) 	&= \sum_{n=-\infty}^\infty C_n e^{i \frac{2\pi n}{L} x} , \\
C_n 	&= \frac{1}{L} \int_{\text{one period}} \dd x' f(x') e^{-i \frac{2\pi n}{L} x'} . \nonumber
\end{align}
(I have derived this in our linear algebra discussion.) The Fourier series can be viewed as the discrete analog of the Fourier transform. In fact, one way to go from the Fourier series to the Fourier transform, is to take the infinite box limit $L \to \infty$. Just as the meaning of the Fourier transform is the decomposition of some wave profile into its continuous infinity of wave modes, the Fourier series can be viewed as the discrete analog of that. One example is that of waves propagating on a guitar or violin string -- the string (of length $L$) is tied down at the end points, so the amplitude of the wave $\psi$ has to vanish there
\begin{align}
\psi(x=0) = \psi(x=L) = 0 .
\end{align}
Even though the Fourier series is supposed to represent the profile $\psi$ of a periodic function, there is nothing to stop us from imagining duplicating our guitar/violin string infinite number of times. Then, the decomposition in \eqref{FourierSeries} applies, and is simply the superposition of possible vibrational modes allowed on the string itself.
\begin{myP}
\qquad (From Riley et al.) Find the Fourier series representation of the Dirac comb, i.e., find the $\{C_n\}$ in
\begin{align}
\sum_{n=-\infty}^\infty \delta(x+nL) = \sum_{n=-\infty}^{\infty} C_n e^{i\frac{2\pi n}{L} x}, \qquad x \in \mathbb{R} .
\end{align}
Then prove the {\it Poisson summation formula}; where for an arbitrary function $f(x)$ and its Fourier transform $\widetilde{f}$,
\begin{align}
\sum_{n=-\infty}^\infty f(x+nL) 
= \frac{1}{L} \sum_{n=-\infty}^{\infty} \widetilde{f}\left(\frac{2\pi n}{L}\right) e^{i\frac{2\pi n}{L} x} .
\end{align}
Hint: Note that
\begin{align}
f(x+nL) = \int_{-\infty}^{+\infty} \dd x' f(x') \delta(x-x'+nL) .
\end{align} \qed
\end{myP}
\begin{myP}
{\it Gibbs phenomenon} \qquad The Fourier series of a discontinuous function suffers from what is known as the Gibbs phenomenon -- near the discontinuity, the Fourier series does not fit the actual function very well. As a simple example, consider the periodic function $f(x)$ where within a period $x \in[0,L)$,
\begin{align}
\label{GibbsPhenomenon_Example}
f(x) 	&= -1, \qquad -L/2 \leq x \leq 0 \\
		&= 1, \qquad 0 \leq x \leq L/2 .
\end{align}
Find its Fourier series representation
\begin{align}
f(x) &= \sum_{n=-\infty}^\infty C_n e^{i \frac{2\pi n}{L} x} .
\end{align}
Since this is an odd function, you should find that the series becomes a sum over sines -- cosine is an even function -- which in turn means you can rewrite the summation as one only over positive integers $n$. Truncate this sum at $N=20$ and $N=50$, namely
\begin{align}
f_N(x) \equiv \sum_{n=-N}^N C_n e^{i \frac{2\pi n}{L} x} ,
\end{align}
and find a computer program to plot $f_N(x)$ as well as $f(x)$ in eq. \eqref{GibbsPhenomenon_Example}. You should see the $f_N(x)$ over/undershooting the $f(x)$ near the latter's discontinuities, even for very large $N \gg 1$.\footnote{See \S 5.7 of \href{http://www.physics.miami.edu/~nearing/mathmethods/mathematical_methods-one.pdf}{James Nearing's Math Methods book} for a pedagogical discussion of how to estimate both the location and magnitude of the (first) maximum overshoot.} \qed
\end{myP}

\newpage
\section{Advanced Calculus: Special Techniques and Asymptotic Expansions}
\label{Chapter_SpecialTechniquesAndAsymptotics}
Integration is usually much harder than differentiation. Any function $f(x)$ you can build out of powers, logs, trigonometric functions, etc., can usually be readily differentiated.\footnote{The ease of differentiation ceases once you start dealing with ``special functions"; see, for e.g., \href{http://blog.wolfram.com/2016/05/16/new-derivatives-of-the-bessel-functions-have-been-discovered-with-the-help-of-the-wolfram-language/}{here} for a discussion on how to differentiate the Bessel function $J_\nu(z)$ with respect to its order $\nu$.} But to integrate a function in closed form you have to know another function $g(x)$ whose derivative yields $f(x)$; that's the essential content of the fundamental theorem of calculus.
\begin{align}
\int f(x) \dd x \stackrel{?}{=} \int g'(x) \dd x = g(x) + \text{constant}
\end{align}
Here, I will discuss integration techniques that I feel are not commonly found in standard treatments of calculus. Among them, some techniques will show how to extract approximate answers from integrals. This is, in fact, a good place to highlight the importance of approximation techniques in physics. For example, most of the predictions from quantum field theory -- our fundamental framework to describe elementary particle interactions at the highest energies/smallest distances -- is based on perturbation theory.

\subsection{Gaussian integrals}

As a start, let us consider the following ``Gaussian" integral:
\begin{align}
I_G(a) \equiv \int_{-\infty}^{+\infty} e^{-a x^2} \dd x, 
\end{align}
where Re$(a) > 0$. (Why is this restriction necessary?) Let us suppose that $a > 0$ for now. Then, we may consider squaring the integral, i.e., the 2-dimensional (2D) case:
\begin{align}
(I_{G}(a))^2 = \int_{-\infty}^{+\infty} \int_{-\infty}^{+\infty} e^{-a x^2} e^{-a y^2} \dd x \dd y  .
\end{align}
You might think ``doubling" the problem is only going to make it harder, not easier. But let us now view $(x,y)$ as Cartesian coordinates on the 2D plane and proceed to change to polar coordinates, $(x,y) = r(\cos\phi,\sin\phi)$; this yields $\dd x \dd y = \dd\phi \dd r\cdot r$.
\begin{align}
\left(I_{G}(a)\right)^2 
= \int_{-\infty}^{+\infty} e^{-a (x^2+y^2)} \dd x \dd y 
= \int_0^{2\pi} \dd\phi \int_{0}^{+\infty} \dd r \cdot r e^{-a r^2}
\end{align}
The integral over $\phi$ is straightforward; whereas the radial one now contains an additional $r$ in the integrand -- this is exactly what makes the integral do-able.
\begin{align}
\left(I_{G}(a)\right)^2
&= 2\pi \int_{0}^{+\infty} \dd r \frac{1}{-2a} \partial_r e^{-a r^2} \nonumber \\
&= \left[ \frac{-\pi}{a} e^{-a r^2} \right]_{r=0}^{r=\infty} = \frac{\pi}{a}
\end{align}
Because $e^{-a x^2}$ is a positive number if $a$ is positive, we know that $I_G(a>0)$ must be a positive number too. Since $\left(I_{G}(a)\right)^2 = \pi/a$ the Gaussian integral itself is just the positive square root 
\begin{align}
\label{Integral_Gaussian}
\int_{-\infty}^{+\infty} e^{-a x^2} \dd x = \sqrt{\frac{\pi}{a}}, \qquad \text{Re}(a) > 0 .
\end{align}
Because both sides of eq. \eqref{Integral_Gaussian} can be differentiated readily with respect to $a$ (for $a \neq 0$), by analytic continuation, even though we started out assuming $a$ is positive, we may now relax that assumption and only impose Re$(a) > 0$. If you are uncomfortable with this analytic continuation argument, you can also tackle the integral directly. Suppose $a = \rho e^{i\delta}$, with $\rho > 0$ and $-\pi/2 < \delta < \pi/2$. Then we may rotate the contour for the $x$ integration from $x \in (-\infty,+\infty)$ to the contour $C$ defined by $z \equiv e^{-i\delta/2} \xi$, where $\xi \in (-\infty,+\infty)$. (The 2 arcs at infinity contribute nothing to the integral -- can you prove it?)
\begin{align}
I_G(a) 
&= \int_{\xi=-\infty}^{\xi=+\infty} e^{-\rho e^{i\delta} (e^{-i\delta/2} \xi)^2} \dd (e^{-i\delta/2} \xi) \nonumber\\
&= \frac{1}{e^{i\delta/2}} \int_{\xi=-\infty}^{\xi=+\infty} e^{-\rho \xi^2} \dd \xi
\end{align}
At this point, since $\rho>0$ we may refer to our result for $I_G(a>0)$ and conclude
\begin{align}
I_G(a) = \frac{1}{e^{i\delta/2}} \sqrt{\frac{\pi}{\rho}} = \sqrt{\frac{\pi}{\rho e^{i\delta}}} = \sqrt{\frac{\pi}{a}} ,
\qquad\qquad
-\frac{\pi}{2} < \text{arg}[a] < \frac{\pi}{2} .
\end{align}
\begin{myP}
\qquad Compute, for Re$(a) > 0$,
{\allowdisplaybreaks\begin{align}
\int_0^{+\infty} e^{-a x^2} \dd x,& 			\qquad \text{ for Re$(a) > 0$} \\
\int_{-\infty}^{+\infty} e^{-a x^2} x^n \dd x,& \qquad \text{ for $n$ odd} \\
\int_{-\infty}^{+\infty} e^{-a x^2} x^n \dd x,& \qquad \text{ for $n$ even} \\
\int_{0}^{+\infty} e^{-a x^2} x^\beta \dd x,&	\qquad \text{ for Re$(\beta) > -1$} 
\end{align}}
Hint: For the very last integral, consider the change of variables $x' \equiv \sqrt{a} x$, and refer to eq. 5.2.1 of the NIST page \href{http://dlmf.nist.gov/5.2}{here}. \qed
\end{myP}
\begin{myP}
\qquad There are many applications of the Gaussian integral in physics. Here, we give an application in geometry, and calculate the solid angle in $D$ spatial dimensions. In $D$-space, the solid angle $\Omega_{D-1}$ subtended by a sphere of radius $r$ is defined through the relation
\begin{align}
\text{Surface area of sphere} \equiv \Omega_{D-1} \cdot r^{D-1} .
\end{align}
Since $r$ is the only length scale in the problem, and since area in $D$-space has to scale as [Length$^{D-1}$], we see that $\Omega_{D-1}$ is independent of the radius $r$. Moreover, the volume of a spherical shell of radius $r$ and thickness $\dd r$ must be the area of the sphere times $\dd r$. Now, argue that the $D$ dimensional integral in spherical coordinates becomes
\begin{align}
\label{Integral_SolidAngle_StepI}
\left(I_G(a=1)\right)^D = \int_{\mathbb{R}^D} \dd^D \vec{x} e^{-\vec{x}^2} = \Omega_{D-1} \int_0^\infty \dd r \cdot r^{D-1} e^{-r^2} .
\end{align}
Next, evaluate $\left(I_G(a=1)\right)^D$ directly. Then use the results of the previous problem to compute the last equality of eq. \eqref{Integral_SolidAngle_StepI}. At this point you should arrive at
\begin{align}
\Omega_{D-1} = \frac{2\pi^{D/2}}{\Gamma(D/2)},
\end{align}
where $\Gamma$ is the Gamma function. \qed
\end{myP}

\subsection{Complexification}

Sometimes complexifying the integral makes it easier. Here's a simple example from Matthews and Walker \cite{MatthewsWalker}.
\begin{align}
I = \int_{0}^{\infty} \dd x e^{-ax} \cos(\lambda x), \qquad a > 0, \ \lambda \in\mathbb{R}.
\end{align}
If we regard $\cos(\lambda x)$ as the real part of $e^{i\lambda x}$,
\begin{align}
I &= \text{Re}\int_{0}^{\infty} \dd x e^{-(a - i\lambda )x} \nonumber\\
&= \text{Re} \left[\frac{e^{-(a - i\lambda )x}}{-(a - i\lambda )}\right]_{x=0}^{x=\infty}  \nonumber\\
&= \text{Re} \frac{1}{a - i\lambda} = \text{Re} \frac{a + i \lambda}{a^2 + \lambda^2} = \frac{a}{a^2 + \lambda^2} 
\end{align}
\begin{myP}
	\qquad What is
	\begin{align}
	\int_{0}^{\infty} \dd x e^{-ax} \sin(\lambda x), \qquad a > 0, \ \lambda \in\mathbb{R} ?
	\end{align}
\end{myP}

\subsection{Differentiation under the integral sign (Leibniz's theorem)}

Differentiation under the integral sign, or Leibniz's theorem, is the result
\begin{align}
\label{Integral_LeibnizRule}
\frac{\dd}{\dd z} \int_{a(z)}^{b(z)} \dd s F\left(z,s\right)
= b'(z) F\left(z,b(z)\right) - a'(z) F\left(z,a(z)\right) + \int_{a(z)}^{b(z)} \dd s \frac{\partial F\left(z,s\right)}{\partial z} .
\end{align}
\begin{myP}
\qquad By using the limit definition of the derivative, i.e.,
\begin{align}
\frac{\dd}{\dd z} H(z) = \lim_{\delta \to 0}  \frac{H(z+\delta) - H(z)}{\delta} ,
\end{align}
argue the validity of eq. \eqref{Integral_LeibnizRule}. \qed
\end{myP}
Why this result is useful for integration can be illustrated by some examples. The art involves creative insertion of some auxiliary parameter $\alpha$ in the integrand. Let's start with
\begin{align}
\Gamma(n+1) = \int_0^\infty \dd t t^n e^{-t} , \qquad \text{ $n$ a positive integer} .
\end{align}
For Re$(n) > -1$ this is in fact the definition of the Gamma function. We introduce the parameter as follows
\begin{align}
I_n(\alpha) = \int_0^\infty \dd t t^n e^{-\alpha t}, \qquad \alpha > 0,
\end{align}
and notice
\begin{align}
I_n(\alpha) 
&= (-\partial_\alpha)^n \int_0^\infty \dd t e^{-\alpha t} = (-\partial_\alpha)^n \frac{1}{\alpha} \nonumber\\
&= (-)^n (-1) (-2) \dots (-n) \alpha^{-1-n} = n! \alpha^{-1-n}
\end{align}
By setting $\alpha=1$, we see that the Gamma function $\Gamma(z)$ evaluated at integer values of $z$ returns the factorial.
\begin{align}
\Gamma(n+1) = I_n(\alpha=1) = n! .
\end{align}
Next, we consider a trickier example:
\begin{align}
\int_{-\infty}^{\infty} \frac{\sin(x)}{x} \dd x .
\end{align}
This can be evaluated via a contour integral. But here we do so by introducing a $\alpha \in \mathbb{R}$,
\begin{align}
I(\alpha) \equiv \int_{-\infty}^{\infty} \frac{\sin(\alpha x)}{x} \dd x .
\end{align}
Observe that the integral is odd with respect to $\alpha$, $I(-\alpha) = -I(\alpha)$. Differentiating once,
\begin{align}
I'(\alpha) = \int_{-\infty}^{\infty} \cos(\alpha x) \dd x = \int_{-\infty}^{\infty} e^{i\alpha x} \dd x = 2\pi \delta(\alpha).
\end{align}
($\cos(\alpha x)$ can be replaced with $e^{i\alpha x}$ because the $i \sin(\alpha x)$ portion integrates to zero.) Remember the derivative of the step function $\Theta(\alpha)$ is the Dirac $\delta$-function $\delta(\alpha)$: $\Theta'(z) = \Theta'(-z) = \delta(z)$. Taking into account $I(-\alpha) = -I(\alpha)$, we can now deduce the answer to take the form
\begin{align}
I(\alpha) = \pi \left(\Theta(\alpha) - \Theta(-\alpha)\right) = \pi \text{sgn}(\alpha) ,
\end{align}
There is no integration constant here because it will spoil the property $I(-\alpha) = -I(\alpha)$. What remains is to choose $\alpha=1$,
\begin{align}
I(1) = \int_{-\infty}^{\infty} \frac{\sin(x)}{x} \dd x = \pi .
\end{align}
\begin{myP}
\qquad Evaluate the following integral
\begin{align}
I(\alpha) = \int_0^\pi \ln \left[ 1 - 2 \alpha \cos(x) + \alpha^2 \right] \dd x , \qquad |\alpha| \neq 1,
\end{align}
by differentiating once with respect to $\alpha$, changing variables to $t \equiv \tan(x/2)$, and then using complex analysis. ({\it Do not} copy the solution from Wikipedia!) You may need to consider the cases $|\alpha| > 1$ and $|\alpha| < 1$ separately. \qed
\end{myP}

\subsection{Symmetry}

You may sometimes need to do integrals in higher than one dimension. If it arises from a physical problem, it may exhibit symmetry properties you should definitely exploit. The case of rotational symmetry is a common and important one, and we shall focus on it here. A simple example is as follows. In 3-dimensional (3D) space, we define
\begin{align}
I(\vec{k}) \equiv \int_{\mathbb{S}^2} \frac{\dd\Omega_{\widehat{n}}}{4\pi} e^{i \vec{k} \cdot \widehat{n}} .
\end{align}
The $\int_{\mathbb{S}^2} \dd\Omega$ means we are integrating the unit radial vector $\widehat{n}$ with respect to the solid angles on the sphere; $\vec{k} \cdot \vec{x}$ is just the Euclidean dot product. For example, if we use spherical coordinates, the Cartesian components of the unit vector would be
\begin{align}
\widehat{n} = (\sin\theta \cos\phi, \sin\theta \sin\phi, \cos\theta) ,
\end{align}
and $\dd\Omega = \dd(\cos\theta) \dd\phi$. The key point here is that we have a rotationally invariant integral. In particular, the $(\theta,\phi)$ here are measured with respect to some $(x^1,x^2,x^3)$-axes. If we rotated them to some other (orthonormal) $(x'^1,x'^2,x'^3)$-axes related via some rotation matrix $R^i_{\phantom{i}j}$,
\begin{align}
\widehat{n}^i(\theta,\phi) = R^i_{\phantom{i}j} \widehat{n}'^j(\theta',\phi') ,
\end{align}
where $\det R^i_{\phantom{i}j} = 1$; in matrix notation $\widehat{n} = R \widehat{n}'$ and $R^T R = \mathbb{I}$. Then $\dd(\cos\theta) \dd\phi = \dd \Omega = \dd \Omega' \det R^i_{\phantom{i}j} = \dd\Omega' = \dd(\cos\theta') \dd\phi'$, and
\begin{align}
I(R \vec{k}) 
= \int_{\mathbb{S}^2} \frac{\dd\Omega_{\widehat{n}}}{4\pi} e^{i \vec{k} \cdot (R^T \widehat{n})} 
= \int_{\mathbb{S}^2} \frac{\dd\Omega'_{\widehat{n}'}}{4\pi} e^{i \vec{k} \cdot \widehat{n}'} = I(\vec{k}) .
\end{align}
In other words, because $R$ was an arbitrary rotation matrix, $I(\vec{k}) = I(|\vec{k}|)$; the integral cannot possibly depend on the direction of $\vec{k}$, but only on the magnitude $|\vec{k}|$. That in turn means we may as well pretend $\vec{k}$ points along the $x^3$-axis, so that the dot product $\vec{k} \cdot \widehat{n}'$ only involved the $\cos\theta \equiv \widehat{n}' \cdot \widehat{e}_3$.
\begin{align}
I(|\vec{k}|) 
= \int_{0}^{2\pi} \dd\phi \int_{-1}^{+1} \frac{\dd(\cos\theta)}{4\pi} e^{i |\vec{k}| \cos\theta} 
= \frac{e^{i |\vec{k}|} - e^{-i |\vec{k}|}}{2i |\vec{k}|} .
\end{align}
We arrive at
\begin{align}
\label{Integral_PlaneWaveOverSphere}
\int_{\mathbb{S}^2} \frac{\dd\Omega_{\widehat{n}}}{4\pi} e^{i \vec{k} \cdot \widehat{n}} = \frac{\sin |\vec{k}|}{|\vec{k}|} .
\end{align}
\begin{myP}
\qquad With $\widehat{n}$ denoting the unit radial vector in $3-$space, evaluate
\begin{align}
I(\vec{x}) = \int_{\mathbb{S}^2} \frac{\dd\Omega_{\widehat{n}}}{|\vec{x}-\vec{r}|}, \qquad \vec{r} \equiv r \widehat{n} .
\end{align}
Note that the answer for $|\vec{x}| > |\vec{r}| = r$ differs from that when $|\vec{x}| < |\vec{r}| = r$. Can you explain the physical significance? Hint: This can be viewed as an electrostatics problem. \qed
\end{myP}
\begin{myP}
\qquad A problem that combines both rotational symmetry and the higher dimensional version of ``differentiation under the integral sign" is the (tensorial) integral
\begin{align}
\int_{\mathbb{S}^2} \frac{\dd\Omega}{4\pi} \widehat{n}^{i_1} \widehat{n}^{i_2} \dots \widehat{n}^{i_N} ,
\end{align}
where $N$ is an integer greater than or equal to $1$. The answer for odd $N$ can be understood by asking, how does the integrand and the measure $\dd\Omega_{\widehat{n}}$ transform under a parity flip of the coordinate system, namely under $\widehat{n} \to -\widehat{n}$? What's the answer for even $N$? Hint: consider differentiating eq. \eqref{Integral_PlaneWaveOverSphere} with respect to $k^{i_1}, \dots, k^{i_N}$; how is that related to the Taylor expansion of $(\sin|\vec{k}|)/|\vec{k}|$? \qed
\end{myP}
\begin{myP}
	\qquad Can you generalize eq. \eqref{Integral_PlaneWaveOverSphere} to $D$ spatial dimensions, namely
\begin{align}
\label{Integral_PlaneWaveOverDSphere}
\int_{\mathbb{S}^{D-1}} \dd\Omega_{\widehat{n}} e^{i \vec{k} \cdot \hat{n}} = ?
\end{align}
The $\vec{k}$ is an arbitrary vector in $D$-space and $\widehat{n}$ is the unit radial vector in the same. Hint: Refer to eq. 10.9.4 of the NIST page \href{http://dlmf.nist.gov/10.9}{here}. \qed
\end{myP}
{\it Example from Matthews and Walker \cite{MatthewsWalker}} \qquad Next, we consider the following integral involving two arbitrary vectors $\vec{a}$ and $\vec{k}$ in 3D space.
\begin{align}
I\left(\vec{a},\vec{k}\right) = \int_{\mathbb{S}^2} \dd\Omega_{\widehat{n}} \frac{\vec{a} \cdot \widehat{n}}{1 + \vec{k} \cdot \widehat{n}} 
\end{align}
First, we write it as $\vec{a}$ dotted into a vector integral $\vec{J}$, namely
\begin{align}
I\left(\vec{a},\vec{k}\right) = \vec{a} \cdot \vec{J} , \qquad
\vec{J}\left(\vec{k}\right) \equiv \int_{\mathbb{S}^2} \dd\Omega_{\widehat{n}} \frac{\widehat{n}}{1 + \vec{k} \cdot \widehat{n}} .
\end{align}
Let us now consider replacing $\vec{k}$ with a rotated version of $\vec{k}$. This amounts to replacing $\vec{k} \to R \vec{k}$, where $R$ is an orthogonal $3 \times 3$ matrix of unit determinant, with $R^T R = R R^T = \mathbb{I}$. We shall see that $\vec{J}$ transforms as a vector $\vec{J} \to R \vec{J}$ under this same rotation. This is because $\int\dd\Omega_{\widehat{n}} \to \int\dd\Omega_{\widehat{n}'}$, for $\widehat{n}' \equiv R^T \widehat{n}$, and
\begin{align}
\vec{J}\left(R\vec{k}\right) 
&= \int_{\mathbb{S}^2} \dd\Omega_{\widehat{n}} \frac{R (R^T \widehat{n})}{1 + \vec{k} \cdot \left(R^T \widehat{n}\right)} \nonumber\\
&= R \int_{\mathbb{S}^2} \dd\Omega_{\widehat{n}'} \frac{\widehat{n}'}{1 + \vec{k} \cdot \widehat{n}'} = R \vec{J}(\vec{k}) .
\end{align}
But the only vector that $\vec{J}$ depends on is $\vec{k}$. Therefore the result of $\vec{J}$ has to be some scalar function $f$ times $\vec{k}$.
\begin{align}
\vec{J} = f \cdot \vec{k}, \qquad\Rightarrow\qquad I\left(\vec{a},\vec{k}\right) &= f \vec{a} \cdot \vec{k} .
\end{align}
To calculate $f$ we now dot both sides with $\vec{k}$.
\begin{align}
f = \frac{\vec{J} \cdot \vec{k}}{\vec{k}^2} 
= \frac{1}{\vec{k}^2} \int_{\mathbb{S}^2} \dd\Omega_{\widehat{n}} \frac{\vec{k} \cdot \widehat{n}}{1 + \vec{k} \cdot \widehat{n}}
\end{align}
At this point, the nature of the remaining scalar integral is very similar to the one we've encountered previously. Choosing $\vec{k}$ to point along the $\widehat{e}_3$ axis,
{\allowdisplaybreaks\begin{align}
f 
&= \frac{2\pi}{\vec{k}^2} \int_{-1}^{+1}\dd(\cos\theta) \frac{|\vec{k}| \cos\theta}{1 + |\vec{k}| \cos\theta} \nonumber\\
&= \frac{2\pi}{\vec{k}^2} \int_{-1}^{+1}\dd c \left( 1 - \frac{1}{1 + |\vec{k}| c} \right) 
= \frac{4\pi}{\vec{k}^2} \left( 1 - \frac{1}{2 |\vec{k}|} \ln \left(\frac{1+|\vec{k}|}{1-|\vec{k}|}\right) \right) .
\end{align}}
Therefore,
\begin{align}
\int_{\mathbb{S}^2} \dd\Omega_{\widehat{n}} \frac{\vec{a} \cdot \widehat{n}}{1 + \vec{k} \cdot \widehat{n}} 
= \frac{4\pi \left(\vec{k} \cdot \vec{a}\right)}{\vec{k}^2} \left( 1 - \frac{1}{2 |\vec{k}|} \ln \left(\frac{1+|\vec{k}|}{1-|\vec{k}|}\right) \right) .
\end{align}
This technique of reducing tensor integrals into scalar ones find applications even in quantum field theory calculations.
\begin{myP}
\qquad Calculate 
\begin{align}
A^{ij}(\vec{a}) \equiv \int \frac{\dd^3 k}{(2\pi)^3} \frac{k^i k^j}{\vec{k}^2 + (\vec{k}\cdot\vec{a})^4},
\end{align}
where $\vec{a}$ is some (dimensionless) vector in 3D Euclidean space. Do so by first arguing that this integral transforms as a tensor in $D$-space under rotations. In other words, if $R^i_{\phantom{i}j}$ is a rotation matrix, under the rotation
\begin{align}
a^i \to R^i_{\phantom{i}j} a^j,
\end{align}
we have
\begin{align}
A^{ij}(R^k_{\phantom{k}l} a^l) = R^i_{\phantom{i}l} R^j_{\phantom{j}k} A^{kl}(\vec{a}) .
\end{align}
Hint: The only rank-2 tensors available here are $\delta^{ij}$ and $a^i a^j$, so we must have
\begin{align}
A^{ij} = f_1 \delta^{ij} + f_2 a^i a^j .
\end{align}
To find $f_{1,2}$ take the trace and also consider $A^{ij} a_i a_j$. \qed
\end{myP}

\subsection{Asymptotic expansion of integrals}

\footnote{The material in this section is partly based on Chapter 3 of Matthews and Walker's ``{\it Mathematical Methods of Physics}" \cite{MatthewsWalker}; and the latter portions are heavily based on Chapter 6 of Bender and Orszag's ``{\it Advanced mathematical methods for scientists and engineers}" \cite{BenderOrszag}.}Many solutions to physical problems, say arising from some differential equations, can be expressed as integrals. Moreover the ``special functions" of mathematical physics, whose properties are well studied -- Bessel, Legendre, hypergeometric, etc. -- all have integral representations. Often we wish to study these functions when their arguments are either very small or very large, and it is then useful to have techniques to extract an answer from these integrals in such limits. This topic is known as the ``asymptotic expansion of integrals".

\subsubsection{Integration-by-parts (IBP)}

In this section we will discuss how to use integration-by-parts (IBP) to approximate integrals.

Previously we evaluated 
\begin{align}
\frac{2}{\sqrt{\pi}} \int_0^{+\infty} e^{-t^2} \dd t = 1 .
\end{align}
The erf function is defined as
\begin{align}
\text{erf}(x) \equiv \frac{2}{\sqrt{\pi}} \int_{0}^{x} \dd t e^{-t^2} .
\end{align}
Its small argument limit can be obtained by Taylor expansion,
\begin{align}
\text{erf}(x \ll 1) 
&= \frac{2}{\sqrt{\pi}} \int_{0}^{x} \dd t \left( 1 - t^2 + \frac{t^4}{2!} - \frac{t^6}{3!} + \dots \right) \nonumber\\
&= \frac{2}{\sqrt{\pi}} \left( x - \frac{x^3}{3} + \frac{t^5}{10} - \frac{t^7}{42} + \dots \right) .
\end{align}
But what about its large argument limit erf$(x \gg 1)$? We may write
\begin{align}
\text{erf}(x) 
&= \frac{2}{\sqrt{\pi}} \left(\int_{0}^{\infty} \dd t - \int_{x}^{\infty} \dd t\right) e^{-t^2} \nonumber\\
&= 1 - \frac{2}{\sqrt{\pi}} I(x), \qquad\qquad I(x) \equiv \int_{x}^{\infty} \dd t e^{-t^2} .
\end{align}
Integration-by-parts may be employed as follows.
{\allowdisplaybreaks\begin{align}
\label{AsymptoticSeries_Erf_Steps}
I(x) &= \int_{x}^{\infty} \dd t \frac{1}{-2t} \partial_t e^{-t^2} 
		= \left[\frac{e^{-t^2}}{-2t}\right]_{t=x}^{t=\infty} - \int_{x}^{\infty} \dd t \partial_t \left(\frac{1}{-2t}\right) e^{-t^2} \nonumber\\
&= \frac{e^{-x^2}}{2x} - \int_{x}^{\infty} \dd t \frac{e^{-t^2}}{2t^2} 
		= \frac{e^{-x^2}}{2x} - \int_{x}^{\infty} \dd t \frac{1}{2t^2(-2t)} \partial_t e^{-t^2} \\
&= \frac{e^{-x^2}}{2x} - \frac{e^{-x^2}}{4x^3} + \int_{x}^{\infty} \dd t \frac{3}{4t^4} e^{-t^2} \nonumber
\end{align}}
\begin{myP}
\qquad After $n$ integration by parts,
\begin{align}
\label{AsymptoticSeries_Erf}
\int_{x}^{\infty} \dd t e^{-t^2}
= e^{-x^2} \sum_{\ell=1}^{n} (-)^{\ell-1} \frac{1 \cdot 3 \cdot 5 \dots (2\ell-3)}{2^\ell x^{2\ell-1}}
- (-)^n \frac{1 \cdot 3 \cdot 5 \dots (2n-1)}{2^n} \int_x^\infty \dd t \frac{e^{-t^2}}{t^{2n}} .
\end{align}
This result can be found in Matthew and Walker, but can you prove it more systematically by mathematical induction? For a fixed $x$, find the $n$ such that the next term generated by integration-by-parts is larger than the previous term. This series does not converge -- why? \qquad \qed
\end{myP}

If we drop the remainder integral in eq. \eqref{AsymptoticSeries_Erf}, the resulting series does not converge as $n \to \infty$. However, for large $x \gg 1$, it is not difficult to argue that the first few terms do offer an excellent approximation, since each subsequent term is suppressed relative to the previous by a $1/x$ factor.\footnote{In fact, as observed by Matthews and Walker \cite{MatthewsWalker}, since this is an oscillating series, the optimal $n$ to truncate the series is the one right before the smallest.} 
\begin{myP}
	\qquad Using integration-by-parts, develop a large $x \gg 1$ expansion for
\begin{align}
I(x) \equiv \int_x^\infty \dd t \frac{\sin(t)}{t} .
\end{align} 
Hint: Consider instead $\int_x^\infty \dd t \frac{\exp(it)}{t}$. \qed
\end{myP}
% For example, after one integration-by-parts, we may compare the magnitudes of the leading order answer with that of the remainder integral:
%\begin{align}
%\frac{e^{-x^2}}{2x} \qquad \text{ vs.} \qquad 
%\int_{x}^{\infty} \dd t \frac{e^{-t^2}}{2t^2} < \frac{1}{2x^2} \int_{x}^{\infty} \dd t e^{-t^2} = \frac{I(x)}{2x^2} .
%\end{align}
%If the leading behavior of $I(x \gg 1)$ is $e^{-x^2}/(2x)$, then the remainder integral is suppressed by a further $1/x$.
{\bf What is an asymptotic series?} \qquad A Taylor expansion of say $e^x$
\begin{align}
e^x = 1 + x + \frac{x}{2!} + \frac{x^3}{3!} + \dots 
\end{align}
converges for all $|x|$. In fact, for a fixed $|x|$, we know summing up more terms of the series
\begin{align}
\sum_{\ell=0}^{N} \frac{x^\ell}{\ell !} ,
\end{align}
-- the larger $N$ we go -- the closer to the actual value of $e^x$ we would get.

An asymptotic series of the sort we have encountered above, and will be doing so below, is a series of the sort
\begin{align}
S_N(x) = A_0 + \frac{A_1}{x} + \frac{A_2}{x^2} + \dots + \frac{A_N}{x^N} .
\end{align}
For a fixed $|x|$ the series oftentimes diverges as we sum up more and more terms ($N \to \infty$). However, for a fixed $N$, it can usually be argued that as $x \to +\infty$ the $S_N(x)$ becomes an increasingly better approximation to the object we derived it from in the first place. 

As Matthews and Walker \cite{MatthewsWalker} further explains:\begin{quotation}
	``\dots {\it an asymptotic series may be added, multiplied, and integrated to obtain the asymptotic series for the corresponding sum, product and integrals of the corresponding functions. Also, the asymptotic series of a given function is unique, but \dots An asymptotic series does not specify a function uniquely.}" 
\end{quotation}

\subsubsection{Laplace's Method, Method of Stationary Phase, Steepest Descent}

{\bf Exponential suppression} \qquad The asymptotic methods we are about to encounter in this section rely on the fact that, the integrals we are computing really receive most of their contribution from a small region of the integration region. Outside of the relevant region the integrand itself is highly exponentially suppressed -- a basic illustration of this is
\begin{align}
I(x) = \int_{0}^x e^{-t} = 1-e^{-x}.
\end{align}
As $x\to\infty$ we have $I(\infty)=1$. Even though it takes an infinite range of integration to obtain $1$, we see that most of the contribution ($\gg 99 \%$) comes from $t=0$ to $t \sim \mathcal{O}(10)$. For example, $e^{-5} \approx 6.7 \times 10^{-3}$ and $e^{-10} \approx 4.5 \times 10^{-5}$. You may also think about evaluating this integral numerically; what this shows is that it is not necessary to sample your integrand out to very large $t$ to get an accurate answer.\footnote{In the Fourier transform section I pointed out how, if you merely need to resolve the coarser features of your wave profile, then provided the short wavelength modes do not have very large amplitudes, only the coefficients of the modes with longer wavelengths need to be known accurately. Here, we shall see some integrals only require us to know their integrands in a small region, if all we need is an approximate (but oftentimes highly accurate) answer. This is a good rule of thumb to keep in mind when tackling difficult, apparently complicated, problems in physics: focus on the most relevant contributions to the final answer, and often this will simplify the problem-solving process.}

{\bf Laplace's Method} \qquad We now turn to integrals of the form
\begin{align}
\label{LaplaceMethod_I}
I(x) = \int_{a}^{b} f(t) e^{x \phi(t)} \dd t
\end{align}
where both $f$ and $\phi$ are real. (There is no need to ever consider the complex $f$ case since it can always be split into real and imaginary parts.) We will consider the $x \to +\infty$ limit and try to extract the leading order behavior of the integral.

The main strategy goes roughly as follows. Find the location of the maximum of $\phi(t)$ -- say it is at $t=c$. This can occur in between the limits of integration $a < c < b$ or at one of the end points $c=a$ or $c=b$. As long as $f(c) \neq 0$, we may expand both $f(t)$ and $\phi(t)$ around $t=c$. For simplicity we display the case where $a < c < b$:
\begin{align}
\label{LaplaceMethod_II}
I(x) \sim e^{x \phi(c)} \int_{c-\kappa}^{c+\kappa} (f(c) + (t-c) f'(c) + \dots) \exp\left(x \left\{ \frac{\phi^{(p)}(c)}{p!}(t-c)^p + \dots \right\}\right)\dd t,
\end{align}
where we have assumed the first non-zero derivative of $\phi$ is at the $p$th order, and $\kappa$ is some small number ($\kappa < |b-a|$) such that the expansion can be justified, because the errors incurred from switching from $\int_{a}^{b} \to \int_{c-\kappa}^{c+\kappa}$ are exponentially suppressed. (Since $\phi(t=c)$ is maximum, $\phi'(c)$ is usually -- but not always! -- zero.) Then, term by term, these integrals, oftentimes after a change of variables, can be tackled using the Gamma function integral representation
\begin{align}
\label{GammaFunction_IntegralRep}
\Gamma(z) \equiv \int_{0}^{\infty} t^{z-1} e^{-t} \dd t , \qquad \text{Re}(z) > 0,
\end{align}
by extending the former's limits to infinity, $\int_{c-\kappa}^{c+\kappa} \to \int_{-\infty}^{+\infty}$. This last step, like the expansion in eq. \eqref{LaplaceMethod_II}, is usually justified because the errors incurred are again exponentially small.

{\it Examples} \qquad The first example, where $\phi'(c) \neq 0$, is related to the integral representation of the parabolic cylinder function; for Re$(\nu)>0$,
\begin{align}
I(x) = \int_{0}^{100} t^{\nu-1} e^{-t^2/2} e^{-xt} \dd t.
\end{align}
Here, $\phi(t)=-t$ and its maximum is at the lower limit of integration. For large $t$ the integrand is exponentially suppressed, and we expect the contribution to arise mainly for $t \in [0, \text{a few})$. In this region we may Taylor expand $e^{-t^2/2}$. Term-by-term, we may then extend the upper limit of integration to infinity, provided we can justify the errors incurred are small enough for $x \gg 1$.
{\allowdisplaybreaks\begin{align}
I(x \to \infty) 
&\sim \int_{0}^{\infty} t^{\nu-1} \left( 1-\frac{t^2}{2} + \dots \right) e^{-xt} \dd t \nonumber\\
&= \int_{0}^{\infty} \frac{(xt)^{\nu-1}}{x^{\nu-1}} \left( 1-\frac{(xt)^2}{2x^2} + \dots \right) e^{-(xt)} \frac{\dd (xt)}{x} \nonumber\\
&= \frac{\Gamma(\nu)}{x^\nu} \left( 1 + \mathcal{O}\left(x^{-2}\right) \right) . 
\end{align}}
The second example is
{\allowdisplaybreaks\begin{align}
I(x\to\infty) &= \int_{0}^{88} \frac{\exp(-x \cosh(t))}{\sqrt{\sinh(t)}} \dd t \nonumber\\
&\sim \int_{0}^{\infty} \frac{\exp\left( -x\left\{ 1 + \frac{t^2}{2} + \dots \right\} \right)}{\sqrt{t}\sqrt{1 + t^2/6 + \dots}} \dd t \nonumber\\
&\sim e^{-x} \int_{0}^{\infty} \frac{(x/2)^{1/4} \exp\left(-(\sqrt{x/2}t)^2\right)}{\sqrt{\sqrt{x/2}t}} \frac{\dd (\sqrt{x/2}t)}{\sqrt{x/2}}  .
\end{align}}
To obtain higher order corrections to this integral, we would have to be expand both the exp and the square root in the denominator. But the $t^2/2 + \dots$ comes multiplied with a $x$ whereas the denominator is $x$-independent, so you'd need to make sure to keep enough terms to ensure you have captured all the contributions to the next- and next-to-next leading corrections, etc. We will be content with just the dominant behavior: we put $z \equiv t^2 \Rightarrow \dd z = 2 t \dd t = 2\sqrt{z} \dd t$.
\begin{align}
\int_{0}^{88} \frac{\exp(-x \cosh(t))}{\sqrt{\sinh(t)}} \dd t
&\sim \frac{e^{-x}}{(x/2)^{1/4}} \int_{0}^{\infty} z^{\left(1-\frac{1}{4}-\frac{1}{2}\right)-1} e^{-z} \frac{\dd z}{2} \nonumber\\
&= e^{-x} \frac{\Gamma(1/4)}{2^{3/4} x^{1/4}} .
\end{align}
In both examples, the integrand really behaves very differently from the first few terms of its expanded version for $t \gg 1$, but the main point here is -- it doesn't matter! The error incurred, for very large $x$, is exponentially suppressed anyway. If you care deeply about rigor, you may have to prove this assertion on a case-by-case basis; see Example 7 and 8 of Bender \& Orszag's Chapter 6 \cite{BenderOrszag} for careful discussions of two specific integrals.

{\it Stirling's formula} \qquad Can Laplace's method apply to obtain a large $x \gg 1$ limit representation of the Gamma function itself?
\begin{align}
\Gamma(x) = \int_{0}^{\infty} t^{x-1} e^{-t} \dd t = \int_{0}^{\infty} e^{(x-1)\ln(t)} e^{-t} \dd t 
\end{align}
It does not appear so because here $\phi(t)=\ln(t)$ and the maximum is at $t=\infty$. Actually, the maximum of the exponent is at
\begin{align}
\frac{\dd}{\dd t}\left((x-1)\ln(t) - t\right) = \frac{x-1}{t} -1 = 0 \qquad \Rightarrow \qquad t = x-1 .
\end{align}
Re-scale $t \to (x-1) t$:
\begin{align}
\Gamma(x) = (x-1) e^{(x-1)\ln(x-1)} \int_{0}^{\infty} e^{(x-1) (\ln(t) - t)} \dd t .
\end{align}
Comparison with eq. \eqref{LaplaceMethod_I} tells us $\phi(t) = \ln(t)-t$ and $f(t)=1$. We may now expand the exponent about its maximum at $1$:
\begin{align}
\ln(t)-t = -1 -\frac{(t-1)^2}{2} + \frac{(t-1)^3}{3} + \dots .
\end{align}
This means 
\begin{align}
\Gamma(x) 
&\sim \sqrt{\frac{2}{x-1}} (x-1)^{x} e^{-(x-1)} \\
&\qquad \times \int_{-\infty}^{+\infty} \exp\left( -\left(\sqrt{x-1}\frac{t-1}{\sqrt{2}}\right)^2 + \mathcal{O}((t-1)^3) \right) \dd (\sqrt{x-1}t/\sqrt{2}) . \nonumber
\end{align}
Noting $x-1 \approx x$ for large $x$; we arrive at Stirling's formula,
\begin{align}
\Gamma(x \to \infty) \sim \sqrt{\frac{2\pi}{x}} \frac{x^x}{e^x} .
\end{align}
\begin{myP}
\qquad What is the leading behavior of
\begin{align}
I(x) \equiv \int_{0}^{50.12345+e^{\sqrt{2}}+\pi^{\sqrt{e}}} e^{-x \cdot t^\pi} \sqrt{1 + \sqrt{t}} \dd t
\end{align}
in the limit $x \to +\infty$? And, how does the first correction scale with $x$? \qed
\end{myP}
\begin{myP}
\qquad What is the leading behavior of 
\begin{align}
I(x) = \int_{-\pi/2}^{\pi/2} \frac{e^{-x \cos(t)^2}}{(\cos(t))^p} \dd t,
\end{align} 
for $0 \leq p < 1$, in the limit $x \to +\infty$? Note that there are two maximums of $\phi(t)$ here. \qed
\end{myP}
{\bf Method of Stationary Phase} \qquad We now consider the case where the exponent is purely imaginary, 
\begin{align}
\label{StationaryPhase_I}
I(x) = \int_{a}^{b} f(t) e^{i x \phi(t)} \dd t .
\end{align}
Here, both $f$ and $\phi$ are real. As we did previously, we will consider the $x \to +\infty$ limit and try to extract the leading order behavior of the integral. 

What will be very useful, to this end, is the following lemma.
\begin{quotation}
The {\it Riemann-Lebesgue lemma} states that $I(x\to\infty)$ in eq. \eqref{StationaryPhase_I} goes to zero provided: (I) $\int_{a}^{b}|f(t)| \dd t < \infty$; (II) $\phi(t)$ is continuously differentiable; and (III) $\phi(t)$ is not constant over a finite range within $t \in [a,b]$.
\end{quotation}
We will not prove this result, but it is heuristically very plausible: as long as $\phi(t)$ is not constant, the $e^{i x \phi(t)}$ fluctuates wildly as $x\to+\infty$ on the $t \in [a,b]$ interval. For large enough $x$, $f(t)$ will be roughly constant over `each period' of $e^{i x \phi(t)}$, which in turn means $f(t) e^{i x \phi(t)}$ will integrate to zero over this same `period'.

{\it Case I: $\phi(t)$ has no turning points} \qquad The first implication of the Riemann-Lebesgue lemma is that, if $\phi'(t)$ is not zero anywhere within $t \in [a,b]$; and as long as $f(t)/\phi'(t)$ is smooth enough within $t \in [a,b]$ and exists on the end points; then we can use integration-by-parts to show that the integral in eq. \eqref{StationaryPhase_I} has to scale as $1/x$ as $x \to \infty$.
\begin{align}
I(x) 
&= \int_{a}^{b} \frac{f(t)}{ix \phi'(t)} \frac{\dd}{\dd t} e^{i x \phi(t)} \dd t \nonumber\\
&= \frac{1}{ix} \left\{ \left[ \frac{f(t)}{\phi'(t)} e^{i x \phi(t)} \right]_{a}^{b}
- \int_{a}^{b} e^{i x \phi(t)} \frac{\dd}{\dd t} \left(\frac{f(t)}{\phi'(t)}\right) \dd t  \right\} .
\end{align}
The integral on the second line within the curly brackets is one where Riemann-Lebesgue applies. Therefore it goes to zero relative to the (boundary) term preceding it, as $x\to\infty$. Therefore what remains is
\begin{align}
\int_{a}^{b} f(t) e^{i x \phi(t)} \dd t \sim \frac{1}{ix} \left[ \frac{f(t)}{\phi'(t)} e^{i x \phi(t)} \right]_{a}^{b}, 
\qquad x \to +\infty, \qquad
\phi'(a \leq t \leq b) \neq 0.
\end{align}
{\it Case II: $\phi(c)$ has at least one turning point} \qquad If there is at least one point where the phase is stationary, $\phi'(a \leq c \leq b)=0$, then provided $f(c)\neq 0$, we shall see that the dominant behavior of the integral in eq. \eqref{StationaryPhase_I} scales as $1/x^{1/p}$, where $p$ is the lowest order derivative of $\phi$ that is non-zero at $t=c$. Because $1/p < 1$, the $1/x$ behavior we found above is sub-dominant to $1/x^{1/p}$ -- hence the need to analyze the two cases separately.

Let us, for simplicity, assume the stationary point is at $a$, the lower limit. We shall discover the leading behavior to be
\begin{align}
\label{StationaryPhase_II}
\int_{a}^{b} f(t) e^{i x \phi(t)} \dd t \sim f(a) \exp\left( ix\phi(a) \pm i \frac{\pi}{2p} \right) \frac{\Gamma(1/p)}{p} \left(\frac{p!}{x |\phi^{(p)}(a)|}\right)^{1/p} ,
\end{align}
where $\phi^{(p)}(a)$ is first non-vanishing derivative of $\phi(t)$ at the stationary point $t=a$; while the $+$ sign is to be chosen if $\phi^{(p)}(a)>0$ and $-$ if $\phi^{(p)}(a)<0$.

To understand eq. \eqref{StationaryPhase_II}, we decompose the integral into
\begin{align}
I(x) &= \int_{a}^{a+\kappa} f(t) e^{i x \phi(t)} \dd t + \int_{a+\kappa}^b f(t) e^{i x \phi(t)} \dd t .
\end{align}
The second integral scales as $1/x$, as already discussed, since we assume there are no stationary points there. The first integral, which we shall denote as $S(x)$, may be expanded in the following way provided $\kappa$ is chosen appropriately:
\begin{align}
S(x) = \int_{a}^{a+\kappa} (f(a) + \dots) e^{ix\phi(a)} \exp\left(\frac{ix}{p!} (t-a)^p \phi^{(p)}(a) + \dots \right) \dd t .
\end{align}
To convert the oscillating exp into a real, dampened one, let us rotate our contour. Around $t=a$, we may change variables to $t-a \equiv \rho e^{i\theta} \Rightarrow (t-a)^p = \rho^p e^{ip\theta} = i\rho^p$ (i.e., $\theta=\pi/(2p)$) if $\phi^{(p)}(a)>0$; and $(t-a)^p = \rho^p e^{ip\theta} = -i\rho^p$ (i.e., $\theta=-\pi/(2p)$) if $\phi^{(p)}(a)<0$. Since our stationary point is at the lower limit, this is for $\rho > 0$.\footnote{If $p$ is even, and if the stationary point is not one of the end points, observe that we can choose $\theta = \pm(\pi/(2p) + \pi) \Rightarrow e^{ip\theta} = \pm i$ for the $\rho < 0$ portion of the contour -- i.e., run a straight line rotated by $\theta$ through the stationary point -- and the final result would simply be twice of eq. \eqref{StationaryPhase_II}.}
\begin{align}
&S(x \to \infty) \nonumber\\
&\sim f(a) e^{ix\phi(a)} e^{\pm i\pi/(2p)} \int_0^{+\infty} \exp\left(-\frac{x}{p!} |\phi^{(p)}(a)| \rho^p \right) \frac{\dd (\rho^p)}{p \cdot \rho^{p-1}} \\
&\sim f(a) e^{ix\phi(a)} \frac{e^{\pm i\pi/(2p)}}{p ( \frac{x}{p!} |\phi^{(p)}(a)| )^{1/p}} \int_0^{+\infty} \left(\frac{x}{p!} |\phi^{(p)}(a)| s\right)^{\frac{1}{p}-1}\exp\left(-\frac{x}{p!} |\phi^{(p)}(a)| s \right) \dd \left(\frac{x}{p!} |\phi^{(p)}(a)| s\right) . \nonumber
\end{align}
This establishes the result in eq. \eqref{StationaryPhase_II}.
\begin{myP}
\qquad Starting from the following integral representation of the Bessel function
\begin{align}
J_n(x) = \frac{1}{\pi} \int_{0}^{\pi} \cos\left( n\theta - x \sin\theta \right) \dd\theta
\end{align}
where $n = 0,1,2,3,\dots$, show that the leading behavior as $x \to +\infty$ is
\begin{align}
J_n(x) \sim \sqrt{\frac{2}{\pi x}} \cos\left( x - \frac{n \pi}{2} - \frac{\pi}{4}\right) .
\end{align}
Hint: Express the cosine as the real part of an exponential. Note the stationary point is two-sided, but it is fairly straightforward to deform the contour appropriately.
\end{myP}
{\bf Method of Steepest Descent} \qquad We now allow our exponent to be complex.
\begin{align}
I(x) = \int_C f(t) e^{x u(t)} e^{i x v(t)} \dd t ,
\end{align}
The $f$, $u$ and $v$ are real; $C$ is some contour on the complex $t$ plane; and as before we will study the $x \to \infty$ limit. We will assume $u+iv$ forms an analytic function of $t$.

The method of steepest descent is the strategy to deform the contour $C$ to some $C'$ such that it lies on a constant-phase path -- where the imaginary part of the exponent does not change along it.
\begin{align}
I(x) = e^{i x v} \int_{C'} f(t) e^{x u(t)} \dd t 
\end{align}
One reason for doing so is that the constant phase contour also coincides with the steepest descent one of the real part of the exponent -- unless the contour passes through a saddle point, where more than one steepest descent paths can intersect. Along a steepest descent path, Laplace's method can then be employed to obtain an asymptotic series. 

To understand this further we recall that the gradient is perpendicular to the lines of constant potential, i.e., the gradient points along the curves of most rapid change. Assuming $u+iv$ is an analytic function, and denoting $t=x+iy$ (for $x$ and $y$ real), the Cauchy-Riemann equations they obey
\begin{align}
\partial_x u = \partial_y v, \qquad \partial_y u = -\partial_x v 
\end{align}
means the dot product of their gradients is zero:
\begin{align}
\vec{\nabla} u \cdot \vec{\nabla} v
= \partial_x u \partial_x v + \partial_y u \partial_y v
= \partial_y v \partial_x v - \partial_x v \partial_y v = 0 .
\end{align}
To sum:
\begin{quotation}
A constant phase line -- namely, the contour line where $v$ is constant -- is necessarily perpendicular to $\vec{\nabla} v$. But since $\vec{\nabla} u \cdot \vec{\nabla} v = 0$ in the relevant region of the 2D complex $(t=x+iy)$-plane where $u(t)+iv(t)$ is assumed to be analytic, a constant phase line must therefore be (anti)parallel to $\vec{\nabla} u$, the direction of most rapid change of the real amplitude $e^{x u}$.
\end{quotation}
We will examine the following simple example:
\begin{align}
I(x) = \int_{0}^{1} \ln(t) e^{ixt} \dd t.
\end{align}
We deform the contour $\int_{0}^{1}$ so it becomes the sum of the straight lines $C_1$, $C_2$ and $C_3$. $C_1$ runs from $t=0$ along the positive imaginary axis to infinity. $C_2$ runs horizontally from $i\infty$ to $i\infty+1$. Then $C_3$ runs from $i\infty+1$ back down to $1$. There is no contribution from $C_2$ because the integrand there is $\ln(i\infty) e^{-x\infty}$, which is zero for positive $x$.
\begin{align}
I(x) 
&= i\int_{0}^{\infty} \ln(it) e^{-xt} \dd t - i\int_{0}^{\infty} \ln(1+it) e^{ix(1+it)} \dd t  \nonumber\\
&= i\int_{0}^{\infty} \ln(it) e^{-xt} \dd t - i e^{ix} \int_{0}^{\infty} \ln(1+it) e^{-xt} \dd t  .
\end{align}
Notice the exponents in both integrands have now zero (and therefore constant) phases.
\begin{align}
I(x) 
&= i\int_{0}^{\infty} \ln(i(xt)/x) e^{-(xt)} \frac{\dd (xt)}{x} - i e^{ix} \int_{0}^{\infty} \ln(1+i(xt)/x) e^{-(xt)} \frac{\dd (xt)}{x}  \nonumber\\
&= i\int_{0}^{\infty} (\ln(z)-\ln(x)+i\pi/2) e^{-z} \frac{\dd z}{x} - i e^{ix} \int_{0}^{\infty} \left( i\frac{z}{x} + \mathcal{O}(x^{-2}) \right) e^{-z} \frac{\dd z}{x} .
\end{align}
The only integral that remains unfamiliar is the first one
\begin{align}
\int_{0}^{\infty} e^{-z} \ln(z) 
= \left.\frac{\partial}{\partial \mu}\right\vert_{\mu=1} \int_{0}^{\infty} e^{-z} e^{(\mu-1)\ln(z)} 
&= \left.\frac{\partial}{\partial \mu}\right\vert_{\mu=1} \int_{0}^{\infty} e^{-z} z^{\mu-1} \nonumber\\
&= \Gamma'(1) = -\gamma_\text{E}
\end{align}
The $\gamma_\text{E} = 0.577216 \dots$ is known as the Euler-Mascheroni constant. At this point,
\begin{align}
\int_{0}^{1} \ln(t) e^{ixt} \dd t
\sim \frac{i}{x} \left(-\gamma_\text{E} - \ln(x) + i\frac{\pi}{2} - \frac{i e^{ix}}{x} + \mathcal{O}(x^{-2}) \right), \qquad x \to +\infty .
\end{align}
\begin{myP}
\qquad Perform an asymptotic expansion of
\begin{align}
I(k) \equiv \int_{-1}^{+1} e^{ikx^2} \dd x
\end{align}
using the steepest descent method. Hint: Find the point $t=t_0$ on the real line where the phase is stationary. Then deform the integration contour such that it passes through $t_0$ and has a stationary phase everywhere. Can you also tackle $I(k)$ using integration-by-parts? \qed
\end{myP}

\subsection{JWKB solution to $-\epsilon^2 \psi''(x) + U(x) \psi(x) = 0$, for $0 < \epsilon \ll 1$}

Many physicists encounter for the first time the following Jeffreys-Wentzel-Kramers-Brillouin (JWKB; aka WKB) method and its higher dimensional generalization, when solving the Schr\"{o}dinger equation -- and are told that the approximation amounts to the semi-classical limit where Planck's constant tends to zero, $\hbar \to 0$. Here, I want to highlight its general nature: it is not just applicable to quantum mechanical problems but oftentimes finds relevance when the wavelength of the solution at hand can be regarded as `small' compared to the other length scales in the physical setup. The statement that electromagnetic waves in curved spacetimes or non-trivial media propagate predominantly on the null cone in the (effective) geometry, is in fact an example of such a `short wavelength' approximation.

We will focus on the 1D case. Many physical problems reduce to the following 2nd order linear ordinary differential equation (ODE):
\begin{align}
\label{JWKB_MainODE}
-\epsilon^2 \psi''(x) + U(x) \psi(x) = 0 ,
\end{align}
where $\epsilon$ is a ``small" (usually fictitious) parameter. This second order ODE is very general because both the Schr\"{o}dinger and the (frequency space) Klein-Gordon equation with some potential reduces to this form. (Also recall that the first derivative terms in all second order ODEs may be removed via a redefinition of $\psi$.) The main goal of this section is to obtain its approximate solutions.

We will use the ansatz
\begin{align*}
\psi(x) = \sum_{\ell = 0}^\infty \epsilon^\ell \alpha_\ell(x) e^{i S(x)/\epsilon} .
\end{align*}
Plugging this into our ODE, we obtain
\begin{align}
\label{JWKB_MainEqn}
0 = \sum_{\ell = 0}^\infty 
\epsilon ^{\ell} \left( \alpha_\ell(x) \left(S'(x)^2+U(x)\right) -i\left(\alpha_{\ell-1}(x) S''(x)+2 S'(x) \alpha_{\ell-1}'(x)\right) - \alpha_{\ell-2}''(x) \right)
\end{align}
with the understanding that $\alpha_{-2}(x) = \alpha_{-1}(x) = 0$. We need to set the coefficients of $\epsilon^\ell$ to zero. The first two terms ($\ell=0,1$) give us solutions to $S(x)$ and $\alpha_0(x)$.
\begin{align*}
0 &= a_0 \left( S'(x)^2 + U(x) \right)  \qquad \Rightarrow \qquad S_\pm(x) = \sigma_0 \pm i \int^x \dd x' \sqrt{U(x')}; \ \sigma_0 = \text{const.} \\
0 &= -i\epsilon \left( 2 \alpha_0'(x) S'(x) + \alpha_0(x) S''(x) \right), \qquad \Rightarrow \qquad \alpha_0(x) = \frac{C_0}{U(x)^{1/4}}
\end{align*}
(While the solutions $S_\pm(x)$ contains two possible signs, the $\pm$ in $S'$ and $S''$ factors out of the second equation and thus $\alpha_0$ does not have two possible signs.) 
\begin{myP}
{\it Recursion relation for higher order terms} \qquad By considering the $\ell \geq 2$ terms in eq. \eqref{JWKB_MainEqn}, show that there is a recursion relation between $\alpha_{\ell}(x)$ and $\alpha_{\ell+1}(x)$. Can you use them to deduce the following two linearly independent JWKB solutions?
\begin{align}
0 &= -\epsilon^2 \psi_\pm''(x) + U(x) \psi_\pm(x) \\
\label{JWB_Solution}
\psi_\pm(x) &= \frac{1}{U(x)^{1/4}} \exp\left[ \mp \frac{1}{\epsilon} \int^x \dd x' \sqrt{U(x')} \right] \sum_{\ell = 0}^\infty \epsilon^\ell Q_{(\ell|\pm)}(x), \\
Q_{(\ell|\pm)}(x) &= \pm \frac{1}{2} \int^x \frac{\dd x'}{U(x')^{1/4}} \frac{\dd^2}{\dd x'^2} \left( \frac{Q_{(\ell-1|\pm)}(x')}{U(x')^{1/4}} \right), \qquad Q_{(0|\pm)}(x) \equiv 1
\end{align}
To lowest order
\begin{align}
\psi_\pm(x) &= \frac{1}{U^{1/4}(x)} \exp\left[ \mp \frac{1}{\epsilon} \int^x \dd x' \sqrt{U[x']} \right] \left( 1 + \mathcal{O}[\epsilon] \right) .
\end{align} 
Note: in these solutions, the $\sqrt{\cdot}$ and $\sqrt[4]{\cdot}$ are positive roots. \qed
\end{myP}
{\bf JWKB Counts Derivatives} \qquad In terms of the $Q_{(n)}$s we see that the JWKB method is really an approximation that works whenever each dimensionless derivative $\dd/\dd x$ acting on some power of $U(x)$ yields a smaller quantity, i.e., roughly speaking $\dd \ln U(x)/\dd x \sim \epsilon \ll 1$; this small derivative approximation is related to the short wavelength approximation. Also notice from the exponential $\exp[i S/\epsilon] \sim \exp[\pm (i/\epsilon) \int \sqrt{-U}]$ that the $1/\epsilon$ indicates an integral (namely, an inverse derivative). To sum: 
\begin{quotation}
	The ficticious parameter $\epsilon \ll 1$ in the JWKB solution of $-\epsilon^2 \psi'' + U \psi = 0$ counts the number of derivatives; whereas $1/\epsilon$ is an integral. The JWKB approximation works well whenever each additional dimensionless derivative acting on some power of $U$ yields a smaller and smaller quantity.
\end{quotation}
{\bf Breakdown and connection formulas} \qquad There is an important aspect of JWKB that I plan to discuss in detail in a future version of these lecture notes. From the $1/\sqrt[4]{U(x)}$ prefactor of the solution in eq. \eqref{JWB_Solution}, we see the approximation breaks down at $x=x_0$ whenever $U(x_0)=0$. The JWKB solutions on either side of $x=x_0$ then need to be joined by matching onto a valid solution in the region $x \sim x_0$. One common approach is to replace $U$ with its first non-vanishing derivative, $U(x) \to ((x-x_0)^n/n!) U^{(n)}(x_0)$; if $n=1$, the corresponding solutions to the 2nd order ODE are Airy functions -- see, for e.g., Sakurai's {\it Modern Quantum Mechanics} for a discussion. Another approach, which can be found in Matthews and Walker \cite{MatthewsWalker}, is to complexify the JWKB solutions, perform analytic continuation, and match them on the complex plane.

\newpage

\section{Differential Geometry of Curved Spaces}
\label{Chapter_DifferentialGeometry_CurvedSpaces}
\subsection{Preliminaries, Tangent Vectors, Metric, and Curvature}

Being fluent in the mathematics of differential geometry is mandatory if you wish to understand Einstein's General Relativity, humanity's current theory of gravity. But it also gives you a coherent framework to understand the multi-variable calculus you have learned, and will allow you to generalize it readily to dimensions other than the 3 spatial ones you are familiar with. In this section I will provide a practical introduction to differential geometry, and will show you how to recover from it what you have encountered in 2D/3D vector calculus. My goal here is that you will understand the subject well enough to perform concrete calculations, without worrying too much about the more abstract notions like, for e.g., what a manifold is.

I will assume you have an intuitive sense of what space means -- after all, we live in it! Spacetime is simply space with an extra time dimension appended to it, although the notion of `distance' in spacetime is a bit more subtle than that in space alone. To specify the (local) geometry of a space or spacetime means we need to understand how to express distances in terms of the coordinates we are using. For example, in Cartesian coordinates $(x,y,z)$ and by invoking Pythagoras' theorem, the square of the distance $(\dd\ell)^2$ between $(x,y,z)$ and $(x+\dd x,y+\dd y,z+\dd z)$ in flat (aka Euclidean) space is
\begin{align}
\label{DifferentialGeometry_3DFlatEuclideanMetric_Cartesian}
(\dd \ell)^2 = (\dd x)^2 + (\dd y)^2 + (\dd z)^2 .
\end{align}
\footnote{In 4-dimensional flat space{\it time}, with time $t$ in addition to the three spatial coordinates $\{x,y,z\}$, the infinitesimal distance is given by a modified form of Pythagoras' theorem: $\dd s^2 \equiv (\dd t)^2 - (\dd x)^2 - (\dd y)^2 - (\dd z)^2$. (The opposite sign convention, i.e., $\dd s^2 \equiv -(\dd t)^2 + (\dd x)^2 + (\dd y)^2 + (\dd z)^2$, is also equally valid.) Why the ``time" part of the distance differs in sign from the ``space" part of the metric would lead us to a discussion of the underlying Lorentz symmetry. Because I wish to postpone the latter for the moment, I will develop differential geometry for curved spaces, not curved spacetimes. Despite this restriction, rest assured most of the subsequent formulas do carry over to curved spacetimes by simply replacing Latin/English alphabets with Greek ones -- see the ``Conventions" paragraph below.}A significant amount of machinery in differential geometry involves understanding how to employ arbitrary coordinate systems -- and switching between different ones. For instance, we may convert the Cartesian coordinates flat space of eq. \eqref{DifferentialGeometry_3DFlatEuclideanMetric_Cartesian} into spherical coordinates,
\begin{align}
(x,y,z) \equiv r \left(\sin\theta \cdot \cos\phi, \sin\theta \cdot \sin\phi,\cos\theta \right) ,
\end{align}
and find 
\begin{align}
\label{DifferentialGeometry_3DFlatEuclideanMetric_Spherical}
(\dd \ell)^2 = \dd r^2 + r^2 (\dd\theta^2 + \sin(\theta)^2 \dd\phi^2 ) .
\end{align}
The geometries in eq. \eqref{DifferentialGeometry_3DFlatEuclideanMetric_Cartesian} and eq. \eqref{DifferentialGeometry_3DFlatEuclideanMetric_Spherical} are exactly the same. All we have done is to express them in different coordinate systems. 

{\bf Conventions} \qquad This is a good place to (re-)introduce the Einstein summation convention and the index convection. First, instead of $(x,y,z)$, we can instead use $x^i \equiv (x^1,x^2,x^3)$; here, the superscript does not mean we are raising $x$ to the first, second and third powers. A derivative with respect to the $i$th coordinate is $\partial_i \equiv \partial/\partial x^i$. The advantage of such a notation is its compactness: we can say we are using coordinates $\{x^i\}$, where $i \in \{1,2,3\}$.\footnote{It is common to use the English alphabets to denote space coordinates and Greek letters to denote spacetime ones. We will adopt this convention in these notes, but note that it is not a universal one; so be sure to check the notation of the book you are reading.} Not only that, we can employ Einstein's summation convention, which says all repeated indices are automatically summed over their relevant range. For example, eq. \eqref{DifferentialGeometry_3DFlatEuclideanMetric_Cartesian} now reads:
\begin{align}
(\dd x^1)^2 + (\dd x^2)^2 + (\dd x^3)^2 =
\delta_{ij} \dd x^i \dd x^j \equiv \sum_{1 \leq i,j \leq 3} \delta_{ij} \dd x^i \dd x^j .
\end{align}
(We say the indices of the $\{\dd x^i\}$ are being contracted with that of $\delta_{ij}$.) The symbol $\delta_{ij}$ is known as the Kronecker delta, defined as
\begin{align}
\label{DifferentialGeometry_KroneckerDelta}
\delta_{ij} &= 1, \qquad i = j , \\
			&= 0, \qquad i \neq j .
\end{align}
Of course, $\delta_{ij}$ is simply the $ij$ component of the identity matrix. Already, we can see $\delta_{ij}$ can be readily defined in an arbitrary $D$ dimensional space, by allowing $i,j$ to run from $1$ through $D$. With these conventions, we can re-express the change of variables from eq. \eqref{DifferentialGeometry_3DFlatEuclideanMetric_Cartesian} and eq. \eqref{DifferentialGeometry_3DFlatEuclideanMetric_Spherical} as follows. First write 
\begin{align}
\label{CartesianToSpherical}
\xi^i \equiv (r \geq 0,0 \leq \theta \leq \pi,0 \leq \phi < 2\pi) .
\end{align}
Then \eqref{DifferentialGeometry_3DFlatEuclideanMetric_Cartesian} becomes
\begin{align}
\label{DifferentialGeometry_3DFlatEuclideanMetric_CartesianToSpherical}
\delta_{ij} \dd x^i \dd x^j 
= \delta_{ab} \frac{\partial x^a}{\partial \xi^i} \frac{\partial x^b}{\partial \xi^j} \dd \xi^i \dd \xi^j 
= \frac{\partial \vec{x}}{\partial \xi^i} \cdot \frac{\partial \vec{x}}{\partial \xi^j} \dd \xi^i \dd \xi^j ,
\end{align}
where in the second equality we have, for convenience, expressed the contraction with the Kronecker delta as an ordinary (vector calculus) dot product. At this point, let us notice, if we call the coefficients of the quadratic form $g_{ij}$; for example, $\delta_{ij} \dd x^i \dd x^j \equiv g_{ij} \dd x^i \dd x^j$, we have
\begin{align}
g_{i'j'}(\vec{\xi}) 
= \frac{\partial \vec{x}}{\partial \xi^i} \cdot \frac{\partial \vec{x}}{\partial \xi^j} , 
\end{align}
where the primes on the indices are there to remind us this is not $g_{ij}(\vec{x}) = \delta_{ij}$, the components written in the Cartesian coordinates, but rather the ones written in spherical coordinates. In fact, what we are finding in eq. \eqref{DifferentialGeometry_3DFlatEuclideanMetric_CartesianToSpherical} is
\begin{align}
g_{i'j'}(\vec{\xi}) 
= g_{ab}(\vec{x}) \frac{\partial x^a}{\partial \xi^i} \frac{\partial x^b}{\partial \xi^j} .
\end{align}
Let's proceed to work out the above dot products out. Firstly,
\begin{align}
\frac{\partial\vec{x}}{\partial r}
&= \left(\sin\theta \cdot \cos\phi, \sin\theta \cdot \sin\phi, \cos\theta \right) , \\
\frac{\partial\vec{x}}{\partial \theta}
&= r \left(\cos\theta \cdot \cos\phi, \cos\theta \cdot \sin\phi, -\sin\theta \right) , \\
\frac{\partial\vec{x}}{\partial \phi}
&= r \left(-\sin\theta \cdot \sin\phi, \sin\theta \cdot \cos\phi, 0 \right) .
\end{align}
A direct calculation should return the results
\begin{align}
g_{r\theta} = g_{\theta r} = \frac{\partial\vec{x}}{\partial r} \cdot \frac{\partial\vec{x}}{\partial \theta} 	= 0, \qquad\qquad
g_{r\phi} = g_{\phi r} = \frac{\partial\vec{x}}{\partial r} \cdot \frac{\partial\vec{x}}{\partial \phi}	= 0, \qquad\qquad
g_{\theta\phi} 	= g_{\phi\theta} = \frac{\partial\vec{x}}{\partial \theta} \cdot \frac{\partial\vec{x}}{\partial \phi} = 0 ;
\end{align}
and
\begin{align}
g_{rr} 
&= \frac{\partial\vec{x}}{\partial r} \cdot \frac{\partial\vec{x}}{\partial r} 
\equiv \left(\frac{\partial\vec{x}}{\partial r} \cdot \frac{\partial\vec{x}}{\partial r}\right)^2 
= 1 , \\
g_{\theta\theta}
&= \left( \frac{\partial\vec{x}}{\partial\theta} \right)^2 = r^2 , \\
g_{\phi\phi}
&= \left( \frac{\partial\vec{x}}{\partial\phi} \right)^2 = r^2 \sin^2(\theta) .
\end{align}
Altogether, these yield eq. \eqref{DifferentialGeometry_3DFlatEuclideanMetric_Spherical}.

{\bf Tangent vectors} \qquad In Euclidean space, we may define vectors by drawing a directed straight line between one point to another. In curved space, the notion of a `straight line' is not straightforward, and as such we no longer try to implement such a definition of a vector. Instead, the notion of tangent vectors, and their higher rank tensor generalizations, now play central roles in curved spacetime geometry and physics. Imagine, for instance, a thin layer of water flowing over an undulating 2D surface -- an example of a tangent vector on a curved space is provided by the velocity of an infinitesimal volume within the flow.

More generally, let $\vec{x}(\lambda)$ denote the trajectory swept out by an infinitesimal volume of fluid as a function of (fictitious) time $\lambda$, transversing through a $(D \geq 2)-$dimensional space. (The $\vec{x}$ need not be Cartesian coordinates.) We may then define the tangent vector $v^i(\lambda) \equiv \dd\vec{x}(\lambda)/\dd \lambda$. Conversely, given a vector field $v^i(\vec{x})$ -- a $(D \geq 2)-$component object defined at every point in space -- we may find a trajectory $\vec{x}(\lambda)$ such that $\dd\vec{x}/\dd\lambda = v^i(\vec{x}(\lambda))$. (This amounts to integrating an ODE, and in this context is why $\vec{x}(\lambda)$ is called the {\it integral curve of $v^i$}.) In other words, tangent vectors do fit the mental picture that the name suggests, as `little arrows' based at each point in space, describing the local `velocity' of some (perhaps fictitious) flow.

You may readily check that tangent vectors at a given point $p$ in space do indeed form a vector space. However, we have written the components $v^i$ but did not explain what their basis vectors were. Geometrically speaking, $v$ tells us in what direction and how quickly to move away from the point $p$. This can be formalized by recognizing that the number of independent directions that one can move away from $p$ corresponds to the number of independent partial derivatives on some arbitrary (scalar) function defined on the curved space; namely $\partial_i f(\vec{x})$ for $i=1,2,\dots,D$, where $\{x^i\}$ are the coordinates used. Furthermore, the set of $\{\partial_i\}$ do span a vector space, based at $p$. We would thus say that any tangent vector $v$ is a superposition of partial derivatives:
\begin{align}
v 
= v^i(\vec{x}) \frac{\partial}{\partial x^i} 
\equiv v^i(x^1,x^2,\dots,x^D) \frac{\partial}{\partial x^i}
\equiv v^i \partial_i .
\end{align}
As already alluded to, given these components $\{v^i\}$, the vector $v$ can be thought of as the velocity with respect to some (fictitious) time $\lambda$ by solving the ordinary differential equation $v^i = \dd x^i(\lambda)/\dd \lambda$. We may now see this more explicitly; $v^i \partial_i f(\vec{x})$ is the time derivative of $f$ along the integral curve of $\vec{v}$ because
\begin{align}
v^i \partial_i f\left( \vec{x}(\lambda) \right) 
= \frac{\dd x^i}{\dd \lambda} \partial_i f(\vec{x}) 
= \frac{\dd f(\lambda)}{\dd \lambda} . 
\end{align}
To sum: the $\{\partial_i\}$ are the basis kets based at a given point $p$ in the curved space, allowing us to enumerate all the independent directions along which we may compute the `time derivative' of $f$ at the same point $p$.

{\bf General spatial metric} \qquad In a generic curved space, the square of the infinitesimal distance between the neighboring points $\vec{x}$ and $\vec{x} + \dd\vec{x}$, which we will continue to denote as $(\dd \ell)^2$, is no longer given by eq. \eqref{DifferentialGeometry_3DFlatEuclideanMetric_Cartesian} -- because we cannot expect Pythagoras' theorem to apply. But by scaling arguments it should still be quadratic in the infinitesimal distances $\{\dd x^i\}$. The most general of such expression is 
\begin{align}
\label{DifferentialGeometry_GenericSpatialMetric}
(\dd \ell)^2 = g_{ij}(\vec{x}) \dd x^i \dd x^j .
\end{align}
Since it measures distances, $g_{ij}$ needs to be real. It is also symmetric, since any antisymmetric portion would drop out of the summation in eq. \eqref{DifferentialGeometry_GenericSpatialMetric} anyway. (Why?) Finally, because we are discussing curved spaces for now, $g_{ij}$ needs to have strictly positive eigenvalues.

Additionally, given $g_{ij}$, we can proceed to define the inverse metric $g^{ij}$ in any coordinate system, as the matrix inverse of $g_{ij}$: 
\begin{align}
g^{ij} g_{jl} \equiv \delta^i_l .
\end{align}
Everything else in a differential geometric calculation follows from the curved metric in eq. \eqref{DifferentialGeometry_GenericSpatialMetric}, once it is specified for a given setup:\footnote{As with most physics texts on differential geometry, we will ignore torsion.} the ensuing Christoffel symbols, Riemann/Ricci tensors, covariant derivatives/curl/divergence; what defines straight lines; parallel transportation; etc. 

\noindent{\bf Distances} \qquad If you are given a path $\vec{x}(\lambda_1 \leq \lambda \leq \lambda_2)$ between the points $\vec{x}(\lambda_1) = \vec{x}_1$ and $\vec{x}(\lambda_2) = \vec{x}_2$, then the distance swept out by this path is given by the integral
\begin{align}
\label{DifferentialGeometry_LengthIntegral}
\ell 
= \int_{\vec{x}(\lambda_1 \leq \lambda \leq \lambda_2)} \sqrt{ g_{ij}\left( \vec{x}(\lambda) \right) \dd x^i \dd x^j}
= \int_{\lambda_1}^{\lambda_2} \dd\lambda \sqrt{ g_{ij}\left( \vec{x}(\lambda) \right) \frac{\dd x^i(\lambda)}{\dd\lambda} \frac{\dd x^j(\lambda)}{\dd \lambda} } .
\end{align}
\begin{myP}
	\qquad Show that this definition of distance is invariant under change of the parameter $\lambda$, as long as the transformation is orientation preserving. That is, suppose we replace $\lambda \to \lambda(\lambda')$ and thus $\dd\lambda = (\dd\lambda/\dd\lambda') \dd\lambda'$ -- then as long as $\dd \lambda/\dd \lambda' > 0$, we have
	\begin{align}
	\ell
	= \int_{\lambda'_1}^{\lambda'_2} \dd\lambda' 
	\sqrt{ g_{ij}\left( \vec{x}(\lambda') \right) \frac{\dd x^i(\lambda')}{\dd\lambda'} \frac{\dd x^j(\lambda')}{\dd \lambda'} } ,
	\end{align}
	where $\lambda(\lambda'_{1,2}) = \lambda_{1,2}$. % Consider
	%\begin{align}
	%\dd\lambda \sqrt{g_{ij} \frac{\dd x^i}{\dd \lambda} \frac{\dd x^j}{\dd \lambda}} 
	%= \dd\lambda' \left(\frac{\dd\lambda'(\lambda)}{\dd \lambda}\right)^{-1} \sqrt{g_{ij} \frac{\dd x^i}{\dd \lambda} \frac{\dd x^j}{\dd \lambda}} .
	%\end{align}
	Why can we always choose $\lambda$ such that
	\begin{align}
	\sqrt{ g_{ij}\left( \vec{x}(\lambda) \right) \frac{\dd x^i(\lambda)}{\dd\lambda} \frac{\dd x^j(\lambda)}{\dd \lambda} } = \text{constant},
	\end{align}
	i.e., the square root factor can be made constant along the entire path linking $\vec{x}_1$ to $\vec{x}_2$? Hint: Up to a re-scaling and a 1D translation, this amounts using the path length itself as the parameter $\lambda$.
\end{myP}
{\bf Kets and Bras} \qquad Earlier, while discussing tangent vectors, we stated that the $\{ \partial_i \}$ are the ket's, the basis tangent vectors at a given point in space. The infinitesimal distances $\{\dd x^i\}$ can now, in turn, be thought of as the basis dual vectors (the bra's) -- through the definition
\begin{align}
\label{DifferentialGeometry_Braket}
\braket{\dd x^i}{\partial_j} = \delta^i_j .
\end{align}
Why this is a useful perspective is due to the following. Let us consider an infinitesimal variation of our arbitrary function at $\vec{x}$:
\begin{align}
\label{DifferentialGeometry_df}
\dd f = \partial_i f(\vec{x}) \dd x^i .
\end{align}
Then, given a vector field $v$, we can employ eq. \eqref{DifferentialGeometry_Braket} to construct the derivative of the latter along the former, at some point $\vec{x}$, by
\begin{align}
\braket{\dd f}{v} = v^j \partial_i f(\vec{x}) \braket{\dd x^i}{\partial_j} = v^i \partial_i f(\vec{x}) .
\end{align}
What about the inner products $\braket{\dd x^i}{\dd x^j}$ and $\braket{\partial_i}{\partial_j}$? They are
\begin{align}
\braket{\dd x^i}{\dd x^j} = g^{ij} \qquad \text{ and } \qquad \braket{\partial_i}{\partial_j} = g_{ij} .
\end{align}
This is because
\begin{align}
g_{ij} \ket{\dd x^j} \equiv \ket{\partial_i} 
\qquad \Leftrightarrow \qquad 
g_{ij} \bra{\dd x^j} \equiv \bra{\partial_i} ;
\end{align}
or, equivalently,
\begin{align}
\ket{\dd x^j} \equiv g^{ij} \ket{\partial_i} 
\qquad \Leftrightarrow \qquad 
\bra{\dd x^j} \equiv g^{ij} \bra{\partial_i} .
\end{align}
In other words, 
\begin{quotation}
	At a given point in a curved space, one may define two different vector spaces -- one spanned by the basis tangent vectors $\{\ket{\partial_i}\}$ and another by its dual `bras' $\{ \ket{\dd x^i} \}$. Moreover, these two vector spaces are connected through the metric $g_{ij}$ and its inverse.
\end{quotation}
{\bf Parallel transport and Curvature} \qquad Roughly speaking, a curved space is one where the usual rules of Euclidean (flat) space no longer apply. For example, Pythagoras' theorem does not hold; and the sum of the angles of an extended triangle is not $\pi$.

The quantitative criteria to distinguish a curved space from a flat one, is to parallel transport a tangent vector $v^i(\vec{x})$ around a closed loop on a coordinate grid. If, upon bringing it back to the same location $\vec{x}$, the tangent vector is the same one we started with -- {\it for all possible coordinate loops} -- then the space is flat. Otherwise the space is curved. In particular, if you parallel transport a vector around an infinitesimal closed loop formed by a pair of `y-coordinate' and `z-coordinate' lines, starting from any one of its corners, and if the resulting vector is compared with original one, you would find that the difference is proportional to the Riemann curvature tensor $R^i_{\phantom{i}jkl}$. Specifically, suppose $v^i$ is parallel transported along a parallelogram, from $\vec{x}$ to $\vec{x}+\dd\vec{y}$; then to $\vec{x}+\dd\vec{y}+\dd\vec{z}$; then to $\vec{x}+\dd\vec{z}$; then back to $\vec{x}$. Then, denoting the end result as $v'^i$, we would find that
\begin{align}
v'^i - v^i \propto R^i_{\phantom{i}jkl} v^j \dd y^k \dd z^l .
\end{align}
Therefore, whether or not a geometry is locally curved is determined by this tensor. Of course, we have not defined what parallel transport actually is; to do so requires knowing the covariant derivative -- but let us first turn to a simple example where our intuition still holds.

{\it $2-$sphere as an example} \qquad A common textbook example of a curved space is that of a $2-$sphere of some fixed radius, sitting in 3D flat space, parametrized by the usual spherical coordinates $(0 \leq \theta \leq \pi,0 \leq \phi < 2\pi)$.\footnote{Any curved space can in fact always be viewed as a curved surface residing in a higher dimensional flat space.} Start at the north pole with the tangent vector $v = \partial_\theta$ pointing towards the equator with azimuthal direction $\phi = \phi_0$. Let us parallel transport $v$ along itself, i.e., with $\phi = \phi_0$ fixed, until we reach the equator itself. At this point, the vector is perpendicular to the equator, pointing towards the South pole. Next, we parallel transport $v$ along the equator from $\phi = \phi_0$ to some other longitude $\phi = \phi'_0$; here, $v$ is still perpendicular to the equator, and still pointing towards the South pole. Finally, we parallel transport it back to the North pole, along the $\phi = \phi'_0$ line. Back at the North pole, $v$ now points along the $\phi = \phi'_0$ longitude line and no longer along the original $\phi = \phi_0$ line. Therefore, $v$ does not return to itself after parallel transport around a closed loop: the $2-$sphere is a curved surface. This same example also provides us a triangle whose sum of its internal angles is $\pi + |\phi_0 - \phi'_0| > \pi$.\footnote{The $2-$sphere has positive curvature; whereas a saddle has negative curvature, and would support a triangle whose angles add up to less than $\pi$. In a very similar spirit, the Cosmic Microwave Background (CMB) sky contains hot and cold spots, whose angular size provide evidence that we reside in a spatially flat universe. See the Wilkinson Microwave Anisotropy Probe (WMAP) pages \href{https://map.gsfc.nasa.gov/mission/sgoals_parameters_geom.html}{here} and \href{https://map.gsfc.nasa.gov/media/030639/index.html}{here}.} Finally, notice in this 2-sphere example, the question of what a straight line means -- let alone using it to define a vector, as one might do in flat space -- does not produce a clear answer.

{\it Comparing tangent vectors at different places} \qquad That tangent vectors cannot, in general, be parallel transported in a curved space also tells us comparing tangent vectors based at different locations is not a straightforward procedure, especially compared to the situation in flat Euclidean space. This is because, if $\vec{v}(\vec{x})$ is to be compared to $\vec{w}(\vec{x}')$ by parallel transporting $\vec{v}(\vec{x})$ to $\vec{x}'$; different results will be obtained by simply choosing different paths to get from $\vec{x}$ to $\vec{x}'$.

{\bf Intrinsic vs extrinsic curvature} \qquad A 2D cylinder (embedded in 3D flat space) formed by rolling up a flat rectangular piece of paper has a surface that is {\it intrinsically} flat -- the Riemann tensor is zero everywhere because the intrinsic geometry of the surface is the same flat metric before the paper was rolled up. However, the paper as viewed by an ambient 3D observer does have an {\it extrinsic} curvature due to its cylindrical shape. To characterize extrinsic curvature mathematically, one would erect a vector perpendicular to the surface in question and parallel transport it along this same surface: the latter is flat if the vector remains parallel; otherwise it is curved. In curved spacetimes, when this vector refers to the flow of time and is perpendicular to some spatial surface, the extrinsic curvature also describes its time evolution.

\subsection{Locally Flat Coordinates \& Symmetries, Infinitesimal Volumes, General Tensors, Orthonormal Basis}

\noindent{\bf Locally flat coordinates\footnote{Also known as Riemann normal coordinates.} and symmetries} \qquad It is a mathematical fact that, given some fixed point $y^i_0$ on the curved space, one can find coordinates $y^i$ such that locally the metric does become flat: 
\begin{align}
\label{DifferentialGeometry_EP}
\lim_{\vec{y} \to \vec{y}_0} g_{ij}(\vec{y}) 
= \delta_{ij} + g_2 \cdot R_{ikjl}(\vec{y}_0) \ (y-y_0)^k (y-y_0)^l + \dots ,
\end{align}
with a similar result for curved spacetimes. In this ``locally flat" coordinate system, the first corrections to the flat Euclidean metric is quadratic in the displacement vector $\vec{y}-\vec{y}_0$, and $R_{ik jl}(\vec{y}_0)$ is the Riemann tensor -- which is the chief measure of curvature -- evaluated at $\vec{y}_0$. (The $g_2$ is just a numerical constant, whose precise value is not important for our discussion.) In a curved space{\it time}, that geometry can always be viewed as locally flat is why the mathematics you are encountering here is the appropriate framework for reconciling gravity as a force, Einstein's equivalence principle, and the Lorentz symmetry of Special Relativity.

Note that under spatial rotations $\{ \widehat{R}^i_{\phantom{i}j} \}$, which obeys $\widehat{R}^a_{\phantom{a}i} \widehat{R}^b_{\phantom{b}j} \delta_{ab} = \delta_{ij}$, if we define in Euclidean space the following change-of-Cartesian coordinates (from $\vec{x}$ to $\vec{x}'$)
\begin{align}
x^i \equiv \widehat{R}^i_{\phantom{i}j} x'^j ;
\end{align}
the flat metric would retain the same form
\begin{align}
\delta_{ij} \dd x^i \dd x^j 
= \delta_{ab} \widehat{R}^a_{\phantom{a}i} \widehat{R}^b_{\phantom{b}j} \dd x'^i \dd x'^j 
= \delta_{ij} \dd x'^i \dd x'^j .
\end{align}
A similar calculation would tell us flat Euclidean space is invariant under parity flips, i.e., $x'^k \equiv -x^k$ for some fixed $k$, as well as spatial translations $\vec{x}' \equiv \vec{x} + \vec{a}$, for constant $\vec{a}$. To sum:
\begin{quotation}
	At a given point in a curved space, it is always possible to find a coordinate system -- i.e., a geometric viewpoint/`frame' -- such that the space is flat up to distances of $\mathcal{O}(1/|\max R_{ijlk}(\vec{y}_0)|^{1/2})$, and hence `locally' invariant under rotations, translations, and reflections.
\end{quotation}
This is why it took a while before humanity came to recognize we live on the curved surface of the (approximately spherical) Earth: locally, the Earth's surface looks flat!

\noindent{\bf Coordinate-transforming the metric} \qquad Note that, in the context of eq. \eqref{DifferentialGeometry_GenericSpatialMetric}, $\vec{x}$ is not a vector in Euclidean space, but rather another way of denoting $x^a$ without introducing too many dummy indices $\{a,b,\dots,i,j,\dots\}$. Also, $x^i$ in eq. \eqref{DifferentialGeometry_GenericSpatialMetric} are not necessary Cartesian coordinates, but can be completely arbitrary. The metric $g_{ij}(\vec{x})$ can viewed as a $3 \times 3$ (or $D \times D$, in $D$ dimensions) matrix of functions of $\vec{x}$, telling us how the notion of distance vary as one moves about in the space. Just as we were able to translate from Cartesian coordinates to spherical ones in Euclidean 3-space, in this generic curved space, we can change from $\vec{x}$ to $\vec{\xi}$, i.e., one arbitrary coordinate system to another, so that
\begin{align}
g_{ij}\left( \vec{x} \right) \dd x^i \dd x^j 
= g_{ij}\left( \vec{x}(\vec{\xi}) \right)
	\frac{\partial x^i(\vec{\xi})}{\partial \xi^a} \frac{\partial x^j(\vec{\xi})}{\partial \xi^b} \dd \xi^a \dd \xi^b 
\equiv g_{ab}( \vec{\xi} ) \dd \xi^a \dd \xi^b  .
\end{align}
We can attribute all the coordinate transformation to how it affects the components of the metric:
\begin{align}
\label{DifferentialGeometry_GenericSpatialMetric_CoordinateTransformation}
g_{ab}( \vec{\xi} ) 
= g_{ij}\left( \vec{x}(\vec{\xi}) \right)
\frac{\partial x^i(\vec{\xi})}{\partial \xi^a} \frac{\partial x^j(\vec{\xi})}{\partial \xi^b}  .
\end{align}
The left hand side are the metric components in $\vec{\xi}$ coordinates. The right hand side consists of the Jacobians $\partial x/\partial \xi$ contracted with the metric components in $\vec{x}$ coordinates -- but now with the $\vec{x}$ replaced with $\vec{x}(\vec{\xi})$, their corresponding expressions in terms of $\vec{\xi}$.

\noindent{\bf Inverse metric} \qquad Previously, we defined $g^{ij}$ to be the matrix inverse of the metric tensor $g_{ij}$. We can also view $g^{ij}$ as components of the tensor
\begin{align}
g^{ij}(\vec{x}) \partial_i \otimes \partial_j ,
\end{align}
where we have now used $\otimes$ to indicate we are taking the tensor product of the partial derivatives $\partial_i$ and $\partial_j$. In $g_{ij}\left( \vec{x} \right) \dd x^i \dd x^j$ we really should also have $\dd x^i \otimes \dd x^j$, but I prefer to stick with the more intuitive idea that the metric (with lower indices) is the sum of squares of distances. Just as we know how $\dd x^i$ transforms under $\vec{x} \to \vec{x}(\vec{\xi})$, we also can work out how the partial derivatives transform.
\begin{align}
g^{ij}(\vec{x}) \frac{\partial}{\partial x^i} \otimes \frac{\partial}{\partial x^j} 
= g^{ab}\left( \vec{x}(\vec{\xi}) \right) \frac{\partial \xi^i}{\partial x^a} \frac{\partial \xi^j}{\partial x^b} 
\frac{\partial}{\partial \xi^i} \otimes \frac{\partial}{\partial \xi^j}
\end{align}
In terms of its components, we can read off their transformation rules:
\begin{align}
g^{ij}(\vec{\xi})  
= g^{ab}\left( \vec{x}(\vec{\xi}) \right) \frac{\partial \xi^i}{\partial x^a} \frac{\partial \xi^j}{\partial x^b} .
\end{align}
The left hand side is the inverse metric written in the $\vec{\xi}$ coordinate system, whereas the right hand side involves the inverse metric written in the $\vec{x}$ coordinate system -- contracted with two Jacobian's $\partial \xi/\partial x$ -- except all the $\vec{x}$ are replaced with the expressions $\vec{x}(\vec{\xi})$ in terms of $\vec{\xi}$. 

A technical point: here and below, the Jacobian $\partial x^a(\vec{\xi})/\partial \xi^j$ can be calculated in terms of $\vec{\xi}$ by direct differentiation if we have defined $\vec{x}$ in terms of $\vec{\xi}$, namely $\vec{x}(\vec{\xi})$. But the Jacobian $(\partial \xi^i/\partial x^a)$ in terms of $\vec{\xi}$ requires a matrix inversion. For, by the chain rule,
\begin{align}
\label{DifferentialGeometry_Jacobians}
\frac{\partial x^i}{\partial \xi^l} \frac{\partial \xi^l}{\partial x^j} = \frac{\partial x^i}{\partial x^j}
= \delta^i_j, \qquad\text{ and }\qquad
\frac{\partial \xi^i}{\partial x^l} \frac{\partial x^l}{\partial \xi^j} = \frac{\partial \xi^i}{\partial \xi^j}
= \delta^i_j .
\end{align}
In other words, given $\vec{x} \to \vec{x}(\vec{\xi})$, we can compute $\mathcal{J}^a_{\phantom{a}i} \equiv \partial x^a/\partial \xi^i$ in terms of $\vec{\xi}$, with $a$ being the row number and $i$ as the column number. Then find the inverse, i.e., $(\mathcal{J}^{-1})^a_{\phantom{a}i}$ and identify it with $\partial \xi^a/\partial x^i$ in terms of $\vec{\xi}$.

{\bf General tensor} \qquad A {\it scalar} $\varphi$ is an object with no indices that transforms as
\begin{align}
\label{DifferentialGeometry_ScalarDef}
\varphi( \vec{\xi} ) = \varphi\left(\vec{x}(\vec{\xi})\right) .
\end{align}
That is, take $\varphi(\vec{x})$ and simply replace $\vec{x} \to \vec{x}(\vec{\xi})$ to obtain $\varphi(\vec{\xi})$.

A {\it vector} $v^i(\vec{x}) \partial_i$ transforms as, by the chain rule,
\begin{align}
\label{DifferentialGeometry_VectorDef}
v^i(\vec{x}) \frac{\partial}{\partial x^i} 
= v^i(\vec{x}(\vec{\xi})) \frac{\partial \xi^j}{\partial x^i} \frac{\partial}{\partial \xi^j} 
\equiv v^j(\vec{\xi}) \frac{\partial}{\partial \xi^j}
\end{align}
If we attribute all the transformations to the components, the components in the $\vec{x}$-coordinate system $v^i(\vec{x})$ is related to those in the $\vec{y}$-coordinate system $v^i(\vec{\xi})$ through the relation
\begin{align}
v^i(\vec{\xi}) = v^i(\vec{x}(\vec{\xi})) \frac{\partial \xi^j}{\partial x^i} .
\end{align}
Similarly, a {\it 1-form} $A_i \dd x^i$ transforms, by the chain rule,
\begin{align}
\label{DifferentialGeometry_1FormDef}
A_i(\vec{x}) \dd x^i
= A_i(\vec{x}(\vec{\xi})) \frac{\partial x^i}{\partial \xi^j} \dd \xi^j 
\equiv A_j(\vec{\xi}) \dd \xi^j.
\end{align}
If we again attribute all the coordinate transformations to the components; the ones in the $\vec{x}$-system $A_i(\vec{x})$ is related to the ones in the $\vec{\xi}$-system $A_i(\vec{\xi})$ through
\begin{align}
A_j(\vec{\xi}) = A_i(\vec{x}(\vec{\xi})) \frac{\partial x^i}{\partial \xi^j} .
\end{align}
By taking tensor products of $\{\ket{\partial_i}\}$ and $\{\bra{\dd x^i}\}$, we may define a {\it rank $\binom{N}{M}$ tensor} $T$ as an object with $N$ ``upper indices" and $M$ ``lower indices" that transforms as
\begin{align}
\label{DifferentialGeometry_GenericTensor_CoordinateTransformation}
T^{i_1 i_2 \dots i_N}_{\phantom{i_1 i_2 \dots i_N} j_i j_2 \dots j_M}(\vec{\xi})
= T^{a_1 a_2 \dots a_N}_{\phantom{a_1 a_2 \dots a_N} b_i b_2 \dots b_M}\left( \vec{x}(\vec{\xi}) \right)
\frac{\partial \xi^{i_1}}{\partial x^{a_1}} \dots \frac{\partial \xi^{i_N}}{\partial x^{a_N}} 
\frac{\partial x^{b_1}}{\partial \xi^{j_1}} \dots \frac{\partial x^{b_M}}{\partial \xi^{j_M}} .
\end{align}
The left hand side are the tensor components in $\vec{\xi}$ coordinates and the right hand side are the Jacobians $\partial x/\partial \xi$ and $\partial \xi/\partial x$ contracted with the tensor components in $\vec{x}$ coordinates -- but now with the $\vec{x}$ replaced with $\vec{x}(\vec{\xi})$, their corresponding expressions in terms of $\vec{\xi}$. This multi-indexed object should be viewed as the components of 
\begin{align}
\label{DifferentialGeometry_GenericTensor}
T^{i_1 i_2 \dots i_N}_{\phantom{i_1 i_2 \dots i_N} j_i j_2 \dots j_M}(\vec{x})
\ket{\frac{\partial}{\partial x^{i_1}}} \otimes \dots \otimes \ket{\frac{\partial}{\partial x^{i_N}}} \otimes
\bra{\dd x^{j_1}} \otimes \dots \otimes \bra{\dd x^{j_M}} .
\end{align}
\footnote{Strictly speaking, when discussing the metric and its inverse above, we should also have respectively expressed them as $g_{ij} \bra{\dd x^i} \otimes \bra{\dd x^j}$ and $g^{ij} \ket{\partial_i} \otimes \ket{\partial_j}$, with the appropriate bras and kets enveloping the $\{\dd x^i\}$ and $\{\partial_i\}$. We did not do so because we wanted to highlight the geometric interpretation of $g_{ij} \dd x^i \dd x^j$ as the square of the distance between $\vec{x}$ and $\vec{x} + \dd \vec{x}$, where the notion of $\dd x^i$ as (a component of) an infinitesimal `vector' -- as opposed to being a 1-form -- is, in our opinion, more useful for building the reader's geometric intuition. \newline\indent It may help the physicist reader to think of a scalar field in eq. \eqref{DifferentialGeometry_ScalarDef} as an observable, such as the temperature $T(\vec{x})$ of the 2D undulating surface mentioned above. If you were provided such an expression for $T(\vec{x})$, together with an accompanying definition for the coordinate system $\vec{x}$; then, to convert this same temperature field to a different coordinate system (say, $\vec{\xi}$) one would, in fact, do $T(\vec{\xi}) \equiv T(\vec{x}(\vec{\xi}))$, because you'd want $\vec{\xi}$ to refer to the same point in space as $\vec{x} = \vec{x}(\vec{\xi})$. For a general tensor in eq. \eqref{DifferentialGeometry_GenericTensor}, the tensor components $T^{i_1 i_2 \dots i_N}_{\phantom{i_1 i_2 \dots i_N} j_i j_2 \dots j_M}$ may then be regarding as scalars describing some weighted superposition of the tensor product of basis vectors and 1-forms. Its transformation rules in eq. \eqref{DifferentialGeometry_GenericTensor_CoordinateTransformation} are really a shorthand for the lazy physicist who does not want to carry the basis vectors/1-forms around in his/her calculations.}Above, we only considered $T$ with all upper indices followed by all lower indices. Suppose we had $T^{i\phantom{j}k}_{\phantom{i}j}$; it is the components of
\begin{align}
T^{i\phantom{j}k}_{\phantom{i}j}(\vec{x}) \ket{\partial_i} \otimes \bra{\dd x^j} \otimes \ket{\partial_k} .
\end{align}
{\bf Raising and lowering tensor indices} \qquad The indices on a tensor are moved -- from upper to lower, or vice versa -- using the metric tensor. For example,
\begin{align}
T^{m_1 \dots m_a \phantom{i} n_1 \dots n_b}_{\phantom{m_1 \dots m_a}i} 
&= g_{ij} T^{m_1 \dots m_a j n_1 \dots n_b} , \\
T_{m_1 \dots m_a \phantom{i} n_1 \dots n_b}^{\phantom{m_1 \dots m_a}i} 
&= g^{ij} T_{m_1 \dots m_a j n_1 \dots n_b} .
\end{align}
Because upper indices transform oppositely from lower indices -- see eq. \eqref{DifferentialGeometry_Jacobians} -- when we contract a upper and lower index, it now transforms as a scalar. For example,
\begin{align}
	A^i_{\phantom{i}l}(\vec{\xi}) B^{lj}(\vec{\xi})
	&= \frac{\partial \xi^i}{\partial x^m} A^m_{\phantom{m}a}\left( \vec{x}(\vec{\xi}) \right) \frac{\partial x^a}{\partial \xi^l}
	\frac{\partial \xi^l}{\partial x^c} B^{cn}\left( \vec{x}(\vec{\xi}) \right) \frac{\partial \xi^j}{\partial x^n} \nonumber\\
	&= \frac{\partial \xi^i}{\partial x^m} \frac{\partial \xi^j}{\partial x^n}
	A^m_{\phantom{m}c}\left( \vec{x}(\vec{\xi}) \right) B^{cn}\left( \vec{x}(\vec{\xi}) \right) .
\end{align}
{\bf General covariance} \qquad Tensors are ubiquitous in physics: the electric and magnetic fields can be packaged into one Faraday tensor $F_{\mu\nu}$; the energy-momentum-shear-stress tensor of matter $T_{\mu\nu}$ is what sources the curved geometry of spacetime in Einstein's theory of General Relativity; etc. The coordinate transformation rules in eq. \eqref{DifferentialGeometry_GenericTensor_CoordinateTransformation} that defines a tensor is actually the statement that, the mathematical description of the physical world (the tensors themselves in eq. \eqref{DifferentialGeometry_GenericTensor}) should not depend on the coordinate system employed. Any expression or equation with physical meaning -- i.e., it yields quantities that can in principle be measured -- must be put in a form that is generally covariant: either a scalar or tensor under coordinate transformations.\footnote{You may also demand your equations/quantities to be tensors/scalars under group transformations.} An example is, it makes no sense to assert that your new-found law of physics depends on $g^{11}$, the $11$ component of the inverse metric -- for, in what coordinate system is this law expressed in? What happens when we use a different coordinate system to describe the outcome of some experiment designed to test this law?

Another aspect of general covariance is that, although tensor equations should hold in any coordinate system -- if you suspect that two tensors quantities are actually equal, say
\begin{align}
S^{i_1 i_2 \dots } = T^{i_1 i_2 \dots } ,
\end{align}
it suffices to find one coordinate system to prove this equality. It is not necessary to prove this by using abstract indices/coordinates because, as long as the coordinate transformations are invertible, then once we have verified the equality in one system, the proof in any other follows immediately once the required transformations are specified. One common application of this observation is to apply the fact mentioned around eq. \eqref{DifferentialGeometry_EP}, that at any given point in a curved space(time), one can always choose coordinates where the metric there is flat. You will often find this ``locally flat" coordinate system simplifies calculations -- and perhaps even aids in gaining some intuition about the relevant physics, since the expressions usually reduce to their more familiar counterparts in flat space. A simple but important example of this brings us to the next concept: what is the curved analog of the infinitesimal volume, which we would usually write as $\dd^D x$ in Cartesian coordinates?

{\bf Determinant of metric and the infinitesimal volume} \qquad The determinant of the metric transforms as
\begin{align}
\det g_{ij}(\vec{\xi}) = \det \left[ g_{ab}\left( \vec{x}(\vec{\xi}) \right) \frac{\partial x^a}{\partial \xi^i} \frac{\partial x^b}{\partial \xi^j} \right] .
\end{align}
Using the properties $\det A \cdot B = \det A \det B$ and $\det A^T = \det A$, for any two square matrices $A$ and $B$,
\begin{align}
\det g_{ij}(\vec{\xi}) = \left(\det \frac{\partial x^a(\vec{\xi})}{\partial \xi^b}\right)^2 \det g_{ij}\left( \vec{x}(\vec{\xi}) \right) .
\end{align}
The square root of the determinant of the metric is often denoted as $\sqrt{|g|}$. It transforms as
\begin{align}
\label{DifferentialGeometry_SqureRootDetg_CoordinateTransformation}
\sqrt{\left| g(\vec{\xi}) \right|} = \sqrt{\left\vert g\left( \vec{x}(\vec{\xi}) \right) \right\vert} 
\left\vert \det \frac{\partial x^a(\vec{\xi})}{\partial \xi^b} \right\vert .
\end{align}
We have previously noted that, given any point $\vec{x}_0$ in the curved space, we can always choose local coordinates $\{\vec{x}\}$ such that the metric there is flat. This means at $\vec{x}_0$ the infinitesimal volume of space is $\dd^D \vec{x}$ and $\det g_{ij}(\vec{x}_0) = 1$. Recall from multi-variable calculus that, whenever we transform $\vec{x} \to \vec{x}(\vec{\xi})$, the integration measure would correspondingly transform as
\begin{align}
\dd^D \vec{x} = \dd^D \vec{\xi} \left\vert\det \frac{\partial x^i}{\partial \xi^a} \right\vert,
\end{align}
where $\partial x^i/\partial \xi^a$ is the Jacobian matrix with row number $i$ and column number $a$. Comparing this multi-variable calculus result to eq. \eqref{DifferentialGeometry_SqureRootDetg_CoordinateTransformation} specialized to our metric in terms of $\{\vec{x}\}$ but evaluated at $\vec{x}_0$, we see the determinant of the Jacobian {\it is} in fact the square root of the determinant of the metric in some other coordinates $\vec{\xi}$,
\begin{align}
\sqrt{\left| g(\vec{\xi}) \right|} 
= \left( \sqrt{\left\vert g\left( \vec{x}(\vec{\xi})\right) \right\vert} 
\left\vert \det \frac{\partial x^i(\vec{\xi})}{\partial \xi^a} \right\vert \right)_{\vec{x}=\vec{x}_0} 
= \left\vert \det \frac{\partial x^i(\vec{\xi})}{\partial \xi^a} \right\vert_{\vec{x}=\vec{x}_0} .
\end{align}
In flat space and by employing Cartesian coordinates $\{\vec{x}\}$, the infinitesimal volume (at some location $\vec{x}=\vec{x}_0$) is $\dd^D \vec{x}$. What is its curved analog? What we have just shown is that, by going from $\vec{\xi}$ to a locally flat coordinate system $\{\vec{x}\}$,
\begin{align}
\label{DifferentialGeometry_LocallyFlatCartesianToCurvilinear}
\dd^D \vec{x} 
= \dd^D \vec{\xi} \left\vert \det \frac{\partial x^i(\vec{\xi})}{\partial \xi^a} \right\vert_{\vec{x}=\vec{x}_0}
= \dd^D \vec{\xi} \sqrt{|g(\vec{\xi})|} .
\end{align}
However, since $\vec{x}_0$ was an arbitrary point in our curved space, we have argued that, in a general coordinate system $\vec{\xi}$, the infinitesimal volume is given by
\begin{align}
\label{DifferentialGeometry_InfinitesimalVolume}
\dd^D \vec{\xi} \sqrt{\left| g(\vec{\xi}) \right|} \equiv \dd\xi^1 \dots \dd\xi^D \sqrt{\left| g(\vec{\xi}) \right|} .
\end{align}
\begin{myP}
\qquad Upon an orientation preserving change of coordinates $\vec{y} \to \vec{y}(\vec{\xi})$, where $\det \partial y/\partial \xi > 0$, show that
\begin{align}
\dd^D \vec{y} \sqrt{\left| g(\vec{y}) \right|} = \dd^D \vec{\xi} \sqrt{\left| g(\vec{\xi}) \right|} .
\end{align}
Therefore calling $\dd^D \vec{x} \sqrt{|g(\vec{x})|}$ an infinitesimal volume is a generally covariant statement.

Note: $g(\vec{y})$ is the determinant of the metric written in the $\vec{y}$ coordinate system; whereas $g(\vec{\xi})$ is that of the metric written in the $\vec{\xi}$ coordinate system. The latter is {\it not} the same as the determinant of the metric written in the $\vec{y}$-coordinates, with $\vec{y}$ replaced with $\vec{y}(\vec{\xi})$; i.e., be careful that the determinant is not a scalar. \qed
\end{myP}
{\bf Volume integrals} \qquad If $\varphi(\vec{x})$ is some scalar quantity, finding its volume integral within some domain $\mathfrak{D}$ in a generally covariant way can be now carried out using the infinitesimal volume we have uncovered; it reads
\begin{align}
I \equiv \int_{\mathfrak{D}} \dd^D \vec{x}\sqrt{|g(\vec{x})|} \varphi(\vec{x}) .
\end{align}
In other words, $I$ is the same result no matter what coordinates we used to compute the integral on the right hand side.
\begin{myP}
{\it Spherical coordinates in $D$ space dimensions.} \qquad In $D$ space dimensions, we may denote the $D$-th unit vector as $\widehat{e}_D$; and $\widehat{n}_{D-1}$ as the unit radial vector, parametrized by the angles $\{0 \leq \theta^1 < 2\pi,0 \leq \theta^2 \leq \pi,\dots,0 \leq \theta^{D-2} \leq \pi\}$, in the plane perpendicular to $\widehat{e}_D$. Let $r \equiv |\vec{x}|$ and $\widehat{n}_D$ be the unit radial vector in the $D$ space. Any vector $\vec{x}$ in this space can thus be expressed as
\begin{align}
\vec{x} = r \widehat{n}\left( \vec{\theta} \right) 
= r \cos(\theta^{D-1}) \widehat{e}_D + r \sin(\theta^{D-1}) \widehat{n}_{D-1}, \qquad 0 \leq \theta^{D-1} \leq \pi .
\end{align}
(Can you see why this is nothing but the Gram-Schmidt process?) Just like in the 3D case, $r \cos(\theta^{D-1})$ is the projection of $\vec{x}$ along the $\widehat{e}_D$ direction; while $r \sin(\theta^{D-1})$ is that along the radial direction in the plane perpendicular to $\widehat{e}_D$.

First show that the Cartesian metric $\delta_{ij}$ in $D$-space transforms to
\begin{align}
(\dd\ell)^2 
= \dd r^2 + r^2 \dd\Omega_D^2
= \dd r^2 + r^2 \left( (\dd\theta^{D-1})^2 + (\sin\theta^{D-1})^2 \dd\Omega_{D-1}^2 \right),
\end{align}
where $\dd\Omega_N^2$ is the square of the infinitesimal solid angle in $N$ spatial dimensions, and is given by
\begin{align}
\dd\Omega_N^2 
\equiv \sum_{\text{I},\text{J}=1}^{N-1} \Omega^{(N)}_{\text{I}\text{J}} \dd\theta^\text{I} \dd\theta^\text{J} , \qquad\qquad
\Omega^{(N)}_{\text{IJ}} 
\equiv \sum_{i,j=1}^{N} \delta_{ij} \frac{\partial \widehat{n}^i_N}{\partial \theta^\text{I}} \frac{\partial \widehat{n}^j_N}{\partial \theta^\text{J}}  .
\end{align}
Proceed to argue that the full $D$-metric in spherical coordinates is
\begin{align}
\dd\ell^2 
&= \dd r^2 + r^2 \left((\dd\theta^{D-1})^2 
+ \sum_{\text{I}=2}^{D-1} s_{D-1}^2 \dots s_{D-\text{I}+1}^2 (\dd\theta^{D-\text{I}})^2\right) , \\
\theta^1 &\in [0,2\pi), \qquad\qquad \theta^{2}, \dots, \theta^{D-1} \in [0,\pi] .
\end{align}
(Here, $s_\text{I} \equiv \sin\theta^\text{I}$.) Show that the determinant of the angular metric $\Omega^{(N)}_{\text{IJ}}$ obeys a recursion relation
\begin{align}
\det \Omega^{(N)}_{\text{IJ}} = \left(\sin\theta^{N-1}\right)^{2(N-2)} \cdot \det \Omega^{(N-1)}_{\text{IJ}} .
\end{align}
Explain why this implies there is a recursion relation between the infinitesimal solid angle in $D$ space and that in $(D-1)$ space. Moreover, show that the integration volume measure $\dd^D \vec{x}$ in Cartesian coordinates then becomes, in spherical coordinates,
\begin{align}
\dd^D \vec{x} = \dd r \cdot r^{D-1} \cdot \dd \theta^1 \dots \dd\theta^{D-1} \left(\sin\theta^{D-1}\right)^{D-2} \sqrt{\det \Omega^{(D-1)}_{\text{IJ}}} . 
\end{align} \qed
\end{myP}
\begin{myP}
\qquad Let $x^i$ be Cartesian coordinates and 
\begin{align}
\xi^i \equiv (r,\theta,\phi)
\end{align}
be the usual spherical coordinates; see eq. \eqref{CartesianToSpherical}. Calculate $\partial \xi^i/\partial x^a$ in terms of $\vec{\xi}$ and thereby, from the flat metric $\delta^{ij}$ in Cartesian coordinates, find the inverse metric $g^{ij}(\vec{\xi})$ in the spherical coordinate system. \qed
\end{myP}
{\bf Symmetries (aka isometries) and infinitesimal displacements} \qquad In some Cartesian coordinates $\{ x^i \}$ the flat space metric is $\delta_{ij} \dd x^i \dd x^j$. Suppose we chose a different set of axes for new Cartesian coordinates $\{ x'^i \}$, the metric will still take the same form, namely $\delta_{ij} \dd x'^i \dd x'^j$. Likewise, on a $2$-sphere the metric is $\dd\theta^2 + (\sin\theta)^2 \dd\phi^2$ with a given choice of axes for the 3D space the sphere is embedded in; upon any rotation to a new axis, so the new angles are now $(\theta',\phi')$, the $2$-sphere metric is still of the same form $\dd\theta'^2 + (\sin\theta')^2 \dd\phi'^2$. All we have to do, in both cases, is swap the symbols $\vec{x} \to \vec{x}'$ and $(\theta,\phi) \to (\theta',\phi')$. The reason why we can simply swap symbols to express the same geometry in different coordinate systems, is because of the symmetries present: for flat space and the $2$-sphere, the geometries are respectively indistinguishable under translation/rotation and rotation about its center.

Motivated by this observation that geometries enjoying symmetries (aka isometries) retain their {\it form} under an active coordinate transformation -- one that corresponds to an actual displacement from one location to another\footnote{As opposed to a passive coordinate transformation, which is one where a different set of coordinates are used to describe the same location in the geometry.} -- we now consider a infinitesimal coordinate transformation as follows. Starting from $\vec{x}$, we define a new set of coordinates $\vec{x}'$ through an infinitesimal vector $\vec{\xi}(\vec{x})$,
\begin{align}
\vec{x}' \equiv \vec{x} - \vec{\xi}(\vec{x}) .
\end{align}
(The $-$ sign is for technical convenience.) One interpretation of this definition is that of an active coordinate transformation -- given some location $\vec{x}$, we now move to a point $\vec{x}'$ that is displaced infinitesimally far away, with the displacement itself described by $-\vec{\xi}(\vec{x})$. On the other hand, since $\vec{\xi}$ is assumed to be ``small," we may replace in the above equation, $\vec{\xi}(\vec{x})$ with $\vec{\xi}(\vec{x}') \equiv \vec{\xi}(\vec{x} \to \vec{x}')$. This is because the error incurred would be of $\mathcal{O}(\xi^2)$.
\begin{align}
\vec{x} = \vec{x}'+\vec{\xi}(\vec{x}') + \mathcal{O}(\xi^2) \qquad \Rightarrow \qquad
\frac{\partial x^i}{\partial x'^a} = \delta^i_a + \partial_{a'} \xi^i(\vec{x}') + \mathcal{O}(\xi \partial \xi)
\end{align}
How does this change our metric?
{\allowdisplaybreaks\begin{align}
g_{ij}\left(\vec{x}\right)\dd x^i \dd x^j
&= g_{ij}\left(\vec{x}'+\vec{\xi}(\vec{x}')+\dots\right)\left( \delta^i_a + \partial_{a'} \xi^i + \dots \right)\left( \delta^j_b + \partial_{b'} \xi^j + \dots\right)\dd x'^a \dd x'^b\nonumber\\
&= \left(g_{ij}\left(\vec{x}'\right) + \xi^c \partial_{c'} g_{ij}(\vec{x}') + \dots\right) \left( \delta^i_a + \partial_{a'} \xi^i + \dots\right)\left( \delta^j_b + \partial_{b'} \xi^j + \dots\right) \dd x'^a \dd x'^b \nonumber\\
&= \left( g_{ij}(\vec{x}') + \delta_\xi g_{ij}(\vec{x}') + \mathcal{O}(\xi^2) \right) \dd x'^i \dd x'^j ,
\end{align}}
where
\begin{align}
\label{DifferentialGeometry_LieDerivativeOfMetric}
\delta_\xi g_{ij}(\vec{x}') \equiv \xi^c(\vec{x}') \frac{\partial g_{ij}(\vec{x}')}{\partial x'^c} + g_{ia}(\vec{x}') \frac{\partial \xi^a(\vec{x}')}{\partial x'^j} + g_{ja}(\vec{x}') \frac{\partial \xi^a(\vec{x}')}{\partial x'^i} .
\end{align}
At this point, we see that if the geometry enjoys a symmetry along the entire curve whose tangent vector is $\vec{\xi}$, then it must retain its form $g_{ij}(\vec{x}) \dd x^i \dd x^j = g_{ij}(\vec{x}') \dd x'^i \dd x'^j$ and therefore,\footnote{We reiterate, by the same {\it form}, we mean $g_{ij}(\vec{x})$ and $g_{ij}(\vec{x}')$ are the same functions if we treat $\vec{x}$ and $\vec{x}'$ as dummy variables. For example, $g_{33}(r,\theta) = (r\sin\theta)^2$ and $g_{3'3'}(r',\theta') = (r'\sin\theta')^2$ in the $2$-sphere metric.}
\begin{align}
\label{DifferentialGeometry_KillingEqn_v1}
\delta_\xi g_{ij} = 0 , \qquad \text{(isometry along $\vec{\xi}$)} .
\end{align}
\footnote{$\delta_\xi g_{ij}$ is known as the Lie derivative of the metric along $\xi$, and is commonly denoted as $(\pounds_\xi g)_{ij}$.}Conversely, if $\delta_\xi g_{ij} = 0$ everywhere in space, then starting from some point $\vec{x}$, we can make incremental displacements along the curve whose tangent vector is $\vec{\xi}$, and therefore find that the metric retain its form along its entirety. Now, a vector $\vec{\xi}$ that satisfies $\delta_\xi g_{ij} = 0$ is called a Killing vector. We may then summarize: \begin{quotation}
A geometry enjoys an isometry along $\vec{\xi}$ if and only if $\vec{\xi}$ is a Killing vector satisfying eq. \eqref{DifferentialGeometry_KillingEqn_v1} everywhere in space.
\end{quotation}
\begin{myP}
\qquad Can you justify the statement: ``If the metric $g_{ij}$ is independent of one of the coordinates, say $x^k$, then $\partial_k$ is a Killing vector of the geometry"? \qed
\end{myP}
{\bf Orthonormal frame} \qquad So far, we have been writing tensors in the coordinate basis -- the basis vectors of our tensors are formed out of tensor products of $\{ \dd x^i \}$ and $\{ \partial_i \}$. To interpret components of tensors, however, we need them written in an orthonormal basis. This amounts to using a uniform set of measuring sticks on all axes, i.e., a local set of (non-coordinate) Cartesian axes where one ``tick mark" on each axis translates to the same length.

As an example, suppose we wish to describe some fluid's velocity $v^x \partial_x + v^y \partial_y$ on a 2 dimensional flat space. In Cartesian coordinates $v^x(x,y)$ and $v^y(x,y)$ describe the velocity at some point $\vec{\xi} = (x,y)$ flowing in the $x$- and $y$-directions respectively. Suppose we used polar coordinates, however,
\begin{align}
\xi^i = r (\cos\phi, \sin\phi) .
\end{align}
The metric would read
\begin{align}
(\dd\ell)^2 = \dd r^2 + r^2 \dd\phi^2 .
\end{align}
The velocity now reads $v^r(\vec{\xi}) \partial_r + v^\phi(\vec{\xi}) \partial_\phi$, where $v^r(\vec{\xi})$ has an interpretation of ``rate of flow in the radial direction". However, notice the dimensions of the $v^\phi$ is not even the same as that of $v^r$; if $v^r$ were of [Length/Time], then $v^\phi$ is of [1/Time]. At this point we recall -- just as $\dd r$ (which is dual to $\partial_r$) can be interpreted as an infinitesimal length in the radial direction, the arc length $r \dd\phi$ (which is dual to $(1/r) \partial_\phi$) is the corresponding one in the perpendicular azimuthal direction. Using these as a guide, we would now express the velocity at $\vec{\xi}$ as
\begin{align}
v = v^r \frac{\partial}{\partial r} + (r \cdot v^\phi) \left(\frac{1}{r}\frac{\partial}{\partial \phi}\right) ,
\end{align}
so that now $v^{\widehat{\phi}} \equiv r \cdot v^\phi$ may be interpreted as the velocity in the azimuthal direction.

More formally, given a (real, symmetric) metric $g_{ij}$ we may always find a orthogonal transformation $O^a_{\phantom{a}i}$ that diagonalizes it; and by absorbing into this transformation the eigenvalues of the metric, the orthonormal frame fields emerge:
{\allowdisplaybreaks\begin{align}
\label{DifferentialGeometry_OrthonormalFrame}
g_{ij} \dd x^i \dd x^j 
&= \sum_{a,b} \left( O^a_{\phantom{a}i} \cdot \lambda_a \delta_{ab} \cdot O^b_{\phantom{b}j} \right)\dd x^i \dd x^j \nonumber\\
&= \sum_{a,b} \left( \sqrt{\lambda_a} O^a_{\phantom{a}i} \cdot \delta_{ab} \cdot \sqrt{\lambda_b} O^b_{\phantom{b}j} \right)\dd x^i \dd x^j \nonumber\\
&= \left( \delta_{ab} \varepsilon^{\widehat{a}}_{\phantom{\widehat{a}}i} \varepsilon^{\widehat{b}}_{\phantom{\widehat{b}}j} \right) \dd x^i \dd x^j 
= \delta_{ab} \left(\varepsilon^{\widehat{a}}_{\phantom{\widehat{a}}i} \dd x^i \right)\left(\varepsilon^{\widehat{b}}_{\phantom{\widehat{b}}j} \dd x^j\right), \\
\varepsilon^{\widehat{a}}_{\phantom{\widehat{a}}i} &\equiv \sqrt{\lambda_a} O^a_{\phantom{a}i}, \qquad\qquad
\text{(No sum over $a$.)} .
\end{align}}
In the first equality, we have exploited the fact that any real symmetric matrix $g_{ij}$ can be diagonalized by an appropriate orthogonal matrix $O^a_{\phantom{a}i}$, with real eigenvalues $\{\lambda_a\}$; in the second we have exploited the assumption that we are working in Riemannian spaces, where all eigenvalues of the metric are positive,\footnote{As opposed to semi-Riemannian/Lorentzian spaces, where the eigenvalue associated with the `time' direction has a different sign from the rest.} to take the positive square roots of the eigenvalues; in the third we have defined the orthonormal frame vector fields as $\varepsilon^{\widehat{a}}_{\phantom{\widehat{a}}i} = \sqrt{\lambda_a} O^a_{\phantom{a}i}$, with no sum over $a$. Finally, from eq. \eqref{DifferentialGeometry_OrthonormalFrame} and by defining the infinitesimal lengths $\varepsilon^{\widehat{a}} \equiv \varepsilon^{\widehat{a}}_{\phantom{\widehat{a}}i} \dd x^i$, we arrive at the following curved space parallel to Pythagoras' theorem in flat space:
\begin{align}
\left(\dd\ell\right)^2 = g_{ij} \dd x^i \dd x^j 
= \left( \varepsilon^{\widehat{1}} \right)^2 + \left( \varepsilon^{\widehat{2}} \right)^2 
+ \dots + \left( \varepsilon^{\widehat{D}} \right)^2 .
\end{align}
The metric components are now%That the eigenvalues are always positive, so that their positive square roots can be taken, is part of the definition of curved geometries with Euclidean signature.
\begin{align}
g_{ij} = \delta_{ab} \varepsilon^{\widehat{a}}_{\phantom{\widehat{a}}i} \varepsilon^{\widehat{b}}_{\phantom{\widehat{b}}j} .
\end{align}
Whereas the metric determinant reads
\begin{align}
\label{DifferentialGeometry_detg_OrthonormalFrame}
\det g_{ij} = \left(\det \varepsilon^{\widehat{a}}_{\phantom{\widehat{a}}i}\right)^2 .
\end{align}
We say the metric on the right hand side of eq. \eqref{DifferentialGeometry_OrthonormalFrame} is written in an orthonormal frame, because in this basis $\{ \varepsilon^{\widehat{a}}_{\phantom{\widehat{a}}i} \dd x^i | a = 1,2,\dots,D \}$, the metric components are identical to the flat Cartesian ones. We have put a $\widehat{\cdot}$ over the $a$-index, to distinguish from the $i$-index, because the latter transforms as a tensor
\begin{align}
\label{DifferentialGeometry_OrthonormalFrame_CT}
\varepsilon^{\widehat{a}}_{\phantom{\widehat{a}}i}(\vec{\xi})
= \varepsilon^{\widehat{a}}_{\phantom{\widehat{a}}j}\left( \vec{x}(\vec{\xi}) \right) \frac{\partial x^j(\vec{\xi})}{\partial \xi^i} .
\end{align}
This also implies the $i$-index can be moved using the metric; for example
\begin{align}
\varepsilon^{\widehat{a} i}(\vec{x}) \equiv g^{ij}(\vec{x}) \varepsilon^{\widehat{a}}_{\phantom{\widehat{a}} j}(\vec{x}) .
\end{align}
The $\widehat{a}$ index does not transform under coordinate transformations. But it can be rotated by an orthogonal matrix $R^{\widehat{a}}_{\phantom{\widehat{a}} \widehat{b}}(\vec{\xi})$, which itself can depend on the space coordinates, while keeping the metric in eq. \eqref{DifferentialGeometry_OrthonormalFrame} the same object. By orthogonal matrix, we mean any $R$ that obeys
\begin{align}
\widehat{R}^{\widehat{a}}_{\phantom{\widehat{a}} \widehat{c}} \delta_{ab} \widehat{R}^{\widehat{b}}_{\phantom{\widehat{b}} \widehat{f}} &= \delta_{cf} \\
\widehat{R}^T \widehat{R} &= \mathbb{I} .
\end{align}
Upon the replacement
\begin{align}
\label{DifferentialGeometry_OrthonormalFrame_SODT}
\varepsilon^{\widehat{a}}_{\phantom{\widehat{a}}i}(\vec{x}) \to
\widehat{R}^{\widehat{a}}_{\phantom{\widehat{a}} \widehat{b}}(\vec{x}) \varepsilon^{\widehat{b}}_{\phantom{\widehat{b}}i}(\vec{x}) ,
\end{align}
we have
\begin{align}
g_{ij} \dd x^i \dd x^j 
&\to \left(\delta_{ab} \widehat{R}^{\widehat{a}}_{\phantom{\widehat{a}} \widehat{c}} \widehat{R}^{\widehat{a}}_{\phantom{\widehat{b}} \widehat{f}}\right)
\varepsilon^{\widehat{c}}_{\phantom{\widehat{c}}i} \varepsilon^{\widehat{f}}_{\phantom{\widehat{f}}j} \dd x^i \dd x^j 
= g_{ij} \dd x^i \dd x^j .
\end{align}
The interpretation of eq. \eqref{DifferentialGeometry_OrthonormalFrame_SODT} is that the choice of local Cartesian-like (non-coordinate) axes are not unique; just as the Cartesian coordinate system in flat space can be redefined through a rotation $R$ obeying $R^T R = \mathbb{I}$, these local axes can also be rotated freely. It is a consequence of this O$_D$ symmetry that upper and lower orthonormal frame indices actually transform the same way. We begin by demanding that rank-1 tensors in an orthonormal frame transform as
\begin{align}
V^{\widehat{a}'} = \widehat{R}^{\widehat{a}}_{\phantom{\widehat{a}} \widehat{c}} V^{\widehat{c}}, \qquad
V_{\widehat{a}'} = (\widehat{R}^{-1})_{\phantom{\widehat{f}}\widehat{a}}^{\widehat{f}} V_{\widehat{f}}
\end{align}
so that
\begin{align}
V^{\widehat{a}'} V_{\widehat{a}'} = V^{\widehat{a}} V_{\widehat{a}} .
\end{align}
But $\widehat{R}^T \widehat{R} = \mathbb{I}$ means $\widehat{R}^{-1} = \widehat{R}^T$ and thus the $a$th row and $c$th column of the inverse, namely $(\widehat{R}^{-1})_{\phantom{\widehat{a}}\widehat{c}}^{\widehat{a}}$, is equal to the $c$th row and $a$th column of $\widehat{R}$ itself: $(\widehat{R}^{-1})^{\widehat{a}}_{\phantom{\widehat{a}}\widehat{c}} = \widehat{R}^{\widehat{c}}_{\phantom{\widehat{c}}\widehat{a}}$.
\begin{align}
V_{\widehat{a}'} = \sum_f \widehat{R}^{\widehat{a}}_{\phantom{\widehat{a}}\widehat{f}} V_{\widehat{f}} .
\end{align}
In other words, $V_{\widehat{a}}$ transforms just like $V^{\widehat{a}}$.

To sum, we have shown that the orthonormal frame index is moved by the Kronecker delta; $V^{\widehat{a}'} = V_{\widehat{a}'}$ for any vector written in an orthonormal frame, and in particular,
\begin{align}
\varepsilon^{\widehat{a}}_{\phantom{\widehat{a}}i}(\vec{x}) 
= \delta^{ab} \varepsilon_{\widehat{b} i}(\vec{x}) = \varepsilon_{\widehat{a} i}(\vec{x}) .
\end{align}
Next, we also demonstrate that these vector fields are indeed of unit length.
\begin{align}
\label{DifferentialGeometry_OrthonormalFrame_Completeness}
\varepsilon^{\widehat{f}}_{\phantom{\widehat{f}}j} \varepsilon^{\widehat{b}j} 
= \varepsilon^{\widehat{f}}_{\phantom{\widehat{f}}j}  \varepsilon^{\widehat{b}}_{\phantom{\widehat{b}}k} g^{jk} = \delta^{fb} , \\
\varepsilon_{\widehat{f}}^{\phantom{\widehat{f}}j} \varepsilon_{\widehat{b}j} 
= \varepsilon_{\widehat{f}}^{\phantom{\widehat{f}}j} \varepsilon_{\widehat{b}}^{\phantom{\widehat{b}}k} g_{jk} = \delta_{fb} .
\end{align}
To understand this we begin with the diagonalization of the metric, $\delta_{cf} \varepsilon^{\widehat{c}}_{\phantom{\widehat{c}}i} \varepsilon^{\widehat{f}}_{\phantom{\widehat{f}}j} = g_{ij}$. Contracting both sides with the orthonormal frame vector $\varepsilon^{\widehat{b}j}$,
\begin{align}
\delta_{cf} \varepsilon^{\widehat{c}}_{\phantom{\widehat{c}}i} \varepsilon^{\widehat{f}}_{\phantom{\widehat{f}}j} \varepsilon^{\widehat{b}j} 
&= \varepsilon^{\widehat{b}}_{\phantom{\widehat{b}}i} , \\
(\varepsilon^{\widehat{b}j} \varepsilon_{\widehat{f}j}) \varepsilon^{\widehat{f}}_{\phantom{\widehat{f}} i} 
&= \varepsilon^{\widehat{b}}_{\phantom{\widehat{b}}i} .
\end{align}
If we let $M$ denote the matrix $M^b_{\phantom{b}f} \equiv (\varepsilon^{\widehat{b}j} \varepsilon_{\widehat{f}j})$, then we have $i=1,2,\dots,D$ matrix equations $M \cdot \varepsilon_i = \varepsilon_i$. As long as the determinant of $g_{ab}$ is non-zero, then $\{\varepsilon_i\}$ are linearly independent vectors spanning $\mathbb{R}^D$ (see eq. \eqref{DifferentialGeometry_detg_OrthonormalFrame}). Since every $\varepsilon_i$ is an eigenvector of $M$ with eigenvalue one, that means $M = \mathbb{I}$, and we have proved eq. \eqref{DifferentialGeometry_OrthonormalFrame_Completeness}.

To summarize,
\begin{align}
\label{DifferentialGeometry_CoordinateVsOrthonormalFrame_Summary}
g_{ij} = \delta_{ab} \varepsilon^{\widehat{a}}_{\phantom{\widehat{a}}i} \varepsilon^{\widehat{b}}_{\phantom{\widehat{b}}j},\qquad
g^{ij} = \delta^{ab} \varepsilon_{\widehat{a}}^{\phantom{\widehat{a}}i} \varepsilon_{\widehat{b}}^{\phantom{\widehat{b}}j}, \nonumber\\
\delta_{ab} = g_{ij} \varepsilon_{\widehat{a}}^{\phantom{\widehat{a}}i} \varepsilon_{\widehat{b}}^{\phantom{\widehat{b}}j}, \qquad
\delta^{ab} = g^{ij} \varepsilon^{\widehat{a}}_{\phantom{\widehat{a}}i} \varepsilon^{\widehat{b}}_{\phantom{\widehat{b}}j} .
\end{align}
Now, any tensor with written in a coordinate basis can be converted to one in an orthonormal basis by contracting with the orthonormal frame fields $\varepsilon^{\widehat{a}}_{\phantom{\widehat{a}}i}$ in eq. \eqref{DifferentialGeometry_OrthonormalFrame}. For example, the velocity field in an orthonormal frame is
\begin{align}
v^{\widehat{a}} = \varepsilon^{\widehat{a}}_{\phantom{\widehat{a}}i} v^i .
\end{align}
For the two dimension example above,
\begin{align}
(\dd r)^2 + (r \dd\phi)^2 = \delta_{rr} (\dd r)^2 + \delta_{\phi\phi} (r \dd\phi)^2 ,
\end{align}
allowing us to read off the only non-zero components of the orthonormal frame fields are
\begin{align}
\varepsilon^{\widehat{r}}_{\phantom{\widehat{r}}r} = 1, \qquad
\varepsilon^{\widehat{\phi}}_{\phantom{\widehat{\phi}}\phi} = r ;
\end{align}
which in turn implies
\begin{align}
v^{\widehat{r}} = \varepsilon^{\widehat{r}}_{\phantom{\widehat{r}}r} v^r = v^r, \qquad 
v^{\widehat{\phi}} = \varepsilon^{\widehat{\phi}}_{\phantom{\widehat{\phi}}\phi} v^\phi = r \ v^\phi .
\end{align}
More generally, what we are doing here is really switching from writing the same tensor in coordinates basis $\{ \dd x^i \}$ and $\{ \partial_i \}$ to an orthonormal basis $\{ \epud{\widehat{a}}{i} \dd x^i \}$ and $\{ \epdu{\widehat{a}}{i} \partial_i \}$. For example,
\begin{align}
\label{DifferentialGeometry_OrthonormalFrameBasis_Eg}
T_{ijk}^{\phantom{ijk}l} \bra{\dd x^i} \otimes \bra{\dd x^j} \otimes \bra{\dd x^k} \otimes \ket{\partial_l}
&= T_{\widehat{i}\widehat{j}\widehat{k}}^{\phantom{\widehat{i}\widehat{j}\widehat{k}}\widehat{l}} 
\bra{\varepsilon^{\widehat{i}}} \otimes \bra{\varepsilon^{\widehat{j}}} \otimes \bra{\varepsilon^{\widehat{k}}} \otimes \ket{\varepsilon_{\widehat{l}}} \\
\label{DifferentialGeometry_OrthonormalFrameBasis}
\varepsilon^{\widehat{i}} &\equiv \epud{\widehat{i}}{a} \dd x^a \qquad\qquad
\varepsilon_{\widehat{i}} \equiv \epdu{\widehat{i}}{a} \partial_a .
\end{align} 
Even though the physical dimension of the whole tensor $[T]$ is necessarily consistent, because the $\{ \dd x^i \}$ and $\{ \partial_i \}$ do not have the same dimensions -- compare, for e.g., $\dd r$ versus $\dd\theta$ in spherical coordinates -- the components of tensors in a coordinate basis do not all have the same dimensions, making their interpretation difficult. By using orthonormal frame fields as defined in eq. \eqref{DifferentialGeometry_OrthonormalFrameBasis}, we see that
\begin{align}
\sum_a \left( \varepsilon^{\widehat{a}} \right)^2 
&= \delta_{ab} \epud{\widehat{a}}{i} \epud{\widehat{b}}{j} \dd x^i \dd x^j
= g_{ij} \dd x^i \dd x^j \\
\left[ \varepsilon^{\widehat{a}} \right] &= \text{Length} ;
\end{align} 
and
\begin{align}
\sum_a \left( \varepsilon_{\widehat{a}} \right)^2 
&= \delta^{ab} \epdu{\widehat{a}}{i} \epdu{\widehat{b}}{j} \partial_i \partial_j
= g^{ij} \partial_i \partial_j \\
\left[ \varepsilon_{\widehat{a}} \right] &= 1/\text{Length} ;
\end{align} 
which in turn implies, for instance, the consistency of the physical dimensions of the orthonormal components $T_{\widehat{i}\widehat{j}\widehat{k}}^{\phantom{\widehat{i}\widehat{j}\widehat{k}}\widehat{l}}$ in eq. \eqref{DifferentialGeometry_OrthonormalFrameBasis_Eg}:
\begin{align}
[T_{\widehat{i}\widehat{j}\widehat{k}}^{\phantom{\widehat{i}\widehat{j}\widehat{k}}\widehat{l}}] 
[\varepsilon^{\widehat{i}}]^3 [\varepsilon_{\widehat{l}}] &= [T] , \\
\left[ T_{\widehat{i}\widehat{j}\widehat{k}}^{\phantom{\widehat{i}\widehat{j}\widehat{k}}\widehat{l}} \right] 
&= \frac{[T]}{\text{Length}^2} .
\end{align}
\begin{myP}
\qquad Find the orthonormal frame fields $\{\varepsilon^{\widehat{a}}_{\phantom{\widehat{a}}i}\}$ in 3-dimensional Cartesian, Spherical and Cylindrical coordinate systems. Hint: Just like the 2D case above, by packaging the metric $g_{ij} \dd x^i \dd x^j$ appropriately, you can read off the frame fields without further work. \qquad \qed
\end{myP}
{\bf (Curved) Dot Product} \qquad So far we have viewed the metric $(\dd\ell)^2$ as the square of the distance between $\vec{x}$ and $\vec{x} + \dd\vec{x}$, generalizing Pythagoras' theorem in flat space. The generalization of the dot product between two (tangent) vectors $U$ and $V$ at some location $\vec{x}$ is
\begin{align}
U(\vec{x}) \cdot V(\vec{x}) \equiv g_{ij}(\vec{x}) U^i(\vec{x}) V^j(\vec{x}) .
\end{align}
That this is in fact the analogy of the dot product in Euclidean space can be readily seen by going to the orthonormal frame:
\begin{align}
U(\vec{x}) \cdot V(\vec{x}) = \delta_{ij} U^{\widehat{i}}(\vec{x}) V^{\widehat{j}}(\vec{x}) .
\end{align}
\noindent{\bf Line integral} \qquad The line integral that occurs in 3D vector calculus, is commonly written as $\int \vec{A} \cdot \dd \vec{x}$. While the dot product notation is very convenient and oftentimes quite intuitive, there is an implicit assumption that the underlying coordinate system is Cartesian in flat space. The integrand that actually transforms covariantly is the tensor $A_i \dd x^i$, where the $\{x^i\}$ are no longer necessarily Cartesian. The line integral itself then consists of integrating this over a prescribed path $\vec{x}(\lambda_1 \leq \lambda \leq \lambda_2)$, namely
\begin{align}
\int_{\vec{x}(\lambda_1 \leq \lambda \leq \lambda_2)} A_i \dd x^i 
= \int_{\lambda_1}^{\lambda_2} A_i\left( \vec{x}(\lambda) \right) \frac{\dd x^i(\lambda)}{\dd\lambda}\dd \lambda .
\end{align}

\subsection{Covariant derivatives, Parallel Transport, Levi-Civita, Hodge Dual}

{\bf Covariant Derivative} \qquad How do we take derivatives of tensors in such a way that we get back a tensor in return? To start, let us see that the partial derivative of a tensor is not a tensor. Consider
\begin{align}
\label{DifferentialGeometry_PartialD}
\frac{\partial T_j(\vec{\xi})}{\partial \xi^i}
&= \frac{\partial x^a}{\partial \xi^i} \frac{\partial}{\partial x^a} \left( T_b\left( \vec{x}(\vec{\xi}) \right) \frac{\partial x^b}{\partial \xi^j} \right) \nonumber \\
&= \frac{\partial x^a}{\partial \xi^i} \frac{\partial x^b}{\partial \xi^j} \frac{\partial T_b\left( \vec{x}(\vec{\xi}) \right)}{\partial x^a}
+ \frac{\partial^2 x^b}{\partial \xi^j \partial \xi^i} T_b\left( \vec{x}(\vec{\xi}) \right) .
\end{align}
The second derivative $\partial^2 x^b/\partial \xi^i \partial \xi^j$ term is what spoils the coordinate transformation rule we desire. To fix this, we introduce the concept of the covariant derivative $\nabla$, which is built out of the partial derivative and the Christoffel symbols $\Gamma^i_{\phantom{i}jk}$, which in turn is built out of the metric tensor,
\begin{align}
\label{DifferentialGeometry_ChristoffelSymbols}
\Gamma^i_{\phantom{i}jk} = \frac{1}{2} g^{il} \left( \partial_j g_{kl} + \partial_k g_{jl} - \partial_l g_{jk} \right) .
\end{align}
Notice the Christoffel symbol is symmetric in its lower indices: $\Gamma^i_{\phantom{i}jk} = \Gamma^i_{\phantom{i}kj}$.

For a scalar $\varphi$ the covariant derivative is just the partial derivative
\begin{align}
\nabla_i \varphi = \partial_i \varphi .
\end{align}
For a $\binom{0}{1}$ or $\binom{1}{0}$ tensor, its covariant derivative reads
\begin{align}
\nabla_i T_j &= \partial_i T_j - \Gamma^l_{\phantom{l}ij} T_l , \\
\nabla_i T^j &= \partial_i T^j + \Gamma^j_{\phantom{l}il} T^l .
\end{align}
Under $\vec{x} \to \vec{x}(\vec{\xi})$, we have,
\begin{align}
\nabla_{\xi^i} \varphi(\vec{\xi}) &= \frac{\partial x^a}{\partial \xi^i} \nabla_{x^a} \varphi\left( \vec{x}(\vec{\xi}) \right)\\
\nabla_{\xi^i} T_j(\xi) &= \frac{\partial x^a}{\partial \xi^i} \frac{\partial x^b}{\partial \xi^j}  
			\nabla_{x^a} T_b\left( \vec{x}(\vec{\xi}) \right) .
\end{align}
For a general $\binom{N}{M}$ tensor, we have
\begin{align}
\nabla_k T^{i_1 i_2 \dots i_N}_{\phantom{i_1 i_2 \dots i_N} j_i j_2 \dots j_M}
&= \partial_k T^{i_1 i_2 \dots i_N}_{\phantom{i_1 i_2 \dots i_N} j_i j_2 \dots j_M} \\
&+ \Gamma^{i_1}_{\phantom{i_1}kl} T^{l i_2 \dots i_N}_{\phantom{i_1 i_2 \dots i_N} j_i j_2 \dots j_M}
+ \Gamma^{i_2}_{\phantom{i_2}kl} T^{i_1 l \dots i_N}_{\phantom{i_1 i_2 \dots i_N} j_i j_2 \dots j_M} + \dots
+ \Gamma^{i_N}_{\phantom{i_N}kl} T^{i_1 \dots i_{N-1} l}_{\phantom{i_1 i_2 \dots i_N} j_i j_2 \dots j_M} \nonumber\\
&
- \Gamma^l_{\phantom{l}k j_1} T^{i_1 \dots i_N}_{\phantom{i_1 i_2 \dots i_N} l j_2 \dots j_M}
- \Gamma^l_{\phantom{l}k j_2} T^{i_1 \dots i_N}_{\phantom{i_1 i_2 \dots i_N} j_1 l \dots j_M} - \dots
- \Gamma^l_{\phantom{l}k j_M} T^{i_1 \dots i_N}_{\phantom{i_1 i_2 \dots i_N} j_1 \dots j_{M-1} l} . \nonumber
\end{align}
\footnote{The semi-colon is sometimes also used to denote the covariant derivative. For example, $\nabla_l \nabla_i T^{jk} \equiv T^{jk}_{\phantom{jk};il}$.}By using eq. \eqref{DifferentialGeometry_PartialD} we may infer how the Christoffel symbols themselves must transform -- they are not tensors. Firstly,
\begin{align}
\label{DifferentialGeometry_ChristoffelSymbols_TI}
\nabla_{\xi^i} T_j(\vec{\xi}) 
&= \partial_{\xi^i} T_j(\vec{\xi}) - \Gamma^l_{\phantom{l}ij}(\vec{\xi}) T_l(\vec{\xi}) \nonumber\\
&= \frac{\partial x^a}{\partial \xi^i} \frac{\partial x^b}{\partial \xi^j} \partial_{x^a} T_b\left( \vec{x}(\vec{\xi}) \right)
+ \left(\frac{\partial^2 x^b}{\partial \xi^j \partial \xi^i} - \Gamma^l_{\phantom{l}ij}(\vec{\xi}) \frac{\partial x^b(\vec{\xi})}{\partial \xi^l}\right) T_b\left( \vec{x}(\vec{\xi}) \right)
\end{align}
On the other hand,
\begin{align}
\label{DifferentialGeometry_ChristoffelSymbols_TII}
\nabla_{\xi^i} T_j(\vec{\xi}) 
&= \frac{\partial x^a}{\partial \xi^i} \frac{\partial x^b}{\partial \xi^j} \nabla_{x^a} T_b\left( \vec{x}(\vec{\xi}) \right) \nonumber\\
&= \frac{\partial x^a}{\partial \xi^i} \frac{\partial x^b}{\partial \xi^j} \left\{
\partial_{x^a} T_b\left( \vec{x}(\vec{\xi}) \right) - \Gamma^l_{\phantom{l}ab}\left(\vec{x}(\vec{\xi})\right) T_l\left( \vec{x}(\vec{\xi}) \right)
 \right\}
\end{align}
Comparing equations \eqref{DifferentialGeometry_ChristoffelSymbols_TI} and \eqref{DifferentialGeometry_ChristoffelSymbols_TII} leads us to relate the Christoffel symbol written in $\vec{\xi}$ coordinates $\Gamma^l_{\phantom{m}ij}(\vec{\xi})$ and that written in $\vec{x}$ coordinates $\Gamma^l_{\phantom{m}ij}(\vec{x})$.
\begin{align}
\label{DifferentialGeometry_ChristoffelSymbols_Tranformation}
\Gamma^l_{\phantom{m}ij}(\vec{\xi})  
= \Gamma^k_{\phantom{l}mn}\left(\vec{x}(\vec{\xi}) \right)
		\frac{\partial \xi^l}{\partial x^k(\vec{\xi})} \frac{\partial x^m(\vec{\xi})}{\partial \xi^i} \frac{\partial x^n(\vec{\xi})}{\partial \xi^j} 
		+ \frac{\partial \xi^l}{\partial x^k(\vec{\xi})} \frac{\partial^2 x^k(\vec{\xi})}{\partial \xi^j \partial \xi^i} .
\end{align}
On the right hand side, all $\vec{x}$ have been replaced with $\vec{x}(\vec{\xi})$.\footnote{We note in passing that in gauge theory -- which encompasses humanity's current description of the non-gravitational forces (electromagnetic-weak (SU$_2)_\text{left-handed fermions} \times (U_1)_\text{hypercharge}$ and strong nuclear (SU$_3)_\text{color}$) -- the fundamental fields there $\{A_{\phantom{b}\mu}^b\}$ transforms (in a group theory sense) in a very similar fashion as the Christoffel symbols do (under a coordinate transformation) in eq. \eqref{DifferentialGeometry_ChristoffelSymbols_Tranformation}.}

The covariant derivative, like its partial derivative counterpart, obeys the product rule. Suppressing the indices, if $T_1$ and $T_2$ are both tensors, we have
\begin{align}
\label{DifferentialGeometry_CovariantD_ProductRule}
\nabla \left( T_1 T_2 \right) = (\nabla T_1) T_2 + T_1 (\nabla T_2) .
\end{align}
As you will see below, the metric is parallel transported in all directions,
\begin{align}
\nabla_i g_{jk} = \nabla_i g^{jk} = 0.
\end{align}
Combined with the product rule in eq. \eqref{DifferentialGeometry_CovariantD_ProductRule}, this means when raising and lowering of indices of a covariant derivative of a tensor, the metric may be passed in and out of the $\nabla$. For example,
\begin{align}
g_{ia} \nabla_j T^{k a l} &= \nabla_j g_{ia} \cdot T^{k a l} + g_{ia} \nabla_j T^{k a l} = \nabla_j (g_{ia} T^{k a l}) \nonumber\\
&= \nabla_j T^{k \phantom{i} l}_{\phantom{k} i} .
\end{align}
\noindent{\it Remark} \qquad I have introduced the Christoffel symbol here by showing how it allows us to define a derivative operator on a tensor that returns a tensor. I should mention here that, alternatively, it is also possible to view $\Gamma^i_{\phantom{i}jk}$ as ``rotation matrices," describing the failure of parallel transporting the basis bras $\{\bra{\dd x^i}\}$ and kets $\{\ket{\partial_i}\}$ as they are moved from one point in space to a neighboring point infinitesimally far away. Specifically,
\begin{align}
\nabla_i \bra{\dd x^j} = -\Gamma^j_{\phantom{l}ik} \bra{\dd x^k} \qquad \text{   and   } \qquad
\nabla_i \ket{\partial_j} = \Gamma^l_{\phantom{l}ij} \ket{\partial_l} .
\end{align}
Within this perspective, the tensor components are scalars. The product rule then yields, for instance,
\begin{align}
\nabla_i \left( V_a \bra{\dd x^a} \right)
&= (\nabla_i V_a) \bra{\dd x^a} + V_a \nabla_i \bra{\dd x^a} \nonumber\\
&= (\partial_i V_j - V_a \Gamma^a_{\phantom{a}ij}) \bra{\dd x^j}  .
\end{align}
{\bf Riemann and Ricci tensors} \qquad I will not use them very much in the rest of our discussion in this section (\S\eqref{Chapter_DifferentialGeometry_CurvedSpaces}), but I should still highlight that the Riemann and Ricci tensors are fundamental to describing curvature. The Riemann tensor is built out of the Christoffel symbols via
\begin{align}
\label{DifferentialGeometry_Riemann}
R^{i}_{\phantom{i}jkl} 
= \partial_{k} \Gamma^i_{\phantom{i}lj} - \partial_{l} \Gamma^i_{\phantom{i}kj} 
	+ \Gamma^i_{\phantom{i}s k} \Gamma^s_{\phantom{s}l j} - \Gamma^i_{\phantom{i}s l} \Gamma^s_{\phantom{s}k j} .
\end{align}
The failure of parallel transport of some vector $V^i$ around an infinitesimally small loop, is characterized by
\begin{align}
\label{DifferentialGeometry_CommutatorOfNabla_IofII}
[\nabla_k, \nabla_l] V^i &\equiv (\nabla_k \nabla_l - \nabla_l \nabla_k) V^i = R^{i}_{\phantom{i}jkl} V^j ,\\
\label{DifferentialGeometry_CommutatorOfNabla_IIofII}
[\nabla_k, \nabla_l] V_j &\equiv (\nabla_k \nabla_l - \nabla_l \nabla_k) V_j = -R^{i}_{\phantom{i}jkl} V_i .
\end{align}
The generalization to higher rank tensors is
\begin{align}
[\nabla_i,\nabla_j] T^{k_1 \dots k_N}_{\phantom{k_1 \dots k_N} l_1 \dots l_M}
&= R^{k_1}_{\phantom{k_1}a ij} T^{a k_2 \dots k_N}_{\phantom{a k_2 \dots k_N} l_1 \dots l_M}
+ R^{k_2}_{\phantom{k_2}a ij} T^{k_1 a k_3 \dots k_N}_{\phantom{k_1 a k_3 \dots k_N} l_1 \dots l_M}
+ \dots 
+ R^{k_N}_{\phantom{k_N}a ij} T^{k_1 \dots k_{N-1} a}_{\phantom{k_1 \dots k_{N-1} a} l_1 \dots l_M} \nonumber\\
&- R^{a}_{\phantom{a}l_1 ij} T^{k_1 \dots k_N}_{\phantom{k_1 \dots k_N} a l_2 \dots l_M}
- R^{a}_{\phantom{a}l_2 ij} T^{k_1 \dots k_N}_{\phantom{k_1 \dots k_N} l_1 a l_3 \dots l_M}
- \dots
- R^{a}_{\phantom{a}l_M ij} T^{k_1 \dots k_N}_{\phantom{k_1 \dots k_N} l_1 \dots l_{M-1} a} .
\end{align}
The Riemann tensor obeys the following symmetries.
\begin{align}
R_{ij ab} = R_{ab ij}, \qquad R_{ij ab} = -R_{ji ab}, \qquad R_{ab ij} = -R_{ab ji} .
\end{align}
The Riemann tensor also obeys the Bianchi identities\footnote{The symbol $[\dots]$ means the indices within it are fully anti-symmetrized; in particular, $T_{[ijk]} = T_{ijk} - T_{ikj} - T_{jik} + T_{jki} - T_{kji} + T_{kij}$. We will have more to say about this operation later on.}
\begin{align}
\label{DifferentialGeometry_BianchiIdentities}
R^i_{\phantom{i}[j kl]} 
= \nabla_{[i} R^{jk}_{\phantom{jk}lm]}
= 0 .
\end{align}
In $D$ dimensions, the Riemann tensor has $D^2(D^2-1)/12$ algebraically independent components. In particular, in $D=1$ dimension, space is always flat because $R_{11 11}=-R_{11 11}=0$.

The Ricci tensor is defined as the non-trivial contraction of a pair of the Riemann tensor's indices.
\begin{align}
\label{DifferentialGeometry_RicciTensor}
R_{jl} \equiv R^{i}_{\phantom{i}jil} .
\end{align}
It is symmetric
\begin{align}
R_{ij} = R_{ji} .
\end{align}
Finally the Ricci scalar results from a contraction of the Ricci tensor's indices.
\begin{align}
\label{DifferentialGeometry_RicciScalar}
\mathcal{R} \equiv g^{jl} R_{jl} .
\end{align}
Contracting eq. \eqref{DifferentialGeometry_BianchiIdentities} appropriately yields the Bianchi identities involving the Ricci tensor and scalar
\begin{align}
\label{DifferentialGeometry_ContractedBianchiIdentities}
\nabla^i \left( R_{ij} - \frac{g_{ij}}{2} \mathcal{R} \right) = 0 .
\end{align}
This is a good place to pause and state, the Christoffel symbols in eq. \eqref{DifferentialGeometry_ChristoffelSymbols}, covariant derivatives, and the Riemann/Ricci tensors, etc., are in general very tedious to compute. If you ever have to do so on a regular basis, say for research, I highly recommend familiarizing yourself with one of the various software packages available that could do them for you.

{\bf Geodesics} \qquad Recall the distance integral in eq. \eqref{DifferentialGeometry_LengthIntegral}. If you wish to determine the shortest path (aka geodesic) between some given pair of points $\vec{x}_1$ and $\vec{x}_2$, you will need to minimize eq. \eqref{DifferentialGeometry_LengthIntegral}. This is a ``calculus of variation" problem. The argument runs as follows. Suppose you found the path $\vec{z}(\lambda)$ that yields the shortest $\ell$. Then, if you consider a slight variation $\delta \vec{z}$ of the path, namely consider
\begin{align}
\vec{x}(\lambda) = \vec{z}(\lambda) + \delta\vec{z}(\lambda),
\end{align}
we must find the contribution to $\ell$ at first order in $\delta \vec{z}$ to be zero. This is analogous to the vanishing of the first derivatives of a function at its minimum.\footnote{There is some smoothness condition being assumed here. For instance, the tip of the pyramid (or a cone) is the maximum height achieved, but the derivative slightly away from the tip is negative in all directions.} In other words, in the integrand of eq. \eqref{DifferentialGeometry_LengthIntegral} we must replace
\begin{align}
g_{ij}\left( \vec{x}(\lambda) \right) 
&\to g_{ij}\left( \vec{z}(\lambda) + \delta \vec{z}(\lambda) \right)
= g_{ij}\left( \vec{z}(\lambda) \right) + \delta z^k(\lambda) \frac{\partial g_{ij}\left( \vec{z}(\lambda) \right)}{\partial z^k} + \mathcal{O}(\delta z^2) \\
\frac{\dd x^i(\lambda)}{\dd\lambda} &\to \frac{\dd z^i(\lambda)}{\dd\lambda} + \frac{\dd \delta z^i(\lambda)}{\dd\lambda} .
\end{align}
Since $\delta \vec{z}$ was arbitrary, at first order, its coefficient within the integrand must vanish. If we further specialize to affine parameters $\lambda$ such that 
\begin{align}
\sqrt{g_{ij} (\dd z^i/\dd \lambda) (\dd z^j/\dd \lambda)} = \text{constant along the entire path $\vec{z}(\lambda)$},
\end{align}
then one would arrive at the following second order non-linear ODE. Minimizing the distance $\ell$ between $\vec{x}_1$ and $\vec{x}_2$ leads to the shortest path $\vec{z}(\lambda)$ ($\equiv$ geodesic) obeying:
\begin{align}
\label{DifferentialGeometry_GeodesicEquation}
0 &= \frac{\dd^2 z^i}{\dd \lambda^2} + \Gamma^i_{\phantom{i}jk}\left(g_{ab}(\vec{z})\right) \frac{\dd z^j}{\dd \lambda} \frac{\dd z^k}{\dd \lambda} ,
\end{align}
with the boundary conditions
\begin{align}
\label{DifferentialGeometry_GeodesicEquation_BC}
\vec{z}(\lambda_1) = \vec{x}_1, \qquad \vec{z}(\lambda_2) = \vec{x}_2 .
\end{align}
The converse is also true, in that -- if the geodesic equation in eq. \eqref{DifferentialGeometry_GeodesicEquation} holds, then $g_{ij}$ $(\dd z^i/\dd \lambda) (\dd z^j/\dd \lambda)$ is a constant along the entire geodesic. Denoting $\ddot{z}^i \equiv \dd^2 z^i/\dd \lambda^2$ and $\dot{z}^i \equiv \dd z^i/\dd \lambda$,
\begin{align}
\frac{\dd}{\dd \lambda} \left( g_{ij} \dot{z}^i \dot{z}^j \right)
&= 2 \ddot{z}^i \dot{z}^j g_{ij} + \dot{z}^k \partial_k g_{ij} \dot{z}^i \dot{z}^j \nonumber\\
&= 2 \ddot{z}^i \dot{z}^j g_{ij} + \dot{z}^k \dot{z}^i \dot{z}^j\left( \partial_k g_{ij} + \partial_i g_{kj} - \partial_j g_{ik} \right) 
\end{align}
Note that the last two terms inside the parenthesis of the second equality cancels. The reason for inserting them is because the expression contained within the parenthesis is related to the Christoffel symbol; keeping in mind eq. \eqref{DifferentialGeometry_ChristoffelSymbols},
\begin{align}
\frac{\dd}{\dd \lambda} \left( g_{ij} \dot{z}^i \dot{z}^j \right)
&= 2 \dot{z}^i \left\{ \ddot{z}^j g_{ij} + \dot{z}^k \dot{z}^j g_{il} \frac{g^{lm}}{2} \left( \partial_k g_{jm} + \partial_j g_{km} - \partial_m g_{jk} \right) \right\} \nonumber\\
&= 2 g_{il} \dot{z}^i \left\{ \ddot{z}^l + \dot{z}^k \dot{z}^j \Gamma^l_{\phantom{l}kj} \right\} = 0 .
\end{align}
The last equality follows because the expression in the $\{ \dots \}$ is the left hand side of eq. \eqref{DifferentialGeometry_GeodesicEquation}. This constancy of $g_{ij}$ $(\dd z^i/\dd \lambda) (\dd z^j/\dd \lambda)$ is useful for solving the geodesic equation itself.
\begin{myP}
{\bf Noether's theorem for Lagrangian mechanics} \qquad Show that the affine parameter form of the geodesic \eqref{DifferentialGeometry_GeodesicEquation} follows from demanding the following integral be extremized:
\begin{align}
\label{DifferentialGeometry_SyngeWorldFunction}
\ell^2
= (\lambda_2-\lambda_1) \int_{\lambda_1}^{\lambda_2} \dd\lambda
g_{ij}\left( \vec{z}(\lambda) \right) \frac{\dd z^i}{\dd\lambda} \frac{\dd z^j}{\dd \lambda} .
\end{align}
(In the General Relativity literature, $\ell^2/2$ (half of eq. \eqref{DifferentialGeometry_SyngeWorldFunction}) is known as Synge's world function.) That is, show that eq. \eqref{DifferentialGeometry_GeodesicEquation} follows from applying the Euler-Lagrange equations to the Lagrangian
\begin{align}
\label{DifferentialGeometry_GeodesicLagrangian}
L \equiv \frac{1}{2} g_{ij} \dot{z}^i \dot{z}^j, 
\qquad\qquad
\dot{z}^i \equiv \frac{\dd z^i}{\dd \lambda} .
\end{align}
Now argue that the Hamiltonian $H$ is equal to the Lagrangian $L$. Can you prove that $H$, and therefore $L$, is a constant of motion? Moreover, if the geodesic equation \eqref{DifferentialGeometry_GeodesicEquation} is satisfied by $z^\mu(\lambda)$, argue that the integral in eq. \eqref{DifferentialGeometry_SyngeWorldFunction} yields the square of the geodesic distance between $\vec{x}_1 \equiv \vec{z}(\lambda_1)$ and $\vec{x}_2 \equiv \vec{z}(\lambda_2)$?

{\it Conserved quantities from symmetries} \qquad Finally, suppose $\partial_k$ is a Killing vector. Explain why 
\begin{align}
\frac{\partial L}{\partial\dot{z}^k} = \text{constant}.
\end{align}
This is an example of Noether's theorem. For example, in flat Euclidean space, since the metric in Cartesian coordinates is a constant $\delta_{ij}$, all the $\{ \partial_i \vert i=1,2,\dots,D \}$ are Killing vectors. Therefore, from $L=(1/2)\delta_{ij} \dot{z}^i \dot{z}^j$, and we have
\begin{align}
\frac{\dd}{\dd\lambda} \frac{\dd z^i}{\dd \lambda} = 0 
\qquad \Rightarrow \qquad 
\frac{\dd z^i}{\dd \lambda} = \text{constant}.
\end{align}
This is, in fact, the statement that the center of mass of an isolated system obeying Newtonian mechanics moves with a constant velocity. By re-writing the Euclidean metric in spherical coordinates, provide the proper definition of angular momentum (about the $D-$axis) and proceed to prove that it is conserved.

{\it Geodesics on a $2-$sphere} \qquad How many geodesics are there joining any two points on the $2-$sphere? How many geodesics are there joining the North Pole and South Pole? Solve the geodesic equation (cf. eq. \eqref{DifferentialGeometry_GeodesicEquation}) on the unit $2-$sphere described by
\begin{align}
\dd\ell^2 = \dd\theta^2 + \sin(\theta)^2 \dd\phi^2 .
\end{align}
Explain how your answer would change if the sphere were of radius $R$ instead. Hint: To solve the geodesic equation it helps to exploit the spherical symmetry of the problem; for e.g., what are the geodesics emanating from the North Pole? Then transform the answer to the more general case.
\end{myP}
{\bf Christoffel symbols from Lagrangian} \qquad As an example of how the action principle in eq. \eqref{DifferentialGeometry_SyngeWorldFunction} allows us to extract the Christoffel symbols, let us consider the following $D-$dimensional metric:
\begin{align}
\dd\ell^2 \equiv a(\vec{x})^2 \dd\vec{x}\cdot\dd\vec{x} ,
\end{align}
where $a(\vec{x})$ is an arbitrary function. The Lagrangian in eq. \eqref{DifferentialGeometry_GeodesicLagrangian} is now
\begin{align}
L = \frac{1}{2} a^2 \delta_{ij} \dot{z}^i \dot{z}^j, \qquad\qquad
\dot{z}^i \equiv \frac{\dd z^i}{\dd\lambda} .
\end{align}
Applying the Euler-Lagrange equations,
{\allowdisplaybreaks\begin{align}
\frac{\dd}{\dd\lambda} \frac{\partial L}{\partial \dot{z}^i} - \frac{\partial L}{\partial z^i} 	&= 0 \\
\frac{\dd}{\dd\lambda} \left( a^2 \dot{z}^i \right) - a \partial_i a \dot{\vec{z}}^2 			&= 0 \\
2 a \dot{z}^j \partial_j a \ \dot{z}^i + a^2 \ddot{z}^i - a \partial_i a \dot{\vec{z}}^2 		&= 0 \\
\ddot{z}^i + \left( \frac{\partial_j a}{a} \delta_l^i + \frac{\partial_l a}{a} \delta_j^i - \frac{\partial_i a}{a} \delta_{lj} \right) \dot{z}^l \dot{z}^j 
= \ddot{z}^i + \Gamma^i_{\phantom{i}lj} \dot{z}^l \dot{z}^j&= 0 .
\end{align}}
Using $\{\dots\}$ to indicate symmetrization of the indices, we have derived
\begin{align}
\Gamma^i_{\phantom{i}lj} 
&= \frac{1}{a}\left( \partial_{\{j} a \delta_{l\}}^i - \partial_i a \delta_{lj} \right) \nonumber\\
&= \left( \delta^k_{\{j} \delta_{l\}}^i - \delta^{ki} \delta_{lj} \right) \partial_k \ln a .
\end{align}
\begin{myP}
\qquad It is always possible to find a coordinate system with coordinates $\vec{y}$ such that, as $\vec{y} \to \vec{y}_0$, the Christoffel symbols vanish
\begin{align}
\Gamma^k_{\phantom{k}ij}(\vec{y}_0) = 0.
\end{align}
Can you demonstrate why this is true from the equivalence principle encoded in eq. \eqref{DifferentialGeometry_EP}? Hint: it is important that, locally, the first deviation from flat space is quadratic in the displacement vector $(y-y_0)^i$. \qed
\end{myP}
{\it Remark} \qquad That there is always an orthonormal frame where the metric is flat -- recall eq. \eqref{DifferentialGeometry_OrthonormalFrame} -- as well as the existence of a locally flat coordinate system, is why the measure of curvature, in particular the Riemann tensor in eq. \eqref{DifferentialGeometry_Riemann}, depends on gradients (second derivatives) of the metric.
\begin{myP}
\qquad Why do the Christoffel symbols take on the form in eq. \eqref{DifferentialGeometry_ChristoffelSymbols}? It comes from assuming that the Christoffel symbol obeys the symmetry $\Gamma^i_{\phantom{i}jk} = \Gamma^i_{\phantom{i}kj}$ -- this is the torsion-free condition -- and demanding that the covariant derivative of a metric is a zero tensor,
\begin{align}
\nabla_i g_{jk} = 0 .
\end{align}
This can be expanded as
\begin{align}
\nabla_i g_{jk} = 0 = \partial_i g_{jk} - \Gamma^l_{\phantom{l}ij} g_{lk} - \Gamma^l_{\phantom{l}ik} g_{jl} .
\end{align}
Expand also $\nabla_j g_{ki}$ and $\nabla_k g_{ij}$, and show that
\begin{align}
2 \Gamma^l_{\phantom{l}ij} g_{lk} = \partial_i g_{jk} + \partial_j g_{ik} - \partial_k g_{ij} .
\end{align}
Divide both sides by $2$ and contract both sides with $g^{km}$ to obtain $\Gamma^m_{\phantom{m}ij}$ in eq. \eqref{DifferentialGeometry_ChristoffelSymbols}. \qquad \qed
\end{myP}
\begin{myP}
\qquad Can you show that the $\delta_\xi g_{ij}$ in eq. \eqref{DifferentialGeometry_LieDerivativeOfMetric} can be re-written in a more covariant looking expression
\begin{align}
\delta_\xi g_{ij}(\vec{x}') = \nabla_i \xi_j + \nabla_j \xi_i ?
\end{align}
$\delta_\xi g_{ij} = \nabla_i \xi_j + \nabla_j \xi_i = 0$ is known as Killing's equation,\footnote{The maximum number of linearly independent Killing vectors in $D$ dimensions is $D(D+1)/2$. See Chapter 13 of Weinberg's {\it Gravitation and Cosmology} for a discussion.} and a vector that satisfies Killing's equation is called a Killing vector. Showing that $\delta_\xi g_{ij}$ is a tensor indicate such a characterization of symmetry is a generally covariant statement.

Hint: Convert all partial derivatives into covariant ones by adding/subtracting Christoffel symbols appropriately; for instance $\partial_a \xi^i = \nabla_a \xi^i - \Gamma^i_{\phantom{i}ab} \xi^b$. \qed
\end{myP}
\begin{myP}
\qquad Argue that, if a tensor $T^{i_1 i_2 \dots i_N}$ is zero in some coordinate system, it must be zero in any other coordinate system. \qquad \qed
\end{myP}
\begin{myP}
\qquad Prove that the tensor $T_{i_1}^{\phantom{i_1}i_2 \dots i_N}$ is zero if and only if the corresponding tensor $T_{i_1 i_2 \dots i_N}$ is zero. Then, using the product rule, explain why $\nabla_i g_{jk}=0$ implies $\nabla_i g^{jk}=0$. Hint: start with $\nabla_i (g_{aj} g_{bk} g^{jk})$. \qquad \qed
\end{myP}
\begin{myP}
\qquad Calculate the Christoffel symbols of the 3-dimensional Euclidean metric in Cartesian coordinates $\delta_{ij}$. Then calculate the Christoffel symbols for the same space, but in spherical coordinates: $(\dd\ell)^2 = \dd r^2 + r^2 (\dd\theta^2 + (\sin\theta)^2 \dd\phi^2)$. To start you off, the non-zero components of the metric are
\begin{align}
g_{rr} &=1, \qquad g_{\theta\theta} = r^2, \qquad g_{\phi\phi} = r^2 (\sin\theta)^2 ; \\
g^{rr} &=1, \qquad g^{\theta\theta} = r^{-2}, \qquad g^{\phi\phi} = \frac{1}{r^2 (\sin\theta)^2} .
\end{align}
Also derive the Christoffel symbols in spherical coordinates from their Cartesian counterparts using eq. \eqref{DifferentialGeometry_ChristoffelSymbols_Tranformation}. This lets you cross-check your results; you should also feel free to use software to help. Partial answer: the non-zero components in spherical coordinates are
\begin{align}
\Gamma^r_{\phantom{r}\theta\theta} 		&= -r, \qquad \Gamma^r_{\phantom{r}\phi\phi} = -r (\sin\theta)^2 , \\
\Gamma^\theta_{\phantom{\theta}r\theta} &= \Gamma^\theta_{\phantom{\theta}\theta r} = \frac{1}{r}, \qquad 
			\Gamma^\theta_{\phantom{\theta}\phi\phi} = -\cos\theta \cdot \sin\theta , \\ 
\Gamma^\phi_{\phantom{\phi}r\phi} 		&= \Gamma^\phi_{\phantom{\phi}\phi r} = \frac{1}{r}, \qquad 
\Gamma^\phi_{\phantom{\phi}\theta\phi} = \Gamma^\phi_{\phantom{\phi}\phi\theta} = \cot\theta . 
\end{align}
To provide an example, let us calculate the Christoffel symbols of 2D flat space written in cylindrical coordinates $\xi^i \equiv (r,\phi)$,
\begin{align}
\dd\ell^2 = \dd r^2 + r^2 \dd\phi, \qquad r \geq 0, \ \phi \in [0,2\pi) .
\end{align}
This means the non-zero components of the metric are 
\begin{align}
g_{rr} = 1, \qquad g_{\phi\phi} = r^2,\qquad
g^{rr} = 1, \qquad g^{\phi\phi} = r^{-2} .
\end{align}
Keeping the diagonal nature of the metric in mind, let us start with
\begin{align}
\Gamma^r_{\phantom{r}ij} 
&= \frac{1}{2} g^{rk} \left(\partial_i g_{jk} + \partial_j g_{ik} - \partial_k g_{ij} \right) 
= \frac{1}{2} g^{rr} \left(\partial_i g_{jr} + \partial_j g_{ir} - \partial_r g_{ij} \right) \nonumber\\
&= \frac{1}{2} \left( \delta^r_j \partial_i g_{rr} + \delta^r_i \partial_j g_{rr} - \delta^\phi_i \delta^\phi_j \partial_r r^2 \right) 
= - \delta^\phi_i \delta^\phi_j r .
\end{align}
In the third equality we have used the fact that the only $g_{ij}$ that depends on $r$ (and therefore yield a non-zero $r$-derivative) is $g_{\phi\phi}$. Now for the 
\begin{align}
\Gamma^\phi_{\phantom{r}ij} 
&= \frac{1}{2} g^{\phi\phi} \left(\partial_i g_{j \phi} + \partial_j g_{i \phi} - \partial_\phi g_{ij} \right) \nonumber\\
&= \frac{1}{2 r^2} \left( \delta^\phi_j \partial_i g_{\phi\phi} + \delta^\phi_i \partial_j g_{\phi\phi} \right) 
= \frac{1}{2 r^2} \left( \delta^\phi_j \delta_i^r \partial_r r^2 + \delta^\phi_i \delta_j^r \partial_r r^2 \right) \nonumber\\
&= \frac{1}{r} \left( \delta^\phi_j \delta_i^r + \delta^\phi_i \delta_j^r \right) . 
\end{align}
If we had started from Cartesian coordinates $x^i$,
\begin{align}
x^i = r(\cos\phi,\sin\phi) ,
\end{align}
we know the Christoffel symbols in Cartesian coordinates are all zero, since the metric components are constant. If we wish to use  eq. \eqref{DifferentialGeometry_ChristoffelSymbols_Tranformation} to calculate the Christoffel symbols in $(r,\phi)$, the first term on the right hand side is zero and what we need are the $\partial x/\partial \xi$ and $\partial^2 x/\partial \xi \partial \xi$ matrices. The first derivative matrices are
\begin{align}
\frac{\partial x^i}{\partial \xi^j} 
&= \left[\begin{array}{cc}
\cos\phi 		& -r\sin\phi \\
\sin\phi 		& r\cos\phi \\
\end{array}
\right]^i_{\phantom{i}j} \\
\frac{\partial \xi^i}{\partial x^j} = \left(\left(\frac{\partial x}{\partial \xi}\right)^{-1}\right)^i_{\phantom{i}j}
&= \left[\begin{array}{cc}
\cos\phi 			& \sin\phi \\
-r^{-1}\sin\phi 	& r^{-1} \cos\phi \\
\end{array}
\right]^i_{\phantom{i}j} ,
\end{align}
whereas the second derivative matrices are
\begin{align}
\frac{\partial^2 x^1}{\partial \xi^i \xi^j} 
&= \left[\begin{array}{cc}
0 			& -\sin\phi \\
-\sin\phi 	& -r\cos\phi 
\end{array}
\right] \\
\frac{\partial^2 x^2}{\partial \xi^i \xi^j} 
&= \left[\begin{array}{cc}
0			& \cos\phi \\
\cos\phi 	& -r\sin\phi 
\end{array}
\right] .
\end{align}
Therefore, from eq. \eqref{DifferentialGeometry_ChristoffelSymbols_Tranformation},
\begin{align}
\Gamma^r_{\phantom{i}ij}(r,\phi)
&= \frac{\partial r}{\partial x^k} \frac{\partial x^k}{\partial \xi^i \partial \xi^j} \\
&= \cos\phi \cdot \left[\begin{array}{cc}
0 			& -\sin\phi \\
-\sin\phi 	& -r\cos\phi 
\end{array}
\right] + \sin\phi \cdot 
\left[\begin{array}{cc}
0			& \cos\phi \\
\cos\phi 	& -r\sin\phi 
\end{array}
\right]
= \left[\begin{array}{cc}
0	& 0 \\
0	& -r 
\end{array}
\right] . \nonumber
\end{align}
Similarly,
\begin{align}
\Gamma^\phi_{\phantom{i}ij}(r,\phi)
&= \frac{\partial \phi}{\partial x^k} \frac{\partial x^k}{\partial \xi^i \partial \xi^j} \\
&= -r^{-1}\sin\phi \left[\begin{array}{cc}
0 			& -\sin\phi \\
-\sin\phi 	& -r\cos\phi 
\end{array}
\right] + r^{-1} \cos\phi \left[\begin{array}{cc}
0			& \cos\phi \\
\cos\phi 	& -r\sin\phi 
\end{array}
\right] = \left[\begin{array}{cc}
0			& r^{-1} \\
r^{-1} 	& 0
\end{array}
\right] . \nonumber
\end{align} \qed
\end{myP}
\noindent{\bf Parallel transport} \qquad Let $v^i$ be a (tangent) vector field and $T^{j_1 \dots j_N}$ be some tensor. (Here, the placement of indices on the $T$ is not important, but we will assume for convenience, all of them are upper indices.) We say that the tensor $T$ is invariant under parallel transport along the vector $v$ when
\begin{align}
v^i  \nabla_i T^{j_1 \dots j_N} = 0.
\end{align}
\begin{myP}
\qquad As an example, let's calculate the Christoffel symbols of the metric on the 2-sphere with unit radius, 
\begin{align}
(\dd\ell)^2 = \dd\theta^2 + (\sin\theta)^2 \dd\phi^2 .
\end{align}
Do not calculate from scratch -- remember you have already computed the Christoffel symbols in 3D Euclidean space. How do you extract the $2$-sphere Christoffel symbols from that calculation?

In the coordinate system $(\theta,\phi)$, define the vector $v^i = (v^\theta,v^\phi) = (1,0)$, i.e., $v = \partial_\theta$. This is the vector tangent to the sphere, at a given location $(0 \leq \theta \leq \pi,0 \leq \phi < 2 \pi)$ on the sphere, such that it points away from the North and towards the South pole, along a constant longitude line. Show that it is parallel transported along itself, as quantified by the statement
\begin{align}
v^i \nabla_i v^j = \nabla_\theta v^j = 0 .
\end{align}
Also calculate $\nabla_\phi v^j$; comment on the result at $\theta = \pi/2$. Hint: recall our earlier 2-sphere discussion, where we considered parallel transporting a tangent vector from the North pole to the equator, along the equator, then back up to the North pole. \qquad \qed
\end{myP}

\noindent{\bf Variation of the metric \& divergence of tensors} \qquad If we perturb the metric slightly
\begin{align}
g_{ij} \to g_{ij} + h_{ij} ,
\end{align}
where the components of $h_{ij}$ are to be viewed as ``small", the inverse metric will become
\begin{align}
g^{ij} \to g^{ij} - h^{ij} + h^{i k} h_k^{\phantom{k} j} + \mathcal{O}\left(h^3\right) ,
\end{align}
then the square root of the determinant of the metric will change as
\begin{align}
\label{DifferentialGeometry_VariationOfRootg}
\sqrt{|g|} \to \sqrt{|g|} \left( 1 + \frac{1}{2} g^{ab} h_{ab} + \mathcal{O}(h^2) \right) .
\end{align}
\begin{myP}
\qquad Use the matrix identity, where for any square matrix $X$,
\begin{align}
\det e^X = e^{\Tr{X}} ,
\end{align}
\footnote{See, for e.g., Theorem 3.10 of \href{http://arxiv.org/abs/math-ph/0005032}{arXiv: math-ph/0005032}.}to prove eq. \eqref{DifferentialGeometry_VariationOfRootg}. (The Tr $X$ means the trace of the matrix $X$ -- sum over its diagonal terms.) Hint: Start with $\det (g_{ij} + h_{ij}) = \det (g_{ij}) \cdot \det(\delta^i_j + h^i_{\phantom{i}j})$, with $h^i_{\phantom{i}j} \equiv g^{ik} h_{kj}$. Then massage $\delta^i_j + h^i_{\phantom{i}j} = \exp(\ln(\delta^i_j + h^i_{\phantom{i}j}))$. \qed
\end{myP}
\begin{myP}
\qquad Use eq. \eqref{DifferentialGeometry_VariationOfRootg} and the definition of the Christoffel symbol to show that
\begin{align}
\label{DifferentialGeometry_PartialDonRootg}
\partial_i \ln \sqrt{|g|} = \frac{1}{2} g^{ab} \partial_i g_{ab} = \Gamma^s_{\phantom{s}i s} . 
\end{align} \qed
\end{myP}
\begin{myP}
{\bf Divergence of tensors.} \qquad Verify the following formulas for the divergence of a vector $V^i$, a fully antisymmetric rank-$(N \leq D)$ tensor $F^{i_1 i_2 \dots i_N}$ and a symmetric tensor $S^{ij} = S^{ji}$,
{\allowdisplaybreaks\begin{align}
\label{DifferentialGeometry_DivergenceOfVector}
\nabla_i V^i  &= \frac{\partial_i \left( \sqrt{|g|} V^i \right)}{\sqrt{|g|}}, \\
\label{DifferentialGeometry_DivergenceOfMaxwellF}
\nabla_j F^{j i_2 \dots i_N} &= \frac{\partial_j \left( \sqrt{|g|} F^{j i_2 \dots i_N} \right)}{\sqrt{|g|}}, \\
\label{DifferentialGeometry_DivergenceOfSymmetricS}
\nabla_i S^{ij}  &= \frac{\partial_i \left( \sqrt{|g|} S^{ij} \right)}{\sqrt{|g|}} + \Gamma^j_{\phantom{j}ab} S^{ab} .
\end{align}}
Note that, fully antisymmetric means, swapping any pair of indices costs a minus sign,
\begin{align}
F^{i_1 \dots i_{a-1} i_a i_{a+1} \dots i_{b-1} i_b i_{b+1} \dots i_N}
= -F^{i_1 \dots i_{a-1} i_b i_{a+1} \dots i_{b-1} i_a i_{b+1} \dots i_N} .
\end{align}
Comment on how these expressions, equations \eqref{DifferentialGeometry_DivergenceOfVector}-\eqref{DifferentialGeometry_DivergenceOfSymmetricS}, transform under a coordinate transformation, i.e., $\vec{x} \to \vec{x}(\vec{\xi})$. \qed
\end{myP}
{\bf Gradient of a scalar} \qquad It is worth highlighting that the gradient of a scalar, with upper indices, depends on the metric; whereas the covariant derivative on the same scalar, with lower indices, does not.
\begin{align}
\label{DifferentialGeometry_Gradient}
\nabla^i \varphi = g^{ij} \nabla_j \varphi = g^{ij} \partial_j \varphi .
\end{align}
This means, even in flat space, $\nabla^i \varphi$ is not always equal to $\nabla_i \varphi$. (They are equal in Cartesian coordinates.) For instance, in spherical coordinates $(r,\theta,\phi)$, where
\begin{align}
g^{ij} = \text{diag}(1,r^{-2},r^{-2} (\sin\theta)^{-2}) ;
\end{align}
the gradient of a scalar is
\begin{align}
\nabla^i \varphi = \left( \partial_r \varphi, r^{-2} \partial_\theta \varphi, r^{-2} (\sin\theta)^{-2} \partial_\phi \varphi\right) .
\end{align}
while the same object with lower indices is simply
\begin{align}
\nabla_i \varphi = \left( \partial_r \varphi, \partial_\theta \varphi, \partial_\phi \varphi\right) .
\end{align}
{\bf Divergence of a vector} \qquad The divergence of a vector $V^i$ is
\begin{align}
\nabla_i V^i = \nabla^i V_i .
\end{align}
{\bf Laplacian of a scalar} \qquad The Laplacian of a scalar $\psi$ can be thought of as the divergence of its gradient. In 3D vector calculus you would write is as $\vec{\nabla}^2$ but in curved spaces we may also write it as $\Box$ or $\nabla_i \nabla^i$:
\begin{align}
\label{DifferentialGeometry_Laplacian}
\Box \psi \equiv \vec{\nabla}^2 \psi = \nabla_i \nabla^i \psi = g^{ij} \nabla_i \nabla_j \psi .
\end{align}
\begin{myP}
\qquad Show that the Laplacian of a scalar can be written more explicitly in terms of the determinant of the metric and the inverse metric as
\begin{align}
\label{DifferentialGeometry_Laplacian_Detg}
\Box \psi \equiv \nabla_i \nabla^i \psi = \frac{1}{\sqrt{|g|}} \partial_i \left( \sqrt{|g|} g^{ij} \partial_j \psi \right) .
\end{align}
Hint: Start with the expansion $\nabla_i \nabla^i \psi = \partial_i \nabla^i \psi + \Gamma^i_{\phantom{i}ij} \nabla^j \psi$. \qed
\end{myP}
{\bf Levi-Civita Tensor} \qquad We have just seen how to write the divergence in any curved or flat space. We will now see that the curl from vector calculus also has a differential geometric formulation as an antisymmetric tensor, which will allow us to generalize the former to not only curved spaces but also arbitrary dimensions greater than $2$. But first, we have to introduce the Levi-Civita tensor, and with it, the Hodge dual.

In $D$ spatial dimensions we first define a Levi-Civita {\it symbol}
\begin{align}
\epsilon_{i_1 i_2 \dots i_{D-1} i_D} .
\end{align}
It is defined by the following properties.
\begin{itemize}
\item It is completely antisymmetric in its indices. This means swapping any of the indices $i_a \leftrightarrow i_b$ (for $a \neq b$) will return
\begin{align}
\epsilon_{i_1 i_2 \dots i_{a-1} i_a i_{a+1} \dots i_{b-1} i_b i_{b+1} \dots i_{D-1} i_D} 
= - \epsilon_{i_1 i_2 \dots i_{a-1} i_b i_{a+1} \dots i_{b-1} i_a i_{b+1} \dots i_{D-1} i_D} .
\end{align}
\item For a given ordering of the $D$ distinct coordinates $\{ x^i | i=1,2,3,\dots,D \}$, $\epsilon_{1 2 3 \dots D} \equiv 1$. Below, we will have more to say about this choice.
\end{itemize}
These are sufficient to define every component of the Levi-Civita symbol. From the first definition, if any of the $D$ indices are the same, say $i_a = i_b$, then the Levi-Civita symbol returns zero. (Why?) From the second definition, when all the indices are distinct, $\epsilon_{i_1 i_2 \dots i_{D-1} i_D}$ is a $+1$ if it takes even number of swaps to go from $\{ 1,\dots,D \}$ to $\{ i_1,\dots,i_D \}$; and is a $-1$ if it takes an odd number of swaps to do the same.

For example, in the (perhaps familiar) 3 dimensional case, in Cartesian coordinates $(x^1,x^2,x^3)$,
\begin{align}
1 = \epsilon_{123} = - \epsilon_{213} = -\epsilon_{321} = -\epsilon_{132} = \epsilon_{231} = \epsilon_{312} .
\end{align}
%Or, in spherical coordinates $(r,\theta,\phi)$,
%\begin{align}
%1 = \epsilon_{r\theta\phi} = - \epsilon_{\theta r\phi} = -\epsilon_{\phi\theta r} = -\epsilon_{r\phi\theta} = \epsilon_{\theta\phi r} = \epsilon_{\phi r \theta} .
%\end{align}
The Levi-Civita {\it tensor} $\widetilde{\epsilon}_{i_1 \dots i_D}$ is defined as
\begin{align}
\label{DifferentialGeometry_LeviCivitaTensor_Def}
\widetilde{\epsilon}_{i_1 i_2 \dots i_D} \equiv \sqrt{|g|} \epsilon_{i_1 i_2 \dots i_D} .
\end{align}
Let us understand why it is a (pseudo-)tensor. Because the Levi-Civita {\it symbol} is just a multi-index array of $\pm 1$ and $0$, it does not change under coordinate transformations. Equation \eqref{DifferentialGeometry_SqureRootDetg_CoordinateTransformation} then implies
\begin{align}
\label{DifferentialGeometry_LeviCivitaTensor_Proof_1}
\sqrt{|g(\vec{\xi})|} \epsilon_{a_1 a_2 \dots a_D} 
= \sqrt{\left\vert g\left( \vec{x}(\vec{\xi}) \right) \right\vert} \left\vert \det \frac{\partial x^i(\vec{\xi})}{\partial \xi^j} \right\vert \epsilon_{a_1 a_2 \dots a_D} .
\end{align}
On the right hand side, $\left\vert g\left(\vec{x}(\vec{\xi}) \right) \right\vert$ is the absolute value of the determinant of $g_{ij}$ written in the coordinates $\vec{x}$ but with $\vec{x}$ replaced with $\vec{x}(\vec{\xi})$. 

If $\widetilde{\epsilon}_{i_1 i_2 \dots i_D}$ were a tensor, on the other hand, it must obey eq. \eqref{DifferentialGeometry_GenericTensor_CoordinateTransformation},
\begin{align}
\label{DifferentialGeometry_LeviCivitaTensor_Proof_2}
\sqrt{|g(\vec{\xi})|} \epsilon_{a_1 a_2 \dots a_D} 
&\stackrel{?}{=} \sqrt{\left\vert g\left(\vec{x}(\vec{\xi}) \right) \right\vert} 
\epsilon_{i_1 \dots i_D} \frac{\partial x^{i_1}}{\partial \xi^{a_1}} \dots \frac{\partial x^{i_D}}{\partial \xi^{a_D}}, \nonumber\\
&= \sqrt{\left\vert g\left(\vec{x}(\vec{\xi}) \right) \right\vert} 
\left(\det \frac{\partial x^i}{\partial \xi^j}\right) \epsilon_{a_1 \dots a_D} ,
\end{align}
where in the second line we have recalled the co-factor expansion determinant of any matrix $M$,
\begin{align}
\label{DifferentialGeometry_CofactorExpansionDet}
\epsilon_{a_1 \dots a_D} \det M 
&= \epsilon_{i_1 \dots i_D} M^{i_1}_{\phantom{i_1}a_1} \dots M^{i_D}_{\phantom{i_D}a_D} .
\end{align}
Comparing equations \eqref{DifferentialGeometry_LeviCivitaTensor_Proof_1} and \eqref{DifferentialGeometry_LeviCivitaTensor_Proof_2} tells us the Levi-Civita $\widetilde{\epsilon}_{a_1 \dots a_D}$ transforms as a tensor for {\it orientation-preserving} coordinate transformations, namely for all coordinate transformations obeying
\begin{align}
\det \frac{\partial x^i}{\partial \xi^j}
= \epsilon_{i_1 i_2 \dots i_D} \frac{\partial x^{i_1}}{\partial \xi^1} \frac{\partial x^{i_2}}{\partial \xi^2} \dots \frac{\partial x^{i_D}}{\partial \xi^D}
> 0 .
\end{align}
{\it Parity flips} \qquad This restriction on the sign of the determinant of the Jacobian means the Levi-Civita tensor is invariant under ``parity", and is why I call it a pseudo-tensor. Parity flips are transformations that reverse the orientation of some coordinate axis, say $\xi^i \equiv -x^i$ (for some fixed $i$) and $\xi^j = x^j$ for $j \neq i$. For the Levi-Civita tensor,
\begin{align}
\sqrt{g(\vec{x})} \epsilon_{i_1 \dots i_D}
= \sqrt{g(\vec{\xi})} \left\vert \det \text{diag}[1,\dots,1,\underbrace{-1}_{i \text{th component}},1,\dots,1] \right\vert \epsilon_{i_1 \dots i_D} 
= \sqrt{g(\vec{\xi})} \epsilon_{i_1 \dots i_D} ; 
\end{align}
whereas, under the usual rules of coordinate transformations (eq. \eqref{DifferentialGeometry_GenericTensor_CoordinateTransformation}) we would have expected a `true' tensor $T_{i_1 \dots i_D}$ to behave, for instance, as
\begin{align}
T_{(1)(2)\dots(i-1)(i)(i+1)\dots(D)}(\vec{x}) \frac{\partial x^i}{\partial \xi^i}
= -T_{(1)(2)\dots(i-1)(i)(i+1)\dots(D)}(\vec{\xi}) .
\end{align}
{\it Orientation of coordinate system} \qquad What is orientation? It is the choice of how one orders the coordinates in use, say $(x^1,x^2,\dots,x^D)$, together with the convention that $\epsilon_{12 \dots D} \equiv 1$. 

In 2D flat spacetime, for example, we may choose the `right-handed' $(x^1,x^2)$ as Cartesian coordinates, $\epsilon_{12} \equiv 1$, and obtain the infinitesimal volume $\dd^2\vec{x} = \dd x^1 \dd x^2$. We can switch to cylindrical coordinates
\begin{align}
\vec{x}(\vec{\xi}) = r(\cos\phi,\sin\phi) .
\end{align}
so that
\begin{align}
\frac{\partial x^i}{\partial r} 		= (\cos\phi,\sin\phi) , \qquad
\frac{\partial x^i}{\partial \phi} 		= r(-\sin\phi,\cos\phi) , \qquad r \geq 0, \ \phi\in [0,2\pi) .
\end{align}
If we ordered $(\xi^1,\xi^2) = (r,\phi)$, we would have
\begin{align}
\epsilon_{i_1 i_2} \frac{\partial x^{i_1}}{\partial r} \frac{\partial x^{i_2}}{\partial \phi}
= \det \left[
\begin{array}{cc}
\cos\phi	& -r\sin \phi \\
\sin\phi	& r \cos\phi
\end{array}
\right] = r (\cos\phi)^2 + r (\sin\phi)^2 = r .
\end{align}
If we instead ordered $(\xi^1,\xi^2) = (\phi,r)$, we would have
\begin{align}
\epsilon_{i_1 i_2} \frac{\partial x^{i_1}}{\partial \phi} \frac{\partial x^{i_2}}{\partial r}
= \det \left[
\begin{array}{cc}
-r\sin \phi	& \cos\phi \\
r \cos\phi	& \sin\phi	
\end{array}
\right] = -r (\sin\phi)^2 - r (\cos\phi)^2 = -r .
\end{align}
We can see that going from $(x^1,x^2)$ to $(\xi^1,\xi^2) \equiv (r,\phi)$ is orientation preserving; and we should also choose $\epsilon_{r\phi} \equiv 1$.\footnote{We have gone from a `right-handed' coordinate system $(x^1,x^2)$ to a `right-handed' $(r,\phi)$; we could also have gone from a `left-handed' one $(x^2,x^1)$ to a `left-handed' $(\phi,r)$ and this would still be orientation-preserving.}
\begin{myP}
\qquad By going from Cartesian coordinates $(x^1,x^2,x^3)$ to spherical ones,
\begin{align}
\vec{x}(\vec{\xi}) = r(\sin\theta \cos\phi, \sin\theta \sin\phi, \cos\theta) ,
\end{align}
determine what is the orientation preserving ordering of the coordinates of $\vec{\xi}$, and is $\epsilon_{r\theta\phi}$ equal $+1$ or $-1$? \qed
\end{myP}
{\it Infinitesimal volume re-visited} \qquad The infinitesimal volume we encountered earlier can really be written as
\begin{align}
\dd(\text{vol.}) 
= \dd^D \vec{x} \sqrt{|g(\vec{x})|} \epsilon_{12 \dots D} 
= \dd^D \vec{x} \sqrt{|g(\vec{x})|} ,
\end{align}
so that under a coordinate transformation $\vec{x} \to \vec{x}(\vec{\xi})$, the necessarily positive infinitesimal volume written in $\vec{x}$ transforms into another positive infinitesimal volume, but written in $\vec{\xi}$:
\begin{align}
\dd^D \vec{x} \sqrt{|g(\vec{x})|} \epsilon_{12 \dots D}  
= \dd^D \vec{\xi} \sqrt{ \left\vert g( \vec{\xi} ) \right\vert } \epsilon_{12 \dots D} .
\end{align}
Below, we will see that $\dd^D \vec{x} \sqrt{|g(\vec{x})|}$ in modern integration theory is viewed as a differential $D-$form.
\begin{myP}
\qquad We may consider the infinitesimal volume in 3D flat space in Cartesian coordinates
\begin{align}
\dd(\text{vol.}) = \dd x^1 \dd x^2 \dd x^3 .
\end{align}
Now, let us switch to spherical coordinates $\vec{\xi}$, with the ordering in the previous problem. Show that it is given by
\begin{align}
\dd x^1 \dd x^2 \dd x^3 = \dd^3\vec{\xi} \sqrt{|g(\vec{\xi})|}, \qquad
\sqrt{|g(\vec{\xi})|} = \epsilon_{i_1 i_2 i_3} \frac{\partial x^{i_1}}{\partial \xi^1} \frac{\partial x^{i_2}}{\partial \xi^2} \frac{\partial x^{i_3}}{\partial \xi^3} .
\end{align} 
Can you compare $\sqrt{|g(\vec{\xi})|}$ with the volume of the parallelepiped formed by $\partial_{\xi^1} x^i$, $\partial_{\xi^2} x^i$ and $\partial_{\xi^3} x^i$?\footnote{Because of the existence of locally flat coordinates $\{y^i\}$, the interpretation of $\sqrt{|g(\xi)|}$ as the volume of parallelepiped formed by $\{ \partial_{\xi^1} y^i, \dots, \partial_{\xi^D} y^i \}$ actually holds very generally.} \qed
\end{myP}
{\it Cross-Product in Flat 3D, Right-hand rule} \qquad Notice the notion of orientation in 3D is closely tied to the ``right-hand rule" in vector calculus. Let $\vec{X}$ and $\vec{Y}$ be vectors in Euclidean 3-space. In Cartesian coordinates, where $g_{ij} = \delta_{ij}$, you may check that their cross product is
\begin{align}
\left(\vec{X} \times \vec{Y}\right)^k = \epsilon^{ijk} X^i Y^j. 
\end{align}
For example, if $\vec{X}$ is parallel to the positive $x^1$ axis and $\vec{Y}$ parallel to the positive $x^2$-axis, so that $\vec{X}=|\vec{X}|(1,0,0)$ and $\vec{Y}=|\vec{Y}|(0,1,0)$, the cross product reads
\begin{align}
\left(\vec{X} \times \vec{Y}\right)^k \to |\vec{X}||\vec{Y}| \epsilon^{12 k} = |\vec{X}||\vec{Y}| \delta^k_3 ,
\end{align}
i.e., it is parallel to the positive $x^3$ axis. (Remember $k$ cannot be either $1$ or $2$ because $\epsilon^{ijk}$ is fully antisymmetric.) If we had chosen $\epsilon_{123} = \epsilon^{123} \equiv -1$, then the cross product would become the ``left-hand rule". Below, I will continue to point out, where appropriate, how this issue of orientation arises in differential geometry.
\begin{myP}
\qquad Show that the Levi-Civita tensor with all upper indices is given by
\begin{align}
\label{DifferentialGeometry_LeviCivita_UpperIndices}
\widetilde{\epsilon}^{i_1 i_2 \dots i_D} = \frac{\text{sgn } \det (g_{ab})}{\sqrt{|g|}} \epsilon_{i_1 i_2 \dots i_D} .
\end{align}
In curved spaces, the sign of the $\det g_{ab} = 1$; whereas in curved spacetimes it depends on the signature used for the flat metric.\footnote{See eq. \eqref{DifferentialGeometry_detg_OrthonormalFrame} to understand why the sign of the determinant of the metric is always determined by the sign of the determinant of its flat counterpart.} Hint: Raise the indices by contracting with inverse metrics, then recall the cofactor expansion definition of the determinant. \qed
\end{myP}
\begin{myP}
\qquad Show that the covariant derivative of the Levi-Civita tensor is zero.
\begin{align}
\nabla_j \widetilde{\epsilon}_{i_1 i_2 \dots i_D} = 0 .
\end{align}
(Hint: Start by expanding the covariant derivative in terms of Christoffel symbols; then go through some combinatoric reasoning or invoke the equivalence principle.) From this, explain why the following equalities are true; for some vector $V$,
\begin{align}
\nabla_j \left( \widetilde{\epsilon}^{i_1 i_2 \dots i_{D-2} jk} V_k \right)
= \widetilde{\epsilon}^{i_1 i_2 \dots i_{D-2} jk} \nabla_j V_k
= \widetilde{\epsilon}^{i_1 i_2 \dots i_{D-2} jk} \partial_j V_k .
\end{align}
Why is $\nabla_i V_j - \nabla_j V_i = \partial_i V_j - \partial_j V_i$ for any $V_i$? Hint: expand the covariant derivatives in terms of the partial derivatives and the Christoffel symbols. \qed
\end{myP}
\noindent{\bf Combinatorics} \qquad This is an appropriate place to state how to actually construct a fully antisymmetric tensor from a given tensor $T_{i_1 \dots i_N}$. Denoting $\Pi(i_1 \dots i_N)$ to be a permutation of the indices $\{i_1 \dots i_N\}$, the antisymmetrization procedure is given by
\begin{align}
\label{DifferentialGeometry_Antisymmetrize}
T_{[i_1 \dots i_N]} 
&= \sum_{\text{permutations $\Pi$ of }\{i_1,i_2,\dots,i_N\}}^{N!} \sigma_\Pi \cdot T_{\Pi(i_1 \dots i_N)} \\
&= \sum_{\text{even permutations $\Pi$ of }\{i_1,i_2,\dots,i_N\}} T_{\Pi(i_1 \dots i_N)} 
- \sum_{\text{odd permutations $\Pi$ of }\{i_1,i_2,\dots,i_N\}} T_{\Pi(i_1 \dots i_N)} . \nonumber
\end{align}
In words: for a rank$-N$ tensor, $T_{[i_1 \dots i_N]}$ consists of a sum of $N!$ terms. The first is $T_{i_1 \dots i_N}$. Each and every other term consists of $T$ with its indices permuted over all the $N!-1$ distinct remaining possibilities, multiplied by $\sigma_\Pi = +1$ if it took even number of index swaps to get to the given permutation, and $\sigma_\Pi = -1$ if it took an odd number of swaps. (The $\sigma_\Pi$ is often called the sign of the permutation $\Pi$.) For example,
\begin{align}
T_{[ij]} = T_{ij} - T_{ji},\qquad
T_{[ijk]} = T_{ijk}-T_{ikj}-T_{jik}+T_{jki}+T_{kij}-T_{kji} .
\end{align}
Can you see why eq. \eqref{DifferentialGeometry_Antisymmetrize} yields a fully antisymmetric object? Consider any pair of distinct indices, say $i_a$ and $i_b$, for $1 \leq (a \neq b) \leq N$. Since the sum on its right hand side contains every permutation (multiplied by the sign) -- we may group the terms in the sum of eq. \eqref{DifferentialGeometry_Antisymmetrize} into pairs, say $\sigma_{\Pi_\ell} T_{j_1 \dots i_a \dots i_b \dots j_N} - \sigma_{\Pi_\ell} T_{j_1 \dots i_b \dots i_a \dots j_N}$. That is, for a given term $\sigma_{\Pi_\ell} T_{j_1 \dots i_a \dots i_b \dots j_N}$ there must be a counterpart with $i_a \leftrightarrow i_b$ swapped, multipled by a minus sign, because -- if the first term involved even (odd) number of swaps to get to, then the second must have involved an odd (even) number. If we now considered swapping $i_a \leftrightarrow i_b$ in every term in the sum on the right hand side of eq. \eqref{DifferentialGeometry_Antisymmetrize},
\begin{align}
T_{[i_1 \dots i_a \dots i_b \dots i_N]} 
&= \sigma_{\Pi_\ell} T_{j_1 \dots i_a \dots i_b \dots j_N} - \sigma_{\Pi_\ell} T_{j_1 \dots i_b \dots i_a \dots j_N} + \dots , \\
T_{[i_1 \dots i_b \dots i_a \dots i_N]} 
&= -\left(\sigma_{\Pi_\ell} T_{j_1 \dots i_a \dots i_b \dots j_N} - \sigma_{\Pi_\ell} T_{j_1 \dots i_b \dots i_a \dots j_N} + \dots\right) .
\end{align}
%\begin{myP}
%	If $T_{i_1 \dots i_N}$ is fully antisymmetric, explain why
%	\begin{align}
%	T_{[i_1 \dots i_N]} = N! T_{i_1 \dots i_N} .
%	\end{align} \qed
%\end{myP}
\begin{myP}
\qquad Given $T_{i_1 i_2 \dots i_N}$, how do we construct a fully symmetric object from it, i.e., such that swapping any two indices returns the same object? \qed
\end{myP}
\begin{myP}
\qquad If the Levi-Civita symbol is subject to the convention $\epsilon_{12\dots D} \equiv 1$, explain why it is equivalent to the following expansion in Kronecker $\delta$s.
\begin{align}
\label{DifferentialGeometry_LeviCivitaSymbol_KroneckerExpansion}
\epsilon_{i_1 i_2 \dots i_D} = \delta^1_{[i_1} \delta^2_{i_2} \dots \delta^{D-1}_{i_{D-1}} \delta^D_{i_D]} 
\end{align}
Can you also explain why the following is true?
\begin{align}
\label{DifferentialGeometry_LeviCivitaSymbol_CofactorExpansionDet}
\epsilon_{a_1 a_2 \dots a_{D-1} a_D} \det A 
= \epsilon_{i_1 i_2 \dots i_{D-1} i_D} A^{i_1}_{\phantom{i_1}a_1} A^{i_2}_{\phantom{i_2}a_2} \dots A^{i_{D-1}}_{\phantom{i_{D-1}}a_{D-1}} A^{i_D}_{\phantom{i_D}a_D} 
\end{align}
\end{myP}
\begin{myP}
\qquad Argue that
\begin{align}
\label{DifferentialGeometry_Combinatorics_I}
T_{[i_1 \dots i_N]} 
= T_{[i_1 \dots i_{N-1}] i_N} 
- T_{[i_N i_2 \dots i_{N-1}] i_1} 
&- T_{[i_1 i_N i_3 \dots i_{N-1}] i_2} \\
- T_{[i_1 i_2 i_N i_4 \dots i_{N-1}] i_3} 
&- \dots - T_{[i_1 \dots i_{N-2} i_N] i_{N-1}} . \nonumber
\end{align} \qed
\end{myP}
{\bf Product of Levi-Civita tensors} \qquad The product of two Levi-Civita tensors will be important for the discussions to come. We have
\begin{align}
\label{DifferentialGeometry_LeviCivita_Product}
\widetilde{\epsilon}^{i_1 \dots i_N k_1 \dots k_{D-N}} \widetilde{\epsilon}_{j_1 \dots j_N k_1 \dots k_{D-N}} 
	&= \text{sgn}\det (g_{ab}) \cdot A_N \delta^{i_1}_{[j_1} \dots \delta^{i_N}_{j_N]}, \qquad 1 \leq N \leq D, \\
\widetilde{\epsilon}^{k_1 \dots k_{D}} \widetilde{\epsilon}_{k_1 \dots k_{D}} 
	&= \text{sgn}\det (g_{ab}) \cdot A_0, \qquad A_{N \geq 0} \equiv (D-N)! .
\end{align}
(Remember $0! = 1! = 1$; also, $\delta^{i_1}_{[j_1} \dots \delta^{i_N}_{j_N]} = \delta^{[i_1}_{j_1} \dots \delta^{i_N]}_{j_N}$.) Let us first understand why there are a bunch of Kronecker deltas on the right hand side, starting from the $N=D$ case -- where no indices are contracted. 
\begin{align}
\label{DifferentialGeometry_LeviCivita_ProductNoContract}
\text{sgn}\det (g_{ab}) \widetilde{\epsilon}^{i_1 \dots i_D} \widetilde{\epsilon}_{j_1 \dots j_D} 
= \epsilon_{i_1 \dots i_D} \epsilon_{j_1 \dots j_D} 
= \delta^{i_1}_{[j_1} \dots \delta^{i_D}_{j_D]} 
\end{align}
(This means $A_D=1$.) The first equality follows from eq. \eqref{DifferentialGeometry_LeviCivita_UpperIndices}. The second may seem a bit surprising, because the indices $\{i_1,\dots,i_D\}$ are attached to a completely different $\widetilde{\epsilon}$ tensor from the $\{j_1,\dots,j_D\}$. However, if we manipulate
\begin{align}
\delta^{i_1}_{[j_1} \dots \delta^{i_D}_{j_D]} 
= \delta^{i_1}_{[1} \dots \delta^{i_D}_{D]} \sigma_j 
= \delta^{1}_{[1} \dots \delta^{D}_{D]} \sigma_i \sigma_j
= \sigma_i \sigma_j 
= \epsilon_{i_1 \dots i_D} \epsilon_{j_1 \dots j_D} ,
\end{align}
where $\sigma_i = 1$ if it took even number of swaps to re-arrange $\{i_1,\dots,i_D\}$ to $\{1,\dots,D\}$ and $\sigma_i=-1$ if it took odd number of swaps; similarly, $\sigma_j = 1$ if it took even number of swaps to re-arrange $\{j_1,\dots,j_D\}$ to $\{1,\dots,D\}$ and $\sigma_j=-1$ if it took odd number of swaps. But $\sigma_i$ is precisely the Levi-Civita {\it symbol} $\epsilon_{i_1 \dots i_D}$ and likewise $\sigma_j = \epsilon_{j_1 \dots j_D}$. The $(\geq 1)$-contractions between the $\widetilde{\epsilon}$s can, in principle, be obtained by contracting the right hand side of \eqref{DifferentialGeometry_LeviCivita_ProductNoContract}. Because one contraction of the $(N+1)$ Kronecker deltas have to return $N$ Kronecker deltas, by induction, we now see why the right hand side of eq. \eqref{DifferentialGeometry_LeviCivita_Product} takes the form it does for any $N$. 

What remains is to figure out the actual value of $A_N$. We will do so recursively, by finding a relationship between $A_N$ and $A_{N-1}$. We will then calculate $A_1$ and use it to generate all the higher $A_N$s. Starting from eq. \eqref{DifferentialGeometry_LeviCivita_Product}, and employing eq. \eqref{DifferentialGeometry_Combinatorics_I},
\begin{align}
&\widetilde{\epsilon}^{i_1 \dots i_{N-1} \sigma k_1 \dots k_{D-N}} \widetilde{\epsilon}_{j_1 \dots j_{N-1} \sigma k_1 \dots k_{D-N}} 
= A_N \delta^{i_1}_{[j_1} \dots \delta^{i_{N-1}}_{j_{N-1}} \delta^{\sigma}_{\sigma]}  \\
&= A_N \left(
\delta^{i_1}_{[j_1} \dots \delta^{i_{N-1}}_{j_{N-1}]} \delta^{\sigma}_{\sigma} 
- \delta^{i_1}_{[\sigma} \delta^{i_2}_{j_2} \dots \delta^{i_{N-1}}_{j_{N-1}]} \delta^{\sigma}_{j_1}
- \delta^{i_1}_{[j_1} \delta^{i_2}_{\sigma} \delta^{i_3}_{j_3} \dots \delta^{i_{N-1}}_{j_{N-1}]} \delta^{\sigma}_{j_2}
- \dots - \delta^{i_1}_{[j_1} \dots \delta^{i_{N-2}}_{j_{N-2}} \delta^{i_{N-1}}_{\sigma]} \delta^{\sigma}_{j_{N-1}}
\right)  \nonumber\\
&= A_N \cdot (D - (N-1)) \delta^{i_1}_{[j_1} \dots \delta^{i_{N-1}}_{j_{N-1}]} \equiv A_{N-1} \delta^{i_1}_{[j_1} \dots \delta^{i_{N-1}}_{j_{N-1}]} . \nonumber
\end{align}
(The last equality is a definition, because $A_{N-1}$ {\it is} the coefficient of $\delta^{i_1}_{[j_1} \dots \delta^{i_{N-1}}_{j_{N-1}]}$.) We have the relationship
\begin{align}
A_N = \frac{A_{N-1}}{D - (N-1)} .
\end{align}
If we contract every index, we have to sum over all the $D!$ (non-zero components of the Levi-Civita symbol)$^2$,
\begin{align}
\label{A0}
\widetilde{\epsilon}^{i_1 \dots i_D} \widetilde{\epsilon}_{i_1 \dots i_D} 
= \text{sgn}\det (g_{ab}) \cdot \sum_{i_1,\dots,i_D} (\epsilon_{i_1 \dots i_D})^2
= \text{sgn}\det (g_{ab}) \cdot D!
\end{align}
That means $A_0 = D!$. If we contracted every index but one,
\begin{align}
\widetilde{\epsilon}^{i k_1 \dots k_D} \widetilde{\epsilon}_{j k_1 \dots k_D} &= \text{sgn}\det (g_{ab}) A_1 \delta^i_j .
\end{align}
Contracting the $i$ and $j$ indices, and invoking eq. \eqref{A0},
\begin{align}
\text{sgn}\det (g_{ab}) \cdot D! &= \text{sgn}\det (g_{ab}) A_1 \cdot D \qquad \Rightarrow \qquad A_1 = (D-1)!. 
\end{align}
That means we may use $A_1$ (or, actually, $A_0$) to generate all other $A_{N \geq 0}$s,
{\allowdisplaybreaks\begin{align}
A_N &= \frac{A_{N-1}}{(D-(N-1))} = \frac{1}{D-(N-1)} \frac{A_{N-2}}{D-(N-2)} = \dots \nonumber\\
&= \frac{A_1}{(D-1)(D-2)(D-3) \dots (D-(N-1))} = \frac{(D-1)!}{(D-1)(D-2)(D-3) \dots (D-(N-1))}  \nonumber\\
&= \frac{(D-1)(D-2)(D-3) \dots (D-(N-1))(D-N)(D-(N+1)) \dots 3 \cdot 2 \cdot 1}{(D-1)(D-2)(D-3) \dots (D-(N-1))} \nonumber\\
&= (D-N)!.
\end{align}}
Note that $0! = 1$, so $A_D=1$ as we have found earlier.
\begin{myP}
{\bf Matrix determinants revisited} \qquad Explain why the cofactor expansion definition of a square matrix in eq. \eqref{DetOfMatrix} can also be expressed as
\begin{align}
\det A = \epsilon^{i_1 i_2 \dots i_{D-1} i_D} A^{1}_{\phantom{1}i_1} A^{2}_{\phantom{2}i_2} \dots A^{D-1}_{\phantom{D-1}i_{D-1}} A^{D}_{\phantom{D}i_D} 
\end{align}
provided we define $\epsilon^{i_1 i_2 \dots i_{D-1} i_D}$ in the same way we defined its lower index counterpart, including $\epsilon^{123 \dots D} \equiv 1$. That is, why can we cofactor expand about either the rows or the columns of a matrix, to obtain its determinant? What does that tell us about the relation $\det A^T = \det A$? Can you also prove, using our result for the product of two Levi-Civita symbols, that $\det(A \cdot B) = (\det A)(\det B)$? \qed
\end{myP}
\begin{myP}
\qquad In 3D vector calculus, the curl of a gradient of a scalar is zero -- how would you express that using the $\widetilde{\epsilon}$ tensor? What about the statement that the divergence of a curl of a vector field is zero? Can you also derive, using the $\widetilde{\epsilon}$ tensor in Cartesian coordinates and eq. \eqref{DifferentialGeometry_LeviCivita_Product}, the 3D vector cross product identity
\begin{align}
\vec{A} \times (\vec{B} \times \vec{C}) = (\vec{A} \cdot \vec{C}) \vec{B} - (\vec{A} \cdot \vec{B}) \vec{C} ? 
\end{align} \qed
\end{myP}
{\bf Hodge dual} \qquad We are now ready to define the Hodge dual. Given a fully antisymmetric rank-$N$ tensor $T_{i_1 \dots i_N}$, its Hodge dual -- which I shall denote as $\widetilde{T}^{j_1 \dots j_{D-N}}$ -- is a fully antisymmetric rank-$(D-N)$ tensor whose components are
\begin{align}
\label{DifferentialGeometry_HodgeDual}
\widetilde{T}^{j_1 \dots j_{D-N}} \equiv \frac{1}{N!} \widetilde{\epsilon}^{j_1 \dots j_{D-N} i_1 \dots i_N} T_{i_1 \dots i_N} .
\end{align}
\begin{quotation}
{\it Invertible} \qquad Note that the Hodge dual is an invertible operation, as long as we are dealing with fully antisymmetric tensors, in that given $\widetilde{T}^{j_1 \dots j_{D-N}}$ we can recover $T_{i_1 \dots i_N}$ and vice versa.\footnote{The fully antisymmetric property is crucial here: any symmetric portion of a tensor contracted with the Levi-Civita tensor would be lost. For example, an arbitrary rank-2 tensor can always be decomposed as $T_{ij} = (1/2) T_{\{ij\}} + (1/2) T_{[ij]}$; then, $\widetilde{\epsilon}^{i_1 \dots i_{D-2} jk} T_{jk} = \widetilde{\epsilon}^{i_1 \dots i_{D-2} jk} ( (1/2) T_{\{jk\}} + (1/2) T_{[jk]} ) = (1/2) \widetilde{\epsilon}^{i_1 \dots i_{D-2} jk} T_{[jk]}$. The symmetric part is lost because $\widetilde{\epsilon}^{i_1 \dots i_{D-2} jk} T_{\{jk\}} = -\widetilde{\epsilon}^{i_1 \dots i_{D-2} kj} T_{\{kj\}}$.} All you have to do is contract both sides with the Levi-Civita tensor, namely
\begin{align}
\label{DifferentialGeometry_HodgeDualInverse}
T_{i_1 \dots i_N} 
= \frac{(-)^{N(D-N)}}{(D-N)!} \widetilde{\epsilon}_{j_1 \dots j_{D-N} i_1 \dots i_N} \widetilde{T}^{j_1 \dots j_{D-N}} .
\end{align}
In other words $\widetilde{T}^{j_1 \dots j_{D-N}}$ and $T_{i_1 \dots i_N}$ contain the same amount of information.
\end{quotation}
\begin{myP}
\qquad Using eq. \eqref{DifferentialGeometry_LeviCivita_Product}, verify the proportionality constant $(-)^{N(D-N)}/(D-N)!$ in the inverse Hodge dual of eq. \eqref{DifferentialGeometry_HodgeDualInverse}, and thereby prove that the Hodge dual is indeed invertible for fully antisymmetric tensors. \qed
\end{myP}
{\bf Curl} \qquad The curl of a vector field $A_i$ can now either be defined as the antisymmetric rank-2 tensor
\begin{align}
\label{DifferentialGeometry_MagneticField}
F_{ij} \equiv \partial_{[i} A_{j]}
\end{align}
or its rank-$(D-2)$ Hodge dual
\begin{align}
\label{DifferentialGeometry_MagneticField_HodgeDual}
\widetilde{F}^{i_1 i_2 \dots i_{D-2}} \equiv \frac{1}{2} \widetilde{\epsilon}^{i_1 i_2 \dots i_{D-2} jk} \partial_{[j} A_{k]} .
\end{align}
($D=3$)-dimensional space is a special case where both the original vector field $A^i$ and the Hodge dual $\widetilde{F}^{i}$ are rank-$1$ tensors. This is usually how electromagnetism is taught: that in 3D the magnetic field is a vector arising from the curl of the vector potential $A_i$:
\begin{align}
\label{DifferentialGeometry_MagneticField_3D}
B^k 
= \frac{1}{2} \widetilde{\epsilon}^{ijk} \partial_{[j} A_{k]} 
= \widetilde{\epsilon}^{ijk} \partial_{j} A_{k} .
\end{align}
In particular, when we specialize to 3D flat space with Cartesian coordinates:
\begin{align}
\left(\vec{\nabla} \times \vec{A}\right)^i &= \epsilon^{ijk} \partial_j A_k , \qquad \text{(Flat 3D Cartesian)}. \\
\left(\vec{\nabla} \times \vec{A}\right)^1 &= \epsilon^{1 23} \partial_2 A_3 + \epsilon^{1 32} \partial_3 A_2 = \partial_2 A_3 - \partial_3 A_2, \qquad \text{etc.}
\end{align}
By setting $i=1,2,3$ we can recover the usual definition of the curl in 3D vector calculus. But you may have noticed from equations \eqref{DifferentialGeometry_MagneticField} and \eqref{DifferentialGeometry_MagneticField_HodgeDual}, in any other dimension, that the magnetic field is really not a (rank$-1$) vector but should be viewed either as a rank$-2$ curl or a rank$-(D-2)$ Hodge dual of this curl. \qquad \qed

{\it Divergence versus Curl} \qquad We can extend the definition of a curl of a vector field to that of a rank$-N$ fully antisymmetric $B_{i_1 \dots i_N}$ as
\begin{align}
\nabla_{[\sigma} B_{i_1 \dots i_N]} = \partial_{[\sigma} B_{i_1 \dots i_N]} .
\end{align}
(Can you explain why the $\nabla$ can be replaced with $\partial$?) With the Levi-Civita tensor, we can convert the curl of an antisymmetric tensor into the divergence of its dual,
\begin{align}
\label{DifferentialGeometry_DivVsCurl}
\nabla_\sigma \widetilde{B}^{j_1 \dots j_{D-N-1} \sigma}
&= \frac{1}{N!} \widetilde{\epsilon}^{j_1 \dots j_{D-N-1} \sigma i_1 \dots i_N} \nabla_{\sigma} B_{i_1 \dots i_N} \\
&= (N+1) \cdot \widetilde{\epsilon}^{j_1 \dots j_{D-N-1} \sigma i_1 \dots i_N} \partial_{[\sigma} B_{i_1 \dots i_N]} .
\end{align}
\begin{myP}
\qquad Show, by contracting both sides of eq. \eqref{DifferentialGeometry_MagneticField_3D} with an appropriate $\widetilde{\epsilon}$-tensor, that
\begin{align}
\widetilde{\epsilon}_{ijk} B^k = 2 \partial_{[i} A_{j]} .
\end{align}
Assume $\text{sgn } \det (g_{ab}) = 1$. \qed
\end{myP}
\begin{myP}
\qquad In $D$-dimensional space, is the Hodge dual of a rank-$D$ fully antisymmetric tensor $F_{i_1 \dots i_D}$ invertible? Hint: If $F_{i_1 \dots i_D}$ is fully antisymmetric, how many independent components does it have? Can you use that observation to relate $\widetilde{F}$ and $F_{i_1 \dots i_D}$ in
\begin{align}
\widetilde{F} \equiv \frac{1}{D!} \widetilde{\epsilon}^{i_1 \dots i_D} F_{i_1 \dots i_D} ?
\end{align} \qed
\end{myP}
\begin{myP}
{\bf Curl, divergence and all that} \qquad The electromagnetism textbook by J.D.Jackson contains on its very last page explicit forms of the gradient and Laplacian of a scalar as well as divergence and curl of a vector -- in Cartesian, cylindrical, and spherical coordinates in 3-dimensional flat space. Can you derive them with differential geometric techniques? Note that the vectors there are expressed in an orthonormal basis.

{\it Cartesian coordinates} \qquad In Cartesian coordinates $\{ x^1,x^2,x^3 \} \in \mathbb{R}^3$, we have the metric
\begin{align}
\dd \ell^2 = \delta_{ij} \dd x^i \dd x^j .
\end{align}
Show that the gradient of a scalar $\psi$ is
\begin{align}
\vec{\nabla} \psi 
= (\partial_1 \psi, \partial_2 \psi, \partial_3 \psi)
= (\partial^1 \psi, \partial^2 \psi, \partial^3 \psi);
\end{align}
the Laplacian of a scalar $\psi$ is
\begin{align}
\nabla_i \nabla^i \psi 
= \delta^{ij} \partial_i \partial_j \psi 
= \left( \partial_1^2 + \partial_2^2 + \partial_3^2 \right) \psi ;
\end{align}
the divergence of a vector $A$ is
\begin{align}
\nabla_i A^i = \partial_i A^i ;
\end{align}
and the curl of a vector $A$ is
\begin{align}
(\vec{\nabla} \times \vec{A})^i = \epsilon^{ijk} \partial_j A_k .
\end{align}
{\it Cylindrical coordinates} \qquad In cylindrical coordinates $\{\rho \geq 0, 0 \leq \phi < 2\pi, z \in \mathbb{R} \}$, employ the following parametrization for the Cartesian components of the 3D Euclidean coordinate vector
\begin{align}
\vec{x} = \left( \rho \cos\phi, \rho \sin\phi, z \right)
\end{align}
to argue that the flat metric is translated from $g_{ij} = \delta_{ij}$ to
\begin{align}
\dd \ell^2 = \dd \rho^2 + \rho^2 \dd\phi^2 + \dd z^2 .
\end{align}
Show that the gradient of a scalar $\psi$ is
\begin{align}
\nabla^{\widehat{\rho}} \psi = \partial_\rho \psi, \qquad
\nabla^{\widehat{\phi}} \psi = \frac{1}{\rho}\partial_\phi \psi, \qquad
\nabla^{\widehat{z}} \psi = \partial_z \psi ;
\end{align}
the Laplacian of a scalar $\psi$ is
\begin{align}
\nabla_i \nabla^i \psi = \frac{1}{\rho} \partial_\rho \left( \rho \partial_\rho \psi \right) 
+ \frac{1}{\rho^2} \partial_\phi^2 \psi + \partial_z^2 \psi ;
\end{align}
the divergence of a vector $A$ is
\begin{align}
\nabla_i A^i = 
\frac{1}{\rho} \left(\partial_\rho \left( \rho A^{\widehat{\rho}} \right) + \partial_\phi A^{\widehat{\phi}} \right)
+ \partial_z A^{\widehat{z}} ;
\end{align}
and the curl of a vector $A$ is
\begin{align}
\widetilde{\epsilon}^{\widehat{\rho} jk} \partial_j A_k = \frac{1}{\rho} \partial_\phi A^{\widehat{z}} - \partial_z A^{\widehat{\phi}}, \qquad
\widetilde{\epsilon}^{\widehat{\phi} jk} \partial_j A_k = \partial_z A^{\widehat{\rho}} - \partial_\rho A^{\widehat{z}}, \nonumber\\
\widetilde{\epsilon}^{\widehat{z} jk} \partial_j A_k = \frac{1}{\rho} \left(\partial_\rho \left( \rho A^{\widehat{\phi}} \right) - \partial_\phi A^{\widehat{\rho}} \right) .
\end{align}
{\it Spherical coordinates} \qquad In spherical coordinates $\{ r \geq 0, 0 \leq \theta \leq \pi, 0 \leq \phi < 2\pi\}$ the Cartesian components of the 3D Euclidean coordinate vector reads
\begin{align}
\vec{x} = \left( r \sin(\theta) \cos(\phi), r \sin(\theta) \sin(\phi), r \cos(\theta) \right) .
\end{align}
Show that the flat metric is now
\begin{align}
\dd \ell^2 = \dd r^2 + r^2 \left( \dd\theta^2 + (\sin\theta)^2 \dd\phi^2 \right) ;
\end{align}
the gradient of a scalar $\psi$ is
\begin{align}
\nabla^{\widehat{r}} \psi = \partial_r \psi, \qquad
\nabla^{\widehat{\theta}} \psi = \frac{1}{r}\partial_\theta \psi, \qquad
\nabla^{\widehat{\phi}} \psi = \frac{1}{r \sin\theta} \partial_\phi \psi ;
\end{align}
the Laplacian of a scalar $\psi$ is
\begin{align}
\nabla_i \nabla^i \psi = \frac{1}{r^2} \partial_r \left( r^2 \partial_r \psi \right) 
+ \frac{1}{r^2 \sin\theta} \partial_\theta \left(\sin\theta \cdot \partial_\theta \psi\right) 
+ \frac{1}{r^2 (\sin\theta)^2} \partial_\phi^2 \psi ;
\end{align} 
the divergence of a vector $A$ reads
\begin{align}
\nabla_i A^i = 
\frac{1}{r^2} \partial_r \left( r^2 A^{\widehat{r}} \right) + \frac{1}{r \sin\theta} \partial_\theta \left( \sin\theta \cdot A^{\widehat{\theta}} \right)
+ \frac{1}{r \sin\theta} \partial_\phi A^{\widehat{\phi}} ;
\end{align}
and the curl of a vector $A$ is given by
\begin{align}
\widetilde{\epsilon}^{\widehat{r} jk} \partial_j A_k 
	= \frac{1}{r \sin\theta} \left(\partial_\theta (\sin\theta \cdot A^{\widehat{\phi}}) - \partial_\phi A^{\widehat{\theta}}\right), \qquad
\widetilde{\epsilon}^{\widehat{\theta} jk} \partial_j A_k 
	= \frac{1}{r \sin\theta} \partial_\phi A^{\widehat{r}} - \frac{1}{r} \partial_r (r A^{\widehat{\phi}} ), \nonumber\\
\widetilde{\epsilon}^{\widehat{\phi} jk} \partial_j A_k 
	= \frac{1}{r} \left(\partial_r \left( r A^{\widehat{\theta}} \right) - \partial_\theta A^{\widehat{r}} \right).
\end{align} \qed
\end{myP}

\subsection{Hypersurfaces}

\subsubsection{Induced Metrics}

There are many physical and mathematical problems where we wish to study some $(N < D)$-dimensional (hyper)surface residing (aka embedded) in a $D$ dimensional ambient space. One way to describe this surface is to first endow it with $N$ coordinates $\{\xi^\text{I} | \text{I} = 1,2,\dots,N\}$, whose indices we will denote with capital letters to distinguish from the $D$ coordinates $\{x^i\}$ parametrizing the ambient space. Then the position of the point $\vec{\xi}$ on this hypersurface in the ambient perspective is given by $\vec{x}(\vec{\xi})$. Distances on this hypersurface can be measured using the ambient metric by restricting the latter on the former, i.e.,
\begin{align}
\label{DifferentialGeometry_InducedMetric}
g_{ij} \dd x^i \dd x^j 
&\to g_{ij}\left(\vec{x}(\vec{\xi})\right) \frac{\partial x^i(\vec{\xi})}{\partial \xi^\text{I}} \frac{\partial x^j(\vec{\xi})}{\partial \xi^\text{J}} 
\dd \xi^\text{I} \dd \xi^\text{J} 
\equiv H_\text{IJ}(\vec{\xi}) \dd \xi^\text{I} \dd \xi^\text{J} .
\end{align}
The $H_\text{IJ}$ is the (induced) metric on the hypersurface.\footnote{The Lorentzian signature of curved space{\it time}s, as opposed to the Euclidean one in curved spaces, complicates the study of hypersurfaces in the former. One has to distinguish between timelike, spacelike and null surfaces. For a pedagogical discussion see Eric Poisson's {\it A Relativist's Toolkit} -- in fact, much of the material in this section is heavily based on its Chapter 3. Note, however, it is not necessary to know General Relativity to study hypersurfaces in curved spacetimes.}

Observe that the $N$ vectors
\begin{align}
\left. \left\{ \frac{\partial x^i}{\partial \xi^\text{I}} \partial_i \right\vert \text{I} = 1,2,\dots,N \right\} ,
\end{align}
are tangent to this hypersurface. They form a basis set of tangent vectors at a given point $\vec{x}(\vec{\xi})$, but from the ambient $D$-dimensional perspective. On the other hand, the $\partial/\partial \xi^\text{I}$ themselves form a basis set of tangent vectors, from the perspective of an observer confined to live on this hypersurface.

{\bf Example} \qquad A simple example is provided by the $2$-sphere of radius $R$ embedded in 3D flat space. We already know that it can be parametrized by two angles $\xi^\text{I} \equiv (0 \leq \theta \leq \pi, 0 \leq \phi < 2\pi)$, such that from the ambient perspective, the sphere is described by
\begin{align}
x^i(\vec{\xi}) = R (\sin\theta \cos\phi, \sin\theta \sin\phi, \cos\theta) , \qquad \text{(Cartesian components)}.
\end{align}
(Remember $R$ is a fixed quantity here.) The induced metric on the sphere itself, according to eq. \eqref{DifferentialGeometry_InducedMetric}, will lead us to the expected result
\begin{align}
\label{DifferentialGeometry_InducedMetric_2Sphere}
H_\text{IJ}(\vec{\xi}) \dd \xi^\text{I} \dd \xi^\text{J} = R^2 \left( \dd\theta^2 + (\sin\theta)^2 \dd\phi^2 \right) .
\end{align}
{\bf Area of 2D surface in 3D flat space} \qquad A common vector calculus problem is to give some function $f(x,y)$ of two variables, where $x$ and $y$ are to be interpreted as Cartesian coordinates on a flat plane; then proceed to ask what its area is for some specified domain on the $(x,y)$-plane. We see such a problem can be phrased as a differential geometric one. First, we view $f$ as the $z$ coordinate of some hypersurface embedded in 3-dimensional flat space, so that
\begin{align}
X^i \equiv (x,y,z) = (x,y,f(x,y)) .
\end{align}
The tangent vectors $(\partial X^i/\partial \xi^\text{I})$ are
\begin{align}
\frac{\partial X^i}{\partial x} = \left(1,0,\partial_x f\right), \qquad
\frac{\partial X^i}{\partial y} = \left(0,1,\partial_y f\right) .
\end{align}
The induced metric, according to eq. \eqref{DifferentialGeometry_InducedMetric}, is given by
\begin{align}
\label{DifferentialGeometry_Areaoff(x,y)}
H_\text{IJ}(\vec{\xi}) \dd \xi^\text{I} \dd \xi^\text{J} 
&= \delta_{ij} \left( \frac{\partial X^i}{\partial x} \frac{\partial X^j}{\partial x} (\dd x)^2
+ \frac{\partial X^i}{\partial y} \frac{\partial X^j}{\partial y} (\dd y)^2
+ 2 \frac{\partial X^i}{\partial x} \frac{\partial X^j}{\partial y} \dd x \dd y
\right) , \nonumber\\
H_\text{IJ}(\vec{\xi}) &\stackrel{\cdot}{=} \left[
\begin{array}{cc}
1 + (\partial_x f)^2 			& \partial_x f \partial_y f \\
\partial_x f \partial_y f		& 1 + (\partial_y f)^2 
\end{array}
\right] , \qquad \xi^\text{I} \equiv (x,y),
\end{align}
where on the second line the ``$\stackrel{\cdot}{=}$" means it is ``represented by" the matrix to its right -- the first row corresponds, from left to right, to the $xx$, $xy$ components; the second row $yx$ and $yy$ components. Recall that the infinitesimal volume ($=$ 2D area) is given in any coordinate system $\vec{\xi}$ by $\dd^2 \xi \sqrt{\det H_{\text{IJ}}(\vec{\xi})}$. That means from taking the $\det$ of eq. \eqref{DifferentialGeometry_Areaoff(x,y)}, if the domain on $(x,y)$ is denoted as $\mathfrak{D}$, the corresponding area swept out by $f$ is given by the 2D integral
\begin{align}
\label{DifferentialGeometry_AreaOf2DSurface_I}
\int_{\mathfrak{D}} \dd x \dd y \sqrt{\det H_{\text{IJ}}(x,y)}
&= \int_{\mathfrak{D}} \dd x \dd y \sqrt{(1 + (\partial_x f)^2)(1 + (\partial_y f)^2) - (\partial_x f \partial_y f)^2} \nonumber\\
&= \int_{\mathfrak{D}} \dd x \dd y \sqrt{1 + (\partial_x f(x,y))^2 + (\partial_y f(x,y))^2} .
\end{align}
{\bf Differential Forms and Volume} \qquad Although we have not (and shall not) employ differential forms very much, it is very much part of modern integration theory. One no longer writes $\int\dd^3\vec{x} f(\vec{x})$, for instance, but rather
\begin{align}
\int f(\vec{x}) \dd x^1 \wedge \dd x^2 \wedge \dd x^3 .
\end{align}
More generally, whenever the following $N-$form occur under an integral sign, we have the definition
\begin{align}
\underbrace{\dd x^1 \wedge \dd x^2 \wedge \dots \wedge \dd x^{N-1} \wedge \dd x^N}_{\text{(Differential form notation)}}
\equiv \underbrace{\dd^N \vec{x}}_{\text{Physicists' colloquial math-speak}} .
\end{align}
(Here $N \leq D$, where $D$ is the dimension of space.) This needs to be supplemented with the constraint that it is a fully antisymmetric object:
\begin{align}
\label{DifferentialGeometry_DifferentialForms_Antisymmetry}
\dd x^{i_1} \wedge \dd x^{i_2} \wedge \dots \wedge \dd x^{i_{N-1}} \wedge \dd x^{i_N}
= \epsilon_{i_1 \dots i_N} \dd x^{1} \wedge \dd x^{2} \wedge \dots \wedge \dd x^{N-1} \wedge \dd x^N .
\end{align}
The superposition of rank-$(N \leq D)$ differential forms spanned by $\{ (1/N!) F_{i_1 \dots i_N} \dd x^{i_1} \wedge \dots \wedge \dd x^{i_N}\}$, for arbitrary but fully antisymmetric $\{ F_{i_1 \dots i_N} \}$, forms a vector space.

Why differential forms are fundamental to integration theory is because, it is this antisymmetry that allows its proper definition as the volume spanned by an $N-$parallelepiped. For one, the antisymmetric nature of forms is responsible for the Jacobian upon a change-of-variables $\vec{x}(\vec{y})$ familiar from multi-variable calculus -- using eq. \eqref{DifferentialGeometry_DifferentialForms_Antisymmetry}:
\begin{align}
\dd x^{1} \wedge \dd x^{2} \wedge \dots \wedge \dd x^{N-1} \wedge \dd x^{N}
&= \frac{\partial x^1}{\partial y^{i_1}} \frac{\partial x^2}{\partial y^{i_2}} \dots \frac{\partial x^N}{\partial y^{i_N}} \dd y^{i_1} \wedge \dd y^{i_2} \wedge \dots \wedge \dd y^{i_{N-1}} \wedge \dd y^{i_N} \nonumber\\
&= \frac{\partial x^1}{\partial y^{i_1}} \frac{\partial x^2}{\partial y^{i_2}} \dots \frac{\partial x^N}{\partial y^{i_N}} \epsilon^{i_1 \dots i_N} \dd y^{1} \wedge \dd y^{2} \wedge \dots \wedge \dd y^{N-1} \wedge \dd y^{N} \nonumber\\
&= \left(\det \frac{\partial x^a}{\partial y^b}\right) \dd y^{1} \wedge \dd y^{2} \wedge \dots \wedge \dd y^{N-1} \wedge \dd y^{N} .
\end{align}
In a $(D \geq 2)-$dimensional flat space, you might be familiar with the statement that $D$ linearly independent vectors define a $D-$parallelepiped. Its volume, in turn, is computed through the determinant of the matrix whose columns (or rows) are these vectors. If we now consider the $(N \leq D)-$form built out of $N$ scalar fields $\{ \Phi^\text{I} \vert \text{I}=1,2,\dots,N \}$, i.e.,
\begin{align}
\dd \Phi^{1} \wedge \dots \wedge \dd \Phi^{N} ,
\end{align}
let us see how it defines an infinitesimal $N-$volume by generalizing the notion of volume-as-determinants.\footnote{These scalar fields $\{ \Phi^\text{I} \}$ can also be thought of as coordinates parametrizing some $N-$dimensional sub-space of the ambient $D-$dimensional space.} Focusing on the $N=2$ case, if $\vec{v} \equiv (p_1 \dd x^1,\dots,p_D \dd x^D)$ and $\vec{w} \equiv (q_1 \dd x^1,\dots,q_D \dd x^D)$ are two linearly independent vectors formed from $p_i = \partial_i \Phi^1$ and $q_i = \partial_i \Phi^2$, then 
\begin{align}
\dd \Phi^1 \wedge \dd \Phi^2 = (p_i \dd x^i) \wedge (q_j \dd x^j) = p_i q_j \dd x^i \wedge \dd x^j
\end{align}
is in fact the 2D area spanned by the parallelepiped defined by $\vec{v}$ and $\vec{w}$. For, since $\dd \Phi^1 \wedge \dd \Phi^2$ is a coordinate scalar, we may choose a locally flat coordinate system $\{y^i\}$ such that $p_i$ and $q_i$ lie on the $(1,2)-$plane; i.e., $p_{i > 2}=q_{i>2}=0$ and
\begin{align}
\dd \Phi^1 \wedge \dd \Phi^2 
= (p_i \dd y^i) \wedge (q_j \dd y^j) 
&= p_1 q_2 \dd y^1 \wedge \dd y^2 + p_2 q_1 \dd y^2 \wedge \dd y^1 \nonumber\\
&= (p_1 q_2 - p_2 q_1) \dd dx^1 \dd x^2
= \det\left[ 
\begin{array}{cc}
\vec{v} & \vec{w} 
\end{array}
\right] ;
\end{align}
where now
\begin{align}
\vec{v} &= \left(\partial_1 \Phi^1 \dd y^1, \partial_2 \Phi^1 \dd y^2, \vec{0} \right) , \\
\vec{w} &= \left(\partial_1 \Phi^2 \dd y^1, \partial_2 \Phi^2 \dd y^2, \vec{0} \right) .
\end{align}
This argument can be readily extended to higher $2 < N \leq D$.

\subsubsection{Fluxes, Gauss-Stokes' theorems, Poincar\'{e} lemma}

\noindent{\bf Normal to hypersurface}\qquad Suppose the hypersurface is $(D-1)$ dimensional, sitting in a $D$ dimensional ambient space. Then it could also be described by first identifying a scalar function of the ambient space $f(\vec{x})$ such that some constant-$f$ surface coincides with the hypersurface,
\begin{align}
f(\vec{x}) = C \equiv \text{constant} .
\end{align}
For example, a $2$-sphere of radius $R$ can be defined in Cartesian coordinates $\vec{x}$ as
\begin{align}
\label{DifferentialGeometry_2SphereDef}
f(\vec{x})=R^2, \qquad\text{ where }\qquad f(\vec{x})=\vec{x}^2 .
\end{align} 
Given the function $f$, we now show that $\dd f = 0$ can be used to define a unit normal $n^i$ through
\begin{align}
\label{DifferentialGeometry_UnitNormal}
n^i \equiv \frac{\nabla^i f}{\sqrt{\nabla^j f \nabla_j f}} = \frac{g^{ik} \partial_k f}{\sqrt{g^{lm} \nabla_l f \nabla_m f}} .
\end{align}
That $n^i$ is of unit length can be checked by a direct calculation. For $n^i$ to be normal to the hypersurface means, when dotted into the latter's tangent vectors from our previous discussion, it returns zero:
\begin{align}
\left.\frac{\partial x^i(\vec{\xi})}{\partial \xi^\text{I}} \partial_i f(\vec{x}) \right\vert_{\text{on hypersurface}}
= \frac{\partial}{\partial \xi^\text{I}} f\left( \vec{x}(\vec{\xi}) \right) = \partial_\text{I} f( \vec{\xi} ) = 0 .
\end{align}
The second and third equalities constitute just a re-statement that $f$ is constant on our hypersurface. Using $n^i$ we can also write down the induced metric on the hypersurface as
\begin{align}
\label{DifferentialGeometry_InducedMetric_Normal}
H_{ij} = g_{ij} - n_i n_j .
\end{align}
This makes sense as an induced metric on the hypersurface of one lower dimension than that of the ambient $D$-space, because $H_{ij}$ is itself orthogonal to $n^i$:
\begin{align}
H_{ij} n^j = \left(g_{ij} - n_i n_j\right) n^j = n_i - n_i = 0.
\end{align}
Any other vector $u$ dotted into the metric will have its $n$-component subtracted out:
\begin{align}
H^{i}_{\phantom{i}j} u^j = \left( \delta^i_{\phantom{i}j} - n^i n_j\right) u^j = u^i - n^i (n_j u^j) .
\end{align}
\begin{myP}
\qquad For the $2$-sphere in 3-dimensional flat space, defined by eq. \eqref{DifferentialGeometry_2SphereDef}, calculate the components of the induced metric $H_{ij}$ in eq. \eqref{DifferentialGeometry_InducedMetric_Normal} and compare it that in eq. \eqref{DifferentialGeometry_InducedMetric_2Sphere}. Hint: compute $\dd \sqrt{\vec{x}^2}$ in terms of $\{\dd x^i\}$ and exploit the constraint $\vec{x}^2 = R^2$; then consider what is the $-(n_i \dd x^i)^2$ occurring in $H_{ij} \dd x^i \dd x^j$, when written in spherical coordinates? \qed
\end{myP}
\begin{myP}
\qquad Consider some $2$-dimensional surface parametrized by $\xi^\text{I} = (\sigma,\rho)$, whose trajectory in $D$-dimensional flat space is provided by the Cartesian coordinates $\vec{x}(\sigma,\rho)$. What is the formula analogous to eq. \eqref{DifferentialGeometry_AreaOf2DSurface_I}, which yields the area of this 2D surface over some domain $\mathfrak{D}$ on the $(\sigma,\rho)$ plane? Hint: First ask, ``what is the 2D induced metric?" Answer:
\begin{align}
\text{Area} = \int_{\mathfrak{D}} \dd \sigma \dd \rho 
\sqrt{(\partial_\sigma \vec{x})^2 (\partial_\rho \vec{x})^2 - (\partial_\sigma \vec{x} \cdot \partial_\rho \vec{x})^2},
\qquad (\partial_\text{I} \vec{x})^2 \equiv \partial_\text{I} x^i \partial_\text{I} x^j \delta_{ij} .
\end{align} 
(This is not too far from the Nambu-Goto action of string theory.) \qed
\end{myP}
{\bf Directed surface elements} \qquad What is the analog of $\dd \vec{\text{(Area)}}$ from vector calculus? This question is important for the discussion of the curved version of Gauss' theorem, as well as the description of fluxes -- rate of flow of, say, a fluid -- across surface areas. If we have a $(D-1)$ dimensional hypersurface with induced metric $H_\text{IJ}(\xi^\text{K})$, determinant $H \equiv \det H_\text{IJ}$, and a unit normal $n^i$ to it, then the answer is
\begin{align}
\label{DifferentialGeometry_DirectedArea_v1}
\dd^{D-1}\Sigma_i &\equiv \dd^{D-1}\vec{\xi} \sqrt{|H(\vec{\xi})|} n_i\left( \vec{x}(\vec{\xi}) \right) \\
\label{DifferentialGeometry_DirectedArea_v2}
&= \dd^{D-1}\vec{\xi} \ \widetilde{\epsilon}_{i j_1 j_2 \dots j_{D-1}}\left( \vec{x}(\vec{\xi}) \right) 
\frac{\partial x^{j_1}(\vec{\xi})}{\partial \xi^1} \frac{\partial x^{j_2}(\vec{\xi})}{\partial \xi^2} \dots \frac{\partial x^{j_{D-1}}(\vec{\xi})}{\partial \xi^{D-1}} .
\end{align}
The difference between equations \eqref{DifferentialGeometry_DirectedArea_v1} and \eqref{DifferentialGeometry_DirectedArea_v2} is that the first requires knowing the normal vector beforehand, while the second description is purely intrinsic to the hypersurface and can be computed once its parametrization $\vec{x}(\vec{\xi})$ is provided. Also be aware that the choice of orientation of the $\{\xi^\text{I}\}$ should be consistent with that of the ambient $\{ \vec{x} \}$ and the infinitesimal volume $\dd^D\vec{x} \sqrt{|g|} \epsilon_{12\dots D}$.

The $\dd^{D-1}\xi \sqrt{|H|}$ is the (scalar) infinitesimal area ($= (D-1)$-volume) and $n_i$ provides the direction. The second equality requires justification. Let's define $\{\mathcal{E}_\text{I}^{\phantom{\text{I}}i} | \text{I} = 1,2,3,\dots,D-1\}$ to be the $(D-1)$ vector fields
\begin{align}
\mathcal{E}_\text{I}^{\phantom{\text{I}}i}(\vec{\xi}) \equiv \frac{\partial x^i(\vec{\xi})}{\partial \xi^\text{I}} .
\end{align}
\begin{myP}
\qquad Show that the tensor in eq. \eqref{DifferentialGeometry_DirectedArea_v2},
\begin{align}
\widetilde{n}_i \equiv \widetilde{\epsilon}_{i j_1 j_2 \dots j_{D-1}} \mathcal{E}_\text{1}^{\phantom{\text{1}}j_1} \dots \mathcal{E}_{D-1}^{\phantom{D-1}j_{D-1}}
\end{align}
is orthogonal to all the $(D-1)$ vectors $\{\mathcal{E}_\text{I}^{\phantom{\text{I}}i} \}$. Since $n_i$ is the sole remaining direction in the $D$ space, $\widetilde{n}_i$ must be proportional to $n_i$
\begin{align}
\widetilde{n}_i = \varphi \cdot n_i .
\end{align}
To find $\varphi$ we merely have to dot both sides with $n^i$,
\begin{align}
\label{DifferentialGeometry_DirectedArea_StepI}
\varphi(\vec{\xi}) 
= \sqrt{|g(\vec{x}(\vec{\xi}))|} \epsilon_{i j_1 j_2 \dots j_{D-1}} n^i 
\frac{\partial x^{j_1}(\vec{\xi})}{\partial \xi^\text{1}} \dots \frac{\partial x^{j_{D-1}}(\vec{\xi})}{\partial \xi^{D-1}} .
\end{align}
Given a point of the surface $\vec{x}(\vec{\xi})$ we can always choose the coordinates $\vec{x}$ of the ambient space such that, at least in a neighborhood of this point, $x^1$ refers to the direction orthogonal to the surface and the $\{x^2,x^3,\dots,x^{D}\}$ lie on the surface itself. Argue that, in this coordinate system, eq. \eqref{DifferentialGeometry_UnitNormal} becomes
\begin{align}
n^i = \frac{g^{(i)(1)}}{\sqrt{g^{(1)(1)}}} ,
\end{align}
and therefore eq. \eqref{DifferentialGeometry_DirectedArea_StepI} reads
\begin{align}
\label{DifferentialGeometry_DirectedArea_StepII}
\varphi(\vec{\xi}) = \sqrt{|g(\vec{x}(\vec{\xi}))|} \sqrt{g^{(1)(1)}} .
\end{align}
Cramer's rule (cf. \eqref{DetOfMatrix_CramersRule}) from matrix algebra reads: the $ij$ component (the $i$th row and $j$th column) of the inverse of a matrix $(A^{-1})_{ij}$ is $((-)^{i+j}/\det A)$ times the determinant of $A$ with the $j$th row and $i$th column removed. Use this and the definition of the induced metric to conclude that 
\begin{align}
\varphi(\vec{\xi}) = \sqrt{|H(\vec{\xi})|} ,
\end{align}
thereby proving the equality of equations \eqref{DifferentialGeometry_DirectedArea_v1} and \eqref{DifferentialGeometry_DirectedArea_v2}.  \qed
\end{myP}
{\bf Gauss' theorem} \qquad We are now ready to state (without proof) {\it Gauss' theorem}. In 3D vector calculus, Gauss tells us the volume integral, over some domain $\mathfrak{D}$, of the divergence of a vector field is equal to the flux of the same vector field across the boundary $\partial \mathfrak{D}$ of the domain. Exactly the same statement applies in a $D$ dimensional ambient curved space with some closed $(D-1)$ dimensional hypersurface that defines $\partial \mathfrak{D}$. \begin{quotation}
	Let $V^i$ be an arbitrary vector field, and let $\vec{x}(\vec{\xi})$ describe this closed boundary surface so that it has an (outward) directed surface element $\dd^{D-1} \Sigma_i$ given by equations \eqref{DifferentialGeometry_DirectedArea_v1} and \eqref{DifferentialGeometry_DirectedArea_v2}. Then
\begin{align}
\label{DifferentialGeometry_GaussTheorem}
\int_{\mathfrak{D}} \dd^D x \sqrt{|g(\vec{x})|} \nabla_i V^i(\vec{x}) = \int_{\partial \mathfrak{D}} \dd^{D-1} \Sigma_i V^i\left( \vec{x}(\vec{\xi}) \right)  .
\end{align}
\end{quotation}
{\it Flux} \qquad Just as in 3D vector calculus, the $\dd^{D-1} \Sigma_i V^i$ can be viewed as the flux of some fluid described by $V^i$ across an infinitesimal element of the hypersurface $\partial \mathfrak{D}$. 

\noindent{\it Remark} \qquad Gauss' theorem is not terribly surprising if you recognize the integrand as a total derivative,
\begin{align}
\sqrt{|g|}\nabla_i V^i = \partial_i (\sqrt{|g|} V^i) 
\end{align}
(recall eq. \eqref{DifferentialGeometry_DivergenceOfVector}) and therefore it should integrate to become a surface term ($\equiv (D-1)$-dimensional integral). The right hand side of eq. \eqref{DifferentialGeometry_GaussTheorem} merely makes this surface integral explicit, in terms of the coordinates $\vec{\xi}$ describing the boundary $\partial \mathfrak{D}$.
%The proof of Gauss' theorem proceeds by using a coordinate system $(x^1,x_\perp^\text{I})$, where the first coordinate is the radial one, i.e., the region $\mathfrak{D}$ is foliated by constant $x^1$ surfaces. Let $x^1=0$ be the origin and $x^1=1$ be the boundary $\partial \mathfrak{D}$. The left hand side of eq. \eqref{DifferentialGeometry_GaussTheorem} is
%{\allowdisplaybreaks\begin{align}
%& \int_0^1 \dd x^1 \int\dd^{D-1} x_\perp \left\{ \partial_{\text{I}} \left(\sqrt{|g|} V^{\text{I}}(\vec{x})\right) + \partial_1 \left(\sqrt{|g|} V^1(\vec{x})\right) \right\} \nonumber\\
%\label{DifferentialGeometry_GaussTheorem_Proof_I}
%&= \int_0^1 \dd x^1 \int\dd^{D-1} x_\perp \partial_{\text{I}} \left(\sqrt{|g|} V^{\text{I}}(\vec{x})\right) + 
%\int_0^1 \dd x^1 \frac{\dd}{\dd x^1} \int \dd^{D-1} x_\perp \left(\sqrt{|g|} V^1(\vec{x})\right) 
%\end{align}}
%The first term of eq. \eqref{DifferentialGeometry_GaussTheorem_Proof_I} is zero because, for a fixed $x^1$ surface, the $x_\perp$-integral is one over a closed $(D-1)$ dimensional hypersurface embedded in $D$ space.

\noindent{\it Closed surface} \qquad Note that if you apply Gauss' theorem eq. \eqref{DifferentialGeometry_GaussTheorem}, on a closed surface such as the sphere, the result is immediately zero. A closed surface is one where there are no boundaries. (For the $2$-sphere, imagine starting with the Northern Hemisphere; the boundary is then the equator. By moving this boundary south-wards, i.e., from one latitude line to the next, until it vanishes at the South Pole -- our boundary-less surface becomes the $2$-sphere.) Since there are no boundaries, the right hand side of eq. \eqref{DifferentialGeometry_GaussTheorem} is automatically zero.
\begin{myP}
\qquad We may see this directly for the $2$-sphere case. The metric on the $2$-sphere of radius $R$ is
\begin{align}
\dd\ell^2 = R^2 (\dd\theta^2 + (\sin\theta)^2 \dd\phi^2), \qquad \theta\in[0,\pi], \ \phi\in[0,2\pi).
\end{align}
Let $V^i$ be an arbitrary smooth vector field on the $2$-sphere. Show explicitly -- namely, do the integral -- that
\begin{align}
\int_{\mathbb{S}^2} \dd^2 x \sqrt{|g(\vec{x})|}\nabla_i V^i = 0.
\end{align}
Hint: For the $\phi$-integral, remember that $\phi=0$ and $\phi=2\pi$ refer to the same point, for a fixed $\theta$.  \qed
\end{myP}
\begin{myP}
{\bf Hudge dual formulation of Gauss' theorem in $D$-space.} \qquad Let us consider the Hodge dual of the vector field in eq. \eqref{DifferentialGeometry_GaussTheorem},
\begin{align}
\widetilde{V}_{i_1 \dots i_{D-1}} \equiv \widetilde{\epsilon}_{i_1 \dots i_{D-1} j} V^j .
\end{align}
First show that
\begin{align}
\widetilde{\epsilon}^{j i_1 \dots i_{D-1}} \nabla_j \widetilde{V}_{i_1 \dots i_{D-1}} 
\propto \partial_{[1} \widetilde{V}_{23\dots D]} 
\propto \nabla_i V^i .
\end{align}
(Find the proportionality factors.) Then deduce the dual formulation of Gauss' theorem, namely, the relationship between
\begin{align}
\label{DifferentialGeometry_GaussTheorem_DualForm}
\int_{\mathfrak{D}} \dd^D x \partial_{[1} \widetilde{V}_{23\dots D]} 
\qquad \text{   and   }
\qquad
\int_{\partial\mathfrak{D}} \dd^{D-1} \xi \widetilde{V}_{i_1 \dots i_{D-1}}\left( \vec{x}(\vec{\xi}) \right)
\frac{\partial x^{i_1}(\vec{\xi})}{\partial \xi^1} \cdots \frac{\partial x^{i_{D-1}}(\vec{\xi})}{\partial \xi^{D-1}} .
\end{align}
The $\widetilde{V}_{i_1 \dots i_{D-1}} \partial_{\xi^1} x^{i_1} \dots \partial_{\xi^{D-1}} x^{i_{D-1}}$ can be viewed as the original tensor $\widetilde{V}_{i_1 \dots i_{D-1}}$, but projected onto the boundary $\partial \mathfrak{D}$. 

In passing, I should point out, what you have shown in eq. \eqref{DifferentialGeometry_GaussTheorem_DualForm} can be written in a compact manner using differential forms notation:
\begin{align}
\int_{\mathfrak{D}} \dd \widetilde{V} = \int_{\partial\mathfrak{D}} \widetilde{V} ,
\end{align} 
by viewing the fully antisymmetric object $\widetilde{V}$ as a differential $(D-1)$-form.  \qed
\end{myP}
{\it Coulomb potential} \qquad A basic application of Gauss' theorem is the derivation of the (spherically symmetric) Coulomb potential of a unit point charge in $D$ spatial dimensions, satisfying
\begin{align}
\nabla_i \nabla^i \psi = -\delta^{(D)}(\vec{x}-\vec{x}')
\end{align}
in flat space. Let us consider as domain $\mathfrak{D}$ the sphere of radius $r$ centered at the point charge at $\vec{x}'$. Using spherical coordinates, $\vec{x} = r \widehat{n}(\vec{\xi})$, where $\widehat{n}$ is the unit radial vector emanating from $\vec{x}'$, the induced metric on the boundary $\partial\mathfrak{D}$ is simply the metric of the $(D-1)$-sphere. We now identify in eq. \eqref{DifferentialGeometry_GaussTheorem} $V^i = \nabla^i \psi$. The normal vector is simply $n^i \partial_i = \partial_r$, and so Gauss' law using eq. \eqref{DifferentialGeometry_DirectedArea_v1} reads
\begin{align}
-1 &= \int_{\mathbb{S}^{D-1}} \dd^{D-1} \vec{\xi} \sqrt{|H|} r^{D-1} \partial_r \psi(r) .
\end{align}
The $\int_{\mathbb{S}^{D-1}} \dd^{D-1} \vec{\xi} \sqrt{|H|} = 2\pi^{D/2}/\Gamma(D/2)$ is simply the solid angle subtended by the $(D-1)$-sphere ($\equiv$ volume of the $(D-1)$-sphere of unit radius). So at this point we have
\begin{align}
\partial_r \psi(r) = -\frac{\Gamma(D/2)}{2 \pi^{D/2} r^{D-1}} \qquad \Rightarrow \qquad 
\psi(r) = \frac{\Gamma(D/2)}{4 ((D-2)/2) \pi^{D/2} r^{D-2}} = \frac{\Gamma(\frac{D}{2}-1)}{4 \pi^{D/2} r^{D-2}} .
\end{align}
I have used the Gamma-function identity $\Gamma(z)z = \Gamma(z+1)$. Replacing $r \to |\vec{x}-\vec{x}'|$, we conclude that the Coulomb potential due to a unit strength electric charge is
\begin{align}
\psi(\vec{x}) = \frac{\Gamma(\frac{D}{2}-1)}{4 \pi^{D/2} |\vec{x}-\vec{x}'|^{D-2}} .
\end{align}
It is instructive to also use Gauss' law using eq. \eqref{DifferentialGeometry_DirectedArea_v2}.
\begin{align}
-1 
&= \int_{\mathbb{S}^{D-1}} \dd^{D-1} \vec{\xi} \epsilon_{i_1 \dots i_{D-1} j} \frac{\partial x^{i_1}}{\partial \xi^1} \cdots \frac{\partial x^{i_{D-1}}}{\partial \xi^{D-1}} g^{jk}(\vec{x}(\vec{\xi})) \partial_k \psi(r\equiv\sqrt{\vec{x}^2}) .
\end{align}
On the surface of the sphere, we have the completeness relation (cf. \eqref{CompletenessRelation}):
\begin{align}
g^{jk}(\vec{x}(\vec{\xi})) 
= \delta^{\text{IJ}} \frac{\partial x^j}{\partial \xi^\text{I}} \frac{\partial x^k}{\partial \xi^\text{J}} 
+ \frac{\partial x^j}{\partial r} \frac{\partial x^k}{\partial r} .
\end{align}
(This is also the coordinate transformation for the inverse metric from Cartesian to Spherical coordinates.) At this point,
\begin{align}
-1 
&= \int_{\mathbb{S}^{D-1}} \dd^{D-1} \vec{\xi} 
\epsilon_{i_1 \dots i_{D-1} j} \frac{\partial x^{i_1}}{\partial \xi^1} \cdots \frac{\partial x^{i_{D-1}}}{\partial \xi^{D-1}} \left(\delta^{\text{IJ}} \frac{\partial x^j}{\partial \xi^\text{I}} \frac{\partial x^k}{\partial \xi^\text{J}} 
+ \frac{\partial x^j}{\partial r} \frac{\partial x^k}{\partial r}\right) \partial_k \psi(r\equiv\sqrt{\vec{x}^2}) \nonumber\\
&= \int_{\mathbb{S}^{D-1}} \dd^{D-1} \vec{\xi} 
\epsilon_{i_1 \dots i_{D-1} j} \frac{\partial x^{i_1}}{\partial \xi^1} \cdots \frac{\partial x^{i_{D-1}}}{\partial \xi^{D-1}} \frac{\partial x^j}{\partial r} \left( \frac{\partial x^k}{\partial r} \partial_k \psi(r\equiv\sqrt{\vec{x}^2}) \right) .
\end{align}
The Levi-Civita symbol contracted with the Jacobians can now be recognized as simply the determinant of the $D$-dimensional metric written in spherical coordinates $\sqrt{|g(r,\vec{\xi})|}$. (Note the determinant is positive because of the way we ordered our coordinates.) That is in fact equal to $\sqrt{|H(r,\vec{\xi})|}$ because $g_{rr}=1$. Whereas $(\partial x^k/\partial r) \partial_k \psi = \partial_r \psi$. We have therefore recovered the previous result using eq. \eqref{DifferentialGeometry_DirectedArea_v1}.

{\bf Tensor elements} \qquad Suppose we have a $(N < D)$-dimensional domain $\mathfrak{D}$ parametrized by $\{\vec{x}(\xi^\text{I}) | \text{I}=1,2,\dots,N \}$ whose boundary $\partial \mathfrak{D}$ is parametrized by $\{\vec{x}(\theta^\mathfrak{A}) | \mathfrak{A}=1,2,\dots,N-1 \}$. We may define a $(D-N)$-tensor element that generalizes the one in eq. \eqref{DifferentialGeometry_DirectedArea_v2}
\begin{align}
\label{DifferentialGeometry_TensorArea_v1}
\dd^N \Sigma_{i_1 \dots i_{D-N}}
\equiv \dd^N \xi \ \widetilde{\epsilon}_{i_1 \dots i_{D-N} j_1 j_2 \dots j_N}\left( \vec{x}(\vec{\xi}) \right) 
\frac{\partial x^{j_1}(\vec{\xi})}{\partial \xi^1} \frac{\partial x^{j_2}(\vec{\xi})}{\partial \xi^2} \dots \frac{\partial x^{j_N}(\vec{\xi})}{\partial \xi^N} .
\end{align}
We may further define the boundary surface element
\begin{align}
\label{DifferentialGeometry_TensorArea_v2}
\dd^{N-1} \Sigma_{i_1 \dots i_{D-N} k}
\equiv \dd^{N-1} \theta \ \widetilde{\epsilon}_{i_1 \dots i_{D-N} k j_1 \dots j_{N-1}}\left( \vec{x}(\vec{\theta}) \right) 
\frac{\partial x^{j_1}(\vec{\theta})}{\partial \theta^1} \frac{\partial x^{j_2}(\vec{\theta})}{\partial \theta^2} \dots \frac{\partial x^{j_{N-1}}(\vec{\theta})}{\partial \theta^{N-1}} .
\end{align}
\begin{quotation}
	{\bf Stokes' theorem}\footnote{Just like for the Gauss' theorem case, in equations \eqref{DifferentialGeometry_TensorArea_v1} and \eqref{DifferentialGeometry_TensorArea_v2}, the $\vec{\xi}$ and $\vec{\theta}$ coordinate systems need to be defined with orientations consistent with the ambient $\dd^D \vec{x} \sqrt{|g(\vec{x})|} \epsilon_{12\dots D}$ one.} \qquad Stokes' theorem is the assertion that, in a $(N < D)$-dimensional simply connected subregion $\mathfrak{D}$ of some $D$-dimensional ambient space, the divergence of a fully antisymmetric rank $(D-N+1)$ tensor field $B^{i_1 \dots i_{D-N} k}$ integrated over the domain $\mathfrak{D}$ can also be expressed as the integral of $B^{i_1 \dots i_{D-N} k}$ over its boundary $\partial \mathfrak{D}$. Namely,
\begin{align}
\label{DifferentialGeometry_StokesTheorem}
\int_{\mathfrak{D}} \dd^N \Sigma_{i_1 \dots i_{D-N}} \nabla_k B^{i_1 \dots i_{D-N} k}
		= \frac{1}{D-N+1} \int_{\partial\mathfrak{D}} \dd^{N-1} \Sigma_{i_1 \dots i_{D-N} k} B^{i_1 \dots i_{D-N} k}, \\
 N < D, \ B^{[i_1 \dots i_{D-N} k]} = (D-N+1)! B^{i_1 \dots i_{D-N} k}. \nonumber
\end{align}
\end{quotation}
\begin{myP}
{\bf Hodge dual formulation of Stokes' theorem.} \qquad Define
\begin{align}
\widetilde{B}_{j_1 \dots j_{N-1}} 
\equiv \frac{1}{(D-N+1)!} \widetilde{\epsilon}_{j_1 \dots j_{N-1} i_1 \dots i_{D-N} k} B^{i_1 \dots i_{D-N} k} .
\end{align}
Can you convert eq. \eqref{DifferentialGeometry_StokesTheorem} into a relationship between
\begin{align}
\int_{\mathfrak{D}} \dd^N \vec{\xi} \partial_{[i_1} \widetilde{B}_{i_2 \dots i_N]} \frac{\partial x^{i_1}}{\partial \xi^1} \dots \frac{\partial x^{i_N}}{\partial \xi^N} \qquad \text{   and   } \qquad
\int_{\partial \mathfrak{D}} \dd^{N-1} \vec{\theta} \widetilde{B}_{i_1 \dots i_{N-1}} \frac{\partial x^{i_1}}{\partial \theta^1} \dots \frac{\partial x^{i_{N-1}}}{\partial \theta^{N-1}} ?
\end{align}
Furthermore, explain why the Jacobians can be ``brought inside the derivative".
\begin{align}
\partial_{[i_1} \widetilde{B}_{i_2 \dots i_N]} \frac{\partial x^{i_1}}{\partial \xi^1} \dots \frac{\partial x^{i_N}}{\partial \xi^N}
= \frac{\partial x^{i_1}}{\partial \xi^{[1}} \partial_{|i_1|} \left( \frac{\partial x^{i_2}}{\partial \xi^2} \dots \frac{\partial x^{i_N}}{\partial \xi^{N]}} \widetilde{B}_{i_2 \dots i_N} \right) .
\end{align}
The $|\cdot|$ around $i_1$ indicate it is {\it not} to be part of the anti-symmetrization; only do so for the $\xi$-indices.

Like for Gauss' theorem, we point out that -- by viewing $\widetilde{B}_{j_1 \dots j_{N-1}}$ as components of a $(N-1)$-form, Stokes' theorem in eq. \eqref{DifferentialGeometry_StokesTheorem} reduces to the simple expression
\begin{align}
\int_{\mathfrak{D}} \dd \widetilde{B} = \int_{\partial\mathfrak{D}} \widetilde{B} .
\end{align} \qed
\end{myP}
{\it Relation to 3D vector calculus} \qquad Stokes' theorem in vector calculus states that the flux of the curl of a vector field over some 2D domain $\mathfrak{D}$ sitting in the ambient 3D space, is equal to the line integral of the same vector field along the boundary $\partial \mathfrak{D}$ of the domain. Because eq. \eqref{DifferentialGeometry_StokesTheorem} may not appear, at first sight, to be related to the Stokes' theorem from 3D vector calculus, we shall work it out in some detail.
\begin{myP}
\qquad Consider some 2D hypersurface $\mathfrak{D}$ residing in a 3D curved space. For simplicity, let us foliate $\mathfrak{D}$ with constant $\rho$ surfaces; let the other coordinate be $\phi$, so $\vec{x}(0 \leq \rho \leq \rho_>,0 \leq \phi \leq 2\pi)$ describes a given point on $\mathfrak{D}$ and the boundary $\partial \mathfrak{D}$ is given by the closed loop $\vec{x}(\rho = \rho_>,0 \leq \phi \leq 2\pi)$. Let 
\begin{align}
B^{ik} \equiv \widetilde{\epsilon}^{i k j} A_j  
\end{align}
for some vector field $A^j$. This implies in Cartesian coordinates,
\begin{align}
\nabla_k B^{ik} = \left(\vec{\nabla} \times \vec{A}\right)^i .
\end{align}
Denote $\vec{\xi} = (\rho,\phi)$. Show that Stokes' theorem in eq. \eqref{DifferentialGeometry_StokesTheorem} reduces to the $N=2$ vector calculus case:
\begin{align}
\label{DifferentialGeometry_StokesTheorem_2D}
\int_0^{\rho_>} \dd\rho \int_{0}^{2\pi} \dd\phi \sqrt{|H(\vec{\xi})|} \vec{n} \cdot \left( \vec{\nabla} \times \vec{A} \right)
= \int_{0}^{2\pi} \dd\phi \frac{\partial\vec{x}(\rho_>,\phi)}{\partial\phi} \cdot \vec{A}(\vec{x}(\rho_>,\phi)) .
\end{align}
where the unit normal vector is given by
\begin{align}
\vec{n} = \frac{ (\partial \vec{x}(\vec{\xi})/\partial \rho) \times (\partial \vec{x}(\vec{\xi})/\partial\phi) }{\left\vert (\partial \vec{x}(\vec{\xi})/\partial \rho) \times (\partial \vec{x}(\vec{\xi})/\partial\phi) \right\vert} .
\end{align}
Of course, once you've verified Stokes' theorem for a particular coordinate system, you know by general covariance it holds in any coordinate system, i.e.,
\begin{align}
\label{DifferentialGeometry_StokesTheorem_2D_GenerallyCovariant}
\int_{\mathfrak{D}} \dd^2 \xi \sqrt{|H(\vec{\xi})|} n_i \widetilde{\epsilon}^{ijk} \partial_j A_k = \int_{\partial\mathfrak{D}} A_i \dd x^i .
\end{align}
{\it Step-by-step guide:} \qquad Start with eq. \eqref{DifferentialGeometry_DirectedArea_v2}, and show that in a Cartesian basis,
\begin{align}
\dd^2 \Sigma_i = \dd^2 \xi \left( \frac{\partial \vec{x}}{\partial \rho} \times \frac{\partial \vec{x}}{\partial \phi} \right)^i .
\end{align}
The induced metric on the 2D domain $\mathfrak{D}$ is
\begin{align}
H_\text{IJ} = \delta_{ij} \partial_\text{I} x^i \partial_\text{J} x^j .
\end{align}
Work out its determinant. Then work out
\begin{align}
\left\vert (\partial\vec{x}/\partial\rho) \times (\partial\vec{x}/\partial\phi) \right\vert^2
\end{align}
using the identity
\begin{align}
\label{DifferentialGeometry_ProductOfEpsilons_3D}
\widetilde{\epsilon}^{ijk} \widetilde{\epsilon}_{lmk} = \delta^i_l \delta^j_m - \delta^i_m \delta^j_l .
\end{align}
Can you thus relate $\sqrt{|H(\vec{\xi})|}$ to $\left\vert (\partial\vec{x}/\partial\rho) \times (\partial\vec{x}/\partial\phi) \right\vert$, and thereby verify the left hand side of eq. \eqref{DifferentialGeometry_StokesTheorem} yields the left hand side of \eqref{DifferentialGeometry_StokesTheorem_2D}? 

For the right hand side of eq. \eqref{DifferentialGeometry_StokesTheorem_2D}, begin by arguing that the boundary (line) element in eq. \eqref{DifferentialGeometry_TensorArea_v2} becomes
\begin{align}
\dd \Sigma_{k i} = \dd\phi \ \widetilde{\epsilon}_{k i j} \frac{\partial x^j}{\partial \phi} .
\end{align}
Then use $\widetilde{\epsilon}^{i j_1 j_2} \widetilde{\epsilon}_{k j_1 j_2} = 2 \delta^i_k$ to then show that the right hand side of eq. \eqref{DifferentialGeometry_StokesTheorem} is now that of eq. \eqref{DifferentialGeometry_StokesTheorem_2D}. \qquad \qed
\end{myP}
\begin{myP}
\qquad Discuss how the tensor element in eq. \eqref{DifferentialGeometry_TensorArea_v1} transforms under a change of hypersurface coordinates $\vec{\xi} \to \vec{\xi}(\vec{\xi}')$. Do the same for the tensor element in eq. \eqref{DifferentialGeometry_TensorArea_v2}: how does it transforms under a change of hypersurface coordinates $\vec{\theta} \to \vec{\theta}(\vec{\theta}')$? \qquad \qed
\end{myP}
%\begin{myP}
%Consider the following scalar field configuration on the $3$-sphere, with radius $R$ and angular coordinates $\xi^i \equiv \{ \psi,\theta \in [0,\pi], \phi \in [0,2\pi) \}$).
%\begin{align}
%\Psi(\psi) = -\frac{\cot(\psi)}{4\pi} .
%\end{align} 
%By a direct calculation, show that
%\begin{align}
%\nabla_i \nabla^i \psi = 0
%\end{align}
%almost everywhere. Then use Stokes' theorem to deduce that 
%\begin{align}
%-\nabla_i \nabla^i \psi = \frac{\delta^{(3)}(\vec{\xi}-\vec{\xi}_+)}{\sqrt[4]{|g(\vec{\xi}) \ g(\vec{\xi}_+)|}} - \frac{\delta^{(3)}(\vec{\xi}-\vec{\xi}_-)}{\sqrt[4]{|g(\vec{\xi}) \ g(\vec{\xi}_-)|}}, 
%\end{align}
%where $\xi_-^i = (\phi,\theta,\psi=0)$ and $\xi_-^i = (\phi,\theta,\psi=\pi)$. This is a special case of Poisson's equation; the interpretation here is that the field $\Psi$ is sourced by a negative unit strength point charge/mass at the North pole ($\psi = 0$) and a positive unit strength point charge/mass at the South pole ($\psi = \pi$).

%Hint: Start by considering the following volume integral over the ``cap" defined by $0 \leq \theta \leq \theta_0$, where $0 < \theta_0 < \pi$.
%\begin{align}
%J \equiv \int_{0 \leq \theta \leq \theta_0} \dd^3 \vec{\xi} \sqrt{|g(\vec{\xi})|} (-\nabla_i \nabla^i \Psi(\vec{\xi})) 
%\end{align}
%By using Stokes' theorem, show that $J$ does not depend on $\theta_0$. This single calculation should allow you to infer the amplitudes of both charges, one at the North %and South poles -- why? \qed
%\end{myP}
{\bf Poincar\'{e} Lemma} \qquad In 3D vector calculus you have learned that a vector $\vec{B}$ is divergence-less everywhere in space iff it is the curl of another vector $\vec{A}$. 
\begin{align}
\vec{\nabla} \cdot \vec{B} = 0 \qquad \Leftrightarrow \qquad \vec{B} = \vec{\nabla} \times \vec{A}  .
\end{align}
And, the curl of a vector $\vec{B}$ is zero everywhere in space iff it is the gradient of scalar $\psi$.
\begin{align}
\vec{\nabla} \times \vec{B} = 0 \qquad \Leftrightarrow \qquad \vec{B} = \vec{\nabla} \psi .
\end{align}
Here, we shall see that these statements are special cases of the following.
\begin{quotation}
{\it Poincar\'{e} lemma} \qquad In an arbitrary $D$ dimensional curved space, let $B_{i_1 \dots i_N}(\vec{x})$ be a fully antisymmetric rank-$N$ tensor field, with $N \leq D$. Then, everywhere within a simply connected region of space,
\begin{align}
B_{i_1 \dots i_N} = \partial_{[i_1} C_{i_2 \dots i_N]} ,
\end{align}
-- i.e., $B$ is the ``curl" of a fully antisymmetric rank-$(N-1)$ tensor $C$ -- if and only if
\begin{align}
\partial_{[j} B_{i_1 \dots i_N]} = 0 .
\end{align}
In differential form notation, by treating $C$ as a $(N-1)$-form and $B$ as a $N$-form, Poincar\'{e} would read: throughout a simply connected region of space,
\begin{align}
\dd B = 0 \text{   iff   } B = \dd C .
\end{align}
\end{quotation}
{\it Example I: Electromagnetism} \qquad Let us recover the 3D vector calculus statement above, that the divergence-less nature of the magnetic field is equivalent to it being the curl of some vector field. Consider the dual of the magnetic field $B^i$:
\begin{align}
\widetilde{B}^{ij} \equiv \widetilde{\epsilon}^{ij k} B_k .
\end{align}
The Poincar\'{e} Lemma says $\widetilde{B}_{ij} = \partial_{[i} A_{j]}$ if and only if $\partial_{[k} \widetilde{B}_{ij]} = 0$ everywhere in space. We shall proceed to take the dual of these two conditions. Via eq. \eqref{DifferentialGeometry_LeviCivita_Product}, the first is equivalent to
\begin{align}
\widetilde{\epsilon}^{k ij} \widetilde{B}_{ij} 
&= \widetilde{\epsilon}^{k ij} \partial_{[i} A_{j]} , \nonumber\\
&= 2 \widetilde{\epsilon}^{k ij} \partial_{i} A_{j} .
\end{align}
On the other hand, employing eq. \eqref{DifferentialGeometry_LeviCivita_Product},
\begin{align}
\widetilde{\epsilon}^{k ij} \widetilde{B}_{ij} 
&= \widetilde{\epsilon}^{k ij} \widetilde{\epsilon}_{ijl} B^l = 2 B^k ;
\end{align}
and therefore $\vec{B}$ is the curl of $A_i$:
\begin{align}
B^k = \widetilde{\epsilon}^{k ij} \partial_{i} A_{j}
\end{align}
While the latter condition $\dd \widetilde{B} = 0$ is, again utilizing eq. \eqref{DifferentialGeometry_LeviCivita_Product}, equivalent to
\begin{align}
0 &= \widetilde{\epsilon}^{kij} \partial_{k} \widetilde{B}_{ij} \nonumber\\
&= \widetilde{\epsilon}_{kij} \widetilde{\epsilon}^{ijl} \nabla_k B_l
= 2 \nabla_l B^l  . 
\end{align}
That is, the divergence of $\vec{B}$ is zero.

{\it Example II} \qquad A simple application is that of the line integral
\begin{align}
I(\vec{x},\vec{x}';\mathfrak{P}) \equiv \int_{\mathfrak{P}} A_i \dd x^i ,
\end{align}
where $\mathfrak{P}$ is some path in $D$-space joining $\vec{x}'$ to $\vec{x}$. Poincar\'{e} tells us, if $\partial_{[i} A_{j]} = 0$ everywhere in space, then $A_i = \partial_i \varphi$, the $A_i$ is a gradient of a scalar $\varphi$. Then $A_i \dd x^i = \partial_i \varphi \dd x^i = \dd \varphi$, and the integral itself is actually path independent -- it depends only on the end points:
\begin{align}
\int_{\vec{x}'}^{\vec{x}} A_i \dd x^i = \int_{\mathfrak{P}} \dd \varphi = \varphi(\vec{x}) - \varphi(\vec{x}'),
\qquad \text{whenever $\partial_{[i} A_{j]} = 0$} .
\end{align}
\begin{myP}
\qquad Make a similar translation, from the Poincar\'{e} Lemma, to the 3D vector calculus statement that a vector $B$ is curl-less if and only if it is a pure gradient everywhere. \qed
\end{myP}
\begin{myP}
\qquad Consider the vector potential, written in 3D Cartesian coordinates,
\begin{align}
A_i \dd x^i = \frac{x^1 \dd x^2 - x^2 \dd x^1}{(x^1)^2 + (x^2)^2} .
\end{align}
Can you calculate
\begin{align}
F_{ij} = \partial_{[i} A_{j]} ?
\end{align}
Consider a 2D surface whose boundary $\partial \mathfrak{D}$ circle around the $(0,0,-\infty < x^3 < +\infty)$ line once. Can you use Stokes' theorem to show that
\begin{align}
F_{ij} = 2\pi \epsilon_{ij3} \delta(x^1) \delta(x^2) ?
\end{align}
Hint: Convert from Cartesian to polar coordinates $(x,y,z) = (r \cos\phi,r \sin\phi,z)$; the line integral on the right hand side of eq. \eqref{DifferentialGeometry_StokesTheorem_2D_GenerallyCovariant} should simplify considerably. This problem illustrates the subtlety regarding the ``simply connected" requirement of the Poincar\'{e} lemma. The magnetic field $F_{ij}$ here describes that of a highly localized solenoid lying along the $z$-axis; its corresponding vector potential is a pure gradient in any simply connected $3-$volume not containing the $z$-axis, but it is no longer a pure gradient in say a solid torus region encircling (but still not containing) it. \qed
\end{myP}
 
\newpage

\section{Differential Geometry In Curved Spacetimes}
\label{Chapter_DifferentialGeometry_CurvedSpacetimes}

We now move on to differential geometry in curved spacetimes. I assume the reader is familiar with basic elements of Special Relativity and with the discussion in \S \eqref{Chapter_DifferentialGeometry_CurvedSpaces} -- in many instances, I will simply bring over the results from there to the curved spacetime context. In \S \eqref{Chapter_PoincareLorentz} I discuss Lorentz/Poincar\'{e} symmetry in flat spacetime, since it is fundamental to both Special and General Relativity. I then cover curved spacetime differential geometry proper from \S \eqref{Chapter_DifferentialGeometry_CurvedSpacetimes_C1} through \S \eqref{Chapter_DifferentialGeometry_CurvedSpacetimes_C3}, focusing on issues not well developed in \S \eqref{Chapter_DifferentialGeometry_CurvedSpaces}. These three sections, together with \S \eqref{Chapter_DifferentialGeometry_CurvedSpaces}, are intended to form the first portion -- the {\it kinematics} of curved space(time)s part\footnote{As opposed to the dynamics of spacetime, which involves studying General Relativity, Einstein's field equations for the metric, and its applications.} -- of a course on gravitation. Following that, \S \eqref{Chapter_DifferentialGeometry_CurvedSpacetimes_C4} contains somewhat specialized content regarding the expansion of geometric quantities off some fixed `background' geometry; and finally, in \S \eqref{Chapter_DifferentialGeometry_ConformalTransformations} we compile conformal transformation properties of geometric objects. 

\subsection{Poincar\'{e} and Lorentz symmetry}
\label{Chapter_PoincareLorentz}
% All differential geometry operators and notions -- at least in these notes -- follow from the metric!
% Re-write physics in spacetime-covariant form.
% What is `covariance'? Problem: Analogy with Non-Abelian Gauge Theory.
% Elements of classical field theory in curved spacetime.
% Noether's theorem

% Define notation! d, D, etc.

% Time dilation, length contraction, 

Poincar\'{e} and Lorentz symmetries play fundamental roles in our understanding of both classical relativistic physics and quantum theories of elementary particle interactions. In this section, we shall study it in some detail.

The metric of flat spacetime is, in Cartesian coordinates $\{x^\mu\}$,
\begin{align}
\label{Metric_Minkowski}
\dd s^2 		&\equiv \eta_{\mu\nu} \dd x^\mu \dd x^\nu, \\
\label{FlatMetric}
\eta_{\mu\nu}	&\equiv \text{diag}[1,-1,\dots,-1] .
\end{align}
Strictly speaking we should be writing eq. \eqref{DifferentialGeometry_MinkowskiMetric} in the `dimensionally-correct' form
\begin{align}
\dd s^2 = c^2 \dd t^2 - \dd\vec{x}\cdot\dd\vec{x} ;
\end{align}
where $c$ is the speed of light and $[\dd s^2]=[\text{Length}^2]$. %The derivatives ought to read
%\begin{align}
%\partial_0 \equiv \frac{1}{c}\frac{\partial}{\partial t}, 
%\qquad \text{ and } \qquad
%\partial_i \equiv \frac{\partial}{\partial x^i} .
%\end{align}
However, as explained in \S \eqref{Chapter_DimensionalAnalysis}, since the speed of light shows up frequently in relativity and gravitational physics, it is often advantageous to set $c=1$, which in turn means all speeds are measured using $c$ as the base unit. ($v=0.23$ would mean $v=0.23c$, for instance.) We shall do so throughout this section.

Notice too, we have switched from Latin/English alphabets in \S \eqref{Chapter_DifferentialGeometry_CurvedSpaces}, say $i,j,k,\dots \in \{ 1,2,3,\dots,D \}$ to Greek ones $\mu,\nu, \dots \in \{ 0,1,2,\dots,D \equiv d-1 \}$; the former run over the spatial coordinates while the latter over time ($0$th) and space $(1,\dots,D)$. Also note that the opposite `mostly plus' sign convention $\eta_{\mu\nu} = \text{diag}[-1,+1,\dots,+1]$ is equally valid and, in fact, more popular in the contemporary physics literature.

We shall define Poincar\'{e} transformations\footnote{Poincar\'{e} transformations are also sometimes known as inhomogeneous Lorentz transformations.} $x(x')$ to be the set of all coordinate transformations that leave the flat spacetime metric invariant:
\begin{align}
\label{PoincareSymmetry_Def}
\dd s^2 = \eta_{\mu\nu} \dd x^\mu \dd x^\nu = \eta_{\alpha'\beta'} \dd x'^\alpha \dd x'^\beta .
\end{align}
As we will now proceed to demonstrate, the most general invertible Poincar\'{e} transformation is
\begin{align}
\label{PoincareTransformation}
x^\mu &= a^\mu + \Lambda^\mu_{\phantom{\mu}\nu} x'^\nu , 
\end{align}
where $a^\mu$ is a constant vector describing a spacetime translation; and $\Lambda^\mu_{\phantom{\mu}\nu}$ is an arbitrary (spacetime-constant) Lorentz transformation, which in turn is defined as one that leaves $\eta_{\mu\nu}$ invariant in the following manner:
\begin{align}
\label{LorentzTransformation}
\Lambda^\mu_{\phantom{\mu}\alpha} \Lambda^\nu_{\phantom{\nu}\beta} \eta_{\mu\nu} = \eta_{\alpha\beta} .
\end{align}
{\it Derivation of eq. \eqref{PoincareSymmetry_Def}}\footnote{This argument can be found in Weinberg \cite{Weinberg:1972kfs}.} \qquad Now, under a coordinate transformation, eq. \eqref{PoincareSymmetry_Def} reads
\begin{align}
\label{PoincareSymmetry_Proof_Step1}
\eta_{\mu\nu} \dd x^\mu \dd x^\nu 
= \eta_{\mu\nu} \frac{\partial x^\mu}{\partial x'^\alpha} \frac{\partial x^\nu}{\partial x'^\beta} \dd x'^\alpha \dd x'^\beta 
= \eta_{\alpha'\beta'} \dd x'^\alpha \dd x'^\beta .  
\end{align}
Let us differentiate both sides of eq. \eqref{PoincareSymmetry_Proof_Step1} with respect to $x'^\sigma$. 
\begin{align}
\label{PoincareSymmetry_Proof_Step2}
\eta_{\mu\nu} \frac{\partial^2 x^\mu}{\partial x'^\sigma \partial x'^\alpha} \frac{\partial x^\nu}{\partial x'^\beta} 
+ \eta_{\mu\nu} \frac{\partial x^\mu}{\partial x'^\alpha} \frac{\partial^2 x^\nu}{\partial x'^\sigma \partial x'^\beta} 
&= 0 .
\end{align}
Next, consider symmetrizing $\sigma\alpha$ and anti-symmetrizing $\sigma\beta$.
\begin{align}
\label{PoincareSymmetry_Proof_Step2A}
2 \eta_{\mu\nu} \frac{\partial^2 x^\mu}{\partial x'^\sigma \partial x'^\alpha} \frac{\partial x^\nu}{\partial x'^\beta} 
+ \eta_{\mu\nu} \frac{\partial x^\mu}{\partial x'^\alpha} \frac{\partial^2 x^\nu}{\partial x'^\sigma \partial x'^\beta}
+ \eta_{\mu\nu} \frac{\partial x^\mu}{\partial x'^\sigma} \frac{\partial^2 x^\nu}{\partial x'^\alpha \partial x'^\beta}  
&= 0 \\
\label{PoincareSymmetry_Proof_Step2B}
\eta_{\mu\nu} \frac{\partial^2 x^\mu}{\partial x'^\sigma \partial x'^\alpha} \frac{\partial x^\nu}{\partial x'^\beta} 
- \eta_{\mu\nu} \frac{\partial^2 x^\mu}{\partial x'^\beta \partial x'^\alpha} \frac{\partial x^\nu}{\partial x'^\sigma} 
&= 0 
\end{align}
Since partial derivatives commute, the second term from the left of eq. \eqref{PoincareSymmetry_Proof_Step2} vanishes upon anti-symmetrization of $\sigma\beta$. Adding equations \eqref{PoincareSymmetry_Proof_Step2A} and \eqref{PoincareSymmetry_Proof_Step2B} hands us
\begin{align}
\label{PoincareSymmetry_Proof_Step2C}
3 \eta_{\mu\nu} \frac{\partial^2 x^\mu}{\partial x'^\sigma \partial x'^\alpha} \frac{\partial x^\nu}{\partial x'^\beta} 
+ \eta_{\mu\nu} \frac{\partial x^\mu}{\partial x'^\alpha} \frac{\partial^2 x^\nu}{\partial x'^\sigma \partial x'^\beta}
= 0 .
\end{align}
Finally, subtracting eq. \eqref{PoincareSymmetry_Proof_Step2} from eq. \eqref{PoincareSymmetry_Proof_Step2C} produces
\begin{align}
2 \eta_{\mu\nu} \frac{\partial^2 x^\mu}{\partial x'^\sigma \partial x'^\alpha} \frac{\partial x^\nu}{\partial x'^\beta} = 0 .
\end{align}
Because we have assumed Poincar\'{e} transformations are invertible, we may contract both sides with $\partial x'^\beta/\partial x^\kappa$.
\begin{align}
\label{PoincareSymmetry_Proof_Step3}
\eta_{\mu\nu} 
\frac{\partial^2 x^\mu}{\partial x'^\sigma \partial x'^\alpha} 
\frac{\partial x^\nu}{\partial x'^\beta} 
\frac{\partial x'^\beta}{\partial x^\kappa} 
= \eta_{\mu\nu} 
\frac{\partial^2 x^\mu}{\partial x'^\sigma \partial x'^\alpha} \delta^\nu_\kappa
= 0 .
\end{align}
Finally, we contract both sides with $\eta^{\kappa\rho}$:
\begin{align}
\label{PoincareSymmetry_Proof_Step4}
\eta_{\mu'\kappa'} \eta^{\kappa'\rho} \frac{\partial^2 x^\mu}{\partial x'^\sigma \partial x'^\alpha} 
= \frac{\partial^2 x^\rho}{\partial x'^\sigma \partial x'^\alpha} 
= 0 .
\end{align}
In words: since the second $x'$-derivative of $x$ has to vanish, the transformation from $x$ to $x'$ can at most go linearly as $x'$; it cannot involve higher powers of $x'$. This implies the form in eq. \eqref{PoincareTransformation}. Plugging eq. \eqref{PoincareTransformation}  the latter into eq. \eqref{PoincareSymmetry_Proof_Step1}, we recover the necessary definition of the Lorentz transformation in eq. \eqref{LorentzTransformation}.
\begin{quotation}
	The most general invertible coordinate transformations that leave the Cartesian Minkowski metric invariant involve the (spacetime-constant) Lorentz transformations $\{ \Lambda^\mu_{\phantom{\mu}\alpha} \}$ of eq \eqref{LorentzTransformation} plus constant spacetime translations.
\end{quotation}
{\bf (Homogeneous) Lorentz Transformations form a Group} \qquad If $\Lambda^\mu_{\phantom{\mu}\alpha}$ and $\Lambda'^\mu_{\phantom{'\mu}\alpha}$ denotes different Lorentz transformations, then notice the composition
\begin{align}
\label{LorentzGroupOperation}
\Lambda''^\mu_{\phantom{''\mu}\alpha} \equiv \Lambda^\mu_{\phantom{\mu}\sigma} \Lambda'^\sigma_{\phantom{'\sigma}\alpha} 
\end{align}
is also a Lorentz transformation. For, keeping in mind the fundamental definition in eq. \eqref{LorentzTransformation}, we may directly compute
\begin{align}
\Lambda''^\mu_{\phantom{''\mu}\alpha} \Lambda''^\nu_{\phantom{''\nu}\beta} \eta_{\mu\nu}
&= 
\Lambda^\mu_{\phantom{\mu}\sigma} \Lambda'^\sigma_{\phantom{'\sigma}\alpha} 
\Lambda^\nu_{\phantom{\nu}\rho} \Lambda'^\rho_{\phantom{'\rho}\beta} \eta_{\mu\nu} \nonumber\\
&= \Lambda'^\sigma_{\phantom{'\sigma}\alpha} \Lambda'^\rho_{\phantom{'\rho}\beta} \eta_{\sigma\rho} = \eta_{\alpha\beta} .
\end{align}
To summarize:
\begin{quotation}
	The set of all Lorentz transformations $\{ \Lambda^\mu_{\phantom{\mu}\alpha} \}$ satisfying eq. \eqref{LorentzTransformation}, together with the composition law in eq. \eqref{LorentzGroupOperation} for defining successive Lorentz transformations, form a {\it Group}.
\end{quotation}
{\it Proof} \qquad Let $\Lambda^\mu_{\phantom{\mu}\alpha}$, $\Lambda'^\mu_{\phantom{'\mu}\alpha}$ and $\Lambda''^\mu_{\phantom{''\mu}\alpha}$ denote distinct Lorentz transformations. 
\begin{itemize}
	\item {\it Closure} \qquad Above, we have just verified that applying successive Lorentz transformations yields another Lorentz transformation; for e.g., $\Lambda^\mu_{\phantom{\mu}\sigma} \Lambda'^\sigma_{\phantom{'\sigma}\nu}$ and $\Lambda^\mu_{\phantom{\mu}\sigma} \Lambda'^\sigma_{\phantom{'\sigma}\rho} \Lambda''^\rho_{\phantom{''\rho}\nu}$ are Lorentz transformations.
	\item {\it Associativity} \qquad Because applying successive Lorentz transformations amount to matrix multiplication, and since the latter is associative, that means Lorentz transformations are associative: 
	\begin{align}
	\Lambda \cdot \Lambda' \cdot \Lambda'' = \Lambda \cdot (\Lambda' \cdot \Lambda'') = (\Lambda \cdot \Lambda') \cdot \Lambda''.
	\end{align}
	\item {\it Identity} \qquad $\delta^\mu_{\phantom{\mu}\alpha}$ is the identity Lorentz transformation:
	\begin{align}
	\delta^\mu_{\phantom{\mu}\sigma} \Lambda^\sigma_{\phantom{\sigma}\nu} = \Lambda^\mu_{\phantom{\mu}\sigma} \delta^\sigma_{\phantom{\sigma}\nu} = \Lambda^\mu_{\phantom{\mu}\nu} ,
	\end{align}
	and
	\begin{align}
	\delta^\mu_{\phantom{\mu}\alpha} \delta^\nu_{\phantom{\nu}\beta} \eta_{\mu\nu} = \eta_{\alpha\beta} .
	\end{align}
	\item {\it Inverse} \qquad Let us take the determinant of both sides of eq. \eqref{LorentzTransformation} -- by viewing the latter as matrix multiplication, we have $\Lambda^T \cdot \eta \cdot \Lambda = \eta$, which in turn means
	\begin{align}
	\label{LorentzTransformation_Det}
	(\det \Lambda)^2 = 1 \qquad \Rightarrow \qquad \det \Lambda = \pm 1 .
	\end{align} 
	Here, we have recalled $\det A^T = \det A$ for any square matrix $A$. Since the determinant of $\Lambda$ is strictly non-zero, what eq. \eqref{LorentzTransformation_Det} teaches us is that $\Lambda$ is always invertible: $\Lambda^{-1}$ is guaranteed to exist. What remains is to check that, if $\Lambda$ is a Lorentz transformation, so is $\Lambda^{-1}$. Starting with the matrix form of eq. \eqref{LorentzTransformation}, and utilizing $(\Lambda^{-1})^T = (\Lambda^T)^{-1}$,
	\begin{align}
	\Lambda^T \eta \Lambda &= \eta \\
	(\Lambda^T)^{-1} \Lambda^T \eta \Lambda \Lambda^{-1} &= (\Lambda^T)^{-1} \cdot \eta \cdot \Lambda^{-1} \\
	\eta &= (\Lambda^{-1})^T \cdot \eta \cdot \Lambda^{-1} .
	\end{align}
\end{itemize}
{\bf Lorentzian `inner product' is preserved} \qquad That $\Lambda$ is a Lorentz transformation means it is a linear operator that preserves the Lorentzian inner product. For suppose $v$ and $w$ are arbitrary vectors, the inner product of $v' \equiv \Lambda v$ and $w' \equiv \Lambda w$ is that between $v$ and $w$.
\begin{align}
v' \cdot w' 
\equiv \eta_{\alpha\beta} v'^\alpha w'^\beta
&= \eta_{\alpha\beta} \Lambda^\alpha_{\phantom{\alpha}\mu} \Lambda^\beta_{\phantom{\beta}\nu} v^\mu w^\nu \\
&= \eta_{\mu\nu} v^\mu w^\nu = v \cdot w .
\end{align}
This is very much analogous to rotations in $\mathbb{R}^D$ being the linear transformations that preserve the Euclidean inner product between spatial vectors: $\vec{v} \cdot \vec{w} = \vec{v}' \cdot \vec{w}'$ for all $\widehat{R}^T \widehat{R} = \mathbb{I}_{D \times D}$, where $\vec{v}' \equiv \widehat{R} \vec{v}$ and $\vec{w}' \equiv \widehat{R} \vec{w}$. 
\begin{myP}
	{\bf 4D Lorentz Group and SL$_{2,\mathbb{C}}$}\qquad Define $\{\sigma^\mu\}$ to be the basis set of $2 \times 2$ complex matrices formed by the $2 \times 2$ identity matrix together with the Pauli matrices, namely 
	\begin{align}
	\label{PauliMatrices}
	\sigma^0 \equiv
	\left[\begin{array}{cc}
	1 & 0 \\
	0 & 1
	\end{array}\right], \qquad
	\sigma^1 \equiv
	\left[\begin{array}{cc}
	0 & 1 \\
	1 & 0
	\end{array}\right], \qquad
	\sigma^2 \equiv
	\left[\begin{array}{cc}
	0 	& -i \\
	i 	& 0
	\end{array}\right], \qquad
	\sigma^3 \equiv
	\left[\begin{array}{cc}
	1 	& 0 	\\
	0 	& -1
	\end{array}\right] .
	\end{align}
	Now let $p_\mu \equiv (p_0,p_1,p_2,p_3)$ be a 4-component collection of real numbers, and verify that
	\begin{align}
	\det p_\mu \sigma^\mu = \eta^{\mu\nu} p_\mu p_\nu \equiv p^2 .
	\end{align}
	Next, consider the following transformation, 
	\begin{align}
	p_\mu \sigma^\mu \to L^\dagger \cdot p_\mu \sigma^\mu \cdot L ,
	\end{align} 
	where $L$ is some arbitrary $2 \times 2$ complex matrix. (This transformation preserves the Hermitian nature of $p_\mu \sigma^\mu$ for real $p_\mu$.) Then consider taking their determinant:
	\begin{align}
	\det [p_\mu \sigma^\mu] \to \det \left[ L^\dagger \cdot p_\mu \sigma^\mu \cdot L \right]
	\end{align}
	What property must $L$ obey in order that this leaves the determinant invariant, i.e.,
	\begin{align}
	\label{SL2C}
	\det [p_\mu \sigma^\mu] = \det \left[ L^\dagger \cdot p_\mu \sigma^\mu \cdot L \right] = p^2 ?
	\end{align}
	Argue that the set of all $L$'s obeying eq. \eqref{SL2C}, with $|\det L| = 1$ (this is the `S'$\equiv$`special' in SL$_{2,\mathbb{C}}$), forms a group. \qed
\end{myP}
We wish to study in some detail what the most general form $\Lambda^\mu_{\phantom{\mu}\alpha}$ may take. To this end, we shall do so by examining how it acts on some arbitrary vector field $v^\mu$. Even though this section deals with Minkowski spacetime, this $v^\mu$ may also be viewed as a vector in a curved spacetime written in an orthonormal basis.

{\bf Rotations} \qquad Let us recall that any spatial vector $v^i$ may be rotated to point along the $1-$axis while preserving its Euclidean length. That is, there is always a $\widehat{R}$, obeying $\widehat{R}^T \widehat{R} = \mathbb{I}$ such that
\begin{align}
\widehat{R}^i_{\phantom{i}j} v^j \dot{=} \pm |\vec{v}| (1,0,\dots,0)^T  ,
\qquad\qquad
|\vec{v}| \equiv \sqrt{\delta_{ij} v^i v^j} . 
\end{align}
\footnote{This $\widehat{R}$ is not unique: for example, by choosing another rotation matrix $\widehat{R}''$ that only rotates the space orthogonal to $v^i$, $\widehat{R} \widehat{R}'' \vec{v}$ and $\widehat{R} \vec{v}$ both yield the same result.}Conversely, since $\widehat{R}$ is necessarily invertible, any spatial vector $v^i$ can be obtained by rotating it from $|\vec{v}|(1,\vec{0}^T)$. Moreover, in $D+1$ notation, these rotation matrices can be written as
\begin{align}
\widehat{R}^\mu_{\phantom{\mu}\nu}
&\dot{=} \left[\begin{array}{cc}
1 		& \vec{0}^T \\
\vec{0}	& \widehat{R}^i_{\phantom{i}j} 
\end{array}\right] \\
\widehat{R}^0_{\phantom{0}\nu} v^\nu &= v^0, \\
\widehat{R}^i_{\phantom{i}\nu} v^\nu &= \widehat{R}^i_{\phantom{i}j} v^j = (\pm |\vec{v}|,0,\dots,0)^T . 
\end{align}
These considerations tell us, if we wish to study Lorentz transformations that are {\it not} rotations, we may reduce their study to the $(1+1)$D case. To see this, we first observe that
\begin{align}
\Lambda \left[ \begin{array}{c}
v^0 \\ v^1 \\ \vdots \\ v^D 
\end{array} \right]
= \Lambda \left[ \begin{array}{cc}
1 		& \vec{0}^T 	\\
\vec{0}	& \widehat{R}
\end{array}\right] \left[ \begin{array}{c}
v^0 \\ \pm |\vec{v}| \\ \vec{0}
\end{array}\right] .
\end{align}
And if the result of this matrix multiplication yields non-zero spatial components, namely $(v'^0,v'^1,\dots,v'^D)^T$, we may again find a rotation matrix $\widehat{R}'$ such that
\begin{align}
\Lambda \left[ \begin{array}{c}
v^0 \\ v^1 \\ \vdots \\ v^D 
\end{array} \right]
= \left[ \begin{array}{c}
v'^0 \\ v'^1 \\ \vdots \\ v'^D 
\end{array} \right]
= \left[ \begin{array}{cc}
1 		& \vec{0}^T 	\\
\vec{0}	& \widehat{R}'
\end{array}\right] \left[ \begin{array}{c}
v'^0 \\ \pm |\vec{v}'| \\ \vec{0}
\end{array}\right] .
\end{align}
At this point, we have reduced our study of Lorentz transformations to
\begin{align}
\left[ \begin{array}{cc}
1 		& \vec{0}^T 	\\
\vec{0}	& \widehat{R}'^T
\end{array}\right] \Lambda
\left[ \begin{array}{cc}
1 		& \vec{0}^T 	\\
\vec{0}	& \widehat{R}
\end{array}\right] \left[ \begin{array}{c}
v^0 \\ v^1 \\ \vec{0}
\end{array}\right]
\equiv \Lambda' \left[ \begin{array}{c}
v^0 \\ v^1 \\ \vec{0}
\end{array}\right]
= \left[ \begin{array}{c}
v'^0 \\ v'^1 \\ \vec{0}
\end{array}\right] .
\end{align}
Because $\Lambda$ was arbitrary so is $\Lambda'$, since one can be gotten from another via rotations.

{\bf Time Reversal \& Parity Flips} \qquad Suppose the time component of the vector $v^\mu$ were negative ($v^0 < 0$), we may write it as
\begin{align}
\left[\begin{array}{c}
-|v^0| \\ \vec{v}
\end{array}\right]
= \widehat{T} \left[\begin{array}{c}
|v^0| \\ \vec{v}
\end{array}\right] , \qquad\qquad
\widehat{T} \equiv \left[\begin{array}{cc}
-1 		& \vec{0}^T \\
\vec{0}	& \mathbb{I}_{D \times D}
\end{array}\right] ;
\end{align}
where $\widehat{T}$ is the time reversal matrix since it reverses the sign of the time component of the vector. You may readily check that $\widehat{T}$ itself is a Lorentz transformation in that it satisfies $\widehat{T}^T \eta \widehat{T} = \eta$.
\begin{myP}
	{\bf Parity flip of the $i$th axis} \qquad Suppose we wish to flip the sign of the $i$th spatial component of the vector, namely $v^i \to -v^i$. You can probably guess, this may be implemented via the diagonal matrix with all entries set to unity, except the $i$th component -- which is set instead to $-1$.
	\begin{align}
	\,_{i}\widehat{P}^\mu_{\phantom{\mu}\nu} v^\nu 	&= v^\mu, \qquad\qquad \mu \neq i, 	\\
	\,_{i}\widehat{P}^i_{\phantom{i}\nu} v^\nu 		&= -v^i,							 \\
	\,_{i}\widehat{P} &\equiv \text{diag}[1,1,\dots,1,\underbrace{-1}_{(i+1)\text{th component}},1,\dots,1] .
	\end{align}
	Define the rotation matrix $\widehat{R}^\mu_{\phantom{\mu}\nu}$ such that it leaves all the axes orthogonal to the 1st and $i$th invariant, namely
	\begin{align}
	\widehat{R}^\mu_{\phantom{\mu}\nu} \widehat{e}_\ell^\nu &= \widehat{e}_\ell^\nu , \\
	\widehat{e}_\ell^\mu &\equiv \delta_\ell^\mu, \qquad\qquad
	\ell \neq 1,i ;
	\end{align}
	while rotating the $(1,i)$-plane clockwise by $\pi/2$:
	\begin{align}
	\widehat{R} \cdot \widehat{e}_1 = -\widehat{e}_i , \qquad\qquad
	\widehat{R} \cdot \widehat{e}_i = +\widehat{e}_1 .
	\end{align} 
	Now argue that
	\begin{align}
	\,_{i}\widehat{P} = \widehat{R}^T \cdot \,_{1}\widehat{P} \cdot \widehat{R} .
	\end{align} 
	Is $\,_{i}\widehat{P}$ a Lorentz transformation? \qed
\end{myP}
{\bf Lorentz Boosts} \qquad As already discussed, we may focus on the 2D case to elucidate the form of the most general Lorentz boost. This is the transformations that would mix time and space components, and yet leave the metric of spacetime $\eta_{\mu\nu} = \text{diag}[1,-1]$ invariant. (Neither time reversal, parity flips, nor spatial rotations mix time and space.) This is what revolutionized humanity's understanding of spacetime at the beginning of the 1900's: inspired by the fact that the speed of light is the same in all inertial frames, Einstein discovered {\it Special Relativity}, that the space and time coordinates of one frame have to become intertwined when being translated to those in another frame. We will turn this around later when discussing Maxwell's equations: the constancy of the speed of light in all inertial frames is in fact a consequence of the Lorentz covariance of the former.
\begin{myP}
	We wish to find a $2 \times 2$ matrix $\Lambda$ that obeys $\Lambda^T \cdot \eta \cdot \Lambda = \eta$, where $\eta_{\mu\nu} = \text{diag}[1,-1]$. By examining the diagonal terms of $\Lambda^T \cdot \eta \cdot \Lambda = \eta$, show that
	\begin{align}
	\Lambda \dot{=} \left[ \begin{array}{cc}
	\sigma_1 \cosh(\xi_1)	& \sigma_2 \sinh(\xi_2) \\
	\sigma_3 \sinh(\xi_1)	& \sigma_4 \cosh(\xi_2) 
	\end{array} \right] ,
	\end{align}
	where the $\sigma_{1,2,3,4}$ are either $+1$ or $-1$; altogether, there are 16 choices of signs. (Hint: $x^2-y^2 = c^2$, for constant $c$, describes a hyperbola on the $(x,y)$ plane.) From the off diagonal terms of $\Lambda^T \cdot \eta \cdot \Lambda = \eta$, argue that either $\xi_1 = \xi_2 \equiv \xi$ or $\xi_1 = -\xi_2 \equiv \xi$. Then explain why, if $\Lambda^0_{\phantom{0}0}$ were not positive, we can always multiply it by a time reversal matrix to render it so; and likewise $\Lambda^1_{\phantom{1}1}$ can always be rendered positive by multiplying it by a parity flip. By requiring $\Lambda^0_{\phantom{0}0}$ and $\Lambda^1_{\phantom{1}1}$ be both positive, therefore, prove that the resulting 2D Lorentz boost is
	\begin{align}
	\label{LorentzTransformation_2D_General}
	\Lambda^\mu_{\phantom{\mu}\nu}(\xi)
	= \left[ \begin{array}{cc}
	\cosh(\xi) & \sinh(\xi) \\
	\sinh(\xi) & \cosh(\xi)
	\end{array} \right] .
	\end{align}
	This $\xi$ is known as {\it rapidity}. In 2D, the rotation matrix is
	\begin{align}
	\label{Rotation_2D}
	\widehat{R}^i_{\phantom{i}j}(\theta)
	= \left[ \begin{array}{cc}
	\cos(\theta) 	& -\sin(\theta) \\
	\sin(\theta)	& \cos(\theta)
	\end{array} \right] ;
	\end{align}
	and therefore rapidity $\xi$ is to the Lorentz boost in eq. \eqref{LorentzTransformation_2D_General} what the angle $\theta$ is to rotation $\widehat{R}^i_{\phantom{i}j}(\theta)$ in eq. \eqref{Rotation_2D}. 
	
	\begin{quotation}
		{\bf 2D Lorentz Group:} In (1+1)D, the continuous boost in $\Lambda^\mu_{\phantom{\mu}\nu}(\xi)$ in eq. \eqref{LorentzTransformation_2D_General}, the continuous rotation $\widehat{R}^i_{\phantom{i}j}(\theta)$ in eq. \eqref{Rotation_2D}; and the discrete time reversal and spatial reflection operators
		\begin{align}
		\widehat{T} = \left[ 
		\begin{array}{cc}
		-1 	& 0 \\
		0	& 1
		\end{array}
		\right]
		\qquad \text{ and } \qquad
		\widehat{P} = \left[ 
		\begin{array}{cc}
		1 	& 0 \\
		0	& -1
		\end{array}
		\right] ;
		\end{align}
		altogether form the full set of Lorentz transformations -- i.e., all solutions to eq. \eqref{LorentzT_Def} consist of products of these 4 matrices.
	\end{quotation} \qed
\end{myP}
To understand the meaning of the rapidity $\xi$, let us consider applying it to an arbitrary 2D vector $U^\mu$.
\begin{align}
\label{LorentzBoost_2D}
U' \equiv \Lambda \cdot U = \left[\begin{array}{c}
U^0 \cosh(\xi) + U^1 \sinh(\xi) \\
U^1 \cosh(\xi) + U^0 \sinh(\xi) 
\end{array}\right] .
\end{align}
{\bf Lorentz Boost: Timelike case} \qquad Suppose $U$ were timelike, $U^2 > 0 \Rightarrow (U^0)^2 > (U^1)^2 \Rightarrow |U^0/U^1| > 1$. Then it is not possible to find a finite $\xi$ such that $U'^0=0$, because that would amount to solving $\tanh(\xi) = -U^0/U^1$ but $\tanh$ lies between $-1$ and $+1$ while $-U^0/U^1$ is either less than $-1$ or greater than $+1$. On the other hand, it does mean we may solve for $\xi$ that would set the spatial component to zero: $\tanh(\xi) = -U^1/U^0$. Recall that tangent vectors may be interpreted as the derivative of the spacetime coordinates with respect to some parameter $\lambda$, namely $U^\mu \equiv \dd x^\mu/\dd \lambda$. Therefore 
\begin{align}
\frac{U^1}{U^0} = \frac{\dd x^1}{\dd \lambda} \frac{\dd \lambda}{\dd x^0} = \frac{\dd x^1}{\dd x^0} \equiv v 
\end{align}
is the velocity associated with $U^\mu$ in the frame $\{ x^\mu \}$. Starting from $\tanh(\xi)=-v$, some algebra would then hand us (cf. eq. \eqref{LorentzTransformation_2D_General})
\begin{align}
\cosh(\xi) &= \gamma \equiv \frac{1}{\sqrt{1-v^2}}, \\
\sinh(\xi) &= -\gamma \cdot v = -\frac{v}{\sqrt{1-v^2}} , \\
\label{LorentzTransformation_2DBoost}
\Lambda^\mu_{\phantom{\mu}\nu}
&= \left[ \begin{array}{cc}
\gamma 			& -\gamma \cdot v \\
-\gamma \cdot v & \gamma
\end{array} \right] .
\end{align}
This in turn yields
\begin{align}
U' = \left( \text{sgn}(U^0) \sqrt{\eta_{\mu\nu} U^\mu U^\nu}, 0 \right)^T ;
\end{align}
leading us to interpret the $\Lambda^\mu_{\phantom{\mu}\nu}$ we have found in eq. \eqref{LorentzTransformation_2DBoost} as the boost that bring observers to the frame where the flow associated with $U^\mu$ is `at rest'. (Note that, if $U^\mu = \dd x^\mu/\dd\tau$, where $\tau$ is proper time, then $\eta_{\mu\nu} U^\mu U^\nu=1$.) % Moreover, if we interpret $U^\mu$ and $U'^\mu$ as vectors joining a pair of spacetime points, but in their respective reference frames, we may take the ratio of their time components:
%\begin{align}
%\frac{U'^0}{U^0} = \sqrt{1-v^2} < 1 .
%\end{align}

As an important aside, we may generalize the two-dimensional Lorentz boost in eq. \eqref{LorentzTransformation_2DBoost} to $D-$dimensions. One way to do it, is to simply append to the 2D Lorentz-boost matrix a $(D-2) \times (D-2)$ identity matrix (that leaves the $2-$ through $D-$spatial components unaltered) in a block diagonal form:
\begin{align}
\label{LorentzBoost_2DAppended}
\Lambda^\mu_{\phantom{\mu}\nu} \stackrel{?}{=}
\left[ \begin{array}{ccc}
\gamma			&	-\gamma \cdot v 	&	0 \\
-\gamma \cdot v	& 	\gamma				&	0 \\
0				& 	0					& \mathbb{I}_{(D-2) \times (D-2)}
\end{array}\right] .
\end{align}
But this is not doing much: we are still only boosting in the $1-$direction. What if we wish to boost in $v^i$ direction, where $v^i$ is now some arbitrary spatial vector? To this end, we may promote the $(0,1)$ and $(1,0)$ components of eq. \eqref{LorentzTransformation_2DBoost} to the spatial vectors $\Lambda^0_{\phantom{0}i}$ and $\Lambda^i_{\phantom{i}0}$ parallel to $v^i$. Whereas the $(1,1)$ component of eq. \eqref{LorentzTransformation_2DBoost} is to be viewed as acting on the 1D space parallel to $v^i$, namely the operator $v^i v^j/\vec{v}^2$. (As a check: When $v^i = v(1,\vec{0})$, $v^i v^j/\vec{v}^2 = \delta_1^i \delta^j_1$.) The identity operator acting on the orthogonal $(D-2) \times (D-2)$ space, i.e., the analog of $\mathbb{I}_{(D-2) \times (D-2)}$ in eq. \eqref{LorentzBoost_2DAppended}, is $\Pi^{ij} = \delta^{ij} - v^i v^j/\vec{v}^2$. (Notice: $\Pi^{ij} v^j = (\delta^{ij} - v^i v^j/\vec{v}^2) v^j = 0$.) Altogether, the Lorentz boost in the $v^i$ direction is given by
\begin{align}
\label{LorentzBoost_along_v}
\LTud{\mu}{\nu}(\vec{v}) \dot{=}
\left[
\begin{array}{cc}
\gamma 		& - \gamma v^i \\
-\gamma v^i	& \gamma \frac{v^i v^j}{\vec{v}^2} + \left( \delta^{ij} - \frac{v^i v^j}{\vec{v}^2} \right)
\end{array} \right] , \qquad\qquad 
\vec{v}^2 \equiv \delta_{ab} v^a b^b .
\end{align}
It may be worthwhile to phrase this discussion in terms of the Cartesian coordinates $\{ x^\mu \}$ and $\{ x'^\mu \}$ parametrizing the two inertial frames. What we have shown is that the Lorentz boost in eq. \eqref{LorentzBoost_along_v} describes
\begin{align}
U'^\mu 	&= \LTud{\mu}{\nu}(\vec{v}) U^\nu, \qquad\qquad \\
U^\mu	&= \frac{\dd x^\mu}{\dd \lambda} , \qquad\qquad
U'^\mu 	
= \frac{\dd x'^\mu}{\dd \lambda} 
= \left( \text{sgn}(U^0) \sqrt{\eta_{\mu\nu} U^\mu U^\nu}, 0 \right)^T .
\end{align}
$\lambda$ is the intrinsic 1D coordinate parametrizing the worldlines, and by definition does not alter under Lorentz boost. The above statement is therefore equivalent to
\begin{align}
\dd x'^\mu 	&= \LTud{\mu}{\nu}(\vec{v}) \dd x^\nu , \\
\label{LorentzBoost_Frames}
x'^\mu 	&= \LTud{\mu}{\nu}(\vec{v}) x^\nu + a^\mu ,
\end{align}
where the spacetime translation $a^\mu$ shows up here as integration constants. 
\begin{myP}
	{\bf Lorentz boost in $(D+1)-$dimensions} \qquad If $v^\mu \equiv (1,v^i)$, check via a direction calculation that the $\LTud{\mu}{\nu}$ in eq. \eqref{LorentzBoost_along_v} produces a $\LTud{\mu}{\nu} v^\nu$ that has no non-trivial spatial components. Also check that eq. \eqref{LorentzBoost_along_v} is, in fact, a Lorentz transformation. What is $\LTud{\mu}{\sigma}(\vec{v}) \LTud{\sigma}{\nu}(-\vec{v})$?
\end{myP}
{\bf Lorentz Boost: Spacelike case} \qquad Suppose $U$ were spacelike, $U^2 < 0 \Rightarrow (U^0)^2 < (U^1)^2 \Rightarrow |U^1/U^0| = |\dd x^1/\dd x^0| \equiv |v| > 1$. Then, recalling eq. \eqref{LorentzBoost_2D}, it is not possible to find a finite $\xi$ such that $U'^1=0$, because that would amount to solving $\tanh(\xi)=-U^1/U^0$, but $\tanh$ lies between $-1$ and $+1$ whereas $-U^1/U^0 = -v$ is either less than $-1$ or greater than $+1$. On the other hand, it is certainly possible to have $U'^0=0$. Simply do $\tanh(\xi)=-U^0/U^1=-1/v$. Similar algebra to the timelike case then hands us 
\begin{align}
\cosh(\xi) 	&= \left(1-v^{-2}\right)^{-1/2} = \frac{|v|}{\sqrt{v^2-1}}, 		\\
\sinh(\xi) 	&= -(1/v)\left(1-v^{-2}\right)^{-1/2} = - \frac{\text{sgn}(v)}{\sqrt{v^2-1}}, 	\\
U' 			&= \left( 0,\text{sgn}(v) \sqrt{-\eta_{\mu\nu} U^\mu U^\nu} \right)^T ,\qquad\qquad v \equiv \frac{U^1}{U^0} .
\end{align}
We may interpret $U'^\mu$ and $U^\mu$ as infinitesimal vectors joining the same pair of spacetime points but in their respective frames. Specifically, $U'^\mu$ are the components in the frame where the pair lies on the same constant-time surface $(U'^0=0)$. While $U^\mu$ are the components in a boosted frame. %Furthermore, because $0 < v^{-2} < 1$, 
%\begin{align}
%\frac{U'^1}{U^1} = \sqrt{1-(U^0/U^1)^2} = \sqrt{1-v^{-2}} < 1.
%\end{align}
%In a frame moving with respect to the one where the two points lie on a constant time surface, the spatial length is contracted.

{\bf Lorentz Boost: Null (aka lightlike) case} \qquad If $U$ were null, that means $(U^0)^2 = (U^1)^2$, which in turn means
\begin{align}
U^\mu = \omega(1,\pm 1)
\end{align}
for some real number $\omega$. Upon a Lorentz boost, eq. \eqref{LorentzBoost_2D} tells us
\begin{align}
U' \equiv \Lambda \cdot U 
= \omega \left[\begin{array}{c}
\cosh(\xi) \pm \sinh(\xi) \\
\sinh(\xi) \pm \cosh(\xi) 
\end{array}\right] .
\end{align}
As we shall see below, if $U^\mu$ describes the $d-$momentum of a photon, so that $|\omega|$ is its frequency in the un-boosted frame, the $U'^0/U^0 = \cosh(\xi) \pm \sinh(\xi)$ describes the photon's red- or blue-shift in the boosted frame. Notice it is not possible to set either the time nor the space component to zero, unless $\xi \to \pm \infty$.
\begin{quotation}
	{\bf Summary} \qquad Our analysis of the group of matrices $\{ \Lambda \}$ obeying $\LTud{\alpha}{\mu} \LTud{\beta}{\nu} \eta_{\alpha\beta} = \eta_{\mu\nu}$ reveals that these Lorentz transformations consists of: time reversals, parity flips, spatial rotations and Lorentz boosts. A timelike vector can always be Lorentz-boosted so that all its spatial components are zero; while a spacelike vector can always be Lorentz-boosted so that its time component is zero.
\end{quotation}
\begin{myP}
	{\bf Null, spacelike vs. timelike} \qquad Do null vectors form a vector space? Simiarly, do spacelike or timelike vectors form a vector space? \qed
\end{myP}
\begin{myP}
	{\bf Determinants and discontinuities} \qquad What are the determinants of the time reversal $\widehat{T}$ and parity flips $\{\,_{i}\widehat{P}\}$ matrices? What is the determinant of the Lorentz boost matrix in eq. \eqref{LorentzTransformation_2D_General}? Hint: Your answers should tells us, as long as the determinants of Lorentz transformations are real, Lorentz transformations involving odd number of time-reversals and/or parity flips cannot be continuously connected to the identity transformation. Whereas, when the rapidity $\xi$ and rotation angle $\theta$ are set to zero, Lorentz boosts and rotations respectively become the identity in a continuous manner. \qed
\end{myP}
\begin{myP}
	{\bf Non-singular Coordinate transformations form a group} \qquad Let us verify explicitly that the Jacobians associated with general non-singular coordinate transformations form a group. Specifically, let us consider transforming from the coordinate system $x^\alpha$ to $y^\mu$, and assume $x^\alpha$ in terms of $y^\mu$ has been provided (i.e., $x^\alpha(y^\mu)$ is known). We may also proceed to consider transforming to a third coordinate system, from $y^\mu$ to $z^\kappa$.
	\begin{itemize}
		\item {\it Closure} \qquad Denote the Jacobian as, for e.g., $\mathcal{J}^\alpha_{\phantom{\alpha}\mu}[x \to y] \equiv \partial x^\alpha/\partial y^\mu$. If we define the group operation as simply that of matrix multiplication, verify that
		\begin{align}
		\mathcal{J}^\alpha_{\phantom{\alpha}\sigma}[x \to y] \mathcal{J}^\sigma_{\phantom{\alpha}\nu}[y \to z]
		= \mathcal{J}^\alpha_{\phantom{\alpha}\nu}[x \to z] .
		\end{align}
		%observe that
		%\begin{align}
		%\mathcal{J}^\alpha_{\phantom{\alpha}\sigma}[x \to y] \mathcal{J}^\sigma_{\phantom{\sigma}\nu}[y \to z]
		%&= \frac{\partial x^\alpha}{\partial y^\sigma} \frac{\partial y^\sigma}{\partial z^\nu} \\
		%&= \frac{\partial x^\alpha}{\partial z^\nu} \equiv \mathcal{J}^\alpha_{\phantom{\alpha}\nu}[x \to z] . \nonumber
		%\end{align}
		In words: multiplying the transformation matrix bringing us from $x$ to $y$ followed by that from $y$ to $z$, yields the Jacobian that brings us from $x$ directly to $z$. This composition law is what we would need, if the group operation is to implement coordinate transformations.
		\item {\it Associativity} \qquad Explain why the composition law for Jacobians is associative.
		\item {\it Identity} \qquad What is the identity Jacobian? What is the most general coordinate transformation it corresponds to?
		\item {\it Inverse} \qquad By non-singular, we mean $\det \mathcal{J}^\alpha_{\phantom{\alpha}\mu} \neq 0$. What does this imply about the existence of the inverse $(\mathcal{J}^{-1})^\alpha_{\phantom{\alpha}\mu}$?
	\end{itemize}	   \qed
\end{myP}

\subsection{Constancy of $c$; Orthonormal Frames; Timelike, Spacelike vs. Null Vectors; Gravitational Time Dilation}
\label{Chapter_DifferentialGeometry_CurvedSpacetimes_C1}
{\bf Flat Spacetimes} \qquad Cartesian coordinates play a basic but special role in interpreting physics in both flat Euclidean space $\delta_{ij}$ and flat Minkowski spacetime $\eta_{\mu\nu}$: they parametrize time durations and spatial distances in orthogonal directions -- i.e., every increasing tick mark along a given Cartesian axis corresponds directly to a measurement of increasing length or time in that direction. This is generically not so, say, for coordinates in curved space(time) because the notion of what constitutes a `straight line' is significantly more subtle there; or even spherical coordinates $(r \geq 0, 0 \leq \theta \leq \pi, 0 \leq \phi < 2\pi)$ in flat 3D space -- for the latter, only the radial coordinate $r$ corresponds to actual distance (from the origin).

We will therefore begin in flat spacetime written in Cartesian coordinates $\{x^\mu \equiv (t,\vec{x})\}$. Flat spacetime is also otherwise known as {\it Minkowski} spacetime, and the `square' of the distance between $x^\mu$ and $x^\mu + \dd x^\mu$, is given by
\begin{align}
\dd s^2 = \eta_{\mu\nu} \dd x^\mu \dd x^\nu 
\label{DifferentialGeometry_MinkowskiMetric}
&= (\dd x^0)^2 - \dd\vec{x}\cdot\dd\vec{x} \nonumber\\
&= (\dd t)^2 - \delta_{ij} \dd x^i \dd x^j ;
\end{align}
where the Minkowski metric tensor reads
\begin{align}
\label{MinkowskiMetric_etamunu}
\eta_{\mu\nu} \dot{=} \text{diag}[1,-1,\dots,-1] .
\end{align}
{\bf Constancy of $c$} \qquad One of the primary motivations that led Einstein to recognize eq. \eqref{DifferentialGeometry_MinkowskiMetric} as the proper geometric setting to describe physics, is the realization that the speed of light $c$ is constant in all inertial frames. In modern physics, the latter is viewed as a consequence of spacetime translation and Lorentz symmetry, as well as the null character of the trajectories swept out by photons. That is, for transformation matrices $\{ \Lambda \}$ satisfying
\begin{align}
\label{LorentzT_Def}
\LTud{\alpha}{\mu} \LTud{\beta}{\nu} \eta_{\alpha\beta} = \eta_{\mu\nu} ,
\end{align}
and constant vectors $\{ a^\mu \}$ we have
\begin{align}
\eta_{\mu\nu} \dd x^\mu \dd x^\nu &= \eta_{\mu\nu} \dd x'^\mu \dd x'^\nu 
\end{align}
whenever
\begin{align}
x^\alpha &= \LTud{\alpha}{\mu} x'^\mu + a^\alpha .
\end{align}
The physical interpretation is that the frames parametrized by $\{ x^\mu = (t,\vec{x}) \}$ and $\{ x'^\mu = (t',\vec{x}') \}$ are {\it inertial} frames: compact bodies with no external forces acting on them will sweep out geodesics $\dd^2 x^\mu/\dd \tau^2 = 0 = \dd^2 x'^\mu/\dd \tau'^2$, where the proper times $\tau$ and $\tau'$ are defined through the relations $\dd\tau = \dd t \sqrt{1-(\dd\vec{x}/\dd t)^2}$ and $\dd\tau' = \dd t' \sqrt{1-(\dd\vec{x}'/\dd t')^2}$. To interpret physical phenomenon taking place in one frame from the other frame's perspective, one would first have to figure out how to translate between $x$ and $x'$.

Let $x^\mu$ be the spacetime Cartesian coordinates of a single photon; in a different Lorentz frame it has Cartesian coordinates $x'^\mu$. Invoking its null character, namely $\dd s^2 = 0$ -- which holds in any inertial frame -- we have $(\dd x^0)^2 = \dd\vec{x}\cdot\dd\vec{x}$ and $(\dd x'^0)^2 = \dd\vec{x}'\cdot\dd\vec{x}'$. This in turn tells us the speeds in both frames is unity:
\begin{align}
\label{SpeedOfLightIsOne}
\frac{|\dd \vec{x}|}{\dd x^0} = \frac{|\dd \vec{x}'|}{\dd x'^0} = 1 .
\end{align}
A more thorough and hence deeper justification would be to recognize, it is the sign difference between the `time' part and the `space' part of the metric in eq. \eqref{DifferentialGeometry_MinkowskiMetric} -- together with its Lorentz invariance -- that gives rise to the wave equations obeyed by the photon. Equation \eqref{SpeedOfLightIsOne} then follows as a consequence.

{\bf Curved Spacetime, Spacetime Volume \& Orthonormal Basis} \qquad The generalization of the `distance-squared' between $x^\mu$ to $x^\mu + \dd x^\mu$, from the Minkowski to the curved case, is the following ``line element":
\begin{align}
\label{CurvedSpacetime_Metric}
\dd s^2 = g_{\mu\nu}(x) \dd x^\mu \dd x^\nu ,
\end{align}
where $x$ is simply shorthand for the spacetime coordinates $\{x^\mu\}$, which we emphasize may no longer be Cartesian. We also need to demand that $g_{\mu\nu}$ be real, symmetric, and has 1 positive eigenvalue associated with the one `time' coordinate and $(d-1)$ negative ones for the spatial coordinates. The infinitesimal spacetime volume continues to take the form
\begin{align}
\label{CurvedSpacetime_Volume}
\dd(\text{vol.}) = \dd^d x \sqrt{|g(x)|} ,
\end{align}
where $|g(x)| = |\det g_{\mu\nu}(x)|$ is now the absolute value of the determinant of the metric $g_{\mu\nu}$.

Just like the curved space case, to interpret physics in the neighborhood of some spacetime location $x^\mu$, we introduce an orthonormal basis $\{ \dbeinud{\mu}{\alpha} \}$ through the `diagonalization' process:
\begin{align}
g_{\mu\nu}(x) = \eta_{\alpha\beta} \dbeinud{\alpha}{\mu}(x) \dbeinud{\beta}{\nu}(x) .
\end{align}
By defining $\varepsilon^{\widehat{\alpha}} \equiv \dbeinud{\alpha}{\mu} \dd x^\mu$, the analog to achieving a Cartesian-like expression for the spacetime metric is
\begin{align}
\dd s^2 
= \left(\varepsilon^{\widehat{0}}\right)^2 - \sum_{i=1}^{D} \left(\varepsilon^{\widehat{i}}\right)^2 
= \eta_{\mu\nu} \varepsilon^{\widehat{\mu}} \varepsilon^{\widehat{\nu}} .
\end{align}
This means under a local Lorentz transformation -- i.e., for all 
\begin{align}
\LTud{\mu}{\alpha}(x) \LTud{\nu}{\beta}(x) \eta_{\mu\nu} 
&= \eta_{\alpha\beta} , \\
\varepsilon'^{\widehat{\mu}}(x) 
&= \LTud{\mu}{\alpha}(x) \varepsilon'^{\widehat{\alpha}}(x)
\end{align}
-- the metric remains the same:
\begin{align}
\dd s^2 
= \eta_{\mu\nu} \varepsilon^{\widehat{\mu}} \varepsilon^{\widehat{\nu}} 
= \eta_{\mu\nu} \varepsilon'^{\widehat{\mu}} \varepsilon'^{\widehat{\nu}} .
\end{align}
By viewing $\widehat{\varepsilon}$ as the matrix with the $\alpha$th row and $\mu$th column given by $\dbeinud{\alpha}{\mu}$, the determinant of the metric $g_{\mu\nu}$ can be written as
\begin{align}
\det g_{\mu\nu}(x)
= \left(\det \widehat{\varepsilon}\right)^2 \det \eta_{\mu\nu} .
\end{align}
The infinitesimal spacetime volume in eq. \eqref{CurvedSpacetime_Volume} now can be expressed as
\begin{align}
\dd^d x \sqrt{|g(x)|} 
&= \dd^d x \det \widehat{\varepsilon} \\
&= \varepsilon^{\widehat{0}} \wedge \varepsilon^{\widehat{1}} \wedge \dots \wedge \varepsilon^{\widehat{d-1}} .
\end{align}
The second equality follows because
\begin{align}
\varepsilon^{\widehat{0}} \wedge \dots \wedge \varepsilon^{\widehat{d-1}} 
&= \dbeinud{0}{\mu_1} \dd x^{\mu_1} \wedge \dots \wedge \dbeinud{0}{\mu_d} \dd x^{\mu_d} \nonumber\\
&= \epsilon_{\mu_1 \dots \mu_d} \dbeinud{0}{\mu_1} \dots \dbeinud{d-1}{\mu_d} \dd x^{0} \wedge \dots \wedge \dd x^{d-1} = (\det \widehat{\varepsilon}) \dd^d x .
\end{align}
Of course, that $g_{\mu\nu}$ may be `diagonalized' follows from the fact that $g_{\mu\nu}$ is a real symmetric matrix:
\begin{align}
\label{CurvedSpacetime_MetricDiagonalization}
g_{\mu\nu} 
= \sum_{\alpha,\beta} O^\alpha_{\phantom{\alpha}\mu} \lambda_\alpha \eta_{\alpha\beta} O^\beta_{\phantom{\beta}\nu} 
= \sum_{\alpha,\beta} \dbeinud{\alpha}{\mu} \eta_{\alpha\beta} \dbeinud{\beta}{\nu} ,
\end{align}
where all $\{ \lambda_\alpha \}$ are positive by assumption, so we may take their positive root:
\begin{align}
\label{CurvedSpacetime_dBeins}
\dbeinud{\alpha}{\mu} = \sqrt{\lambda_\alpha} O^\alpha_{\phantom{\alpha}\mu} , \qquad\qquad 
\{\lambda_\alpha > 0\} , \qquad\qquad
(\text{No sum over $\alpha$}).
\end{align}
That $\dbeinud{0}{\mu}$ acts as `standard clock' and $\{ \dbeinud{i}{\mu} \vert i = 1,2,\dots,D \}$ act as `standard rulers' is because they are of unit length:
\begin{align}
g^{\mu\nu} \dbeinud{\alpha}{\mu} \dbeinud{\beta}{\nu} = \eta^{\alpha\beta} .
\end{align}
The $\widehat{\cdot}$ on the index indicates it is to be moved with the flat metric, namely
\begin{align}
\dbeinud{\alpha}{\mu} = \eta^{\alpha\beta} \dbeindd{\beta}{\mu}
\qquad \text{ and } \qquad
\dbeindd{\alpha}{\mu} = \eta_{\alpha\beta} \dbeinud{\beta}{\mu} ;
\end{align}
while the spacetime index is to be moved with the spacetime metric
\begin{align}
\dbeinuu{\alpha}{\mu} = g^{\mu\nu} \dbeinud{\alpha}{\nu}
\qquad \text{ and } \qquad
\dbeinud{\alpha}{\mu} = g_{\mu\nu} \dbeinuu{\alpha}{\nu} .
\end{align}
In other words, we view $\dbeindu{\alpha}{\mu}$ as the $\mu$th spacetime component of the $\alpha$th vector field in the basis set $\{ \dbeindu{\alpha}{\mu} \vert \alpha = 0,1,2,\dots,D \equiv d-1 \}$. We may elaborate on the interpretation that $\{ \dbeinud{\alpha}{\mu} \}$ act as `standard clock/rulers' as follows. For a test (scalar) function $f(x)$ defined throughout spacetime, the rate of change of $f$ along $\varepsilon_{\widehat{0}}$ is
\begin{align}
\braket{\dd f}{\varepsilon_{\widehat{0}}}
= \dbeindu{0}{\mu} \partial_\mu f \equiv \frac{\dd f}{\dd y^0} ;
\end{align}
whereas that along $\varepsilon_{\widehat{i}}$ is
\begin{align}
\braket{\dd f}{\varepsilon_{\widehat{i}}} 
= \dbeindu{i}{\mu} \partial_\mu f \equiv \frac{\dd f}{\dd y^i} ;
\end{align}
where $y^0$ and $\{y^i\}$ are to be viewed as `time' and `spatial' parameters along the integral curves of $\{ \dbeindu{\mu}{\alpha} \}$. That these are Cartesian-like can now be expressed as
\begin{align}
\label{dbeins_orthonormality}
\braket{\frac{\dd}{\dd y^\mu}}{\frac{\dd}{\dd y^\nu}} 
= \dbeindu{\mu}{\alpha} \dbeindu{\nu}{\beta} \braket{\partial_\alpha}{\partial_\beta}
= \dbeindu{\mu}{\alpha} \dbeindu{\nu}{\beta} g_{\alpha\beta}
= \eta_{\mu\nu} .
\end{align}
It is worth reiterating that the first equalities of eq. \eqref{CurvedSpacetime_MetricDiagonalization} are really assumptions, in that the definitions of curved spaces include assuming all the eigenvalues of the metric are positive whereas that of curved spacetimes include assuming all but one eigenvalue is negative.\footnote{In $d-$spacetime dimensions, with our sign convention in place, if there were $n$ `time' directions and $(d-n)$ `spatial' ones, then this carries with it the assumption that $g_{\mu\nu}$ has $n$ positive eigenvalues and $(d-n)$ negative ones.}

Note that the $\{ \dd/\dd y^\mu \}$ in eq. \eqref{dbeins_orthonormality} do not, generically, commute. For instance, acting on a scalar function,
\begin{align}
\left[\frac{\dd}{\dd y^\mu}, \frac{\dd}{\dd y^\nu}\right] f(x) 
&= \left(\frac{\dd}{\dd y^\mu}\frac{\dd}{\dd y^\nu} - \frac{\dd}{\dd y^\nu}\frac{\dd}{\dd y^\mu}\right) f(x) \\
&= \left( \dbeindu{\mu}{\alpha} \partial_\alpha \dbeindu{\nu}{\beta} - \dbeindu{\nu}{\alpha} \partial_\alpha \dbeindu{\mu}{\beta} \right) \partial_\beta f(x) 
\neq 0 .
\end{align}
A theorem in differential geometry -- see, for instance, Schutz \cite{Schutz_GeometricalMethods} for a pedagogical discussion -- tells us: 
\begin{quotation}
	A set of $1 < N \leq d$ vector fields $\{ \dd/\dd \xi^\mu \}$ form a coordinate basis in the $N-$dimensional space(time) they inhabit, if and only if they commute.
\end{quotation}
When $N = d$, and if $[\dd/\dd y^\mu,\dd/\dd y^\nu]=0$ in eq. \eqref{dbeins_orthonormality}, we would not only have found coordinates $\{ y^\mu \}$ for our spacetime, we would have found this spacetime is a flat one.

It is perhaps important to clarify what a coordinate system is. In 2D, for instance, if we had $[\dd/\dd y^0,\dd/\dd y^1] \neq 0$, this means it is not possible to vary the `coordinate' $y^0$ (i.e., along the integral curve of $\dd/\dd y^0$) without holding the `coordinate' $y^1$ fixed; or, it is not possible to hold $y^0$ fixed while moving along the integral curve of $\dd/\dd y^1$.
\begin{myP}
	{\bf Example: Schutz \cite{Schutz_GeometricalMethods} Exercise 2.1} \qquad In 2D flat space, starting from Cartesian coordinates $x^i$, we may convert to cylindrical coordinates
	\begin{align}
	(x^1,x^2) = r(\cos\phi,\sin\phi) .
	\end{align}
	The pair of vector fields $(\partial_r,\partial_\phi)$ do form a coordinate basis -- it is possible to hold $r$ fixed while going along the integral curve of $\partial_\phi$ and vice versa. However, show via a direct calculation that the following commutator involving the unit vector fields $\widehat{r}$ and $\widehat{\phi}$ is not zero:
	\begin{align}
	\left[\widehat{r} , \widehat{\phi}\right] f(r,\phi) \neq 0 ;
	\end{align}
	where
	\begin{align}
	\widehat{r} 	&\equiv \cos(\phi) \partial_{x^1} + \sin(\phi) \partial_{x^2} , \\
	\widehat{\phi} 	&\equiv -\sin(\phi) \partial_{x^1} + \cos(\phi) \partial_{x^2} .
	\end{align} 
	Therefore $\widehat{r}$ and $\widehat{\phi}$ do not form a coordinate basis. \qed
\end{myP}
{\bf Timelike, Spacelike, and Null Distances/Vectors} \qquad A fundamental difference between (curved) space versus spacetime, is that the former involves strictly positive distances while the latter -- because of the $\eta_{00} = +1$ for orthonormal `time' versus $\eta_{ii} = -1$ for the $i$th orthonormal space component -- involves positive, zero, and negative distances.

With our `mostly minus' sign convention (cf. eq. \eqref{DifferentialGeometry_MinkowskiMetric}), a vector $v^\mu$ is:
\begin{itemize}
	\item {\it Time-like} if $v^2 \equiv \eta_{\mu\nu} v^{\widehat{\mu}} v^{\widehat{\nu}} > 0$. We have seen in \S \eqref{Chapter_PoincareLorentz}: if $v^2 > 0$, it is always possible to find a Lorentz transformation $\Lambda$ (cf. eq. \eqref{LorentzT_Def}) such that $\Lambda^\mu_{\phantom{\mu}\alpha} v^{\widehat{\alpha}} = (v'^{\widehat{0}},\vec{0})$. In flat spacetime, if $\dd s^2 = \eta_{\mu\nu} \dd x^\mu \dd x^\nu > 0$ then this result indicates it is always possible to find an inertial frame where $\dd s^2 = \dd t'^2$: hence the phrase `timelike'. 
	
	More generally, for a timelike trajectory $z^\mu(\lambda)$ in curved spacetime -- i.e., $g_{\mu\nu} (\dd z^\mu/\dd\lambda) (\dd z^\nu/\dd\lambda) > 0$, we may identify
	\begin{align}
	\dd \tau \equiv \dd\lambda \sqrt{g_{\mu\nu}(z(\lambda)) \frac{\dd z^\mu}{\dd\lambda} \frac{\dd z^\nu}{\dd\lambda}} 
	\end{align}
	as the (infinitesimal) {\it proper time}, the time read by the watch of an observer whose worldline is $z^\mu(\lambda)$. (As a check: when $g_{\mu\nu}=\eta_{\mu\nu}$ and the observer is at rest, namely $\dd\vec{z}=0$, then $\dd\tau = \dd t$.) Using orthonormal frame fields in eq. \eqref{CurvedSpacetime_MetricDiagonalization},
	\begin{align}
	\dd \tau = \dd\lambda \sqrt{\eta_{\alpha\beta} \frac{\dd z^{\widehat{\alpha}}}{\dd\lambda} \frac{\dd z^{\widehat{\beta}}}{\dd\lambda}} , \qquad\qquad
	\frac{\dd z^{\widehat{\alpha}}}{\dd\lambda} \equiv \dbeinud{\alpha}{\mu} \frac{\dd z^{\mu}}{\dd\lambda} .
	\end{align}
	Furthermore, since $v^{\widehat{\mu}} \equiv \dd z^{{\widehat{\mu}}}/\dd \lambda$ is assumed to be timelike, it must be possible to find a local Lorentz transformation $\LTud{\mu}{\nu}(z)$ such that $\LTud{\mu}{\nu} v^{\widehat{\nu}} = (v'^{\widehat{0}},\vec{0})$; assuming $\dd\lambda > 0$,
	\begin{align}
	\dd \tau 
	&= \dd\lambda \sqrt{\eta_{\mu\nu} \LTud{\mu}{\alpha} \LTud{\nu}{\beta} \frac{\dd z^{\widehat{\alpha}}}{\dd\lambda} \frac{\dd z^{\widehat{\beta}}}{\dd\lambda}} , \nonumber\\
	&= \dd\lambda \sqrt{\left(\frac{\dd z'^{\widehat{0}}}{\dd\lambda}\right)^2}
	= |\dd z'^{\widehat{0}}| .
	\end{align}
	\item {\it Space-like} if $v^2 \equiv \eta_{\mu\nu} v^{\widehat{\mu}} v^{\widehat{\nu}} < 0$. We have seen in \S \eqref{Chapter_PoincareLorentz}: if $v^2 < 0$, it is always possible to find a Lorentz transformation $\Lambda$ such that $\Lambda^\mu_{\phantom{\mu}\alpha} v^{\widehat{\alpha}} = (0,v'^{\widehat{i}})$. In flat spacetime, if $\dd s^2 = \eta_{\mu\nu} \dd x^\mu \dd x^\nu < 0$ then this result indicates it is always possible to find an inertial frame where $\dd s^2 = -\dd \vec{x}'^2$: hence the phrase `spacelike'.
	
	More generally, for a spacelike trajectory $z^\mu(\lambda)$ in curved spacetime -- i.e., $g_{\mu\nu} (\dd z^\mu/\dd\lambda) (\dd z^\nu/\dd\lambda) < 0$, we may identify
	\begin{align}
	\dd \ell \equiv \dd\lambda \sqrt{\left\vert g_{\mu\nu}(z(\lambda)) \frac{\dd z^\mu}{\dd\lambda} \frac{\dd z^\nu}{\dd\lambda} \right\vert} 
	\end{align}
	as the (infinitesimal) {\it proper length}, the distance read off some measuring rod whose trajectory is $z^\mu(\lambda)$. (As a check: when $g_{\mu\nu}=\eta_{\mu\nu}$ and $\dd t = 0$, i.e., the rod is lying on the constant$-t$ surface, then $\dd\ell = |\dd\vec{x}\cdot\dd\vec{x}|^{1/2}$.) Using the orthonormal frame fields in eq. \eqref{CurvedSpacetime_MetricDiagonalization},
	\begin{align}
	\dd \ell = \dd\lambda \sqrt{\left\vert \eta_{\alpha\beta} \frac{\dd z^{\widehat{\alpha}}}{\dd\lambda} \frac{\dd z^{\widehat{\beta}}}{\dd\lambda} \right\vert} , \qquad\qquad
	\frac{\dd z^{\widehat{\alpha}}}{\dd\lambda} \equiv \dbeinud{\alpha}{\mu} \frac{\dd z^{\mu}}{\dd\lambda} .
	\end{align}
	Furthermore, since $v^{\widehat{\mu}} \equiv \dd z^{{\widehat{\mu}}}/\dd \lambda$ is assumed to be spacelike, it must be possible to find a local Lorentz transformation $\LTud{\mu}{\nu}(z)$ such that $\LTud{\mu}{\nu} v^{\widehat{\nu}} = (0,v'^{\widehat{i}})$; assuming $\dd\lambda > 0$,
	\begin{align}
	\dd \ell
	&= \dd\lambda \sqrt{\eta_{\mu\nu} \LTud{\mu}{\alpha} \LTud{\nu}{\beta} \frac{\dd z^{\widehat{\alpha}}}{\dd\lambda} \frac{\dd z^{\widehat{\beta}}}{\dd\lambda}}
	= \left\vert \dd \vec{z}' \right\vert ; \\
	\dd\vec{z}'^{\widehat{i}}
	&\equiv \LTud{i}{\mu} \dbeinud{\mu}{\nu} \dd z^\nu .
	\end{align}
	\item {\it Null} if $v^2 \equiv \eta_{\mu\nu} v^{\widehat{\mu}} v^{\widehat{\nu}} = 0$. We have already seen, in flat spacetime, if $\dd s^2 = \eta_{\mu\nu} \dd x^\mu \dd x^\nu = 0$ then $|\dd\vec{x}|/\dd x^0 = |\dd\vec{x}'|/\dd x'^0 = 1$ in all inertial frames.
\end{itemize}
It is physically important to reiterate: one of the reasons why it is important to make such a distinction between vectors, is because it is {\it not possible} to find a Lorentz transformation that would linearly transform one of the above three types of vectors into another different type -- for e.g., it is not possible to Lorentz transform a null vector into a time-like one (a photon has no `rest frame'); or a time-like vector into a space-like one; etc. This is because their Lorentzian `norm-squared'
\begin{align}
v^2 \equiv \eta_{\mu\nu} v^{\widehat{\mu}} v^{\widehat{\nu}}
= \eta_{\alpha\beta} \LTud{\alpha}{\mu} \LTud{\beta}{\nu} v^{\widehat{\mu}} v^{\widehat{\nu}} 
= \eta_{\alpha\beta} v'^{\widehat{\alpha}} v'^{\widehat{\beta}} 
\end{align}
has to be invariant under all Lorentz transformations $v'^{\widehat{\alpha}} \equiv \LTud{\alpha}{\mu} v^{\widehat{\mu}}$. This in turn teaches us: if $v^2$ were positive, it has to remain so; likewise, if it were zero or negative, a Lorentz transformation cannot alter this attribute.
\begin{myP}
	{\bf Orthonormal Frames in Kerr-Schild Spacetimes} \qquad A special class of geometries, known as {\it Kerr-Schild} spacetimes, take the following form.
	\begin{align}
	g_{\mu\nu} = \gb_{\mu\nu} + H k_\mu k_\nu 
	\end{align}
	Many of the known black hole spacetimes can be put in this form; and in such a context, $\gb_{\mu\nu}$ usually refers to flat or de Sitter spacetime.\footnote{See Gibbons et al. \cite{Gibbons:2004uw} arXiv: \href{https://arxiv.org/abs/hep-th/0404008}{hep-th/0404008}. The special property of Kerr-Schild coordinates is that Einstein's equations become {\it linear} in these coordinates.} The $k_\mu$ is null with respect to $\gb_{\mu\nu}$, i.e.,
	\begin{align}
	\gb_{\alpha\beta} k^\alpha k^\beta = 0,
	\end{align}
	and we shall move its indices with $\gb_{\mu\nu}$. 
	
	Verify that the inverse metric is
	\begin{align}
	g^{\mu\nu} = \gb^{\mu\nu} - H k^\mu k^\nu, 
	\end{align}
	where $\gb^{\mu\sigma}$ is the inverse of $\gb_{\mu\sigma}$, namely $\gb^{\mu\sigma} \gb_{\sigma\nu} \equiv \delta^\mu_{\phantom{\mu}\nu}$. Then, verify that the orthonormal frame fields are
	\begin{align}
	\dbeinud{\alpha}{\mu} = \delta^\alpha_{\phantom{\alpha}\mu} + \frac{1}{2} H k^\alpha k_\mu .
	\end{align}
	Can you explain why $k^\mu$ is also null with respect to the full metric $g_{\mu\nu}$? \qed
\end{myP}
{\bf Proper times and Gravitational Time Dilation} \qquad Consider two observers sweeping out their respective timelike worldlines in spacetime, $y^\mu(\lambda)$ and $z^\mu(\lambda)$. If we use the time coordinate of the geometry to parameterize their trajectories, their proper times -- i.e., the time read by their watches -- are given by
\begin{align}
\dd\tau_y &\equiv \dd t \sqrt{g_{\mu\nu}(y(t)) \dot{y}^\mu \dot{y}^\nu}, \qquad\qquad 
\dot{y}^\mu \equiv \frac{\dd y^\mu}{\dd t} ; \\
\dd\tau_z &\equiv \dd t \sqrt{g_{\mu\nu}(z(t)) \dot{z}^\mu \dot{z}^\nu}, \qquad\qquad 
\dot{z}^\mu \equiv \frac{\dd z^\mu}{\dd t} .
\end{align} 
In flat spacetime, clocks that are synchronized in one frame are no longer synchronized in a different frame -- chronology is not a Lorentz invariant. We see that, in curved spacetime, the infinitesimal {\it passage} of proper time measured by observers at the same `coordinate time' $t$ depends on their spacetime locations:
\begin{align}
\label{DifferentialGeometry_ProperTimeRatios}
\frac{\dd \tau_y}{\dd \tau_z} 
= \sqrt{\frac{g_{\mu\nu}(y(t)) \dot{y}^\mu \dot{y}^\nu}{g_{\alpha\beta}(z(t)) \dot{y}^\alpha \dot{y}^\beta}} .
\end{align}
Physically speaking, eq. \eqref{DifferentialGeometry_ProperTimeRatios} does not, in general, yield the ratio of proper times measured by observers at two different locations. (Drawing a spacetime diagram here helps.) To do so, one would have to specify the trajectories of both $y^\mu(\lambda_1 \leq \lambda \leq \lambda_2)$ and $z^\mu(\lambda'_1 \leq \lambda' \leq \lambda'_2)$, before the integrals $\Delta \tau_1 \equiv \int_{\lambda_1}^{\lambda_2} \dd\lambda \sqrt{g_{\mu\nu} \dot{y}^\mu \dot{y}^\nu}$ and $\Delta \tau_2 \equiv \int_{\lambda'_1}^{\lambda'_2} \dd\lambda' \sqrt{g_{\mu\nu} \dot{z}^\mu \dot{z}^\nu}$ are evaluated and compared.
\begin{myP}
	\label{Problem_GravitationalTimeDilation}
	{\bf Example} \qquad The spacetime geometry around the Earth itself can be approximated by the line element
	\begin{align}
	\label{Schwarzschild_Earth}
	\dd s^2 = \left(1-\frac{r_{s,E}}{r}\right) \dd t^2 - \frac{\dd r^2}{1 - r_{s,E}/r} - r^2 \left(\dd\theta^2 + \sin(\theta)^2 \dd\phi^2 \right) ,
	\end{align}
	where $t$ is the time coordinate and $(r,\theta,\phi)$ are analogs of the spherical coordinates. Whereas $r_{s,E}$ is known as the Schwarzschild radius of the Earth, and depends on the Earth's mass $M_\text{E}$ through the expression
	\begin{align}
	\label{SchwarzschildRadius_Earth}
	r_{s,E} \equiv 2 \GN M_\text{E} .
	\end{align}
	Find the $4-$beins of the geometry in eq. \eqref{Schwarzschild_Earth}. Then find the numerical value of $r_{s,E}$ in eq. \eqref{SchwarzschildRadius_Earth} and take the ratio $r_{s,E}/R_\text{E}$, where $R_\text{E}$ is the radius of the Earth. Explain why this means we may -- for practical purposes -- expand the metric in eq. \eqref{SchwarzschildRadius_Earth} as
	\begin{align}
	\label{Schwarzschild_Earth_TaylorExpansion}
	\dd s^2 = \left(1-\frac{r_{s,E}}{r}\right) \dd t^2 &- \dd r^2 \left( 1 + \frac{r_{s,E}}{r} + \left(\frac{r_{s,E}}{r}\right)^2 + \left(\frac{r_{s,E}}{r}\right)^3 + \dots \right) \nonumber\\
	&- r^2 \left(\dd\theta^2 + \sin(\theta)^2 \dd\phi^2 \right) .
	\end{align}	
	Since we are not in flat spacetime, the $(t,r,\theta,\phi)$ are no longer subject to the same interpretation. However, use your computation of $r_{s,E}/R_\text{E}$ to {\it estimate} the error incurred if we do continue to interpret $t$ and $r$ as though they measured time and radial distances, with respect to a frame centered at the Earth's core.
	
	Consider placing one clock at the base of the Taipei 101 tower and another at its tip. Denoting the time elapsed at the base of the tower as $\Delta \tau_\text{B}$; that at the tip as $\Delta \tau_\text{T}$; and assuming for simplicity the Earth is a perfect sphere -- show that eq. \eqref{DifferentialGeometry_ProperTimeRatios} translates to
	\begin{align}
	\label{EarthTimeDilation}
	\frac{\Delta \tau_\text{B}}{\Delta \tau_\text{T}} 
	= \sqrt{\frac{g_{00}(R_\text{E})}{g_{00}(R_\text{E}+h_{101})}}
	\approx 1 + \frac{1}{2} \left( \frac{r_{s,E}}{R_\text{E}+h_{101}} - \frac{r_{s,E}}{R_\text{E}} \right) .
	\end{align} 
	Here, $R_{\text{E}}$ is the radius of the Earth and $h_{101}$ is the height of the Taipei 101 tower. Notice the right hand side is related to the difference in the Newtonian gravitational potentials at the top and bottom of the tower. 
	
	In actuality, both clocks are in motion, since the Earth is rotating. Can you estimate what is the error incurred from assuming they are at rest? First arrive at eq. \eqref{EarthTimeDilation} analytically, then plug in the relevant numbers to compute the numerical value of $\Delta \tau_\text{B}/\Delta \tau_\text{T}$. Does the clock at the base of Taipei 101 or that on its tip tick more slowly? 
	
	This gravitational time dilation is an effect that needs to be accounted for when setting up a network of Global Positioning Satellites (GPS); for details, see Ashby \cite{Ashby:2003vja}. \qed 
\end{myP}

\subsection{Connections, Curvature, Geodesics, Isometries}
\label{Chapter_DifferentialGeometry_CurvedSpacetimes_C2}
{\bf Connections \& Christoffel Symbols} \qquad The partial derivative on a scalar $\varphi$ is a rank-1 tensor, so we shall simply define the covariant derivative acting on $\varphi$ to be
\begin{align}
\nabla_\alpha \varphi = \partial_\alpha \varphi .
\end{align}
Because the partial derivative itself cannot yield a tensor once it acts on tensor, we need to introduce a connection $\Christ{\mu}{\alpha}{\beta}$, i.e., 
\begin{align}
\label{Connection_CovD}
\nabla_\sigma V^\mu
&= \partial_\sigma V^\mu + \Gamma^\mu_{\phantom{\mu}\sigma\rho} V^\rho .
\end{align}
Under a coordinate transformation of the partial derivatives and $V^\mu$, say going from $x$ to $x'$,
\begin{align}
\label{Connection_StepI}
\partial_\sigma V^\mu + \Gamma^\mu_{\phantom{\mu}\sigma\rho} V^\rho
&= \frac{\partial x'^\lambda}{\partial x^\sigma} \frac{\partial x^\mu}{\partial x'^\nu} \partial_{\lambda'} V^{\nu'} 
+ \left(\frac{\partial x'^\lambda}{\partial x^\sigma} \frac{\partial^2 x^\mu}{\partial x'^\lambda x'^\nu} 
+ \Gamma^\mu_{\phantom{\mu}\sigma\rho} \frac{\partial x^\rho}{\partial x'^\nu}\right) V^{\nu'} .
\end{align}
On the other hand, if $\nabla_\sigma V^\mu$ were to transform as a tensor,
\begin{align}
\label{Connection_StepII}
\partial_\sigma V^\mu + \Gamma^\mu_{\phantom{\mu}\sigma\rho} V^\rho
&= \frac{\partial x'^\lambda}{\partial x^\sigma} \frac{\partial x^\mu}{\partial x'^\nu} \partial_{\lambda'} V^{\nu'} 
+ \frac{\partial x'^\lambda}{\partial x^\sigma} \frac{\partial x^\mu}{\partial x'^\tau} \Gamma^{\tau'}_{\phantom{\mu'}\lambda'\nu'} V^{\nu'} .
\end{align}
\footnote{All un-primed indices represent tensor components in the $x$-system; while all primed indices those in the $x'$ system.}Since $V^{\nu'}$ is an arbitrary vector, we may read off its coefficient on the right hand sides of equations \eqref{Connection_StepI} and \eqref{Connection_StepII}, and deduce the connection has to transform as
\begin{align}
\frac{\partial x'^\lambda}{\partial x^\sigma} \frac{\partial^2 x^\mu}{\partial x'^\lambda x'^\nu} 
+ \Gamma^\mu_{\phantom{\mu}\sigma\rho}(x) \frac{\partial x^\rho}{\partial x'^\nu}
= \frac{\partial x'^\lambda}{\partial x^\sigma} \frac{\partial x^\mu}{\partial x'^\tau} \Gamma^{\tau'}_{\phantom{\mu'}\lambda'\nu'}(x') .
\end{align}
Moving all the Jacobians onto the connection written in the $\{ x^\mu \}$ frame,
\begin{align}
\label{Connection_CoordinateTransformation}
\Gamma^{\tau'}_{\phantom{\mu'}\kappa'\nu'}(x')
= \frac{\partial x'^\tau}{\partial x^\mu} \frac{\partial^2 x^\mu}{\partial x'^\kappa x'^\nu} 
+ \frac{\partial x'^\tau}{\partial x^\mu} \Gamma^\mu_{\phantom{\mu}\sigma\rho}(x) \frac{\partial x^\sigma}{\partial x'^\kappa} \frac{\partial x^\rho}{\partial x'^\nu} .
\end{align}
All connections have to satisfy this non-tensorial transformation law. On the other hand, if we found an object that transforms according to eq. \eqref{Connection_CoordinateTransformation}, and if one employs it in eq. \eqref{Connection_CovD}, then the resulting $\nabla_\alpha V^\mu$ would transform as a tensor. 

{\it Product rule} \qquad For physical applications, because covariant derivatives should reduce to partial derivatives in flat Cartesian coordinates, it is natural to require the former to obey the usual product rule. For any two tensors $T_1$ and $T_2$, and suppressing all indices,
\begin{align}
\nabla(T_1 T_2) = (\nabla T_1) T_2 + T_1 (\nabla T_2) .
\end{align}
\begin{myP}
	Let us take the covariant derivative of a 1-form:
	\begin{align}
	\nabla_\alpha V_\mu = \partial_\alpha V_\mu + \Gamma'^\sigma_{\phantom{\sigma}\alpha\mu} V_\sigma .
	\end{align}
	Can you prove that this connection is negative of the vector one in eq. \eqref{Connection_CovD}?
	\begin{align}
	\label{Connection_1Form_vs_Vector}
	\Gamma'^\sigma_{\phantom{\sigma}\alpha\mu} = - \Gamma^\sigma_{\phantom{\sigma}\alpha\mu} ,
	\end{align}
	where $\Gamma^\sigma_{\phantom{\sigma}\alpha\mu}$ is the connection in eq. \eqref{Connection_CovD} -- if we define the covariant derivative of a scalar to be simply the partial derivative acting on the same, i.e.,
	\begin{align}
	\nabla_\alpha \left( V^\mu W_\mu \right) = \partial_\alpha \left( V^\mu W_\mu \right) ?
	\end{align}
	You should assume the product rule holds, namely $\nabla_\alpha \left( V^\mu W_\mu \right) = \left( \nabla_\alpha V^\mu  \right) W_\mu + V^\mu \left( \nabla_\alpha W_\mu \right)$. Expand these covariant derivatives in terms of the connections and argue why this leads to eq. \eqref{Connection_1Form_vs_Vector}. \qed
\end{myP}
Suppose we found two such connections, $\,_{(1)}\Gamma^{\tau}_{\phantom{\mu}\kappa\nu}(x)$ and $\,_{(2)}\Gamma^{\tau}_{\phantom{\mu}\kappa\nu}(x)$. Notice their difference does transform as a tensor because the first term on the right hand side involving the Hessian $\partial^2 x/\partial x' \partial x'$ cancels out:
\begin{align}
\,_{(1)}\Gamma^{\tau'}_{\phantom{\mu'}\kappa'\nu'}(x') - \,_{(2)}\Gamma^{\tau'}_{\phantom{\mu'}\kappa'\nu'}(x')
&= \frac{\partial x'^\tau}{\partial x^\mu} \left(\,_{(1)}\Gamma^\mu_{\phantom{\mu}\sigma\rho}(x) - \,_{(2)}\Gamma^\mu_{\phantom{\mu}\sigma\rho}(x)\right) \frac{\partial x^\sigma}{\partial x'^\kappa} \frac{\partial x^\rho}{\partial x'^\nu} .
\end{align}
Now, any connection can be decomposed into its symmetric and antisymmetric parts in the following sense:
\begin{align}
\Christ{\mu}{\alpha}{\beta}
= \frac{1}{2} \Christ{\mu}{\{\alpha}{\beta\}} + \frac{1}{2} \Christ{\mu}{[\alpha}{\beta]} .
\end{align}
This is, of course, mere tautology. However, let us denote 
\begin{align}
\,_{(1)}\Gamma^{\mu}_{\phantom{\mu}\alpha\beta} \equiv \frac{1}{2} \Gamma^{\mu}_{\phantom{\mu}\alpha\beta} 
\qquad \text{ and } \qquad
\,_{(2)}\Gamma^{\mu}_{\phantom{\mu}\alpha\beta} \equiv \frac{1}{2} \Gamma^{\mu}_{\phantom{\mu}\beta\alpha} ;
\end{align}
so that 
\begin{align}
\frac{1}{2} \Gamma^{\mu}_{\phantom{\mu}[\alpha\beta]} = \,_{(1)}\Gamma^{\mu}_{\phantom{\mu}\alpha\beta} - \,_{(2)}\Gamma^{\mu}_{\phantom{\mu}\alpha\beta}
\equiv T^{\mu}_{\phantom{\mu}\alpha\beta} .
\end{align}
We then see that this anti-symmetric part of the connection is in fact a tensor. It is the symmetric part $(1/2)\Christ{\mu}{\{\alpha}{\beta\}}$ that does not transform as a tensor. {\it For the rest of these notes, by $\Christ{\mu}{\alpha}{\beta}$ we shall always mean a symmetric connection.} This means our covariant derivative would now read
\begin{align}
\nabla_\alpha V^\mu 
= \partial_\alpha V^\mu + \Christ{\mu}{\alpha}{\beta} V^\beta + T^\mu_{\phantom{\mu}\alpha\beta} V^\beta .
\end{align}
As is common within the physics literature, we proceed to set to zero the torsion term: $T^\mu_{\phantom{\mu}\alpha\beta} \to 0$. If we further impose the metric compatibility condition,
\begin{align}
\label{MetricCompatibility}
\nabla_\mu g_{\alpha\beta} = 0 ,
\end{align}
then we have already seen in \S \eqref{Chapter_DifferentialGeometry_CurvedSpaces} this implies
\begin{align}
\label{Christoffel_Def}
\Gamma^\mu_{\phantom{\mu}\alpha\beta}
= \frac{1}{2} g^{\mu\sigma} \left( \partial_\alpha g_{\beta\sigma} + \partial_\beta g_{\alpha\sigma} - \partial_\sigma g_{\alpha\beta} \right) .
\end{align}
{\bf Parallel Transport \& Riemann Tensor} \qquad Along a curve $z^\mu(\lambda)$ such that one end is $z^\mu(\lambda=\lambda_1)=x'^\mu$ and the other end is $z^\mu(\lambda=\lambda_2)=x^\mu$, we may parallel transport some vector $V^\alpha$ from $x'$ to $x$ by exponentiating the covariant derivative along $z^\mu(\lambda)$. If $V^\alpha(x' \to x)$ is the result of this parallel transport, we have
\begin{align}
\label{DifferentialGeometry_ParallelTransportExp}
V^\alpha(x' \to x) = e^{(\lambda_2-\lambda_1) \dot{z}^\mu(\lambda_1) \nabla_\mu} V^\alpha(x') .
\end{align}
This is the covariant derivative analog of the Taylor expansion of a scalar function -- where, translation by a constant spacetime vector $a^\mu$ may be implemented as
\begin{align}
f(x^\mu+a^\mu) = \exp\left( a^\nu \partial_\nu \right) f(x^\mu) .
\end{align}
To elucidate the definition of geometric curvature as the failure of tensors to remain invariant under parallel transport, we may now attempt to parallel transport a vector $V^\alpha$ around a closed parallelogram defined by the tangent vectors $A$ and $B$. We shall soon see how the Riemann tensor itself emerges from such an analysis.

Let the 4 sides of this parallelogram have infinitesimal affine parameter length $\epsilon$. We will now start from one of its 4 corners, which we will denote as $x$. $V^\alpha$ will be parallel transported from $x$ to $x + \epsilon A$; then to $x + \epsilon A + \epsilon B$; then to $x + \epsilon A + \epsilon B - \epsilon A = x + \epsilon B$; and finally back to $x + \epsilon B - \epsilon B = x$. Let us first work out the parallel transport along the `side' $A$ using eq. \eqref{DifferentialGeometry_ParallelTransportExp}. Denoting $\nabla_A \equiv A^\mu \nabla_\mu$, $\nabla_B \equiv B^\mu \nabla_\mu$, etc.,
\begin{align}
V^\alpha(x \to x+\epsilon A) 
&= \exp(\epsilon \nabla_A) V^\alpha(x),  \nonumber\\
&= V^\alpha(x) + \epsilon \nabla_A V^\alpha(x) + \frac{\epsilon^2}{2} \nabla_A^2 V^\alpha(x) + \mathcal{O}\left(\epsilon^3\right) .
\end{align}
We then parallel transport this result from $x + \epsilon A$ to $x + \epsilon A + \epsilon B$.
{\allowdisplaybreaks\begin{align}
	&V^\alpha(x \to x+\epsilon A \to x+\epsilon A + \epsilon B)  \nonumber\\
	&= \exp(\epsilon \nabla_B) \exp(\epsilon \nabla_A) V^\alpha(x), \nonumber\\
	&= V^\alpha(x) + \epsilon \nabla_A V^\alpha(x) + \frac{\epsilon^2}{2} \nabla_A^2 V^\alpha(x) \nonumber\\
	&\qquad + \epsilon \nabla_B V^\alpha(x) + \epsilon^2 \nabla_B \nabla_A V^\alpha(x) \nonumber\\
	&\qquad + \frac{\epsilon^2}{2} \nabla_B^2 V^\alpha(x) + \mathcal{O}\left(\epsilon^3\right) \nonumber\\
	&= V^\alpha(x) + \epsilon \left( \nabla_A + \nabla_B \right) V^\alpha(x)
	+ \frac{\epsilon^2}{2} \left( \nabla_A^2 + \nabla_B^2 + 2 \nabla_B \nabla_A \right) V^\alpha(x) + \mathcal{O}\left(\epsilon^3\right) .
	\end{align}}
Pressing on, we now parallel transport this result from $x + \epsilon A + \epsilon B$ to $x + \epsilon B$.
{\allowdisplaybreaks\begin{align}
	&V^\alpha(x \to x+\epsilon A \to x+\epsilon A + \epsilon B \to x+\epsilon B)  \nonumber\\
	&= \exp(-\epsilon \nabla_A) \exp(\epsilon \nabla_B) \exp(\epsilon \nabla_A) V^\alpha(x), \nonumber\\
	&= V^\alpha(x) + \epsilon \left( \nabla_A + \nabla_B \right) V^\alpha(x)
	+ \frac{\epsilon^2}{2} \left( \nabla_A^2 + \nabla_B^2 + 2 \nabla_B \nabla_A \right) V^\alpha(x) \nonumber\\
	&\qquad
	- \epsilon \nabla_A V^\alpha(x) - \epsilon^2 \left( \nabla_A^2 + \nabla_A \nabla_B \right) V^\alpha(x) \nonumber\\
	&\qquad + \frac{\epsilon^2}{2} \nabla_A^2 V^\alpha(x) + \mathcal{O}\left(\epsilon^3\right) \nonumber\\
	&= V^\alpha(x) + \epsilon \nabla_B V^\alpha(x)
	+ \epsilon^2 \left( \frac{1}{2} \nabla_B^2 + \nabla_B \nabla_A - \nabla_A \nabla_B \right) V^\alpha(x) 
	+ \mathcal{O}\left(\epsilon^3\right)  . 
	\end{align}}
Finally, we parallel transport this back to $x + \epsilon B - \epsilon B = x$.
{\allowdisplaybreaks\begin{align}
	&V^\alpha(x \to x+\epsilon A \to x+\epsilon A + \epsilon B \to x+\epsilon B \to x)  \nonumber\\
	&= \exp(-\epsilon \nabla_B) \exp(-\epsilon \nabla_A) \exp(\epsilon \nabla_B) \exp(\epsilon \nabla_A) V^\alpha(x), \nonumber\\
	&= V^\alpha(x) + \epsilon \nabla_B V^\alpha(x)
	+ \epsilon^2 \left( \frac{1}{2} \nabla_B^2 + \nabla_B \nabla_A - \nabla_A \nabla_B \right) V^\alpha(x) \nonumber\\
	&\qquad
	- \epsilon \nabla_B V^\alpha(x) - \epsilon^2 \nabla_B^2 V^\alpha(x) \nonumber\\
	&\qquad
	+ \frac{\epsilon^2}{2} \nabla_B^2 V^\alpha(x) + \mathcal{O}\left(\epsilon^3\right)  \nonumber\\
	&= V^\alpha(x) 
	+ \epsilon^2 \left( \nabla_B \nabla_A - \nabla_A \nabla_B \right) V^\alpha(x) 
	+ \mathcal{O}\left(\epsilon^3\right)  .
	\end{align}}
We have arrived at the central characterization of {\it local} geometric curvature. By parallel transporting a vector around an infinitesimal parallelogram, we see the parallel transported vector differs from the original one by the commutator of covariant derivatives with respect to the two tangent vectors defining the parallelogram. In the same vein, their difference is also proportional to the area of this parallogram, i.e., it scales as $\mathcal{O}\left(\epsilon^2\right)$ for infinitesimal $\epsilon$.
\begin{align}
\label{ParallelTransport_ClosedLoop_epsilonAB}
V^\alpha(x \to x+\epsilon A \to x+\epsilon A + \epsilon B \to x+\epsilon B \to x) &- V^\alpha(x) \\
&= \epsilon^2 \left[\nabla_B, \nabla_A \right] V^\alpha(x) + \mathcal{O}\left( \epsilon^3 \right), \nonumber\\
\left[\nabla_B, \nabla_A \right] &\equiv \nabla_B \nabla_A - \nabla_A \nabla_B .
\end{align}
We shall proceed to calculate the commutator in a coordinate basis.
\begin{align}
\label{CommutatorOfCovD_I}
[\nabla_A, \nabla_B] V^\mu
&\equiv A^\sigma \nabla_\sigma\left( B^\rho\nabla_\rho V^\mu \right)
- B^\sigma\nabla_\sigma \left( A^\rho\nabla_\rho V^\mu \right) \nonumber\\
&= \left(A^\sigma \nabla_\sigma B^\rho - B^\sigma\nabla_\sigma A^\rho\right) \nabla_\rho V^\mu 
+ A^\sigma B^\rho [\nabla_\sigma,\nabla_\rho] V^\mu .
\end{align}
Let us tackle the two groups separately. Firstly,
{\allowdisplaybreaks\begin{align}
	[A,B]^\rho \nabla_\rho V^\mu
	&\equiv \left(A^\sigma \nabla_\sigma B^\rho - B^\sigma\nabla_\sigma A^\rho\right) \nabla_\rho V^\mu \nonumber\\
	&= \left( A^\sigma \partial_\sigma B^\rho + \Christ{\rho}{\sigma}{\lambda} A^\sigma B^\lambda
	- B^\sigma \partial_\sigma A^\rho - \Christ{\rho}{\sigma}{\lambda} B^\sigma A^\lambda \right) \nabla_\rho V^\mu \nonumber\\
	\label{CommutatorOfCovD_II}
	&= \left( A^\sigma \partial_\sigma B^\rho - B^\sigma \partial_\sigma A^\rho \right) \nabla_\rho V^\mu .
	\end{align}}
Next, we need $A^\sigma B^\rho [\nabla_\sigma, \nabla_\rho] V^\mu = A^\sigma B^\rho (\nabla_\sigma \nabla_\rho - \nabla_\rho \nabla_\sigma) V^\mu$. The first term is 
{\allowdisplaybreaks\begin{align}
	A^\sigma B^\rho \nabla_\sigma \nabla_\rho V^\mu 
	&= A^\sigma B^\rho \left( \partial_\sigma \nabla_\rho V^\mu 
	- \Christ{\lambda}{\sigma}{\rho} \nabla_\lambda V^\mu
	+ \Christ{\mu}{\sigma}{\lambda} \nabla_\rho V^\lambda \right) \nonumber\\
	&= A^\sigma B^\rho \left( \partial_\sigma \left(\partial_\rho V^\mu + \Christ{\mu}{\rho}{\lambda} V^\lambda \right) 
	- \Christ{\lambda}{\sigma}{\rho} \left(\partial_\lambda V^\mu + \Christ{\mu}{\lambda}{\omega} V^\omega \right)
	+ \Christ{\mu}{\sigma}{\lambda} \left(\partial_\rho V^\lambda + \Christ{\lambda}{\rho}{\omega} V^\omega \right) \right) \nonumber\\
	&= A^\sigma B^\rho \Big\{ \partial_\sigma \partial_\rho V^\mu + \partial_\sigma \Christ{\mu}{\rho}{\lambda} V^\lambda 
	+ \Christ{\mu}{\rho}{\lambda} \partial_\sigma V^\lambda  
	- \Christ{\lambda}{\sigma}{\rho} \left(\partial_\lambda V^\mu + \Christ{\mu}{\lambda}{\omega} V^\omega \right) \nonumber\\
	&\qquad\qquad
	+ \Christ{\mu}{\sigma}{\lambda} \left(\partial_\rho V^\lambda + \Christ{\lambda}{\rho}{\omega} V^\omega \right) \Big\} .
	\end{align}}
Swapping $(\sigma \leftrightarrow \rho)$ within the parenthesis $\{ \dots \}$ and subtract the two results, we gather
\begin{align}
A^\sigma B^\rho [\nabla_\sigma, \nabla_\rho] V^\mu
&= A^\sigma B^\rho \Big\{ \partial_{[\sigma} \Christ{\mu}{\rho]}{\lambda} V^\lambda 
+ \Christ{\mu}{\lambda}{[\rho} \partial_{\sigma]} V^\lambda  
- \Christ{\lambda}{[\sigma}{\rho]} \left(\partial_\lambda V^\mu + \Christ{\mu}{\lambda}{\omega} V^\omega \right) \nonumber\\
&\qquad\qquad\qquad\qquad
+ \Christ{\mu}{\lambda}{[\sigma} \partial_{\rho]} V^\lambda + \Christ{\mu}{\lambda}{[\sigma} \Christ{\lambda}{\rho]}{\omega} V^\omega \Big\} \\
\label{CommutatorOfCovD_III}
&= A^\sigma B^\rho \left( \partial_{[\sigma} \Christ{\mu}{\rho]}{\omega}  
+ \Christ{\mu}{\lambda}{[\sigma} \Christ{\lambda}{\rho]}{\omega} \right) V^\omega .
\end{align}
Notice we have used the symmetry of the Christoffel symbols $\Christ{\mu}{\alpha}{\beta} = \Christ{\mu}{\beta}{\alpha}$ to arrive at this result. Since $A$ and $B$ are arbitrary, let us observe that the commutator of covariant derivatives acting on a vector field is not a different operator, but rather an algebraic operation:
\begin{align}
\label{CommutatorOfCovD_Vector}
[\nabla_\mu,\nabla_\nu] V^\alpha 	
&= R^\alpha_{\phantom{\alpha}\beta \mu\nu} V^\beta , \\
\label{Riemann_Def}
R^\alpha_{\phantom{\alpha}\beta \mu\nu}	
&\equiv \partial_{[\mu} \Christ{\alpha}{\nu]}{\beta} + \Christ{\alpha}{\sigma}{[\mu} \Christ{\sigma}{\nu]}{\beta} \\
&= \partial_{\mu} \Christ{\alpha}{\nu}{\beta} - \partial_{\nu} \Christ{\alpha}{\mu}{\beta}
+ \Christ{\alpha}{\sigma}{\mu} \Christ{\sigma}{\nu}{\beta} 
- \Christ{\alpha}{\sigma}{\nu} \Christ{\sigma}{\mu}{\beta} .
\end{align}
Inserting the results in equations \eqref{CommutatorOfCovD_II} and \eqref{CommutatorOfCovD_III} into eq. \eqref{CommutatorOfCovD_I} -- we gather, for arbitrary vector fields $A$ and $B$:
\begin{align}
\label{Riemann}
\left( [\nabla_A, \nabla_B] - \nabla_{[A,B]} \right)V^\mu
= R^\mu_{\phantom{\mu}\nu \alpha\beta} V^\nu A^\alpha B^\beta .
\end{align}
Moreover, we may return to eq. \eqref{ParallelTransport_ClosedLoop_epsilonAB} and re-express it as
\begin{align}
\label{ParallelTransport_FailureRiemann}
&V^\alpha(x \to x+\epsilon A \to x+\epsilon A + \epsilon B \to x+\epsilon B \to x) - V^\alpha(x) \\
&\qquad\qquad\qquad\qquad
= \epsilon^2 \left( R^\alpha_{\phantom{\alpha}\beta \mu\nu}(x) V^\beta(x) B^\mu(x) A^\nu(x) + \nabla_{[B,A]} V^\alpha(x) \right)
+ \mathcal{O}\left( \epsilon^3 \right) .
\end{align}
When $A = \partial_\mu$ and $B = \partial_\nu$ are coordinate basis vectors themselves, $[A,B] = [\partial_\mu,\partial_\nu]=0$, and eq. \eqref{Riemann} then coincides with eq. \eqref{CommutatorOfCovD_Vector}. Earlier, we have already mentioned: if $[A,B]=0$, the vector fields $A$ and $B$ can be integrated to form a local 2D coordinate system; while if $[A,B] \neq 0$, they cannot form a good coordinate system. Hence the failure of parallel transport invariance due to the $\nabla_{[A,B]}$ term in eq. \eqref{ParallelTransport_FailureRiemann} is really a measure of the coordinate-worthiness of $A$ and $B$; whereas it is the Riemann tensor term that appears to tell us something about the intrinsic local curvature of the geometry itself.
\begin{myP}
	{\bf Symmetries of the Riemann tensor} \qquad Explain why, if a tensor $\Sigma_{\alpha\beta}$ is antisymmetric in one coordinate system, it has to be anti-symmetric in any other coordinate system. Similarly, explain why, if $\Sigma_{\alpha\beta}$ is symmetric in one coordinate system, it has to be symmetric in any other coordinate system. Compute the Riemann tensor in a locally flat coordinate system and show that
	\begin{align}
	\label{Riemann_FNC}
	R_{\alpha\beta \mu\nu} 
	= \frac{1}{2} \left( \partial_{\beta} \partial_{[\mu} g_{\nu]\alpha} - \partial_{\alpha} \partial_{[\mu} g_{\nu]\beta} \right) .
	\end{align}
	From this result, argue that Riemann has the following symmetries:
	\begin{align}
	\label{Riemann_Symmetries_I}
	R_{\mu\nu \alpha\beta} = R_{\alpha\beta \mu\nu} ,
	\qquad\qquad
	R_{\mu\nu \alpha\beta} = -R_{\nu\mu \alpha\beta} ,
	\qquad\qquad
	R_{\mu\nu \alpha\beta} = -R_{\mu\nu \beta\alpha} .
	\end{align} 
	This indicates the components of the Riemann tensor are not all independent. Below, we shall see there are additional differential relations (aka ``Bianchi identities") between various components of the Riemann tensor.
	
	Finally, use these symmetries to show that
	\begin{align}
	\label{CommutatorOfCovD_1Form}
	[\nabla_\alpha, \nabla_\beta] V_\nu = - R^\mu_{\phantom{\mu}\nu \alpha\beta} V_\mu .
	\end{align} 
	Hint: Start with $[\nabla_\alpha, \nabla_\beta](g_{\nu\sigma} V^\sigma)$. \qed
\end{myP}
{\bf Ricci tensor and scalar} \qquad Because of the symmetries of Riemann in eq. \eqref{Riemann_Symmetries_I}, we have $g^{\alpha\beta} R_{\alpha\beta \mu\nu} = -g^{\alpha\beta} R_{\beta\alpha \mu\nu} = -g^{\beta\alpha} R_{\beta\alpha \mu\nu} = 0$; and likewise, $R_{\alpha\beta\mu}^{\phantom{\alpha\beta\mu}\mu} = 0$. In fact, the Ricci tensor is defined as the sole distinct and non-zero contraction of Riemann:
\begin{align}
R_{\mu\nu} \equiv R^\sigma_{\phantom{\sigma}\mu\sigma\nu} .
\end{align}
This is a symmetric tensor, $R_{\mu\nu} = R_{\nu\mu}$, because of eq. \eqref{Riemann_Symmetries_I}; for,
\begin{align}
R_{\mu\nu} 
= g^{\sigma\rho} R_{\sigma\mu\rho\nu} 
= g^{\rho\sigma} R_{\rho\nu \sigma\mu} = R_{\nu\mu} .
\end{align}
Its contraction yields the Ricci scalar
\begin{align}
\mathcal{R} \equiv g^{\mu\nu} R_{\mu\nu} .
\end{align}
\begin{myP} {\bf Commutator of covariant derivatives on higher rank tensor} \qquad Prove that
	\begin{align}
	\label{Riemann_CovD}
	&[\nabla_\mu,\nabla_\nu] T^{\alpha_1 \dots \alpha_N}_{\phantom{\alpha_1 \dots \alpha_N} \beta_1 \dots \beta_M} \nonumber\\
	&= R^{\alpha_1}_{\phantom{\alpha_1}\sigma \mu\nu} T^{\sigma \alpha_2 \dots \alpha_N}_{\phantom{\sigma \alpha_2 \dots \alpha_N} \beta_1 \dots \beta_M}
	+ R^{\alpha_2}_{\phantom{\alpha_2}\sigma \mu\nu} T^{\alpha_1 \sigma \alpha_3 \dots \alpha_N}_{\phantom{\alpha_1 \sigma \alpha_3 \dots \alpha_N} \beta_1 \dots \beta_M}
	+ \dots 
	+ R^{\alpha_N}_{\phantom{\alpha_N}\sigma \mu\nu} T^{\alpha_1 \dots \alpha_{N-1} \sigma}_{\phantom{\alpha_1 \dots \alpha_{N-1} \sigma} \beta_1 \dots \beta_M} \nonumber\\
	&- R^{\sigma}_{\phantom{\sigma}\beta_1 \mu\nu} T^{\alpha_1 \dots \alpha_N}_{\phantom{\alpha_1 \dots \alpha_N} \sigma \beta_2 \dots \beta_M}
	- R^{\sigma}_{\phantom{\sigma}\beta_2 \mu\nu} T^{\alpha_1 \dots \alpha_N}_{\phantom{\alpha_1 \dots \alpha_N} \beta_1 \sigma \beta_3 \dots \beta_M}
	- \dots
	- R^{\sigma}_{\phantom{\sigma}\beta_M \mu\nu} T^{\alpha_1 \dots \alpha_N}_{\phantom{\alpha_1 \dots \alpha_N} \beta_1 \dots \beta_{M-1} \sigma} .
	\end{align} 
	Also verify that 
	\begin{align}
	[\nabla_\alpha,\nabla_\beta]\varphi = 0,
	\end{align}
	where $\varphi$ is a scalar. \qed
\end{myP}
\begin{myP}
	{\bf Differential Bianchi identities I} \qquad Show that
	\begin{align}
	\label{Riemann_Bianchi_1}
	R^\mu_{\phantom{\mu} [\alpha\beta\delta]} = 0 .
	\end{align} 
	Hint: Use the Riemann tensor expressed in an FNC system. \qed
\end{myP}
\begin{myP}
	{\bf Differential Bianchi identities II} \qquad If $[A,B] \equiv AB-BA$, can you show that the differential operator
	\begin{align}
	[\nabla_\alpha, [\nabla_\beta,\nabla_\delta]]
	+ [\nabla_\beta, [\nabla_\delta,\nabla_\alpha]]
	+ [\nabla_\delta, [\nabla_\alpha,\nabla_\beta]]
	\end{align}
	is actually zero? (Hint: Just expand out the commutators.) Why does that imply
	\begin{align}
	\label{Riemann_Bianchi_2}
	\nabla_{[\alpha} R^{\mu\nu}_{\phantom{\mu\nu}\beta\delta]} = 0 ?
	\end{align} 
	Using this result, show that
	\begin{align}
	\label{Bianchi_Contracted_I}
	\nabla_\sigma R^{\sigma\beta}_{\phantom{\sigma\beta}\mu\nu} 
	= \nabla_{[\mu} R^\beta_{\phantom{\beta}\nu]} .
	\end{align}
	The {\it Einstein tensor} is defined as
	\begin{align}
	G_{\mu\nu} \equiv R_{\mu\nu} - \frac{1}{2} g_{\mu\nu} \mathcal{R} .
	\end{align}
	From eq. \eqref{Bianchi_Contracted_I} can you show the divergence-less property of the Einstein tensor, i.e.,
	\begin{align}
	\label{EinsteinBianchi}
	\nabla^\mu G_{\mu\nu} = \nabla^\mu \left( R_{\mu\nu} - \frac{1}{2} g_{\mu\nu} \mathcal{R} \right) = 0 ?
	\end{align} 
	This will be an important property when discussing Einstein's equations for General Relativity. \qed
\end{myP}
{\bf Geodesics} \qquad As already noted, even in flat spacetime, $\dd s^2$ is not positive-definite (cf. \eqref{DifferentialGeometry_MinkowskiMetric}), unlike its purely spatial counterpart. Therefore, when computing the distance along a line in spacetime $z^\mu(\lambda)$, with boundary values $z(\lambda_1) \equiv x'$ and $z(\lambda_2) \equiv x$, we need to take the square root of its absolute value:
\begin{align}
s 
= \int_{\lambda_1}^{\lambda_2} \left\vert g_{\mu\nu}\left( z(\lambda) \right) \frac{\dd z^\mu(\lambda)}{\dd\lambda} \frac{\dd z^\nu(\lambda)}{\dd\lambda} \right\vert^{1/2} \dd\lambda .
\end{align}
A geodesic in curved spacetime that joins two points $x$ and $x'$ is a path that extremizes the distance between them. Using an affine parameter to describe the geodesic, i.e., using a $\lambda$ such that $\sqrt{|g_{\mu\nu} \dot{z}^\mu \dot{z}^\nu|} =$ constant, this amounts to imposing the principle of stationary action on Synge's world function:
\begin{align}
\label{SyngeWorldFunction}
\sigma(x,x') 
&\equiv \frac{1}{2} (\lambda_2 - \lambda_1) \int_{\lambda_1}^{\lambda_2} g_{\alpha\beta}\left(z(\lambda)\right) \frac{\dd z^\alpha}{\dd \lambda} \frac{\dd z^\beta}{\dd \lambda} \dd\lambda , \\
z^\mu(\lambda_1) &= x'^\mu, \qquad z^\mu(\lambda_2) = x^\mu .
\end{align}
When evaluated on geodesics, eq. \eqref{SyngeWorldFunction} is half the square of the geodesic distance between $x$ and $x'$. The curved spactime geodesic equation in affine-parameter form which follows from eq. \eqref{SyngeWorldFunction}, is
\begin{align}
\label{GeodesicEquation}
\frac{D^2 z^\mu}{\dd \lambda^2}
\equiv \frac{\dd^2 z^\mu}{\dd \lambda^2} 
+ \Gamma^{\mu}_{\phantom{\mu}\alpha\beta} \frac{\dd z^\alpha}{\dd \lambda} \frac{\dd z^\beta}{\dd \lambda}
= 0 .
\end{align}
The Lagragian associated with eq. \eqref{SyngeWorldFunction}, 
\begin{align}
\label{GeodesicLagrangian}
L_{\text{g}} \equiv \frac{1}{2} g_{\mu\nu}(z(\lambda)) \dot{z}^\mu \dot{z}^\nu, \qquad\qquad
\dot{z}^\mu \equiv \frac{\dd z^\mu}{\dd \lambda} ,
\end{align} 
not only oftentimes provides a more efficient means of computing the Christoffel symbols, it is a constant of motion. Unlike the curved space case, however, this Lagrangian $L_{\text{g}}$ can now be positive, zero, or negative.
\begin{itemize}
	\item If $\dot{z}^\mu$ is timelike, then by choosing the affine parameter to be proper time $\dd\lambda \sqrt{g_{\mu\nu}\dot{z}^\mu \dot{z}^\nu} = \dd\tau$, we see that the Lagrangian is then set to $L_{\text{g}} = 1/2$. 
	\item If $\dot{z}^\mu$ is spacelike, then by choosing the affine parameter to be proper length $\dd\lambda \sqrt{| g_{\mu\nu}\dot{z}^\mu \dot{z}^\nu |} = \dd\ell$, we see that the Lagrangian is then set to $L_{\text{g}} = -1/2$.
	\item If $\dot{z}^\mu$ is null, then the Lagrangian is zero: $L_{\text{g}}=0$. 
\end{itemize}
{\it Formal solution to geodesic equation} \qquad We may re-write eq. \eqref{GeodesicEquation} into an integral equation by simply integrating both sides with respect to the affine parameter $\lambda$:
\begin{align}
\label{GeodesicEquation_IntegralForm}
v^\mu(\lambda)
&= v^\mu(\lambda_1) - \int_{z(\lambda_1)}^{z(\lambda)} \Gamma^{\mu}_{\phantom{\mu}\alpha\beta} v^\alpha \dd z^\beta ; 
\end{align}
where $v^\mu \equiv \dd z^\mu/\dd \lambda$; the lower limit is $\lambda=\lambda_1$; and we have left the upper limit indefinite. The integral on the right hand side can be viewed as an integral operator acting on the tangent vector at $v^\alpha(z(\lambda))$. By iterating this equation infinite number of times -- akin to the Born series expansion in quantum mechanics -- it is possible to arrive at a formal (as opposed to explicit) solution to the geodesic equation.
\begin{myP}
	{\bf Synge's World Function In Minkowski} \qquad Verify that Synge's world function (cf. \eqref{SyngeWorldFunction}) in Minkowski spacetime is
	\begin{align}
	\sbar(x,x') &= \frac{1}{2}(x-x')^2 \equiv \frac{1}{2} \eta_{\mu\nu} (x-x')^\mu (x-x')^\nu , \\
	(x-x')^\mu 	&\equiv x^\mu - x'^\mu .
	\end{align}
	Hint: If we denote the geodesic $z^\mu(0 \leq \lambda \leq 1)$ joining $x'$ to $x$ in Minkowski spacetime, verify that the solution is
	\begin{align}
	z^\mu(0 \leq \lambda \leq 1) = x'^\mu + \lambda (x-x')^\mu .
	\end{align} \qed
\end{myP}
\begin{myP}
	Show that eq. \eqref{GeodesicEquation} takes the same form under re-scaling and constant shifts of the parameter $\lambda$. That is, if
	\begin{align}
	\lambda = a \lambda' + b ,
	\end{align}
	for constants $a$ and $b$, then eq. \eqref{GeodesicEquation} becomes
	\begin{align}
	\frac{D^2 z^\mu}{\dd \lambda'^2}
	\equiv \frac{\dd^2 z^\mu}{\dd \lambda'^2} 
	+ \Gamma^{\mu}_{\phantom{\mu}\alpha\beta} \frac{\dd z^\alpha}{\dd \lambda'} \frac{\dd z^\beta}{\dd \lambda'}
	= 0 .
	\end{align} 
	For the timelike and spacelike cases, this is telling us that proper time and proper length are respectively only defined up to an overall re-scaling and an additive shift. In other words, both the base units and its `zero' may be altered at will. \qed
\end{myP}
\begin{myP}
	Let $v^\mu(x)$ be a vector field defined throughout a given spacetime. Show that the geodesic equation \eqref{GeodesicEquation} follows from
	\begin{align}
	\label{GeodesicEquation_VectorFieldForm}
	v^\sigma \nabla_\sigma v^\mu = 0 ,
	\end{align}
	i.e., $v^\mu$ is parallel transported along itself -- provided we recall the `velocity flow' interpretation of a vector field:
	\begin{align}
	v^\mu\left(z(s)\right) = \frac{\dd z^\mu}{\dd s} .
	\end{align}
	{\it Parallel transport preserves norm-squared} \qquad The metric compatibility condition in eq. \eqref{MetricCompatibility} obeyed by the covariant derivative $\nabla_\alpha$ can be thought of as the requirement that the norm-squared $v^2 \equiv g_{\mu\nu} v^\mu v^\nu$ of a geodesic vector ($v^\mu$ subject to eq. \eqref{GeodesicEquation_VectorFieldForm}) be preserved under parallel transport. Can you explain this statement using the appropriate equations?
	
	{\it Non-affine form of geodesic equation} \qquad Suppose instead 
	\begin{align}
	\label{GeodesicEquation_VectorFieldForm_NonAffineParameter}
	v^\sigma \nabla_\sigma v^\mu = \kappa v^\mu .
	\end{align}
	This is the more general form of the geodesic equation, where the parameter $\lambda$ is not an affine one. Nonetheless, by considering the quantity $v^\sigma \nabla_\sigma (v^\mu/(v_\nu v^\nu)^p)$, for some real number $p$, show how eq. \eqref{GeodesicEquation_VectorFieldForm_NonAffineParameter} can be transformed into the form in eq. \eqref{GeodesicEquation_VectorFieldForm}; that is, identify an appropriate $v'^\mu$ such that
	\begin{align}
	v'^\sigma \nabla_\sigma v'^\mu = 0 .
	\end{align}
	You should comment on how this re-scaling fails when $v^\mu$ is null. 
	
	Starting from the finite distance integral
	\begin{align}
	s &\equiv \int_{\lambda_1}^{\lambda_2} \dd\lambda \sqrt{|g_{\mu\nu}(z(\lambda)) \dot{z}^\mu \dot{z}^\nu|},
	\qquad\qquad \dot{z}^\mu \equiv \frac{\dd z^\mu}{\dd\lambda} , \\
	z^\mu(\lambda_1) &= x', \qquad\qquad z^\mu(\lambda_2) = x ;
	\end{align}
	show that demanding $s$ be extremized leads to the non-affine geodesic equation
	\begin{align}
	\label{GeodesicEquation_NonAffine}
	\ddot{z}^\mu + \Christ{\mu}{\alpha}{\beta} \dot{z}^\alpha \dot{z}^\beta 
	= \dot{z}^\mu \frac{\dd}{\dd\lambda} \ln\sqrt{g_{\alpha\beta} \dot{z}^\alpha \dot{z}^\beta} .
	\end{align}\qed
\end{myP}
\begin{myP}
	{\bf Null Geodesics \& Weyl Transformations} \qquad Suppose two geometries $g_{\mu\nu}$ and $\gb_{\mu\nu}$ are related via a Weyl transformation
	\begin{align}
	g_{\mu\nu}(x) = \Omega(x)^2 \gb_{\mu\nu}(x) .
	\end{align}
	Consider the null geodesic equation in the geometry $g_{\mu\nu}(x)$,
	\begin{align}
	k'^\sigma \nabla_\sigma k'^\mu = 0,
	\qquad\qquad
	g_{\mu\nu} k'^\mu k'^\nu = 0
	\end{align}
	where $\nabla$ is the covariant derivative with respect to $g_{\mu\nu}$; as well as the null geodesic equation in $\gb_{\mu\nu}(x)$,
	\begin{align}
	k^\sigma \Db_\sigma k^\mu = 0, 
	\qquad\qquad
	\gb_{\mu\nu} k^\mu k^\nu = 0 ;
	\end{align}
	where $\Db$ is the covariant derivative with respect to $\gb_{\mu\nu}$. Show that
	\begin{align}
	k^\mu = \Omega^2 \cdot k'^\mu .
	\end{align}
	Hint: First show that the Christoffel symbol $\overline{\Gamma}^\mu_{\phantom{\mu}\alpha\beta}[\gb]$ built solely out of $\gb_{\mu\nu}$ is related to $\Christ{\mu}{\alpha}{\beta}[g]$ built out of $g_{\mu\nu}$ through the relation
	\begin{align}
	\Christ{\mu}{\alpha}{\beta}[g]
	= \bar{\Gamma}^\mu_{\phantom{\mu}\alpha\beta}[\gb]
	+ \delta_{\{\beta}^\mu \Db_{\alpha\}} \ln \Omega - \gb_{\alpha\beta} \Db^\mu \ln \Omega .
	\end{align}
	Then remember to use the constraint $g_{\mu\nu} k'^\mu k'^\nu = 0 = \gb_{\mu\nu} k^\mu k^\nu$. 
	
	A spacetime is said to be conformally flat if it takes the form
	\begin{align}
	g_{\mu\nu}(x) = \Omega(x)^2 \eta_{\mu\nu} .
	\end{align}
	Solve the null geodesic equation explicitly in such a spacetime. \qed
\end{myP}
\begin{myP}
	{\bf Light Deflection Due To Static Mass Monopole in 4D} \qquad In General Relativity the weak field metric generated by an isolated system, of total mass $M$, is dominated by its mass monopole and hence goes as $1/r$ (i.e., its Newtonian potential)
	\begin{align}
	\label{StaticNewtonian_Metric}
	g_{\mu\nu} 
	= \eta_{\mu\nu} + 2\Phi \delta_{\mu\nu}
	= \eta_{\mu\nu} - \frac{r_s}{r} \delta_{\mu\nu} ,
	\end{align}
	where we assume $|\Phi| = r_s/r \ll 1$ and
	\begin{align}
	r_s \equiv 2 \GN M .
	\end{align} 
	Now, the metric of an isolated static non-rotating black hole -- i.e., the Schwarzschild black hole -- in isotropic coordinates is
	\begin{align}
	\label{Schwarzschild_Isotropic}
	\dd s^2 = \left( \frac{1 - \frac{r_s}{4r}}{1 + \frac{r_s}{4r}} \right)^2 \dd t^2 
	- \left( 1 + \frac{r_s}{4 r} \right)^4 \dd\vec{x}\cdot\dd\vec{x}, \qquad\qquad
	r \equiv \sqrt{\vec{x}\cdot\vec{x}} .
	\end{align}
	The $r_s \equiv 2 \GN M$ here is the Schwarzschild radius; any object falling behind $r < r_s$ will not be able to return to the $r > r_s$ region unless it is able to travel faster than light.
	
	Expand this metric in eq. \eqref{Schwarzschild_Isotropic} up to first order $r_s/r$ and verify this yields eq. \eqref{StaticNewtonian_Metric}. We may therefore identify eq. \eqref{StaticNewtonian_Metric} as either the metric due to the monopole moment of some static mass density $\rho(\vec{x})$ or the far field limit $r_s/r \ll 1$ of the Schwarzschild black hole.
	
	{\it Statement of Problem:} \qquad Now consider shooting a beam of light from afar, and by solving the appropriate null geodesic equations, figure out how much angular deflection $\Delta \varphi$ it suffers due to the presence of a mass monopole. Express the answer in terms of the coordinate radius of closest approach $r_0$.
	
	{\it Hints}: First, write down the affine-parameter form of the Lagrangian $L_g$ for geodesic motion in eq. \eqref{StaticNewtonian_Metric} in spherical coordinates
	\begin{align}
	\vec{x} = r \left( \sin(\theta) \cos(\phi), \sin(\theta) \sin(\phi), \cos(\theta) \right) .
	\end{align}
	Because of the spherical symmetry of the problem, we may always assume that all geodesic motion takes place on the equatorial plane:
	\begin{align}
	\label{StaticNewtonian_Equatorial}
	\theta = \frac{\pi}{2} .
	\end{align}
	Proceed to argue one may always choose the affine parameter $\lambda$ such that
	\begin{align}
	\label{StaticNewtonian_EnergyConservation}
	\dot{t} = 1 + \frac{r_s}{r} \left( 1 - \frac{r_s}{r} \right)^{-1} ;
	\end{align}
	such that when $r_s \to 0$, the coordinate time $t$ becomes proper time. Next, show that angular momentum conservation $-\partial L_g/\partial \dot{\phi} \equiv \ell$ (constant) yields 
	\begin{align}
	\label{StaticNewtonian_AngularMomentumConservation}
	\dot{\phi} = \frac{\ell}{r^2} \left( 1 + \frac{r_s}{r} \right)^{-1} . 
	\end{align}
	We are primarily interested in the trajectory as a function of angle, so we may eliminate all $\dot{r} \equiv \dd r/\dd\lambda$ as
	\begin{align}
	\label{StaticNewtonian_rdot}
	\dot{r} = \frac{\dd \phi}{\dd \lambda} r'(\phi) = \frac{\ell}{r^2} \left( 1 + \frac{r_s}{r} \right)^{-1} r'(\phi) ,
	\end{align}
	where eq. \eqref{StaticNewtonian_AngularMomentumConservation} was employed in the second equality. At this point, by utilizing equations \eqref{StaticNewtonian_Equatorial}, \eqref{StaticNewtonian_EnergyConservation}, \eqref{StaticNewtonian_AngularMomentumConservation} and \eqref{StaticNewtonian_rdot}, verify that the geodesic Lagrangian now takes the form
	\begin{align}
	\label{StaticNewtonian_Lag}
	L_g = \frac{1}{2} \left( 
	\frac{r}{r-r_s}
	- \frac{\ell^2}{r^2(1 + r_s/r)}\left( 1 + \left(\frac{r'(\phi)}{r}\right)^2 \right) \right) .
	\end{align} 
	Remember that null geodesics render $L_g = 0$. If $r_0$ is the coordinate radius of closest approach, which we shall assume is appreciably larger than the Schwarzschild radius $r_0 \gg r_s$, that means $r'(\phi) = 0$ when $r=r_0$. Show that
	\begin{align}
	\ell = r_0 \sqrt{\frac{r_0 + r_s}{r_0 - r_s}} .
	\end{align}
	Working to first order in $r_s$, proceed to show that
	\begin{align}
	\frac{\dd\phi}{\dd r} 
	= \frac{1}{\sqrt{r^2 - r_0^2}} \left( \frac{r_0}{r} + \frac{r_s}{r + r_0} \right) 
	+ \mathcal{O}\left( r_s^2 \right) .
	\end{align}
	By integrating from infinity $r=\infty$ to closest approach $r=r_0$ and then out to infinity again $r=\infty$, show that the angular deflection is
	\begin{align}
	\Delta \varphi = \frac{2 r_s}{r_0} .
	\end{align}
	Even though $r_0$ is the coordinate radius of closest approach, in a weakly curved spacetime dominated by the monopole moment of the central object, estimate the error incurred if we set $r_0$ to be the {\it physical} radius of closest approach. What is the angular deflection due to the Sun, if a beam of light were to just graze its surface?
	
	Note that, if the photon were undeflected, the total change in angle $(\int_{r=\infty}^{r_0} \dd r + \int_{r_0}^{\infty} \dd r) (\dd \phi/\dd r)$ would be $\pi$. Therefore, the total {\it deflection} angle is
	\begin{align}
	\Delta \varphi = 2 \left\vert \int_{r=\infty}^{r_0} \frac{\dd\phi}{\dd r} \dd r \right\vert - \pi .
	\end{align}
	For further help on this problem, consult \S 8.5 of Weinberg \cite{Weinberg:1972kfs}. \qed
\end{myP}

\subsection{Equivalence Principles \& Geometry-Induced Tidal Forces}
\label{Chapter_DifferentialGeometry_CurvedSpacetimes_C3}
{\bf Weak Equivalence Principle, ``Free-Fall" \& Gravity as a Non-Force} \qquad The universal nature of gravitation -- how it appears to act in the same way upon all material bodies independent of their internal composition -- is known as the Weak Equivalence Principle. Within non-relativistic physics, the acceleration of some mass $M_1$ located at $\vec{x}_1$, due to the Newtonian gravitational `force' exerted by some other mass $M_2$ at $\vec{x}_2$, is given by
\begin{align}
\label{NewtonianGravity}
M_1 \frac{\dd^2 \vec{x}_1}{\dd t^2}
= -\widehat{n} \frac{\GN M_1 M_2}{|\vec{x}_1 - \vec{x}_2|^2} , \qquad\qquad
\widehat{n} \equiv \frac{\vec{x}_1 - \vec{x}_2}{|\vec{x}_1 - \vec{x}_2|}.
\end{align}
Strictly speaking the $M_1$ on the left hand side is the `inertial mass', a characterization of the resistance -- so to speak -- of any material body to being accelerated by an external force. While the $M_1$ on the right hand side is the `gravitational mass', describing the strength to which the material body interacts with the gravitational `force'. Viewed from this perspective, the equivalence principle is the assertion that the inertial and gravitational masses are the same, so that the resulting motion does not depend on them:
\begin{align}
\frac{\dd^2 \vec{x}_1}{\dd t^2} = -\widehat{n} \frac{\GN M_2}{|\vec{x}_1 - \vec{x}_2|^2} .
\end{align}
Similarly, the acceleration of body 2 due to the gravitational force exerted by body 1 is independent of $M_2$:
\begin{align}
\frac{\dd^2 \vec{x}_2}{\dd t^2} = +\widehat{n} \frac{\GN M_1}{|\vec{x}_1 - \vec{x}_2|^2} .
\end{align}
This Weak Equivalence Principle\footnote{See Will \cite{Will:2014kxa} arXiv: \href{https://arxiv.org/abs/1403.7377}{1403.7377} for a review on experimental tests of various versions of the Equivalence Principle and other aspects of General Relativity.} is one of the primary motivations that led Einstein to recognize gravitation as the manifestation of curved spacetime. The reason why inertial mass appears to be equal to its gravitational counterpart, is because material bodies now follow (timelike) geodesics $z^\mu(\tau)$ in curved spacetimes: 
\begin{align}
a^\mu
\equiv \frac{D^2 z^\mu}{\dd \tau^2}
\equiv \frac{\dd^2 z^\mu}{\dd \tau^2} + \Christ{\mu}{\alpha}{\beta} \frac{\dd z^\alpha}{\dd\tau} \frac{\dd z^\beta}{\dd\tau}
= 0; \qquad\qquad
g_{\mu\nu}\left(z(\lambda)\right) \frac{\dd z^\mu}{\dd \tau} \frac{\dd z^\nu}{\dd \tau} > 0 ;
\end{align}
so that their motion only depends on the curved geometry itself and does not depend on their own mass.\footnote{If there {\it were} an external non-gravitational force $f^\mu$, then the covariant Newton's second law for a system of mass $M$ would read: $M D^2 z^\mu/\dd \tau^2 = f^\mu$.} From this point of view, gravity is no longer a force.

Note that, strictly speaking, this ``gravity-induced-dynamics-as-geodesics" is actually an idealization that applies for material bodies with no internal structure and whose proper sizes are very small compared to the length scale(s) associated with the geometric curvature itself. In reality, all physical systems have internal structure -- non-trivial quadrupole moments, spin/rotation, etc. -- and may furthermore be large enough that their full dynamics require detailed analysis to understand properly.

{\it Newton vs. Einstein} \qquad Observe that the Newtonian gravity of eq. \eqref{NewtonianGravity} in an instantaneous force, in that the force on body $1$ due to body $2$ (or, vice versa) changes immediately when body $2$ starts changing its position $\vec{x}_2$ -- even though it is located at a finite distance away. However, Special Relativity tells us there ought to be an ultimate speed limit in Nature, i.e., no physical effect/information can travel faster than $c$. This apparent inconsistency between Newtonian gravity and Einstein's Special Relativity is of course a driving motivation that led Einstein to General Relativity. As we shall see shortly, by postulating that the effects of gravitation are in fact the result of residing in a curved spacetime, the Lorentz symmetry responsible for Special Relativity is recovered in any local ``freely-falling" frame.

{\it Massless particles} \qquad Finally, this dynamics-as-geodesics also led Einstein to realize -- if gravitation does indeed apply universally -- that massless particles such as photons, i.e., electromagnetic waves, must also be influenced by the gravitational field too. This is a significant departure from Newton's law of gravity in eq. \eqref{NewtonianGravity}, which may lead one to suspect otherwise, since $M_\text{photon} = 0$. It is possible to justify this statement in detail, but we shall simply assert here -- to leading order in the JWKB approximation, photons in fact sweep out {\it null} geodesics $z^\mu(\lambda)$ in curved spacetimes:
\begin{align}
a^\mu \equiv \frac{D^2 z^\mu}{\dd \lambda^2} = 0, \qquad\qquad
g_{\mu\nu}\left(z(\lambda)\right) \frac{\dd z^\mu}{\dd \lambda} \frac{\dd z^\nu}{\dd \lambda} = 0 .
\end{align}
{\bf Locally flat coordinates, Einstein Equivalence Principle \& Symmetries} \qquad We now come to one of the most important features of curved spacetimes. In the neighborhood of a timelike geodesic $y^\mu=(s,\vec{y})$, one may choose {\it Fermi normal coordinates} $x^\mu \equiv (s,\vec{x})$ such that spacetime appears flat up to distances of $\mathcal{O}(1/|\max R_{\mu\nu \alpha\beta}(y=(s,\vec{y}))|^{1/2})$; namely, $g_{\mu\nu} = \eta_{\mu\nu}$ plus corrections that begin at quadratic order in the displacement $\vec{x}-\vec{y}$:
\begin{align}
\label{FermiNormalCoordinateExpansion_00}
g_{00}(x) &= 1 - R_{0a0b}(s) \cdot (x^a-y^a) (x^b-y^b) + \mathcal{O}\left((x-y)^3\right) , \\
\label{FermiNormalCoordinateExpansion_0i}
g_{0i}(x) &= - \frac{2}{3} R_{0aib}(s)\cdot (x^a-y^a) (x^b-y^b) + \mathcal{O}\left((x-y)^3\right) , \\
\label{FermiNormalCoordinateExpansion_ij}
g_{ij}(x) &= \eta_{ij} - \frac{1}{3} R_{iajb}(s) \cdot (x^a-y^a) (x^b-y^b) + \mathcal{O}\left((x-y)^3\right) .  
\end{align}
Here $x^0 = s$ is the time coordinate, and is also the proper time of the observer with the trajectory $y^\mu(s) = (s,\vec{y})$. (The $\vec{y}$ are fixed spatial coordinates; i.e., they do not depend on $s$.) Suppose you were placed inside a closed box, so you cannot tell what's outside. Then provided the box is small enough, you will not be able to distinguish between being in ``free-fall" in a gravitational field versus being in a completely empty Minkowski spacetime.

As already alluded to in the ``Newton vs. Einstein" discussion above, just as the rotation and translation symmetries of flat Euclidean space carried over to a small enough region of curved spaces -- the FNC expansion of equations \eqref{FermiNormalCoordinateExpansion_00} through \eqref{FermiNormalCoordinateExpansion_ij} indicates that, within the spacetime neighborhood of a freely-falling observer, any curved spacetime is Lorentz and spacetime-translation symmetric. To sum:
\begin{quotation}
	Physically speaking, in a freely falling frame $\{x^\mu\}$ -- i.e., centered along a timelike geodesic at $x=y$ -- physics in a curved spacetime is the same as that in flat Minkowski spacetime up to corrections that go at least as 
	\begin{align}
	\epsilon_\text{E} \equiv \frac{\text{Length or inverse mass scale of system}}{\text{Length scale of the spacetime geometric curvature}} .
	\end{align}
\end{quotation}
This is the essence of the equivalence principle that lead Einstein to recognize curved spacetime to be the setting to formulate his General Theory of Relativity. As a simple example, the geodesic $y^\mu$ itself obeys the free-particle version of Newton's 2nd law: $\dd^2 y^\mu/\dd s^2 = 0$.
\begin{myP}
	Verify that the coefficients in front of the Riemann tensor in equations \eqref{FermiNormalCoordinateExpansion_00}, \eqref{FermiNormalCoordinateExpansion_0i} and \eqref{FermiNormalCoordinateExpansion_ij} are independent of the spacetime dimension. That is, starting with 
	\begin{align}
	\label{FermiNormalCoordinateExpansion_00A}
	g_{00}(x) &= 1 - A \cdot R_{0a0b}(s) \cdot (x-y)^a (x-y)^b + \mathcal{O}\left((x-y)^3\right) , \\
	\label{FermiNormalCoordinateExpansion_0iB}
	g_{0i}(x) &= - B \cdot R_{0aib}(s)\cdot (x-y)^a (x-y)^b + \mathcal{O}\left((x-y)^3\right) , \\
	\label{FermiNormalCoordinateExpansion_ijC}
	g_{ij}(x) &= \eta_{ij} - C \cdot R_{iajb}(s) \cdot (x-y)^a (x-y)^b + \mathcal{O}\left((x-y)^3\right) ,  
	\end{align}
	where $A,B,C$ are unknown constants, compute the Riemann tensor at $x=y$. \qed
\end{myP}
\begin{myP}
	\label{Problem_StaticNewtonianSpacetimes_NewtonianGravity}
	{\bf Gravitational force in a weak gravitational field} \qquad Consider the following metric:
	\begin{align}
	\label{StaticNewtonian}
	g_{\mu\nu}(t,\vec{x}) = \eta_{\mu\nu} + 2 \Phi(\vec{x}) \delta_{\mu\nu} ,
	\end{align}
	where $\Phi(\vec{x})$ is time-independent. Assume this is a weak gravitational field, in that $|\Phi| \ll 1$ everywhere in spacetime, and there are no non-gravitational forces. (Linearized General Relativity reduce to the familiar Poisson equation $\vec{\nabla}^2 \Phi = 4\pi\GN \rho$, where $\rho(\vec{x})$ is the mass/energy density of matter.) Starting from the non-affine form of the action principle
	\begin{align}
	-M s 
	&= -M \int_{t_1}^{t_2} \dd t \sqrt{ g_{\mu\nu} \dot{z}^\mu \dot{z}^\nu }, \qquad\qquad 
	\dot{z}^\mu \equiv \frac{\dd z^\mu}{\dd t} \nonumber\\
	&= -M \int_{t_1}^{t_2} \dd t \sqrt{ 1 - \vec{v}^2 + 2 \Phi (1 + \vec{v}^2) }, 
	\qquad\qquad
	\vec{v}^2 \equiv \delta_{ij} \dot{z}^i \dot{z}^j ;
	\end{align}
	expand this action to lowest order in $\vec{v}^2$ and $\Phi$ and work out the geodesic equation of a `test mass' $M$ sweeping out some worldline $z^\mu$ in such a spacetime. (You should find something very familiar from Classical Mechanics.) Show that, in this non-relativistic limit, Newton's law of gravitation is recovered:
	\begin{align}
	\frac{\dd^2 z^i}{\dd t^2} = -\partial_i \Phi .
	\end{align}
	We see that, in the weakly curved spacetime of eq. \eqref{StaticNewtonian}, $\Phi$ may indeed be identified as the Newtonian potential. \qed
\end{myP}
{\bf Geodesic Deviation \& Tidal Forces} \qquad We now turn to the derivation of the {\it geodesic deviation} equation. Consider two geodesics that are infinitesimally close-by. Let both of them be parametrized by $\lambda$, so that we may connect one geodesic to the other at the same $\lambda$ via an infinitesimal vector $\xi^\mu$. We will denote the tangent vector to one of geodesics to be $U^\mu$, such that
\begin{align}
U^\sigma \nabla_\sigma U^\mu = 0 .
\end{align}
Furthermore, we will assume that $[U,\xi]=0$, i.e., $U$ and $\xi$ may be integrated to form a 2D coordinate system in the neighborhood of this pair of geodesics. Then
\begin{align}
\label{GeodesicDeviation}
U^\alpha U^\beta \nabla_\alpha \nabla_\beta \xi^\mu 
= \nabla_U \nabla_U \xi^\mu
= -R^\mu_{\phantom{\mu}\nu \alpha\beta} U^\nu \xi^\alpha U^\beta .
\end{align}
As its name suggests, this equation tells us how the deviation vector $\xi^\mu$ joining two infinitesimally displaced geodesics is accelerated by the presence of spacetime curvature through the Riemann tensor. If spacetime were flat, the acceleration will be zero: two initially parallel geodesics will remain so.

For a macroscopic system, if $U^\mu$ is a timelike vector tangent to, say, the geodesic trajectory of its center-of-mass, the geodesic deviation equation \eqref{GeodesicDeviation} then describes {\it tidal forces} acting on it. In other words, the relative acceleration between the `particles' that comprise the system -- induced by spacetime curvature -- would compete with the system's internal forces.\footnote{The first gravitational wave detectors were in fact based on measuring the tidal squeezing and stretching of solid bars of aluminum. They are known as ``Weber bars", named after their inventor Joseph Weber.}

{\it Derivation of eq. \eqref{GeodesicDeviation}} \quad Starting with the geodesic equation $U^\sigma \nabla_\sigma U^\mu = 0$, we may take its derivative along $\xi$.
{\allowdisplaybreaks\begin{align}
	\xi^\alpha \nabla_\alpha \left( U^\beta \nabla_\beta U^\mu \right) &= 0 , \nonumber\\
	\left(\xi^\alpha \nabla_\alpha U^\beta - U^\alpha \nabla_\alpha \xi^\beta \right) \nabla_\beta U^\mu
	+ U^\beta \nabla_\beta \xi^\alpha \nabla_\alpha U^\mu
	+ \xi^\alpha U^\beta \nabla_\alpha \nabla_\beta U^\mu &= 0 \nonumber\\
	[\xi,U]^\beta \nabla_\beta U^\mu
	+ U^\beta \nabla_\beta (\xi^\alpha \nabla_\alpha U^\mu)
	- U^\beta \xi^\alpha \nabla_\beta \nabla_\alpha U^\mu
	+ \xi^\alpha U^\beta \nabla_\alpha \nabla_\beta U^\mu &= 0 \nonumber\\
	U^\beta \nabla_\beta (U^\alpha \nabla_\alpha \xi^\mu)
	&= -\xi^\alpha U^\beta [\nabla_\alpha,\nabla_\beta] U^\mu \nonumber\\
	U^\beta \nabla_\beta (U^\alpha \nabla_\alpha \xi^\mu)
	&= -\xi^\alpha U^\beta R^\mu_{\phantom{\mu}\nu \alpha\beta} U^\nu . \nonumber
	\end{align}}
We have repeatedly used $[\xi,U]=0$ to state, for example, $\nabla_U \xi^\rho = U^\sigma \nabla_\sigma \xi^\rho = \xi^\sigma \nabla_\sigma U^\rho = \nabla_\xi U^\rho$. It is also possible to use a more elegant notation to arrive at eq. \eqref{GeodesicDeviation}.
{\allowdisplaybreaks\begin{align}
	\nabla_U U^\mu &= 0 \\
	\nabla_\xi \nabla_U U^\mu &= 0 \\
	\nabla_U \underbrace{\nabla_\xi U^\mu}_{= \nabla_U \xi^\mu} + \left[ \nabla_\xi,\nabla_U \right] U^\mu &= 0 \\
	\nabla_U \nabla_U \xi^\mu &= -R^\mu_{\phantom{\mu}\nu\alpha\beta} U^\nu \xi^\alpha U^\beta 
	\end{align}}
On the last line, we have exploited the assumption that $[U,\xi]=0$ to say $\left[ \nabla_\xi,\nabla_U \right] U^\mu = (\left[ \nabla_\xi,\nabla_U \right] - \nabla_{[\xi,U]}) U^\mu$ -- recall eq. \eqref{Riemann}. 
\begin{myP}
	{\bf Geodesic Deviation \& FNC} \qquad Argue that all the Christoffel symbols $\Christ{\alpha}{\mu}{\nu}$ evaluated along the free-falling geodesic in equations \eqref{FermiNormalCoordinateExpansion_00}-\eqref{FermiNormalCoordinateExpansion_ij}, namely when $x=y$, vanish. Then argue that all the time derivatives of the Christoffel symbols vanish along $y$ too: $\partial^{n \geq 1}_s \Christ{\alpha}{\mu}{\nu}=0$. Why does this imply, denoting $U^\mu \equiv \dd y^\mu/\dd s$, the geodesic equation
	\begin{align}
	U^\nu \nabla_\nu U^\mu = \frac{\dd U^\mu}{\dd s} = 0 ?
	\end{align}
	Next, evaluate the geodesic deviation equation in these Fermi Normal Coordinates (FNC) system. Specifically, show that
	\begin{align}
	\label{GeodesicDeviation_FNC}
	U^\alpha U^\beta \nabla_\alpha \nabla_\beta \xi^\mu = \frac{\dd^2 \xi^\mu}{\dd s^2} 
	= - R^\mu_{\phantom{\mu}0 \nu 0} \xi^\nu . 
	\end{align}
	Why does this imply, {\it if} the deviation vector is purely spatial at a given $s=s_0$, specifically $\xi^0(s_0)=\dd \xi^0/\dd s_0=0$, then it remains so for all time? \qed
\end{myP}

\begin{myP}
	{\bf Tidal forces due to mass monopole of isolated body} \qquad In this problem we will consider sprinkling test masses initially at rest on the surface of an imaginary sphere of very small radius $r_\epsilon$, whose center is located far from that of a static isolated body whose stress tensor is dominated by its mass density $\rho(\vec{x})$. We will examine how these test masses will respond to the gravitational tidal forces exerted by $\rho$.
	
	Show that the vector field
	\begin{align}
	U^\mu(t,\vec{x}) \equiv \delta^\mu_0 (1-\Phi(\vec{x})) - t \delta^\mu_i \partial_i \Phi(\vec{x})
	\end{align}
	is a timelike geodesic up to linear order in the Newtonian potential $\Phi$. This $U^\mu$ may be viewed as the tangent vector to the worldline of the observer who was released from rest in the $(t,\vec{x})$ coordinate system at $t=0$. (To ensure this remains a valid perturbative solution we shall also assume $t/r \ll 1$.) Let $\xi^\mu = (\xi^0,\vec{\xi})$ be the deviation vector whose spatial components we wish to interpret as the small displacement vector joining the center of the imaginary sphere to its surface. Use the above $U^\alpha$ to show that -- up to first order in $\Phi$ -- the right hand sides of its geodesic deviation equations are
	\begin{align}
	U^\alpha U^\beta \nabla_\alpha \nabla_\beta \xi^0 &= 0 , \\
	U^\alpha U^\beta \nabla_\alpha \nabla_\beta \xi^i &= R_{i0 j0} \xi^j ;
	\end{align} 
	where the linearized Riemann tensor reads
	\begin{align}
	R_{i0 j0} = -\partial_i \partial_j \Phi(\vec{x}) .
	\end{align}
	Assuming that the monopole contribution dominates,
	\begin{align}
	\Phi(\vec{x}) \approx \Phi(r) = -\frac{\GN M}{r} = -\frac{r_s}{2r} ,
	\end{align}
	show that these tidal forces have strengths that scale as $1/r^3$ as opposed to the $1/r^2$ forces of Newtonian gravity itself -- specifically, you should find
	\begin{align}
	R_{i0 j0} 
	\approx -\left( \delta^{ij} - \widehat{r}^i \widehat{r}^j \right) \frac{\Phi'(r)}{r}
	- \widehat{r}^i \widehat{r}^j \Phi''(r) , 
	\qquad\qquad
	\widehat{r}^i \equiv \frac{x^i}{r} , 
	\end{align}
	so that the result follows simply from counting the powers of $1/r$ from $\Phi'(r)/r$ and $\Phi''(r)$. By setting $\vec{\xi}$ to be (anti-)parallel and perpendicular to the radial direction $\widehat{r}$, argue that the test masses lying on the radial line emanating from the body centered at $\vec{x}=\vec{0}$ will be {\it stretched apart} while the test masses lying on the plane perpendicular to $\widehat{r}$ will be {\it squeezed together}. (Hint: You should be able to see that $\delta^{ij} - \widehat{r}^i \widehat{r}^j$ is the Euclidean space orthogonal to $\widehat{r}$.) 
	
	The shape of the Earth's ocean tides can be analyzed in this manner by viewing the Earth as `falling' in the gravitational fields of the Moon and the Sun.  \qed
\end{myP}
\noindent{\bf Interlude} \qquad Let us pause to summarize the physics we have revealed thus far.
\begin{quotation}
	In a curved spacetime, the collective motion of a system of mass $M$ sweeps out a timelike geodesic -- recall equations \eqref{GeodesicEquation}, \eqref{GeodesicEquation_VectorFieldForm}, and \eqref{GeodesicEquation_NonAffine} -- whose dynamics is actually independent of $M$ as long as its internal structure can be neglected. In the co-moving frame of an observer situated within this same system, physical laws appear to be the same as that in Minkowski spacetime up to distances of order $1/|\max R_{\widehat{\alpha}\widehat{\beta}\widehat{\mu}\widehat{\nu}}|^{1/2}$. However, once the finite size of the physical system is taken into account, one would find tidal forces exerted upon it due to spacetime curvature itself -- this is described by the geodesic deviation eq. \eqref{GeodesicDeviation_FNC}. 
\end{quotation}
{\bf Killing Vectors} \qquad A geometry is said to enjoy an isometry -- or, symmetry -- when we perform the following infinitesimal displacement
\begin{align}
\label{GaugeTransformation_Coordinates}
x^\mu \to x^\mu + \xi^\mu(x)
\end{align}
and find that the geometry is unchanged 
\begin{align}
g_{\mu\nu}(x) \to g_{\mu\nu}(x) + \mathcal{O}\left( \xi^2 \right) .
\end{align}
Generically, under the infinitesimal transformation of eq. \eqref{GaugeTransformation_Coordinates},
\begin{align}
g_{\mu\nu}(x) \to g_{\mu\nu}(x) + \nabla_\mu \xi_{\nu} + \nabla_\nu \xi_{\mu} .
\end{align}
where
\begin{align}
\label{KillingEquation_LHS_CoordinateForm}
\nabla_{\{\mu} \xi_{\nu\}}
= \xi^\sigma \partial_\sigma g_{\mu\nu} + g_{\sigma\{\mu} \partial_{\nu\}} \xi^\sigma .
\end{align}
If an isometry exists along the integral curve of $\xi^\mu$, it has to obey Killing's equation
\begin{align}
\label{KillingEquation}
\nabla_{\{\mu} \xi_{\nu\}}
= \xi^\sigma \partial_\sigma g_{\mu\nu} + \partial_{\{\mu} \xi^\sigma g_{\nu\}\sigma}
= 0 .
\end{align}
In fact, by exponentiating the infinitesimal coordinate transformation, it is possible to show that -- if $\xi^\mu$ is a Killing vector (i.e., it satisfies eq. \eqref{KillingEquation}), then an isometry exists along its integral curve. In other words,
\begin{quotation}
	A spacetime geometry enjoys an isometry (aka symmetry) along the integral curve of $\xi^\mu$ iff it obeys $\nabla_{\{\mu} \xi_{\nu\}} = \nabla_\mu \xi_{\nu} + \nabla_\nu \xi_{\mu} = 0$. 
\end{quotation}
In a $d-$dimensional spacetime, there are at most $d(d+1)/2$ Killing vectors. A spacetime that has $d(d+1)/2$ Killing vectors is called {\it maximally symmetric}. (See Weinberg \cite{Weinberg:1972kfs} for a discussion.)
\begin{myP}
	{\bf Conserved quantities along geodesics} \qquad If $p_\mu$ denotes the `momentum' variable of a geodesic
	\begin{align}
	p_\mu \equiv \frac{\partial L_{\text{g}}}{\partial \dot{z}^\mu} ,
	\end{align}
	where $L_\text{g}$ is defined in eq. \eqref{GeodesicLagrangian}, and if $\xi^\mu$ is a Killing vector of the same geometry $\nabla_{\{\alpha} \xi_{\beta\}} = 0$, show that 
	\begin{align}
	\xi^\mu(z(\lambda)) p_\mu(\lambda) 
	\end{align}
	is a constant along the geodesic $z^\mu(\lambda)$. 
	
	The vector field version of this result goes as follows. 
	\begin{quotation}
		If the geodesic equation $v^\sigma \nabla_\sigma v^\mu = 0$ holds, and if $\xi^\mu$ is a Killing vector, then $\xi_\nu v^\nu$ is conserved along the integral curve of $v^\mu$.
	\end{quotation} 
	Can you demonstrate the validity of this statement? \qed
\end{myP}
{\it Second Derivatives of Killing Vectors} \qquad Now let us also consider the second derivatives of $\xi^\mu$. In particular, we will now explain why
\begin{align}
\label{KillingEquation_2ndD}
\nabla_\alpha \nabla_\beta \xi_\delta = R^\lambda_{\phantom{\lambda}\alpha\beta\delta} \xi_\lambda .
\end{align}
Consider
\begin{align}
0 &= \nabla_\delta \nabla_{\{\alpha} \xi_{\beta\}} \\
&= [\nabla_\delta, \nabla_\alpha] \xi_\beta + \nabla_\alpha \nabla_\delta \xi_\beta + [\nabla_\delta, \nabla_\beta] \xi_\alpha + \nabla_\beta \nabla_\delta \xi_\alpha \\
&= -R^\lambda_{\phantom{\lambda}\beta \delta\alpha} \xi_\lambda - \nabla_\alpha \nabla_\beta \xi_\delta - R^\lambda_{\phantom{\lambda}\alpha \delta\beta} \xi_\lambda - \nabla_\beta \nabla_\alpha \xi_\delta 
\end{align}
Because Bianchi says $0 = R^\lambda_{\phantom{\lambda}[\alpha\beta \delta]} \Rightarrow R^\lambda_{\phantom{\lambda}\alpha\beta \delta} = R^\lambda_{\phantom{\lambda}\beta\alpha\delta} + R^\lambda_{\phantom{\lambda}\delta\beta\alpha}$.
{\allowdisplaybreaks\begin{align}
	0 &= -R^\lambda_{\phantom{\lambda}\beta \delta\alpha} \xi_\lambda - \nabla_\alpha \nabla_\beta \xi_\delta + \left( R^\lambda_{\phantom{\lambda}\beta\alpha\delta} + R^\lambda_{\phantom{\lambda}\delta\beta\alpha} \right) \xi_\lambda - \nabla_\beta \nabla_\alpha \xi_\delta \\
	0 &= -2 R^\lambda_{\phantom{\lambda}\beta \delta\alpha} \xi_\lambda - \nabla_{\{\beta} \nabla_{\alpha\}} \xi_\delta - [\nabla_\beta,\nabla_\alpha] \xi_\delta \\
	0 &= -2 R^\lambda_{\phantom{\lambda}\beta \delta\alpha} \xi_\lambda - 2 \nabla_\beta \nabla_\alpha \xi_\delta
	\end{align}}
This proves eq. \eqref{KillingEquation_2ndD}.

{\it Commutators of Killing Vectors} \qquad Next, we will show that 
\begin{quotation}
	The commutator of 2 Killing vectors is also a Killing vector.
\end{quotation}
Let $U$ and $V$ be Killing vectors. If $\xi \equiv [U,V]$, we need to verify that
\begin{align}
\nabla_{\{\alpha} \xi_{\beta\}} 
= \nabla_{\{\alpha} [U,V]_{\beta\}} = 0 .
\end{align}
More explicitly, let us compute:
{\allowdisplaybreaks\begin{align*}
	&\nabla_\alpha (U^\mu \nabla_\mu V_\beta - V^\mu \nabla_\mu U_\beta) + (\alpha\leftrightarrow\beta) \\
	&= \nabla_\alpha U^\mu \nabla_\mu V_\beta - \nabla_\alpha V^\mu \nabla_\mu U_\beta
	+ U^\mu \nabla_\alpha \nabla_\mu V_\beta - V^\mu \nabla_\alpha \nabla_\mu U_\beta + (\alpha\leftrightarrow\beta) \\
	&= -\nabla_\mu U_\alpha \nabla^\mu V_\beta + \nabla_\mu V_\alpha \nabla^\mu U_\beta
	+ U^\mu \nabla_{[\alpha} \nabla_{\mu]} V_\beta + U^\mu \nabla_\mu \nabla_\alpha V_\beta 
	- V^\mu \nabla_{[\alpha} \nabla_{\mu]} U_\beta - V^\mu \nabla_\mu \nabla_\alpha U_\beta + (\alpha\leftrightarrow\beta) \\
	&= -U^\mu R^\sigma_{\phantom{\sigma}\beta \alpha\mu} V_\sigma + V^\mu R^\sigma_{\phantom{\sigma}\beta \alpha\mu} U_\sigma + (\alpha\leftrightarrow\beta) \\
	&= -U^{[\mu} V^{\sigma]} R_{\sigma\{\beta \alpha\}\mu} = 0 .
	\end{align*}}
The $(\alpha \leftrightarrow \beta)$ means we are taking all the terms preceding it and swapping $\alpha \leftrightarrow \beta$. Moreover, we have repeatedly used the Killing equations $\nabla_\alpha U_\beta = -\nabla_\beta U_\alpha$ and $\nabla_\alpha V_\beta = -\nabla_\beta V_\alpha$.
\begin{myP}
	{\bf Killing Vectors in Minkowski} \qquad In Minkowski spacetime $g_{\mu\nu} = \eta_{\mu\nu}$, with Cartesian coordinates $\{ x^\mu \}$, use eq. \eqref{KillingEquation_2ndD} to argue that the most general Killing vector takes the form
	\begin{align}
	\xi_{\mu} = \ell_\mu + \omega_{\mu\nu} x^\nu ,
	\end{align}
	for constant $\ell_\mu$ and $\omega_{\mu\nu}$. (Hint: Think about Taylor expansions.) Then use the Killing equation \eqref{KillingEquation} to infer that
	\begin{align}
	\omega_{\mu\nu} = -\omega_{\nu\mu} .
	\end{align}
	The $\ell_\mu$ corresponds to infinitesimal spacetime translation and the $\omega_{\mu\nu}$ to infinitesimal Lorentz boosts and rotations. Explain why this implies the following are the Killing vectors of flat spacetime:
	\begin{align}
	\label{Killing_MinkowskiTranslation}
	\partial_\mu \qquad\qquad \text{(Generators of spacetime translations)}
	\end{align}
	and
	\begin{align}
	\label{Killing_MinkowskiRotationBoosts}
	x^{[\mu} \partial^{\nu]} \qquad\qquad \text{(Generators of Lorentz boosts or rotations)} .
	\end{align} 
	There are $d$ distinct $\partial_\mu$'s and (due to their antisymmetry) $(1/2)(d^2-d)$ distinct $x^{[\mu} \partial^{\nu]}$'s. Therefore there are a total of $d(d+1)/2$ Killing vectors in Minkowski -- i.e., it is maximally symmetric. \qed
\end{myP}
It might be instructive to check our understanding of rotation and boosts against the 2D case we have worked out earlier via different means. Up to first order in the rotation angle $\theta$, the 2D rotation matrix in eq. \eqref{Rotation_2D} reads 
\begin{align}
\widehat{R}^i_{\phantom{i}j}(\theta)
= \left[ \begin{array}{cc}
1 		& -\theta \\
\theta	& 1
\end{array} \right] + \mathcal{O}\left( \theta^2 \right) .
\end{align} 
In other words, $\widehat{R}^i_{\phantom{i}j}(\theta) = \delta_{ij} - \theta \epsilon_{ij}$, where $\epsilon_{ij}$ is the Levi-Civita symbol in 2D with $\epsilon_{12} \equiv 1$. Applying a rotation of the 2D Cartesian coordinates $x^i$ upon a test (scalar) function $f$,
\begin{align}
f(x^i) 
\to f\left(\widehat{R}^i_{\phantom{i}j} x^j\right) 
&= f\left( x^i - \theta \epsilon_{ij} x^j + \mathcal{O}\left( \theta^2 \right) \right) \\
&= f(\vec{x}) - \theta \epsilon_{ij} x^j \partial_i f(\vec{x}) + \mathcal{O}\left( \theta^2 \right) .
\end{align}
Since $\theta$ is arbitrary, the basic differential operator that implements an infinitesimal rotation of the coordinate system on any Minkowski scalar is
\begin{align}
-\epsilon_{ij} x^j \partial_i = x^1 \partial_2 - x^2 \partial_1 .
\end{align}
This is the 2D version of eq. \eqref{Killing_MinkowskiRotationBoosts} for rotations. As for 2D Lorentz boosts, eq. \eqref{LorentzTransformation_2D_General} tells us
\begin{align}
\Lambda^\mu_{\phantom{\mu}\nu}(\xi)
= \left[ \begin{array}{cc}
1 	& \xi \\
\xi & 1
\end{array} \right]
+ \mathcal{O}\left( \xi^2 \right) .
\end{align}
(This $\xi$ is known as {\it rapidity}.)  Here, we have $\Lambda^\mu_{\phantom{\mu}\nu} = \delta^\mu_{\phantom{\mu}\nu} + \xi \cdot \epsilon^\mu_{\phantom{\mu}\nu}$, where $\epsilon_{\mu\nu}$ is the Levi-Civita tensor in 2D Minkowski with $\epsilon_{01} \equiv 1$. Therefore, to implement an infinitesimal Lorentz boost on the Cartesian coordinates within a test (scalar) function $f(x^\mu)$, we do
\begin{align}
f(x^\mu) \to f\left( \LTud{\mu}{\nu} x^\nu \right)
&= f\left( x^\mu + \xi \epsilon^\mu_{\phantom{\mu}\nu} x^\nu + \mathcal{O}\left(\xi^2\right) \right) \\
&= f(x) - \xi \epsilon_{\nu\mu} x^\nu \partial^\mu f(x) + \mathcal{O}\left(\xi^2\right)  .
\end{align}
Since $\xi$ is arbitrary, to implement a Lorentz boost of the coordinate system on any Minkowski scalar, the appropriate differential operator is
\begin{align}
\epsilon_{\mu\nu} x^\mu \partial^\nu = x^0 \partial^1 - x^1 \partial^0 ;
\end{align}
which again is encoded within eq. \eqref{Killing_MinkowskiRotationBoosts}.
\begin{myP}
	{\bf Co-moving Observers \& Rulers In Cosmology} \qquad We live in a universe that, at the very largest length scales, is described by the following spatially flat Friedmann-Lema\^{i}tre-Robertson-Walker (FLRW) metric
	\begin{align}
	\dd s^2 = \dd t^2 - a(t)^2 \dd\vec{x}\cdot\dd\vec{x} ;
	\end{align}
	where $a(t)$ describes the relative size of the universe. Enumerate as many constants-of-motion as possible of this geometry. (Hint: Focus on the spatial part of the metric and try to draw a connection with the previous problem.)
	
	In this cosmological context, a co-moving observer is one that does not move spatially, i.e., $\dd\vec{x}=0$. Solve the geodesic swept out by such an observer. 
	
	Galaxies $A$ and $B$ are respectively located at $\vec{x}$ and $\vec{x}'$ at a fixed cosmic time $t$. What is their spatial distance on this constant $t$ slice of spacetime? \qed
\end{myP}
\begin{myP}
	{\bf Killing identities involving Ricci} \qquad Prove the following results. If $\xi^\mu$ is a Killing vector and $R_{\alpha\beta}$ and $\mathcal{R}$ are the Ricci tensor and scalar respectively, then
	\begin{align}
	\label{Killing_Ricci}
	\xi^\alpha \nabla^\beta R_{\alpha\beta} = 0 
	\qquad \text{ and } \qquad  
	\xi^\alpha \nabla_\alpha \mathcal{R} 	= 0 .
	\end{align}
	Hints: First use eq. \eqref{KillingEquation_2ndD} to show that
	\begin{align}
	\Box \xi_\delta &= -R^{\lambda}_{\phantom{\lambda}\delta} \xi_\lambda , \\
	\Box 			&\equiv g^{\alpha\beta} \nabla_\alpha \nabla_\beta = \nabla_\alpha \nabla^\alpha .
	\end{align} 
	Then take the divergence on both sides. Argue why $\xi^\alpha \nabla^\beta R_{\alpha\beta} = \nabla^\beta (\xi^\alpha R_{\alpha\beta})$. You may also need to employ the Einstein tensor Bianchi identity $\nabla^\mu G_{\mu\nu} = 0$ to infer that $\xi^\alpha \nabla_\alpha \mathcal{R}=0$. \qed
\end{myP}
\begin{myP}
	In $d$ spacetime dimensions, show that
	\begin{align}
	\label{dJep}
	\partial_{[\alpha_1} J^\mu \widetilde{\epsilon}_{\alpha_2 \dots \alpha_d] \mu}
	\end{align}
	is proportional to $\nabla_\sigma J^\sigma$. What is the proportionality factor? (This discussion provides a differential forms based language to write $\dd^d x \sqrt{|g|} \nabla_\sigma J^\sigma$.) If $\nabla_\sigma J^\sigma = 0$, what does the Poincar\'{e} lemma tell us about eq. \eqref{dJep}? Find the dual of your result and argue there must an antisymmetric tensor $\Sigma^{\mu\nu}$ such that 
	\begin{align}
	J^\mu = \nabla_\nu \Sigma^{\mu\nu} .
	\end{align} \qed
\end{myP}
\begin{myP}
	{\bf Gauge-covariant derivative} \qquad Let $\psi$ be a vector under {\it group} transformations. By this we mean that, if $\psi^{\check{a}}$ corresponds to the $a$th component of $\psi$, then given some matrix $U^{\check{a}}_{\phantom{\check{a}}\check{b}}$, $\psi$ transforms as
	\begin{align}
	\label{DifferentialGeometry_GaugeTransformation_Matter}
	\psi^{\check{a}'} = U^{\check{a}'}_{\phantom{\check{a}'}\check{b}} \psi^{\check{b}} \qquad\qquad 
	\left( \text{or, } \psi' = U \psi \right) .
	\end{align}
	Compare eq. \eqref{DifferentialGeometry_GaugeTransformation_Matter} to how a spacetime vector transforms under coordinate transformations:
	\begin{align}
	V^{\mu'}(x') = \mathcal{J}^{\mu'}_{\phantom{\mu'}\sigma} V^\sigma(x),
	\qquad\qquad
	\mathcal{J}^\mu_{\phantom{\mu}\sigma} \equiv \frac{\partial x'^\mu}{\partial x^\sigma}  .
	\end{align}
	Now, let us consider taking the gauge-covariant derivative $\check{D}$ of $\psi$ such that it still transforms `covariantly' under group transformations, namely
	\begin{align}
	\label{DifferentialGeometry_GaugeTransformation_MatterDerivative}
	\check{D}_\alpha \psi' = \check{D}_\alpha (U \psi) = U (\check{D}_\alpha \psi) .
	\end{align}
	%Or, in component form,
	%\begin{align}
	%\check{D}_\alpha \psi^{\check{a}'} = U^{\check{a}}_{\phantom{\check{a}}\check{b}} \check{D}_\alpha \psi^{\check{b}} .
	%\end{align}
	Crucially: 
	\begin{quotation}
		{\it We shall now demand that the gauge-covariant derivative transforms covariantly -- eq. \eqref{DifferentialGeometry_GaugeTransformation_MatterDerivative} holds -- even when the group transformation $U(x)$ depends on spacetime coordinates.}
	\end{quotation}
	First check that, the spacetime-covariant derivative cannot be equal to the gauge-covariant derivative in general, i.e.,
	\begin{align}
	\nabla_\alpha \psi' \neq \check{D}_\alpha \psi' ,
	\end{align}
	by showing that eq. \eqref{DifferentialGeometry_GaugeTransformation_MatterDerivative} is not satisfied. 
	
	Just as the spacetime-covariant derivative was built from the partial derivative by adding a Christoffel symbol, $\nabla = \partial + \Gamma$, we may build a gauge-covariant derivative by adding to the spacetime-covariant derivative a {\it gauge potential}:
	\begin{align}
	\label{DifferentialGeometry_GaugeCovariantD}
	(\check{D}_\mu)^{\check{a}}_{\phantom{\check{a}}\check{b}} \equiv \delta^a_b \nabla_\mu + (A_\mu)^{\check{a}}_{\phantom{\check{a}}\check{b}} .
	\end{align}
	Or, in gauge-index-free notation,
	\begin{align}
	\label{DifferentialGeometry_GaugeCovariantD_ColorIndexFree}
	\check{D}_\mu \equiv \nabla_\mu + A_\mu .
	\end{align}
	With the definition in eq. \eqref{DifferentialGeometry_GaugeCovariantD}, how must the gauge potential $A_\mu$ (or, equivalently, $(A_\mu)^{\check{a}}_{\phantom{\check{a}}\check{b}}$) transform so that eq. \eqref{DifferentialGeometry_GaugeTransformation_MatterDerivative} is satisfied? Compare the answer to the transformation properties of the Christoffel symbol in eq. \eqref{Connection_CoordinateTransformation}. (Since the answer can be found in most Quantum Field Theory textbooks, make sure you verify the covariance explicitly!)
	
	{\it Bonus}: Here, we have treated $\psi$ as a spacetime scalar and the gauge-covariant derivative $\check{D}_\alpha$ itself as a scalar under group transformations. Can you generalize the analysis here to the higher-rank tensor case? \qed
\end{myP}

\subsection{Special Topic 1: Gravitational Perturbation Theory}
\label{Chapter_DifferentialGeometry_CurvedSpacetimes_C4}
Carrying out perturbation theory about some fixed `background' geometry $\gb_{\mu\nu}$ has important physical applications. As such, in this section, we will in fact proceed to set up a general and systematic perturbation theory involving the metric:
\begin{align}
\label{PerturbedMetric}
g_{\mu\nu} = \gb_{\mu\nu} + h_{\mu\nu} ,
\end{align}
where $\gb_{\mu\nu}$ is an arbitrary `background' metric and $h_{\mu\nu}$ is a small deviation. I will also take the opportunity to discuss the transformation properties of $h_{\mu\nu}$ under infinitesimal coordinate transformations, i.e., the gauge transformations of gravitons.

{\bf Metric inverse, Determinant} \qquad Whenever performing a perturbative analysis, we shall agree to move all tensor indices -- including that of $h_{\mu\nu}$ -- with the $\gb_{\alpha\beta}$. For example,
\begin{align}
h^{\alpha}_{\phantom{\alpha}\beta} \equiv \gb^{\alpha\sigma} h_{\sigma\beta} , 
\qquad \text{ and } \qquad 
h^{\alpha\beta} \equiv \gb^{\alpha\sigma} \gb^{\beta\rho} h_{\sigma\rho} .
\end{align}
With this convention in place, let us note that the inverse metric is a geometric series. Firstly,
\begin{align}
\label{PerturbedMetric_ToInverse}
g_{\mu\nu} 
= \gb_{\mu\sigma} \left( \delta^\sigma_\nu + h^\sigma_{\phantom{\sigma}\nu} \right) 
\dot{\equiv} \gb \cdot \left( \mathbb{I} + {\sf h} \right) .
\end{align}
(Here, ${\sf h}$ is a matrix, whose $\mu$th row and $\nu$th column is $h^\mu_{\phantom{\mu}\nu} \equiv \gb^{\mu\sigma} h_{\sigma\nu}$.) Remember that, for invertible matrices $A$ and $B$, we have $(A \cdot B)^{-1} = B^{-1} A^{-1}$. Therefore
\begin{align}
g^{-1} = \left( \mathbb{I} + {\sf h} \right)^{-1} \cdot \gb^{-1} .
\end{align}
If we were dealing with numbers instead of matrices, the geometric series $1/(1+z) = \sum_{\ell=0}^{\infty} (-)^\ell z^\ell$ may come to mind. You may directly verify that this prescription, in fact, still works.
\begin{align}
g^{\mu\nu} 
&= \left( \delta^\mu_{\phantom{\mu}\lambda} + \sum_{\ell=1}^{\infty} (-)^\ell h^{\mu}_{\phantom{\mu}\sigma_1} h^{\sigma_1}_{\phantom{\sigma_1}\sigma_2} \dots h^{\sigma_{\ell-2}}_{\phantom{\sigma_{\ell-2}}\sigma_{\ell-1}} h^{\sigma_{\ell-1}}_{\phantom{\sigma_{\ell-1}}\lambda} \right) \gb^{\lambda\nu} \\
\label{PerturbedMetric_Inverseg}
&= \gb^{\mu\nu} + \sum_{\ell=1}^{\infty} (-)^\ell h^{\mu}_{\phantom{\mu}\sigma_1} h^{\sigma_1}_{\phantom{\sigma_1}\sigma_2} \dots h^{\sigma_{\ell-2}}_{\phantom{\sigma_{\ell-2}}\sigma_{\ell-1}} h^{\sigma_{\ell-1}\nu} \\
&= \gb^{\mu\nu} - h^{\mu\nu}
+ h^{\mu}_{\phantom{\mu}\sigma_1} h^{\sigma_1 \nu}
- h^{\mu}_{\phantom{\mu}\sigma_1} h^{\sigma_1}_{\phantom{\mu}\sigma_2} h^{\sigma_2 \nu} 
+ \dots .
\end{align}
The square root of the determinant of the metric can be computed order-by-order in perturbation theory via the following formula. For any matrix $A$,
\begin{align}
\label{detA_to_Exp}
\det A = \exp\left[ \Tr{\ln A} \right] ,
\end{align}
where Tr is the matrix trace; for e.g., $\Tr{{\sf h}} = h^\sigma_{\phantom{\sigma}\sigma}$. Taking the determinant of both sides of eq. \eqref{PerturbedMetric_ToInverse}, and using the property $\det[A \cdot B] = \det A \cdot \det B$,
\begin{align}
\det g_{\alpha\beta} = \det \gb_{\alpha\beta} \cdot \det\left[ \mathbb{I} + {\sf h} \right] ,
\end{align}
so that eq. \eqref{detA_to_Exp} can be employed to state
\begin{align}
\sqrt{|g|} &= \sqrt{|\gb|} \cdot \exp\left[ \frac{1}{2} \Tr{\ln [\mathbb{I} + {\sf h}]} \right] . 
\end{align}
The first few terms read
\begin{align}
\sqrt{|g|}
&= \sqrt{|\gb|} \bigg( 1 + \frac{1}{2} h 
+ \frac{1}{8} h^2 - \frac{1}{4} h^{\sigma\rho} h_{\sigma\rho} \nonumber \\
&\qquad \qquad 
+ \frac{1}{48} h^3 - \frac{1}{8} h \cdot h^{\sigma\rho} h_{\sigma\rho} + \frac{1}{6} h^{\sigma\rho} h_{\rho\kappa} h^{\kappa}_{\phantom{\kappa}\sigma} + \mathcal{O}[h^{4}] \bigg) \\
h &\equiv h^{\sigma}_{\phantom{\sigma}\sigma} .
\end{align}
{\bf Covariance, Covariant Derivatives, Geometric Tensors} \qquad Under a coordinate transformation $x \equiv x(x')$, the full metric of course transforms as a tensor. The full metric $g_{\alpha'\beta'}$ in this new $x'$ coordinate system reads 
\begin{align}
\label{PerturbedMetric_Covariance}
g_{\alpha'\beta'}(x')
= \left( \gb_{\mu\nu}(x(x')) + h_{\mu\nu}(x(x'))\right) \frac{\partial x^\mu}{\partial x'^\alpha} \frac{\partial x^\nu}{\partial x'^\beta} .
\end{align}
If we define the `background metric' to transform covariantly; namely
\begin{align}
\label{PerturbedMetric_BgCovariance}
\gb_{\alpha'\beta'}(x') \equiv \gb_{\mu\nu}(x(x')) \frac{\partial x^\mu}{\partial x'^\alpha} \frac{\partial x^\nu}{\partial x'^\beta} ;
\end{align}
then, from eq. \eqref{PerturbedMetric_Covariance}, the perturbation itself can be treated as a tensor
\begin{align}
\label{PerturbedMetric_hCovariance}
h_{\alpha'\beta'}(x')
= h_{\mu\nu}(x(x')) \frac{\partial x^\mu}{\partial x'^\alpha} \frac{\partial x^\nu}{\partial x'^\beta} .
\end{align}
These will now guide us to construct the geometric tensors -- the full Riemann tensor, Ricci tensor and Ricci scalar -- using the covariant derivative $\Db$ with respect to the `background metric' $\gb_{\mu\nu}$ and its associated geometric tensors. Let's begin by considering this background covariant derivative acting on the full metric in eq. \eqref{PerturbedMetric}:
\begin{align}
\label{PerturbedMetric_Db_IofII}
\Db_\alpha g_{\mu\nu} 
= \Db_\alpha \left(\gb_{\mu\nu} + h_{\mu\nu}\right)
= \Db_\alpha h_{\mu\nu} .
\end{align}
On the other hand, the usual rules of covariant differentiation tell us
\begin{align}
\label{PerturbedMetric_Db_IIofII}
\Db_\alpha g_{\mu\nu} 
= \partial_\alpha g_{\mu\nu} 
- \overline{\Gamma}^\sigma_{\phantom{\sigma}\alpha \mu} g_{\sigma\nu}
- \overline{\Gamma}^\sigma_{\phantom{\sigma}\alpha \nu} g_{\mu\sigma} ;
\end{align}
where the Christoffel symbols here are built out of the `background metric',
\begin{align}
\overline{\Gamma}^\sigma_{\phantom{\sigma}\alpha \mu}
= \frac{1}{2} \gb^{\sigma\lambda} \left( \partial_\alpha \gb_{\mu \lambda} + \partial_\mu \gb_{\alpha \lambda} - \partial_\lambda \gb_{\mu\alpha} \right) .
\end{align}
\begin{myP}
	\label{Problem_GravityPT_ChristoffelSymbols}
	{\bf Relation between `background' and `full' Christoffel} \qquad Show that equations \eqref{PerturbedMetric_Db_IofII} and \eqref{PerturbedMetric_Db_IIofII} can be used to deduce that the full Christoffel symbol
	\begin{align}
	\label{PerturbedMetric_FullChristoffelDef}
	\Gamma^\alpha_{\phantom{\alpha}\mu\nu}[g]
	= \frac{1}{2} g^{\alpha \sigma} \left( \partial_\mu g_{\nu\sigma} + \partial_\nu g_{\mu\sigma} - \partial_\sigma g_{\mu\nu} \right)
	\end{align}
	can be related to that of its background counterpart through the relation
	\begin{align}
	\label{PerturbedMetric_FullChristoffel}
	\Gamma^\alpha_{\phantom{\alpha}\mu\nu}[g]
	&= \overline{\Gamma}^\alpha_{\phantom{\alpha}\mu\nu}[\gb] + \delta \Gamma^\alpha_{\phantom{\alpha}\mu\nu} .
	\end{align}
	Here,
	\begin{align}
	\label{PerturbedMetric_PerturbedChristoffel}
	\delta \Gamma^\alpha_{\phantom{\alpha}\mu\nu}
	&\equiv \frac{1}{2} g^{\alpha\sigma} H_{\sigma \mu\nu} , \\
	H_{\sigma \mu\nu}
	&\equiv \Db_\mu h_{\nu \sigma} + \Db_\nu h_{\mu \sigma} - \Db_\sigma h_{\mu\nu} .
	\end{align}
	Notice the difference between the `full' and `background' Christoffel symbols, namely $\Gamma^\mu_{\phantom{\mu}\alpha\beta} - \overline{\Gamma}^\mu_{\phantom{\mu}\alpha\beta}$, is a tensor. \qed
\end{myP}
\begin{myP} {\bf Geometric tensors} \qquad With the result in eq. \eqref{PerturbedMetric_FullChristoffel}, show that for an arbitrary 1-form $V_\beta$,
	\begin{align}
	\nabla_\alpha V_\beta
	= \Db_\alpha V_\beta - \delta\Gamma^\sigma_{\phantom{\sigma} \alpha\beta} V_\sigma .
	\end{align}
	Use this to compute $[\nabla_\alpha,\nabla_\beta] V_\lambda$ and proceed to show that the exact Riemann tensor is
	\begin{align}
	\label{PerturbedMetric_FullRiemann}
	R^{\alpha}_{\phantom{\alpha}\beta \mu \nu}[g]
	&= \bar{R}^{\alpha}_{\phantom{\alpha}\beta \mu \nu}[\gb] + \delta R^{\alpha}_{\phantom{\alpha}\beta \mu \nu} , \\
	\delta R^{\alpha}_{\phantom{\alpha}\beta \mu \nu} 
	&\equiv \Db_{[\mu} \delta\Gamma^\alpha_{\phantom{\alpha}\nu]\beta} 
	+ \delta\Gamma^\alpha_{\phantom{\alpha}\sigma[\mu} \delta\Gamma^\sigma_{\phantom{\sigma}\nu]\beta} \\
	&= \frac{1}{2} \Db_{\mu} \left( g^{\alpha \lambda} H_{\lambda \nu \beta}
	\right) - \frac{1}{2} \Db_{\nu} \left( g^{\alpha \lambda} H_{\lambda \mu \beta}
	\right) + \frac{1}{4} g^{\alpha \lambda} g^{\sigma \rho } \left( H_{\lambda \mu
		\sigma} H_{\rho \beta \nu} - H_{\lambda \nu \sigma} H_{\rho \beta \mu} \right) ,
	\end{align}
	where $\bar{R}^{\alpha}_{\phantom{\alpha}\beta \mu \nu}[\gb]$ is the Riemann tensor built entirely out of the background metric $\gb_{\alpha \lambda}$. \qed
\end{myP}	
	From eq. \eqref{PerturbedMetric_FullRiemann}, the Ricci tensor and scalars can be written down:
	\begin{align}
	\label{PerturbedMetric_FullRicci}
	R_{\mu\nu}[g] = R^{\sigma}_{\phantom{\sigma}\mu \sigma \nu} 
	\qquad \text{ and } \qquad
	\mathcal{R}[g] = g^{\mu\nu} R_{\mu\nu} .
	\end{align}
	From these formulas, perturbation theory can now be carried out. The primary reason why these geometric tensors admit an infinite series is because of the geometric series of the full inverse metric eq. \eqref{PerturbedMetric_Inverseg}. I find it helpful to remember, when one multiplies two infinite series which do not have negative powers of the expansion object $h_{\mu\nu}$, the terms that contain precisely $n$ powers of $h_{\mu\nu}$ is a discrete convolution: for instance, such an $n$th order piece of the Ricci scalar is
	\begin{align}
	\delta_n \mathcal{R}
	= \sum_{\ell=0}^{n} \delta_\ell g^{\mu\nu} \delta_{n-\ell} R_{\mu\nu} ,
	\end{align}
	where $\delta_\ell g^{\mu\nu}$ is the piece of the full inverse metric containing exactly $\ell$ powers of $h_{\mu\nu}$ and $\delta_{n-\ell} R_{\mu\nu}$ is that containing precisely $n-\ell$ powers of the same.
\begin{myP}
	{\bf Linearized geometric tensors} \qquad The Riemann tensor that contains up to one power of $h_{\mu\nu}$ can be obtained readily from eq. \eqref{PerturbedMetric_FullRiemann}. The $H^2$ terms begin at order $h^2$, so we may drop them; and since $H$ is already linear in $h$, the $g^{-1}$ contracted into it can be set to the background metric.
	\begin{align}
	\label{PerturbedMetric_LinearizedRiemann}
	R^{\alpha}_{\phantom{\alpha}\beta \mu \nu}[g]
	&= \bar{R}^{\alpha}_{\phantom{\alpha}\beta \mu \nu}[\gb] 
	+ \frac{1}{2} \Db_{[\mu} \left( \Db_{\nu]} h_{\beta}^{\phantom{\beta}\alpha} + \Db_{|\beta|} h_{\nu]}^{\phantom{\nu}\alpha} - \Db^\alpha h_{\nu]\beta} \right) + \mathcal{O}(h^2) \\
	&= \bar{R}^{\alpha}_{\phantom{\alpha}\beta \mu \nu}[\gb] 
	+ \frac{1}{2} \left( [\Db_{\mu}, \Db_\nu] h_{\beta}^{\phantom{\beta}\alpha} + \Db_{\mu} \Db_\beta h_{\nu}^{\phantom{\nu}\alpha} - \Db_{\nu} \Db_\beta h_{\mu}^{\phantom{\mu}\alpha} - \Db_{\mu} \Db^\alpha h_{\nu\beta} + \Db_{\nu} \Db^\alpha h_{\mu\beta} \right) + \mathcal{O}(h^2) . \nonumber
	\end{align}
	(The $|\beta|$ on the first line indicates the $\beta$ is not to be antisymmetrized.) Starting from the linearized Riemann tensor in eq. \eqref{PerturbedMetric_LinearizedRiemann}, let us work out the linearized Ricci tensor, Ricci scalar, and Einstein tensor.
	
	Specifically, show that one contraction of eq. \eqref{PerturbedMetric_LinearizedRiemann} yields the linearized Ricci tensor:
	\begin{align}
	R_{\beta\nu}			&= \overline{R}_{\beta\nu} + \delta_1 R_{\beta\nu} + \mathcal{O}(h^2) , \\
	\delta_1 R_{\beta\nu} 	&\equiv \frac{1}{2} \left( \Db^{\mu} \Db_{\{\beta} h_{\nu\}\mu} -
	\Db_{\nu} \Db_{\beta} h - \Db^\mu \Db_{\mu} h_{\beta\nu} \right) .
	\end{align}
	Contracting this Ricci tensor result with the full inverse metric, verify that the linearized Ricci scalar is
	\begin{align}
	\mathcal{R} 			&= \overline{\mathcal{R}} + \delta_1 \mathcal{R} + \mathcal{O}(h^2), \\
	\delta_1 \mathcal{R} 	&\equiv -h^{\beta\nu} \bar{R}_{\beta\nu} + \left( \Db^{\mu} \Db^\nu - \gb^{\mu\nu} \Db^\sigma \Db_\sigma \right) h_{\mu\nu} .
	\end{align}
	Now, let us define the variable $\bar{h}_{\mu\nu}$ through the relation
	\begin{align}
	\label{PerturbedMetric_htobarh}
	h_{\mu\nu} \equiv \bar{h}_{\mu\nu} - \frac{\gb_{\mu\nu}}{d-2} \bar{h}, \qquad\qquad
	\bar{h} \equiv \bar{h}^\sigma_{\phantom{\sigma}\sigma} .
	\end{align}
	First explain why this is equivalent to
	\begin{align}
	\label{PerturbedMetric_barhtoh}
	\bar{h}_{\mu\nu} = h_{\mu\nu} - \frac{\gb_{\mu\nu}}{2} h .
	\end{align}
	(Hint: First calculate the trace of $\bar{h}$ in terms of $h$.) In (3+1)D this $\bar{h}_{\mu\nu}$ is often dubbed the ``trace-reversed" perturbation -- can you see why? Then show that the linearized Einstein tensor is
	\begin{align}
	\label{PerturbedMetric_Einstein01}
	G_{\mu\nu} = \bar{G}_{\mu\nu}[\gb] + \delta_1 G_{\mu\nu} + \mathcal{O}(\overline{h}^2) ,
	\end{align} 
	where
	\begin{align}
	\label{PerturbedMetric_Einstein1}
	\delta_1 G_{\mu\nu} 
	&\equiv -\frac{1}{2} \left( \overline{\Box} \bar{h}_{\mu\nu} + \gb_{\mu\nu} \Db_\sigma \Db_\rho \bar{h}^{\sigma\rho} - \Db_{\{\mu} \Db^{\sigma} \bar{h}_{\nu\}\sigma} \right) \nonumber\\
	&\qquad\qquad
	+ \frac{1}{2} \left( \gb_{\mu\nu} \bar{h}^{\rho\sigma} \bar{R}_{\rho\sigma} + \bar{h}_{\{\mu}^{\phantom{\{\mu}\sigma} \bar{R}_{\nu\}\sigma} - \bar{h}_{\mu\nu} \bar{\mathcal{R}} - 2 \bar{h}^{\rho\sigma} \bar{R}_{\mu\rho \nu\sigma} \right) .
	\end{align}
	Cosmology, Kerr/Schwarzschild black holes, and Minkowski spacetimes are three physically important geometries. This result may be used to study linear perturbations about them. \qed
\end{myP}
{\it Second order Ricci} \qquad For later purposes, we collect the second order Ricci tensor -- see, for e.g., equation 35.58b of \cite{MTW}:\footnote{I have checked that eq. \eqref{PerturbedMetric_RicciTensor_O2} is consistent with the output from {\sf xAct} \cite{xAct}.}
\begin{align}
\label{PerturbedMetric_RicciTensor_O2}
\delta_2 R_{\mu\nu}
&= \frac{1}{2} \Bigg\{ 
\frac{1}{2} \Db_{\mu} h_{\alpha\beta} \Db_\nu h^{\alpha\beta}
+ h^{\alpha\beta} \left( \Db_{\nu} \Db_{\mu} h_{\alpha\beta} + \Db_{\beta} \Db_{\alpha} h_{\mu\nu} - \Db_{\beta} \Db_{\nu} h_{\mu\alpha} - \Db_{\beta} \Db_{\mu} h_{\nu\alpha} \right)  \\
&\qquad\qquad
+ \Db^{\beta} h^{\alpha}_{\phantom{\alpha}\nu} \left( \Db_{\beta} h_{\mu\alpha} - \Db_{\alpha} h_{\mu\beta} \right)
- \Db_\beta \left( h^{\alpha\beta} - \frac{1}{2} \gb^{\alpha\beta} h \right)
\left( \Db_{\{\nu} h_{\mu\}\alpha} - \Db_\alpha h_{\mu\nu} \right) 
\Bigg\} . \nonumber
\end{align}
{\bf Gauge transformations: Infinitesimal Coordinate Transformations} \qquad In the above discussion, we regarded the `background metric' as a tensor. As a consequence, the metric perturbation $h_{\mu\nu}$ was also a tensor. However, since it is the full metric that enters any generally covariant calculation, it really is the combination $\gb_{\mu\nu} + h_{\mu\nu}$ that transforms as a tensor. As we will now explore, when the coordinate transformation
\begin{align}
\label{CoordinateTransformation_Infinitesimal}
x^\mu = x'^\mu + \xi^\mu(x')  
\end{align}
is infinitesimal, in that $\xi^\mu$ is small in the same sense that $h_{\mu\nu}$ is small, we may instead attribute all the ensuing coordinate transformations to a transformation of $h_{\mu\nu}$ alone. This will allow us to view `small' coordinate transformations as gauge transformations, and will also be important for the discussion of the linearized Einstein's equations. 

In what follows, we shall view the $x$ and $x'$ in eq. \eqref{CoordinateTransformation_Infinitesimal} as referring to the same spacetime point, but expressed within infinitesimally different coordinate systems. Now, transforming from $x$ to $x'$,
{\allowdisplaybreaks\begin{align}
\dd s^2 
&= g_{\mu\nu}(x) \dd x^\mu \dd x^\nu \\
&= \left( \gb_{\mu\nu}(x'+\xi) + h_{\mu\nu}(x'+\xi) \right) 
\left( \dd x'^\mu + \partial_{\alpha'} \xi^\mu \dd x'^\alpha \right) \left( \dd x'^\nu + \partial_{\beta'} \xi^\nu \dd x'^\beta \right) \nonumber \\
&= \left( \gb_{\mu\nu}(x') + \xi^\sigma \partial_{\sigma'} \gb_{\mu\nu}(x') + h_{\mu\nu}(x') + \mathcal{O}\left(\xi^2,\xi \partial h \right) \right) 
\left( \dd x'^\mu + \partial_{\alpha'} \xi^\mu \dd x'^\alpha \right) \left( \dd x'^\nu + \partial_{\beta'} \xi^\nu \dd x'^\beta \right) \nonumber \\
&= \left(\gb_{\mu\nu}(x') 
+ \xi^\sigma(x') \partial_{\sigma'} \gb_{\mu\nu}(x') 
+ \gb_{\sigma\{\mu}(x') \partial_{\nu'\}} \xi^\sigma(x') 
+ h_{\mu\nu}(x') + \mathcal{O}\left(\xi^2,\xi \partial h \right) \right) \dd x'^\mu \dd x'^\nu . \nonumber
\end{align}}
This teaches us that, the infinitesimal coordinate transformation of eq. \eqref{CoordinateTransformation_Infinitesimal} amounts to keeping the background metric fixed, but shifting
\begin{align}
h_{\mu\nu}(x)
\to h_{\mu\nu}(x) + \xi^\sigma(x) \partial_{\sigma} \gb_{\mu\nu}(x) + \gb_{\sigma\{\mu}(x) \partial_{\nu\}} \xi^\sigma(x) ,
\end{align}
followed by replacing
\begin{align}
x^\mu \to x'^\mu 
\qquad \text{ and } \qquad 
\partial_\mu \equiv \frac{\partial}{\partial x^\mu} \to \frac{\partial}{\partial x'^\mu} \equiv \partial_{\mu'} .
\end{align}
However, since $x$ and $x'$ refer to the same point in spacetime,\footnote{We had, earlier, encountered very similar mathematical manipulations while considering the geometric symmetries that left the metric in the same form upon an active coordinate transformation -- an actual displacement from one point to another infinitesimally close by. Here, we are doing a passive coordinate transformation, where $x$ and $x'$ describe the same point in spacetime, but using infinitesimally different coordinate systems.} it is customary within the contemporary physics literature to drop the primes and simply phrase the coordinate transformation as replacement rules:
\begin{align}
\label{PerturbedMetric_GaugeTransformation_Coords}
x^\mu				&\to x^\mu + \xi^\mu(x) , \\
\label{PerturbedMetric_GaugeTransformation_Bg}
\gb_{\mu\nu}(x)		&\to \gb_{\mu\nu}(x) , \\
\label{PerturbedMetric_GaugeTransformation_h}
h_{\mu\nu}(x) 		&\to h_{\mu\nu}(x) + \Db_{\{\mu} \xi_{\nu\}}(x) ;
\end{align}
where we have recognized 
\begin{align}
\xi^\sigma \partial_{\sigma} \gb_{\mu\nu} + \gb_{\sigma\{\mu} \partial_{\nu\}} \xi^\sigma = \Db_{\{\mu} \xi_{\nu\}} .
\end{align}
\begin{myP}
	\label{Problem_CoordinateGaugeTransformation}
	{\bf Gauge transformations of a tensor} \qquad Consider perturbing a spacetime tensor
	\begin{align}
	\label{PerturbedTensor}
	T^{\mu_1 \dots \mu_N}_{\phantom{\mu_1 \dots \mu_N} \nu_1 \dots \nu_M}
	\equiv \overline{T}^{\mu_1 \dots \mu_N}_{\phantom{\mu_1 \dots \mu_N} \nu_1 \dots \nu_M}
	+ \delta T^{\mu_1 \dots \mu_N}_{\phantom{\mu_1 \dots \mu_N} \nu_1 \dots \nu_M} ,
	\end{align}
	where $\delta T^{\mu_1 \dots \mu_N}_{\phantom{\mu_1 \dots \mu_N} \nu_1 \dots \nu_M}$ is small in the same sense that $\xi^\alpha$ and $h_{\mu\nu}$ are small. Perform the infinitesimal coordinate transformation in eq. \eqref{CoordinateTransformation_Infinitesimal} on the tensor in eq. \eqref{PerturbedTensor} and attribute all the transformations to the $\delta T^{\mu_1 \dots \mu_N}_{\phantom{\mu_1 \dots \mu_N} \nu_1 \dots \nu_M}$. Write down the ensuing gauge transformation, in direct analogy to eq. \eqref{PerturbedMetric_GaugeTransformation_h}. Can you write it in a generally covariant form? Then justify the statement: 
	\begin{quotation}
		``If the background tensor is zero, the perturbed tensor is gauge-invariant at first order in coordinate transformations."
	\end{quotation} \qed
\end{myP}

\subsection{Special Topic 2: Conformal/Weyl Transformations}
\label{Chapter_DifferentialGeometry_ConformalTransformations}
In this section, we collect for the reader's reference, the conformal transformation properties of various geometric objects. We shall define a conformal transformation on a metric to be a change of the geometry by an overall spacetime dependent scale. That is,
\begin{align}
\label{ConformalTransformation_Metric}
g_{\mu\nu}(x) \equiv \Omega^2(x) \bar{g}_{\mu\nu}(x) .
\end{align}
The inverse metric is
\begin{align}
g^{\mu\nu}(x) = \Omega(x)^{-2} \gb^{\mu\nu}(x),
\qquad\qquad
\gb^{\mu\sigma} \gb_{\sigma\nu} \equiv \delta^\mu_\nu .
\end{align}
We shall now enumerate how the geometric objects/operations built out of $g_{\mu\nu}$ is related to that built out of $\gb_{\mu\nu}$. In what follows, all indices on barred tensors are raised and lowered with $\gb^{\mu\nu}$ and $\bar{g}_{\mu\nu}$ while all indices on un-barred tensors are raised/lowered with $g^{\mu\nu}$ and $g_{\mu\nu}$; the covariant derivative $\nabla$ is with respect to $g_{\mu\nu}$ while the $\overline{\nabla}$ is with respect to $\gb_{\mu\nu}$.

{\bf Metric Determinant} \qquad Since 
\begin{align}
\det g_{\mu\nu} = \det \left( \Omega^2 \gb_{\mu\nu} \right) = \Omega^{2d} \det \gb_{\mu\nu} ,
\end{align}
we must also have
\begin{align}
|g|^{1/2} = \Omega^d |\bar{g}|^{1/2} .
\end{align}
{\bf Scalar Gradients} \qquad The scalar gradient with a lower index is just a partial derivative. Therefore
\begin{align}
\nabla_\mu \varphi = \overline{\nabla}_\mu \varphi = \partial_\mu \varphi .
\end{align}
while $\nabla^\mu \varphi = g^{\mu\nu} \nabla_\nu \varphi = \Omega^{-2} \bar{g}^{\mu\nu} \overline{\nabla}_\nu \varphi$, so
\begin{align}
\nabla^\mu \varphi = \Omega^{-2} \overline{\nabla}^\mu \varphi .
\end{align}
{\bf Scalar Wave Operator} \qquad The wave operator $\Box$ in the geometry $g_{\mu\nu}$ is defined as
\begin{align}
\Box \equiv g^{\mu\nu} \nabla_\mu \nabla_\nu = \nabla_\mu \nabla^\mu .
\end{align}
By a direct calculation, the wave operator $\Box$ with respect to $g_{\mu\nu}$ acting on a scalar $\psi$ is
\begin{align}
\Box \varphi 
= \frac{1}{\Omega^2} \left( \frac{d-2}{\Omega} \overline{\nabla}_\mu \Omega \cdot \overline{\nabla}^\mu \varphi + \overline{\Box} \varphi \right) ,
\end{align}
where $\overline{\Box}$ is the wave operator with respect to $\gb_{\mu\nu}$. We also have 
\begin{align}
\Box \left( \Omega^s \psi \right) 
&= \frac{1}{\Omega^2} 
\Big\{ \left( s \Omega^{s-1} \overline{\Box} \Omega + s \left( d + s-3 \right) \Omega^{s-2} \overline{\nabla}_\mu \Omega \overline{\nabla}^\mu \Omega \right) \psi \nonumber\\
&\qquad\qquad
+ \left( 2 s + d - 2 \right) \Omega^{s-1} \overline{\nabla}_\mu \Omega \overline{\nabla}^\mu \psi 
+ \Omega^s \overline{\Box} \psi
\Big\} .
\end{align}
{\bf Scalar Field Action} \qquad In $d$ dimensional spacetime, the following action involving the scalar $\varphi$ and Ricci scalar $\mathcal{R}[g]$,
\begin{align}
S[\varphi] \equiv \int \dd^d x \sqrt{|g|} \frac{1}{2} 
\left( g^{\alpha\beta} \nabla_\alpha \varphi \nabla_\beta \varphi + \frac{d-2}{4(d-1)} \mathcal{R} \varphi^2 \right),
\end{align}
is invariant -- up to surface terms -- under the simultaneous replacements
\begin{align}
g_{\alpha\beta} \to \Omega^2 g_{\alpha\beta}, \qquad
g^{\alpha\beta} &\to \Omega^{-2} g^{\alpha\beta}, \qquad
\sqrt{|g|} \to \Omega^d \sqrt{|g|}, \\
\varphi &\to \Omega^{1-\frac{d}{2}} \varphi .
\end{align}
The jargon here is that $\varphi$ transforms covariantly under conformal transformations, with weight $s = 1-(d/2)$. We see in two dimensions, $d=2$, a minimally coupled massless scalar theory automatically enjoys conformal/Weyl symmetry.

\noindent{\bf Christoffel Symbols} \qquad A direct calculation shows:
\begin{align}
\label{ConformalTransformation_Christoffel}
\Gamma^\mu_{\alpha\beta}[g]
&= \overline{\Gamma}^\mu_{\alpha\beta}[\gb] + \left( \partial_{\{\alpha} \ln \Omega \right) \delta_{\beta\}}^\mu - \gb_{\alpha\beta} \gb^{\mu\nu} \left( \partial_\nu \ln \Omega \right) \\
&= \overline{\Gamma}^\mu_{\alpha\beta}[\gb] + \left( \Db_{\{\alpha} \ln \Omega \right) \delta_{\beta\}}^\mu - \gb_{\alpha\beta} \Db^\mu \ln \Omega .
\end{align}
{\bf Riemann Tensor} \qquad By viewing the difference between $g_{\mu\nu}$ and $\gb_{\mu\nu}$ as a `perturbation',
\begin{align}
g_{\mu\nu} - \gb_{\mu\nu} = \left(\Omega^2-1\right) \gb_{\mu\nu} \equiv h_{\mu\nu} ,
\end{align}
we may employ the results in \S \eqref{Chapter_DifferentialGeometry_CurvedSpacetimes_C4}. In particular, eq. \eqref{PerturbedMetric_FullRiemann} may be used to infer that the Riemann tensor is
\begin{align}
\label{ConformalTransformation_Riemann}
R^\alpha_{\phantom{\alpha}\beta \mu\nu}[g]
= \bar{R}^\alpha_{\phantom{\alpha}\beta \mu\nu}[\gb]
&+ \Db_\beta \Db_{[\mu} \ln\Omega \delta_{\nu]}^\alpha - \gb_{\beta[\nu} \Db_{\mu]} \Db^\alpha \ln\Omega \nonumber \\
&
+ \delta^\alpha_{[\mu} \Db_{\nu]} \ln \Omega \Db_{\beta} \ln\Omega + \Db^\alpha \ln\Omega \Db_{[\mu} \ln\Omega \gb_{\nu]\beta} 
+ \left(\Db\ln \Omega\right)^2 \gb_{\beta[\mu} \delta_{\nu]}^\alpha .
\end{align}
{\bf Ricci Tensor} \qquad In turn, the Ricci tensor is
\begin{align}
\label{ConformalTransformation_RicciTensor}
R_{\beta\nu}[g]
= \bar{R}_{\beta\nu}[\gb]
&+ (2-d) \Db_\beta \Db_\nu \ln\Omega - \gb_{\beta\nu} \overline{\Box} \ln\Omega \\
&+ (d-2) \left( \Db_\beta \ln \Omega \Db_\nu \ln \Omega - \gb_{\beta\nu} \left(\Db \ln \Omega\right)^2 \right) .
\end{align}
{\bf Ricci Scalar} \qquad Contracting the Ricci tensor with $g^{\beta\nu} = \Omega^{-2}\gb^{\beta\nu}$, we conclude 
\begin{align}
\label{ConformalTransformation_RicciScalar}
\mathcal{R}[g]
= \Omega^{-2} \left(
\overline{\mathcal{R}}[\gb] + 2(1-d) \overline{\Box} \ln\Omega + (d-2)(1-d) \left( \Db \ln \Omega \right)^2
\right)
\end{align}
{\bf Weyl Tensor} \qquad The Weyl tensor, for spacetime dimensions greater than two ($d > 2$), is defined to be the completely trace-free portion of the Riemann tensor:
\begin{align}
\label{WeylTensor}
C_{\mu\nu \alpha\beta}
&\equiv
R_{\mu\nu \alpha\beta}
- \frac{1}{d-2} \left( R_{\alpha[\mu} g_{\nu]\beta} - R_{\beta[\mu} g_{\nu]\alpha} \right)
+ \frac{ g_{\mu[\alpha} g_{\beta]\nu} }{ (d-2)(d-1) } \mathcal{R}[g] .
\end{align}
By a direct calculation, one may verify $C_{\mu\nu \alpha\beta}$ has the same index-symmetries as $R_{\mu\nu \alpha\beta}$ and is indeed completely traceless: $g^{\mu\alpha} C_{\mu\nu \alpha\beta} = 0$. Using equations \eqref{ConformalTransformation_Metric}, \eqref{ConformalTransformation_Riemann}, \eqref{ConformalTransformation_RicciTensor}, and \eqref{ConformalTransformation_RicciScalar}, one may then deduce the Weyl tensor with one upper index is {\it invariant} under conformal transformations:
\begin{align}
C^\mu_{\phantom{\mu}\nu \alpha\beta}[g] = C^\mu_{\phantom{\mu}\nu \alpha\beta}[\gb] .
\end{align}
If we lower the index $\mu$ on both sides,
\begin{align}
C_{\mu\nu \alpha\beta}[g] = \Omega^2 C_{\mu\nu \alpha\beta}[\gb] .
\end{align}
{\bf Einstein Tensor} \qquad From equations \eqref{ConformalTransformation_Metric}, \eqref{ConformalTransformation_RicciTensor} and \eqref{ConformalTransformation_RicciScalar}, we may also compute the transformation of the Einstein tensor $G_{\beta\nu} \equiv R_{\beta\nu} -(g_{\beta\nu}/2) \mathcal{R}$.
\begin{align}
G_{\beta\nu}[g] 
&= \overline{G}_{\beta\nu}[\gb] 
+ (2-d) \left(\Db_\beta \Db_\nu \ln\Omega - \gb_{\beta\nu} \overline{\Box} \ln\Omega \right) \nonumber\\
&\qquad\qquad
+ (d-2) \left( \Db_\beta \ln \Omega \Db_\nu \ln \Omega - \gb_{\beta\nu} \frac{3-d}{2} \left(\Db \ln \Omega\right)^2 \right) 
\end{align}
Notice the Einstein tensor is invariant under constant conformal transformations: $G_{\beta\nu}[g] = \overline{G}_{\beta\nu}[\gb]$ whenever $\partial_\mu \Omega = 0$.

To reiterate: on the right-hand-sides of these expressions for the Riemann tensor, Ricci tensor and scalar, all indices are raised and lowered with $\gb$; for example, $(\Db A)^2 \equiv \gb^{\sigma\tau} \Db_\sigma A \Db_\tau A$ and $\Db^\alpha A \equiv \gb^{\alpha\lambda} \Db_\lambda A$. The $R^\alpha_{\phantom{\alpha}\beta \mu\nu}[g]$ is built out of the metric $g_{\alpha\beta}$ but the $\bar{R}^\alpha_{\phantom{\alpha}\beta \mu\nu}[\gb]$ is built entirely out of $\gb_{\mu\nu}$, etc.

\newpage

\section{Linear Partial Differential Equations (PDEs)}
\label{Chapter_LinearPDE}
A partial differential equation (PDE) is a differential equation involving more than one variable. Much of fundamental physics -- electromagnetism, quantum mechanics, gravitation and more -- involves PDEs. We will first examine Poisson's equation, and introduce the concept of the Green's function, in order to solve it. Because the Laplacian $\vec{\nabla}^2$ will feature a central role in our study of PDEs, we will study its eigenfunctions/values in various contexts. Then we will use their spectra to tackle the heat/diffusion equation via an initial value formulation. In the final sections we will study the wave equation in flat spacetime, and study various routes to obtain its solutions, both in position/real spacetime and in Fourier space.

\subsection{Laplacians and Poisson's Equation}

\subsubsection{Poisson's equation, uniqueness of solutions}

Poisson's equation in $D$-space is defined to be
\begin{align}
\label{PoissonEquation}
-\vec{\nabla}^2 \psi(\vec{x}) = J(\vec{x}),
\end{align}
where $J$ is to be interpreted as some given mass/charge density that sources the Newtonian/electric potential $\psi$. The most physically relevant case is in 3D; if we use Cartesian coordinates, Poisson's equation reads
\begin{align}
-\vec{\nabla}^2 \psi(\vec{x}) 
= -\left(\frac{\partial^2 \psi}{\partial (x^1)^2} + \frac{\partial^2 \psi}{\partial (x^2)^2} + \frac{\partial^2 \psi}{\partial (x^3)^2}\right) = J(\vec{x}) .
\end{align} 
We will soon see how to solve eq. \eqref{PoissonEquation} by first solving for the inverse of the negative Laplacian ($\equiv$ Green's function).

\noindent{\bf Uniqueness of solution} \qquad We begin by showing that the solution of Poisson's equation (eq. \eqref{PoissonEquation}) in some domain $\mathfrak{D}$ is unique once $\psi$ is specified on the boundary of the domain $\partial \mathfrak{D}$. As we shall see, this theorem holds even in curved spaces. If it is the normal derivative $n^i \nabla_i \psi$ that is specified on the boundary $\partial \mathfrak{D}$, then $\psi$ is unique up to an additive constant.

The proof goes by contradiction. Suppose there were two distinct solutions, $\psi_1$ and $\psi_2$. Let us start with the integral
\begin{align}
\label{PoissonEquation_UniquenessProof_I}
I \equiv \int_{\mathfrak{D}} \dd^D \vec{x} \sqrt{|g|} \nabla_i \Psi^\dagger \nabla^i \Psi \geq 0 .
\end{align}
That this is greater or equal to zero, even in curved spaces, can be seen by writing the gradients in an orthonormal frame (cf. eq. \eqref{DifferentialGeometry_OrthonormalFrame}), where $g^{ij} = \varepsilon_{\widehat{a}}^{\phantom{\widehat{a}}i} \varepsilon_{\widehat{b}}^{\phantom{\widehat{b}}j} \delta^{ab}$.\footnote{Expressing the gradients in an orthonormal frame is, in fact, the primary additional ingredient to this proof, when compared to the flat space case. Moreover, notice this proof relies on the Euclidean (positive definite) nature of the metric.} The $\sqrt{|g|}$ is always positive, since it describes volume, whereas $\nabla_i \Psi \nabla^i \Psi$ is really a sum of squares.
\begin{align}
\sqrt{|g|} \delta^{ab} \nabla_{\widehat{a}} \Psi^\dagger \nabla_{\widehat{b}} \Psi 
= \sqrt{|g|} \sum_a \left\vert\nabla_{\widehat{a}} \Psi\right\vert^2 \geq 0 .
\end{align}
We may now integrate-by-parts eq. \eqref{PoissonEquation_UniquenessProof_I} and use the curved space Gauss' theorem in eq. \eqref{DifferentialGeometry_GaussTheorem}.
\begin{align}
\label{PoissonEquation_UniquenessProof_II}
I = \int_{\partial \mathfrak{D}} \dd^{D-1}\Sigma_i \cdot \Psi^\dagger \nabla^i \Psi - \int_{\mathfrak{D}} \dd^D \vec{x} \sqrt{|g|} \cdot \Psi^\dagger \nabla_i \nabla^i \Psi .
\end{align}
Remember from eq. \eqref{DifferentialGeometry_DirectedArea_v1} that $\dd^{D-1}\Sigma_i \nabla^i \Psi = \dd^{D-1}\vec{\xi} \sqrt{|H(\vec{\xi})|} n^i \nabla_i \Psi$, where $n^i$ is the unit (outward) normal to the boundary $\partial \mathfrak{D}$. If either $\psi(\partial \mathfrak{D})$ or $n^i \partial_i \psi(\partial \mathfrak{D})$ is specified, therefore, the first term on the right hand side of eq. \eqref{PoissonEquation_UniquenessProof_II} is zero -- since $\Psi(\partial \mathfrak{D}) = \psi_1(\partial \mathfrak{D}) - \psi_2(\partial \mathfrak{D})$ and $n^i \partial_i \Psi(\partial \mathfrak{D}) = n^i \partial_i \psi_1(\partial \mathfrak{D}) - n^i \partial_i \psi_2(\partial \mathfrak{D})$. The seccond term is zero too, since
\begin{align}
-\nabla_i \nabla^i \Psi = -\nabla_i \nabla^i(\psi_1 - \psi_2) = J-J = 0.
\end{align}
But we have just witnessed how $I$ is itself the integral, over the domain, of the sum of squares of $| \nabla_{\widehat{a}} \Psi |$. The only way summing squares of something is zero is that something is identically zero.
\begin{align}
\nabla_{\widehat{a}} \Psi = \varepsilon_{\widehat{a}}^{\phantom{\widehat{a}}i} \partial_i \Psi = 0, \qquad \text{(everywhere in $\mathfrak{D}$)} .
\end{align}
Viewing the $\varepsilon_{\widehat{a}}^{\phantom{\widehat{a}}i}$ as a vector field, so $\nabla_{\widehat{a}} \Psi$ is the derivative of $\Psi$ in the $a$th direction, this translates to the conclusion that $\Psi = \psi_1 - \psi_2$ is constant in every direction, all the way up to the boundary; i.e., $\psi_1$ and $\psi_2$ can at most differ by an additive constant. If the normal derivative $n^i \nabla_i \psi(\partial \mathfrak{D})$ were specified, so that $n^i \nabla_i \Psi = 0$ there, then $\psi_1(\vec{x}) - \psi_2(\vec{x}) =$ non-zero constant can still yield the same normal derivative. However, if instead $\psi(\partial \mathfrak{D})$ were specified on the boundary, $\Psi(\partial\mathfrak{D})=0$ there, and must therefore be zero everywhere in $\mathfrak{D}$. In other words $\psi_1 = \psi_2$, and there cannot be more than 1 distinct solution. This completes the proof.

\subsubsection{(Negative) Laplacian as a Hermitian operator}

We will now demonstrate that the negative Laplacian in some domain $\mathfrak{D}$ can be viewed as a Hermitian operator, if its eigenfunctions obey 
\begin{align}
\label{Laplacian_Dirichlet}
\{\psi_\lambda(\partial\mathfrak{D}) = 0 \} \qquad \text{(Dirichlet)}
\end{align} 
or 
\begin{align}
\label{Laplacian_Neumann}
\{n^i \nabla_i \psi_\lambda(\partial\mathfrak{D}) = 0 \} \qquad \text{(Neumann)} ,
\end{align}
or if there are {\it no boundaries}.\footnote{In this chapter on PDEs we will focus mainly on Dirichlet (and occasionally, Neumann) boundary conditions. There are plenty of other possible boundary conditions, of course.} The steps we will take here are very similar to those in the uniqueness proof above. Firstly, by Hermitian we mean the negative Laplacian enjoys the property that
\begin{align}
I \equiv \int_{\mathfrak{D}} \dd^D \vec{x} \sqrt{|g(\vec{x})|} \psi_1^\dagger(\vec{x}) \left(-\vec{\nabla}^2_{\vec{x}} \psi_2(\vec{x}) \right)
= \int_{\mathfrak{D}} \dd^D \vec{x} \sqrt{|g(\vec{x})|} \left(-\vec{\nabla}^2_{\vec{x}} \psi_1^\dagger(\vec{x})\right) \psi_2(\vec{x}) ,
\end{align}
for any functions $\psi_{1,2}(\vec{x})$ spanned by the eigenfunctions of $-\vec{\nabla}^2$, and therefore satisfy the same boundary conditions. We begin on the left hand side and again employ the curved space Gauss' theorem in eq. \eqref{DifferentialGeometry_GaussTheorem}.
\begin{align}
I 
&= \int_{\partial \mathfrak{D}} \dd^{D-1} \Sigma_i \psi_1^\dagger \left(- \nabla^i \psi_2 \right)
+ \int_{\mathfrak{D}} \dd^D \vec{x} \sqrt{|g|} \nabla_i \psi_1^\dagger \nabla^i \psi_2, \nonumber\\
&= \int_{\partial \mathfrak{D}} \dd^{D-1} \Sigma_i \left\{ \psi_1^\dagger \left(- \nabla^i \psi_2 \right) + \left(\nabla^i \psi_1^\dagger\right) \psi_2 \right\}
+ \int_{\mathfrak{D}} \dd^D \vec{x} \sqrt{|g|} \left(-\nabla^i\nabla_i \psi_1^\dagger\right) \psi_2,
\end{align}
We see that, if either $\psi_{1,2}(\partial\mathfrak{D})=0$, or $n^i \nabla_i \psi_{1,2}(\partial\mathfrak{D})= 0$, the surface integrals vanish, and the Hermitian nature of the Laplacian is established. 
%In case the second condition is not as apparent, let us work it out more explicitly; the key term is
%\begin{align}
%\int_{\partial \mathfrak{D}} \dd^{D-1} \Sigma_i \left\{ \psi_1^\dagger \left(- \nabla^i \psi_2 \right) + \left(\nabla^i \psi_1^\dagger\right) \psi_2 \right\}
%= \int_{\partial \mathfrak{D}} \dd^{D-1} \vec{\xi} \sqrt{|H(\vec{\xi})|} \alpha \left\{ - \psi_1^\dagger \psi_2 + \psi_1^\dagger \psi_2 \right\} = 0 ,
%\end{align}
%where $\dd^{D-1} \vec{\xi} \sqrt{|H(\vec{\xi})|}$ denotes the volume/area element on the boundary $\partial\mathfrak{D}$. Finally, if there are no boundaries to begin with, there will be no surface terms to speak of and the Laplacian is thus automatically Hermitian.

\noindent{\bf Non-negative eigenvalues} \qquad Let us understand the bounds on the spectrum of the negative Laplacian subject to the Dirichlet (eq. \eqref{Laplacian_Dirichlet}) or Neumann boundary (eq. \eqref{Laplacian_Neumann}) conditions, or when there are no boundaries. Let $\psi_\lambda$ be an eigenfunction obeying
\begin{align}
\label{Laplacian_HermitianNonNegativeProof_I}
- \vec{\nabla}^2 \psi_\lambda = \lambda \psi_\lambda .
\end{align}
We have previously argued that 
\begin{align}
\label{Laplacian_HermitianNonNegativeProof_II}
I' = \int_{\mathfrak{D}} \dd^D \vec{x} \sqrt{|g|} \nabla_i \psi_\lambda^\dagger \nabla^i \psi_\lambda 
\end{align}
is strictly non-negative. If we integrate-by-parts,
\begin{align}
I' &= 
\int_{\partial\mathfrak{D}} \dd^{D-1}\Sigma_i \psi_\lambda^\dagger \nabla^i \psi_\lambda 
		+ \int_{\mathfrak{D}} \dd^D \vec{x} \sqrt{|g|} \psi_\lambda^\dagger \left(-\nabla_i\nabla^i \psi_\lambda\right) \geq 0 . 
\end{align}
If there are no boundaries -- for example, if $\mathfrak{D}$ is a $(n \geq 2)$-sphere (usually denoted as $\mathbb{S}^n$) -- there will be no surface terms; if there are boundaries but the eigenfunctions obey either Dirichlet conditions in eq. \eqref{Laplacian_Dirichlet} or Neumann conditions in eq. \eqref{Laplacian_Neumann}, the surface terms will vanish. In all three cases, we see that the corresponding eigenvalues $\{\lambda\}$ are strictly non-negative, since $\int_{\mathfrak{D}} \dd^D \vec{x} \sqrt{|g|} |\psi_\lambda|^2 \geq 0$:
\begin{align}
I' = \lambda \int_{\mathfrak{D}} \dd^D \vec{x} \sqrt{|g|} |\psi_\lambda|^2 \geq 0 .
\end{align}
%For orthonormal eigenfunctions obeying the Neumann boundary conditions in eq. \eqref{Laplacian_Robin},
%\begin{align}
%\lambda \int_{\mathfrak{D}} \dd^D \vec{x} \sqrt{|g|} |\psi_\lambda|^2 
%		&\geq -\alpha \int_{\partial\mathfrak{D}} \dd^D \vec{\xi} \sqrt {|H(\vec{\xi})|}  |\psi_\lambda|^2 \\
%\lambda &\geq -\alpha \frac{\int_{\partial\mathfrak{D}} \dd^D \vec{\xi} \sqrt{|H(\vec{\xi})|}  |\psi_\lambda|^2}{\int_{\mathfrak{D}} \dd^D \vec{x} \sqrt{|g|} %|\psi_\lambda|^2}
%\end{align}
\begin{myP}
\qquad Instead of Dirichlet or Neumann boundary conditions, let us allow for mixed (aka Robin) boundary conditions, namely
\begin{align}
\label{RobinBC}
\alpha \cdot \psi + \beta \cdot n^i \nabla_i \psi = 0 
\end{align}
on the boundary $\partial\mathfrak{D}$. Show that the negative Laplacian is Hermitian if we impose
\begin{align}
\frac{\alpha}{\alpha^*} = \frac{\beta}{\beta^*} .
\end{align}
In particular, if $\alpha$ and $\beta$ are both real, imposing eq. \eqref{RobinBC} automatically yields a Hermitian Laplacian. \qed
\end{myP}

\subsubsection{Inverse of the negative Laplacian: Green's function and reciprocity}

Given the Dirichlet boundary condition in eq. \eqref{Laplacian_Dirichlet}, i.e., $\{ \psi_\lambda(\partial\mathfrak{D}) = 0 \}$, we will now understand how to solve Poisson's equation, through the inverse of the negative Laplacian. Roughly speaking,
\begin{align}
-\vec{\nabla}^2 \psi = J \qquad \Rightarrow \qquad \psi = \left( -\vec{\nabla}^2 \right)^{-1} J .
\end{align}
(The actual formula, in a finite domain, will be a tad more complicated, but here we are merely motivating the reason for defining $G$.) Since, given any Hermitian operator
\begin{align}
H = \sum_\lambda \lambda \ketbra{\lambda}{\lambda}, \qquad \{ \lambda \in \mathbb{R} \} ,
\end{align}
its inverse is 
\begin{align}
H^{-1} = \sum_\lambda \frac{\ketbra{\lambda}{\lambda}}{\lambda}, \qquad \{ \lambda \in \mathbb{R} \} ;
\end{align}
we see that the inverse of the negative Laplacian in the position space representation is the following mode expansion involving its eigenfunctions $\{ \psi_\lambda \}$.
\begin{align}
\label{GreensFunction_Laplacian_ModeSum}
G(\vec{x},\vec{x}') 
&= \braOket{\vec{x}}{\frac{1}{-\vec{\nabla}^2}}{\vec{x}'} 
= \sum_{\lambda} \frac{\psi_\lambda(\vec{x}) \psi_\lambda(\vec{x}')^\dagger}{\lambda}, \\
-\vec{\nabla}^2 \psi_\lambda &= \lambda \psi_\lambda, \qquad \psi_\lambda(\vec{x}) \equiv \braket{\vec{x}}{\lambda} .
\end{align}
(The summation sign is schematic; it can involve either (or both) a discrete sum or/and an integral over a continuum.) Since the mode functions are subject to $\{ \psi_\lambda(\partial\mathfrak{D}) = 0 \}$, the Green's function itself also obeys Dirichlet boundary conditions:
\begin{align}
\label{GreensFunction_Laplacian_DirichletBC}
G(\vec{x}\in\mathfrak{D},\vec{x}') = G(\vec{x},\vec{x}'\in\mathfrak{D}) = 0 .
\end{align}
The Green's function $G$ satisfies the PDE
\begin{align}
\label{GreensFunction_Laplacian_PDE}
-\vec{\nabla}^2_{\vec{x}} G(\vec{x},\vec{x}') 
= -\vec{\nabla}^2_{\vec{x}'} G(\vec{x},\vec{x}') = \frac{\delta^{(D)}(\vec{x}-\vec{x}')}{\sqrt[4]{|g(\vec{x}) g(\vec{x}'))|}} ,
\end{align}
because the negative Laplacian is Hermitian and thus its eigenfunctions obey the following completeness relation (cf. \eqref{CompletenessRelation})
\begin{align}
\label{GreensFunction_Laplacian_Completeness}
\sum_{\lambda} \psi_\lambda(\vec{x})^\dagger \psi_\lambda(\vec{x}') = \frac{\delta^{(D)}(\vec{x}-\vec{x}')}{\sqrt[4]{|g(\vec{x}) g(\vec{x}'))|}} .
\end{align}
Eq. \eqref{GreensFunction_Laplacian_PDE} follows from $-\vec{\nabla}^2 \psi_\lambda = \lambda \psi_\lambda$ and
\begin{align}
-\vec{\nabla}^2_{\vec{x}} G(\vec{x},\vec{x}') 
&= \sum_{\lambda} \frac{-\vec{\nabla}^2_{\vec{x}} \psi_\lambda(\vec{x}) \psi_\lambda(\vec{x}')^\dagger}{\lambda} 
		= \sum_{\lambda} \psi_\lambda(\vec{x}) \psi_\lambda(\vec{x}')^\dagger , \\
-\vec{\nabla}^2_{\vec{x}'} G(\vec{x},\vec{x}') 
&= \sum_{\lambda} \frac{\psi_\lambda(\vec{x}) (-\vec{\nabla}^2_{\vec{x}'} \psi_\lambda(\vec{x}')^\dagger)}{\lambda} 
		= \sum_{\lambda} \psi_\lambda(\vec{x}) \psi_\lambda(\vec{x}')^\dagger .
\end{align}
Because the $\delta^{(D)}$-functions on the right hand side of eq. \eqref{GreensFunction_Laplacian_PDE} is the (position representation) of the identity operator, the Green's function itself is really the inverse of the negative Laplacian. 

{\bf Physically speaking} these $\delta$-functions also lend eq. \eqref{GreensFunction_Laplacian_PDE} to the interpretation that the Green's function is the field at $\vec{x}$ produced by a point source at $\vec{x}'$. Therefore, the Green's function of the negative Laplacian {\it is} the gravitational/electric potential produced by a unit strength point charge/mass.

{\bf Isolated zero eigenvalue implies non-existence of inverse} \qquad Within a finite domain $\mathfrak{D}$, we see that the Neumann boundary conditions $\{ n^i \nabla_i \psi_\lambda(\partial\mathfrak{D}) = 0 \}$ imply there must be a zero eigenvalue; for, the $\psi_0 =$ constant is the corresponding eigenvector, whose normal derivative on the boundary is zero:
\begin{align}
-\vec{\nabla}^2 \psi_0 = -\frac{\partial_i \left( \sqrt{|g|} g^{ij} \partial_j \psi_0 \right)}{\sqrt{|g|}} = 0 \cdot \psi_0 .
\end{align}
As long as this is an isolated zero -- i.e., there are no eigenvalues continuously connected to $\lambda = 0$ -- this mode will contribute a discrete term in the mode sum of eq. \eqref{GreensFunction_Laplacian_ModeSum} that yields a $1/0$ infinity. That is, the inverse of the Laplacian does not make sense if there is an isolated zero mode.\footnote{In the infinite flat $\mathbb{R}^{D}$ case below, we will see the $\{ \exp(i\vec{k}\cdot\vec{x}) \}$ are the eigenfunctions and hence there is also a zero mode, gotten by setting $\vec{k} \to \vec{0}$. However the inverse does exist because the mode sum of eq. \eqref{GreensFunction_Laplacian_ModeSum} is really an integral, and the integration measure $\dd^D\vec{k}$ ensures convergence of the integral.}

{\bf Discontinuous first derivatives} \qquad Because it may not be apparent from the mode expansion in eq. \eqref{GreensFunction_Laplacian_ModeSum}, it is worth highlighting that the Green's function must contain discontinuous first derivatives as $\vec{x} \to \vec{x}'$ in order to yield, from a second order Laplacian, $\delta$-functions on the right hand side of eq. \eqref{GreensFunction_Laplacian_PDE}. For Green's functions in a finite domain $\mathfrak{D}$, there are potentially additional discontinuities when both $\vec{x}$ and $\vec{x}'$ are near the boundary of the domain $\partial \mathfrak{D}$.

{\bf Flat $\mathbb{R}^D$ and Method of Images} \qquad An example is provided by the eigenfunctions of the negative Laplacian in infinite $D$-space.
\begin{align}
\psi_{\vec{k}}(\vec{x}) 							= \frac{e^{i \vec{k} \cdot \vec{x}}}{(2\pi)^{D/2}} , \qquad 
-\vec{\nabla}_{\vec{x}}^2 \psi_{\vec{k}}(\vec{x}) 	= \vec{k}^2 \psi_{\vec{k}}(\vec{x}).
\end{align}
Because we know the integral representation of the $\delta$-function, eq. \eqref{GreensFunction_Laplacian_Completeness} now reads
\begin{align}
\int_{\mathbb{R}^D} \frac{\dd^D \vec{k}}{(2\pi)^D} e^{i \vec{k} \cdot (\vec{x}-\vec{x}')} = \delta^{(D)}(\vec{x}-\vec{x}') .
\end{align}
Through eq. \eqref{GreensFunction_Laplacian_ModeSum}, we may write down the integral representation of the inverse of the negative Laplacian in Euclidean $D$-space.
\begin{align}
\label{GreensFunctionLaplacian_ModeExpansion_}
G(\vec{x},\vec{x}') 
= \int_{\mathbb{R}^D} \frac{\dd^D \vec{k}}{(2\pi)^D} \frac{e^{i\vec{k}\cdot(\vec{x}-\vec{x}')}}{\vec{k}^2} 
= \frac{\Gamma\left(\frac{D}{2}-1\right)}{4\pi^{D/2} |\vec{x}-\vec{x}'|^{D-2}}.
\end{align}
Now, one way to think about the Green's function $G_D(\mathfrak{D})$ of the negative Laplacian in a finite domain $\mathfrak{D}$ of flat space is to view it as the sum of its counterpart in infinite $\mathbb{R}^D$ plus a term that is a homogeneous solution $H_D(\mathfrak{D})$ in the finite domain $\mathfrak{D}$, such that the desired boundary conditions are achieved on $\partial \mathfrak{D}$. Namely,
\begin{align}
G_D(\vec{x},\vec{x}';\mathfrak{D}) 
	&= \frac{\Gamma\left(\frac{D}{2}-1\right)}{4\pi^{D/2} |\vec{x}-\vec{x}'|^{D-2}} + H(\vec{x},\vec{x}';\mathfrak{D}) , \nonumber\\
-\vec{\nabla}_{\vec{x}}^2 G_D(\vec{x},\vec{x}';\mathfrak{D}) = -\vec{\nabla}_{\vec{x}'}^2 G_D(\vec{x},\vec{x}';\mathfrak{D}) 
	&= \delta^{(D)}\left(\vec{x}-\vec{x}'\right), \qquad \text{(Cartesian coordinates)} \nonumber\\
-\vec{\nabla}_{\vec{x}}^2 H_D(\vec{x},\vec{x}';\mathfrak{D}) 
	&= -\vec{\nabla}_{\vec{x}'}^2 H_D(\vec{x},\vec{x}';\mathfrak{D}) = 0 , \qquad \vec{x},\vec{x}'\in\mathfrak{D} .
\end{align}
If Dirichlet boundary conditions are desired, we would demand
\begin{align}
\frac{\Gamma\left(\frac{D}{2}-1\right)}{4\pi^{D/2} |\vec{x}-\vec{x}'|^{D-2}} + H(\vec{x},\vec{x}';\mathfrak{D})  = 0
\end{align}
whenever $\vec{x} \in \partial\mathfrak{D}$ or $\vec{x}' \in \partial\mathfrak{D}$.

The {\it method of images}, which you will likely learn about in an electromagnetism course, is a special case of such a strategy of solving the Green's function. We will illustrate it through the following example. Suppose we wish to solve the Green's function in a half-infinite space, i.e., for $x^D \geq 0$ only, but let the rest of the $\{x^1,\dots,x^{D-1}\}$ run over the real line. We further want the boundary condition
\begin{align}
G_D(x^D=0) = G_D(x'^D=0) = 0 .
\end{align}
The strategy is to notice that the infinite plane that is equidistant between one positive and one negative point mass/charge has zero potential, so if we wish to solve the Green's function (the potential of the positive unit mass) on the half plane, we place a negative unit mass on the opposite side of the boundary at $x^D = 0$. Since the solution to Poisson's equation is unique, the solution for $x^D \geq 0$ is therefore
\begin{align}
\label{MethodOfImages_Eg}
G_D(\vec{x},\vec{x}';\mathfrak{D}) 
&= \frac{\Gamma\left(\frac{D}{2}-1\right)}{4\pi^{D/2} |\vec{x}-\vec{x}'|^{D-2}}
		- \frac{\Gamma\left(\frac{D}{2}-1\right)}{4\pi^{D/2} |\vec{\xi}|^{D-2}} , \\
|\vec{\xi}| &\equiv \sqrt{\sum_{j=1}^{D-1} (x^j-x'^j)^2 + (x^D + x'^D)^2 }, \qquad x^D,x'^D \geq 0 . \nonumber
\end{align}
Mathematically speaking, when the negative Laplacian is applied to the second term in eq. \eqref{MethodOfImages_Eg}, it yields $\prod_{j=1}^{D-1} \delta(x^j - x'^j) \delta(x^D + x'^D)$, but since $x^D, x'^D \geq 0$, the very last $\delta$-function can be set to zero. Hence, the second term is a homogeneous solution when attention is restricted to $x^D \geq 0$.

%\qquad Explain why eq. \eqref{GreensFunction_Laplacian_Completeness} means eq. \eqref{GreensFunction_Laplacian_PDE} is satisfied. Explain why the Green's function in eq. \eqref{GreensFunction_Laplacian_ModeSum} obeys the appropriate boundary conditions, either
%\begin{align}
%\label{GreensFunction_Laplacian_BC}
%G\left( \vec{x}\in\partial\mathfrak{D},\vec{x}' \right) = G\left( \vec{x}, \vec{x}' \in \partial\mathfrak{D} \right) = 0 \qquad \text{   or   } \qquad \nonumber\\
%\left. n^i(\vec{x}) \nabla_{\vec{x}^i} G\left( \vec{x}, \vec{x}' \right) \right\vert_{\vec{x}\in\partial\mathfrak{D}}
%= \left. n^i(\vec{x}') \nabla_{\vec{x}'^i} G\left( \vec{x}, \vec{x}' \right)\right\vert_{\vec{x}'\in\partial\mathfrak{D}} = 0 .
%\end{align}
{\bf Reciprocity} \qquad We will also now show that the Green's function itself is a Hermitian object, in that
\begin{align}
G(\vec{x},\vec{x}')^\dagger = G(\vec{x}',\vec{x}) = G(\vec{x},\vec{x}') .
\end{align}
The first equality follows from the real positive nature of the eigenvalues, as well as the mode expansion in eq. \eqref{GreensFunction_Laplacian_ModeSum}
\begin{align}
G(\vec{x},\vec{x}')^* = \sum_{\lambda} \frac{\psi_\lambda(\vec{x}') \psi_\lambda(\vec{x})^\dagger}{\lambda} = G(\vec{x}',\vec{x}) .
\end{align}
The second requires considering the sort of integrals we have been examining in this section.
\begin{align}
I(x,x') \equiv \int_{\mathfrak{D}} \dd^D \vec{x}'' \sqrt{|g(\vec{x}'')|} \left\{ 
G\left( \vec{x},\vec{x}'' \right) (-\vec{\nabla}_{\vec{x}''}^2) G\left( \vec{x}',\vec{x}'' \right) - G\left( \vec{x}',\vec{x}'' \right) (-\vec{\nabla}_{\vec{x}''}^2) G\left( \vec{x},\vec{x}'' \right)
\right\} .
\end{align}
Using the PDE obeyed by $G$,
\begin{align}
I(x,x') = G(\vec{x},\vec{x}')  - G(\vec{x}',\vec{x}) .
\end{align}
We may integrate-by-parts too.
\begin{align}
I(x,x') &= \int_{\partial \mathfrak{D}} \dd^{D-1}\Sigma_{i''} \left\{ 
G(\vec{x},\vec{x}'') (-\nabla^{i''}) G(\vec{x}',\vec{x}'') - G(\vec{x}',\vec{x}'') (-\nabla^{i''}) G(\vec{x},\vec{x}'')
\right\} \nonumber\\
&+ \int \dd^D \vec{x}'' \sqrt{|g(\vec{x}'')|} \left\{ 
\nabla_{i''} G(\vec{x},\vec{x}'') \nabla^{i''} G(\vec{x}',\vec{x}'') - \nabla_{i''} G(\vec{x}',\vec{x}'') \nabla^{i''} G(\vec{x},\vec{x}'')
\right\} .
\end{align}
The terms in the last line cancel. Moreover, for precisely the same boundary conditions that make the negative Laplacian Hermitian, we see the surface terms have to vanish too. Therefore $I(x,x') = 0 = G(\vec{x},\vec{x}')  - G(\vec{x}',\vec{x})$, and we have established the reciprocity of the Green's function.

\subsubsection{Kirchhoff integral theorem and Dirichlet boundary conditions}

Within a finite domain $\mathfrak{D}$ we will now understand why the choice of boundary conditions that makes the negative Laplacian a Hermitian operator, is intimately tied to the type of boundary conditions imposed in solving Poisson's equation eq. \eqref{PoissonEquation}. 

Suppose we have specified the field on the boundary $\psi(\partial\mathfrak{D})$. To solve Poisson's equation $-\vec{\nabla}^2 \psi = J$, we will start by imposing Dirichlet boundary conditions on the eigenfunctions of the Laplacian, i.e., $\{\psi_\lambda(\partial \mathfrak{D}) = 0\}$, so that the resulting Green's function obey eq. \eqref{GreensFunction_Laplacian_DirichletBC}. The solution to Poisson's equation within the domain $\mathfrak{D}$ can now be solved in terms of $G$, the source $J$, and its boundary values $\psi(\partial \mathfrak{D})$ through the following Kirchhoff integral representation:
\begin{align}
\label{GreensFunction_Laplacian_KirchhoffRep_PsiGiven}
\psi(\vec{x}) 
= \int_{\mathfrak{D}} \dd^D \vec{x}' \sqrt{|g(\vec{x}')|} G(\vec{x},\vec{x}') J(\vec{x}')
	- \int_{\partial\mathfrak{D}} \dd^{D-1} \Sigma_{i'} \nabla^{i'} G(\vec{x},\vec{x}') \psi(\vec{x}')  .
\end{align}
If there are no boundaries, then the boundary integral terms in eq. \eqref{GreensFunction_Laplacian_KirchhoffRep_PsiGiven} are zero. Similarly, if the boundaries are infinitely far away, the same boundary terms can usually be assumed to vanish, provided the fields involved decay sufficiently quickly at large distances. Physically, the first term can be interpreted to be the $\psi$ directly due to $J$ the source (the particular solution). Whereas the surface integral terms are independent of $J$ and therefore the homogeneous solutions. 

{\it Derivation of eq. \eqref{GreensFunction_Laplacian_KirchhoffRep_PsiGiven}} \qquad Let us now consider the following integral
\begin{align}
\label{GreensFunction_Laplacian_Kirchhoff_I}
I(\vec{x} \in \mathfrak{D}) \equiv \int_{\mathfrak{D}} \dd^D \vec{x}' \sqrt{|g(\vec{x}')|} \left\{ 
G(\vec{x},\vec{x}') \left(-\vec{\nabla}_{\vec{x}'}^2 \psi(\vec{x}') \right)
- \left(-\vec{\nabla}_{\vec{x}'}^2 G(\vec{x},\vec{x}') \right) \psi(\vec{x}') 
\right\}
\end{align}
If we use the equations \eqref{PoissonEquation} and \eqref{GreensFunction_Laplacian_PDE} obeyed by $\psi$ and $G$ respectively, we obtain immediately
\begin{align}
\label{GreensFunction_Laplacian_Kirchhoff_II}
I(\vec{x}) &= \int_{\mathfrak{D}} \dd^D \vec{x}' \sqrt{|g(\vec{x}')|} G(\vec{x},\vec{x}') J(\vec{x}')
- \psi(\vec{x}) .
\end{align}
On the other hand, we may integrate-by-parts, 
\begin{align}
\label{GreensFunction_Laplacian_Kirchhoff_III}
I(\vec{x}) &= \int_{\partial\mathfrak{D}} \dd^{D-1} \Sigma_{i'} \left\{ 
G(\vec{x},\vec{x}') \left(-\nabla^{i'} \psi(\vec{x}') \right)
- \left(-\nabla^{i'} G(\vec{x},\vec{x}') \right) \psi(\vec{x}') 
\right\} \nonumber\\
&+ \int_{\mathfrak{D}} \dd^D \vec{x}' \sqrt{|g(\vec{x}')|} \left\{ 
\nabla_{i'} G(\vec{x},\vec{x}') \nabla^{i'} \psi(\vec{x}') 
- \nabla^{i'} G(\vec{x},\vec{x}') \nabla_{i'} \psi(\vec{x}') 
\right\} .
\end{align}
The second line cancels. Combining equations \eqref{GreensFunction_Laplacian_Kirchhoff_II} and \eqref{GreensFunction_Laplacian_Kirchhoff_III} then hands us the following Kirchhoff representation:
\begin{align}
\label{GreensFunction_Laplacian_KirchhoffRep_0}
\psi(\vec{x} \in \mathfrak{D})
= \int_{\partial\mathfrak{D}} \dd^{D-1} \Sigma_{i'} & \left\{
G(\vec{x},\vec{x}') \left(\nabla^{i'} \psi(\vec{x}') \right)
-\left(\nabla^{i'} G(\vec{x},\vec{x}') \right) \psi(\vec{x}') 
\right\} \nonumber\\
&+ \int_{\mathfrak{D}} \dd^D \vec{x}' \sqrt{|g(\vec{x}')|} G(\vec{x},\vec{x}') J(\vec{x}') .
\end{align}
\footnote{I have put a prime on the index in $\nabla^{i'}$ to indicate the covariant derivative is with respect to $\vec{x}'$.}If we recall the Dirichlet boundary conditions obeyed by the Green's function $G(\vec{x},\vec{x}')$ (eq. \eqref{GreensFunction_Laplacian_DirichletBC}), the first term on the right hand side of the first line drops out and we obtain eq. \eqref{GreensFunction_Laplacian_KirchhoffRep_PsiGiven}.

%Remember we had proven that the solution $\psi$ of Poisson's equation is unique up to possibly an additive constant, once the field or its normal derivative is specified. The presence of both the $\psi(\partial\mathfrak{D})$ and $n^i \nabla_i \psi(\partial \mathfrak{D})$ in eq. \eqref{GreensFunction_Laplacian_KirchhoffRep_0} appears to over-determine the solution. However, once Dirichlet boundary conditions are placed on the mode functions and thus $G(\vec{x},\vec{x}')$ vanishes once either $\vec{x}$ or $\vec{x}'$ lie on the boundary, this means the first term on the first line of eq. \eqref{GreensFunction_Laplacian_KirchhoffRep_0} is zero. This reduces eq. \eqref{GreensFunction_Laplacian_KirchhoffRep_0} to eq. \eqref{GreensFunction_Laplacian_KirchhoffRep_PsiGiven}.
\begin{myP} {\it Dirichlet B.C. Variation Principle}
\qquad In a finite domain (where $\int_{\mathfrak{D}} \dd^D \vec{x} \sqrt{|g|} < \infty$), let all fields vanish on the boundary $\partial\mathfrak{D}$ and denote the smallest non-zero eigenvalue of the negative Laplacian $-\vec{\nabla}^2$ as $\lambda_0$. Let $\psi$ be an arbitrary function obeying the same boundary conditions as the eigenfunctions of $-\vec{\nabla}^2$. For this problem, assume that the spectrum of the negative Laplacian is discrete. Prove that
\begin{align}
\label{VariationalPrincipleForLaplacianSpectrum}
\frac{\int_{\mathfrak{D}} \dd^D \vec{x} \sqrt{|g|} \nabla_i \psi^\dagger \nabla^i \psi}{\int_{\mathfrak{D}} \dd^D \vec{x} \sqrt{|g|} |\psi|^2}
	\geq \lambda_0 .
\end{align}
Just like in quantum mechanics, we have a variational principle for the spectrum of the negative Laplacian in a finite volume curved space: you can exploit any trial complex function $\psi$ that vanishes on $\mathfrak{D}$ to derive an upper bound for the lowest eigenvalue of the negative Laplacian.

Hint: Expand $\psi$ as a superposition of the eigenfunctions of $-\vec{\nabla}^2$. Then integrate-by-parts one of the $\nabla^i$ in the integrand. \qed
\end{myP}
{\it Example} \qquad Suppose, within a finite 1D box, $x \in [0,L]$ we are provided a real field $\psi$ obeying
\begin{align}
\psi(x=0) = \alpha, \qquad \psi(x=L) = \beta 
\end{align}
without any external sources. You can probably solve this 1D Poisson's equation $(-\partial_x^2 \psi = 0)$ right away; it is a straight line:
\begin{align}
\label{1DG_Example_0}
\psi(0 \leq x \leq L) = \alpha + \frac{\beta-\alpha}{L} x .
\end{align}
But let us try to solve it using the methods developed here. First, we recall the orthonormal eigenfunctions of the negative Laplacian with Dirichlet boundary conditions,
\begin{align}
\braket{x}{n} 				&= \sqrt{\frac{2}{L}} \sin\left(\frac{n\pi}{L} x \right), \qquad n \in \{ 1,2,3,\dots \} , \qquad
\sum_{n=1}^\infty \braket{x}{n} \braket{n}{x'} = \delta(x-x'), \nonumber \\
-\partial_x^2 \braket{x}{n} &= \left(\frac{n\pi}{L}\right)^2 \braket{x}{n} .
\end{align}
The mode sum expansion of the Green's function in eq. \eqref{GreensFunction_Laplacian_ModeSum} is
\begin{align}
G(x,x') = \frac{2}{L} \sum_{n = 1}^{\infty} \left(\frac{n \pi}{L}\right)^{-2} \sin\left(\frac{n\pi}{L} x \right) \sin\left(\frac{n\pi}{L} x' \right) .
\end{align}
The $J$ term in eq. \eqref{GreensFunction_Laplacian_KirchhoffRep_PsiGiven} is zero, while the surface integrals really only involve evaluation at $x=0,L$. Do be careful that the normal derivative refers to the outward normal.
\begin{align}
\psi(\vec{x}) &= \partial_{x'} G(x,x'=0) \psi(x'=0) - \partial_{x'} G(x,x'=L) \psi(x'=L) \nonumber\\
&= -\frac{2}{L} \sum_{n = 1}^{\infty} \frac{L}{n \pi} \sin\left(\frac{n\pi}{L} x \right) \left[ \cos\left(\frac{n\pi}{L} x' \right) \psi(x') \right]_{x'=0}^{x'=L} \nonumber\\
\label{1DG_Example_1}
&= -\sum_{n = 1}^{\infty} \frac{2}{n \pi} \sin\left(\frac{n\pi}{L} x \right) \left( (-)^n \cdot \beta - \alpha \right) 
\end{align}
We may check this answer in the following way. Because the solution in eq. \eqref{1DG_Example_1} is odd under $x \to -x$, let us we extend the solution in the following way:
\begin{align}
\psi_\infty(-L \leq x \leq L) 
&= \alpha + \frac{\beta-\alpha}{L} x , \qquad 0 \leq x \leq L , \nonumber\\
&= -\left(\alpha + \frac{\beta-\alpha}{L} x\right) , \qquad -L \leq x < 0 .
\end{align}
We will then extend the definition of $\psi_\infty$ by imposing periodic boundary conditions, $\psi_\infty(x+2L) = \psi_\infty(x)$. This yields the Fourier series
\begin{align}
\psi_\infty(x) 	&= \sum_{\ell = -\infty}^{+\infty} C_\ell e^{i\frac{2\pi \ell}{2 L} x} .
\end{align}
Multiplying both sides by $\exp(-i (\pi n/L)x)$ and integrating over $x \in [-L,L]$.
\begin{align}
C_n 
= \int_{-L}^{L} \psi_\infty(x) e^{-i\frac{\pi n}{L} x} \frac{\dd x}{2L} 
&= \int_{-L}^{L} \psi_\infty(x) \left( \cos\left(\frac{\pi n}{L} x\right) - i \sin\left(\frac{\pi n}{L} x\right) \right) \frac{\dd x}{2L} \nonumber\\
&= -i \int_{0}^{L} \left(\alpha + \frac{\beta-\alpha}{L} x\right) \sin\left( \frac{\pi n}{L} x\right) \frac{\dd x}{L} \nonumber\\
&= \frac{i}{\pi n} \left( (-)^n \beta - \alpha \right) .
\end{align}
Putting this back to into the Fourier series,
\begin{align}
\psi_\infty(x) 	
&= i \sum_{n = 1}^{+\infty} \frac{1}{\pi n} \left\{ \left( (-)^n \beta - \alpha \right) e^{i\frac{\pi n}{L} x} 
- \left( (-)^{-n} \beta - \alpha \right) e^{-i\frac{\pi n}{L} x} \right\} \nonumber\\
&= - \sum_{n = 1}^{+\infty} \frac{2}{\pi n} \left( (-)^n \beta - \alpha \right) \sin \left( \frac{\pi n}{L} x \right) .
\end{align}
Is it not silly to obtain a complicated infinite sum for a solution, when it is really a straight line? The answer is that, while the Green's function/mode sum method here does appear unnecessarily complicated, this mode expansion method is very general and is oftentimes the only known means of solving the problem analytically. % YZ: Should we recall Gibb's phenomenon here?
\begin{myP}
Solve the 2D flat space Poisson equation $-(\partial_x^2 + \partial_y^2) \psi(0 \leq x \leq L_1, 0 \leq y \leq L_2) = 0$, up to quadrature, with the following boundary conditions
\begin{align}
\psi(0,y) = \varphi_1(y), \qquad \psi(L_1,y) = \varphi_2(y), \qquad \psi(x,0) = \rho_1(x), \qquad \psi(x,L_2) = \rho_2(x) .
\end{align}
Write the solution as a mode sum, using the eigenfunctions
\begin{align}
\psi_{m,n}(x,y) \equiv \braket{x,y}{m,n} = \frac{2}{\sqrt{L_1 L_2}} \sin \left( \frac{\pi m}{L_1} x \right) \sin \left( \frac{\pi n}{L_2} y \right) .
\end{align}
Hint: your answer will involve 1D integrals on the 4 boundaries of the rectangle. \qed
\end{myP}

\subsection{Laplacians and their spectra}

Let us recall our discussions from both linear algebra and differential geometry. Given a (Euclidean signature) metric 
\begin{align}
\dd\ell^2 = g_{ij}(\vec{x}) \dd x^i \dd x^j,
\end{align}
the Laplacian acting on a scalar $\psi$ can be written as
\begin{align}
\label{Laplacian_OnScalar}
\vec{\nabla}^2 \psi \equiv \nabla_i \nabla^i \psi = \frac{\partial_i \left( \sqrt{|g|} g^{ij} \partial_j \psi \right)}{\sqrt{|g|}} ,
\end{align}
where $\sqrt{|g|}$ is the square root of the determinant of the metric.

\noindent{\bf Spectra} \qquad Now we turn to the primary goal of this section, to study the eigenvector/value problem
\begin{align}
\label{Laplacian_EigenProblem}
-\vec{\nabla}^2 \psi_\lambda(\vec{x}) = -\vec{\nabla}^2\braket{\vec{x}}{\lambda} = \lambda \braket{\vec{x}}{\lambda} .
\end{align}

\subsubsection{Infinite $\mathbb{R}^D$ in Cartesian coordinates}

In infinite flat Euclidean $D$-space $\mathbb{R}^D$, we have already seen that the plane waves $\{ \exp(i \vec{k} \cdot \vec{x}) \}$ are the eigenvectors of $-\vec{\nabla}^2$ with eigenvalues $\{ k^2 \vert - \infty < k < \infty \}$. This is a coordinate invariant statement, since the $\psi$ and Laplacian in eq. \eqref{Laplacian_EigenProblem} are coordinate scalars. Also notice that the eigenvalue/vector equation \eqref{Laplacian_EigenProblem} is a ``local" PDE in that it is possible to solve it only in the finite neighborhood of $\vec{x}$; it therefore requires appropriate boundary conditions to pin down the correct eigen-solutions. 

In Cartesian coordinates, moreover,
\begin{align}
\label{Laplacian_Spectrum_DCartesian}
\psi_{\vec{k}}(\vec{x}) = e^{i \vec{k} \cdot \vec{x}} = \prod_{j=1}^{D} e^{i k_j x^j} ,
\qquad \vec{k}^2 = \delta^{ij} k_i k_j = \sum_{i=1}^D (k_i)^2 \equiv \vec{k}^2 ,
\end{align}
with completeness relations (cf. \eqref{CompletenessRelation}) given by
\begin{align}
\int_{\mathbb{R}^D} \dd^D \vec{x} \left.\left\langle \vec{k} \right\vert \vec{x} \right\rangle \left\langle \vec{x} \left\vert \vec{k}' \right\rangle\right.
		&= (2\pi)^D \delta^{(D)}\left( \vec{k} - \vec{k}' \right), \\
\int_{\mathbb{R}^D} \frac{\dd^D \vec{k}}{(2\pi)^D} \left\langle \vec{x} \left\vert \vec{k} \right\rangle\right. \left.\left\langle \vec{k} \right\vert \vec{x}' \right\rangle
		&= \delta^{(D)}\left( \vec{x} - \vec{x}' \right) .
\end{align}
{\it Translation symmetry and degeneracy} \qquad For a fixed $1 \leq i \leq D$, notice the translation operator in the $i$th Cartesian direction, namely $-i \partial_j \equiv -i \partial/\partial x^j$ commutes with $-\vec{\nabla}^2$. The translation operators commute amongst themselves too. This is why one can simultaneously diagonalize the Laplacian, and all the $D$ translation operators.
\begin{align}
-i\partial_j \braket{\vec{x}}{k^2} = k_j \braket{\vec{x}}{k^2} 
\end{align}
In fact, we see that the eigenvector of the Laplacian $\ket{k^2}$ can be viewed as a tensor product of the eigenstates of $P_j$.
\begin{align}
\ket{k^2 = \vec{k}^2} &= \ket{k_1} \otimes \ket{k_2} \otimes \dots \otimes \ket{k_D} \\
\braket{\vec{x}}{k^2} 
		&= \left(\bra{x^1} \otimes \dots \otimes \bra{x^D}\right)\left(\ket{k_1} \otimes \dots \otimes \ket{k_D}\right) \nonumber\\
		&= \braket{x^1}{k_1} \braket{x^2}{k_2} \dots \braket{x^D}{k_D} = \prod_{j=1}^{D} e^{i k_j x^j} .
\end{align}
As we have already highlighted in the linear algebra of continuous spaces section, the spectrum of the negative Laplacian admits an infinite fold degeneracy here. Physically speaking we may associate it with the translation symmetry of $\mathbb{R}^D$.

\subsubsection{1 Dimension}

{\bf Infinite Flat Space} \qquad In one dimension, the metric\footnote{One dimensional space(time)s are always flat -- the Riemann tensor is identically zero.} is
\begin{align}
\dd\ell^2 = \dd z^2 ,
\end{align}
for $z \in \mathbb{R}$, and eq. \eqref{Laplacian_Spectrum_DCartesian} reduces to 
\begin{align}
\label{Laplacian_Spectrum_1D}
-\vec{\nabla}^2_1 \psi_k(z) = -\partial_z^2 \psi_k(z) = k^2 \psi_k(z), \qquad \braket{z}{k} \equiv \psi_k(z) = e^{ikz} ;
\end{align}
and their completeness relation (cf. eq. \eqref{CompletenessRelation}) is
\begin{align}
\int_{-\infty}^{\infty} \frac{\dd k}{2\pi} \braket{z}{k} \braket{k}{z'} = \int_{-\infty}^{\infty} \frac{\dd k}{2\pi} e^{ik(z-z')} = \delta(z-z') .
\end{align}
{\bf Periodic infinite space} \qquad If the 1D space obeys periodic boundary conditions, with period $L$, we have instead
\begin{align}
\label{Laplacian_Spectrum_1D_Periodic}
-\vec{\nabla}^2_1 \psi_m(z) = -\partial_z^2 \psi_m(z) = \left(\frac{2\pi m}{L}\right)^2 \psi_m(z), \nonumber\\ 
\braket{z}{m} \equiv \psi_m(z) = L^{-1/2} e^{i\frac{2\pi m}{L} z}, \qquad m = 0,\pm 1, \pm 2,\dots .
\end{align}
The orthonormal eigenvectors obey
\begin{align}
\int_0^L \dd z \braket{m}{z}\braket{z}{m'} = \delta^m_{m'}, \qquad \braket{z}{m} = L^{-1/2} e^{i\frac{2\pi m}{L} z} ;
\end{align}
while their completeness relation reads, for $0 \leq z,z' \leq L$,
\begin{align}
\label{Laplacian_Spectrum_1D_Completeness}
\sum_{m=-\infty}^{\infty} \braket{z}{m} \braket{m}{z'} = \frac{1}{L}\sum_{m=-\infty}^{\infty} e^{\frac{2\pi m}{L} i(z-z')} = \delta(z-z') .
\end{align}
{\bf Unit Circle} \qquad A periodic infinite space can be thought of as a circle, and vice versa. Simply identify $L \equiv 2\pi r$, where $r$ is the radius of the circle as embedded in 2D space. For concreteness we will consider a circle of radius $1$. Then we may write the metric as
\begin{align}
\dd\ell^2 = (\dd\phi)^2, \qquad \phi \in [0,2\pi) .
\end{align}
We may then bring over the results from the previous discussion.
\begin{align}
\label{Laplacian_Spectrum_1D_Circle}
-\vec{\nabla}^2_{\mathbb{S}^1} \psi_m(\phi) = -\partial_\phi^2 \psi_m(\phi) = m^2 \psi_m(\phi), \nonumber\\ 
\braket{\phi}{m} \equiv \psi_m(\phi) = (2\pi)^{-1/2} e^{i m \phi}, \qquad m = 0,\pm 1, \pm 2,\dots .
\end{align}
The orthonormal eigenvectors obey
\begin{align}
\int_0^{2\pi} \dd\phi \braket{m}{\phi} \braket{\phi}{m'} = \delta^m_{m'}, \qquad \braket{\phi}{m} = (2\pi)^{-1/2} e^{im\phi} . 
\end{align}
while their completeness relation reads, for $0 \leq z,z' \leq L$,
\begin{align}
\label{Laplacian_Spectrum_1D_Circle_Completeness}
\sum_{m=-\infty}^{\infty} \braket{\phi}{m} \braket{m}{\phi'} = \frac{1}{2\pi} \sum_{m=-\infty}^{\infty} e^{im(\phi-\phi')} = \delta(\phi-\phi') .
\end{align}
{\it Fourier series re-visited.} \qquad Note that $-i \partial_\phi$ can be thought of as the ``momentum operator" on the unit circle (in the position representation) with eigenvalues $\{m\}$ and corresponding eigenvectors $\{ \braket{\phi}{m} \}$. Namely, if we define
\begin{align}
\braOket{\phi}{P_\phi}{\psi} = -i \partial_\phi \braket{\phi}{\psi}
\end{align}
for any state $\ket{\psi}$, we shall see it is Hermitian. Given arbitrary states $\ket{\psi_{1,2}}$,
\begin{align}
\braOket{\psi_1}{P_\phi}{\psi_2} 
&= \int_{0}^{2\pi} \dd \phi \braket{\psi_1}{\phi}\left(-i\partial_\phi \braket{\phi}{\psi_2}\right) \\
&= \left[ -i\braket{\psi_1}{\phi} \braket{\phi}{\psi_2} \right]_{\phi=0}^{\phi=2\pi}
+ \int_{0}^{2\pi} \dd \phi \left(i\partial_\phi \braket{\psi_1}{\phi}\right) \braket{\phi}{\psi_2} . \nonumber
\end{align}
As long as we are dealing with the space of {\it continuous} functions $\psi_{1,2}(\phi)$ on a circle, the boundary terms must vanish because $\phi=0$ and $\phi=2\pi$ really refer to the same point. Therefore,
\begin{align}
\braOket{\psi_1}{P_\phi}{\psi_2} 
&= \int_{0}^{2\pi} \dd \phi \left(- i\partial_\phi \braket{\phi}{\psi_1} \right)^* \braket{\phi}{\psi_2} 
= \int_{0}^{2\pi} \dd \phi \overline{\braOket{\phi}{P_\phi}{\psi_1}} \braket{\phi}{\psi_2} \nonumber \\
&= \int_{0}^{2\pi} \dd \phi \braOket{\psi_1}{P_\phi^\dagger}{\phi} \braket{\phi}{\psi_2} = \braOket{\psi_1}{P_\phi^\dagger}{\psi_2} . 
\end{align}
We must therefore have
\begin{align}
\braOket{\phi}{e^{-i\theta P_\phi}}{\psi} = e^{-i\theta (-i\partial_\phi)} \braket{\phi}{\psi}
= e^{-\theta \partial_\phi} \braket{\phi}{\psi} = \braket{\phi - \theta}{\psi} .
\end{align}
Any function on a circle can be expanded in the eigenstates of $P_\phi$, which in turn can be expressed through its position representation.
\begin{align}
\ket{\psi} 
		= \sum_{\ell=-\infty}^{+\infty} \ket{m} \braket{m}{\psi} 
		&= \sum_{\ell=-\infty}^{+\infty} \int_{0}^{2\pi} \dd\phi \ket{\phi} \braket{\phi}{m} \braket{m}{\psi} 
		= \sum_{\ell=-\infty}^{+\infty} \int_{0}^{2\pi} \frac{\dd\phi}{\sqrt{2\pi}} \ket{\phi} \braket{m}{\psi} e^{im\phi} , \nonumber\\
\braket{m}{\psi} 
				&= \int_{0}^{2\pi} \dd\phi' \braket{m}{\phi'} \braket{\phi'}{\psi} 
				= \int_{0}^{2\pi} \frac{\dd\phi'}{\sqrt{2\pi}} e^{-i m \phi'} \psi(\phi') .
\end{align}
This is nothing but the Fourier series expansion of $\psi(\phi)$.

\subsubsection{2 Dimensions}

{\bf Flat Space, Cylindrical Coordinates} \qquad The 2D flat metric in cylindrical coordinates reads
\begin{align}
\dd\ell^2 = \dd r^2 + r^2 \dd\phi^2, \qquad r \geq 0, \qquad \phi\in [0,2\pi), \qquad\sqrt{|g|} = r .
\end{align}
The negative Laplacian is therefore
\begin{align}
\label{Laplacian_2D_Cylindrical}
-\vec{\nabla}^2_2 \varphi_k(r,\phi) 
&= -\frac{1}{r}\left( \partial_r \left(r \partial_r \varphi_k \right) + \frac{1}{r} \partial_\phi^2 \varphi_k \right) \\
&= -\left\{ \frac{1}{r} \partial_r \left(r \partial_r \varphi_k \right) + \frac{1}{r^2} \partial_\phi^2 \varphi_k \right\} .
\end{align}
Our goal here is to diagonalize the negative Laplacian in cylindrical coordinates, and re-write the plane wave using its eigenstates. In this case we will in fact tackle the latter and use the results to do the former. To begin, note that the plane wave in 2D cylindrical coordinates is
\begin{align}
\label{2DPlaneWave}
\langle \vec{x} \vert \vec{k} \rangle = \exp(i \vec{k} \cdot \vec{x}) = \exp(i kr \cos(\phi - \phi_k)), \qquad k \equiv |\vec{k}|, \ r \equiv |\vec{x}| ;
\end{align}
because the Cartesian components of $\vec{k}$ and $\vec{x}$ are
\begin{align}
k_i = k \left( \cos\phi_k, \sin\phi_k \right) \qquad
x^i = r \left( \cos\phi, \sin\phi \right) .
\end{align}
We observe that this is a periodic function of the angle $\Delta \phi \equiv \phi - \phi_k$ with period $L=2\pi$, which means it must admit a Fourier series expansion. Referring to equations \eqref{FourierSeries_I} and \eqref{FourierSeries_II},
\begin{align}
\langle \vec{x} \vert \vec{k} \rangle = \sum_{m=-\infty}^{+\infty} \chi_m(kr) \frac{e^{im(\phi-\phi_k)}}{\sqrt{2\pi}} .
\end{align}
and
{\allowdisplaybreaks\begin{align}
\chi_m(kr) 
&= \int_{0}^{2\pi} \frac{\dd\phi''}{\sqrt{2\pi}} e^{ikr \cos\phi''} e^{-im\phi''} \\
&= \sqrt{2\pi} \int_{\phi''=0}^{\phi''=2\pi} \frac{\dd(\phi''+\pi/2)}{2\pi} e^{ikr \cos(\phi''+\pi/2-\pi/2)} e^{-im(\phi''+\pi/2-\pi/2)} \nonumber\\
&= \sqrt{2\pi} \int_{\pi/2}^{5\pi/2} \frac{\dd\phi'}{2\pi} e^{ikr \sin\phi'} e^{-im\phi'} i^m 
= i^m \sqrt{2\pi} \int_{-\pi}^{+\pi} \frac{\dd\phi'}{2\pi} e^{ikr \sin\phi'} e^{-im\phi'} \nonumber
\end{align}}
(In the last line, we have used the fact that the integrand is itself a periodic function of $\phi'$ with period $2\pi$ to change the limits of integration.) As it turns out, the Bessel function $J_m$ admits an integral representation (cf. eq. (10.9.2) of the NIST page \href{http://dlmf.nist.gov/10.9}{here}.)
\begin{align}
\label{BesselJ_IntegralRep}
J_m(z) 		&= \int_{-\pi}^{\pi} \frac{\dd \phi'}{2\pi} e^{iz \sin\phi' - i m \phi'},  \qquad m \in \{0,\pm 1, \pm 2, \dots \} , \\
J_{-m}(z)	&= (-)^m J_m(z) .
\end{align}
As an aside, $J_\nu(z)$ also has a series representation
\begin{align}
J_\nu(z) = \left(\frac{z}{2}\right)^\nu \sum_{k=0}^{\infty} \frac{ (-)^k (z/2)^{2k} }{k! \Gamma(\nu+k+1)} ;
\end{align}
and the large argument asymptotic expansion
\begin{align}
J_{\pm \nu}(z \gg \nu) \sim \sqrt{\frac{2}{\pi z}} \cos\left( z \mp \frac{\pi}{2} \nu - \frac{\pi}{4} \right) .
\end{align}
We have arrived at the result
\begin{align}
\label{2DPlaneWave_CylindricalCoordinates}
\langle \vec{x} \vert \vec{k} \rangle 
&= \exp(i \vec{k} \cdot \vec{x}) = \exp(i kr \cos(\phi - \phi_k)), \qquad k \equiv |\vec{k}|, \ r \equiv |\vec{x}| \nonumber\\
&= \sum_{\ell=-\infty}^{\infty} i^\ell J_\ell(kr) e^{im(\phi-\phi_k)} .
\end{align}
Because the $\{ e^{im\phi} \}$ are basis vectors on the circle of fixed radius $r$, every term in the infinite sum is a linearly independent eigenvector of $-\vec{\nabla}_2^2$. That is, we can now read off the basis eigenvectors of the negative Laplacian in 2D cylindrical coordinates. To obtain orthonormal ones, however, let us calculate their normalization using the following orthogonality relation, written in cylindrical coordinates,
\begin{align}
(2\pi)^2 \frac{\delta(k-k') \delta(\phi_{k}-\phi_{k'})}{\sqrt{kk'}}  
&= \int_{\mathbb{R}^2} \dd^2 x \exp(i(\vec{k}-\vec{k}')\cdot\vec{x}) \\
&= \sum_{m,m'=-\infty}^{+\infty} \int_0^\infty \dd r \cdot r \int_{0}^{2\pi} \dd\phi \cdot i^{m}(-i)^{m'} J_m(kr) J_{m'}(k'r) e^{im(\phi-\phi_k)} e^{-im'(\phi-\phi_{k'})} \nonumber\\
&= (2\pi) \sum_{m=-\infty}^{+\infty} \int_0^\infty \dd r \cdot r J_m(kr) J_{m}(k'r) e^{im(\phi_{k'}-\phi_k)} . \nonumber
\end{align}
We now replace the $\delta(\phi-\phi_k)$ on the left hand side with the completeness relation in eq. \eqref{Laplacian_Spectrum_1D_Completeness}, where now $z = \phi_k$, $z'=\phi_{k'}$ and the period is $L = 2\pi$. Equating the result to the last line then brings us to
\begin{align}
\sum_{m=-\infty}^{+\infty} \frac{\delta(k-k')}{\sqrt{kk'}} e^{im(\phi_k - \phi_{k'})}
&= \sum_{m=-\infty}^{+\infty} \int_0^\infty \dd r \cdot r J_m(kr) J_{m}(k'r) e^{im(\phi_{k'}-\phi_k)} .
\end{align}
The coefficients of each (linearly independent) vector $e^{im(\phi_k - \phi_{k'})}$ on both sides should be the same. This yields the completeness relation of the radial mode functions:
\begin{align}
\label{BesselIntegrals}
\int_0^\infty \dd r \cdot r J_m(kr) J_{m}(k'r) &= \frac{\delta(k-k')}{\sqrt{kk'}}, \\
\int_0^\infty \dd k \cdot k J_m(kr) J_{m}(kr') &= \frac{\delta(r-r')}{\sqrt{rr'}}.
\end{align}
To summarize, we have found, in 2D infinite flat space, that the eigenvectors/values of the negative Laplacian in cylindrical coordinates $(r \geq 0,0\leq \phi < 2\pi)$ are
\begin{align}
\label{Laplacian_Spectrum_2D_Cylindrical}
-\vec{\nabla}_2^2 \braket{r,\phi}{k,m} = k^2 \braket{r,\phi}{k,m}, \qquad
\braket{r,\phi}{k,m} &\equiv J_m(kr) \frac{\exp\left(i m \phi\right)}{\sqrt{2\pi}}, \nonumber\\
m &= 0,\pm 1, \pm 2, \pm 3,\dots .
\end{align}
The eigenvectors are normalized as
\begin{align}
\int_0^\infty \dd r \cdot r \int_{0}^{2\pi} \dd\phi \braket{k,m}{r,\phi} \braket{r,\phi}{k',m'}
= \delta^{m}_{m'} \frac{\delta(k-k')}{\sqrt{kk'}} .
\end{align}
{\it Rotational symmetry and degeneracy} \qquad Note that $-i\partial_\phi$ is the translation operator in the azimuthal direction ($\equiv$ rotation operator), with eigenvalue $m$. The spectrum here is discretely and infinitely degenerate, which can be physically interpreted to be due to the presence of rotational symmetry.

{\it Bessel's equation} \qquad As a check of our analysis here, we may now directly evaluate the 2D negative Laplacian acting on the its eigenvector $\braket{r,\phi}{k,m}$, and see that we are lead to Bessel's equation. Starting from the eigenvector/value equation in \eqref{Laplacian_Spectrum_2D_Cylindrical}, followed by using the explicit expression in eq. \eqref{Laplacian_2D_Cylindrical} and the angular eigenvalue/vector equation $\partial_\phi^2 \exp(im\phi) = -m^2 \exp(im\phi)$, this hands us
\begin{align}
k^2 J_m(kr) &= -\left\{ \frac{1}{r} \partial_r \left(r \partial_r J_m(kr) \right) - \frac{m^2}{r^2} J_m(kr) \right\} .
\end{align}
Let us then re-scale $\rho \equiv k r$, where $k \equiv |\vec{k}|$, so that $\partial_r = k \partial_\rho$.
\begin{align}
\label{BesselEquation}
\rho^2 \cdot J''(\rho) + \rho \cdot J'(\rho) + (\rho^2 - m^2) J(\rho)  = 0
\end{align}
Equation 10.2.1 of the NIST page \href{http://dlmf.nist.gov/10.2}{here} tells us we have indeed arrived at Bessel's equation. Two linearly independent solutions are $J_m(kr)$ and $Y_m(kr)$. However, eq. (10.2.2) of the NIST page \href{http://dlmf.nist.gov/10.2}{here} and eq. (10.8.1) of the NIST page \href{http://dlmf.nist.gov/10.8}{here} tell us, for small argument, $Y_m(z \to 0)$ has at least a log singularity of the form $\ln(z/2)$ and for $m \neq 0$ has also a power law singularity that goes as $1/z^{|m|}$. Whereas, $J_m(z)$ is $(z/2)^{|m|}$ times a power series in the variable $(z/2)^2$, and is not only smooth for small $z$, the power series in fact has an infinite radius of convergence. It makes sense that our plane wave expansion only contains $J_m$ and not $Y_m$ because it is smooth for all $r$.

\begin{myP}
\qquad Explain how you would modify the analysis here, if we were not dealing with an infinite 2D space, but only a wedge of 2D space -- namely, $r \geq 0$ but $0 \leq \phi \leq \phi_0 < 2\pi$. How would you modify the analysis here, if $\phi \in [0,2\pi)$, but now $0 \leq r \leq r_0 < \infty$? You do not need to carry out the calculations in full, but try to be as detailed as you can. Assume Dirichlet boundary conditions. \qed
\end{myP}
\noindent{\bf 2-sphere $\mathbb{S}^2$, Separation-Of-Variables, and the Spherical Harmonics}\footnote{In these notes we focus solely on the spherical harmonics on $\mathbb{S}^2$; for spherical harmonics in arbitrary dimensions, see \href{https://arxiv.org/abs/1205.3548}{ arXiv:1205.3548}.} \qquad The $2$-sphere of radius $R$ can be viewed as a curved surface embedded in 3D flat space parametrized as
\begin{align}
\vec{x}(\vec{\xi}=(\theta,\phi)) = R \left( \sin\theta \ \cos\phi, \sin\theta \ \sin\phi, \cos\theta \right),
\qquad\qquad \vec{x}^2 = R^2 .
\end{align}
For concreteness we will consider the case where $R=1$. Its metric is therefore given by $H_\text{IJ} = \partial_\text{I} x^i \partial_\text{J} x^j \dd \xi^\text{I} \dd \xi^\text{J}$,
\begin{align}
H_\text{IJ} \dd \xi^\text{I} \dd \xi^\text{J} = \dd\theta^2 + (\sin\theta)^2 \dd\phi^2,
\qquad\qquad \sqrt{|H|} = \sin\theta .
\end{align}
(Or, simply take the 3D flat space metric in spherical coordinates, and set $\dd r \to 0$ and $r \to 1$.)

We wish to diagonalize the negative Laplacian on this unit radius $2-$sphere. The relevant eigenvector/value equation is
\begin{align}
-\vec{\nabla}_{\mathbb{S}^2}^2 Y(\theta,\phi) = \nu(\nu+1) Y(\theta,\phi) ,
\end{align}
where $\nu$ for now is some arbitrary positive number.

To do so, we now turn to the {\it separation of variables} technique, which is a method to reduce a PDE into a bunch of ODEs -- and hence more manageable. The main idea is, for highly symmetric problems such as the Laplacian in flat space(time)s or on the $D$-sphere, one postulates that a multi-variable eigenfunction factorizes into a product of functions, each depending only on one variable. If solutions can be found, then we are assured that such an ansatz works.

In the unit radius $2-$sphere case we postulate
\begin{align}
\label{2Sphere_SeparationOfVariables}
Y(\theta,\phi) = \Lambda(\theta) \Phi(\phi) .
\end{align}
First work out the Laplacian explicitly, with $s \equiv \sin\theta$,
\begin{align}
-\left\{ \frac{1}{s} \partial_\theta \left( s \partial_\theta Y \right) + \frac{1}{s^2} \partial_\phi^2 Y \right\} 
= -\left\{ \frac{1}{s} \partial_\theta \left( s \partial_\theta Y \right) + \frac{1}{s^2} \vec{\nabla}_{\mathbb{S}^1}^2 Y \right\} 
= \nu(\nu+1) Y(\theta,\phi) .
\end{align}
We have identified $\vec{\nabla}_{\mathbb{S}^1}^2 = \partial_\phi^2$ to be the Laplacian on the circle, from eq. \eqref{Laplacian_Spectrum_1D_Circle}. This suggests we should choose $\Phi$ to be the eigenvector of $\vec{\nabla}_{\mathbb{S}^1}^2$.
\begin{align}
\Phi(\phi) \propto \exp(im\phi), \qquad m = 0,\pm 1, \pm 2, \dots
\end{align}
Moreover, it will turn out to be very useful to change variables to $c \equiv \cos\theta$, which runs from $-1$ to $+1$ over the range $0 \leq \theta \leq \pi$. Since $s \equiv \sin\theta$ is strictly positive there, we have the positive root $s_\theta = (1-c^2)^{1/2}$ and $\partial_\theta = (\partial c/\partial\theta) \partial_c = -\sin\theta \partial_c = -(1-c^2)^{1/2} \partial_c$.
{\allowdisplaybreaks\begin{align*}
-\left\{ - \partial_c \left( -(1-c^2) \partial_c \Lambda \cdot \Phi \right) + \frac{1}{1-c^2} \Lambda \cdot \partial_\phi^2 \Phi \right\} 
	&= \nu(\nu+1) \Lambda \cdot \Phi \\
\partial_c \left( (1-c^2) \partial_c \Lambda \cdot \Phi \right) + \left(\nu(\nu+1) - \frac{m^2}{1-c^2}\right) \Lambda \cdot \Phi 
	&= 0 
\end{align*}}
Canceling the $\Phi$ from the equation, we now obtain an ODE for the $\Lambda$.
\begin{align}
\partial_c \left( (1-c^2) \partial_c \Lambda \right) + \left(\nu(\nu+1) - \frac{m^2}{1-c^2}\right) \Lambda &= 0 
\end{align}
This is solved -- see eq. 14.2.2 of the NIST page \href{http://dlmf.nist.gov/14.2}{here} -- by the two associated Legendre functions $P_\nu^m(c)$ and $Q_\nu^m(c)$. It turns out, to obtain a solution that does not blow up over the entire range $-1 \leq c \leq +1$, we need to choose $P_\nu^m(c)$, set $\nu \equiv \ell$ to be $0$ or a positive integer, and have $m$ run from $-\ell$ to $\ell$.
\begin{align}
\Lambda \propto P_\ell^m(\cos\theta), \qquad \ell\in\{0,1,2,3,\dots\}, \ m\in\{-\ell,-\ell+1,\dots.\ell-1,\ell\} .
\end{align}
Note that 
\begin{align}
\label{SphericalHarmonics_AzimuthalSymmetry}
P_\ell^0(x) = P_\ell(x),
\end{align}
where $P_\ell(x)$ is the $\ell$th Legendre polynomial. A common phase convention that yields an orthonormal basis set of functions on the $2-$sphere is the following definition for the {\it spherical harmonics}
\begin{align}
-\vec{\nabla}_{\mathbb{S}^2}^2 Y_\ell^m(\theta,\phi) &= \ell(\ell+1) Y_\ell^m(\theta,\phi) , \nonumber\\
\label{SphericalHarmonics}
\braket{\theta,\phi}{\ell,m}
&= Y_\ell^m(\theta,\phi) = \sqrt{\frac{2\ell+1}{4\pi} \frac{(\ell-m)!}{(\ell+m)!}} P_\ell^m(\cos\theta) e^{im\phi}, \nonumber\\
&\ell\in\{0,1,2,3,\dots\}, \ m\in\{-\ell,-\ell+1,\dots.\ell-1,\ell\} .
\end{align}
Spherical harmonics should be viewed as ``waves" on the $2-$sphere, with larger $\ell$ modes describing the higher frequency/shorter wavelength/finer features of the state/function on the sphere. Let us examine the spherical harmonics from $\ell=0,1,2,3$. The $\ell=0$ spherical harmonic is a constant.
\begin{align}
Y_0^0 	&= \frac{1}{\sqrt{4\pi}}
\end{align}
The $\ell=1$ spherical harmonics are:
\begin{align}
Y_1^{-1} 	= \frac{1}{2} \sqrt{\frac{3}{2 \pi }} e^{-i \phi } \sin (\theta ), \qquad
Y_1^0		= \frac{1}{2} \sqrt{\frac{3}{\pi }} \cos (\theta ), \qquad
Y_1^1		= -\frac{1}{2} \sqrt{\frac{3}{2 \pi }} e^{i \phi } \sin (\theta) .
\end{align}
The $\ell=2$ spherical harmonics are:
\begin{align}
Y_2^{-2} 	&= \frac{1}{4} \sqrt{\frac{15}{2 \pi }} e^{-2 i	\phi } \sin ^2(\theta ), \qquad 
Y_2^{-1}	= \frac{1}{2} \sqrt{\frac{15}{2 \pi }} e^{-i \phi } \sin(\theta ) \cos (\theta ), \qquad
Y_2^{0}		= \frac{1}{4} \sqrt{\frac{5}{\pi }} \left(3 \cos ^2(\theta)-1\right), \nonumber\\
Y_2^{1} 	&= -\frac{1}{2} \sqrt{\frac{15}{2 \pi }} e^{i \phi } \sin (\theta ) \cos (\theta), \qquad
Y_2^{2} 	= \frac{1}{4} \sqrt{\frac{15}{2 \pi }} e^{2 i \phi } \sin ^2(\theta ) .
\end{align}
The $\ell=3$ spherical harmonics are:
\begin{align}
Y_3^{-3} &= \frac{1}{8} \sqrt{\frac{35}{\pi }} e^{-3 i \phi } \sin ^3(\theta ), \qquad
Y_3^{-2} = \frac{1}{4} \sqrt{\frac{105}{2 \pi }} e^{-2 i \phi } \sin^2(\theta ) \cos (\theta ), \nonumber\\
Y_3^{-1} &= \frac{1}{8} \sqrt{\frac{21}{\pi }} e^{-i \phi } \sin (\theta) \left(5 \cos ^2(\theta )-1\right), \qquad 
Y_3^{0} = \frac{1}{4} \sqrt{\frac{7}{\pi }} \left(5 \cos ^3(\theta )-3 \cos (\theta )\right), \nonumber\\
Y_3^{1} &= -\frac{1}{8} \sqrt{\frac{21}{\pi }} e^{i \phi } \sin (\theta ) \left(5 \cos ^2(\theta )-1\right), \qquad
Y_3^{2} = \frac{1}{4} \sqrt{\frac{105}{2 \pi }} e^{2 i \phi } \sin^2(\theta ) \cos (\theta ), \nonumber\\
Y_3^{3} &= -\frac{1}{8} \sqrt{\frac{35}{\pi }} e^{3 i \phi } \sin^3(\theta) .
\end{align}
For later purposes, note that the $m=0$ case removes any dependence on the azimuthal angle $\phi$, and in fact returns the Legendre polynomial. 
\begin{align}
\label{SphericalHarmonics_m0}
\braket{\theta,\phi}{\ell,m=0} = Y_\ell^0(\theta,\phi) = \sqrt{\frac{2\ell+1}{4\pi}} P_\ell(\cos\theta) .
\end{align}
Orthonormality and completeness of the spherical harmonics read, respectively,
\begin{align}
\label{SphericalHarmonics_Orthonormality}
\braket{\ell',m'}{\ell,m}
	&= \int_{\mathbb{S}^2} \dd^2\vec{\xi} \sqrt{|H|} \ \overline{Y_{\ell'}^{m'}(\theta,\phi)} Y_\ell^m(\theta,\phi) \nonumber\\
	&= \int_{-1}^{+1} \dd(\cos\theta) \int_{0}^{2\pi} \dd\phi \overline{Y_{\ell'}^{m'}}(\theta,\phi) Y_\ell^m(\theta,\phi)
	= \delta^{\ell'}_{\ell} \delta^{m'}_m ,
\end{align}
and
\begin{align}
\label{SphericalHarmonics_Completeness}
\braket{\theta',\phi'}{\theta,\phi} 
	&= \frac{\delta(\theta'-\theta) \delta(\phi-\phi')}{\sqrt{\sin(\theta) \sin(\theta')}} 
	= \delta\left( \cos(\theta')-\cos(\theta) \right) \delta(\phi-\phi') \nonumber\\
	&= \sum_{\ell=0}^{\infty} \sum_{m=-\ell}^{\ell} \overline{Y_\ell^m(\theta',\phi')} Y_\ell^m(\theta,\phi) .
\end{align}
In 3D flat space, let us write the Cartesian components of the momentum vector $\vec{k}$ and the position vector $\vec{x}$ in spherical coordinates.
\begin{align}
k_i &= k \left( \sin\theta_k \cdot \cos\phi_k, 	\sin\theta_k \cdot \sin\phi_k, 	\cos\theta_k \right) \equiv k \widehat{k} \\
x^i &= r \left( \sin\theta \cdot \cos\phi, 		\sin\theta \cdot \sin\phi, 		\cos\theta \right) \equiv r \widehat{x}
\end{align}
{\it Addition formula} \qquad In terms of these variables we may write down a useful identity involving the spherical harmonics and the Legendre polynomial, usually known as the addition formula.
\begin{align}
\label{SphericalHarmonics_AdditionFormula}
P_\ell\left( \widehat{k}\cdot\widehat{x} \right)
= \frac{4\pi}{2\ell+1} \sum_{m=-\ell}^{+\ell} \overline{Y_\ell^m(\theta,\phi)} Y_\ell^m(\theta_k,\phi_k)
= \frac{4\pi}{2\ell+1} \sum_{m=-\ell}^{+\ell} Y_\ell^m(\theta,\phi) \overline{Y_\ell^m(\theta_k,\phi_k)} ,
\end{align}
where $\widehat{k} \equiv \vec{k}/k$ and $\widehat{x} \equiv \vec{x}/r$. The second equality follows from the first because the Legendre polynomial is real.

For a fixed direction $\widehat{k}$, note that $P_\ell( \widehat{k}\cdot\widehat{x})$ in eq. \eqref{SphericalHarmonics_AdditionFormula} is an eigenvector of the negative Laplacian on the $2-$sphere. For, as we have already noted, the eigenvalue equation $-\vec{\nabla}^2 \psi = \lambda \psi$ is a coordinate scalar. In particular, we may choose coordinates such that $\widehat{k}$ is pointing `North', so that $\widehat{k} \cdot \widehat{x} = \cos\theta$, where $\theta$ is the usual altitude angle. By recalling eq. \eqref{SphericalHarmonics_m0}, we see therefore,
\begin{align}
\label{SphericalHarmonics_kdotx}
-\vec{\nabla}_{\vec{x},\mathbb{S}^2}^2 P_\ell\left( \widehat{k}\cdot\widehat{x} \right) = \ell(\ell+1) P_\ell\left( \widehat{k}\cdot\widehat{x} \right) .
\end{align}
Since $P_\ell(\widehat{k}\cdot\widehat{x})$ is symmetric under the swap $k \leftrightarrow x$, it must also be an eigenvector of the Laplacian with respect to $\vec{k}$,
\begin{align}
-\vec{\nabla}_{\vec{k},\mathbb{S}^2}^2 P_\ell\left( \widehat{k}\cdot\widehat{x} \right) = \ell(\ell+1) P_\ell\left( \widehat{k}\cdot\widehat{x} \right) .
\end{align}
{\it Complex conjugation} \qquad Under complex conjugation, the spherical harmonics obey
\begin{align}
\overline{Y_\ell^m(\theta,\phi)} = (-)^m Y_\ell^{-m}(\theta,\phi) .
\end{align}
{\it Parity} \qquad Under a parity flip, meaning if you compare $Y_\ell^m$ evaluated at the point $(\theta,\phi)$ to the point on the opposite side of the sphere $(\pi-\theta,\phi+\pi)$, we have the relation
\begin{align}
Y_\ell^m(\pi-\theta,\phi+\pi) = (-)^\ell Y_\ell^m(\theta,\phi) .
\end{align}
The odd $\ell$ spherical harmonics are thus odd under parity; whereas the even $\ell$ ones are invariant (i.e., even) under parity.

\noindent{\it Poisson Equation on the $2$-sphere} \qquad Having acquired some familiarity of the spherical harmonics, we can now tackle Poisson's equation
\begin{align}
\label{PoissonEquation_2Sphere}
-\vec{\nabla}^2_{\mathbb{S}^2} \psi(\theta,\phi) = J(\theta,\phi)
\end{align}
on the $2-$sphere. Because the spherical harmonics are complete on the sphere, we may expand both $\psi$ and $J$ in terms of them.
\begin{align}
\psi = \sum_{\ell,m} A_\ell^m Y_\ell^m, \qquad J = \sum_{\ell,m} B_\ell^m Y_\ell^m .
\end{align}
(This means, if $J$ is a given function, then we may calculate $B_\ell^m = \int_{\mathbb{S}^2} \dd^2 \Omega \overline{Y_\ell^m(\theta,\phi)} J(\theta,\phi)$.) Inserting these expansions into eq. \eqref{PoissonEquation_2Sphere}, and recalling the eigenvalue equation $-\vec{\nabla}^2_{\mathbb{S}^2} Y_\ell^m = \ell(\ell+1) Y_\ell^m$,
\begin{align}
\sum_{\ell \neq 0,m} \ell(\ell+1) A_\ell^m Y_\ell^m = \sum_{\ell,m} B_\ell^m Y_\ell^m .
\end{align}
On the left hand side, because the eigenvalue of $Y_0^0$ is zero, there is no longer any $\ell=0$ term. Therefore, we see that for there to be a consistent solution, $J$ itself cannot contain a $\ell=0$ term. (This is intimately related to the fact that the sphere has no boundaries.\footnote{For, suppose there is a solution to $-\vec{\nabla}^2\psi = \chi/(4\pi)$, where $\chi$ is a constant. Let us now integrate both sides over the sphere's surface, and apply the Gauss/Stokes' theorem. On the left hand side we get zero because the sphere has no boundaries. On the right hand side we have $\chi$. This inconsistency means no such solution exist.}) At this point, we may then equate the $\ell > 0$ coefficients of the spherical harmonics on both sides, and deduce
\begin{align}
A_\ell^m = \frac{B_\ell^m}{\ell(\ell+1)}, \qquad \ell > 0.
\end{align}
To summarize, given a $J(\theta,\phi)$ that has no ``zero mode," such that it can be decomposed as
\begin{align}
J(\theta,\phi) = \sum_{\ell=1}^\infty \sum_{m=-\ell}^{\ell} B_\ell^m Y_\ell^m(\theta,\phi)
\quad \Leftrightarrow \quad 
B_\ell^m = \int_{-1}^{+1} \dd(\cos\theta) \int_{0}^{2\pi} \dd\phi \overline{Y_\ell^m(\theta,\phi)} J(\theta,\phi) ,
\end{align}
the solution to \eqref{PoissonEquation_2Sphere} is 
\begin{align}
\psi(\theta,\phi) = \sum_{\ell=1}^{\infty} \sum_{m=-\ell}^{+\ell} \frac{B_\ell^m}{\ell(\ell+1)} Y_\ell^m(\theta,\phi) .
\end{align}

\subsubsection{3 Dimensions}

{\bf Infinite Flat Space, Cylindrical Coordinates} \qquad We now turn to 3D flat space, written in cylindrical coordinates,
\begin{align}
\dd\ell^2 = \dd r^2 + r^2 \dd\phi^2 + \dd z^2, \qquad r\geq 0, \ \phi\in [0,2\pi), \ z \in\mathbb{R} , \qquad
\sqrt{|g|} = r .
\end{align}
Because the negative Laplacian on a scalar is the sum of the 1D and the 2D cylindrical case,
\begin{align}
-\vec{\nabla}^2_3 \psi = -\vec{\nabla}^2_2 \psi - \partial_z^2 \psi ,
\end{align}
we may try the separation-of-variables ansatz involving the product of the eigenvectors of the respective Laplacians.
\begin{align}
\psi(r,\phi,z) = \psi_2(r,\phi) \psi_1(z), \qquad 
\psi_2(r,\phi) \equiv J_m(kr) \frac{e^{im\phi}}{\sqrt{2\pi}}, \qquad 
\psi_1(z) \equiv e^{ik_z z} .
\end{align}
This yields
\begin{align}
-\vec{\nabla}^2 \psi = -\psi_1 \vec{\nabla}^2_{2} \psi_2 - \psi_2 \partial_z^2 \psi_1 = (k^2 + (k_z)^2) \psi ,
\end{align}
To sum, the orthonormal eigenfunctions are
\begin{align}
\braket{r,\phi,z}{k,m,k_z} &= J_m(kr) \frac{e^{im\phi}}{\sqrt{2\pi}} e^{ik_z z} \\
\int_{0}^{2\pi} \dd\phi \int_{0}^{\infty} \dd r r \int_{-\infty}^{+\infty} \dd z \braket{k',m',k'_z}{r,\phi,z} \braket{r,\phi,z}{k,m,k_z}
&= \delta^{m'}_m \frac{\delta(k-k')}{\sqrt{kk'}} \cdot (2\pi) \delta(k'_z - k_z) .
\end{align}
Since we already figured out the 2D plane wave expansion in cylindrical coordinates in eq. \eqref{2DPlaneWave_CylindricalCoordinates}, and since the 3D plane wave is simply the 2D one multiplied by the plane wave in the $z$ direction, i.e., $\exp(i\vec{k}\cdot\vec{x}) = \exp(ikr\cos(\phi-\phi_k)) \exp(ik_z z)$, we may write down the 3D expansion immediately
\begin{align}
\langle \vec{x} \vert \vec{k} \rangle
= \exp(i\vec{k}\cdot\vec{x}) 
= \sum_{\ell=-\infty}^{\infty} i^\ell J_\ell(kr) e^{im(\phi-\phi_k)} e^{ik_z z},
\end{align}
where
\begin{align}
k_i = \left( k \cos\phi_k, k \sin\phi_k, k_z \right), \qquad
x^i = \left( r \cos\phi, r \sin\phi, z \right) .
\end{align}
\noindent{\bf Infinite Flat Space, Spherical Coordinates} \qquad We now turn to 3D flat space written in spherical coordinates,
\begin{align}
\dd\ell^2 = \dd r^2 + r^2 \dd\Omega^2_{\mathbb{S}^2}, \qquad
\dd\Omega^2_{\mathbb{S}^2} \equiv \dd \theta^2 + (\sin\theta)^2 \dd\phi^2 ,  \nonumber\\
r\geq 0, \ \phi\in [0,2\pi), \ \theta\in[0,\pi] , \qquad \sqrt{|g|} = r^2 \sin\theta .
\end{align}
The Laplacian on a scalar is
\begin{align}
\label{Laplacian_3D_Spherical}
\vec{\nabla}^2 \psi = \frac{1}{r^2} \partial_r \left( r^2 \partial_r \psi \right) + \frac{1}{r^2} \vec{\nabla}_{\mathbb{S}^2}^2 \psi  .
\end{align}
where $\vec{\nabla}_{\mathbb{S}^2}^2$ is the Laplacian on a $2-$sphere.

\noindent{\it Plane wave} \qquad With
\begin{align}
k_i &= k \left( \sin(\theta_k) \cos(\phi_k), \sin(\theta_k) \sin(\phi_k), \cos(\theta_k) \right) \equiv k \widehat{k}, \\
x^i &= r \left( \sin(\theta) \cos(\phi), \sin(\theta) \sin(\phi), \cos(\theta) \right) \equiv r \widehat{x} ,
\end{align}
we have
\begin{align}
\langle \vec{x} \vert \vec{k} \rangle 
= \exp(i\vec{k} \cdot \vec{x}) = \exp\left( ikr \widehat{k} \cdot \widehat{x} \right) .
\end{align}
If we view $\widehat{k}$ as the $3-$direction, this means the plane wave has no dependence on the azimuthal angle describing rotation about the $3-$direction. This in turn indicates we should be able to expand $\langle \vec{x} \vert \vec{k} \rangle$ using $P_\ell(\widehat{k} \cdot \vec{x})$.
\begin{align}
\exp\left( ikr \widehat{k} \cdot \widehat{x} \right) = \sum_{\ell=0}^{\infty} \chi_\ell(kr) \sqrt{\frac{2\ell+1}{4\pi}} P_\ell\left( \widehat{k} \cdot \widehat{x} \right) .
\end{align}
For convenience we have used the $Y_\ell^0$ in eq. \eqref{SphericalHarmonics_m0}) as our basis. Exploiting the orthonormality of the spherical harmonics to solve for the expansion coefficients:
\begin{align}
\chi_\ell(kr) 
= 2\pi \int_{-1}^{+1} \dd c e^{ikr c} \overline{Y_\ell^0(\theta,\phi)} 
= \sqrt{(4\pi)(2\ell+1)} \frac{1}{2} \int_{-1}^{+1} \dd c e^{ikr c} P_\ell(c) .
\end{align}
(Even though the integral is over the entire solid angle, the azimuthal integral is trivial and yields $2\pi$ immediately.) At this point we may refer to eq. (10.54.2) of the NIST page \href{http://dlmf.nist.gov/10.54}{here} for the following integral representation of the spherical Bessel function of integer order,
\begin{align}
i^\ell j_\ell(z) = \frac{1}{2} \int_{-1}^{+1} \dd c e^{iz c} P_\ell(c), \qquad \ell = 0,1,2,\dots .
\end{align}
(The spherical Bessel function $j_\ell(z)$ is real when $z$ is positive.) We have arrived at
\begin{align}
\label{3DPlaneWave_Spherical}
\langle \vec{x} \vert \vec{k} \rangle = \exp(i\vec{k} \cdot \vec{x})
&= \sum_{\ell=0}^{\infty} (2\ell+1) i^\ell j_\ell(kr) P_\ell\left( \widehat{k} \cdot \widehat{x} \right) , \qquad\qquad k \equiv |\vec{k}| \\
&= 4\pi \sum_{\ell=0}^{\infty} i^\ell j_\ell(kr) \sum_{m=-\ell}^{+\ell} Y_\ell^m(\theta,\phi) \overline{Y_\ell^m(\theta_k,\phi_k)} ,
\end{align}
where, for the second equality, we have employed the additional formula in eq. \eqref{SphericalHarmonics_AdditionFormula}.

\noindent{\it Spectrum} \qquad Just as we did for the 2D plane wave, we may now read off the eigenfunctions of the 3D flat Laplacian in spherical coordinates. First we compute the normalization.
\begin{align}
\int_{\mathbb{R}^3} \dd^3 \vec{x} \exp(i(\vec{k}-\vec{k}')\cdot\vec{x}) 
		&= (2\pi)^3 \frac{\delta(k-k')}{kk'} \delta\left( \cos(\theta_{k'}) - \cos(\theta_k) \right) \delta\left( \phi_k - \phi_{k'} \right) 
\end{align}
Switching to spherical coordinates within the integral on the left-hand-side,
\begin{align}
&(4\pi)^2 \int_{\mathbb{S}^2} \dd^2 \Omega \int_{0}^{\infty} \dd r r^2 \sum_{\ell,\ell'=0}^{\infty} i^\ell (-i)^{\ell'} j_\ell(kr) j_{\ell'}(k'r) \nonumber\\
&\qquad\qquad\times
\sum_{m=-\ell}^{+\ell} \sum_{m'=-\ell'}^{+\ell'} Y_\ell^m(\theta,\phi) \overline{Y_\ell^m(\theta_k,\phi_k)} Y_{\ell'}^{m'}(\theta_k,\phi_k) \overline{Y_{\ell'}^{m'}(\theta,\phi)} \nonumber\\
&\qquad\qquad = (4\pi)^2 \int_{0}^{\infty} \dd r r^2 \sum_{\ell=0}^{\infty} j_\ell(kr) j_{\ell}(k'r) 
\sum_{m=-\ell}^{+\ell} Y_\ell^m(\theta,\phi) \overline{Y_{\ell}^{m}(\theta,\phi)} .
\end{align}
Comparing the right hand sides of the two preceding equations, and utilizing the completeness relation obeyed by the spherical harmonics,
\begin{align}
& 4 (2\pi)^2 \int_{0}^{\infty} \dd r r^2 \sum_{\ell=0}^{\infty} j_\ell(kr) j_{\ell}(k'r) 
\sum_{m=-\ell}^{+\ell} Y_\ell^m(\theta_k,\phi_k) \overline{Y_{\ell}^{m}(\theta_k,\phi_k)}  \nonumber\\
&\qquad \qquad = (2\pi)^3 \frac{\delta(k-k')}{kk'} \sum_{\ell=0}^{\infty} \sum_{m=-\ell}^{+\ell} Y_\ell^m(\theta_k,\phi_k) \overline{Y_{\ell}^{m}(\theta_k,\phi_k)} .
\end{align}
Therefore it must be that
\begin{align}
\int_{0}^{\infty} \dd r r^2 j_\ell(kr) j_{\ell}(k'r) = \frac{\pi}{2} \frac{\delta(k-k')}{kk'} .
\end{align}
Referring to eq. (10.47.3) of the NIST page \href{http://dlmf.nist.gov/10.47}{here},
\begin{align}
j_\ell(z) = \sqrt{\frac{\pi}{2z}} J_{\ell+\frac{1}{2}}(z) 
\end{align}
we see this is in fact the same result as in eq. \eqref{BesselIntegrals}. 

To sum, we have diagonalized the 3D flat space negative Laplacian in spherical coordinates as follows.
{\allowdisplaybreaks\begin{align}
-\vec{\nabla}^2 \braket{r,\theta,\phi}{k,\ell,m} 	&= k^2 \braket{r,\theta,\phi}{k,\ell,m}, \nonumber \\
\braket{r,\theta,\phi}{k,\ell,m} 					&= \sqrt{\frac{2}{\pi}} j_\ell(kr) Y_\ell^m(\theta,\phi) , \\
\braket{k',\ell',m'}{k,\ell,m} 
	&= \int_{\mathbb{S}^2} \dd^2 \Omega \int_0^\infty \dd r r^2 \braket{k',\ell',m'}{r,\theta,\phi} \braket{r,\theta,\phi}{k,\ell,m} , \nonumber\\
	&= \frac{\delta(k-k')}{kk'} \delta^{\ell'}_\ell \delta^{m'}_m . \nonumber
\end{align}}
\begin{myP}
{\bf Prolate Ellipsoidal Coordinates in 3D Flat Space} \qquad 3D Euclidean space can be foliated by prolate ellipsoids in the following way. Let $\vec{x} \equiv (x^1,x^2,x^3)$ be Cartesian coordinates; $\rho$ be the size of a given prolate ellipsoid; and the angular coordinates $(0 \leq \theta \leq \pi,0 \leq \phi < 2\pi)$ specify a point on its 2D surface. Then,
\begin{align}
\vec{x} &= \left( \sqrt{\rho^2-R^2} \sin\theta \cos\phi, \sqrt{\rho^2-R^2} \sin\theta \sin\phi, \rho \cos\theta \right) ; \\
\rho &\geq R, \qquad (\theta,\phi) \in \mathbb{S}^2 .
\end{align}
Explain the geometric meaning of the constant $R$. Work out the 3D flat metric in prolate ellipsoidal coordinates $(\rho,\theta,\phi)$ and proceed to diagonalize the associated scalar Laplacian $\vec{\nabla}^2 \equiv g^{ij} \nabla_i \nabla_j$. Hint: The spherical harmonics $\{ Y_\ell^m(\theta,\phi) \}$ will turn out to still be very useful here.
\end{myP}

\subsection{Heat/Diffusion Equation}

\subsubsection{Definition, uniqueness of solutions}

We will define the heat or diffusion equation to be the PDE
\begin{align}
\label{HeatDiffusionEquation}
\partial_t \psi\left(t,\vec{x}\right) 
= \sigma \vec{\nabla}^2_{\vec{x}} \psi\left(t,\vec{x}\right) 
= \frac{\sigma}{\sqrt{|g|}} \partial_i \left( \sqrt{|g|} g^{ij} \partial_j \psi \right), \qquad \sigma > 0 ,
\end{align}
where $\vec{\nabla}^2_{\vec{x}}$ is the Laplacian with respect to some metric $g_{ij}(\vec{x})$, which we will assume {\it does not} depend on the time $t$. We will also assume the $\psi(t,\vec{x})$ is specified on the boundary of the domain described by $g_{ij}(\vec{x})$, i.e., it obeys Dirichlet boundary conditions.

The diffusion constant $\sigma$ has dimensions of length if $\vec{\nabla}^2$ is of dimensions 1/[Length$^2$]. We may set $\sigma=1$ and thereby describe all other lengths in the problem in units of $\sigma$. As the heat equation, this PDE describes the temperature distribution as a function of space and time. As the diffusion equation in flat space, it describes the probability density of finding a point particle undergoing (random) Brownian motion. As we shall witness, the solution of eq. \eqref{HeatDiffusionEquation} is aided by the knowledge of the eigenfunctions/values of the Laplacian in question.
\begin{quotation}
\noindent{\bf Uniqueness of solution} \qquad Suppose the following initial conditions are given
\begin{align}
\psi(t=t_0,\vec{x}) = \varphi_0(\vec{x}), 
\end{align}
and suppose the field $\psi$ or its normal derivative is specified on the boundaries $\partial \mathfrak{D}$,
\begin{align}
\psi(t,\vec{x}\in\partial\mathfrak{D}) 								&= \varphi_3(\partial\mathfrak{D}), \qquad \text{(Dirichlet)}, \\
\text{or     } n^i \nabla_i \psi(t,\vec{x}\in\partial\mathfrak{D}) 	&= \varphi_4(\partial\mathfrak{D}), \qquad \text{(Neumann)} ,
\end{align}
where $n^i(\partial \mathfrak{D})$ is the unit outward normal vector. Then, the solution to the heat/diffusion equation in eq. \eqref{HeatDiffusionEquation} is unique.
\end{quotation}
{\it Proof} \qquad Without loss of generality, since our heat/diffusion equation is linear, we may assume the field is real. We then suppose there are two such solutions $\psi_1$ and $\psi_2$; the proof is established if we can show, in fact, that $\psi_1$ has to be equal to $\psi_2$. Note that the difference $\Psi \equiv \psi_1 - \psi_2$ is subject to the initial conditions
\begin{align}
\Psi(t=t_0,\vec{x}) = 0 ,
\end{align} 
and the spatial boundary conditions
\begin{align}
\Psi(t,\vec{x}\in\partial\mathfrak{D}) = 0 \qquad \text{  or  } \qquad n^i \nabla_i \Psi(t,\vec{x}\in\partial\mathfrak{D}) = 0 .
\end{align} 
Let us then consider the following (non-negative) integral
\begin{align}
\label{HeatDiffusionEquation_Uniqueness_I}
\rho(t) \equiv \frac{1}{2} \int_{\mathfrak{D}} \dd^D\vec{x} \sqrt{|g(\vec{x})|} \Psi(t,\vec{x})^2 \geq 0 ,
\end{align}
as well as its time derivative
\begin{align}
\partial_t \rho(t) &= \int_{\mathfrak{D}} \dd^D\vec{x} \sqrt{|g(\vec{x})|} \Psi \dot{\Psi} .
\end{align}
We may use the heat/diffusion equation on the $\dot{\Psi}$ term, and integrate-by-parts one of the gradients on the second term,
\begin{align}
\partial_t \rho(t) 
&= \int_{\mathfrak{D}} \dd^D\vec{x} \sqrt{|g(\vec{x})|} \Psi \vec{\nabla}^2\Psi \nonumber\\
&= \int_{\partial\mathfrak{D}} \dd^{D-1}\vec{\xi} \sqrt{|H(\vec{\xi})|} \Psi n^i \nabla_i \Psi
- \int_{\mathfrak{D}} \dd^D\vec{x} \sqrt{|g(\vec{x})|} \nabla_i \Psi \nabla^i \Psi .
\end{align}
By assumption either $\Psi$ or $n^i \nabla_i \Psi$ is zero on the spatial boundary; therefore the first term on the second line is zero. We have previously argued that the integrand in the second term on the second line is strictly non-negative
\begin{align}
\nabla_i \Psi \nabla^i \Psi = \sum_i (\nabla_{\widehat{i}} \Psi)^2 \geq 0 .
\end{align}
This implies 
\begin{align}
\partial_t \rho(t) &= - \int_{\mathfrak{D}} \dd^D\vec{x} \sqrt{|g(\vec{x})|} \nabla_i \Psi \nabla^i \Psi \leq 0 .
\end{align}
However, the initial conditions $\Psi(t=t_0,\vec{x})=0$ indicate $\rho(t=t_0)=0$ (cf. eq. \eqref{HeatDiffusionEquation_Uniqueness_I}). Moreover, since $\rho(t \geq t_0)$ has to be non-negative from its very definition and since we have just shown its time derivative is non-positive, $\rho(t \geq t_0)$ therefore has to remain zero for all subsequent time $t \geq t_0$; i.e., it cannot decrease below zero. And because $\rho(t)$ is the integral of the square of $\Psi$, the only way it can be zero is $\Psi=0 \Rightarrow \psi_1 = \psi_2$. This establishes the theorem. \qed

\subsubsection{Heat Kernel, Solutions with $\psi( \partial \mathfrak{D} ) = 0$}

In this section we introduce the propagator, otherwise known as the {\it heat kernel}, which will prove to be key to solving the heat/diffusion equation. It is the matrix element
\begin{align}
\label{HeatEquation_Propagator}
K(\vec{x},\vec{x}';s \geq 0) \equiv \braOket{\vec{x}}{e^{s \vec{\nabla}^2}}{\vec{x}'} .
\end{align}
It obeys the heat/diffusion equation
\begin{align}
\label{HeatEquation_PropagatorEqn}
\partial_s K(\vec{x},\vec{x}';s) &= \braOket{\vec{x}}{ \vec{\nabla}^2 e^{s \vec{\nabla}^2}}{\vec{x}'} 
= \braOket{\vec{x}}{ e^{s \vec{\nabla}^2} \vec{\nabla}^2 }{\vec{x}'} \nonumber\\
&= \vec{\nabla}^2_{\vec{x}} K(\vec{x},\vec{x}';s) = \vec{\nabla}^2_{\vec{x}'} K(\vec{x},\vec{x}';s) ,
\end{align}
where we have assumed $\vec{\nabla}^2$ is Hermitian. $K$ also obeys the initial condition
\begin{align}
\label{HeatEquation_PropagatorInitialCondition}
K(\vec{x},\vec{x}';s=0) = \braket{\vec{x}}{\vec{x}'} = \frac{\delta^{(D)}(\vec{x}-\vec{x}')}{\sqrt[4]{g(\vec{x}) g(\vec{x}')}} . 
\end{align}
If we demand the eigenfunctions of $\vec{\nabla}^2$ obey Dirichlet boundary conditions,
\begin{align}
\left\{ \psi_\lambda(\partial \mathfrak{D}) = 0 \left\vert -\vec{\nabla}^2 \psi_\lambda = \lambda \psi_\lambda \right\}\right. ,
\end{align}
then the heat kernel obeys the same boundary conditions.
\begin{align}
\label{HeatKernel_VanishesOnBoundary}
K(\vec{x} \in \partial\mathfrak{D},\vec{x}';s) = K(\vec{x},\vec{x}' \in \partial\mathfrak{D} ;s) = 0 .
\end{align}
To see this we need to perform a {\it mode expansion}. By inserting in eq. \eqref{HeatEquation_PropagatorInitialCondition} a complete set of the eigenstates of $\vec{\nabla}^2$, the heat kernel has an explicit solution
\begin{align}
\label{HeatKernel_ModeExpansion}
K(\vec{x},\vec{x}';s \geq 0) 
= \braOket{\vec{x}}{e^{s\vec{\nabla}^2}}{\vec{x}'}
= \sum_{\lambda} e^{-s\lambda} \braket{\vec{x}}{\lambda} \braket{\lambda}{\vec{x}'} ,
\end{align}
where the sum is schematic: depending on the setup at hand, it can consist of either a sum over discrete eigenvalues and/or an integral over a continuum. In this form, it is manifest the heat kernel vanishes when either $\vec{x}$ or $\vec{x}'$ lies on the boundary $\partial \mathfrak{D}$.

{\bf Initial value problem} \qquad In this section we will focus on solving the initial value problem when the field is itself is zero on the boundary $\partial \mathfrak{D}$ for all relevant times. This will in fact be the case for infinite domains; for example, flat $\mathbb{R}^D$, whose heat kernel we will work out explicitly below. The setup is thus as follows:
\begin{align}
\psi(t=t',\vec{x}) \equiv \braket{\vec{x}}{\psi(t')} \ \text{(given)}, \qquad \psi(t \geq t',\vec{x} \in \mathfrak{D}) = 0 .
\end{align}
Then $\psi(t,\vec{x})$ at any later time $t>t'$ is given by
\begin{align}
\label{HeatDiffusionEquation_GeneralSolution}
\psi(t \geq t',\vec{x}) 
= \braOket{\vec{x}}{e^{(t-t') \vec{\nabla}^2}}{\psi(t')} 
&= \int \dd^D \vec{x}' \sqrt{|g(\vec{x}')|} \braOket{\vec{x}}{e^{(t-t') \vec{\nabla}^2}}{\vec{x}'} \braket{\vec{x}'}{\psi(t')} \nonumber\\
&= \int \dd^D \vec{x}' \sqrt{|g(\vec{x}')|} K(\vec{x},\vec{x}';t-t') \psi(t',\vec{x}') .
\end{align}
That this is the correct solution is because the right hand side obeys the heat/diffusion equation through eq. \eqref{HeatEquation_PropagatorEqn}. As $t \to t'$, we also see from eq. \eqref{HeatEquation_PropagatorInitialCondition} that the initial condition is recovered.
\begin{align}
\psi(t=t',\vec{x}) 
= \braket{\vec{x}}{\psi(t')} 
= \int \dd^D \vec{x}' \sqrt{|g(\vec{x}')|} \frac{\delta^{(D)}(\vec{x}-\vec{x}')}{\sqrt[4]{|g(\vec{x}') g(\vec{x})|}} \psi(t',\vec{x}') = \psi(t',\vec{x}) .
\end{align}
Moreover, since the heat kernel obeys eq. \eqref{HeatKernel_VanishesOnBoundary}, the solution automatically maintains the $\psi(t \geq t',\vec{x} \in \mathfrak{D}) = 0$ boundary condition.

\noindent{\bf Decay times, Asymptotics} \qquad Suppose we begin with some temperature distribution $T(t',\vec{x})$. By expanding it in the eigenfunctions of the Laplacian, let us observe that it is the component along the eigenfunction with the small eigenvalue that dominates the late time temperature distribution. From eq. \eqref{HeatDiffusionEquation_GeneralSolution} and \eqref{HeatKernel_ModeExpansion},
\begin{align}
T(t \geq t',\vec{x}) 
&= \sum_{\lambda} \int \dd^D \vec{x}' \sqrt{|g(\vec{x}')|} \braOket{\vec{x}}{e^{(t-t') \vec{\nabla}^2}}{\lambda} \braket{\lambda}{\vec{x}'} \braket{\vec{x}'}{T(t')} \nonumber\\
&= \sum_{\lambda} e^{-(t-t') \lambda} \braket{\vec{x}}{\lambda} \int \dd^D \vec{x}' \sqrt{|g(\vec{x}')|} \braket{\lambda}{\vec{x}'} \braket{\vec{x}'}{T(t')} \nonumber\\
&= \sum_{\lambda} e^{-(t-t') \lambda} \braket{\vec{x}}{\lambda} \braket{\lambda}{T(t')} .
\end{align}
Remember we have proven that the eigenvalues of the Laplacian are strictly non-positive. That means, as $(t-t') \to \infty$, the dominant temperature distribution is
\begin{align}
\label{HeatDiffusionEquation_Dominant}
T(t-t' \to \infty,\vec{x}) 
&\approx e^{-(t-t') \lambda_\text{min}} \braket{\vec{x}}{\lambda_\text{min}} \int \dd^D \vec{x}' \sqrt{|g(\vec{x}')|} \braket{\lambda_\text{min}}{\vec{x}'} \braket{\vec{x}'}{T(t')}, 
\end{align}
because all the $\lambda > \lambda_\text{min}$ become exponentially suppressed (relative to the $\lambda_\text{min}$ state) due to the presence of $e^{-(t-t') \lambda}$. As long as the minimum eigenvalue $\lambda_\text{min}$ is strictly positive, we see the final temperature is zero. 
\begin{align}
T(t-t' \to \infty,\vec{x}) = 0, \qquad \text{ if }\lambda_\text{min} > 0.
\end{align}
When the minimum eigenvalue is zero, we have
\begin{align}
T(t-t' \to \infty,\vec{x}) 
&\to \braket{\vec{x}}{\lambda=0} \int \dd^D \vec{x}' \sqrt{|g(\vec{x}')|} \braket{\lambda=0}{\vec{x}'} \braket{\vec{x}'}{T(t')}, 
\qquad \text{ if }\lambda_\text{min} = 0 .
\end{align}
The exception to the dominant behavior in eq. \eqref{HeatDiffusionEquation_Dominant} is when there is zero overlap between the initial distribution and that eigenfunction with the smallest eigenvalue, i.e., if
\begin{align}
\int \dd^D \vec{x}' \sqrt{|g(\vec{x}')|} \braket{\lambda_\text{min}}{\vec{x}'} \braket{\vec{x}'}{T(t')} = 0 .
\end{align}
Generically, we may say that, with the passage of time, the component of the initial distribution along the eigenfunction corresponding to the eigenvalue $\lambda$ decays as $1/\lambda$; i.e., when $t-t' = 1/\lambda$, its amplitude falls by $1/e$.

{\it Static limit} \qquad Another way of phrasing the $(t-t') \to \infty$ behavior is that -- since every term in the sum-over-eigenvalues that depends on time decays exponentially, it must be that the late time asymptotic limit is simply the static limit, when the time derivative on the left hand side of eq. \eqref{HeatDiffusionEquation} is zero and we obtain Laplace's equation
\begin{align}
0 = \vec{\nabla}^2 \psi(t \to \infty, \vec{x}) .
\end{align}
{\bf Probability interpretation in flat infinite space} \qquad In the context of the diffusion equation in flat space, because of the $\delta$-functions on the right hand side of eq. \eqref{HeatEquation_PropagatorInitialCondition}, the propagator $K(\vec{x},\vec{x}';t-t')$ itself can be viewed as the probability density ($\equiv$ probability per volume) of finding the Brownian particle -- which was infinitely localized at $\vec{x}'$ at the initial time $t'$ -- at a given location $\vec{x}$ some later time $t>t'$. To support this probability interpretation it has to be that
\begin{align}
\int_{\mathbb{R}^D} \dd^D\vec{x} K(\vec{x},\vec{x}';t-t') = 1.
\end{align}
The integral on the left hand side corresponds to summing the probability of finding the Brownian particle over all space -- that has to be unity, since the particle has to be {\it somewhere}. We can verify this directly, by inserting a complete set of states.
{\allowdisplaybreaks\begin{align}
\int_{\mathbb{R}^D} \dd^D\vec{x} \braOket{\vec{x}}{ e^{ (t-t') \vec{\nabla}^2} }{\vec{x}'}
&= \int_{\mathbb{R}^D} \dd^D \vec{k} \int_{\mathbb{R}^D} \dd^D\vec{x} \braOket{\vec{x}}{e^{(t-t') \vec{\nabla}^2}}{\vec{k}} \braket{\vec{k}}{\vec{x}'} \nonumber\\
&= \int_{\mathbb{R}^D} \dd^D \vec{k} \int_{\mathbb{R}^D} \dd^D\vec{x} e^{-(t-t') \vec{k}^2} \langle \vec{x} \vert \vec{k} \rangle \braket{\vec{k}}{\vec{x}'} \nonumber\\ 
&= \int_{\mathbb{R}^D} \dd^D \vec{k} \int_{\mathbb{R}^D} \dd^D\vec{x} e^{-(t-t') \vec{k}^2} \frac{e^{i\vec{k}\cdot(\vec{x}-\vec{x}')}}{(2\pi)^D} \nonumber\\
&= \int_{\mathbb{R}^D} \dd^D \vec{k} e^{-(t-t') \vec{k}^2} e^{-i\vec{k}\cdot \vec{x}'} \delta^{(D)}(\vec{k}) = 1 . 
\end{align}}
{\bf Heat Kernel in flat space} \qquad In fact, the same technique allow us to obtain the heat kernel in flat $\mathbb{R}^D$.
\begin{align}
\braOket{\vec{x}}{ e^{ (t-t') \vec{\nabla}^2} }{\vec{x}'}
&= \int_{\mathbb{R}^D} \dd^D \vec{k} \braOket{\vec{x}}{e^{(t-t') \vec{\nabla}^2}}{\vec{k}} \braket{\vec{k}}{\vec{x}'} \\
&= \int_{\mathbb{R}^D} \frac{\dd^D \vec{k}}{(2\pi)^D} e^{-(t-t') \vec{k}^2} e^{i\vec{k}\cdot(\vec{x}-\vec{x}')}
= \prod_{j=1}^D \int_{-\infty}^{+\infty} \frac{\dd k_j}{2\pi} e^{-(t-t') (k_j)^2} e^{i k_j (x^j - x'^j)} . \nonumber
\end{align}
We may ``complete the square" in the exponent by considering
\begin{align}
-(t-t') \left(k_j - i \frac{x^j - x'^j}{2(t-t')} \right)^2 
= -(t-t') \left( (k_j)^2 - i k_j \frac{x^j - x'^j}{t-t'} - \left(\frac{x^j - x'^j}{2(t-t')}\right)^2 \right) .
\end{align}
The heat kernel in flat $\mathbb{R}^D$ is therefore
\begin{align}
\label{HeatDiffusionEquation_HeatKernel_FlatSpace}
\braOket{\vec{x}}{ e^{ (t-t') \sigma \vec{\nabla}^2} }{\vec{x}'} = \left( 4 \pi \sigma (t-t') \right)^{-D/2} \exp\left(-\frac{(\vec{x}-\vec{x}')^2}{4 \sigma (t-t')}\right) ,
\qquad t > t' ,
\end{align}
where we have put back the diffusion constant $\sigma$. If you have taken quantum mechanics, you may recognize this result to be very similar to the path integral $\,_\text{H}\braket{\vec{x},t}{\vec{x}',t'}_\text{H}$ of a free particle.
%\begin{myP}
%\qquad Suppose $T(t,\vec{x})$ obeys the heat equation $\partial_t T(t,\vec{x}) = \vec{\nabla}^2 T(t,\vec{x})$. How is the long time asymptotic behavior, the solution $T(t\to+\infty,\vec{x})$ related to that of Laplace's equation $\vec{\nabla}^2 T = 0$?

%Now consider a 1D finite box, where $x \in [0,L]$. Let walls of the box be always held at temperatures
%\begin{align}
%T(x=0) = T_1 \qquad \text{   and   } \qquad T(x=L)=T_2 .
%\end{align}
%Suppose the initial temperature distribution is
%\begin{align}
%T(t=0,x) = T_1 + \frac{x}{L} (T_2-T_1) + A \left( x - L \right) x, \qquad A > T_{1,2} > 0.
%\end{align}
%\begin{enumerate}
%\item Check that $T(t=0,x)$ does satisfy the boundary conditions.
%\item Find the propagator $K$ for the heat/diffusion equation in terms of a mode expansion, and use it to solve for $T(t>0,x)$.
%\item What is the final state $T(t\to+\infty,0 \leq x \leq L)$ of the system? Can you guess the solution without doing (much) work? And, can you compare this guess with your calculation of $T(t>0,x)$?
%\end{enumerate}
%\end{myP}
\subsubsection{Green's functions and initial value formulation in a finite domain}

{\bf Green's function from Heat Kernel} \qquad Given the heat kernel defined with Dirichlet boundary conditions, the associated Green's function is defined as
\begin{align}
G(t-t';\vec{x},\vec{x}') \equiv \Theta(t-t') K(\vec{x},\vec{x}';t-t') ,
\end{align}
where we define $\Theta(s)=1$ for $s \geq 0$ and $\Theta(s)=0$ for $s < 0$. This Green's function $G$ obeys
\begin{align}
\label{HeatDiffusionEquation_GreensFunction_PDE}
\left( \partial_t - \vec{\nabla}_{\vec{x}}^2 \right) G(t-t';\vec{x},\vec{x}') 
= \left( \partial_t - \vec{\nabla}_{\vec{x}'}^2 \right) G(t-t';\vec{x},\vec{x}') 
= \delta(t-t') \frac{\delta^{(D)}(\vec{x}-\vec{x}')}{\sqrt[4]{g(\vec{x}) g(\vec{x}')}} ,
\end{align}
the boundary condition
\begin{align}
\label{HeatDiffusionEquation_GreensFunction_DirichletBC}
G(\tau;\vec{x}\in\partial\mathfrak{D},\vec{x}') = G(\tau;\vec{x},\vec{x}'\in\partial\mathfrak{D}) = 0 ,
\end{align}
as well as the causality condition
\begin{align}
\label{HeatDiffusionEquation_GreensFunction_BC}
G(\tau;\vec{x},\vec{x}') = 0 \qquad \text{ when } \qquad \tau < 0 .
\end{align}
The boundary condition in eq. \eqref{HeatDiffusionEquation_GreensFunction_DirichletBC} follows directly from eq. \eqref{HeatKernel_VanishesOnBoundary}; whereas eq. \eqref{HeatDiffusionEquation_GreensFunction_PDE} follow from a direct calculation
\begin{align}
\left( \partial_t - \vec{\nabla}^2 \right) G(t-t';\vec{x},\vec{x}') 
&= \delta(t-t') K(\vec{x},\vec{x}';t-t') 
+ \Theta(t-t') \left( \partial_t - \vec{\nabla}^2 \right) K(\vec{x},\vec{x}';t-t') \nonumber\\
&= \delta(t-t') \frac{\delta^{(D)}(\vec{x}-\vec{x}')}{\sqrt[4]{g(\vec{x}) g(\vec{x}')}} .
\end{align}
\begin{quotation}
{\bf Initial value problem} \qquad Within a spatial domain $\mathfrak{D}$, suppose the initial field configuration $\psi(t',\vec{x}\in\mathfrak{D})$ is given and suppose its value on the spatial boundary $\partial\mathfrak{D}$ is also provided (i.e., Dirichlet B.C.'s $\psi(t \geq t',\vec{x} \in \partial\mathfrak{D})$ are specified). The unique solution $\psi(t \geq t',\vec{x}\in\mathfrak{D})$ to the heat/diffusion equation \eqref{HeatDiffusionEquation} is
\begin{align}
\label{HeatDiffusionEquation_SolutionG}
\psi(t \geq t',\vec{x}) 
&= \int_{\mathfrak{D}} \dd^D\vec{x}' \sqrt{|g(\vec{x}')|} G(t-t';\vec{x},\vec{x}') \psi(t',\vec{x}') \\
&- \int_{t'}^{t} \dd t'' \int_{\partial\mathfrak{D}} \dd^{D-1}\vec{\xi} \sqrt{|H(\vec{\xi})|} n^{i'} \nabla_{i'} G\left( t-t'';\vec{x},\vec{x}'(\vec{\xi}) \right) \psi\left( t'',\vec{x}'(\vec{\xi}) \right) , \nonumber
\end{align}
where the Green's function $G$ obeys the PDE in eq. \eqref{HeatDiffusionEquation_GreensFunction_PDE} and the boundary conditions in equations \eqref{HeatDiffusionEquation_GreensFunction_DirichletBC} and \eqref{HeatDiffusionEquation_GreensFunction_BC}.
\end{quotation}
{\it Derivation of eq. \eqref{HeatDiffusionEquation_SolutionG}} \qquad We begin by multiplying both sides of eq. \eqref{HeatDiffusionEquation_GreensFunction_PDE} by $\psi(t'',\vec{x}')$ and integrating over both space and time (from $t'$ to infinity).
{\allowdisplaybreaks\begin{align}
	\psi(t \geq t',\vec{x}) 
	&= \int_{t'}^{\infty} \dd t'' \int_{\mathfrak{D}} \dd^D\vec{x}' \sqrt{|g(\vec{x}')|} 
	\left( \partial_{t} - \vec{\nabla}_{\vec{x}'}^2 \right)G(t-t'';\vec{x},\vec{x}') \psi(t'',\vec{x}') \\
	&= \int_{t'}^{\infty} \dd t'' \int_{\mathfrak{D}} \dd^D\vec{x}' \sqrt{|g(\vec{x}')|}
	\left( -\partial_{t''} G \psi + \nabla_{i'} G \nabla^{i'} \psi \right) \nonumber\\
	&\qquad\qquad - \int_{t'}^{\infty} \dd t'' \int_{\partial\mathfrak{D}} \dd^{D-1}\vec{\xi} \sqrt{|H(\vec{\xi})|} n^{i'} \nabla_{i'} G  \psi \nonumber\\
	&= \int_{\mathfrak{D}} \dd^D\vec{x}' \sqrt{|g(\vec{x}')|} 
	\left\{ [-G\psi]_{t''=t'}^{t''=\infty} + \int_{t'}^{\infty} \dd t'' G \left( \partial_{t''} - \vec{\nabla}_{\vec{x}''}^2 \right) \psi \right\} \nonumber\\
	&\qquad\qquad + \int_{t'}^{\infty} \dd t'' \int_{\partial\mathfrak{D}} \dd^{D-1}\vec{\xi} \sqrt{|H(\vec{\xi})|} 
	\left( G \cdot n^{i'} \nabla_{i'} \psi - n^{i'} \nabla_{i'} G \cdot \psi\right) . \nonumber
	\end{align}}
If we impose the boundary condition in eq. \eqref{HeatDiffusionEquation_GreensFunction_BC}, we see that $[-G\psi]_{t''=t'}^{t''=\infty} = G(t-t') \psi(t')$ because the upper limit contains $G(t-\infty)\equiv\lim_{t'\to-\infty}\Theta(t-t') K(\vec{x},\vec{x}';t-t')=0$. The heat/diffusion eq. \eqref{HeatDiffusionEquation} removes the time-integral term on the first line of the last equality. If Dirichlet boundary conditions were chosen, we may choose $G(t-t'';\vec{x},\vec{x}'\in\partial\mathfrak{D})=0$ (i.e., eq. \eqref{HeatDiffusionEquation_GreensFunction_DirichletBC}) and obtain eq. \eqref{HeatDiffusionEquation_SolutionG}. Note that the upper limit of integration in the last line is really $t$, because eq. \eqref{HeatDiffusionEquation_GreensFunction_BC} tells us the Green's function vanishes for $t'' > t$.

\subsubsection{Problems}

\begin{myP}
\qquad In infinite flat $\mathbb{R}^D$, suppose we have some initial probability distribution of finding a Brownian particle, expressed in Cartesian coordinates as
\begin{align}
\psi(t=t_0,\vec{x}) = \left(\frac{\omega}{\pi}\right)^{D/2} \exp\left( -\omega (\vec{x}-\vec{x}_0)^2 \right), \qquad \omega > 0 .
\end{align}
Solve the diffusion equation for $t \geq t_0$.
\end{myP}
\begin{myP}
\qquad Suppose we have some initial temperature distribution $T(t=t_0,\theta,\phi) \equiv T_0(\theta,\phi)$ on a thin spherical shell. This distribution admits some multipole expansion: 
\begin{align}
T_0(\theta,\phi) = \sum_{\ell=0}^\infty \sum_{m=-\ell}^\ell a_\ell^m Y_\ell^m(\theta,\phi), \qquad a_\ell^m \in \mathbb{C} .
\end{align}
The temperature as a function of time obeys the heat/diffusion equation
\begin{align}
\partial_t T(t,\theta,\phi) = \sigma \vec{\nabla}^2 T(t,\theta,\phi), \qquad \sigma > 0 ,
\end{align}
where $\vec{\nabla}^2$ is now the Laplacian on the $2-$sphere. Since $\vec{\nabla}^2$ is dimensionless here, $\sigma$ has units of 1/[Time].
\begin{enumerate}
\item Solve the propagator $K$ for the heat/diffusion equation on the $2-$sphere, in terms of a spherical harmonic $\{Y_\ell^m(\theta,\phi)\}$ expansion.
\item Find the solution for $T(t>t_0,\theta,\phi)$.
\item What is the decay rate of the $\ell$th multipole, i.e., how much time does the $\ell$th term in the multipole sum take to decay in amplitude by $1/e$? Does it depend on both $\ell$ and $m$? And, what is the final equilibrium temperature distribution?
\end{enumerate}	
\end{myP}
\begin{myP}
{\it Inverse of Laplacian from Heat Kernel} \qquad In this problem we want to point out how the Green's function of the Laplacian is related to the heat/diffusion equation. To re-cap, the Green's function itself obeys the $D$-dimensional PDE:
\begin{align}
-\vec{\nabla}^2 G(\vec{x},\vec{x}') = \frac{\delta^{(D)}(\vec{x}-\vec{x}')}{\sqrt[4]{g(\vec{x}) g(\vec{x}')}} .
\end{align}
As already suggested by our previous discussions, the Green's function $G(\vec{x},\vec{x}')$ can be viewed the matrix element of the operator $\widehat{G} \equiv 1/(-\vec{\nabla}^2)$, namely\footnote{The perspective that the Green's function be viewed as an operator acting on some Hilbert space was advocated by theoretical physicist Julian Schwinger.}
\begin{align}
G(\vec{x},\vec{x}') = \braOket{\vec{x}}{\widehat{G}}{\vec{x}'} \equiv \braOket{\vec{x}}{\frac{1}{-\vec{\nabla}^2}}{\vec{x}'}.
\end{align}
The $\vec{\nabla}^2$ is now an abstract operator acting on the Hilbert space spanned by the position eigenkets $\{\ket{\vec{x}}\}$. Because it is Hermitian, we have
\begin{align}
-\vec{\nabla}^2_{\vec{x}} \braOket{\vec{x}}{\frac{1}{-\vec{\nabla}^2}}{\vec{x}'}
= \braOket{\vec{x}}{\frac{-\vec{\nabla}^2}{-\vec{\nabla}^2}}{\vec{x}'} 
= \braket{\vec{x}}{\vec{x}'} = \delta^{(D)}(\vec{x}-\vec{x}') .
\end{align}
Now use the Gamma function identity, for Re$(z)$, Re$(b) > 0$,
\begin{align}
\label{SchwingerTrick}
\frac{1}{b^z} = \frac{1}{\Gamma(z)} \int_0^\infty t^{z-1} e^{-bt} \dd t,
\end{align}
where $\Gamma(z)$ is the Gamma function -- to justify 
\begin{align}
\label{GfromHeatKernel}
G(\vec{x},\vec{x}') 				&= \int_0^\infty \dd t K_G\left(\vec{x},\vec{x}';t\right), \\
K_G\left(\vec{x},\vec{x}';t\right) 	&\equiv \braOket{\vec{x}}{e^{t \vec{\nabla}^2}}{\vec{x}'} . \nonumber
\end{align}
Notice how the integrand itself is the propagator (eq. \eqref{HeatEquation_Propagator}) of the heat/diffusion equation.

We will borrow from our previous linear algebra discussion that $-\vec{\nabla}^2 = \vec{P}^2$, as can be seen from its position space representation. Now proceed to re-write this integral by inserting to both the left and to the right of the operator $e^{t \vec{\nabla}^2}$ the completeness relation in momentum space. Use the fact that $\vec{P}^2 = -\vec{\nabla}^2$ and eq. to deduce
\begin{align}
\label{GreensFunctionLaplacian_IntegralRep}
G(\vec{x},\vec{x}') = \int_0^\infty \dd t \int \frac{\dd^D \vec{k}}{(2\pi)^D} e^{-t\vec{k}^2} e^{i \vec{k} \cdot (\vec{x}-\vec{x}')} .
\end{align}
(Going to momentum space allows you to also justify in what sense the restriction Re$(b) > 0$ of the formula in eq. \eqref{SchwingerTrick} was satisfied.) By appropriately ``completing the square" in the exponent, followed by an application of eq. \eqref{SchwingerTrick}, evaluate this integral to arrive at the Green's function of the Laplacian in $D$ spatial dimensions:
\begin{align}
\label{GreensFunctionLaplacian}
G(\vec{x},\vec{x}') = \braOket{\vec{x}}{\frac{1}{-\vec{\nabla}^2}}{\vec{x}'} 
= \frac{\Gamma\left(\frac{D}{2}-1\right)}{4 \pi^{D/2} |\vec{x}-\vec{x}'|^{D-2}} ,
\end{align}
where $|\vec{x}-\vec{x}'|$ is the Euclidean distance between $\vec{x}$ and $\vec{x}'$. 

Next, can you use eq. 18.12.4 of the NIST page \href{http://dlmf.nist.gov/18.12}{here} to perform an expansion of the Green's function of the negative Laplacian in terms of $r_> \equiv \max(r,r')$, $r_< \equiv \min(r,r')$ and $\widehat{n} \cdot \widehat{n}'$, where $r \equiv |\vec{x}|$, $r' \equiv |\vec{x}'|$, $\widehat{n} \equiv \vec{x}/r$, and $\widehat{n}' \equiv \vec{x}'/r'$? The $D=3$ case reads
\begin{align}
\label{GreensFunction_LegendrePolynomialExpansion}
\frac{1}{4\pi |\vec{x}-\vec{x}'|} 
= (4\pi r_>)^{-1} \sum_{\ell=0}^\infty P_\ell\left(\widehat{n}\cdot\widehat{n}'\right) \left(\frac{r_<}{r_>}\right)^\ell 
= \frac{1}{r_>} \sum_{\ell=0}^\infty \sum_{m=-\ell}^{\ell} \frac{\overline{Y_\ell^m(\widehat{n})} Y_\ell^m(\widehat{n}')}{2\ell+1} \left(\frac{r_<}{r_>}\right)^\ell ,
\end{align}
where the $P_\ell$ are Legendre polynomials and in the second line the addition formula of eq. \eqref{SphericalHarmonics_AdditionFormula} was invoked.

Note that while it is not easy to verify by direct differentiation that eq. \eqref{GreensFunctionLaplacian} is indeed the Green's function $1/(-\vec{\nabla}^2)$, one can do so by performing the integral over $t$ in eq. \eqref{GreensFunctionLaplacian_IntegralRep}, to obtain
\begin{align}
\label{GreensFunctionLaplacian_ModeExpansion}
G(\vec{x},\vec{x}') = \int \frac{\dd^D k}{(2\pi)^D} \frac{e^{i \vec{k} \cdot (\vec{x}-\vec{x}')}}{\vec{k}^2} .
\end{align}
We have already seen this in eq. \eqref{GreensFunctionLaplacian_ModeExpansion_}.

Finally, can you use the relationship between the heat kernel and the Green's function of the Laplacian in eq. \eqref{GfromHeatKernel}, to show how in a finite domain, eq. \eqref{HeatDiffusionEquation_SolutionG} leads to eq. \eqref{GreensFunction_Laplacian_KirchhoffRep_PsiGiven} in the late time $t \to \infty$ limit? (You may assume the smallest eigenvalue of the negative Laplacian is strictly positive; recall eq. \eqref{VariationalPrincipleForLaplacianSpectrum}.) \qed
\end{myP}
%\begin{myP}
%\qquad Use the methods of the previous problem to solve the Helmholtz Green's function $G$, for $m^2 > 0$:
%\begin{align}
%\left(-\vec{\nabla}^2 + m^2\right) G(\vec{x},\vec{x}') = \delta^{(D)}(\vec{x}-\vec{x}') .
%\end{align}
%Hint: refer to 10.32.10 of http://dlmf.nist.gov/10.32. \qed
%\end{myP}
\begin{myP}
\qquad Is it possible to solve for the Green's function of the Laplacian on the $2$-sphere? Use the methods of the last two problems, or simply try to write down the mode sum expansion in eq. \eqref{GreensFunction_Laplacian_ModeSum}, to show that you would obtain a $1/0$ infinity. What is the reason for this apparent pathology? Suppose we could solve
\begin{align}
-\vec{\nabla}^2 G(\vec{x},\vec{x}') = \frac{\delta^{(2)}(\vec{x}-\vec{x}')}{\sqrt[4]{g(\vec{x}) g(\vec{x}')}} .
\end{align}
Perform a volume integral of both sides over the $2-$sphere -- explain the contradiction you get. (Recall the discussion in the differential geometry section.) Hint: Apply the curved space Gauss' law in eq. \eqref{DifferentialGeometry_GaussTheorem} and remember the $2$-sphere is a closed surface.
\end{myP}

\subsection{Massless Scalar Wave Equation (Mostly) In Flat Spacetime $\mathbb{R}^{D,1}$}

\subsubsection{Spacetime metric, uniqueness of Minkowski wave solutions}

\noindent{\bf Spacetime Metric} \qquad In Cartesian coordinates $(t,\vec{x})$, it is possible associate a metric to flat spacetime as follows
\begin{align}
\label{MinkowskiMetric}
\dd s^2 = c^2 \dd t^2 - \dd\vec{x}\cdot\dd\vec{x}
\equiv \eta_{\mu\nu} \dd x^\mu \dd x^\nu, \qquad\qquad x^\mu \equiv (c t,x^i),
\end{align}
where $c$ is the speed of light in vacuum; $\mu \in \{ 0,1,2,\dots,D \}$; and $D$ is still the dimension of space.\footnote{In this section it is important to distinguish Greek $\{ \mu,\nu, \dots \}$ and Latin/English alphabets $\{ a,b,i,j,\dots \}$. The former run over $0$ through $D$, where the $0$th index refers to time and the 1st through $D$th to space. The latter run from $1$ through $D$, and are thus strictly ``spatial" indices. Also, be aware that the opposite sign convention, $\dd s^2 = - \dd t^2 + \dd\vec{x}\cdot\dd\vec{x}$, is commonly used too. For most physical applications both sign conventions are valid; see, however, \cite{Carlip:1988gw}.} We also have defined the flat (Minkowski) spacetime metric
\begin{align}
\eta_{\mu\nu} \equiv \text{diag}\left(1,-1,-1,\dots,-1\right) .
\end{align}
The generalization of eq. \eqref{MinkowskiMetric} to curved spacetime is
\begin{align}
\label{CurvedMetric}
\dd s^2 = g_{\mu\nu}(t,\vec{x}) \dd x^\mu \dd x^\nu, \qquad x^\mu = (c t,x^i).
\end{align}
It is common to use the symbol $\Box$, especially in curved spacetime, to denote the spacetime-Laplacian:
\begin{align}
\Box \psi \equiv \nabla_\mu \nabla^\mu \psi 
= \frac{1}{\sqrt{|g|}} \partial_\mu \left( \sqrt{|g|} g^{\mu\nu} \partial_\nu \psi \right) ,
\end{align}
where $\sqrt{|g|}$ is now the square root of the absolute value of the determinant of the metric $g_{\mu\nu}$. In Minkowski spacetime of eq. \eqref{MinkowskiMetric}, we have $\sqrt{|g|}=1$, $\eta^{\mu\nu} = \eta_{\mu\nu}$, and
\begin{align}
\Box \psi
= \eta^{\mu\nu} \partial_\mu \partial_\nu \psi \equiv \partial^2 \psi 
= \left( c^{-2} \partial_t^2 - \delta^{ij} \partial_i \partial_j\right) \psi  ;
\end{align}
where $\delta^{ij} \partial_i \partial_j = \vec{\nabla}^2$ is the spatial Laplacian in flat Euclidean space. The Minkowski ``dot product" between vectors $u$ and $v$ in Cartesian coordinates is now
\begin{align}
u \cdot v \equiv \eta_{\mu\nu} u^\mu v^\nu = u^0 v^0 - \vec{u} \cdot \vec{v}, \qquad u^2 \equiv (u^0)^2 -\vec{u}^2, \quad \text{etc} .
\end{align}
From here on, $x$, $x'$ and $k$, etc. -- without an arrow over them -- denotes collectively the $D+1$ coordinates of spacetime. Indices of spacetime tensors are moved with $g^{\mu\nu}$ and $g_{\mu\nu}$. For instance,
\begin{align}
u^\mu = g^{\mu\nu} u_\nu, \qquad\qquad u_\mu = g_{\mu\nu} u^\nu .
\end{align}
In the flat spacetime geometry of eq. \eqref{MinkowskiMetric}, written in Cartesian coordinates,
\begin{align}
u^0 = u_0, \qquad\qquad u^i = -u_i .
\end{align}
{\it Indefinite signature} \qquad The subtlety with the metric of spacetime, as opposed to that of space only, is that the ``time" part of the distance in eq. \eqref{MinkowskiMetric} comes with a different sign from the ``space" part of the metric. In curved or flat space, if $\vec{x}$ and $\vec{x}'$ have zero geodesic distance between them, they are really the same point. In curved or flat spacetime, however, $x$ and $x'$ may have zero geodesic distance between them, but they could either refer to the same spacetime point (aka ``event") -- or they could simply be lying on each other's light cone:
\begin{align}
0 = (x-x')^2 = \eta_{\mu\nu} (x^\mu-x'^\mu) (x^\nu-x'^\nu) \qquad \Rightarrow \qquad (t-t')^2 = (\vec{x}-\vec{x}')^2 .
\end{align}
To understand this statement more systematically, let us work out the geodesic distance between any pair of spacetime points in flat spacetime.
\begin{myP}
\qquad In Minkowski spacetime expressed in Cartesian coordinates, the Christoffel symbols are zero. Therefore the geodesic equation in \eqref{DifferentialGeometry_GeodesicEquation} returns the following ``acceleration-is-zero" ODE:
\begin{align}
0 &= \frac{\dd^2 Z^\mu(\lambda)}{\dd \lambda^2} .
\end{align}
Show that the geodesic joining the initial spacetime point $Z^\mu(\lambda=0) = x'^\mu$ to the final location $Z^\mu(\lambda=1) = x^\mu$ is the straight line
\begin{align}
Z^\mu(0 \leq \lambda \leq 1) = x'^\mu + \lambda \ (x^\mu - x'^\mu) .
\end{align}
Use eq. \eqref{DifferentialGeometry_LengthIntegral} to show that half the {\it square} of the geodesic distance between $x'$ and $x$ is
\begin{align}
\bar{\sigma}(x,x') = \frac{1}{2} (x-x')^2 .
\end{align}
$\sbar$ is commonly called Synge's world function in the gravitation literature. \qed
\end{myP}
Some jargon needs to be introduced here. (Drawing a spacetime diagram would help.)
\begin{itemize}
\item When $\sbar > 0$, we say $x$ and $x'$ are timelike separated. If you sit at rest in some inertial frame, then the tangent vector to your world line is $u^\mu = (1,\vec{0})$, and $u = \partial_t$ is a measure of how fast the time on your watch is running. Or, simply think about setting $\dd \vec{x}=0$ in the Minkowski metric: $\dd s^2 \to \dd t^2 > 0$.
\item When $\sbar < 0$, we say $x$ and $x'$ are spacelike separated. If you and your friend sit at rest in the same inertial frame, then at a fixed time $\dd t=0$, the (square of the) spatial distance between the both of you is now given by integrating $\dd s^2 \to -\dd\vec{x}^2 < 0$ between your two locations. 
\item When $\sbar = 0$, we say $x$ and $x'$ are null (or light-like) separated. As already alluded to, in 4 dimensional flat spacetime, light travels strictly on null geodesics $\dd s^2=0$. Consider a coordinate system for spacetime centered at $x'$; then we would say $x$ lies on the light cone of $x'$ (and vice versa).
\end{itemize}
As we will soon discover, the indefinite metric of spacetimes -- as opposed to the positive definite one of space itself -- is what allows for wave solutions, for packets of energy/momentum to travel over space and time. In Minkowski spacetime, we will show below, by solving explicitly the Green's function $G_{D+1}$ of the wave operator, that these waves $\psi$, subject to eq. \eqref{ScalarWaveEquation_Minkowski}, will obey causality: they travel strictly on and/or within the light cone, independent of what the source $J$ is.

{\bf Poincar\'{e} symmetry} \qquad Analogous to how rotations $\{R^i_{\phantom{i}a} \vert \delta_{ij} R^i_{\phantom{i}a} R^j_{\phantom{j}b} = \delta_{ab} \}$ and spatial translations $\{a^i\}$ leave the flat Euclidean metric $\delta_{ij}$ invariant, 
\begin{align}
x^i \to R^i_{\phantom{i}j} x^j + a^i \qquad \Rightarrow \qquad
\delta_{ij} \dd x^i \dd x^j \to \delta_{ij} \dd x^i \dd x^j .
\end{align}
(The $R^i_{\phantom{i}j}$ and $a^i$ are constants.) Lorentz transformations $\{\Lambda^\alpha_{\phantom{\alpha}\mu} \vert \eta_{\alpha\beta} \Lambda^\alpha_{\phantom{\alpha}\mu} \Lambda^\beta_{\phantom{\beta}\nu} = \eta_{\mu\nu} \}$ and spacetime translations $\{ a^\mu \}$ are ones that leave the flat Minkowski metric $\eta_{\mu\nu}$ invariant.
\begin{align}
\label{PoincareTransformations}
x^\alpha \to \Lambda^\alpha_{\phantom{\alpha}\mu} x^\mu + a^\alpha \qquad \Rightarrow \qquad
\eta_{\mu\nu} \dd x^\mu \dd x^\nu \to \eta_{\mu\nu} \dd x^\mu \dd x^\nu .
\end{align}
(The $\Lambda^\alpha_{\phantom{\alpha}\mu}$ and $a^\alpha$ are constants.) This in turn leaves the light cone condition $\dd s^2=0$ invariant -- the speed of light is unity, $|\dd\vec{x}|/\dd t = 1$, in all inertial frames related via eq. \eqref{PoincareTransformations}.

\noindent{\bf Wave Equation In Curved Spacetime} \qquad The wave equation (for a minimally coupled massless scalar) in some spacetime geometry $g_{\mu\nu} \dd x^\mu \dd x^\nu$ is a 2nd order in time PDE that takes the following form:
\begin{align}
\label{ScalarWaveEquation_Curved}
\nabla_\mu \nabla^\mu \psi = \frac{1}{\sqrt{|g|}} \partial_\mu \left( \sqrt{|g|} g^{\mu\nu} \partial_\nu \psi \right) = J(x) ,
\end{align}
where $J$ is some specified external source of $\psi$. 

{\it Minkowski} \qquad We will mainly deal with the case of infinite flat (aka ``Minkowski") spacetime in eq. \eqref{MinkowskiMetric}, where in Cartesian coordinates $x^\mu = (ct,\vec{x})$. This leads us to the wave equation
\begin{align}
\label{ScalarWaveEquation_Minkowski}
\left( \partial_t^2 - c^2 \vec{\nabla}^2_{\vec{x}} \right) \psi(t,\vec{x}) = c^2 J(t,\vec{x}) , \qquad\qquad 
\vec{\nabla}^2_{\vec{x}} \equiv \delta^{ij} \partial_i \partial_j .
\end{align}
Here, $c$ will turn out to be the speed of propagation of the waves themselves. Because it will be the most important speed in this chapter, I will set it to unity, $c=1$.\footnote{This is always a good labor-saving strategy when you solve problems. Understand all the distinct dimensionful quantities in your setup -- pick the most relevant/important length, time, and mass, etc. Then set them to one, so you don't have to carry their symbols around in your calculations. Every other length, time, mass, etc. will now be respectively, expressed as multiples of them. For instance, now that $c=1$, the speed(s) $\{v_i\}$ of the various constituents of the source $J$ measured in some center of mass frame, would be measured in multiples of $c$ -- for instance, ``$v^2 = 0.76$" really means $(v/c)^2 = 0.76$.} We will work mainly in flat infinite spacetime, which means the $\vec{\nabla}^2$ is the Laplacian in flat space. This equation describes a diverse range of phenomenon, from the vibrations of strings to that of spacetime itself.

{\bf 2D Minkowski} \qquad We begin the study of the homogeneous wave equation in 2 dimensions. In Cartesian coordinates $(t,z)$,
\begin{align}
\label{ScalarWaveEquation_Minkowski_2D}
\left( \partial_t^2 - \partial_z^2 \right) \psi(t,z) = 0 .
\end{align}
We see that the solutions are a superposition of either left-moving $\psi(z+t)$ or right-moving waves $\psi(z-t)$, where $\psi$ can be any arbitrary function,
\begin{align}
\left( \partial_t^2 - \partial_z^2 \right) \psi(z \pm t) = (\pm)^2 \psi''(z \pm t) - \psi''(z \pm t) = 0.
\end{align}
{\it Remark} \qquad It is worth highlighting the difference between the nature of the general solutions to 2nd order linear homogeneous ODEs versus those of PDEs such as the wave equation here. In the former, they span a 2 dimensional vector space, whereas the wave equation admits arbitrary functions as general solutions. This is why the study of PDEs involve infinite dimensional (oftentimes continuous) Hilbert spaces.

Let us put back the speed $c$ -- by dimensional analysis we know $[c]$=[Length/Time], so $x \pm c t$ would yield the correct dimensions.
\begin{align}
\psi(t,x) = \psi_L(x+ct) + \psi_R(x-ct) .
\end{align}
These waves move strictly at speed $c$.
\begin{myP}
\qquad Let us define light cone coordinates as $x^\pm \equiv t \pm z$. Write down the Minkowski metric in eq. \eqref{MinkowskiMetric} 
\begin{align}
\dd s^2 = \dd t^2 - \dd z^2
\end{align}
in terms of $x^\pm$ and show by direct integration of eq. \eqref{ScalarWaveEquation_Minkowski_2D} that the most general homogeneous wave solution in 2D is the superposition of left- and right-moving (otherwise arbitrary) profiles. \qed
\end{myP}
\begin{quotation}
\noindent{\bf Uniqueness of Minkowski solutions} \qquad Suppose the following initial conditions are given
\begin{align}
\psi(t=t_0,\vec{x}) = \varphi_0(\vec{x}), \qquad \partial_t \psi(t=t_0,\vec{x}) = \varphi_1(\vec{x}) ;
\end{align}
and suppose the scalar field $\psi$ or its normal derivative is specified on the spatial boundaries $\partial \mathfrak{D}$,
\begin{align}
\psi(t,\vec{x}\in\partial\mathfrak{D}) 								&= \varphi_3(\partial\mathfrak{D}), \qquad \text{(Dirichlet)}, \\
\text{or     } n^i \nabla_i \psi(t,\vec{x}\in\partial\mathfrak{D}) 	&= \varphi_4(\partial\mathfrak{D}), \qquad \text{(Neumann)} ,
\end{align}
where $n^i(\partial \mathfrak{D})$ is the unit outward normal vector. Then, the solution to the wave equation in eq. \eqref{ScalarWaveEquation_Minkowski} is unique.
\end{quotation}
\noindent{\it Proof} \qquad Without loss of generality, since our wave equation is linear, we may assume the scalar field is real. We then suppose there are two such solutions $\psi_1$ and $\psi_2$ obeying the same initial and boundary conditions. The proof is established if we can show, in fact, that $\psi_1$ has to be equal to $\psi_2$. Note that the difference $\Psi \equiv \psi_1 - \psi_2$ is subject to the homogeneous wave equation
\begin{align}
\partial^2 \Psi = \ddot{\Psi} - \vec{\nabla}^2 \Psi = 0 
\end{align}
since the $J$ cancels out when we subtract the wave equations of $\psi_{1,2}$. For similar reasons the $\Psi$ obeys the initial conditions
\begin{align}
\Psi(t=t_0,\vec{x}) = 0 \qquad \text{  and  } \qquad \partial_t \Psi(t=t_0,\vec{x}) = 0,
\end{align} 
and the spatial boundary conditions
\begin{align}
\Psi(t,\vec{x}\in\partial\mathfrak{D}) = 0 \qquad \text{  or  } \qquad n^i \nabla_i \Psi(t,\vec{x}\in\partial\mathfrak{D}) = 0 .
\end{align} 
Let us then consider the following integral
\begin{align}
T^{00}(t) \equiv \frac{1}{2} \int_{\mathfrak{D}} \dd^D\vec{x} 
\left( \dot{\Psi}^2(t,\vec{x}) + \vec{\nabla}\Psi(t,\vec{x}) \cdot \vec{\nabla}\Psi(t,\vec{x}) \right)
\end{align}
\footnote{The integrand, for $\Psi$ obyeing the homogeneous wave equation, is in fact its energy density. Therefore $T^{00}(t)$ is the total energy stored in $\Psi$ at a given time $t$.}as well as its time derivative
\begin{align}
\partial_t T^{00}(t) &= \int_{\mathfrak{D}} \dd^D\vec{x} \left( \dot{\Psi} \ddot{\Psi} + \vec{\nabla}\dot{\Psi} \cdot \vec{\nabla}\Psi \right) .
\end{align}
We may use the homogeneous wave equation on the $\ddot{\Psi}$ term, and integrate-by-parts one of the gradients on the second term,
\begin{align}
\partial_t T^{00}(t) 
&= \int_{\partial \mathfrak{D}} \dd^{D-1} \vec{\xi} \sqrt{|H(\vec{\xi})|} \dot{\Psi} n^i \nabla_i \Psi 
+ \int_{\mathfrak{D}} \dd^D\vec{x} \left( \dot{\Psi} \vec{\nabla}^2 \Psi - \dot{\Psi} \vec{\nabla}^2 \Psi \right) .
\end{align}
By assumption either $\Psi$ or $n^i \nabla_i \Psi$ is zero on the spatial boundary; if it were the former, then $\dot{\Psi}(\partial \mathfrak{D}) = 0$ too. Either way, the surface integral is zero. Therefore the right hand side vanishes and we conclude that $T^{00}$ is actually a constant in time. Together with the initial conditions $\dot{\Psi}(t=t_0,\vec{x})^2 = 0$ and $\Psi(t=t_0,\vec{x}) = 0$ (which implies $(\vec{\nabla}\Psi(t=t_0,\vec{x}))^2 = 0$), we see that $T^{00}(t=t_0)=0$, and therefore has to remain zero for all subsequent time $t \geq t_0$. Moreover, since $T^{00}(t \geq t_0) = 0$ is the integral of the sum of $(D+1)$ positive terms $\{ \dot{\Psi}^2, (\partial_i \Psi)^2 \}$, each term must individually vanish, which in turn implies $\Psi$ must be a constant in both space and time. But, since it is zero at the initial time $t=t_0$, it must be in fact zero for $t \geq t_0$. That means $\psi_1 = \psi_2$. \qed

\noindent{\it Remark} \qquad Armed with the knowledge that the ``initial value problem" for the Minkowski spacetime wave equation has a unique solution, we will see how to actually solve it first in Fourier space and then with the retarded Green's function.

\subsubsection{Waves, Initial value problem, Green's Functions}

\noindent{\bf Dispersion relations, Homogeneous solutions} \qquad You may guess that any function $f(t,\vec{x})$ in flat (Minkowski) spacetime can be Fourier transformed.
\begin{align}
f(t,\vec{x}) = \int_{\mathbb{R}^{D+1}} \frac{\dd^{D+1}k}{(2\pi)^{D+1}} \widetilde{f}(\omega,\vec{k}) e^{-i\omega t} e^{i\vec{k} \cdot \vec{x}} 
\qquad \text{(Not quite \dots)} ,
\end{align}
where
\begin{align}
k^\mu \equiv (\omega,k^i) .
\end{align}
Remember the first component is now the $0$th one; so 
\begin{align}
\label{ScalarWaveEquation_PlaneWaves}
\exp(-ik_\mu x^\mu) = \exp(-i \eta_{\mu\nu} k^\mu x^\mu) = \exp(-i\omega t) \exp(i\vec{k} \cdot \vec{x}) .
\end{align}
Furthermore, these plane waves in eq. \eqref{ScalarWaveEquation_PlaneWaves} obey
\begin{align}
\label{ScalarWaveEquation_PlaneWaves_Eigenvectors}
\partial^2 \exp(-ik_\mu x^\mu) = -k^2 \exp(-ik_\mu x^\mu) , \qquad k^2 \equiv k_\mu k^\mu .
\end{align}
This comes from a direct calculation; note that $\partial_\mu (i k_\alpha x^\alpha) = i k_\alpha \delta^\alpha_\mu = i k_\mu$ and similarly $\partial^\mu (i k_\alpha x^\alpha) = i k^\mu$.
\begin{align}
\partial^2 \exp(-ik_\mu x^\mu) = \partial_\mu \partial^\mu \exp(-ik_\mu x^\mu) = (ik_\mu)(ik^\mu) \exp(-ik_\mu x^\mu) .
\end{align}
Therefore, a particular mode $\widetilde{\psi} e^{-ik_\alpha x^\alpha}$ satisfies the homogeneous scalar wave equation in eq. \eqref{ScalarWaveEquation_Minkowski} with $J=0$ -- provided that
\begin{align}
\label{ScalarWaveEquation_Disperson}
0 = \partial^2 \left( \widetilde{\psi} e^{-ik_\alpha x^\alpha} \right) = -k^2 \widetilde{\psi} e^{-ik_\alpha x^\alpha} \qquad \Rightarrow \qquad
k^2 = 0 \qquad \Rightarrow \qquad \omega^2 = \vec{k}^2 .
\end{align} 
This relationship between the zeroth component of the momentum and its spatial ones, is often known as the {\it dispersion relation}. Moreover, the positive root
\begin{align}
\omega = |\vec{k}|
\end{align}
can be interpreted as saying the energy $\omega$ of the photon -- or, the massless particle associated with $\psi$ obeying eq. \eqref{ScalarWaveEquation_Minkowski} -- is equal to the magnitude of its momentum $\vec{k}$.

Therefore, if $\psi$ satisfies the homogeneous wave equation, the Fourier expansion is actually $D$-dimensional not $(D+1)$ dimensional:
\begin{align}
\label{ScalarWaveEquation_HomogeneousSolution_FourierExpansion_StepI}
\psi(t,\vec{x}) 
= \int_{\mathbb{R}^D} \frac{\dd^D \vec{k}}{(2\pi)^D} \left(\widetilde{A}(\vec{k}) e^{-i|\vec{k}|t} 
		+ \widetilde{B}(\vec{k}) e^{i|\vec{k}|t} \right) e^{i\vec{k}\cdot\vec{x}} .
\end{align}
There are two terms in the parenthesis, one for the positive solution $\omega=+|\vec{k}|$ and one for the negative $\omega=-|\vec{k}|$. For a real scalar field $\psi$, the $\widetilde{A}$ and $\widetilde{B}$ are related.
\begin{align}
\psi(t,\vec{x})^* = \psi(t,\vec{x})
&= \int_{\mathbb{R}^D} \frac{\dd^D \vec{k}}{(2\pi)^D} \left( \widetilde{A}(\vec{k})^* e^{i|\vec{k}|t} 
+ \widetilde{B}(\vec{k})^* e^{-i|\vec{k}|t} \right) e^{-i\vec{k}\cdot\vec{x}} \nonumber\\
\label{ScalarWaveEquation_HomogeneousSolution_FourierExpansion_StepII}
&= \int_{\mathbb{R}^D} \frac{\dd^D \vec{k}}{(2\pi)^D} \left( \widetilde{B}(-\vec{k})^* e^{-i|\vec{k}|t}
+ \widetilde{A}(-\vec{k})^* e^{i|\vec{k}|t} \right) e^{i\vec{k}\cdot\vec{x}} .
\end{align}
Comparing equations \eqref{ScalarWaveEquation_HomogeneousSolution_FourierExpansion_StepI} and \eqref{ScalarWaveEquation_HomogeneousSolution_FourierExpansion_StepII} indicate $\widetilde{A}(-\vec{k})^* = \widetilde{B}(\vec{k}) \Leftrightarrow \widetilde{A}(\vec{k}) = \widetilde{B}(-\vec{k})^*$. Therefore,
\begin{align}
\label{ScalarWaveEquation_HomogeneousSolution_FourierExpansion}
\psi(t,\vec{x}) 
= \int_{\mathbb{R}^D} \frac{\dd^D \vec{k}}{(2\pi)^D} \left(\widetilde{A}(\vec{k}) e^{-i|\vec{k}|t} 
+ \widetilde{A}(-\vec{k})^* e^{i|\vec{k}|t} \right) e^{i\vec{k}\cdot\vec{x}} .
\end{align}
Note that $\widetilde{A}(\vec{k})$ itself, for a fixed $\vec{k}$, has two independent parts -- its real and imaginary portions.\footnote{In quantum field theory, the coefficients $\widetilde{A}(\vec{k})$ and $\widetilde{A}(\vec{k})^*$ of the Fourier expansion in \eqref{ScalarWaveEquation_HomogeneousSolution_FourierExpansion} will become operators obeying appropriate commutation relations.}

{\it Contrast} this homogeneous wave solution against the infinite Euclidean (flat) space case, where $-\vec{\nabla}^2 \psi = 0$ does not admit any solutions that are regular everywhere ($\equiv$ does not blow up anywhere), except the $\psi=$ constant solution.

\noindent{\bf Initial value formulation through mode expansion} \qquad Unlike the heat/diffusion equation, the wave equation is second order in time. We therefore expect that, to obtain a unique solution to the latter, we have to supply both the initial field configuration and its first time derivative (conjugate momentum). It is possible to see it explicitly through the mode expansion in eq. \eqref{ScalarWaveEquation_HomogeneousSolution_FourierExpansion} -- the need for two independent coefficients $\widetilde{A}$ and $\widetilde{A}^*$ to describe the homogeneous solution is intimately tied to the need for two independent initial conditions.

Suppose
\begin{align}
\psi(t=0,\vec{x})=\psi_0(\vec{x}) \qquad \text{ and } \qquad \partial_t \psi(t=0,\vec{x})=\dot{\psi}_0(\vec{x}) ,
\end{align}
where the right hand sides are given functions of space. Then, from eq. \eqref{ScalarWaveEquation_HomogeneousSolution_FourierExpansion},
\begin{align}
\psi_0(\vec{x}) 
= \int_{\mathbb{R}^D} \frac{\dd^D k}{(2\pi)^D} \widetilde{\psi}_0(\vec{k}) e^{i\vec{k}\cdot\vec{x}}
&= \int_{\mathbb{R}^D} \frac{\dd^D k}{(2\pi)^D} \left(\widetilde{A}(\vec{k}) + \widetilde{A}(-\vec{k})^*\right) e^{i\vec{k}\cdot\vec{x}} \nonumber\\
\dot{\psi}_0(\vec{x}) 
= \int_{\mathbb{R}^D} \frac{\dd^D k}{(2\pi)^D} \widetilde{\dot{\psi}}_0(\vec{k}) e^{i\vec{k}\cdot\vec{x}}
&= \int_{\mathbb{R}^D} \frac{\dd^D k}{(2\pi)^D} (-i|\vec{k}|)\left(\widetilde{A}(\vec{k}) - \widetilde{A}(-\vec{k})^* \right)e^{i\vec{k}\cdot\vec{x}} .
\end{align}
We have also assumed that the initial field and its time derivative admits a Fourier expansion. By equating the coefficients of the plane waves,
\begin{align}
\widetilde{\psi}_0(\vec{k}) 							&= \widetilde{A}(\vec{k}) + \widetilde{A}(-\vec{k})^* , \nonumber\\
\frac{i}{|\vec{k}|} \widetilde{\dot{\psi}}_0(\vec{k}) 	&= \widetilde{A}(\vec{k}) - \widetilde{A}(-\vec{k})^* .
\end{align}
Inverting this relationship tells us the $\widetilde{A}(\vec{k})$ and $\widetilde{A}(\vec{k})^*$ are indeed determined by (the Fourier transforms) of the initial conditions:
\begin{align}
\widetilde{A}(\vec{k}) 		&= \frac{1}{2}\left( \widetilde{\psi}_0(\vec{k}) + \frac{i}{|\vec{k}|} \widetilde{\dot{\psi}}_0(\vec{k}) \right) \nonumber\\
\widetilde{A}(-\vec{k})^* 	&= \frac{1}{2}\left( \widetilde{\psi}_0(\vec{k}) - \frac{i}{|\vec{k}|} \dot{\psi}_0(\vec{k}) \right)
\end{align}
In other words, given the initial conditions $\psi(t=0,\vec{x})=\psi_0(\vec{x})$ and $\partial_t \psi(t=0,\vec{x})=\dot{\psi}_0(\vec{x})$, we can evolve the homogeneous wave solution forward/backward in time through their Fourier transforms:
\begin{align}
\label{ScalarWaveEquation_HomogeneousSolution_InitialValueFormulation_Fourier}
\psi(t,\vec{x})
&= \frac{1}{2}\int_{\mathbb{R}^D} \frac{\dd^D \vec{k}}{(2\pi)^D} 
\left\{
\left( \widetilde{\psi}_0(\vec{k}) + \frac{i}{|\vec{k}|} \widetilde{\dot{\psi}}_0(\vec{k}) \right) e^{-i|\vec{k}|t}
+ \left( \widetilde{\psi}_0(\vec{k}) - \frac{i}{|\vec{k}|} \widetilde{\dot{\psi}}_0(\vec{k}) \right) e^{i|\vec{k}|t}
\right\}
e^{i\vec{k}\cdot\vec{x}} \nonumber \\
&= \int_{\mathbb{R}^D} \frac{\dd^D \vec{k}}{(2\pi)^D} 
\left( \widetilde{\psi}_0(\vec{k}) \cos(|\vec{k}|t) + \widetilde{\dot{\psi}}_0(\vec{k}) \frac{\sin(|\vec{k}|t)}{|\vec{k}|} \right)
e^{i\vec{k}\cdot\vec{x}} .
\end{align}
We see that the initial profile contributes to the part of the field even under time reversal $t \to -t$; whereas its initial time derivative contributes to the portion odd under time reversal.

Suppose the initial field configuration and its time derivative were specified at some other time $t_0$ (instead of $0$),
\begin{align}
\psi(t=t_0,\vec{x}) = \psi_0(\vec{x}), \qquad \partial_t \psi(t=t_0,\vec{x}) = \dot{\psi}_0(\vec{x}) .
\end{align}
Because of time-translation symmetry, eq. \eqref{ScalarWaveEquation_HomogeneousSolution_InitialValueFormulation_Fourier} becomes
\begin{align}
\psi(t,\vec{x})
= \int_{\mathbb{R}^D} \frac{\dd^D \vec{k}}{(2\pi)^D} 
\left( \widetilde{\psi}_0(\vec{k}) \cos\left( |\vec{k}|(t-t_0) \right) 
		+ \widetilde{\dot{\psi}}_0(\vec{k}) \frac{\sin\left(|\vec{k}|(t-t_0)\right)}{|\vec{k}|} \right)
e^{i\vec{k}\cdot\vec{x}} .
\end{align}
\begin{myP}
\qquad Let's consider an initial Gaussian wave profile with zero time derivative,
\begin{align}
\psi(t=0,\vec{x}) = \exp(-(\vec{x}/\sigma)^2), \qquad \partial_t \psi(t=0,\vec{x}) = 0.
\end{align}
If $\psi$ satisfies the homogeneous wave equation, what is $\psi(t>0,\vec{x})$? Express the answer as a Fourier integral; the integral itself may be very difficult to evaluate. \qed
\end{myP}
\noindent{\bf Inhomogeneous solution in Fourier space} \qquad If there is a non-zero source $J$, we could try the strategy we employed with the 1D damped driven simple harmonic oscillator: first go to Fourier space and then inverse-transform it back to position spacetime. That is, starting with,
\begin{align}
\label{ScalarWaveEquation_FlatSpacetime}
\partial_x^2 \psi(x) 	&= J(x), \\
\partial_x^2 \int_{\mathbb{R}^{D,1}} \frac{\dd^{D+1}k}{(2\pi)^{D+1}} \widetilde{\psi}(k) e^{-ik_\mu x^\mu} 
						&= \int_{\mathbb{R}^{D,1}} \frac{\dd^{D+1}k}{(2\pi)^{D+1}} \widetilde{J}(k) e^{-ik_\mu x^\mu} \\
\int_{\mathbb{R}^{D,1}} \frac{\dd^{D+1}k}{(2\pi)^{D+1}} (-k^2)\widetilde{\psi}(k) e^{-ik_\mu x^\mu} 
						&= \int_{\mathbb{R}^{D,1}} \frac{\dd^{D+1}k}{(2\pi)^{D+1}} \widetilde{J}(k) e^{-ik_\mu x^\mu}, \qquad k^2 \equiv k_\mu k^\mu .
\end{align}
Because the plane waves $\{ \exp(-ik_\mu x^\mu) \}$ are basis vectors, their coefficients on both sides of the equation must be equal.
\begin{align}
\label{WaveEquation_ParticularSolution_Fourier}
\widetilde{\psi}(k) = -\frac{\widetilde{J}(k)}{k^2} .
\end{align}
The advantage of solving the wave equation in Fourier space is, we see that this is the particular solution for $\psi$ -- the portion that is sourced by $J$. Turn off $J$ and you'd turn off (the inhomogeneous part of) $\psi$.

\noindent{\bf Inhomogeneous solution via Green's function} \qquad We next proceed to transform eq. \eqref{WaveEquation_ParticularSolution_Fourier} back to spacetime.
\begin{align}
\psi(x) 
&= -\int_{\mathbb{R}^{D,1}} \frac{\dd^{D+1}k}{(2\pi)^{D+1}} \frac{\widetilde{J}(k)}{k^2} e^{-i k\cdot x} 
= -\int_{\mathbb{R}^{D,1}} \frac{\dd^{D+1}k}{(2\pi)^{D+1}} \int_{\mathbb{R}^{D,1}} \dd^{D+1} x'' \frac{J(x') e^{ik \cdot x''}}{k^2} e^{-i k\cdot x} \nonumber\\
\label{WaveEquation_ParticularSolution_InverseFourier_I}
&= \int_{\mathbb{R}^{D,1}} \dd^{D+1} x'' \left(\int_{\mathbb{R}^{D,1}} \frac{\dd^{D+1}k}{(2\pi)^{D+1}} \frac{e^{-ik \cdot (x-x'')}}{-k^2}\right) J(x'') 
\end{align}
That is, if we define the Green's function of the wave operator as
\begin{align}
\label{ScalarWaveEquation_GreensFunction_ModeExpansion}
G_{D+1}(x-x') 
&= \int_{\mathbb{R}^{D+1}} \frac{\dd^{D+1}k}{(2\pi)^{D+1}} \frac{e^{-ik_\mu(x-x')^\mu}}{-k^2} \nonumber\\
&= -\int \frac{\dd\omega}{2\pi} \int \frac{\dd^D \vec{k}}{(2\pi)^D} \frac{e^{-i\omega(t-t')} e^{i\vec{k}\cdot(\vec{x}-\vec{x}')}}{\omega^2 - \vec{k}^2} ,
\end{align}
eq. \eqref{WaveEquation_ParticularSolution_InverseFourier_I} translates to
\begin{align}
\label{ScalarWaveEquation_GConvolvedWithJ}
\psi(x) = \int_{\mathbb{R}^{D+1}} \dd^{D+1} x'' G_{D+1}(x-x'') J(x'') .
\end{align}
The Green's function $G_{D+1}(x,x')$ itself satisfies the following PDE:
\begin{align}
\label{ScalarWaveEquation_GreensFunction_PDE}
\partial^2_x G_{D+1}(x,x') = \partial^2_{x'} G_{D+1}(x,x') = \delta^{(D+1)}(x-x') = \delta(t-t') \delta^{(D)}\left( \vec{x} - \vec{x}' \right).
\end{align}
This is why we call it the Green's function. Like its counterpart for the Poisson equation, we can view $G_{D+1}$ as the inverse of the wave operator. A short calculation using the Fourier representation in eq. \eqref{ScalarWaveEquation_GreensFunction_ModeExpansion} will verify eq. \eqref{ScalarWaveEquation_GreensFunction_PDE}. If $\partial^2$ denotes the wave operator with respect to either $x$ or $x'$, and if we recall the eigenvalue equation \eqref{ScalarWaveEquation_PlaneWaves_Eigenvectors} as well as the integral representation of the $\delta$-function,
\begin{align}
\partial^2 G_{D+1}(x-x') 
&= \int_{\mathbb{R}^{D+1}} \frac{\dd^{D+1}k}{(2\pi)^{D+1}} \frac{\partial^2 e^{-ik_\mu(x-x')^\mu}}{-k^2} \nonumber\\
&= \int_{\mathbb{R}^{D+1}} \frac{\dd^{D+1}k}{(2\pi)^{D+1}} \frac{-k^2 e^{-ik_\mu(x-x')^\mu}}{-k^2} = \delta^{(D+1)}(x-x')  .
\end{align}
{\bf Observer and Source, $G_{D+1}$ as a field by a point source} \qquad If we compare $\delta^{(D+1)}(x-x')$ in the wave equation obeyed by the Green's function itself (eq. \eqref{ScalarWaveEquation_GreensFunction_PDE}) with that of an external source $J$ in the wave equation for $\psi$ (eq. \eqref{ScalarWaveEquation_FlatSpacetime}), we see $G_{D+1}(x,x')$ itself admits the interpretation that it is the field observed at the spacetime location $x$ produced by a spacetime point source at $x'$. According to eq. \eqref{ScalarWaveEquation_GConvolvedWithJ}, the $\psi(t,\vec{x})$ is then the superposition of the fields due to all such spacetime points, weighted by the physical source $J$. (For a localized $J$, it sweeps out a world tube in spacetime -- try drawing a spacetime diagram to show how its segments contribute to the signal at a given $x$.)

{\bf Contour prescriptions and causality} \qquad From your experience with the mode sum expansion you may already have guessed that the Green's function for the wave operator $\partial^2$, obeying eq. \eqref{ScalarWaveEquation_GreensFunction_PDE}, admits the mode sum expansion in eq. \eqref{ScalarWaveEquation_GreensFunction_ModeExpansion}. However, you will soon run into a stumbling block if you begin with the $k^0 = \omega$ integral, because the denominator of the second line of eq. \eqref{ScalarWaveEquation_GreensFunction_ModeExpansion} gives rise to two singularities on the real line at $\omega=\pm |\vec{k}|$. To ensure the mode expansion in eq. \eqref{ScalarWaveEquation_GreensFunction_ModeExpansion} is well defined, we would need to append to it an appropriate contour prescription for the $\omega$-integral. It will turn out that, each distinct contour prescription will give rise to a Green's function with distinct causal properties.

On the complex $\omega$-plane, we can choose to avoid the singularities at $\omega= \pm|\vec{k}|$ by 
\begin{enumerate}
\item Making a tiny semi-circular clockwise contour around each of them. This will yield the {\it retarded Green's function} $G_{D+1}^+$, where signals from the source propagate forward in time; observers will see signals only from the past.
\item Making a tiny semi-circular counterclockwise contour around each of them. This will yield the {\it advanced Green's function} $G_{D+1}^-$, where signals from the source propagate backward in time; observers will see signals only from the future.
\item Making a tiny semi-circular counterclockwise contour around $\omega = -|\vec{k}|$ and a clockwise one at $\omega=+|\vec{k}|$. This will yield the Feynman Green's function $G_{D+1,F}$, named after the theoretical physicist Richard P. Feynman. The Feynman Green's function is used heavily in Minkowski spacetime perturbative Quantum Field Theory. Unlike its retarded and advanced cousins -- which are purely real -- the Feynman Green's function is complex. The real part is equal to half the advanced plus half the retarded Green's functions. The imaginary part, in the quantum field theory context, describes particle creation by an external source.
\end{enumerate}
These are just 3 of the most commonly used contour prescriptions -- there are an infinity of others, of course. You may also wonder if there is a heat kernel representation of the Green's function of the Minkowski spacetime wave operator, i.e., the generalization of eq. \eqref{GfromHeatKernel} to ``spacetime Laplacians". The subtlety here is that the eigenvalues of $\partial^2$, the $\{-k^2\}$, are not positive definite; to ensure convergence of the proper time $t$-integral in eq. \eqref{GfromHeatKernel} one would in fact be lead to the Feynman Green's function.

For classical physics, we will focus mainly on the retarded Green's function $G_{D+1}^+$ because it obeys causality -- the cause (the source $J$) precedes the effect (the field it generates). We will see this explicitly once we work out the $G_{D+1}^+$ below, for all $D \geq 1$.

To put the issue of contours on concrete terms, let us tackle the 2 dimensional case. Because the Green's function enjoys the spacetime translation symmetry of the Minkowski spacetime it resides in -- namely, under the simultaneous replacements $x^\mu \to x^\mu + a^\mu$ and $x'^\mu \to x'^\mu + a^\mu$, the Green's function remains the same object -- without loss of generality we may set $x'=0$ in eq. \eqref{ScalarWaveEquation_GreensFunction_ModeExpansion}.
\begin{align}
G_2\left( x^\mu=(t,z) \right) = -\int \frac{\dd\omega}{2\pi} \int \frac{\dd k}{2\pi} \frac{e^{-i\omega t} e^{ikz}}{\omega^2 - k^2} 
\end{align}
If we make the retarded contour choice, which we will denote as $G_2^+$, then if $t<0$ we would close it in the upper half plane (recall $e^{-i(i\infty)(-|t|)} = 0$). Because there are no poles for Im$(\omega) > 0$, we'd get zero. If $t>0$, on the other hand, we will form the closed (clockwise) contour $C$ via the lower half plane, and pick up the resides at both poles. We begin with a partial fractions decomposition of $1/k^2$, followed by applying the residue theorem:
\begin{align}
G_2^+\left( t,z \right) 
&= -i \Theta(t) \oint_C \frac{\dd\omega}{2\pi i} \int \frac{\dd k}{2\pi} e^{-i\omega t} \frac{e^{ikz}}{2k}
\left( \frac{1}{\omega-k} - \frac{1}{\omega+k} \right) \\
&= +i \Theta(t) \int \frac{\dd k}{2\pi} \frac{e^{ikz}}{2k} \left( e^{-ikt} - e^{ikt} \right) \nonumber\\
&= -i \Theta(t) \int \frac{\dd k}{2\pi} \frac{e^{ikz}}{2k} \cdot 2i \sin(kt)
= \Theta(t) \int \frac{\dd k}{2\pi} \frac{e^{ikz}}{k} \sin(kt)
\end{align}
At this point, let us note that, if we replace $z \to -z$,
\begin{align}
G_2^+\left( t,-z \right) 
&= \Theta(t) \int \frac{\dd k}{2\pi} \frac{e^{-ikz}}{k} \cdot \sin(kt) = G_2\left( t,z \right)^* \\
&= \Theta(t) \int \frac{\dd k}{2\pi} \frac{e^{i(-k)z}}{(-k)} \cdot \sin((-k)t) = G_2(t,z) .
\end{align}
Therefore not only is $G_2(t,z)$ real, we can also put an absolute value around the $z$ -- the answer for $G_2$ has to be the same whether $z$ is positive or negative anyway. Using the identity $\cos(a) \sin(b) = (1/2)(\sin(a+b)-\sin(a-b))$,
\begin{align}
G_2^+\left( t,z \right) 
&= \Theta(t) \int \frac{\dd k}{2\pi} \frac{\cos(kz)}{k} \cdot \sin(kt) \qquad \text{($G_2$ is real)} \\
&= \frac{1}{2}\Theta(t) I(t,z) ,
\end{align}
where
\begin{align}
I(t,z) \equiv \int \frac{\dd k}{2\pi} \frac{\sin(k(t+|z|)) + \sin(k(t-|z|))}{k} .
\end{align}
We differentiate once with respect to time and obtain the differential equation
\begin{align}
\label{ScalarWaveEquation_GreensFunction_2D_I}
\partial_t I(t,z) 
&= \int \frac{\dd k}{2\pi} \left(\cos(k(t+|z|)) + \cos(k(t-|z|))\right) \nonumber\\
&= \delta(t+|z|)+\delta(t-|z|) . 
\end{align}
Note that, because sine is an odd function, the integral representation of the $\delta$-function really only involves cosine.
\begin{align}
\int \frac{\dd k}{2\pi} e^{ikz} = \int \frac{\dd k}{2\pi} \cos(kz) = \delta(z) 
\end{align}
We use the distributional identity
\begin{align}
\delta(f(z)) = \sum_{z_i \in \{\text{zeroes of }f(z)\}} \frac{\delta(z-z_i)}{|f'(z_i)|}
\end{align}
to re-express the $\delta$-functions in eq. \eqref{ScalarWaveEquation_GreensFunction_2D_I} as
\begin{align}
\label{deltasigmabar}
\delta(\bar{\sigma}) &= \frac{\delta(t-|z|)}{|t|} + \frac{\delta(t+|z|)}{|t|} 
= \frac{\delta(t-|z|) + \delta(t+|z|)}{|z|} , \qquad 
\bar{\sigma} \equiv \frac{t^2-z^2}{2} , \nonumber\\
\delta(\bar{\sigma}) \cdot |t| &= \delta(t-|z|) + \delta(t+|z|) .
\end{align}
Moreover, because $\partial_t \text{sgn}(t) = \partial_t (\Theta(t)-\Theta(-t)) = 2\delta(t)$,
\begin{align}
\partial_t \left\{ \text{sgn}(t) \Theta(\bar{\sigma}) \right\} 
&= \delta(\bar{\sigma}) \cdot |t| + 2 \delta(t) \Theta(\bar{\sigma}) \nonumber\\
&= \delta(t-|z|) + \delta(t+|z|) + 2 \delta(t) \Theta(\bar{\sigma}) .
\end{align}
In the first equality, the first term contains $|t|$ because sgn$(t) \cdot t = t = |t|$ when $t>0$; and $t<0$, sgn$(t) \cdot t = -t = |t|$. The second term $\delta(t) \Theta(\bar{\sigma}) = \delta(t) \Theta(-z^2/2)$ is zero because we will never set $(t,z)=(0,0)$. The solution to the first order differential equation in eq. \eqref{ScalarWaveEquation_GreensFunction_2D_I} is thus
\begin{align}
I(t,z) &= \text{sgn}(t) \Theta(\bar{\sigma}) + C(z) ,
\end{align}
where $C(z)$ is a time independent but possibly $z$-dependent function. But the following boundary condition says
\begin{align}
I(t=0,z) = \int \frac{\dd k}{2\pi} \frac{\sin(k|z|) - \sin(k|z|)}{k} = 0.
\end{align}
Therefore $C(z)=0$ and we have obtained the solution to $G_2$.
\begin{align}
\label{MinkowskiG_2Plus}
G_2^+\left( x-x' \right) 
&= \frac{1}{2}\Theta(t-t') \Theta(\bar{\sigma}), \qquad \bar{\sigma} \equiv \frac{(t-t')^2-(z-z')^2}{2} = \frac{1}{2}(x-x')^2 .
\end{align}
(The $\Theta(t)$ sets $\text{sgn}(t) = 1$; and we have restored $x \to x-x'$.) While the $\Theta(\bar{\sigma})$ allows the signal due to the spacetime point source at $x'$ to propagate both forward and backward in time -- actually, throughout the interior of the light cone of $x'$ -- the $\Theta(t-t')$ implements retarded boundary conditions: the observer time $t$ always comes after the emission time $t'$. If you carry out a similar analysis for $G_2$ but for the advanced contour, you would find
\begin{align}
\label{MinkowskiG_2Minus}
G_2^-\left( x-x' \right) &= \frac{1}{2}\Theta(t'-t) \Theta(\bar{\sigma}) .
\end{align} 
\begin{myP}
\qquad From its Fourier representation , calculate $G_3^\pm(x-x')$, the retarded and advanced Green's function of the wave operator in 3 dimensional Minkowski spacetime. You should find
\begin{align}
\label{MinkowskiG_3}
G_3^\pm(x-x') = \frac{\Theta(\pm(t-t'))}{\sqrt{2} (2\pi)} \frac{\Theta(\bar{\sigma})}{\sqrt{\bar{\sigma}}} .
\end{align} 
{\it Bonus problem}: Can you perform the Fourier integral in eq. \eqref{ScalarWaveEquation_GreensFunction_ModeExpansion} for all $G_{D+1}$? \qed
\end{myP}
{\bf Green's Functions From Recursion Relations} \qquad With the 2 and 3 dimensional Green's function under our belt, I will now show how we can generate the Green's function of the Minkowski wave operator in all dimensions, just by differentiating $G_{2,3}$. The primary observation that allow us to do so, is that a line source in $(D+2)$ spacetime is a point source in $(D+1)$ dimensions; and a plane source in $(D+2)$ spacetime is a point source in $D$ dimensions.\footnote{I will make this statement precise very soon, by you are encouraged to read H. Soodak and M. S. Tiersten, {\it Wakes and waves in N dimensions}, Am. J. Phys. 61 (395), May 1993, for a pedagogical treatment.}

For this purpose let's set the notation. In $(D+1)$ dimensional flat spacetime, let the spatial coordinates be denoted as $x^i = (\vec{x}_\perp,w^1,w^2)$; and in $(D-1)$ dimensions let the spatial coordinates be the $\vec{x}_\perp$. Then $|\vec{x}-\vec{x}'|$ is a $D$ dimensional Euclidean distance between the observer and source in the former, whereas $|\vec{x}_\perp - \vec{x}'_\perp|$ is the $D-1$ counterpart in the latter.

Starting from the integral representation for $G_{D+1}$ in eq. \eqref{ScalarWaveEquation_GreensFunction_ModeExpansion}, we may integrate with respect to the $D$th spatial coordinate $w^2$:
{\allowdisplaybreaks\begin{align}
\label{MinkowskiG_Recursion_LineSource}
&\int_{-\infty}^{+\infty} \dd w'^2 G_{D+1}(t-t',\vec{x}_\perp-\vec{x}'_\perp,\vec{w}-\vec{w}') \nonumber\\
&= \int_{-\infty}^{+\infty} \dd w'^2 \int_{\mathbb{R}^{D+1}} \frac{\dd\omega \dd^{D-2}k_\perp \dd^2 k_\parallel}{(2\pi)^{D+1}} 
		\frac{e^{-i\omega(t-t')} e^{i\vec{k}_\perp\cdot(\vec{x}_\perp-\vec{x}'_\perp)} e^{ik_\parallel \cdot (\vec{w}-\vec{w}')}}{-\omega^2 + \vec{k}_\perp^2 + \vec{k}_\parallel^2 } \nonumber\\
&= \int_{\mathbb{R}^{D+1}} \frac{\dd\omega \dd^{D-2}k_\perp \dd^2 k_\parallel}{(2\pi)^{D+1}} (2\pi)\delta(k_\parallel^2)
\frac{e^{-i\omega(t-t')} e^{i\vec{k}_\perp\cdot(\vec{x}_\perp-\vec{x}'_\perp)} e^{i k_\parallel^1 (w^1-w'^1)} e^{ik_\parallel^2 w^2} }{-\omega^2 + \vec{k}_\perp^2 + \vec{k}_\parallel^2 } \nonumber\\
&= \int_{\mathbb{R}^{D}} \frac{\dd\omega \dd^{D-2}k_\perp \dd k_\parallel^1}{(2\pi)^{D}} 
\frac{e^{-i\omega(t-t')} e^{i\vec{k}_\perp\cdot(\vec{x}_\perp-\vec{x}'_\perp)} e^{ik^1_\parallel(w^1-w'^1)} }{-\omega^2 + \vec{k}_\perp^2 + (k_\parallel^1)^2 } \nonumber\\
&= G_{D}(t-t',\vec{x}_\perp-\vec{x}'_\perp,w^1-w'^1)  .
\end{align}}
The notation is cumbersome, but the math can be summarized as follows. Integrating $G_{D+1}$ over the $D$th spatial coordinate amounts to discarding the momentum integral with respect to its $D$ component and setting its value to zero everywhere in the integrand. But that is nothing but the integral representation of $G_D$. Moreover, because of translational invariance, we could have integrated with respect to either $w'^2$ or $w^2$. If we compare our integral here with eq. \eqref{ScalarWaveEquation_GConvolvedWithJ}, we may identify $J(x'') = \delta(t''-t') \delta^{(D-2)}(\vec{x}'_\perp - \vec{x}''_\perp) \delta(w^1-w''^1)$, an instantaneous line source of unit strength lying parallel to the $D$th axis, piercing the $(D-1)$ space at $(\vec{x}'_\perp,w'^1)$. 

We may iterate this integral recursion relation once more,
\begin{align}
\int_{\mathbb{R}^2} \dd^2 w G_{D+1}\left( t-t',\vec{x}_\perp-\vec{x}'_\perp,\vec{w}-\vec{w}' \right) 
= G_{D-1}\left( t-t',\vec{x}_\perp-\vec{x}'_\perp \right) . 
\end{align}
This is saying $G_{D-1}$ is sourced by a 2D plane of unit strength, lying in $(D+1)$ spacetime. On the left hand side, we may employ cylindrical coordinates to perform the integral
\begin{align}
2\pi \int_0^\infty \dd \rho \rho G_{D+1}\left( t-t',\sqrt{(\vec{x}_\perp-\vec{x}'_\perp)^2 + \rho^2} \right) 
= G_{D-1}\left( t-t',|\vec{x}_\perp-\vec{x}'_\perp| \right) ,
\end{align}
where we are now highlighting the fact that, the Green's function really has only two arguments: one, the time elapsed $t-t'$ between observation $t$ and emission $t'$; and two, the Euclidean distance between observer and source. (We will see this explicitly very shortly.) For $G_{D+1}$ the relevant Euclidean distance is
\begin{align}
|\vec{x}-\vec{x}'| = \sqrt{(\vec{x}_\perp-\vec{x}'_\perp)^2 + (\vec{w}-\vec{w}')^2} .
\end{align}
A further change of variables
\begin{align}
R' \equiv \sqrt{(\vec{x}_\perp-\vec{x}'_\perp)^2 + \rho^2} \qquad \Rightarrow \qquad
\dd R' = \frac{\rho \dd \rho}{R'} .
\end{align}
This brings us to
\begin{align}
2\pi \int_R^\infty \dd R' R' G_{D+1}(t-t',R') = G_{D-1}(t-t',R) .
\end{align}
At this point we may differentiate both sides with respect to $R$ (see Leibniz's rule for differentiation), to obtain the Green's function in $(D+1)$ dimensions from its counterpart in $(D-1)$ dimensions.
\begin{align}
\label{MinkowskiG_Recursion}
G_{D+1}(t-t',R) = -\frac{1}{2\pi R} \frac{\partial}{\partial R} G_{D-1}(t-t',R) .
\end{align}
The meaning of $R$ on the left hand side is the $D$-space length $|\vec{x}-\vec{x}'|$; on the right hand side it is the $(D-2)$-space length $|\vec{x}_\perp-\vec{x}'_\perp|$.

{\bf Green's Function From Extra Dimensional Line Source} \qquad There is an alternate means of obtaining the integral relation in eq. \eqref{MinkowskiG_Recursion_LineSource}, which was key to deriving eq. \eqref{MinkowskiG_Recursion}. In particular, it does not require explicit use of the Fourier integral representation. Let us postulate that $G_D$ is sourced by a ``line charge" $J(w^2)$ extending in the extra spatial dimension of $\mathbb{R}^{D,1}$.
\begin{align}
G_{D}(t-t',\vec{x}_\perp - \vec{x}'_\perp,w^1-w'^1)
&\stackrel{?}{=} \int_{-\infty}^{+\infty} \dd w'^2 G_{D+1}(t-t',\vec{x}_\perp - \vec{x}'_\perp,\vec{w}-\vec{w}') J(w'^2) 
\end{align}
Applying the wave operator in the $((D-1)+1)$-space on the right hand side, and suppressing arguments of the Green's function whenever convenient,
{\allowdisplaybreaks\begin{align}
&\partial_D^2 \int_{-\infty}^{+\infty} \dd w'^2 G_{D+1} \cdot J \qquad\qquad\qquad \left( \text{where } \partial^2_D \equiv \partial_{t'}^2 - \sum_{i=1}^{D-1} \partial_{i'}^2 \right) \nonumber\\
&\qquad = \int_{-\infty}^{+\infty} \dd w'^2 J(w'^2) \left( \partial_D^2 - \left(\frac{\partial}{\partial w'^2}\right)^2 +  \left(\frac{\partial}{\partial w'^2}\right)^2\right) G_{D+1}(w^2-w'^2) \nonumber\\
&\qquad = \int_{-\infty}^{+\infty} \dd w'^2 J(w'^2) \left( \partial^2_{D+1} + \left(\frac{\partial}{\partial w'^2}\right)^2\right) G_{D+1}(w^2-w'^2) \nonumber\\
&\qquad = \int_{-\infty}^{+\infty} \dd w'^2 J(w'^2) \left( \delta(t-t') \delta^{(D-2)}(\vec{x}_\perp - \vec{x}_\perp') \delta^{(2)}(\vec{w} - \vec{w}') + \left(\frac{\partial}{\partial w'^2}\right)^2 G_{D+1}(w^2-w'^2) \right) \nonumber\\
\label{MinkowskiG_Recursion_LineSource_v2}
&\qquad = \delta^{(D-1)}(x-x') \delta(w^1-w'^1) J(w^2) \nonumber\\
&\qquad\qquad\qquad
	+ \left[ J(w'^2) \frac{\partial G_{D+1}(w^2-w'^2)}{\partial w'^2} \right]_{w'^2=-\infty}^{w'^2=+\infty} 
	- \left[ \frac{\partial J(w'^2)}{\partial w'^2} G_{D+1}(w^2-w'^2) \right]_{w'^2=-\infty}^{w'^2=+\infty} \nonumber \\
&\qquad\qquad\qquad + \int_{-\infty}^{+\infty} \dd w'^2 J''(w'^2) G_{D+1}(w^2-w'^2) .
\end{align}}
That is, we would have verified the $((D-1)+1)$ flat space wave equation is satisfied if only the first term in the final equality survives. Moreover, that it needs to yield the proper $\delta$-function measure, namely $\delta^{(D-1)}(x-x') \delta(w^1-w'^1)$, translates to the boundary condition on $J$:
\begin{align}
\label{MinkowskiG_Recursion_LineSource_v2BC}
J(w^2) = 1 .
\end{align}
That the second and third terms of the final equality of eq. \eqref{MinkowskiG_Recursion_LineSource_v2} are zero, requires knowing causal properties of the Green's function: in particular, because the $w'^2 = \pm \infty$ limits correspond to sources infinitely far away from the observer at $(\vec{x}_\perp,w^1,w^2)$, they must lie outside the observer's light cone, where the Green's function is identically zero. The final term of eq. \eqref{MinkowskiG_Recursion_LineSource_v2} is zero if the source obeys the ODE
\begin{align}
\label{MinkowskiG_Recursion_LineSource_v2ODE}
0 = J''(w'^2) .
\end{align}
The solutions of eq. \eqref{MinkowskiG_Recursion_LineSource_v2ODE}, subject to eq. \eqref{MinkowskiG_Recursion_LineSource_v2BC}, are
\begin{align}
J(w'^2) = 1 \qquad \text{ or } \qquad J(w'^2) = \frac{w'^2}{w^2} .
\end{align}
We have deduced the Green's function in $D+1$ dimensions $G_{D+1}$ may be sourced by a line source of two distinct charge densities extending in the extra spatial dimension of $\mathbb{R}^{D+1,1}$.
\begin{align}
G_{D}(t-t',\vec{x}_\perp-\vec{x}'_\perp,w^1-w'^1)
&= \int_{-\infty}^{+\infty} \dd w'^2 G_{D+1}(t-t',\vec{x}_\perp-\vec{x}'_\perp, \vec{w}-\vec{w}')  \\
&= \int_{-\infty}^{+\infty} \dd w'^2 \frac{w'^2}{w^2} G_{D+1}(t-t',\vec{x}_\perp-\vec{x}'_\perp, \vec{w}-\vec{w}') 
\end{align}
As a reminder, $\vec{x}_\perp$ and $\vec{x}'_\perp$ are $D-1$ dimensional spatial coordinates; whereas $\vec{w}$ and $\vec{w}'$ are two dimensional ones.

{\bf $G_{D+1}^\pm$ in all dimensions, Causal structure of physical signals} \qquad At this point we may gather $G_{2,3}^\pm$ in equations \eqref{MinkowskiG_2Plus}, \eqref{MinkowskiG_2Minus}, and \eqref{MinkowskiG_3} and apply to them the recursion relation in eq. \eqref{MinkowskiG_Recursion} to record the explicit expressions of the retarded $G_{D+1}^+$ and advanced $G_{D+1}^-$ Green's functions in all $(D \geq 2)$ dimensions.\footnote{When eq. \eqref{MinkowskiG_Recursion} applied to $G_{2,3}^\pm$ in equations \eqref{MinkowskiG_2Plus}, \eqref{MinkowskiG_2Minus}, and \eqref{MinkowskiG_3}, note that the $(2\pi R)^{-1}\partial_R$ passes through the $\Theta(\pm(t-t'))$ and because the rest of the $G_{2,3}^\pm$ depends solely on $\sbar$, it becomes $(2\pi R)^{-1}\partial_R = (2\pi)^{-1} \partial_{\sbar}$.}
\begin{itemize}
\item In even dimensional spacetimes, $D+1 = 2+2n$ and $n=0,1,2,3,4,\dots$,
\begin{align}
\label{MinkowskiG_Even}
G_{2+2n}^\pm(x-x') 
= \Theta\left(\pm (t-t')\right) \left( \frac{1}{2\pi} \frac{\partial}{\partial \sbar} \right)^n \frac{\Theta(\sbar)}{2} .
\end{align}
\item In odd dimensional spacetime, $D+1 = 3+2n$ and $n=0,1,2,3,4,\dots$,
\begin{align}
\label{MinkowskiG_Odd}
G_{3+2n}^\pm(x-x') 
= \Theta\left(\pm (t-t')\right) \left( \frac{1}{2\pi} \frac{\partial}{\partial \sbar} \right)^n \left(\frac{\Theta(\sbar)}{2\pi\sqrt{2\bar{\sigma}}}\right).
\end{align}
\end{itemize}
Recall that $\sbar(x,x')$ is half the square of the geodesic distance between the observer at $x$ and point source at $x'$,
\begin{align}
\bar{\sigma} \equiv \frac{1}{2}(x-x')^2 .
\end{align}
Hence, $\Theta(\sbar)$ is unity inside the light cone and zero outside; whereas $\delta(\sbar)$ and its derivatives are non-zero strictly on the light cone. Note that the inside-the-light-cone portion of a signal -- for e.g., the $\Theta(\sbar)$ term of the Green's function -- is known as the tail. Notice too, the $\Theta(\pm (t-t'))$ multiplies an expression that is symmetric under interchange of observer and source ($x \leftrightarrow x'$). For a fixed source at $x'$, we may interpret these coefficients of $\Theta(\pm (t-t'))$ as the symmetric Green's function: the field due to the source at $x'$ travels both backwards and forward in time. The retarded $\Theta(t-t')$ (observer time is later than emission time) selects the future light cone portion of this symmetric signal; while the advanced $\Theta(-(t-t'))$ (observer time earlier than emission time) selects the backward light cone part of it.

As already advertised earlier, because the Green's function of the scalar wave operator in Minkowski is the field generated by a unit strength point source in spacetime -- the field $\psi$ generated by an arbitrary source $J(t,\vec{x})$ obeys causality. By choosing the {\it retarded} Green's function, the field generated by the source propagates on and possibly within the forward light cone of $J$. Specifically, $\psi$ travels strictly on the light cone for even dimensions greater or equal to $4$, because $G_{D+1=2n}$ involves only $\delta(\bar{\sigma})$ and its derivatives. In 2 dimensions, the Green's function is pure tail, and is in fact a constant $1/2$ inside the light cone. In 3 dimensions, the Green's function is also pure tail, going as $\bar{\sigma}^{-1/2}$ inside the light cone. For odd dimensions greater than 3, the Green's function has non-zero contributions from both on and inside the light cone. However, the $\partial_{\sbar}$s occurring within eq. \eqref{MinkowskiG_Odd} can be converted into $\partial_{t'}$s and -- at least for material/timelike $J$ -- integrated-by-parts within the integral in eq. \eqref{ScalarWaveEquation_GConvolvedWithJ} to act on the $J$. The result is that, in all odd dimensional Minkowski spacetimes ($d \geq 3$), physical signals propagate strictly inside the null cone, despite the massless nature of the associated particles.\footnote{Explicit formulas for the electromagnetic and linear gravitational case can be found in appendices A and B of \href{https://arxiv.org/abs/1611.00018}{arXiv: 1611.00018} \cite{Chu:2016ngc}.}

{\bf Comparison to heat equation} \qquad The causal structure of the solutions to the wave equation here can be contrasted against those of the infinite flat space heat equation. Referring to the heat kernel in eq. \eqref{HeatDiffusionEquation_HeatKernel_FlatSpace}, we witness how at initial time $t'$, the field $K$ is infinitely sharply localized at $\vec{x}=\vec{x}'$. However, immediately afterwards, it becomes spread out over all space, with a Gaussian profile peaked at $\vec{x} = \vec{x}'$ -- thereby violating causality. In other words, the ``waves" in the heat/diffusion equation of eq. \eqref{HeatDiffusionEquation} propagates with infinite speed. Physically speaking, we may attribute this property to the fact that time and space are treated asymmetrically both in the heat/diffusion eq. \eqref{HeatDiffusionEquation} itself -- one time derivative versus two derivatives per spatial coordinate -- as well as in the heat kernel solution of eq. \eqref{HeatDiffusionEquation_HeatKernel_FlatSpace}. On the other hand, the symmetric portion of the spacetime Green's functions in equations \eqref{MinkowskiG_Even} and \eqref{MinkowskiG_Odd} depend on spacetime solely through $2 \sbar \equiv (t-t')^2 - (\vec{x}-\vec{x}')^2$, which is invariant under global Poincar\'{e} transformations (cf. eq. \eqref{PoincareTransformations}).

{\bf 4 dimensions: Massless Scalar Field} \qquad We highlight the 4 dimensional retarded case, because it is most relevant to the real world. Using eq. \eqref{deltasigmabar} after we recognize $\Theta'(\bar{\sigma}) = \delta(\bar{\sigma})$, 
\begin{align}
\label{ScalarWaveEquation_GreensFunction_4D}
G_4^+(x-x') = \frac{\delta\left( t-t'-|\vec{x}-\vec{x}'| \right)}{4\pi |\vec{x}-\vec{x}'|} .
\end{align}
The $G_4$ says the point source at $(t',\vec{x}')$ produces a spherical wave that propagates strictly on the light cone $t-t'=|\vec{x}-\vec{x}'|$, with amplitude that falls off as 1/(observer-source spatial distance) $= 1/|\vec{x}-\vec{x}'|$. There is another term involving $\delta\left( t-t'+|\vec{x}-\vec{x}'| \right)$, but for this to be non-zero $t-t' = - |\vec{x}-\vec{x}'| < 0$; this is not allowed by the $\Theta(t-t')$.

The solution to $\psi$ from eq. \eqref{ScalarWaveEquation_GConvolvedWithJ} is now
\begin{align}
\label{ScalarWaveEquation_Solution_4D}
\psi(t,\vec{x}) 
&= \int_{-\infty}^{+\infty} \dd t' \int_{\mathbb{R}^3} \dd^3 \vec{x}' G_4^+(t-t',\vec{x}-\vec{x}') J(t',\vec{x}') \nonumber\\
&= \int_{-\infty}^{+\infty} \dd t' \int_{\mathbb{R}^3} \dd^3 \vec{x}' \frac{\delta\left( t-t'-|\vec{x}-\vec{x}'| \right) J(t',\vec{x}')}{4\pi |\vec{x}-\vec{x}'|} \nonumber\\
&= \int_{\mathbb{R}^3} \dd^3 \vec{x}' \frac{J(t_r,\vec{x}')}{4\pi |\vec{x}-\vec{x}'|}, \qquad t_r \equiv t-|\vec{x}-\vec{x}'| .
\end{align}
The $t_r$ is called retarded time. With $c=1$, the time it takes for a signal traveling at unit speed to travel from $\vec{x}'$ to $\vec{x}$ is $|\vec{x}-\vec{x}'|$, and so at time $t$, what the observer detects at $(t,\vec{x})$ is what the source produced at time $t-|\vec{x}-\vec{x}'|$.

{\it Far Zone \& Non-Relativistic Source} \qquad Let us center the coordinate system so that $\vec{x}=\vec{x}'=\vec{0}$ lies within the body of the source $J$ itself. When the observer is located at very large distances from the source compared to the latter's characteristic size, we may approximate
\begin{align}
|\vec{x}-\vec{x}'| 
&= e^{-x'^j \partial_j} |\vec{x}| \nonumber\\
\label{ScalarWaveEquation_LargeDistanceExpansion}
&= |\vec{x}| - \vec{x}' \cdot \widehat{x} + |\vec{x}| \mathcal{O}\left( \left(\frac{|\vec{x}'|}{|\vec{x}|}\right)^2 \right) ,
\qquad\qquad
\widehat{x} \equiv \frac{x^i}{|\vec{x}|} \\
&= |\vec{x}| - \vec{x}' \cdot \widehat{x} + |\vec{x}'| \mathcal{O}\left( \frac{|\vec{x}'|}{|\vec{x}|} \right)
\end{align}
This leads us from eq. \eqref{ScalarWaveEquation_Solution_4D} to the following far zone scalar solution
\begin{align}
&\psi(t,\vec{x}) 
\label{ScalarWaveEquation_Solution_LargeDistanceExpansion}
= \frac{1}{4\pi |\vec{x}|} \int_{\mathbb{R}^3} \dd^3 \vec{x}' \left\{ 1 + \frac{\vec{x}'}{|\vec{x}|} \cdot \widehat{x} + \mathcal{O}\left( \left(\frac{|\vec{x}'|}{|\vec{x}|}\right)^2 \right) \right\} \\
&\times \left\{ J\left(t-|\vec{x}|,\vec{x}'\right) - \left\{ (\vec{x}' \cdot \widehat{x}) + |\vec{x}'| \mathcal{O}\left( \frac{|\vec{x}'|}{|\vec{x}|} \right) \right\} \partial_t J\left(t-|\vec{x}|,\vec{x}'\right) 
+ \mathcal{O}\left( \left(\vec{x}' \cdot \widehat{x}\right)^2 \right) \ddot{J} + \dots \right\} . \nonumber
\end{align}
We see that the corrections to the leading order term scales as either (characteristic size of source)/(observer-source spatial distance) or (characteristic size of source)/(timescale of source); where the former is from eq. \eqref{ScalarWaveEquation_LargeDistanceExpansion} and the latter from the $(\vec{x}'\cdot\vec{x})\dot{J}$ in eq. \eqref{ScalarWaveEquation_Solution_LargeDistanceExpansion}. Therefore, in the far zone but without assuming the source is non-relativistic,
\begin{align}
\psi(t,\vec{x}) 
&\approx \frac{1}{4\pi |\vec{x}|} \int_{\mathbb{R}^3} \dd^3 \vec{x}' 
J\left( t-|\vec{x}| + \vec{x}'\cdot\widehat{x},\vec{x}' \right) .
\end{align}
But if the source is non-relativistic -- namely (characteristic size of source)/(timescale of source) $\ll 1$ --
\begin{align}
\psi(t,\vec{x}) 			&\approx \frac{\mathcal{A}(t-|\vec{x}|)}{4\pi |\vec{x}|}, \\
\mathcal{A}(t-|\vec{x}|) 	&\equiv \int_{\mathbb{R}^3} \dd^3 \vec{x}' J(t-|\vec{x}|,\vec{x}') .
\end{align}
In the far zone and with a non-relativistic source: the amplitude of the wave falls off with increasing distance as $1/(\text{observer-source spatial distance})$; and the time-dependent portion of the wave $\mathcal{A}(t-|\vec{x}|)$ is consistent with that of an outgoing wave, one emanating from the source $J$.

{\bf 4D photons} \qquad In 4 dimensional flat spacetime, the vector potential of electromagnetism, in the Lorenz gauge
\begin{align}
\partial_\mu A^\mu = 0  \qquad \text{(Cartesian coordinates)},
\end{align}
obeys the wave equation
\begin{align}
\partial^2 A^\mu = J^\mu .
\end{align}
Here, $\partial^2$ is the scalar wave operator, and $J^\mu$ is a conserved electromagnetic current describing the motion of some charge density
\begin{align}
\partial_\mu J^\mu = \partial_t J^t + \partial_i J^i = 0.
\end{align}
The electromagnetic fields are the ``curl" of the vector potential
\begin{align}
F_{\mu\nu} = \partial_\mu A_\nu - \partial_\nu A_\mu .
\end{align}
In particular, for a given inertial frame, the electric $E$ and magnetic $B$ fields are, with $i,j,k \in \{1,2,3\}$,
\begin{align}
E^i &= \partial^i A^0 - \partial^0 A^i = -\partial_i A_0 + \partial_0 A_i = -F_{i0}, \\
B^k &= -\epsilon^{ijk} \partial_i A_j = -\frac{1}{2} \epsilon^{ijk} F_{ij}, \qquad \epsilon^{123} \equiv 1.
\end{align}
{\bf 4D gravitational waves} \qquad In a 4D weakly curved spacetime, the metric can be written as one deviating slightly from Minkowski,
\begin{align}
g_{\mu\nu} = \eta_{\mu\nu} + h_{\mu\nu} \qquad \text{(Cartesian coordinates)},
\end{align}
where the dimensionless components of $h_{\mu\nu}$ are assumed to be much smaller than unity.

The (trace-reversed) graviton
\begin{align}
\bar{h}_{\mu\nu} \equiv h_{\mu\nu} - \frac{1}{2} \eta_{\mu\nu} \eta^{\alpha\beta} h_{\alpha\beta} ,
\end{align}
in the de Donder gauge
\begin{align}
\partial^\mu \bar{h}_{\mu\nu} = \partial_t \bar{h}_{t\nu} - \delta^{ij} \partial_i \bar{h}_{j\nu} = 0,
\end{align}
obeys the wave equation\footnote{The following equation is only approximate; it comes from linearizing Einstein's equations about a flat spacetime background, i.e., where all terms quadratic and higher in $h_{\mu\nu}$ are discarded.}
\begin{align}
\partial^2 \bar{h}_{\mu\nu} = -16 \pi \GN T_{\mu\nu}  \qquad \text{(Cartesian coordinates)}.
\end{align}
(The $\GN$ is the same Newton's constant you see in Newtonian gravity $\sim \GN M_1 M_2/r^2$; both $\bar{h}_{\mu\nu}$ and $T_{\mu\nu}$ are symmetric.) The $T_{\mu\nu}$ is a $4 \times 4$ matrix describing the energy-momentum-shear-stress of matter, and has zero divergence (i.e., it is conserved)
\begin{align}
\partial_\mu T^{\mu\nu} = \partial_t T^{t\nu} + \partial_i T^{i\nu} = 0.
\end{align}
\begin{myP}
{\it Electromagnetic radiation zone} \qquad Using $G_4^+$ in eq. \eqref{ScalarWaveEquation_GreensFunction_4D}, write down the solution of $A^\mu$ in terms of $J^\mu$. Like the scalar case, take the far zone limit. In this problem we wish to study some basic properties of $A^\mu$ in this limit. Throughout this analysis, assume that $J^i$ is sufficiently localized that it vanishes at spatial infinity; and assume $J^i$ is a non-relativistic source.
\begin{enumerate}
\item Using $\partial_t J^t = - \partial_i J^i$, the conservation of the current, show that $A^0$ is independent of time in the far zone limit.
\item Now define the dipole moment as
\begin{align}
I^i(t) \equiv \int_{\mathbb{R}^3} \dd^3 \vec{x}' x'^i J^0(t,\vec{x}') .
\end{align}
Can you show its first time derivative is 
\begin{align}
\dot{I}^i(t) \equiv \frac{\dd I^i(t)}{\dd t} = \int_{\mathbb{R}^3} \dd^3 \vec{x}' J^i(t,\vec{x}') ?
\end{align}
\item From this, we shall infer it is $A^i$ that contains radiative effects. Remember the Poynting vector, which describes the direction and rate of flow of energy/momentum carried by electromagnetic waves, is proportional to $\vec{E} \times \vec{B}$. The energy density $\mathcal{E}$ is proportional to $\vec{E}^2 + \vec{B}^2$. Let's focus on the electric field $E^i$; it has to be non-zero for the Poynting vector to carry energy to infinity.
\begin{align}
E^i = \partial^i A^0 - \partial^0 A^i .
\end{align}
Show that in the far zone, it is the $- \partial^0 A^i$ term that dominates, and in particular
\begin{align}
\label{DipoleRadiation}
E^i &\to -\frac{1}{4\pi |\vec{x}|} \frac{\dd^2 I^i(t-|\vec{x}|)}{\dd t^2}  .
\end{align}
\item {\it Bonus problem:} Can you work out the far zone Poynting vector?
\end{enumerate}
Therefore the electric field energy on a $\dd r$ thick spherical shell centered at the source, is a constant as $r \to \infty$. Moreover it depends on the acceleration of the dipole moment evaluated at retarded time:
\begin{align}
\frac{\dd \mathcal{E}(\text{electric})}{\dd r}
\propto \frac{1}{4\pi} \left(\frac{\dd^2 \vec{I}(t-|\vec{x}|)}{\dd t^2}\right)^2 .
\end{align}
The non-zero acceleration of the dipole moment responsible for electromagnetic radiation indicates work needs to be done pushing around electric charges, i.e., forces are needed to give rise to acceleration. \qed
\end{myP}
\begin{myP}
{\it Gravitational radiation zone} \qquad Can you carry out a similar analysis for gravitational radiation? Using $G_4^+$ in eq. \eqref{ScalarWaveEquation_GreensFunction_4D}, write down the solution of $\bar{h}^{\mu\nu}$ in terms of $T^{\mu\nu}$. Then take the far zone limit. Throughout this analysis, assume that $T^{\mu\nu}$ is sufficiently localized that it vanishes at spatial infinity; and assume $T^{\mu\nu}$ is a non-relativistic source.
\begin{enumerate}
\item Using $\partial_t T^{t\nu} = - \partial_i T^{i\nu}$, the conservation of the stress-tensor, show that $\bar{h}^{\nu 0} = \bar{h}^{0\nu}$ is independent of time in the far zone limit.
\item Now define the quadrupole moment as
\begin{align}
I^{ij}(t) \equiv \int_{\mathbb{R}^3} \dd^3 \vec{x}' x'^i x'^j T^{00}(t,\vec{x}') .
\end{align}
Can you show its second time derivative is 
\begin{align}
\ddot{I}^{ij}(t) \equiv \frac{\dd^2 I^{ij}(t)}{\dd t^2} = 2 \int_{\mathbb{R}^3} \dd^3 \vec{x}' T^{ij}(t,\vec{x}') ?
\end{align}
and from it infer that the (trace-reversed) gravitational wave form in the far zone is proportional to the acceleration of the quadrupole moment evaluated at retarded time:
\begin{align}
\bar{h}^{ij}(t,\vec{x}) \to -\frac{2\GN}{|\vec{x}|} \frac{\dd^2 I^{ij}(t-|\vec{x}|)}{\dd t^2} .
\end{align}
Note that the (trace-reversed) gravitational wave $\bar{h}_{ij}(t,\vec{x})$ can be detected by how it squeezes and stretches arms of a laser interferometer such as \href{https://www.advancedligo.mit.edu/}{aLIGO} and \href{http://www.ego-gw.it/public/about/whatIs.aspx}{VIRGO}. Moreover, the non-zero acceleration of the quadrupole moment responsible for gravitational radiation indicates work needs to be done pushing around matter, i.e., forces are needed to give rise to acceleration.
\end{enumerate} \qed
\end{myP}
\begin{myP}
{\it Waves Around Schwarzschild Black Hole.} \qquad The geometry of a non-rotating black hole is described by
\begin{align}
\dd s^2
= \left( 1 - \frac{r_s}{r} \right) \dd t^2 - \frac{\dd r^2}{1 - \frac{r_s}{r}} - r^2 \left( \dd\theta^2 + \sin(\theta)^2 \dd\phi^2 \right) ,
\end{align}
where $x^\mu = (t \in \mathbb{R}, r \geq 0, 0 \leq \theta \leq \pi, 0 \leq \phi < 2\pi)$, and $r_s$ (proportional to the mass of the black hole itself) is known as the Schwarzschild radius -- nothing can fall inside the black hole ($r < r_s$) and still get out.

Consider the (massless scalar) homogeneous wave equation in this black hole spacetime, namely
\begin{align}
\Box \psi(t,r,\theta,\phi) = \nabla_\mu \nabla^\mu \psi = 0 .
\end{align}
Consider the following separation-of-variables ansatz
\begin{align}
\psi(t,r,\theta,\phi) = \int_{-\infty}^{+\infty} \frac{\dd \omega}{2\pi} e^{-i\omega t} \sum_{\ell=0}^{+\infty} \sum_{m = -\ell}^{+\ell} 
\frac{R_\ell(\omega r_*)}{r} Y_\ell^m(\theta,\phi) ,
\end{align}
where $\{ Y_\ell^m \}$ are the spherical harmonics on the $2$-sphere and the ``tortoise coordinate" is
\begin{align}
r_* \equiv r + r_s \ln\left(\frac{r}{r_s}-1\right) .
\end{align}
Show that the wave equation is reduced to an ordinary differential equation for the $\ell$th radial mode function
\begin{align}
R_\ell''(\xi_*)
+ \left( \frac{\xi_s^2}{\xi^4} +\frac{\left( \ell(\ell+1)-1\right) \xi_s}{\xi^3}-\frac{\ell  (\ell +1)}{\xi^2}+1\right) R_\ell(\xi_*) = 0 ,
\end{align}
where $\xi \equiv \omega r$, $\xi_s \equiv \omega r_s$ and $\xi_* \equiv \omega r_*$. 

An alternative route is to first perform the change-of-variables
\begin{align}
x \equiv 1 - \frac{\xi}{\xi_s} ,
\end{align}
and the change of radial mode function
\begin{align}
\frac{R_\ell(\xi_*)}{r} \equiv \frac{Z_\ell(x)}{\sqrt{x(1-x)}} .
\end{align}
Show that this returns the ODE
\begin{align}
\label{SchwarzschildConfluentHeun}
Z_\ell''(x) + \left(
\frac{1}{4 (x -1)^2} + \frac{1 + 4 \xi_s^2}{4 x ^2} + \xi_s^2 + \frac{2\ell(\ell+1) + 1 - 4 \xi_s^2}{2 x} - \frac{2 \ell(\ell+1) + 1}{2 (x -1)} \right) Z_\ell(x) = 0.
\end{align}
You may use {\sf Mathematica} or similar software to help you with the tedious algebra/differentiation; but make sure you explain the intermediate steps clearly.

The solutions to eq. \eqref{SchwarzschildConfluentHeun} are related to the confluent Heun function. For a recent discussion, see for e.g., \S I of \href{http://arxiv.org/abs/1510.06655}{arXiv: 1510.06655}. The properties of Heun functions are not as well studied as, say, the Bessel functions you have encountered earlier. This is why it is still a subject of active research -- see, for instance, the \href{http://theheunproject.org/}{Heun Project}. \qed
\end{myP}

\subsubsection{4D frequency space, Static limit, Discontinuous first derivatives}

{\bf Wave Equation in Frequency Space} \qquad We begin with eq. \eqref{WaveEquation_ParticularSolution_InverseFourier_I}, and translate it to frequency space.
{\allowdisplaybreaks\begin{align}
\psi(t,\vec{x})
&= \int_{-\infty}^{+\infty} \frac{\dd\omega}{2\pi} \widetilde{\psi}(\omega,\vec{x}) e^{-i\omega t} \nonumber\\
&= \int_{-\infty}^{+\infty} \dd t'' \int_{\mathbb{R}^D} \dd^D \vec{x}'' G_{D+1}(t-t'',\vec{x}-\vec{x}'') \int_{-\infty}^{+\infty} \frac{\dd\omega}{2\pi} \widetilde{J}(\omega,\vec{x}'') e^{-i\omega t''} \nonumber\\
&= \int_{-\infty}^{+\infty} \frac{\dd\omega}{2\pi} \int_{-\infty}^{+\infty} \dd (t-t'') e^{i\omega (t-t'')} e^{-i\omega t} \int_{\mathbb{R}^D} \dd^D \vec{x}'' G_{D+1}(t-t'',\vec{x}-\vec{x}'') \widetilde{J}(\omega,\vec{x}'') \nonumber\\
&= \int_{-\infty}^{+\infty} \frac{\dd\omega}{2\pi} e^{-i\omega t} \int_{\mathbb{R}^D} \dd^D \vec{x}'' \widetilde{G}^+_{D+1}(\omega,\vec{x}-\vec{x}'') \widetilde{J}(\omega,\vec{x}'') .
\end{align}}
Equating the coefficients of $e^{-i\omega t}$ on both sides,
\begin{align}
\label{ScalarWaveEquation_DrivenSHO}
\widetilde{\psi}(\omega,\vec{x}) 
&= \int_{\mathbb{R}^D} \dd^D \vec{x}'' \widetilde{G}^+_{D+1}(\omega,\vec{x}-\vec{x}'') \widetilde{J}(\omega,\vec{x}'') ; \\
\widetilde{G}^+_{D+1}(\omega,\vec{x}-\vec{x}'') 
&\equiv \int_{-\infty}^{+\infty} \dd \tau e^{i\omega\tau} G_{D+1}(\tau,\vec{x}-\vec{x}'') .
\end{align}
Equation \eqref{ScalarWaveEquation_DrivenSHO} tells us that the $\omega$-mode of the source is directly responsible for that of the field $\widetilde{\psi}(\omega,\vec{x})$. This is reminiscent of the driven harmonic oscillator system, except now we have one oscillator per point in space $\vec{x}'$ -- hence the integral over all of them.

{\bf 4D Retarded Green's Function in Frequency Space} \qquad Next, we focus on the $(D+1) = (3+1)$ case, and re-visit the 4D retarded Green's function result in eq. \eqref{ScalarWaveEquation_GreensFunction_4D}, but replace the $\delta$-function with its integral representation. This leads us to $\widetilde{G}^+_4(\omega,\vec{x}-\vec{x}')$, the frequency space representation of the retarded Green's function of the wave operator.
\begin{align}
G_4^+(x-x') 
&= \int_{-\infty}^{+\infty}\frac{\dd\omega}{2\pi}\frac{\exp\left( -i \omega (t-t'-|\vec{x}-\vec{x}'|) \right)}{4\pi |\vec{x}-\vec{x}'|} \nonumber\\
&\equiv \int_{-\infty}^{+\infty}\frac{\dd\omega}{2\pi} e^{-i\omega(t-t')} \widetilde{G}^+_4(\omega,\vec{x}-\vec{x}') ,
\end{align}
where
\begin{align}
\label{GreensFunction_4D_FrequencySpace}
\widetilde{G}^+_4(\omega,\vec{x}-\vec{x}')	&\equiv \frac{\exp\left( i\omega |\vec{x}-\vec{x}'| \right)}{4\pi |\vec{x}-\vec{x}'|} .
\end{align}
As we will see, $\omega$ can be interpreted as the frequency of the source of the waves. In this section we will develop a multipole expansion of the field in frequency space by performing one for the source as well. This will allow us to readily take the non-relativistic/static limit, where the motion of the sources (in some center of mass frame) is much slower than 1.

Because the $(3+1)$-dimensional case of eq. \eqref{ScalarWaveEquation_GreensFunction_PDE} in frequency space reads
\begin{align}
\left(\partial_0^2-\vec{\nabla}^2\right)\int_{-\infty}^{+\infty}\frac{\dd\omega}{2\pi}\frac{\exp\left( -i \omega (t-t'-|\vec{x}-\vec{x}'|) \right)}{4\pi |\vec{x}-\vec{x}'|}
				&= \delta(t-t') \delta^{(3)}\left( \vec{x}-\vec{x}' \right) , \\
\int_{-\infty}^{+\infty}\frac{\dd\omega}{2\pi} e^{-i\omega(t-t')} 
				\left( -\omega^2 - \vec{\nabla}^2 \right)\frac{\exp\left( i \omega |\vec{x}-\vec{x}'| \right)}{4\pi |\vec{x}-\vec{x}'|}
				&= \int_{-\infty}^{+\infty}\frac{\dd\omega}{2\pi} e^{-i \omega(t-t')}  \delta^{(3)}\left( \vec{x}-\vec{x}' \right) ,
\end{align}
-- where $\partial_0^2$ can be either $\partial_t^2$ or $\partial_{t'}^2$; $\vec{\nabla}^2$ can be either $\vec{\nabla}_{\vec{x}}$ or $\vec{\nabla}_{\vec{x}'}$; and we have replaced $\delta(t-t')$ with its integral representation -- we can equate the coefficients of the (linearly independent) functions $\{\exp(-i\omega(t-t'))\}$ on both sides to conclude, for fixed $\omega$, the frequency space Green's function of eq. \eqref{GreensFunction_4D_FrequencySpace} obeys the PDE
\begin{align}
\label{HelmholtzGreensFunction}
\left( -\omega^2 - \vec{\nabla}^2 \right) \widetilde{G}^+_4(\omega,\vec{x}-\vec{x}') &= \delta^{(3)}\left( \vec{x}-\vec{x}' \right) .
\end{align}
{\bf Static Limit Equals Zero Frequency Limit} \qquad In any (curved) spacetime that enjoys time translation symmetry -- which, in particular, means there is some coordinate system where the metric $g_{\mu\nu}(\vec{x})$ depends only on space $\vec{x}$ and not on time $t$ -- we expect the Green's function of the wave operator to reflect the symmetry and take the form $G^+(t-t';\vec{x},\vec{x}')$. Furthermore, the wave operator only involves time through derivatives, i.e., eq. \eqref{ScalarWaveEquation_Curved} now reads
\begin{align}
\label{ScalarWaveEquation_Curved_Stationary}
\nabla_\mu \nabla^\mu G 
&= g^{tt} \partial_t \partial_t G + g^{ti} \partial_t \partial_i G + \frac{\partial_i \left(\sqrt{|g|} g^{ti} \partial_t G\right)}{\sqrt{|g|}} 
+ \frac{1}{\sqrt{|g|}} \partial_i \left( \sqrt{|g|} g^{ij} \partial_j G \right) \nonumber\\
&= \frac{\delta(t-t') \delta^{(D)}\left( \vec{x} - \vec{x}' \right)}{\sqrt[4]{g(\vec{x}) g(\vec{x}')}} ;
\end{align}
since $\sqrt{|g|}$ and $g^{\mu\nu}$ are time-independent. In such a time-translation-symmetric situation, we may perform a frequency transform 
\begin{align}
\widetilde{G}^+(\omega;\vec{x},\vec{x}') 
= \int_{-\infty}^{+\infty} \dd\tau e^{i\omega\tau} G^+\left(\tau;\vec{x},\vec{x}'\right) ,
\end{align}
and note that solving the {\it static} equation
\begin{align}
\label{ScalarWaveEquation_Curved_Static}
\nabla_\mu \nabla^\mu G^{(\text{static})}\left(\vec{x},\vec{x}'\right) 
&= \frac{\partial_i \left( \sqrt{|g(\vec{x})|} g^{ij}(\vec{x}) \partial_{j} G^{(\text{static})}\left(\vec{x},\vec{x}'\right) \right)}{\sqrt{|g(\vec{x})|}} \nonumber\\
&= \frac{\partial_{i'} \left( \sqrt{|g(\vec{x}')|} g^{ij}(\vec{x}') \partial_{j'} G^{(\text{static})}\left(\vec{x},\vec{x}'\right) \right)}{\sqrt{|g(\vec{x}')|}} 
= \frac{\delta^{(D)}(\vec{x}-\vec{x}')}{\sqrt[4]{g(\vec{x}) g(\vec{x}')}} ,
\end{align}
amounts to taking the zero frequency limit of the frequency space retarded Green's function. Note that the static equation still depends on the full $(D+1)$ dimensional metric, but the $\delta$-functions on the right hand side is $D$-dimensional.

The reason is the frequency transform of eq. \eqref{ScalarWaveEquation_Curved_Stationary} replaces $\partial_t \to -i\omega$ and the $\delta(t-t')$ on the right hand side with unity.
\begin{align}
g^{tt} (-i\omega)^2 \widetilde{G} + g^{ti} (-i\omega) \partial_i \widetilde{G} 
+ \frac{\partial_i \left(\sqrt{|g|} g^{ti} (-i\omega) G\right)}{\sqrt{|g|}} 
+ \frac{1}{\sqrt{|g|}} \partial_i \left( \sqrt{|g|} g^{ij} \partial_j \widetilde{G} \right) 
= \frac{\delta^{(D)}\left( \vec{x} - \vec{x}' \right)}{\sqrt[4]{g(\vec{x}) g(\vec{x}')}} 
\end{align}
In the zero frequency limit ($\omega \to 0$) we obtain eq. \eqref{ScalarWaveEquation_Curved_Static}. And since the static limit is the zero frequency limit,
\begin{align}
\label{ScalarWaveEquation_Curved_Static_ChargeMass}
G^\text{(static)}(\vec{x},\vec{x}') 
&= \lim_{\omega \to 0} \int_{-\infty}^{+\infty} \dd\tau e^{i\omega\tau} G^+\left(\tau;\vec{x},\vec{x}'\right) , \\
&= \int_{-\infty}^{+\infty} \dd\tau G^+\left(\tau;\vec{x},\vec{x}'\right) 
= \int_{-\infty}^{+\infty} \dd\tau \int \dd^D\vec{x}''\sqrt{|g(\vec{x}'')|} G^+\left(\tau;\vec{x},\vec{x}''\right) \frac{\delta^{(D)}(\vec{x}'-\vec{x}'')}{\sqrt{|g(\vec{x}') g(\vec{x}'')|}} . \nonumber
\end{align}
This second line has the following interpretation: not only is the static Green's function the zero frequency limit of its frequency space retarded counterpart, it can also be viewed as the field generated by a point ``charge/mass" held still at $\vec{x}'$ from past infinity to future infinity.\footnote{Note, however, that in curved spacetimes, holding still a charge/mass -- ensuring it stays put at $\vec{x}'$ -- requires external forces. For example, holding a mass still in a spherically symmetric gravitational field of a star requires an outward external force, for otherwise the mass will move towards the center of the star.}

{\it 4D Minkowski Example} \qquad We may illustrate our discussion here by examining the 4D Minkowski case. The field generated by a charge/mass held still at $\vec{x}'$ is nothing but the Coulomb/Newtonian potential $1/(4\pi |\vec{x}-\vec{x}'|)$. Since we also know the 4D Minkowski retarded Green's function in eq. \eqref{ScalarWaveEquation_GreensFunction_4D}, we may apply the infinite time integral in eq. \eqref{ScalarWaveEquation_Curved_Static_ChargeMass}.
\begin{align}
G^\text{(static)}(\vec{x},\vec{x}') 
	&= \int_{-\infty}^{+\infty} \dd\tau \frac{\delta(\tau-|\vec{x}-\vec{x}'|)}{4\pi |\vec{x}-\vec{x}'|} = \frac{1}{4\pi |\vec{x}-\vec{x}'|} , \\
-\delta^{ij} \partial_i \partial_j G^\text{(static)}(\vec{x},\vec{x}') 
	&= -\vec{\nabla}^2 G^\text{(static)}(\vec{x},\vec{x}') = \delta^{(3)}(\vec{x}-\vec{x}')  .
\end{align}
On the other hand, we may also take the zero frequency limit of eq. \eqref{GreensFunction_4D_FrequencySpace} to arrive at the same answer.
\begin{align}
\lim_{\omega\to 0} \frac{\exp\left( i\omega |\vec{x}-\vec{x}'| \right)}{4\pi |\vec{x}-\vec{x}'|} = \frac{1}{4\pi |\vec{x}-\vec{x}'|} .
\end{align}
\begin{myP} 
{\it Discontinuous first derivatives of the radial Green's function} \qquad In this problem we will understand the discontinuity in the radial Green's function of the frequency space retarded Green's function in 4D Minkowski spacetime. We begin by switching to spherical coordinates and utilizing the following ansatz
\begin{align}
\label{ScalarWaveEquation_4DFlatGFrequencySpace}
\widetilde{G}^+_4\left( \omega,\vec{x}-\vec{x}' \right) 
&= \sum_{\ell = 0}^{\infty}\widetilde{g}_\ell(r,r') 
		\sum_{m=-\ell}^{\ell} Y_\ell^m(\theta,\phi) Y_\ell^m(\theta',\phi')^* , \nonumber\\
\vec{x} 	= r(\sin\theta \ \cos\phi,\sin\theta &\ \sin\phi, \cos\theta) , \qquad 
\vec{x}' 	= r'(\sin\theta' \ \cos\phi',\sin\theta' \ \sin\phi', \cos\theta') .
\end{align}
Show that this leads to the following ODE(s) for the $\ell$th radial Green's function $\widetilde{g}_\ell$:
\begin{align}
\frac{1}{r^2} \partial_r \left( r^2 \partial_r \widetilde{g}_\ell \right) + \left(\omega^2 - \frac{\ell(\ell+1)}{r^2}\right) \widetilde{g}_\ell 
		&= -\frac{\delta(r-r')}{rr'} , \\
\frac{1}{r'^2} \partial_{r'} \left( r'^2 \partial_{r'} \widetilde{g}_\ell \right) + \left(\omega^2 - \frac{\ell(\ell+1)}{r'^2}\right) \widetilde{g}_\ell 
		&= -\frac{\delta(r-r')}{rr'} .
\end{align}
Because $\widetilde{G}^+_4(\omega,\vec{x}-\vec{x}') = \widetilde{G}^+_4( \omega,\vec{x}'-\vec{x})$, i.e., it is symmetric under the exchange of the spatial coordinates of source and observer, it is reasonable to expect that the radial Green's function is symmetric too: $\widetilde{g}(r,r') = \widetilde{g}(r',r)$. That means the results in \S \eqref{Section_SymmetricG} may be applied here. Show that
\begin{align}
\label{4DMinkowski_RadialGreensFunction}
\widetilde{g}_\ell(r,r') = i\omega j_\ell(\omega r_<) h_\ell^{(1)}(\omega r_>) ,
\end{align}
where $j_\ell(z)$ is the spherical Bessel function and $h_\ell^{(1)}(z)$ is the Hankel function of the first kind. Then check that the static limit in eq. \eqref{3DLaplacianGEquation_StaticLimit} is recovered, by taking the limits $\omega r, \omega r' \to 0$.

Some useful formulas include
\begin{align}
j_\ell(x) = (-x)^\ell \left(\frac{1}{x} \frac{\dd}{\dd x}\right)^\ell \frac{\sin x}{x} , \qquad
h_\ell^{(1)}(x) = -i (-x)^\ell \left(\frac{1}{x} \frac{\dd}{\dd x}\right)^\ell \frac{\exp(ix)}{x} ,
\end{align}
their small argument limits
\begin{align}
\label{SphericalBessel_HankelH1_SmallArguments}
j_\ell(x \ll 1) \to \frac{x^\ell}{(2\ell+1)!!} \left( 1 + \mathcal{O}(x^2) \right), \qquad
h_\ell^{(1)}(x \ll 1) \to -\frac{i(2\ell-1)!!}{x^{\ell+1}} \left( 1 + \mathcal{O}(x) \right),
\end{align}
as well as their large argument limits
\begin{align}
\label{SphericalBessel_HankelH1_LargeArguments}
j_\ell(x \gg 1) \to \frac{1}{x} \sin\left(x - \frac{\pi \ell}{2}\right), \qquad
h_\ell^{(1)}(x \gg 1) \to (-i)^{\ell+1} \frac{e^{ix}}{x} .
\end{align}
Their Wronskian is
\begin{align}
\text{Wr}_z\left( j_\ell(z), h_\ell^{(1)}(z) \right) = \frac{i}{z^2} .
\end{align}
Hints: First explain why
\begin{align}
\widetilde{g}_\ell(r,r') 
		&= A^1_\ell j_\ell(\omega r) j_\ell(\omega r') + A^2_\ell h_\ell^{(1)}(\omega r) h_\ell^{(1)}(\omega r') + \mathcal{G}_\ell(r,r') , \\
\mathcal{G}_\ell(r,r') 
		&\equiv F \left\{ (\chi_\ell-1) j_\ell(\omega r_>) h^{(1)}_\ell(\omega r_<) + \chi_\ell \cdot j_\ell(\omega r_<) h^{(1)}_\ell(\omega r_>) \right\} ,
\end{align}
where $A^{1,2}_\ell$, $F$ and $\chi_\ell$ are constants. Fix $F$ by ensuring the ``jump" in the first $r$-derivative at $r=r'$ yields the correct $\delta$-function measure. Then consider the limits $r \to 0$ and $r \gg r'$. For the latter, note that
\begin{align}
|\vec{x}-\vec{x}'| = e^{-\vec{x}'\cdot\vec{\nabla}_{\vec{x}}} |\vec{x}| = |\vec{x}| \left( 1 - (r'/r) \widehat{n}\cdot\widehat{n}' + \mathcal{O}((r'/r)^2) \right) ,
\end{align} 
where $\widehat{n} \equiv \vec{x}/r$ and $\widehat{n}' \equiv \vec{x}'/r'$. \qed
\end{myP}
We will now proceed to understand the utility of obtaining such a mode expansion of the frequency space Green's function.

\noindent{\bf Localized source(s): Static Multipole Expansion} \qquad In infinite flat $\mathbb{R}^3$, Poisson's equation
\begin{align}
-\vec{\nabla}^2 \psi(\vec{x}) = J(\vec{x})
\end{align}
is solved via the static limit of the 4D retarded Green's function we have been discussing. This static limit is given in eq. \eqref{3DLaplacianGEquation_StaticLimit} in spherical coordinates, which we will now exploit to display its usefulness. In particular, assuming the source $J$ is localized in space, we may now ask: \begin{quotation}
	{\it What is the field generated by $J$ and how does it depend on the details of its interior?}
\end{quotation}
Let the origin of our coordinate system lie at the center of mass of the source $J$, and let $R$ be its maximum radius, i.e., $J(r>R)=0$. Therefore we may replace $r_< \to r'$ and $r_> \to r$ in eq. \eqref{3DLaplacianGEquation_StaticLimit}, and the exact solution to $\psi$ now reads
\begin{align}
\label{ScalarField_4DStatic}
\psi(\vec{x};r>R) 
= \int_{\mathbb{R}^3} \dd^3 \vec{x}' G(\vec{x}-\vec{x}') J(\vec{x}') 
= \sum_{\ell=0}^{\infty} \sum_{m = -\ell}^{+\ell} \frac{\rho_\ell^m}{2\ell+1} \frac{Y_\ell^m(\theta,\phi)}{r^{\ell+1}} ,
\end{align}
where the multipole moments $\{ \rho_\ell^m \}$ are defined 
\begin{align}
\label{Multipoles_Static}
\rho_\ell^m \equiv \int_{\mathbb{S}^2} \dd(\cos\theta') \dd\phi' \int_{0}^{\infty} \dd r' r'^{\ell+2} \ \overline{Y_\ell^m(\theta',\phi')} J(r',\theta',\phi') .
\end{align}
It is worthwhile to highlight the following.
\begin{itemize}
\item The spherical harmonics can be roughly thought of as waves on the $2-$sphere. Therefore, the multipole moments $\rho_\ell^m$ in eq. \eqref{Multipoles_Static} with larger $\ell$ and $m$ values, describe the shorter wavelength/finer features of the interior structure of $J$. (Recall the analogous discussion for Fourier transforms.) 
\item Moreover, since there is a $Y_\ell^m(\theta,\phi)/r^{\ell+1}$ multiplying the $(\ell,m)$-moment of $J$, we see that the finer features of the field detected by the observer at $\vec{x}$ is not only directly sourced by finer features of $J$, it falls off more rapidly with increasing distance from $J$. As the observer moves towards infinity, the dominant part of the field $\psi$ is the monopole which goes as $1/r$ times the total mass/charge of $J$.
\item We see why separation-of-variables is not only a useful mathematical technique to reduce the solution of Green's functions from a PDE to a bunch of ODE's, it was the form of eq. \eqref{3DLaplacianGEquation_StaticLimit} that allowed us to cleanly separate the contribution from the source (the multipoles $\{ \rho_\ell^m \}$) from the form of the field they would generate, at least on a mode-by-mode basis.
\end{itemize}
\noindent{\bf Localized source(s): General Multipole Expansions, Far Zone} \qquad Let us generalize the static case to the fully time dependent one, but in frequency space and in the far zone. By the far zone, we mean the observer is located very far away from the source $J$, at distances (from the center of mass) much further than the typical inverse frequency of $\widetilde{J}$, i.e., mathematically, $\omega r \gg 1$. We begin with eq. \eqref{4DMinkowski_RadialGreensFunction} inserted into eq. \eqref{ScalarWaveEquation_4DFlatGFrequencySpace}.
\begin{align}
\widetilde{G}^+_4\left( \omega,\vec{x}-\vec{x}' \right) 
&= \frac{\exp\left(i\omega|\vec{x}-\vec{x}'|\right)}{4\pi |\vec{x}-\vec{x}'|} \\
&= i\omega \sum_{\ell = 0}^{\infty} j_\ell(\omega r_<) h^{(1)}_\ell(\omega r_>) \sum_{m=-\ell}^{\ell} Y_\ell^m(\theta,\phi) Y_\ell^m(\theta',\phi')^* 
\end{align}
Our far zone assumptions means we may replace the Hankel function in eq. \eqref{4DMinkowski_RadialGreensFunction} with its large argument limit in eq. \eqref{SphericalBessel_HankelH1_LargeArguments}.
\begin{align}
\widetilde{G}^+_4\left( \omega r \gg 1 \right) 
&= \frac{e^{i\omega r}}{r} \left( 1 + \mathcal{O}\left(r^{-1}\right) \right)\sum_{\ell = 0}^{\infty} (-i)^\ell j_\ell(\omega r') \sum_{m=-\ell}^{\ell} Y_\ell^m(\theta,\phi) Y_\ell^m(\theta',\phi')^* .
\end{align}
Applying this limit to the general wave solution in eq. \eqref{ScalarWaveEquation_DrivenSHO},
\begin{align}
\widetilde{\psi}(\omega,\vec{x}) 
	&= \int_{\mathbb{R}^3} \dd^3 \vec{x}'' \widetilde{G}^+_4(\omega,\vec{x}-\vec{x}'') \widetilde{J}(\omega,\vec{x}'') , \\
\widetilde{\psi}(\omega r \gg 1) 
	&\approx \frac{e^{i\omega r}}{r} \sum_{\ell = 0}^{\infty} \sum_{m=-\ell}^{\ell} \frac{Y_\ell^m(\theta,\phi)}{2\ell+1} \Omega_\ell^m(\omega) ,
\end{align}
where now the frequency dependent multipole moments are defined as
\begin{align}
\label{Multipoles_FrequencyDependent}
\Omega_\ell^m(\omega)
&\equiv (2\ell+1) (-i)^\ell \int_{\mathbb{S}^2} \dd(\cos\theta') \dd\phi' \int_{0}^{\infty} \dd r' r'^2 j_\ell(\omega r') \overline{Y_\ell^m(\theta',\phi')} \widetilde{J}(\omega,r',\theta',\phi') .
\end{align}
\noindent{\it Low frequency limit equals slow motion limit} \qquad How are the multipole moments $\{ \rho_\ell^m \}$ in eq. \eqref{Multipoles_Static} (which are pure numbers) related to the frequency dependent ones $\{ \Omega_\ell^m(\omega) \}$ in eq. \eqref{Multipoles_FrequencyDependent}? The answer is that the low frequency limit is the slow-motion/non-relativistic limit. To see this in more detail, we take the $\omega r' \ll 1$ limit, which amounts to the physical assumption that the object described by $J$ is localized so that its maximum radius $R$ (from its center of mass) is much smaller than the inverse frequency. In other words, in units where the speed of light is unity, the characteristic size $R$ of the source $J$ is much smaller than the time scale of its typical time variation. Mathematically, this $\omega r' \ll 1$ limit is achieved by replacing $j_\ell(\omega r')$ with its small argument limit in eq. \eqref{SphericalBessel_HankelH1_SmallArguments}.
\begin{align}
\Omega_\ell^m(\omega R \ll 1)
&\approx \frac{(-i \omega)^\ell}{(2\ell-1)!!} \left( 1 + \mathcal{O}(\omega^2) \right)\int_{\mathbb{S}^2} \dd(\cos\theta') \dd\phi' \int_{0}^{\infty} \dd r' r'^{2+\ell} \overline{Y_\ell^m(\theta',\phi')} \widetilde{J}(\omega,r',\theta',\phi') 
\end{align}
Another way to see this ``small $\omega$ equals slow motion limit" is to ask: what is the real time representation of these $\{ \Omega_\ell^m(\omega R \ll 1) \}$? By recognizing every $-i\omega$ as a $t$-derivative,
\begin{align}
\Omega_\ell^m(t)
&\approx \frac{\partial_t^\ell}{(2\ell-1)!!} 
\int_{-\infty}^{+\infty} \frac{\dd \omega}{2\pi} e^{-i\omega t} \int_{\mathbb{S}^2} \dd(\cos\theta') \dd\phi' \int_{0}^{\infty} \dd r' r'^{2+\ell} \overline{Y_\ell^m(\theta',\phi')} \widetilde{J}(\omega,r',\theta',\phi') , \nonumber\\
&\equiv \frac{\partial_t^\ell \rho_\ell^m(t)}{(2\ell-1)!!} .
\end{align}
We see that the $\omega R \ll 1$ is the slow motion/non-relativistic limit because it is in this limit that time derivatives vanish. This is also why the only $1/r$ piece of the static field in eq. \eqref{ScalarField_4DStatic} comes from the monopole.

{\it Spherical waves in small $\omega$ limit} \qquad In this same limit, we may re-construct the real time scalar field, and witness how it is a superposition of spherical waves $\exp(i\omega (r-t))/r$. The observer detects a field that depends on the time derivatives of the multipole moments evaluated at retarded time $t-r$.
{\allowdisplaybreaks\begin{align}
\psi(t,\vec{x}) 
&= \int_{-\infty}^{+\infty} \frac{\dd \omega}{2\pi} e^{-i\omega t} \widetilde{\psi}(\omega,\vec{x}) \nonumber\\
&\approx \int_{-\infty}^{+\infty} \frac{\dd \omega}{2\pi} \frac{e^{i\omega (r-t)}}{r} \sum_{\ell = 0}^{\infty} \sum_{m=-\ell}^{\ell} \frac{Y_\ell^m(\theta,\phi)}{2\ell+1} \Omega_\ell^m(\omega), \qquad\qquad \text{(Far zone spherical wave expansion)} \nonumber\\
% &= \int_{-\infty}^{+\infty} \frac{\dd \omega}{2\pi} \int_{-\infty}^{+\infty} \dd t' \frac{e^{i\omega (r-(t-t'))}}{r} \sum_{\ell = 0}^{\infty} \sum_{m=-\ell}^{\ell} \frac{Y_\ell^m(\theta,\phi)}{(2\ell+1)!!} \frac{\dd^\ell \rho_\ell^m(t')}{\dd t'^\ell} \nonumber \\
% &= \int_{-\infty}^{+\infty} \dd t' \frac{\delta(r-(t-t'))}{r} \sum_{\ell = 0}^{\infty} \sum_{m=-\ell}^{\ell} \frac{Y_\ell^m(\theta,\phi)}{(2\ell+1)!!} \frac{\dd^\ell \rho_\ell^m(t')}{\dd t'^\ell} \nonumber \\
\label{ScalarWaveEquation_4DSlowMotion}
&\approx \frac{1}{r} \sum_{\ell = 0}^{\infty} \sum_{m=-\ell}^{\ell} \frac{Y_\ell^m(\theta,\phi)}{(2\ell+1)!!} \frac{\dd^\ell \rho_\ell^m(t-r)}{\dd t^\ell}, \qquad\qquad \text{(Slow motion limit)}  .
\end{align}}
\begin{myP}
{\it Far zone in position/real space} \qquad Starting from the exact wave solution in eq. \eqref{ScalarWaveEquation_Solution_4D}, show that the leading $1/r$ portion of the solution -- i.e., the far zone limit -- reads
\begin{align}
\psi(t,\vec{x}) 
\label{ScalarWaveEquation_4DFarZone_TaylorSeries}
&\approx \frac{1}{4\pi r} \int_{\mathbb{R}^3} \dd^3 \vec{x}' \sum_{\ell=0}^{\infty} \frac{\left(\vec{x}'\cdot\widehat{r}\right)^\ell}{\ell !} \partial_t^\ell J\left(t-r,\vec{x}'\right), \qquad\qquad r \equiv |\vec{x}| ; \ \widehat{r} \equiv \frac{\vec{x}}{r} , \\
\label{ScalarWaveEquation_4DFarZone}
&= \frac{1}{4\pi r} \int_{\mathbb{R}^3} \dd^3 \vec{x}' J\left(t-r+\vec{x}'\cdot\widehat{r},\vec{x}'\right), 
\end{align}
where we have placed the origin $\vec{x}=\vec{x}'=\vec{0}$ within the source $J$. In terms of the characteristic time scale $\tau_s$ of the source $J$ and its characteristic spatial extent $r_s$, explain what physical situations would allow only the first few terms of the series in eq. \eqref{ScalarWaveEquation_4DFarZone_TaylorSeries} to be retained. Comment on the relationship between equations \eqref{ScalarWaveEquation_4DSlowMotion} and \eqref{ScalarWaveEquation_4DFarZone_TaylorSeries}.
\end{myP}

\subsubsection{Initial value problem via Kirchhoff representation}

{\bf Massless scalar fields} \qquad Previously we showed how, if we specified the initial conditions for the scalar field $\psi$ -- then via their Fourier transforms -- eq. \eqref{ScalarWaveEquation_HomogeneousSolution_InitialValueFormulation_Fourier} tells us how they will evolve forward in time. Now we will derive an analogous expression that is valid in curved spacetime, using the retarded Green's function $G^+_{D+1}$. To begin, the appropriate generalization of equations \eqref{ScalarWaveEquation_Minkowski} and \eqref{ScalarWaveEquation_GreensFunction_PDE} are
\begin{align}
\label{ScalarWaveEquation_CurvedSpacetime}
\Box_x \psi(x) &= J(x) , \nonumber\\
\Box G^+_{D+1}(x,x') &= \frac{\delta^{(D+1)}(x-x')}{\sqrt[4]{|g(x) g(x')|}} .
\end{align}
The derivation is actually very similar in spirit to the one starting in eq. \eqref{GreensFunction_Laplacian_Kirchhoff_I}. Let us consider some ``cylindrical" domain of spacetime $\mathfrak{D}$ with spatial boundaries $\partial \mathfrak{D}_s$ that can be assumed to be infinitely far away, and ``constant time" hypersurfaces $\partial \mathfrak{D}(t_>)$ (final time $t_>$) and $\partial \mathfrak{D}(t_0)$ (initial time $t_0$). (These constant time hypersurfaces need not correspond to the same time coordinate used in the integration.) We will consider an observer residing (at $x$) within this domain $\mathfrak{D}$.
{\allowdisplaybreaks\begin{align}
\label{GreensFunction_Spacetime_Kirchhoff_I}
I(x \in \mathfrak{D}) 
&\equiv \int_{\mathfrak{D}} \dd^{D+1} x' \sqrt{|g(x')|} \left\{ 
G_{D+1}(x,x') \Box_{x'} \psi(x') - \Box_{x'} G_{D+1}(x,x') \cdot \psi(x') 
\right\} \nonumber\\
&= \int_{\partial\mathfrak{D}} \dd^{D} \Sigma_{\alpha'} \left\{ G_{D+1}(x,x') \nabla^{\alpha'} \psi(x') - \nabla^{\alpha'} G_{D+1}(x,x') \cdot \psi(x') \right\} \\
&- \int_{\mathfrak{D}} \dd^{D+1} x' \sqrt{|g(x')|} \left\{ 
\nabla_{\alpha'} G_{D+1}(x,x') \nabla^{\alpha'} \psi(x') - \nabla_{\alpha'} G_{D+1}(x,x') \nabla^{\alpha'} \psi(x') 
\right\} . \nonumber
\end{align}}
The terms in the very last line cancel. What remains in the second equality is the surface integrals over the spatial boundaries $\partial \mathfrak{D}_s$, and constant time hypersurfaces $\partial \mathfrak{D}(t_>)$ and $\partial \mathfrak{D}(t_0)$ -- where we have used the Gauss' theorem in eq. \eqref{DifferentialGeometry_GaussTheorem}. Here is where there is a significant difference between the curved space setup and the curved spacetime one at hand. By causality, since we have $G^+_{D+1}$ in the integrand, the constant time hypersurface $\partial \mathfrak{D}(t_>)$ cannot contribute to the integral because it lies to the future of $x$. Also, if we assume that $G^+_{D+1}(x,x')$, like its Minkowski counterpart, vanishes outside the past light cone of $x$, then the spatial boundaries at infinity also cannot contribute.\footnote{In curved spacetimes where any pair of points $x$ and $x'$ can be linked by a unique geodesic, this causal structure of $G_{D+1}^+$ can be readily proved for the 4 dimensional case.} (Drawing a spacetime diagram here helps.) If we now proceed to invoke the equations obeyed by $\psi$ and $G_{D+1}$ in eq. \eqref{ScalarWaveEquation_CurvedSpacetime}, what remains is
\begin{align}
&-\psi(x) + \int_{\mathfrak{D}} \dd^{D+1} x' \sqrt{|g(x')|} G_{D+1}(x,x') J(x') \\
&
= -\int_{\partial\mathfrak{D}(t_0)} \dd^{D} \vec{\xi} \sqrt{|H(\vec{\xi})|} \left\{ G_{D+1}\left(x,x'(\vec{\xi})\right) n^{\alpha'} \nabla_{\alpha'} \psi\left(x'(\vec{\xi})\right) - n^{\alpha'} \nabla_{\alpha'} G_{D+1}\left(x,x'(\vec{\xi})\right) \cdot \psi\left(x'(\vec{\xi})\right) \right\} . \nonumber
\end{align}
Here, we have assumed there are $D$ coordinates $\vec{\xi}$ such that $x'^\mu(\vec{\xi})$ parametrizes our initial time hypersurface $\partial \mathfrak{D}(t_0)$. The $\sqrt{|H|}$ is the square root of the determinant of its induced metric. Also, remember in Gauss' theorem (eq. \eqref{DifferentialGeometry_GaussTheorem}), the unit normal vector dotted into the gradient $\nabla_{\alpha'}$ is the {\it outward} one (see equations \eqref{DifferentialGeometry_DirectedArea_v1} and \eqref{DifferentialGeometry_DirectedArea_v2}), which in our case is therefore pointing {\it backward} in time: this is our $-n^{\alpha'}$, we have inserted a negative sign in front so that $n^{\alpha'}$ itself is the unit timelike vector pointing towards the future. With all these clarifications in mind, we gather
\begin{align}
\label{Kirchhoff_CurvedSpace}
&\psi(x;x^0>t_0) 
= \int_{\mathfrak{D}} \dd^{D+1} x' \sqrt{|g(x')|} G_{D+1}(x,x') J(x') \\
&+ \int_{\partial\mathfrak{D}(t_0)} \dd^{D} \vec{\xi} \sqrt{|H(\vec{\xi})|} \left\{ G_{D+1}\left( x,x'(\vec{\xi}) \right) n^{\alpha'} \nabla_{\alpha'} \psi\left( x'(\vec{\xi}) \right) - n^{\alpha'} \nabla_{\alpha'} G_{D+1}\left(x,x'(\vec{\xi})\right) \cdot \psi\left(x'(\vec{\xi})\right) \right\} . \nonumber
\end{align}
In Minkowski spacetime, we may choose $t_0$ to be the constant $t$ surface of $\dd s^2 = \dd t^2 - \dd\vec{x}^2$. Then, expressed in these Cartesian coordinates,
\begin{align}
\label{Kirchhoff_Minkowski}
\psi(t > t_0,\vec{x}) 
&= \int_{t' \geq t_0} \dd t' \int_{\mathbb{R}^{D}} \dd^{D}\vec{x}' G_{D+1}\left(t-t',\vec{x}-\vec{x}'\right) J(t',\vec{x}') \\
&+ \int_{\mathbb{R}^D} \dd^{D}\vec{x}' \left\{ G_{D+1}(t-t_0,\vec{x}-\vec{x}') \partial_{t_0} \psi(t_0,\vec{x}') - \partial_{t_0} G_{D+1}(t-t_0,\vec{x}-\vec{x}') \cdot \psi(t_0,\vec{x}') \right\} . \nonumber
\end{align}
We see in both equations \eqref{Kirchhoff_CurvedSpace} and \eqref{Kirchhoff_Minkowski}, that the time evolution of the field $\psi(x)$ can be solved once the retarded Green's function $G_{D+1}^+$, as well as $\psi$'s initial profile and first time derivative is known at $t_0$. Generically, the field at the observer location $x$ is the integral of the contribution from its initial profile and first time derivative on the $t=t_0$ surface from both on and within the past light cone of $x$. (Even in flat spacetime, while in 4 and higher even dimensional flat spacetime, the field propagates only on the light cone -- in 2 and all odd dimensions, we have seen that scalar waves develop tails.)

Let us also observe that the wave solution in eq. \eqref{ScalarWaveEquation_GConvolvedWithJ} is in fact a special case of eq. \eqref{Kirchhoff_Minkowski}: the initial time surface is the infinite past $t_0 \to -\infty$, upon which it is further assumed the initial field and its time derivatives are trivial -- the signal detected at $x$ can therefore be entirely attributed to $J$.
\begin{myP}
\qquad In 4 dimensional infinite flat spacetime, let the initial conditions for the scalar field be given by
\begin{align}
\psi(t=0,\vec{x}) = e^{i\vec{k}\cdot\vec{x}}, \qquad \partial_t \psi(t=0,\vec{x}) = -i|\vec{k}| e^{i\vec{k}\cdot\vec{x}} .
\end{align}
Use the Kirchhoff representation in eq. \eqref{Kirchhoff_Minkowski} to find $\psi(t>0,\vec{x})$. You can probably guess the final answer, but this is a simple example to show you the Kirchhoff representation really works. \qed
\end{myP}

\subsection{Variational Principle in Field Theory}

You may be familiar with the variational principle -- or, the principle of stationary action -- from classical mechanics. Here, we will write down one for the classical field theories leading to the Poisson and wave equations.

{\bf Poisson equation} \qquad Consider the following {\it action} for the real field $\psi$ sourced by some externally prescribed $J(\vec{x})$.
\begin{align}
S_\text{Poisson}[\psi] \equiv \int_{\mathfrak{D}} \dd^D \vec{x} \sqrt{|g(\vec{x})|} \left( \frac{1}{2} \nabla_i \psi(\vec{x}) \nabla^i \psi(\vec{x}) - \psi(\vec{x}) J(\vec{x}) \right)
\end{align}
We claim that the action $S_\text{Poisson}$ is extremized iff $\psi$ is a solution to Poisson's equation (eq. \eqref{PoissonEquation}), provided the field at the boundary $\partial\mathfrak{D}$ of the domain is specified and fixed.

Given a some field $\bar{\psi}$, not necessarily a solution, let us consider some deviation from it; namely,
\begin{align}
\psi = \bar{\psi} + \delta \psi .
\end{align}
($\delta \psi$ is one field; the $\delta$ is pre-pended as a reminder this is a deviation from $\bar{\psi}$.) A direct calculation yields
{\allowdisplaybreaks\begin{align}
S_\text{Poisson}[\bar{\psi} + \delta \psi] 
&= \int_{\mathfrak{D}} \dd^D \vec{x} \sqrt{|g(\vec{x})|} \left( \frac{1}{2} \nabla_i \bar{\psi} \nabla^i \bar{\psi} - \bar{\psi} J \right) \nonumber\\
&+ \int_{\mathfrak{D}} \dd^D \vec{x} \sqrt{|g(\vec{x})|} \left( \nabla_i \bar{\psi} \nabla^i \delta \psi - J \delta\psi \right) \nonumber\\
&+ \int_{\mathfrak{D}} \dd^D \vec{x} \sqrt{|g(\vec{x})|} \left( \frac{1}{2} \nabla_i \delta\psi \nabla^i \delta\psi \right) .
\end{align}}
We may integrate-by-parts, in the second line, the gradient acting on $\delta \psi$.
\begin{align}
S_\text{Poisson}[\bar{\psi} + \delta \psi] 
&= \int_{\mathfrak{D}} \dd^D \vec{x} \sqrt{|g(\vec{x})|} \left( \frac{1}{2} \nabla_i \bar{\psi} \nabla^i \bar{\psi} - \bar{\psi} J 
+ \frac{1}{2} \nabla_i \delta\psi \nabla^i \delta\psi + \delta\psi \left\{ -\vec{\nabla}^2 \bar{\psi} - J \right\} \right) \nonumber\\
& \qquad \qquad + \int_{\partial \mathfrak{D}} \dd^{D-1}\vec{\xi} \sqrt{|H(\vec{\xi})|} \delta\psi n^i \nabla_i \bar{\psi} 
\end{align}
Provided Dirichlet boundary conditions are specified and not varied, i.e., $\psi( \partial \mathfrak{D} )$ is given, then by definition $\delta\psi( \partial \mathfrak{D} ) = 0$ and the surface term on the second line is zero. Now, suppose Poisson's equation is satisfied by $\bar{\psi}$, then $-\vec{\nabla}^2 \bar{\psi} - J = 0$ and because the remaining quadratic-in-$\delta\psi$ is strictly positive (as argued earlier) we see that any deviation {\it increases} the value of $S_\text{Poisson}$ and therefore the solution $\bar{\psi}$ yields a minimal action.

Conversely, just as we say a (real) function $f(x)$ is extremized at $x=x_0$ when $f'(x_0)=0$, we would say $S_\text{Poisson}$ is extremized by $\bar{\psi}$ if the first-order-in-$\delta \psi$ term
\begin{align}
\int_{\mathfrak{D}} \dd^D \vec{x} \sqrt{|g(\vec{x})|} \delta\psi \left\{ -\vec{\nabla}^2 \bar{\psi} - J \right\} 
\end{align}
vanishes for {\it any} deviation $\delta \psi$. But if this were to vanish for any deviation $\delta \psi(\vec{x})$, the terms in the curly brackets must be zero, and Poisson's equation is satisfied.

{\bf Wave equation in infinite space} \qquad Assuming the fields fall off sufficiently quickly at spatial infinity and suppose the initial $\psi(t_{\text{i}},\vec{x})$ and final $\psi(t_{\text{f}},\vec{x})$ configurations are specified and fixed, we now discuss why the action
\begin{align}
S_\text{Wave} \equiv \int_{t_{\text{i}}}^{t_{\text{f}}} \dd t'' \int_{\mathbb{R}^D} \dd^{D}\vec{x} \sqrt{|g(x)|}
\left\{ \frac{1}{2} \nabla_\mu \psi(t'',\vec{x}) \nabla^\mu \psi(t'',\vec{x}) + J(t'',\vec{x}) \psi(t'',\vec{x}) \right\}
\end{align}
(where $x \equiv (t'',\vec{x})$) is extremized iff the wave equation in eq. \eqref{ScalarWaveEquation_Curved} is satisfied. 

Just as we did for $S_\text{Poisson}$, let us consider adding to some given field $\bar{\psi}$, a deviation $\delta\psi$. That is, we will consider
\begin{align}
\psi = \bar{\psi} + \delta \psi ,
\end{align}
without first assuming $\bar{\psi}$ solves the wave equation. A direct calculation yields
{\allowdisplaybreaks\begin{align}
S_\text{Wave}[\bar{\psi} + \delta \psi] 
&= \int_{t_{\text{i}}}^{t_{\text{f}}} \dd t'' \int_{\mathbb{R}^D} \dd^{D}\vec{x} \sqrt{|g(x)|} \left( \frac{1}{2} \nabla_\mu \bar{\psi} \nabla^\mu \bar{\psi} + \bar{\psi} J \right) \nonumber\\
&+ \int_{t_{\text{i}}}^{t_{\text{f}}} \dd t'' \int_{\mathbb{R}^D} \dd^{D}\vec{x} \sqrt{|g(x)|} \left( \nabla_\mu \bar{\psi} \nabla^\mu \delta \psi + J \delta\psi \right) \nonumber\\
&+ \int_{t_{\text{i}}}^{t_{\text{f}}} \dd t'' \int_{\mathbb{R}^D} \dd^{D}\vec{x} \sqrt{|g(x)|} \left( \frac{1}{2} \nabla_\mu \delta\psi \nabla^\mu \delta\psi \right) .
\end{align}}
We may integrate-by-parts, in the second line, the gradient acting on $\delta \psi$. By assuming that the fields fall off sufficiency quickly at spatial infinity, the remaining surface terms involve the fields at the initial and final time hypersurfaces.
\begin{align}
\label{ScalarWaveEquation_ActionPrinciple_0}
S_\text{Wave}[\bar{\psi} + \delta \psi] 
&= \int_{t_{\text{i}}}^{t_{\text{f}}} \dd t'' \int_{\mathbb{R}^D} \dd^D\vec{x} \sqrt{|g(x)|} \left( \frac{1}{2} \nabla_\mu \bar{\psi} \nabla^\mu \bar{\psi} + \bar{\psi} J 
+ \frac{1}{2} \nabla_\mu \delta\psi \nabla^\mu \delta\psi + \delta\psi \left\{ -\nabla_\mu \nabla^\mu \bar{\psi} + J \right\} \right) \nonumber\\
& \qquad \qquad + \int_{\mathbb{R}^D} \dd^{D}\vec{x} \sqrt{|g(x)|} \delta\psi(t=t_\text{f},\vec{x}) g^{0\mu} \partial_\mu \bar{\psi}(t=t_f,\vec{x}) \nonumber\\
& \qquad \qquad - \int_{\mathbb{R}^D} \dd^{D}\vec{x} \sqrt{|g(x)|} \delta\psi(t=t_\text{i},\vec{x}) g^{0\mu} \partial_\mu \bar{\psi}(t=t_i,\vec{x}) . 
\end{align}
The last two lines come from the time derivative part of
\begin{align}
\int_{t_{\text{i}}}^{t_{\text{f}}} \dd t'' \int_{\mathbb{R}^D} \dd^{D}\vec{x} \sqrt{g(x)} \nabla_\mu \left( \delta \psi \nabla^\mu \bar{\psi} \right) 
&= \int_{t'}^{t} \dd t'' \int_{\mathbb{R}^D} \dd^{D}\vec{x} \partial_\mu \left( \sqrt{g(x)} \delta \psi  g^{\mu\nu} \nabla_\nu \bar{\psi} \right) \nonumber\\
&= \left[ \int_{\mathbb{R}^D} \dd^{D}\vec{x} \sqrt{g(x)} \delta \psi g^{0\nu} \partial_\nu \bar{\psi} \right]_{t''=t_\text{i}}^{t''=t_\text{f}} + \dots
\end{align}
Provided the initial and final field values are specified and not varied, then $\delta \psi(t''=t_\text{i,f}) = 0$ and the surface terms are zero. In eq. \eqref{ScalarWaveEquation_ActionPrinciple_0}, we see that the action is extremized, i.e., when the term
\begin{align}
\int_{t_{\text{i}}}^{t_{\text{f}}} \dd t'' \int_{\mathbb{R}^D} \dd^D\vec{x} \sqrt{|g(x)|} \left( \delta\psi \left\{ -\nabla_\mu \nabla^\mu \bar{\psi} + J \right\} \right) 
\end{align}
is zero for all deviations $\delta\psi$, iff the terms in the curly brackets vanish, and the wave equation eq. \eqref{ScalarWaveEquation_Curved} is satisfied. Note that, unlike the case for $S_\text{Poisson}$, because $\nabla_\mu \psi \nabla^\mu \psi$ may not be positive definite, it is not possible to conclude from this analysis whether all solutions minimize, maximize, or merely extremizes the action $S_\text{Wave}$.

{\bf Why?} \qquad Why bother coming up with an action to describe dynamics, especially if we already have the PDEs governing the fields themselves? Apart from the intellectual interest/curiosity in formulating the same physics in different ways, having an action to describe dynamics usually allows the symmetries of the system to be made more transparent. For instance, all of the currently known fundamental forces and fields in Nature -- the Standard Model (SM) of particle physics and gravitation -- can be phrased as an action principle, and the mathematical symmetries they exhibit played key roles in humanity's attempts to understand them. Furthermore, having an action for a given theory allows it to be quantized readily, through the path integral formulation of quantum field theory due to Richard P. Feynman. In fact, our discussion of the heat kernel in, for e.g. eq. \eqref{HeatKernel_ModeExpansion}, is in fact an example of Norbert Wiener's version of the path integral, which was the precursor of Feynman's.

\subsection{Appendix to linear PDEs discourse: \\ Symmetric Green's Function of a real 2nd Order ODE}
\label{Section_SymmetricG}

\noindent{\bf Setup} \qquad In this section we wish to write down the symmetric Green's function of the most general 2nd order real linear ordinary differential operator $D$, in terms of its homogeneous solutions. We define such as differential operator as
\begin{align}
\label{2ndOrderODE_D}
D_z f(z) \equiv p_2(z) \frac{\dd^2 f(z)}{\dd z^2} + p_1(z) \frac{\dd f(z)}{\dd z} + p_0(z) f(z) , \qquad a \leq z \leq b ,
\end{align}
where $p_{0,1,2}$ are assumed to be smooth real functions and we are assuming the setup at hand is defined within the domain $z \in [a,b]$. By homogeneous solutions $f_{1,2}(z)$, we mean they both obey
\begin{align}
\label{2ndOrderODE_Homogeneous}
D_z f_{1,2}(z) = 0 .
\end{align}
Because this is a 2nd order ODE, we expect two linearly independent solutions $f_{1,2}(z)$. What we wish to solve here is the symmetric Green's function $G(z,z')=G(z',z)$ equation
\begin{align}
\label{2ndOrderODE_GreensFunctionEquation}
D_z G(z,z') = \lambda(z) \delta(z-z'), \qquad \text{ and } \qquad D_{z'} G(z,z') = \lambda(z') \delta(z-z') ,
\end{align}
where $\delta(z-z')$ is the Dirac $\delta$-function and $\lambda$ is a function to be determined. With the Green's function $G(z,z')$ at hand we may proceed to solve the particular solution $f_p(z)$ to the inhomogeneous equation, with some prescribed external source $J$,
\begin{align}
D_z f_p(z) = J(z) \qquad \Rightarrow \qquad f_p(z) = \int_{a}^{b} \frac{\dd z'}{\lambda(z')} G(z,z') J(z') .
\end{align}
Of course, for a given problem, one needs to further impose appropriate boundary conditions to obtain a unique solution. Here, we will simply ask: what's the most general ansatz that would solve eq. \eqref{2ndOrderODE_GreensFunctionEquation} in terms of $f_{1,2}$?

\noindent{\bf Wronskian} \qquad The Wronskian of the two linearly independent solutions, defined to be
\begin{align}
\text{Wr}_z(f_1,f_2) \equiv f_1(z) f_2'(z) - f_1'(z) f_2(z) , \qquad a \leq z \leq b ,
\end{align}
will be an important object in what is to follow. We record the following facts.
\begin{itemize}
	\item If Wr$_z(f_1,f_2) \neq 0$, then $f_{1,2}(z)$ are linearly independent.
	\item The Wronskian itself obeys the 1st order ODE
	\begin{align}
	\label{2ndOrderODE_Wronskian_ODE}
	\frac{\dd}{\dd z} \text{Wr}_z(f_1,f_2) &= - \frac{p_1(z)}{p_2(z)} \text{Wr}_z(f_1,f_2),
	\end{align}
	\footnote{This can be readily proven using eq. \eqref{2ndOrderODE_Homogeneous}.}which immediately implies the Wronskian can be determined, up to an overall multiplicative constant, without the need to know explicitly the pair of homogeneous solutions $f_{1,2}$,
	\begin{align}
	\label{2ndOrderODE_Wronskian_GeneralSolution}
	\text{Wr}_z(f_1,f_2) &= W_0 \exp\left(- \int_b^z \frac{p_1(z'')}{p_2(z'')} \dd z'' \right), \qquad W_0 = \text{constant} .
	\end{align}
	\item If we ``rotate" from one pair of linearly independent solutions $(f_1,f_2)$ to another $(g_1,g_2)$ via a constant invertible matrix $M_\text{I}^{\phantom{\text{I}}\text{J}}$,
	\begin{align}
	f_\text{I}(z) = M_\text{I}^{\phantom{\text{I}}\text{J}} g_\text{J}(z), \qquad 
			\text{I},\text{J} \in \{1,2\}, \ \det M_\text{I}^{\phantom{\text{I}}\text{J}} \neq 0 ;
	\end{align}
	then
	\begin{align}
	\text{Wr}_z(f_1,f_2) = \left(\det M_\text{I}^{\phantom{\text{I}}\text{J}} \right) \text{Wr}_z(g_1,g_2) .
	\end{align}
\end{itemize}
\noindent{\bf Discontinuous first derivative at $z=z'$} \qquad The key observation to solving the symmetric Green's function is that, as long as $z \neq z'$ then the $\delta(z-z')=0$ in eq. \eqref{2ndOrderODE_GreensFunctionEquation}. Therefore $G(z,z')$ has to obey the homogeneous equation
\begin{align}
D_z G(z,z') = D_{z'} G(z,z') = 0, \qquad z \neq z' .
\end{align}
For $z>z'$, if we solve $D_z G = 0$ first,
\begin{align}
G(z,z') = \alpha^\text{I}(z') f_\text{I}(z) ,
\end{align}
i.e., it must be a superposition of the linearly independent solutions $\{f_\text{I}(z)\}$ (in the variable $z$). Because $G(z,z')$ is a function of both $z$ and $z'$, the coefficients of the superposition must depend on $z'$. If we then solve 
\begin{align}
D_{z'} G(z,z') = D_{z'} \alpha^\text{I}(z') f_\text{I}(z) = 0 ,
\end{align}
(for $z \neq z'$), we see that the $\{\alpha^\text{I}(z')\}$ must in turn each be a superposition of the linearly independent solutions in the variable $z'$.
\begin{align}
\alpha^\text{I}(z') = A_>^{\text{I}\text{J}} f_\text{J}(z') .
\end{align} 
(The $\{ A_>^{\text{I}\text{J}} \}$ are now constants, because $\alpha^\text{I}(z')$ has to depend only on $z'$ and not on $z$.) What we have deduced is that $G(z>z')$ is a sum of 4 independent terms:
\begin{align}
G(z > z') = A_>^{\text{I}\text{J}} f_\text{I}(z) f_\text{J}(z') , \qquad \qquad A_>^{\text{I}\text{J}} = \text{constant} .
\end{align}
Similar arguments will tell us, 
\begin{align}
G(z < z') = A_<^{\text{I}\text{J}} f_\text{I}(z) f_\text{J}(z') , \qquad \qquad A_<^{\text{I}\text{J}} = \text{constant} .
\end{align}
This may be summarized as
\begin{align}
\label{2ndOrderODE_G_Ansatz}
G(z,z') = \Theta(z-z') A_>^{\text{I}\text{J}} f_\text{I}(z) f_\text{J}(z') + \Theta(z'-z) A_<^{\text{I}\text{J}} f_\text{I}(z) f_\text{J}(z') .
\end{align}
Now we examine the behavior of $G(z,z')$ near $z=z'$. Suppose $G(z,z')$ is discontinuous at $z=z'$. Then its first derivative there will contain $\delta(z-z')$ and its second derivative will contain $\delta'(z-z')$, and $G$ itself will thus not satisfy the right hand side of eq. \eqref{2ndOrderODE_GreensFunctionEquation}. Therefore we may impose the continuity conditions
\begin{align}
A_<^{\text{I}\text{J}} f_\text{I}(z) f_\text{J}(z) &= A_>^{\text{I}\text{J}} f_\text{I}(z) f_\text{J}(z) , \\
A_<^{11} f_1(z)^2 + A_<^{22} f_2(z)^2 + (A_<^{12} + A_<^{21}) f_1(z) f_2(z)
		&= A_>^{11} f_1(z)^2 + A_>^{22} f_2(z)^2 + (A_>^{12} + A_>^{21}) f_1(z) f_2(z) . \nonumber
\end{align}
Since this must hold for all $a \leq z \leq b$, the coefficients of $f_1(z)^2$, $f_2(z)^2$ and $f_1(z) f_2(z)$ on both sides must be equal,
\begin{align}
\label{2ndOrderODE_ContinuityConditionForG}
A_<^{11} = A_>^{11} \equiv A^1, \qquad
A_<^{22} = A_>^{22} \equiv A^2, \qquad
A_<^{12} + A_<^{21} = A_>^{12} + A_>^{21} .
\end{align}
Now let us integrate $D_z G(z,z') = \lambda(z) \delta(z-z')$ around the neighborhood of $z \approx z'$; i.e., for $0 < \epsilon \ll 1$, and a prime denoting $\partial_z$,
\begin{align}
\int_{z'-\epsilon}^{z'+\epsilon} \dd z \lambda(z) \delta(z-z') 
&= \int_{z'-\epsilon}^{z'+\epsilon} \dd z \left\{ p_2 G'' + p_1 G' + p_0 G  \right\} \nonumber\\
\lambda(z')
&= [p_2 G' + p_1 G]_{z'-\epsilon}^{z'+\epsilon} + \int_{z'-\epsilon}^{z'+\epsilon} \dd z \left\{ - p'_2 G' - p'_1 G + p_0 G  \right\} \nonumber\\
&= [ (p_1(z)- \partial_z p_2(z)) G(z,z') + p_2(z) \partial_z G(z,z') ]_{z=z'-\epsilon}^{z=z'+\epsilon} \\
&\qquad\qquad + \int_{z'-\epsilon}^{z'+\epsilon} \dd z \left\{ p''_2(z) G(z,z') - p'_1(z) G(z,z') + p_0(z) G(z,z') \right\} . \nonumber
\end{align}
Because $p_{0,1,2}(z)$ are smooth and because $G$ is continuous at $z=z'$, as we set $\epsilon \to 0$, only the $G'$ remains on the right hand side.
\begin{align}
\lim_{\epsilon\to 0} \left\{ p_2(z'+\epsilon) \frac{\partial G(z = z'+\epsilon,z')}{\partial z} - p_2(z'-\epsilon) \frac{\partial G(z = z'-\epsilon,z')}{\partial z} \right\} = \lambda(z') 
\end{align}
We can set $z' \pm \epsilon \to z'$ in the $p_2$ because it is smooth; the error incurred would go as $\mathcal{O}(\epsilon)$. We have thus arrived at the following ``jump" condition: the first derivative of the Green's function on either side of $z=z'$ {\it has to be} discontinuous and their difference multiplied by $p_2(z')$ is equal to the function $\lambda(z')$, the measure multiplying the $\delta(z-z')$ in eq. \eqref{2ndOrderODE_GreensFunctionEquation}.
\begin{align}
\label{2ndOrderODE_JumpConditionForG}
p_2(z') \left\{\frac{\partial G(z = z'^+,z')}{\partial z} - \frac{\partial G(z = z'^-,z')}{\partial z} \right\} = \lambda(z') 
\end{align}
This translates to
\begin{align}
p_2(z') \left( A_>^{\text{I}\text{J}} f'_\text{I}(z') f_\text{J}(z') - A_<^{\text{I}\text{J}} f'_\text{I}(z') f_\text{J}(z') \right) &= \lambda(z') .
\end{align}
By taking into account eq. \eqref{2ndOrderODE_ContinuityConditionForG},
\begin{align}
p_2(z') \left( (A_>^{12}-A_<^{12}) f'_1(z') f_2(z') + (A_>^{21}-A_<^{21}) f'_2(z') f_1(z') \right) &= \lambda(z') ,
\end{align}
Since $A_<^{12} + A_<^{21} = A_>^{12} + A_>^{21} \Leftrightarrow A_>^{12} - A_<^{12} = - (A_>^{21} - A_<^{21})$, 
\begin{align}
p_2(z') (A_>^{21}-A_<^{21}) \text{Wr}_{z'}(f_1,f_2) &= \lambda(z') , \nonumber\\
p_2(z') (A_>^{21}-A_<^{21})W_0 \exp\left(- \int_b^{z'} \frac{p_1(z'')}{p_2(z'')} \dd z'' \right) &= \lambda(z') ,
\end{align}
where eq. \eqref{2ndOrderODE_Wronskian_GeneralSolution} was employed in the second line. We see that, given a differential operator $D$ of the form in eq. \eqref{2ndOrderODE_D}, this amounts to solving for the measure $\lambda(z')$: it is fixed, up to an overall multiplicative constant $(A_>^{21}-A_<^{21}) W_0$, by the $p_{1,2}$. (Remember the Wronskian itself is fixed up to an overall constant by $p_{1,2}$; cf. eq. 	\eqref{2ndOrderODE_Wronskian_GeneralSolution}.) Furthermore, note that $A_>^{21}-A_<^{21}$ can be absorbed into the functions $f_{1,2}$, since the latter's normalization has remained arbitrary till now. Thus, we may choose $A_>^{21}-A_<^{21} = 1 = -(A_>^{12}-A_<^{12})$. At this point,
\begin{align}
\label{2ndOrderODE_SymmetricG_I}
G(z,z') 
= A^1 f_1(z) f_1(z') &+ A^2 f_2(z) f_2(z') \nonumber\\
&+ \Theta(z-z') ( (A_<^{12}-1) f_1(z) f_2(z') + A_>^{21} f_2(z) f_1(z')) \nonumber\\
&+ \Theta(z'-z) ( A_<^{12} f_1(z) f_2(z') + (A_>^{21}-1) f_2(z) f_1(z')) .
\end{align}
Because we are seeking a symmetric Green's function, let us also consider
\begin{align}
\label{2ndOrderODE_SymmetricG_II}
G(z',z) 
= A^1 f_1(z') f_1(z) &+ A^2 f_2(z') f_2(z) \nonumber\\
&+ \Theta(z'-z) ( (A_<^{12}-1) f_1(z') f_2(z) + A_>^{21} f_2(z') f_1(z)) \nonumber\\
&+ \Theta(z-z') ( A_<^{12} f_1(z') f_2(z) + (A_>^{21}-1) f_2(z') f_1(z)) .
\end{align}
Comparing the first lines of equations \eqref{2ndOrderODE_SymmetricG_I} and \eqref{2ndOrderODE_SymmetricG_II} tells us the $A^{1,2}$ terms are automatically symmetric; whereas the second line of eq. \eqref{2ndOrderODE_SymmetricG_I} versus the third line of eq. \eqref{2ndOrderODE_SymmetricG_II}, together with the third line of eq. \eqref{2ndOrderODE_SymmetricG_I} versus second line of eq. \eqref{2ndOrderODE_SymmetricG_II}, says the terms involving $A_\lessgtr^{12}$ are symmetric iff $A_<^{12} = A_>^{12} \equiv \chi$. We gather, therefore,
\begin{align}
G(z,z') 					&= A^1 f_1(z) f_1(z') + A^2 f_2(z) f_2(z') + \mathcal{G}(z,z';\chi) , \\
\mathcal{G}(z,z';\chi) 		&\equiv (\chi-1) \left\{ \Theta(z-z') f_1(z) f_2(z') + \Theta(z'-z) f_1(z') f_2(z) \right\} \nonumber\\
			&\qquad\qquad	+ \chi \left\{ \Theta(z-z') f_2(z) f_1(z') + \Theta(z-z') f_2(z') f_1(z) \right\} .
\end{align}
The terms in the curly brackets can be written as $(\chi-1)f_1(z_>) f_2(z_<) + \chi \cdot f_1(z_<) f_2(z_>)$, where $z_>$ is the larger and $z_<$ the smaller of the pair $(z,z')$. Moreover, we see it is these terms that contributes to the `jump' in the first derivative across $z=z'$. The terms involving $A^1$ and $A^2$ are smooth across $z=z'$ provided, of course, the functions $f_{1,2}$ themselves are smooth; they are also homogeneous solutions with respect to both $z$ and $z'$.

\noindent{\bf Summary} \qquad Given any pair of linearly independent solutions to 
\begin{align}
\label{2ndOrderODE_DDef_Summary}
D_z f_{1,2}(z) \equiv p_2(z) \frac{\dd^2 f_{1,2}(z)}{\dd z^2} + p_1(z) \frac{\dd f_{1,2}(z)}{\dd z} + p_0(z) f_{1,2}(z) = 0 , \qquad a \leq z \leq b ,
\end{align}
we may solve the symmetric Green's function equation(s)
{\allowdisplaybreaks\begin{align}
\label{2ndOrderODE_GreensFunctionEquation_Summary}
D_z G(z,z') 	&= p_2(z) 	W_0 \exp\left(- \int_b^z 	\frac{p_1(z'')}{p_2(z'')} \dd z'' \right) \delta(z-z') , \\
D_{z'} G(z,z') 	&= p_2(z') 	W_0 \exp\left(- \int_b^{z'} \frac{p_1(z'')}{p_2(z'')} \dd z'' \right) \delta(z-z') , \\
G(z,z') 		&= G(z',z) , 
\end{align}}
by using the general ansatz
{\allowdisplaybreaks\begin{align}
\label{2ndOrderODE_GreensFunctionAnsatz_Summary}
G(z,z') = G(z',z) 	&= A^1 f_1(z) f_1(z') + A^2 f_2(z) f_2(z') + \mathcal{G}(z,z';\chi) , \\
\mathcal{G}(z,z';\chi) 		&\equiv (\chi-1) f_1(z_>) f_2(z_<) + \chi \ f_2(z_>) f_1(z_<), \\
z_> 				&\equiv \max(z,z'), \quad z_< \equiv \min(z,z') .
\end{align}}
Here $W_0$, $A^{1,2}$, and $\chi$ are arbitrary constants. However, once $W_0$ is chosen, the $f_{1,2}$ needs to be normalized properly to ensure the constant $W_0$ is recovered. Specifically,
\begin{align}
\label{2ndOrderODE_GreensFunctionJump_Summary}
\text{Wr}_z(f_1,f_2) 
&= f_1(z) f_2'(z) - f_1'(z) f_2(z) = \frac{\partial \mathcal{G}(z = z'^+,z')}{\partial z} - \frac{\partial \mathcal{G}(z = z'^-,z')}{\partial z} \nonumber\\
&= W_0 \exp\left(- \int_b^z \frac{p_1(z'')}{p_2(z'')} \dd z'' \right) .
\end{align}
We also reiterate, up to the overall multiplicative constant $W_0$, the right hand side of eq. \eqref{2ndOrderODE_GreensFunctionEquation_Summary} is fixed once the differential operator $D$ (in eq. \eqref{2ndOrderODE_DDef_Summary}) is specified; in particular, one may not always be able to set the right hand side of eq. \eqref{2ndOrderODE_GreensFunctionEquation_Summary} to $\delta(z-z')$.

\noindent{\bf 3D Green's Function of Laplacian} \qquad As an example of the methods described here, let us work out the radial Green's function of the Laplacian in 3D Euclidean space. That is, we shall employ spherical coordinates
\begin{align}
x^i 	&= r (s_\theta c_\phi, s_\theta s_\phi, c_\theta) , \\
x'^i 	&= r' (s_{\theta'} c_{\phi'}, s_{\theta'} s_{\phi'}, c_{\theta'}) ;
\end{align}
and try to solve
\begin{align}
\label{3DLaplacianGEquation}
-\vec{\nabla}_{\vec{x}}^2 G(\vec{x}-\vec{x}') = -\vec{\nabla}_{\vec{x}'}^2 G(\vec{x}-\vec{x}') 
		= \frac{\delta(r-r')}{rr'} \delta(c_\theta - c_{\theta'}) \delta(\phi-\phi') .
\end{align}
Because of the rotation symmetry of the problem -- we know, in fact,
\begin{align}
G\left( \vec{x}-\vec{x}' \right) = \frac{1}{4\pi|\vec{x}-\vec{x}'|}
= (4\pi)^{-1} \left( r^2 + r'^2 - 2 r r' \cos\gamma \right)^{-1/2}
\end{align}
depends on the angular coordinates through the dot product $\cos\gamma \equiv \vec{x}\cdot\vec{x}'/(rr') = \widehat{x} \cdot \widehat{x}'$. This allows us to postulate the ansatz
\begin{align}
\label{3DLaplacianGEquation_FullAnsatz}
G(\vec{x}-\vec{x}') 
= \sum_{\ell=0}^{\infty} \frac{\widetilde{g}_\ell(r,r')}{2\ell+1} \sum_{m = -\ell}^{\ell} Y_\ell^m(\theta,\phi) \overline{Y_\ell^m(\theta',\phi')} .
\end{align}
By applying the Laplacian in spherical coordinates (cf. eq. \eqref{Laplacian_3D_Spherical}) and using the completeness relation for spherical harmonics in eq. \eqref{SphericalHarmonics_Completeness}, eq. \eqref{3DLaplacianGEquation} becomes
\begin{align}
\sum_{\ell=0}^{\infty} \frac{\widetilde{g}_\ell'' + (2/r) \widetilde{g}_\ell' - \ell(\ell+1) r^{-2} \widetilde{g}_\ell}{2\ell+1} & \sum_{m = -\ell}^{\ell} Y_\ell^m(\theta,\phi) \overline{Y_\ell^m(\theta',\phi')} \nonumber\\
&= -\frac{\delta(r-r')}{rr'} \sum_{\ell=0}^{\infty} \sum_{m = -\ell}^{\ell} Y_\ell^m(\theta,\phi) \overline{Y_\ell^m(\theta',\phi')} ,
\end{align}
with each prime representing $\partial_r$. Equating the $(\ell,m)$ term on each side,
\begin{align}
D_r \widetilde{g}_\ell 
\equiv \widetilde{g}_\ell'' + \frac{2}{r} \widetilde{g}_\ell' - \frac{\ell(\ell+1)}{r^2} \widetilde{g}_\ell
= -(2\ell+1)\frac{\delta(r-r')}{rr'} .
\end{align}
We already have the $\delta$-function measure -- it is $-(2\ell+1)/r^2$ -- but it is instructive to check its consistency with the right hand side of \eqref{2ndOrderODE_GreensFunctionEquation_Summary}; here, $p_1(r) = 2/r$ and $p_2(r) = 1$, and
\begin{align}
W_0 \exp\left( - 2\int^r \dd r''/r'' \right) = W_0 e^{-2 \ln r} = W_0 r^{-2} .
\end{align}
Now, the two linearly independent solutions to $D_r f_{1,2}(r) = 0$ are
\begin{align}
f_1(r) = \frac{F_1}{r^{\ell+1}}, \qquad f_2(r) = F_2 r^\ell , \qquad F_{1,2} = \text{constant} .
\end{align}
The radial Green's function must, according to eq. \eqref{2ndOrderODE_GreensFunctionAnsatz_Summary}, take the form
{\allowdisplaybreaks\begin{align}
\label{3DLaplacianGEquation_RadialAnsatz}
\widetilde{g}_\ell(r,r') 	
		&= \frac{A^1_\ell}{(rr')^{\ell+1}} + A^2_\ell (rr')^{\ell} + \mathcal{G}_\ell(r,r') , \\
\mathcal{G}_\ell(r,r') 		
		&\equiv F \left\{ \frac{\chi_\ell - 1}{r_>} \left( \frac{r_<}{r_>} \right)^\ell + \frac{\chi_\ell}{r_<} \left( \frac{r_>}{r_<} \right)^\ell \right\} , \\
r_> 	&\equiv \max(r,r'), \ r_< \equiv \min(r,r') ,
\end{align}}
where $A^{1,2}_\ell$, $F$, and $\chi_\ell$ are constants. (What happened to $F_{1,2}$? Strictly speaking $F_1 F_2$ should multiply $A^{1,2}_\ell$ but since the latter is arbitrary their product(s) may be assimilated into one constant(s); similarly, in $\mathcal{G}_\ell(r,r')$, $F = F_1 F_2$ but since $F_{1,2}$ occurs as a product, we may as well call it a single constant.) To fix $F$, we employ eq. \eqref{2ndOrderODE_GreensFunctionJump_Summary}.
\begin{align}
-\frac{2\ell+1}{r^2} = F \ \text{Wr}_r\left( r^{-\ell-1}, r^\ell \right)
= \frac{\partial \mathcal{G}(r=r'^+)}{\partial r} - \frac{\partial \mathcal{G}(r=r'^-)}{\partial r} .
\end{align}
Carrying out the derivatives explicitly,
\begin{align}
-\frac{2\ell+1}{r^2}
&= F \left\{ \frac{\partial}{\partial r} \left(\frac{1}{r'} \left( \frac{r}{r'} \right)^\ell\right)_{r=r'^-} - \frac{\partial}{\partial r} \left(\frac{1}{r} \left( \frac{r'}{r} \right)^\ell\right)_{r=r'^+} \right\} \nonumber\\
&= F \left\{ \frac{\ell \cdot r^{\ell-1}}{r^{\ell+1}} + \frac{(\ell+1) r^\ell}{r^{\ell+2}} \right\} = F \frac{2\ell+1}{r^2} .
\end{align}
Thus, $F = -1$. We may take the limit $r \to 0$ or $r' \to 0$ and see that the terms involving $A^1_\ell$ and $(\chi_\ell/r_<) (r_>/r_<)^\ell$ in eq. \eqref{3DLaplacianGEquation_RadialAnsatz} will blow up for any $\ell$; while $1/(4\pi|\vec{x}-\vec{x}'|) \to 1/(4\pi r')$ or $\to 1/(4\pi r)$ does not. This implies $A^1_\ell = 0$ and $\chi_\ell = 0$. Next, by considering the limits $r \to \infty$ or $r' \to \infty$, we see that the $A^2_\ell$ term will blow up for $\ell > 0$, whereas, in fact, $1/(4\pi|\vec{x}-\vec{x}'|) \to 0$. Hence $A^2_{\ell > 0} = 0$. Moreover, the $\ell=0$ term involving $A^2_0$ is a constant in space because $Y_{\ell=0}^m = 1/\sqrt{4\pi}$ and does not decay to zero for $r,r' \to \infty$; therefore, $A^2_0 = 0$ too. Equation \eqref{3DLaplacianGEquation_RadialAnsatz} now stands as
\begin{align}
\widetilde{g}_\ell(r,r') = \frac{1}{r_>} \left( \frac{r_<}{r_>} \right)^\ell ,
\end{align}
which in turn means eq. \eqref{3DLaplacianGEquation_FullAnsatz} is
\begin{align}
\label{3DLaplacianGEquation_StaticLimit}
G(\vec{x}-\vec{x}') 
= \frac{1}{4\pi |\vec{x}-\vec{x}'|}
= \frac{1}{r_>}\sum_{\ell=0}^{\infty} \frac{1}{2\ell+1} \left(\frac{r_<}{r_>}\right)^\ell \sum_{m = -\ell}^{\ell} Y_\ell^m(\theta,\phi) \overline{Y_\ell^m(\theta',\phi')} .
\end{align}
If we use the addition formula in eq. \eqref{SphericalHarmonics_AdditionFormula}, we then recover eq. \eqref{GreensFunction_LegendrePolynomialExpansion}.
\begin{myP}
\qquad Can you perform a similar ``jump condition" analysis for the 2D Green's function of the negative Laplacian? Your answer should be proportional to eq. \eqref{Laplacian_2DGreensFunction}. You may assume there are no homogeneous contributions to the answer, i.e., set $A^1=A^2=0$ in eq. 	\eqref{2ndOrderODE_GreensFunctionAnsatz_Summary}. Hint: Start by justifying the ansatz
\begin{align}
G_2(\vec{x}-\vec{x}') = \sum_{\ell=-\infty}^{+\infty} \widetilde{g}_\ell(r,r') e^{i\ell(\phi-\phi')} ,
\end{align}
where $\vec{x} \equiv r(\cos\phi,\sin\phi)$ and $\vec{x}' \equiv r'(\cos\phi',\sin\phi')$. Carry out the jump condition analysis, assuming the radial Green's function $\widetilde{g}_\ell$ is a symmetric one.
\end{myP}

\appendix

\section{Copyleft}

You should feel free to re-distribute these notes, as long as they remain freely available. {\it Please do not} post on-line solutions to the problems I have written here! I do have solutions to some of the problems. If you are using these notes for self-study, write to me and I will e-mail them to you.

\section{Conventions}
\label{Chapter_Conventions}
{\bf Function argument} \qquad There is a notational ambiguity whenever we write ``$f$ is a function of the variable $x$" as $f(x)$. If you did not know $f$ were meant to be a function, what is $f(x+\sin(\theta))$? Is it some number $f$ times $x+\sin\theta$? For this reason, in my personal notes and research papers I reserve square brackets exclusively to denote the argument of functions -- I would always write $f[x+\sin[\theta]]$, for instance. (This is a notation I borrowed from the software {\sf Mathematica}.) However, in these lecture notes I will stick to the usual convention of using parenthesis; but I wish to raise awareness of this imprecision in our mathematical notation.

{\bf Einstein summation and index notation} \qquad Repeated indices are always summed over, unless otherwise stated:
\begin{align}
\xi^i p_i \equiv \sum_i \xi^i p_i .
\end{align}
Often I will remain agnostic about the range of summation, unless absolutely necessary. 

In such contexts when the Einstein summation is in force -- unless otherwise stated -- both the superscript and subscript are enumeration labels. $\xi^i$ is the $i$th component of $(\xi^1,\xi^2,\xi^3,\dots)$, not some variable $\xi$ raised to the $i$th power. The position of the index, whether it is super- or sub-script, usually represents how it transforms under the change of basis or coordinate system used. For instance, instead of calling the 3D Cartesian coordinates $(x,y,z)$, we may now denote them collectively as $x^i$, where $i=1,2,3$. When you rotate your coordinate system $x^i \to R^i_{\phantom{i}j} y^j$, the derivative transforms as $\partial_i \equiv \partial/\partial x^i \to (R^{-1})^j_{\phantom{j}i} \partial_j$.

{\bf Dimensions} \qquad Unless stated explicitly, the number of space dimensions is $D$; it is an arbitrary positive integer greater or equal to one. Unless stated explicitly, the number of spacetime dimensions is $d=D+1$; it is an arbitrary positive integer greater or equal to 2.

{\bf Spatial vs. spacetime indices} \qquad I will employ the common notation that spatial indices are denoted with Latin/English alphabets whereas spacetime ones with Greek letters. Spacetime indices begin with $0$; the $0$th index is in fact time. Spatial indices start at $1$. I will also use the ``mostly minus" convention for the metric; for e.g., the flat spacetime geometry in Cartesian coordinates reads
\begin{align}
\eta_{\mu\nu} = \text{diag} \left[ 1, -1, \dots, -1 \right] ,
\end{align}
where ``diag$[a_1, \dots, a_N]$" refers to the diagonal matrix, whose diagonal elements (from the top left to the bottom right) are respectively $a_1$, $a_2$, $\dots$, $a_N$. Spatial derivatives are $\partial_i \equiv \partial/\partial x^i$; and spacetime ones are $\partial_\mu \equiv \partial/\partial x^\mu$. The scalar wave operator in flat spacetime, in Cartesian coordinates, read
\begin{align}
\partial^2 = \Box = \eta^{\mu\nu} \partial_\mu \partial_\nu .
\end{align}
The Laplacian in flat space, in Cartesian coordinates, read instead
\begin{align}
\vec{\nabla}^2 = \delta^{ij} \partial_i \partial_i ,
\end{align}
where $\delta_{ij}$ is the Kronecker delta, the unit $D \times D$ matrix $\mathbb{I}$:
\begin{align}
\delta_{ij} &= 1, \qquad i = j \nonumber\\
&= 0, \qquad i \neq j .
\end{align}

\section{Physical Constants and Dimensional Analysis}
\label{Chapter_DimensionalAnalysis}
In much of these notes we will set Planck's reduced constant and the speed of light to unity: $\hbar = c = 1$. (In the General Relativity literature, Newton's gravitational constant $\GN$ is also often set to one.) What this means is, we are using $\hbar$ as our base unit for angular momentum; and $c$ for speed. 

Since $[c]$ is Length/Time, setting it to unity means
\begin{center}
	[Length] $=$ [Time] .
\end{center}
In particular, since in SI units $c=299,792,458$ meters/second, we have
\begin{align}
1 \text{ second} = 299,792,458 \text{ meters} , \qquad\qquad (c=1) .
\end{align}
Einstein's $E = m c^2$, once $c=1$, becomes the statement that
\begin{center}
	[Energy] $=$ [Mass]
\end{center}
Because $[\hbar]$ is Energy $\times$ Time, setting it to unity means 
\begin{center}
	[Energy] = [1/Time] .
\end{center}
In SI units, $\hbar \approx 1.0545718 \times 10^{-34}$ Joules second -- hence,
\begin{align}
1 \text{ second} &\approx 1/(1.0545718 \times 10^{-34}\text{ Joules}) \qquad\qquad (\hbar=1) .
\end{align}
Altogether, with $\hbar = c = 1$, we may state 
\begin{quotation}
	[Mass] $=$ [Energy] $=$ [1/Time] $=$ [1/Length] .
\end{quotation}
Physically speaking, the energy-mass and time-length equivalence can be attributed to relativity $(c)$; whereas the (energy/mass)-(time/length)$^{-1}$ equivalence can be attributed to quantum mechanics $(\hbar)$.

High energy physicists prefer to work with eV (or its multiples, such as MeV or GeV); and so it is useful to know the relation $\hbar c \approx 197.326,98$ MeV fm. (fm $=$ femtometer $=10^{-15}$ meters.)
\begin{align}
\label{Meters_to_MeV}
10^{-15} \text{ meters} \approx 1/(197.326,98 \text{ MeV}), \qquad\qquad
(\hbar c = 1) .
\end{align}
Using these `natural units' $\hbar=c=1$ is a very common practice throughout the physics literature. 

One key motivation behind setting to unity physical constants occurring frequently in your physics analysis, is that it allows you to focus on the quantities that are more specific (and hence more important) to the problem at hand. Carrying these physical constants around clutter your calculation, and increases the risk of mistakes due to this additional burden. For instance, in the Bose-Einstein or Fermi-Dirac statistical distribution $1/(\exp(E/(k_B T)) \pm 1)$ -- where $E$, $k_B$ and $T$ are respectively the energy of the particle(s), $k_B$ is the Boltzmann constant, and $T$ is the temperature of the system -- what's physically important is the ratio of the energy scales, $E$ versus $k_B T$. The Boltzmann constant $k_B$ is really a distraction, and ought to be set to one, so that temperature is now measured in units of energy: the cleaner expression now reads $1/(\exp(E/T) \pm 1)$.

Another reason why one may want to set a physical constant to unity is because, it could be such an important benchmark in the problem at hand that it should be employed as a base unit. 

Most down-to-Earth engineering problems may not benefit from using the speed of light $c$ as their basic unit for speed. In non-relativistic astrophysical systems bound by their mutual gravity, however, it turns out that General Relativistic corrections to the Newtonian law of gravity will be akin to a series in $v/c$, where $v$ is the typical speed of the bodies that comprise the system. The expansion parameter then becomes $0 \leq v < 1$ if we set $c=1$ -- i.e., if we measure all speeds relative to $c$ -- which in turn means this `post-Newtonian' expansion is a series in the gravitational potential $\GN M/r$ through the virial theorem (kinetic energy $\sim$ potential energy) $v \sim \sqrt{\GN M/r}$. 

Newton's gravitational constant takes the form
\begin{align}
\GN \approx 6.7086 \times 10^{-39} \hbar c(\text{GeV}/c^2)^{-2} .
\end{align}
Just from this dimensional analysis alone, when $\hbar = c = 1$, one may form a mass-energy scale (`Planck mass') 
\begin{align}
\label{PlanckMass}
\mpl \equiv \frac{1}{\sqrt{32\pi\GN}} .
\end{align}
(The $32\pi$ is for technical convenience.) We will provide further justification below, but this suggests -- since $\mpl$ appears to involve relativity ($c$), quantum mechanics ($\hbar$) and gravitation ($\GN$) -- that the energy scale required to probe quantum aspects of gravity is roughly $\mpl$. Therefore, it may be useful to set $\mpl=1$ in quantum gravity calculations, so that all other energy scales in a given problem, say the quantum amplitude of scattering gravitons, are now measured relative to it.

I recommend the following resource for physical and astrophysical constants, particle physics data, etc.:
\begin{center}
	Particle Data Group: \href{http://pdg.lbl.gov}{http://pdg.lbl.gov} .
\end{center}
\begin{myP}
	Let $\hbar = c = 1$.
	\begin{itemize}
		\item If angular momentum is 3.34, convert it to SI units.
		\item What is the mass of the Sun in MeV? What is its mass in parsec?
		\item If Pluto is orbiting roughly 40 astronomical units from the Sun, how many seconds is this orbital distance?
		\item Work out the Planck mass in eq. \eqref{PlanckMass} in seconds, meters, and GeV.
	\end{itemize} \qed
\end{myP}

\newpage

\section{Acknowledgments}

I wish to thank the following people for pointing out errors/typos, and/or for discussions that lead to new or better presentations, etc.: Jake Leistico, Lin Kuan-Nan, Alec Lovlein, Hadi Papei, Jason Payne, Evan Severson, Leon Tsai and Wei Chun-Yu (Viona).

I thank Tolga Birkandan for pointing me to Sages Math (linked below, \cite{SagesMath}); and Leo Stein for clarifying how {\sf xAct} \cite{xAct} carries out perturbation theory.

I would also like to thank Tanmay Vachaspati for encouraging me to post these lecture notes on the arXiv (@ \href{https://arxiv.org/abs/1701.00776}{1701.00776}).

\section{Last update: \today}

\end{document}